\documentclass[prb,aps,nofootinbib,amssymb,twocolumn,superscriptaddress,10pt]{revtex4-2}
\usepackage[T1]{fontenc}
\usepackage{graphicx}
\usepackage{xcolor}
\usepackage{dcolumn}
\usepackage{comment}
\usepackage{amsmath,amssymb,bbold,bm}
\usepackage{hyperref}
\usepackage{amsfonts}
\usepackage{float,latexsym}
\usepackage{multirow}
\usepackage{color}
\usepackage[normalem]{ulem}
\usepackage{cleveref}
\usepackage{physics}
\usepackage{tabularx}
\usepackage{array}
\usepackage[caption=false]{subfig}
\usepackage{siunitx}
\usepackage{gensymb}
\usepackage{pifont}
\usepackage{placeins}
\usepackage{mathtools}
\usepackage{longtable}
\usepackage{multirow}
\usepackage{nicematrix}
\usepackage{tikz}
\usepackage{xstring}
\usepackage{makecell}
\usepackage{xr}
\usepackage{ifthen}
\usetikzlibrary{decorations.markings}
\usetikzlibrary{positioning}
\usetikzlibrary{arrows.meta}
\usepackage{tikz-feynman}
\usetikzlibrary{calc}
\tikzfeynmanset{warn luatex=false}

\tikzset{my node/.style={circle}, 
	strike through post/.append style={
		decoration={markings, mark=at position 0.75 with {
				\draw[-] ++ (0,-5pt) -- (0,5pt);}
		},postaction={decorate}}
}
\tikzset{my node/.style={circle}, 
	strike through pre/.append style={
		decoration={markings, mark=at position 0.25 with {
				\draw[-] ++ (0,-5pt) -- (0,5pt);}
		},postaction={decorate}}
}
\allowdisplaybreaks

\newcolumntype{C}{>{$} c <{$}}
\let\vec\mathbf

\DeclareMathOperator{\Li}{Li}
\DeclareMathOperator{\sgn}{\mathrm{sgn}}
\DeclareMathAlphabet\mathbfcal{OMS}{cmsy}{b}{n}
\DeclarePairedDelimiterX{\setDef}[2]\{\}{%
  \, #1 \;\delimsize\vert\; #2 \,
}
\definecolor{darkgreen}{cmyk}{0.9,0,0.9,0.5} 

\makeatletter
\def\maketag@@@#1{\hbox{\m@th\normalfont\normalsize#1}}
\makeatother

\crefname{appendix}{Appendix}{Appendices}
\crefname{equation}{Eq.}{Eqs.}
\crefname{figure}{Fig.}{Figs.}
\crefname{table}{Table}{Tables}
\crefname{section}{Section}{Sections}
\crefname{enumi}{Point}{Points}
\creflabelformat{appendix}{[#2#1#3]}
\crefmultiformat{figure}{Figs.~#2#1\xdef\mycreffirstarg{#1}#3}%
 { and~#2#1#3}{, #2#1#3}{ and~#2#1#3}

\makeatletter     
\renewcommand\onecolumngrid{
\do@columngrid{one}{\@ne}%
\def\set@footnotewidth{\onecolumngrid}
\def\footnoterule{\kern-6pt\hrule width 1.5in\kern6pt}%
}

\makeatletter
\AtBeginDocument{\let\LS@rot\@undefined}
\makeatother 

\usepackage{xr}
\makeatletter
\newcommand*{\addFileDependency}[1]{
  \typeout{(#1)}
  \@addtofilelist{#1}
  \IfFileExists{#1}{}{\typeout{No file #1.}}
}
\makeatother

\crefname{appendix}{Appendix}{Appendices}
\crefname{equation}{Eq.}{Eqs.}
\crefname{figure}{Fig.}{Figs.}
\crefname{table}{Table}{Tables}
\crefname{section}{Section}{Sections}
\creflabelformat{appendix}{[#2#1#3]}

\makeatletter
\AtBeginDocument{\let\LS@rot\@undefined}
\makeatother

\makeatletter
\renewcommand\onecolumngrid{\do@columngrid{one}{\@ne}\def\set@footnotewidth{\onecolumngrid}\def\footnoterule{\kern-6pt\hrule width 1.5in\kern6pt}}

\newcommand{\citeHFReview}{ (reviewed in \cref{app:sec:HF_review})~\cite{SON22,YU23a}}
\newcommand{\citeCurrents}{ (see \cref{app:sec:cur_deriv_cont})}
\newcommand{\citeBMReview}{  (reviewed in \cref{app:sec:BM_review})}
\newcommand{\citeHue}{ (see \cref{app:sec:many_body_rev})}
\newcommand{\citeManyBodyReview}{ (reviewed in \cref{app:sec:many_body_rev})~\cite{MAH00,CAL23b}}
\newcommand{\citeCurDeriv}{ (see \cref{app:sec:deriv_moire_cur,app:sec:thermoelectric_response})}
\newcommand{\citeThermoelectric}{ (as detailed in \cref{app:sec:thermoelectric_response})}
\newcommand{\citeThermoelectricRef}{\cref{app:sec:thermoelectric_response}}
\newcommand{\citeCurDerivShown}{ (as shown in \cref{app:sec:deriv_moire_cur})}
\newcommand{\citeSymBrSeebeck}{\cref{app:sec:seebeck_sym_br}}
\newcommand{\citeAsymptotes}{ (see \cref{app:sec:asymptotes})}
\newcommand{\citeAsymptotesRef}{\cref{app:sec:asymptotes}}
\newcommand{\citeSymmetric}{ (see \cref{app:sec:seebeck_sym})}
\newcommand{\citeNoninteracting}{ (see \cref{app:sec:seebeck_non_int})}
\newcommand{\citeComparison}{ (see \cref{app:sec:comparison_with_exp})}

\newcommand{\siSection}{appendix}

\allowdisplaybreaks

\begin{document}
\title{The Thermoelectric Effect and Its Natural Heavy Fermion Explanation in Twisted Bilayer and Trilayer Graphene}
\author{Dumitru C\u{a}lug\u{a}ru}
	\thanks{These authors contributed equally to this work.}
	\affiliation{Department of Physics, Princeton University, Princeton, New Jersey 08544, USA}
	\author{Haoyu Hu}
	\thanks{These authors contributed equally to this work.}
	\affiliation{Donostia International Physics Center (DIPC), Paseo Manuel de Lardizábal. 20018, San Sebastián, Spain}
	\author{Rafael Luque Merino}
	\affiliation{ICFO - Institut de Ciencies Fotoniques, The Barcelona Institute of Science and Technology, Castelldefels, Barcelona, 08860, Spain} 
	\affiliation{Faculty of Physics, Ludwig-Maximilians-University Munich, Munich 80799, Germany}
	\affiliation{Munich Center for Quantum Science and Technology (MCQST), Ludwig-Maximilians-University Munich, Munich 80799, Germany}
	\author{Nicolas Regnault}
	\affiliation{Department of Physics, Princeton University, Princeton, New Jersey 08544, USA}
	\affiliation{Laboratoire de Physique de l’Ecole normale sup\'erieure, ENS, Universit\'e PSL, CNRS, Sorbonne Universit\'e, Universit\'e Paris-Diderot, Sorbonne Paris Cit\'e, 75005 Paris, France}
	\author{Dmitri K.~Efetov}
	\affiliation{Faculty of Physics, Ludwig-Maximilians-University Munich, Munich 80799, Germany}
	\affiliation{Munich Center for Quantum Science and Technology (MCQST), Ludwig-Maximilians-University Munich, Munich 80799, Germany}
	\author{B.~Andrei Bernevig}
	\email{bernevig@princeton.edu}
	\affiliation{Department of Physics, Princeton University, Princeton, New Jersey 08544, USA}
	\affiliation{Donostia International Physics Center, P. Manuel de Lardizabal 4, 20018 Donostia-San Sebastian, Spain}
	\affiliation{IKERBASQUE, Basque Foundation for Science, Bilbao, Spain}

\let\oldaddcontentsline\addcontentsline

\begin{abstract}
We study the interacting transport properties of twisted bilayer graphene (TBG) using the topological heavy-fermion (THF) model. In the THF model, TBG comprises localized, correlated $f$-electrons and itinerant, dispersive $c$-electrons. We focus on the Seebeck coefficient, which quantifies the voltage difference arising from a temperature gradient. We find that the TBG's Seebeck coefficient shows unconventional (strongly-interacting) traits: negative values with sawtooth oscillations at positive fillings, contrasting typical band-theory expectations. This behavior is naturally attributed to the presence of heavy (correlated, short-lived $f$-electrons) and light (dispersive, long-lived $c$-electrons) electronic bands. Their longer lifetime and stronger dispersion lead to a dominant transport contribution from the $c$-electrons. At positive integer fillings, the correlated TBG insulators feature $c$- ($f$-)electron bands on the electron (hole) doping side, leading to an overall negative Seebeck coefficient. Additionally, sawtooth oscillations occur around each integer filling due to gap openings. Our results highlight the essential importance of electron correlations in understanding the transport properties of TBG and, in particular, of the lifetime asymmetry between the two fermionic species (naturally captured by the THF model). Our findings are corroborated by new experiments in both twisted bilayer and trilayer graphene, and show the natural presence of strongly-correlated heavy and light carriers in the system. 
\end{abstract}
\maketitle

\textit{Introduction}.~The flat bands emerging near charge neutrality~\cite{BIS11,SUA10,LOP07} give rise to a wealth of experimentally-probed strongly-correlated~\cite{CAO18,KER19,XIE19,SHA19,JIA19,CHO19,POL19,YAN19,LU19,STE20,SAI20,SER20,CHE20b,WON20,CHO20,NUC20,CHO21,SAI21,LIU21c,PAR21c,WU21a,CAO21,DAS21,TSC21,PIE21,STE21,CHO21a,XIE21d,DAS22,NUC23,YU23c}, superconducting~\cite{CAO18a,YAN19,LU19,STE20,SAI20,DE21a,OH21,TIA23,DI22a,CAL22d}, and exotic~\cite{TOM19,CAO20,ZON20,LIS21,BEN21,LIA21c,ROZ21,SAI21a,LU21,HES21,DIE23,HUB22,GHA22,JAO22,PAU22,GRO22,ZHO23a} phases in twisted bilayer graphene (TBG) and twisted symmetric trilayer graphene (TSTG)~\cite{CHE19,CHE19a,PAR21,HAO21,CAO21b,LI22d,TUR22,LIU22b,KIM22,ZHA22d,ZHA22c,SHE23,KIM23}. In parallel, sustained theoretical efforts have focused on constructing models~\cite{LOP07,SUA10,BIS11,UCH14,WIJ15,DAI16,JAI16,NAM17,EFI18,KAN18,ZOU18,PO19,LIU19a,TAR19,MOR19,LI19,FAN19,KHA19,CAR19a,CAR19,RAD19,KWA20,CAR20,TRI20,WIL20,PAR20,CAR20a,FU20,HUA20a,CAL20,WU21d,REN21,HEJ21,CAL21,BER21,BER21a,WAN21a,RAM21,SHI21,LEI21,CAO21a,SHE21,WU21c,KOS18,LED21,GUE22,DAV22,CLA22,LIN22,SAM22,KAN23b,VAF23,SHI23} for these systems' nontrivial single-particle topology~\cite{ZOU18,HEJ19,AHN19,PO19,SON19,HEJ19a,LIU19a,XIE20,LIA20,SON21}, correlated states~\cite{OCH18,THO18,XU18b,KOS18,PO18a,VEN18,YUA18,DOD18,PAD18,KEN18,RAD18,LIU19,HUA19,WU19,CLA19,KAN19,SEO19,DA19,ANG19,XIE20b,BUL20b,CHA20,REP20,CEA20,ZHA20,KAN20a,BUL20a,CHI20b,SOE20,CHR20,EUG20,WU20,VAF20,XIE21,KAN21,LIU21,DA21,LIU21a,THO21,KWA21a,LIA21,ZHA21,PAR21a,VAF21,KWA21b,CHE21,POT21,XIE21b,XIE21a,LED21,CHA21,KWA21,HOF22,WAG22,CHR22,SON22,BRI22,CAL22d,HON22,ZHA23a,BLA22,XIE23a,KWA23,YU23a,WAN23c,FER20,DAT23,RAI23a}, fractional insulators~\cite{ABO20,LED20,REP20a}, excitations~\cite{WU20,KHA20,VAF20,BER21b,XIE21,KUM21,KAN21,KWA22}, superconducting phases~\cite{GUO18,YUA18,XU18,DOD18,PO18a,LIU18a,VEN18,ISO18,PEL18,KEN18,WU18,GUI18,GON19,HUA19,ROY19,WU19a,WU19,YOU19,CLA19,LIA19,HU19a,JUL20,XIE20,CHI20a,LOP20,KON20,CHR20,WAN21,KHA21,LEW21,FER21,QIN21,PHO21,CHO21d,LAK21,CHO21c,LI22c,FIS22,YU22,SCA22,CHR22,CHA22,KWA22,WAG23a,GON23,WAG23,WAN24}, and experimental response~\cite{MOO13,LIU20b,PAD20,GAR20,CAL22d,HON22,KRU23,WAN23b}.
One hurdle in achieving a unified understanding of the rich physics of TBG and TSTG is an apparent contradiction in their phenomenology. On the one hand, the quantum-dot-like behavior probed by scanning-tunneling microscopy experiments~\cite{BEN21,TUR22,KIM22,CAL22d,ZHO23a,KIM23,NUC23}, the Pomeranchuk effect~\cite{ROZ21,SAI21a}, and the formation of ferromagnetic ground states~\cite{SHA19,CHE20b,LIU20a,SAI21,TSC21,LIU22b} point to a localized electronic picture of TBG and TSTG. Conversely, the metalicity~\cite{CAO20}, Chern insulators~\cite{SHA19,LU19,SER20,CHE20b,STE20,CHO20,NUC20,CHO21,SAI21,PAR21c,WU21a,DAS21,TSC21,PIE21,CHO21a}, Dirac-like compressbility~\cite{TOM19,ZON20}, superconductivity~\cite{CAO18a,YAN19,CHE19a,LU19,SHE20,STE20,SAI20,HAO21,CAO21b,DE21a,OH21,TIA23,KIM22,DI22a} seen in transport experiments imply the existence of itinerant electrons. The topological heavy-fermion models for TBG~\cite{SON22,HU23,ZHO24,LAU23,CAL23,CHO23,HU23i,LI23a,RAI23a,WAN24} and TSTG~\cite{YU23a} provide an elegant solution to this dichotomy, wherein the flat bands of TBG and TSTG are realized by the hybridization between localized, strongly-correlated \emph{heavy} ($f$) electrons and itinerant, strongly-dispersive \emph{conduction} ($c$) electrons.

While previous experiments have predominantly highlighted either the localized or itinerant electrons, a definitive experimental signature of both types has been elusive. In this \textit{Letter}, we show that the unconventional thermoelectric transport measured recently in TBG~\cite{MER24} and TSTG~\cite{BAT24} provides clear evidence for the coexistence of the two (heavy and itinerant) electronic species. In the Seebeck effect, an applied thermal gradient $\Delta T$ causes both the electron- and hole-like carriers to diffuse from high to low temperatures. The predominant type of carrier, either electron or hole, determines the sign of the resulting thermoelectric voltage $\Delta V$, with electron-dominant transport yielding a negative Seebeck coefficient $S \equiv - \frac{\Delta V}{\Delta T}$. At low temperatures, the Seebeck coefficient of both TBG and TSTG is approximately antisymmetric around charge-neutrality (due to the approximate particle-hole symmetry) and fully negative with sawtooth oscillations at positive fillings $\nu > 0$. In TBG, we attribute the sawtooth oscillations (which occur around integer filling) to the formation of symmetry-broken correlated insulating states. The latter feature light (heavy) bands arising for doping away from (towards) charge neutrality~\cite{BUL20a,BER21b,KAN21}. The heavy bands, mainly formed by the correlated $f$-electrons (with a large scattering rate), will exhibit a shorter lifetime compared to the light bands formed by the $c$-electrons~\cite{SON22,*[{The acompanying work, }] [{, which focuses on band structure calculations for TBG and TSTG, is a methodolgical companion to the present \textit{Letter}. It was separated for clarity and to provide an more in-depth exposition of the computational methods underlying our findings.}] CAL23b}. This difference in lifetime, combined with the disparity in group velocities between the heavy and light bands, leads to a dominant transport contribution from the electron bands and an overall negative $S$. We supplement our detailed microscopic calculations of the latter with analytical results within a phenomenological two-band model mimicking the charge-one excitation dispersion of the TBG correlated insulators. In particular, we find that the lifetime asymmetry between the electron and hole excitations, which is naturally captured by the THF model via its two fermionic species, is \emph{essential} for the observed fully negative Seebeck coefficient. In TSTG, the presence of an additional Dirac cone~\cite{LI19,KHA19,CAL21,YU23a} does not alter the qualitative behavior of $S$ at low temperatures~\cite{BAT24}. Above the theoretically-computed ordering temperature of TBG ($\SIrange{10}{20}{\kelvin}$)~\cite{RAI23a}, the measured Seebeck coefficient turns positive for $\nu > 0$, while still oscillating around integer fillings. We replicate and theoretically elucidate this behavior using the same THF model in the symmetric phase~\cite{HU23,DAT23,RAI23a}. Overall, our calculations underscore the presence of both heavy and light electronic species within TBG and provide a robust theoretical framework for modeling its transport properties in an interacting setting.
\begin{figure*}[t]
	\centering
	\includegraphics[width=\textwidth]{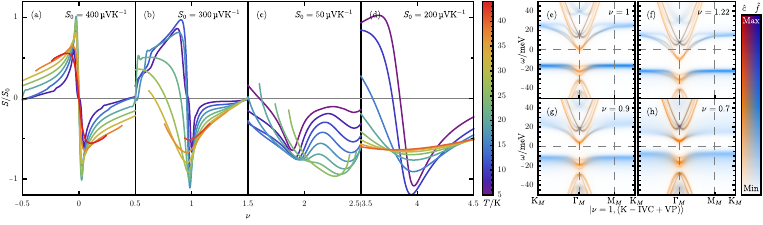}\subfloat{\label{fig:seebeck_sym_br:a}}\subfloat{\label{fig:seebeck_sym_br:b}}\subfloat{\label{fig:seebeck_sym_br:c}}\subfloat{\label{fig:seebeck_sym_br:d}}\subfloat{\label{fig:seebeck_sym_br:e}}\subfloat{\label{fig:seebeck_sym_br:f}}\subfloat{\label{fig:seebeck_sym_br:g}}\subfloat{\label{fig:seebeck_sym_br:h}}\caption{Seebeck coefficient of TBG in its symmetry-broken phases. (a)-(d) display the Seebeck coefficient for the doped $\protect\IfStrEqCase{8}{{1}{\ket{\nu={}4} }
		{2}{\ket{\nu={}3, \mathrm{IVC}}}
		{3}{\ket{\nu={}3, \mathrm{VP}}}
		{4}{\ket{\nu={}2, \mathrm{K-IVC}}}
		{5}{\ket{\nu={}2, \mathrm{VP}}}
		{6}{\ket{\nu={}1, (\mathrm{K-IVC}+\mathrm{VP})}}
		{7}{\ket{\nu={}1, \mathrm{VP}}}
		{8}{\ket{\nu=0, \mathrm{K-IVC}}}
		{9}{\ket{\nu=0, \mathrm{VP}}}
	}
	[nada]
$, $\protect\IfStrEqCase{6}{{1}{\ket{\nu={}4} }
		{2}{\ket{\nu={}3, \mathrm{IVC}}}
		{3}{\ket{\nu={}3, \mathrm{VP}}}
		{4}{\ket{\nu={}2, \mathrm{K-IVC}}}
		{5}{\ket{\nu={}2, \mathrm{VP}}}
		{6}{\ket{\nu={}1, (\mathrm{K-IVC}+\mathrm{VP})}}
		{7}{\ket{\nu={}1, \mathrm{VP}}}
		{8}{\ket{\nu=0, \mathrm{K-IVC}}}
		{9}{\ket{\nu=0, \mathrm{VP}}}
	}
	[nada]
$, $\protect\IfStrEqCase{4}{{1}{\ket{\nu={}4} }
		{2}{\ket{\nu={}3, \mathrm{IVC}}}
		{3}{\ket{\nu={}3, \mathrm{VP}}}
		{4}{\ket{\nu={}2, \mathrm{K-IVC}}}
		{5}{\ket{\nu={}2, \mathrm{VP}}}
		{6}{\ket{\nu={}1, (\mathrm{K-IVC}+\mathrm{VP})}}
		{7}{\ket{\nu={}1, \mathrm{VP}}}
		{8}{\ket{\nu=0, \mathrm{K-IVC}}}
		{9}{\ket{\nu=0, \mathrm{VP}}}
	}
	[nada]
$, and $\protect\IfStrEqCase{1}{{1}{\ket{\nu={}4} }
		{2}{\ket{\nu={}3, \mathrm{IVC}}}
		{3}{\ket{\nu={}3, \mathrm{VP}}}
		{4}{\ket{\nu={}2, \mathrm{K-IVC}}}
		{5}{\ket{\nu={}2, \mathrm{VP}}}
		{6}{\ket{\nu={}1, (\mathrm{K-IVC}+\mathrm{VP})}}
		{7}{\ket{\nu={}1, \mathrm{VP}}}
		{8}{\ket{\nu=0, \mathrm{K-IVC}}}
		{9}{\ket{\nu=0, \mathrm{VP}}}
	}
	[nada]
$ correlated insulators, respectively, for doping up to $\Delta \nu = \pm \frac{1}{2}$ around the integer filling. The Seebeck coefficient scaling factor ($S_0$) is shown in each panel. Line colors represent temperature and, in (a)-(c), only extend for the doping interval in which the symmetry broken solution is stable. (e)-(h) depict the doping-dependent $\vec{k}$-resolved spectral function at $T=\SI{32}{\kelvin}$ for the doped $\protect\IfStrEqCase{6}{{1}{\ket{\nu={}4} }
		{2}{\ket{\nu={}3, \mathrm{IVC}}}
		{3}{\ket{\nu={}3, \mathrm{VP}}}
		{4}{\ket{\nu={}2, \mathrm{K-IVC}}}
		{5}{\ket{\nu={}2, \mathrm{VP}}}
		{6}{\ket{\nu={}1, (\mathrm{K-IVC}+\mathrm{VP})}}
		{7}{\ket{\nu={}1, \mathrm{VP}}}
		{8}{\ket{\nu=0, \mathrm{K-IVC}}}
		{9}{\ket{\nu=0, \mathrm{VP}}}
	}
	[nada]
$ state (filling $\nu$ is indicated in each panel). The hue and intensity correspond, respectively, to the $f$/$c$-electron character and the spectral weight of the excitations\citeHue{}.}
	\label{fig:seebeck_sym_br}
\end{figure*}

\textit{Microscopic theory}.~For brevity, we will focus on TBG; complete 
derivations and results for TSTG are relegated to the appendices. The starting point for a microscopic theory of thermoelectric transport is the particle [$\rho \left( \vec{r}, t\right)$ and $\vec{j} \left( \vec{r}, t \right)$] and thermal [$\rho_Q \left( \vec{r}, t\right)$ and $\vec{j}_Q \left( \vec{r}, t \right)$] density and current operators, which obey the continuity equations $\partial_{t} \rho_{(Q)} \left( \vec{r}, t\right) + \nabla \cdot \vec{j}_{(Q)} \left( \vec{r}, t \right) = 0$~\cite{MAH00}. These operators are derived using Noether's theorem~\cite{PES95,MOR96,PAU03} for generic fermionic theories in the continuum with long-range density-density interactions\citeCurrents{}, such as the Bistritzer-MacDonald (BM) models~\cite{BIS11,LI19,KHA19,BER21a,CAL21} with Coulomb electronic repulsion~\citeBMReview{}. From linear-response theory~\cite{MAH00}, $S$ only depends on the \emph{total} ({\it i.e.}{} zero-momentum) current operators $\vec{J}_{(Q)} \left( t \right) \equiv \int \dd[2]{r} \vec{j}_{(Q)} \left( \vec{r}, t \right)$.

The total current operators (obtained within the interacting BM model) will be projected in the basis of the corresponding THF model\citeHFReview{}. The THF model comprises conduction and heavy electrons, respectively annihilated by $\hat{c}_{\vec{k},a,\eta,s}$ and $\hat{f}_{\vec{k},\alpha,\eta,s}$. Here, $\vec{k}$ labels the moir\'e momentum, $\eta = \pm$, the valley, and $s=\uparrow,\downarrow$, the spin of the corresponding fermions. Additionally, there are four ($1 \leq a \leq 4$) and two ($1 \leq \alpha \leq 2$) orbital degrees of freedom for the $c$- and $f$-electrons~\cite{SON22}. The THF interacting Hamiltonian is given by $H = H_0 + H_{U_1} + H_{U_2} + H_{V} + H_{W} + H_{J} + H_{\tilde{J}} + H_{K}$, where the most important terms are the single-particle Hamiltonian ($H_0$), the onsite Hubbard ($H_{U_1}$) and nearest-neighbor ($H_{U_2}$) $f$-electron repulsion terms, the $c$-electron Coulomb interaction $H_{V}$, and the $f$-$c$ Coulomb ($H_{W}$) and spin-ferromagnetic ($H_J$) interactions. Without moir\'e translation symmetry-breaking, the dynamics of the THF model are encoded in the interacting Matsubara Green's function\citeManyBodyReview{}
\begin{equation}
	\label{app:eqn:mats_gf_def}
	\mathcal{G}_{i \eta s; i' \eta' s'} \left(\tau, \vec{k} \right) = - \left\langle \mathcal{T}_{\tau} \hat{\gamma}_{\vec{k},i,\eta,s} \left( \tau \right) \hat{\gamma}^\dagger_{\vec{k},i',\eta',s'} \left( 0 \right)  \right\rangle,
\end{equation}
where $\hat{\gamma}_{\vec{k},i, \eta, s} \equiv \hat{c}^\dagger_{\vec{k},i,\eta,s}$ ($\hat{\gamma}_{\vec{k},i, \eta, s} \equiv \hat{f}^\dagger_{\vec{k},(i-4),\eta,s}$) for $1 \leq i \leq 4$ ($5 \leq i \leq 6$) and $\mathcal{T}_{\tau}$ enforces the ordering with respect to the imaginary time $\tau$. Both the thermodynamic averaging $\left\langle \dots \right\rangle$ and the imaginary time-evolution $\hat{\gamma}_{\vec{k},i, \eta, s} (\tau) \equiv e^{K \tau}  \hat{\gamma}_{\vec{k},i, \eta, s} e^{- K \tau}$ are performed with the grand canonical Hamiltonian $K \equiv H - \mu \hat{N}$ at inverse temperature $\beta = 1/T$ (with $\mu$ and $\hat{N}$ being, respectively, the chemical potential and the number operator). In Matsubara frequency, the Green's function is obtained using Dyson's equation~\cite{MAH00,CAL23b}
\begin{equation}
	\label{eqn:general_gf_self_energy}
	\mathcal{G}^{-1} \left( i \omega_n, \vec{k} \right) = \left( i\omega_n + \mu \right) \mathbb{1} - h^{\text{MF}} \left( \vec{k} \right) - \Sigma \left( i \omega_n \right),
\end{equation}
with $i \omega_n = \frac{(2n + 1) \pi}{\beta}$ being the fermionic Matsubara frequencies ($n \in \mathbb{Z}$) and $\mathbb{1}$ being the identity matrix. $h^{\text{MF}} \left( \vec{k} \right)$ contains the single-particle THF Hamiltonian matrix, as well as the interaction Hamiltonian decoupled at the Hartree-Fock level. Additionally, the site-diagonal ($\vec{k}$-independent) dynamic $f$-electron self-energy $\Sigma \left( i \omega_n \right)$ stemming from the $H_{U_1} + H_{U_2}$ ($H_{U_1}$) interaction is computed at second-order in perturbation theory~\cite{SCH90} (using dynamical mean-field theory~\cite{GEO96}) in the symmetry-broken (symmetric) phase~\cite{CAL23b}. By including $\Sigma \left( i \omega_n \right)$, we account for the carriers' finite lifetime, in addition to their dispersion. Both $h^{\text{MF}} \left( \vec{k} \right)$ and $\Sigma \left( i \omega_n \right)$ are computed self-consistently for various symmetric and symmetry-broken phases, fillings $\nu$, and temperatures $T$~\cite{CAL23b}. 

In the THF model basis, the imaginary time-evolved total particle and thermal current operators are given by
\begin{align}
	\vec{J} \left(- i \tau \right) =&  \sum_{\substack{i,i' \\ \vec{k}, \eta, s}}  \partial_{\vec{k}} h_{ii'}^{\eta} \left( \vec{k} \right) \hat{\gamma}^\dagger_{\vec{k},i,\eta,s} (\tau) \hat{\gamma}_{\vec{k},i',\eta,s} (\tau), \label{eqn:generic_particle_current} \\
	\vec{J}_Q \left(- i \tau \right) =& \frac{1}{2} \sum_{\substack{i,i' \\ \vec{k}, \eta, s}}  \partial_{\vec{k}} h_{ii'}^{\eta} \left( \vec{k} \right) \left(  \partial_{\tau} \hat{\gamma}^\dagger_{\vec{k},i,\eta,s} (\tau) \hat{\gamma}_{\vec{k},i',\eta,s} (\tau) \right. \nonumber \\
	& \left.  - \hat{\gamma}^\dagger_{\vec{k},i,\eta,s} (\tau) \partial_{\tau} \hat{\gamma}_{\vec{k},i',\eta,s} (\tau) \right), \label{eqn:generic_energy_current}
\end{align}
respectively, where $h^{\eta} \left( \vec{k} \right)$ is the single-particle THF Hamiltonian matrix\citeCurDeriv{}. Upon projecting from the BM to the THF model bases, additional nonabelian Berry connection contributions arise, but are much smaller in magnitude than $\norm{\partial_{\vec{k}} h^{\eta} \left( \vec{k} \right)}$ and can be ignored\citeCurDerivShown{}. An additional quartic term (in the fermion operators and their time-derivatives) arises for the thermal current operator, but can again be ignored as it accounts for no more than a few percent of the quadratic one.

Microscopically, the Seebeck coefficient can be obtained from the ratio of two current-current correlators~\cite{LUT64,MAH00,PAU03}
\begin{equation}
	\label{eqn:seebeck_definition}
	S = \frac{\beta}{e} \frac{L_{12}^{xx}}{L_{11}^{xx}},
\end{equation} 
where $e$ is the electronic charge. The static thermoelectric linear response coefficients $L_{11}^{xx}$ and $L_{12}^{xx}$ are obtained from the dynamic ones
\begin{align}
	L^{\alpha \beta}_{11} \left( i \tilde{\Omega}_n \right) &= \int_{0}^{\beta} \dd{\tau} \frac{e^{i \tilde{\Omega}_n \tau}}{\beta \tilde{\Omega}_n  \Omega}  \left\langle \mathcal{T}_{\tau} J^{\beta} \left( -i\tau \right) J^{\alpha} \left( 0 \right) \right\rangle, \label{eqn:L11_mats} \\
	L^{\alpha \beta}_{12} \left( i \tilde{\Omega}_n \right) &= \int_{0}^{\beta} \dd{\tau} \frac{e^{i \tilde{\Omega}_n \tau}}{\beta \tilde{\Omega}_n  \Omega}  \left\langle \mathcal{T}_{\tau} J_{Q}^{\beta} \left( -i\tau \right) J^{\alpha} \left( 0 \right) \right\rangle, \label{eqn:L12_mats}
\end{align}
by analytically continuing the bosonic Matsubara frequency $i \tilde{\Omega}_n \to \omega + i 0^{+}$, taking the limit $\omega \to 0$, and subsequently extracting the real part~\citeThermoelectric{}. Since both current operators are quadratic in the fermion operators, evaluating the dynamic correlators from \cref{eqn:L11_mats,eqn:L12_mats} entails computing a four-fermion correlation function. Ignoring vertex corrections (an approximation which is valid in the limit of infinite dimensions~\cite{PAU03}), $L^{\alpha \beta}_{11} \left( i \tilde{\Omega}_n \right)$ and $L^{\alpha \beta}_{12} \left( i \tilde{\Omega}_n \right)$ can be readily expressed in terms of the \emph{interacting} THF spectral function
\begin{equation}
	\label{eqn:spectral_function}
	A \left( \omega, \vec{k} \right) \equiv \frac{-1}{2 \pi i} \left( \mathcal{G} \left(\omega + i 0^{+}, \vec{k} \right) - \mathcal{G}^{\dagger} \left(\omega + i 0^{+}, \vec{k} \right) \right),
\end{equation}
obtained from the (analytically continued) Green's function from \cref{app:eqn:mats_gf_def}. We show in \citeThermoelectricRef{} that the static linear response coefficients can be obtained from the same conductivity function 
\begin{equation}
	\label{eqn:def_conduc_func}
	\sigma^{x} \left( \omega \right) =  \frac{1}{\Omega} \sum_{\vec{k}} \Tr \left[ \left( T^{x} \left( \vec{k} \right)  A \left( \omega, \vec{k} \right) \right)^2 \right],
\end{equation}
where $T^{x}_{i \eta s;i' \eta' s'} \left( \vec{k} \right)= \pdv{h^{\eta}_{ii'} \left( \vec{k} \right)}{k^x} \delta_{\eta \eta'} \delta_{s s'}$ is the momentum derivative of the single-particle THF Hamiltonian. Specifically, letting $n_{\mathrm{F}} (\omega) \equiv \left( e^{\beta \omega} + 1  \right)^{-1}$ be the Fermi-Dirac distribution function, we find that
\begin{align}
	L^{xx}_{11} =&  \frac{\pi}{\beta} \int_{-\infty}^{\infty} \dd{\omega} \pdv{n_{\mathrm{F}} \left( \omega \right)}{\omega} \sigma^{x} \left( \omega \right), \label{eqn:L11_static} \\
	L^{xx}_{12} =& - \frac{\pi}{\beta} \int_{-\infty}^{\infty} \dd{\omega} \omega \pdv{n_{\mathrm{F}} \left( \omega \right)}{\omega} \sigma^{x} \left( \omega \right). \label{eqn:L12_static}
\end{align}

\textit{Results in the symmetry-broken phases}.~In \cref{fig:seebeck_sym_br}, we show the numerically obtained Seebeck coefficient in the symmetry-broken phases of TBG. We consider the $\protect\IfStrEqCase{8}{{1}{\ket{\nu={}4} }
		{2}{\ket{\nu={}3, \mathrm{IVC}}}
		{3}{\ket{\nu={}3, \mathrm{VP}}}
		{4}{\ket{\nu={}2, \mathrm{K-IVC}}}
		{5}{\ket{\nu={}2, \mathrm{VP}}}
		{6}{\ket{\nu={}1, (\mathrm{K-IVC}+\mathrm{VP})}}
		{7}{\ket{\nu={}1, \mathrm{VP}}}
		{8}{\ket{\nu=0, \mathrm{K-IVC}}}
		{9}{\ket{\nu=0, \mathrm{VP}}}
	}
	[nada]
$, $\protect\IfStrEqCase{6}{{1}{\ket{\nu={}4} }
		{2}{\ket{\nu={}3, \mathrm{IVC}}}
		{3}{\ket{\nu={}3, \mathrm{VP}}}
		{4}{\ket{\nu={}2, \mathrm{K-IVC}}}
		{5}{\ket{\nu={}2, \mathrm{VP}}}
		{6}{\ket{\nu={}1, (\mathrm{K-IVC}+\mathrm{VP})}}
		{7}{\ket{\nu={}1, \mathrm{VP}}}
		{8}{\ket{\nu=0, \mathrm{K-IVC}}}
		{9}{\ket{\nu=0, \mathrm{VP}}}
	}
	[nada]
$, and $\protect\IfStrEqCase{4}{{1}{\ket{\nu={}4} }
		{2}{\ket{\nu={}3, \mathrm{IVC}}}
		{3}{\ket{\nu={}3, \mathrm{VP}}}
		{4}{\ket{\nu={}2, \mathrm{K-IVC}}}
		{5}{\ket{\nu={}2, \mathrm{VP}}}
		{6}{\ket{\nu={}1, (\mathrm{K-IVC}+\mathrm{VP})}}
		{7}{\ket{\nu={}1, \mathrm{VP}}}
		{8}{\ket{\nu=0, \mathrm{K-IVC}}}
		{9}{\ket{\nu=0, \mathrm{VP}}}
	}
	[nada]
$ correlated insulator states of TBG, together with the band insulator $\protect\IfStrEqCase{1}{{1}{\ket{\nu={}4} }
		{2}{\ket{\nu={}3, \mathrm{IVC}}}
		{3}{\ket{\nu={}3, \mathrm{VP}}}
		{4}{\ket{\nu={}2, \mathrm{K-IVC}}}
		{5}{\ket{\nu={}2, \mathrm{VP}}}
		{6}{\ket{\nu={}1, (\mathrm{K-IVC}+\mathrm{VP})}}
		{7}{\ket{\nu={}1, \mathrm{VP}}}
		{8}{\ket{\nu=0, \mathrm{K-IVC}}}
		{9}{\ket{\nu=0, \mathrm{VP}}}
	}
	[nada]
$. The former three states were shown to be the ground states of the THF model in the absence of strain or relaxation effects~\cite{BUL20a,LIA21,SON22,*[{Strain and relaxation effects can be added systematically, as we show in }] [{.}] HER24a}. We restrict to positive fillings, since $S$ is an odd function of $\nu$ as a result of particle-hole symmetry. Similar results for TSTG with and without displacement field, as well as for other ground state candidates (including the $\nu=3$ ones) are relegated to \citeSymBrSeebeck{}. The solutions at finite doping (up to $\Delta \nu = \pm \frac{1}{2}$ or until the symmetry-broken solution can no longer be stabilized) are obtained incrementally from the integer filled ones~\cite{CAL23b}.  

Around charge neutrality, the Seebeck coefficient shown in \cref{fig:seebeck_sym_br:a} has the conventional shape expected for a particle-hole symmetric semiconductor: negative (positive) for electron (hole) doping. At very low temperatures, the Seebeck coefficient of the $\protect\IfStrEqCase{6}{{1}{\ket{\nu={}4} }
		{2}{\ket{\nu={}3, \mathrm{IVC}}}
		{3}{\ket{\nu={}3, \mathrm{VP}}}
		{4}{\ket{\nu={}2, \mathrm{K-IVC}}}
		{5}{\ket{\nu={}2, \mathrm{VP}}}
		{6}{\ket{\nu={}1, (\mathrm{K-IVC}+\mathrm{VP})}}
		{7}{\ket{\nu={}1, \mathrm{VP}}}
		{8}{\ket{\nu=0, \mathrm{K-IVC}}}
		{9}{\ket{\nu=0, \mathrm{VP}}}
	}
	[nada]
$ and $\protect\IfStrEqCase{1}{{1}{\ket{\nu={}4} }
		{2}{\ket{\nu={}3, \mathrm{IVC}}}
		{3}{\ket{\nu={}3, \mathrm{VP}}}
		{4}{\ket{\nu={}2, \mathrm{K-IVC}}}
		{5}{\ket{\nu={}2, \mathrm{VP}}}
		{6}{\ket{\nu={}1, (\mathrm{K-IVC}+\mathrm{VP})}}
		{7}{\ket{\nu={}1, \mathrm{VP}}}
		{8}{\ket{\nu=0, \mathrm{K-IVC}}}
		{9}{\ket{\nu=0, \mathrm{VP}}}
	}
	[nada]
$ insulators from \cref{fig:seebeck_sym_br:b,fig:seebeck_sym_br:d} also displays the conventional behavior as a function of doping. As the temperature increases in both cases, the positive hole-doping peak shrinks and $S$ turns negative for the entire doping range we consider. Around $\nu = 2$, the Seebeck coefficient shown in \cref{fig:seebeck_sym_br:c} is negative starting from the lowest temperatures we consider. As such, for $T \gtrsim \SI{25}{\kelvin}$, $S$ is \emph{negative} for $\nu \geq 0$ and shows oscillations (negative peaks) around the integer fillings, matching the low-temperature ($T \sim \SI{10}{\kelvin}$) measurements in both TBG~\cite{MER24} and TSTG~\cite{BAT24}. The temperature mismatch between theory and experiment is likely due to an \emph{underestimation} of dynamical fluctuations within second-order perturbation theory. The latter predicts critical temperatures $T^{*} \sim \SI{50}{\kelvin}$ for the transition to a symmetric state, compared to quantum Monte-Carlo simulations which point to $T^{*} \sim \SI{15}{\kelvin}$~\cite{RAI23a}. 

The negative Seebeck coefficient can be understood qualitatively by inspecting $A \left( \omega, \vec{k} \right)$ as a function of doping for the cases with a negative $S$. Taking $\protect\IfStrEqCase{6}{{1}{\ket{\nu={}4} }
		{2}{\ket{\nu={}3, \mathrm{IVC}}}
		{3}{\ket{\nu={}3, \mathrm{VP}}}
		{4}{\ket{\nu={}2, \mathrm{K-IVC}}}
		{5}{\ket{\nu={}2, \mathrm{VP}}}
		{6}{\ket{\nu={}1, (\mathrm{K-IVC}+\mathrm{VP})}}
		{7}{\ket{\nu={}1, \mathrm{VP}}}
		{8}{\ket{\nu=0, \mathrm{K-IVC}}}
		{9}{\ket{\nu=0, \mathrm{VP}}}
	}
	[nada]
$ as an example, \cref{fig:seebeck_sym_br:e} shows that the electron (hole) excitations consist primarily of $c$- ($f$-)fermions. Consequently, under electron doping, dispersive and uncorrelated $c$-electron states become occupied, and the bands are filled rigidly, as seen in \cref{fig:seebeck_sym_br:f}. In contrast, hole doping brings the $f$-electron states close to the chemical potential, and their strong correlation effect leads to a highly incoherent spectral signal. Due to their longer lifetime, the $c$-like electron excitations will dominate transport \emph{even} in the hole-doped cases from \cref{fig:seebeck_sym_br:f,fig:seebeck_sym_br:g}, leading to an overall negative $S$. At lower temperatures (a regime not measured by experiments~\cite{MER24}), the interaction-induced $f$-electron broadening is less significant and Seebeck coefficient does show a positive hole peak.

\begin{figure}[!t]
	\centering
	\includegraphics[width=\columnwidth]{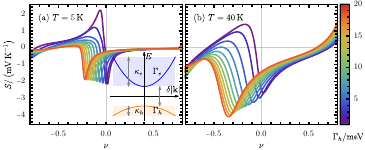}\subfloat{\label{fig:heavy_light:a}}\subfloat{\label{fig:heavy_light:b}}\caption{Seebeck coefficient of the two-band model at $T=\SI{5}{\kelvin}$ (a) and $T=\SI{40}{\kelvin}$ (b), for $\Gamma_e = \SI{1}{\milli\electronvolt}$, $\kappa_h = \SI{-8}{\milli\electronvolt}$, $\kappa_e = \SI{30}{\milli\electronvolt}$, and $\delta = \SI{16}{\milli\electronvolt}$. The filling $\nu$ is measured relative to the half-filled case ($\nu=\pm 1$ for the fully-filled/empty cases). The line color encodes $\Gamma_h$, according to the colormap. Inset in (a) shows the two-band model's schematic dispersion.}
	\label{fig:heavy_light}
\end{figure}

\textit{Analytical heavy-light model}.~In order to provide an analytical understanding and to identify the key ingredient leading to the observed negative Seebeck coefficient~\cite{MER24}, we consider a simple model featuring two quadratic bands, as shown in the inset of \cref{fig:heavy_light:a}\citeAsymptotes{}. The hole and electron bands have bandwidths $\kappa_h = \SI{-8}{\milli\electronvolt}$ and $\kappa_e = \SI{30}{\milli\electronvolt}$, respectively, and are separated by a gap $\delta = \SI{16}{\milli\electronvolt}$ (we use a conventional minus sign for the bandwidth of the hole band). These values are chosen to approximate the charge-one excitation dispersion of the $\protect\IfStrEqCase{6}{{1}{\ket{\nu={}4} }
		{2}{\ket{\nu={}3, \mathrm{IVC}}}
		{3}{\ket{\nu={}3, \mathrm{VP}}}
		{4}{\ket{\nu={}2, \mathrm{K-IVC}}}
		{5}{\ket{\nu={}2, \mathrm{VP}}}
		{6}{\ket{\nu={}1, (\mathrm{K-IVC}+\mathrm{VP})}}
		{7}{\ket{\nu={}1, \mathrm{VP}}}
		{8}{\ket{\nu=0, \mathrm{K-IVC}}}
		{9}{\ket{\nu=0, \mathrm{VP}}}
	}
	[nada]
$ correlated insulator shown in \cref{fig:seebeck_sym_br:e}. We also introduce momentum-independent broadening factors $\Gamma_h$ and $\Gamma_e$ for the hole and electron bands, respectively. Due to the strong dispersion of the charge-one excitation bands of TBG's correlated insulators, we find in \citeAsymptotesRef{} that the quantum-geometric contribution to $S$ are negligible~\cite{KRU23}. Consequently, this two-band framework successfully captures the fundamental characteristics of TBG's quasiparticles, namely their effective masses and lifetimes, thus providing a coherent interpretation of the Seebeck coefficient behavior around TBG's correlated insulating states. 

In what follows, we keep $\Gamma_e = \SI{1}{\milli\electronvolt}$ constant and investigate the effects of varying $\Gamma_h$ in \cref{fig:heavy_light}. At low-temperatures, the Seebeck coefficient shown in \cref{fig:heavy_light:a} is almost particle-hole antisymmetric in the $\Gamma_h = \Gamma_e$ case, despite significant effective mass asymmetry. Analytically, in the low temperature regime and for $\Gamma_{e/h} \ll \kappa_{e/h}$, we can show that the values of the  hole and electron peaks of $S$ are only dependent on the respective quasiparticle lifetime and not on their effective masses $S_{e/h} \approx -\frac{\pi^3}{3 \beta e} \frac{ \sgn\left( \kappa_{e/h} \right) }{2 \Gamma_{e/h}}$. As $\Gamma_h$ is increased, the positive hole peak shrinks: when $\Gamma_h \gg \kappa_h$, the ratio of magnitudes of the two Seebeck peaks is given by $\abs{\frac{S_{h}}{S_{e}}} \approx \frac{2 \abs{\kappa_h}}{\Gamma_h^2 + \kappa_h^2}  \frac{2 \Gamma_e}{\pi}$ (for $\Gamma_h = \SI{20}{\milli\electronvolt}$, $\abs{\frac{S_{h}}{S_{e}}} \sim 0.02$). As temperature is increased in \cref{fig:heavy_light:b}, the $\Gamma_h = \Gamma_e$ case still exhibits a positive hole peak, but for $\Gamma_h \geq \SI{6}{\milli\electronvolt}$, $S$ remains negative even in the hole doping case. This shows that $\Gamma_h \gg \Gamma_e$ is the key ingredient for a fully negative $S$. In the language of the THF model, the measured negative $S$ offers clear evidence - in this natural scenario - that heavy and light electrons (where the former have much \emph{larger} correlation-induced broadening) coexist in TBG. 

\begin{figure}[!t]
	\centering
	\includegraphics[width=\columnwidth]{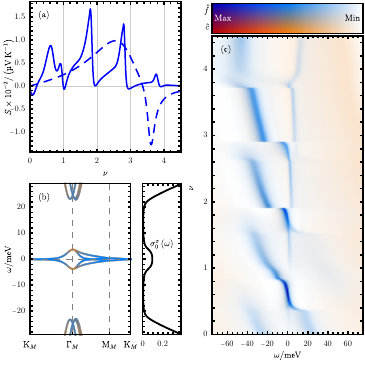}\subfloat{\label{fig:seebeck_sym:a}}\subfloat{\label{fig:seebeck_sym:b}}\subfloat{\label{fig:seebeck_sym:c}}\caption{The symmetric phase of TBG at $T=\SI{15}{\kelvin}$. The continuous (dashed) line in (a) shows the Seebeck coefficient in the symmetric phase (noninteracting limit). (b) illustrates the noninteracting band structure (left) and the noninteracting conductivity function $\sigma^{x}_{0} \left( \omega \right)$ (right) in $\hbar = 1$ units. The band colors indicate the fermionic character: blue (orange) for $f$- ($c$-)electrons. The interacting spectral function is shown in (c) using the same colormap as in \crefrange{fig:seebeck_sym_br:e}{fig:seebeck_sym_br:h}.}
	\label{fig:seebeck_sym}
\end{figure}

\textit{The symmetric phase}.~At larger temperatures, TBG transitions (naturally in the THF model) to a symmetric phase (wherein none of the symmetries of the model are broken)~\cite{RAI23a}. Without an accurate measurement of the transition temperature, we study the Seebeck coefficient in the symmetric state for a range of temperatures\citeSymmetric{}. For $T \lesssim \SI{20}{\kelvin}$, the Seebeck coefficient shown in \cref{fig:seebeck_sym:a} displays sawtooth oscillations around the integer fillings, but has an overall positive offset. The offset can be explained by comparing with the Seebeck coefficient in the noninteracting limit of TBG. For the former, the band structure and the corresponding conductivity function $\sigma^{x}_{0} \left( \omega \right)$ are shown in \cref{fig:seebeck_sym:b}. The Mott formula $S\left( \mu \right) \propto -  \dv{\log \left( \sigma^{x} \left( \mu \right) \right)}{\mu}$ can be used understand the positive offset of $S$ in the noninteracting limit\citeNoninteracting{}. On top of the non-interacting result, interaction-driven gap openings around integer fillings in the symmetric phase~\cite{CAL23b}, which are seen in the spectral function from \cref{fig:seebeck_sym:c}, lead to the observed oscillations of the Seebeck coefficient. The resulting positive $S$ along with its oscillations matches the high-temperature measurements in TBG~\cite{MER24}. The mismatch in temperature scales between theory and experiment can be attributed to strain and relaxation effects, which are unaccounted in our theoretical modeling\citeComparison{}. 

\textit{Conclusion}.~In summary, our investigations reveal that the distinctive thermoelectric response in TBG and TSTG -- especially the fully negative Seebeck coefficient at low temperatures -- serves as compelling evidence for the coexistence of localized heavy electrons and itinerant conduction electrons. Our theoretical model effectively describes this behavior, confirming its suitability for understanding the complex thermoelectric phenomena in these materials. This work lays the groundwork for further studies on transport within an interacting framework for moir\'e graphene systems.

\begin{acknowledgments}
	We thank S. Batlle Porro, P. Stepanov, and F. Koppens for collaborations on the related experimental project~\cite{BAT24} on the thermoelectric response of TSTG. We are also grateful to G. Rai, M. Lindsey, L. Lin, O. Vafek, E. Y. Andrei, S. B\"{u}hler-Paschen, A. Georges, A. Millis, G. Sangiovanni, J. Herzog-Arbeitman, J. Yu, B. Offertaler, and A. O. Denisov for useful discussions. The simulations presented in this article were performed on computational resources managed and supported by Princeton Research Computing, a consortium of groups including the Princeton Institute for Computational Science and Engineering (PICSciE) and the Office of Information Technology's High Performance Computing Center and Visualization Laboratory at Princeton University. D.C. acknowledges the hospitality of the Donostia International Physics Center, at which this work was carried out. B.A.B. was supported by DOE Grant No. DE-SC0016239. D.C. was supported by the European Research Council (ERC) under the European Union’s Horizon 2020 research and innovation program (grant agreement no. 101020833) and by the Simons Investigator Grant No. 404513, the Gordon and Betty Moore Foundation through Grant No. GBMF8685 towards the Princeton theory program, the Gordon and Betty Moore Foundation’s EPiQS Initiative (Grant No. GBMF11070), Office of Naval Research (ONR Grant No. N00014-20-1-2303), Global Collaborative Network Grant at Princeton University, BSF Israel US foundation No. 2018226, NSF-MERSEC (Grant No. MERSEC DMR 2011750). H.H. was supported by the European Research Council (ERC) under the European Union’s Horizon 2020 research and innovation program (Grant Agreement No. 101020833) and the Schmidt Fund Grant. 
\end{acknowledgments}

\renewcommand{\addcontentsline}[3]{}

\let\addcontentsline\oldaddcontentsline

\renewcommand{\thetable}{S\arabic{table}}
\renewcommand{\thefigure}{S\arabic{figure}}
\renewcommand{\theequation}{S\arabic{section}.\arabic{equation}}
\onecolumngrid
\pagebreak
\thispagestyle{empty}
\newpage
\begin{center}
	\textbf{\large Supplementary Information for ''The Thermoelectric Effect and Its Natural Heavy Fermion Explanation in Twisted Bilayer and Trilayer Graphene{}``}\\[.2cm]
\end{center}

\appendix
\renewcommand{\thesection}{\Roman{section}}
\tableofcontents
\let\oldaddcontentsline\addcontentsline
\newpage

\section{Derivation of the particle and energy currents in generic continuum crystalline systems}\label{app:sec:cur_deriv_cont}

In this \siSection{}, we consider a generic fermionic system in the continuum and use Noether's theorem to derive the corresponding particle and energy currents. After introducing the fermionic Hamiltonian in the continuum, we perturb the system with space-time dependent $\mathrm{U} \left(1\right)$ phase rotations and time translations to determine, respectively, the particle and energy currents. We consider generic kinetic terms with arbitrary number of spatial derivatives, as well as generic density-density long-range interactions. While the derivation of the particle current is straightforward, long-range interactions give rise to additional contributions to the energy currents, which we obtain. Finally, we also define the thermal currents, which are useful in the context of thermoelectric transport. Our expressions are consistent with those previously reported in the literature~\cite{PAU03}, which are hereby generalized for arbitrary kinetic energy operators. Additionally, a complete and pedagogical derivation of all the formulae is provided. The results of this \siSection{} will be employed in \cref{app:sec:deriv_moire_cur} to derive the corresponding currents for moir\'e twisted graphene systems.

\subsection{Model}\label{app:sec:cur_deriv_cont:model}
We start by outlining the model of a generic fermionic system in the continuum. To this end, we let $\hat{\psi}^\dagger_{n}\left( \vec{r} \right)$ denote the fermionic field operator in the Schr\"{o}dinger picture that creates an electron of ``flavor'' $n$ at position $\vec{r}$. Throughout this work, \emph{all} creation and annihilation operators will always be denoted with a hat, to differentiate them from complex numbers. The flavor index $n$ of the field operator $\hat{\psi}^\dagger_{n}\left( \vec{r} \right)$ incorporates the various orbital, valley, and spin degrees of freedom of the corresponding fermion. We consider $N_{\psi}$ fermion flavors, such that the flavor index obeys $1 \leq n \leq N_{\psi}$. Using units in which $\hbar = 1$, as will be assumed in the rest of this work, the Lagrangian for a general $d$-dimensional crystalline system can be written as~\cite{PAU03} 
\begin{equation}
	\label{app:eqn:generic_lagrangian}
	\mathcal{L} \left[ \hat{\psi}^\dagger_{}, \hat{\psi}_{} \right] = \frac{i}{2} \sum_{n} \int \dd[d]{r} \left( \hat{\psi}^\dagger_{n} \left( \vec{r}, t \right) \dot{\hat{\psi}}_{n} \left( \vec{r}, t \right) - \dot{\hat{\psi}}^\dagger_{n} \left( \vec{r}, t \right) \hat{\psi}_{n} \left( \vec{r}, t \right)  \right) - \mathcal{H} \left[ \hat{\psi}^\dagger_{}, \hat{\psi}_{} \right],
\end{equation}
where the corresponding Hamiltonian is given by
\begin{align}
	\mathcal{H} \left[ \hat{\psi}^\dagger_{}, \hat{\psi}_{} \right] & = \sum_{n,m} \int \dd[d]{r} \hat{\psi}^\dagger_{n} \left( \vec{r} \right) \left( T_{nm} \left( -i \nabla \right) + U_{nm} \left( \vec{r} \right)  \right) \hat{\psi}_{m} \left( \vec{r} \right) \nonumber \\
	& + \frac{1}{2} \sum_{n,m} \int \dd[d]{r_1} \dd[d]{r_2} V \left( \vec{r}_2 - \vec{r}_1 \right) \hat{\psi}^\dagger_{n} \left( \vec{r}_1 \right) \hat{\psi}_{n} \left( \vec{r}_1 \right) \hat{\psi}^\dagger_{m} \left( \vec{r}_2 \right) \hat{\psi}_{m} \left( \vec{r}_2 \right).
	\label{app:eqn:generic_hamiltonian_non_herm}
\end{align}
Additionally, in \cref{app:eqn:generic_lagrangian} and in what follows throughout this \siSection{}, for any operator $\hat{\mathcal{O}}$ in the Schr\"{o}dinger picture, we define the corresponding operator in the Heisenberg picture according to 
\begin{equation}
	\label{app:eqn:rel_heis_schro}
	\hat{\mathcal{O}} \left( t \right) = e^{i \mathcal{H} \left[ \hat{\psi}^\dagger_{}, \hat{\psi}_{} \right] t} \hat{O} e^{- i \mathcal{H} \left[ \hat{\psi}^\dagger_{}, \hat{\psi}_{} \right] t},
\end{equation}
such that $\hat{\mathcal{O}} = \hat{\mathcal{O}} \left( 0 \right)$. Consequently, the time derivative of the $\hat{\mathcal{O}}$ operator is given by
\begin{equation}
	\label{app:eqn:field_ops_time_deriv}
	\dot{\hat{\mathcal{O}}} \left( t \right) = i \commutator{\mathcal{H} \left[ \hat{\psi}^\dagger_{}, \hat{\psi}_{} \right] }{\hat{\mathcal{O}} \left( t \right)}.
\end{equation}
For example, the fermionic field creation operator in the Heisenberg picture is given by
\begin{equation}
	\hat{\psi}^\dagger_{n} \left( \vec{r}, t \right) = e^{i \mathcal{H} \left[ \hat{\psi}^\dagger_{}, \hat{\psi}_{} \right] t} \hat{\psi}^\dagger_{n} \left( \vec{r} \right) e^{- i \mathcal{H} \left[ \hat{\psi}^\dagger_{}, \hat{\psi}_{} \right] t}.
\end{equation}
In \cref{app:eqn:generic_hamiltonian_non_herm}, $T_{nm} \left( -i \nabla \right)$ denotes the kinetic operator for the field $\hat{\psi}^\dagger_{n} \left( \vec{r} \right)$ ({\it e.g.}{}, $T_{nm} \left( -i \nabla \right) \propto \nabla^2$ for non-relativistic free electron or for a $\vec{k} \cdot \vec{p}$ model of a semiconductor, while $T_{nm} \left( -i \nabla \right) \propto \nabla$ for a massless Dirac fermion or for a $\vec{k} \cdot \vec{p}$ model of graphene), $U_{nm} \left( \vec{r} \right)$ is the crystalline potential, while $V \left( \vec{r} \right)$ is the interaction function corresponding to two electrons separated by a vector $\vec{r}$. In this work, we consider exclusively density-density interactions of the form in \cref{app:eqn:generic_hamiltonian_non_herm} which are identical to the electron-electron interaction in the continuum Bistritzer-MacDonald models introduced in \cref{app:sec:BM_review}. Without loss of generality, we can take $V\left( \vec{r} \right) = V\left( -\vec{r} \right)$. The fermionic field $\hat{\psi}^\dagger_{n} \left( \vec{r} \right)$ obeys the standard (equal-time) anticommutation relations
\begin{equation}
	\begin{split}
		\left\lbrace \hat{\psi}^\dagger_{n} \left( \vec{r} \right), \hat{\psi}_{m} \left( \vec{r}' \right) \right\rbrace &= \delta_{nm} \delta^d \left( \vec{r} - \vec{r}' \right), \\
		\left\lbrace \hat{\psi}^\dagger_{n} \left( \vec{r} \right), \hat{\psi}^\dagger_{m} \left( \vec{r}' \right) \right\rbrace &= 
		\left\lbrace \hat{\psi}_{n} \left( \vec{r} \right), \hat{\psi}_{m} \left( \vec{r}' \right) \right\rbrace = 0.
	\end{split}
\end{equation}
Finally, the action of the system has the usual form 
\begin{equation}
	S \left[ \hat{\psi}^\dagger_{}, \hat{\psi}_{} \right] = \int \dd{t} \mathcal{L} \left[ \hat{\psi}^\dagger_{}, \hat{\psi}_{} \right] .
\end{equation}

Note that the kinetic operator and the crystalline potential satisfy the following Hermiticity conditions
\begin{equation}
	\label{app:eqn:herm_of_kin_and_u}
	T_{nm} \left( -i \nabla \right) = T^{*}_{mn} \left( i \nabla \right),  \quad
	U_{mn} \left( \vec{r} \right) = U^{*}_{nm} \left( \vec{r} \right),
\end{equation}
which follow by requiring that the Hamiltonian $\mathcal{H} \left[ \hat{\psi}^\dagger_{}, \hat{\psi}_{} \right]$ is Hermitian. Using \cref{app:eqn:herm_of_kin_and_u}, as well as integration by parts, one can show that
\begin{align}
	&\sum_{n,m} \int \dd[d]{r} \hat{\psi}^\dagger_{n} \left( \vec{r} \right) T_{nm} \left( -i \nabla \right)  \hat{\psi}_{m} \left( \vec{r} \right) 
	= \sum_{n,m} \int \dd[d]{r} \left( T_{nm} \left( i \nabla \right) \hat{\psi}^\dagger_{n} \left( \vec{r} \right) \right)  \hat{\psi}_{m} \left( \vec{r} \right) \nonumber \\
	=& \sum_{n,m} \int \dd[d]{r} \left( T^{*}_{mn} \left( -i \nabla \right) \hat{\psi}^\dagger_{n} \left( \vec{r} \right) \right)  \hat{\psi}_{m} \left( \vec{r} \right)
	= \sum_{n,m} \int \dd[d]{r} \left( \hat{\psi}^\dagger_{n} \left( \vec{r} \right) T_{nm} \left( -i \nabla \right)  \hat{\psi}_{m} \left( \vec{r} \right) \right)^{\dagger},
\end{align}
which can be used to render the integrand in the first term of \cref{app:eqn:generic_hamiltonian_non_herm} explicitly Hermitian, leading to the following expression of the Hamiltonian 
\begin{align}
	\mathcal{H} \left[ \hat{\psi}^\dagger_{}, \hat{\psi}_{} \right] & = \frac{1}{2} \sum_{n,m} \int \dd[d]{r} \hat{\psi}^\dagger_{n} \left( \vec{r} \right) \left( T_{nm} \left( -i \nabla \right) + U_{nm} \left( \vec{r} \right)  \right) \hat{\psi}_{m} \left( \vec{r} \right) + \text{h.c.} \nonumber \\
	& + \frac{1}{2} \sum_{n,m} \int \dd[d]{r_1} \dd[d]{r_2} V \left( \vec{r}_2 - \vec{r}_1 \right) \hat{\psi}^\dagger_{n} \left( \vec{r}_1 \right) \hat{\psi}_{n} \left( \vec{r}_1 \right) \hat{\psi}^\dagger_{m} \left( \vec{r}_2 \right) \hat{\psi}_{m} \left( \vec{r}_2 \right).
	\label{app:eqn:generic_hamiltonian}
\end{align}
The expression in \cref{app:eqn:generic_hamiltonian}, where ``$+\text{h.c.}$'' denotes the addition of the Hermitian conjugate, is equivalent to the one in \cref{app:eqn:generic_hamiltonian_non_herm}, but has the advantage of having explicitly Hermitian integrands. 

We consider systems with discrete translation symmetry, {\it i.e.}{} for which the crystalline potential is periodic with respect to translations by any lattice vector $\vec{R}$,
\begin{equation}
	\label{app:eqn:translation_generic_potential}
	U_{nm} \left( \vec{r} + \vec{R} \right) = U_{nm} \left( \vec{r} \right),
\end{equation} 
with $\vec{R}$ being an integer linear combination of the primitive lattice vectors $\vec{a}_i$ ($1 \leq i \leq d$),
\begin{equation}
	\vec{R} = \sum_{i=1}^{d} m_i \vec{a}_i, \qq{for} m_i \in \mathbb{Z}.
\end{equation}
As such, it is useful to introduce the following Fourier transformations
\begin{align}
	\hat{\psi}^\dagger_{n} \left( \vec{r} \right) &= \frac{1}{\sqrt{N_0\Omega_0}} \sum_{\vec{G}} \sum_{\vec{k}} e^{-i \left( \vec{k} - \vec{G} \right) \cdot \vec{r}} \hat{a}^\dagger_{\vec{k},\vec{G},n}, \label{app:eqn:ft_psi}\\
	U_{nm} \left( \vec{r} \right) &= \sum_{\vec{G}} U_{nm} \left( \vec{G} \right) e^{-i \vec{G} \cdot \vec{r}}, \label{app:eqn:ft_U} \\
	V \left( \vec{r} \right) &= \frac{1}{N_0 \Omega_0}\sum_{\vec{G}} \sum_{\vec{k}} V \left( \vec{k} + \vec{G} \right) e^{-i \left( \vec{k} + \vec{G} \right) \cdot \vec{r}}, \label{app:eqn:ft_V}
\end{align}
where the summation over $\vec{k}$ is evaluated over the first Brillouin zone (BZ), while the sum over $\vec{G}$ runs over the reciprocal vectors of the crystalline system\footnote{For the summation dummy variables, we will stick to the convention of using upper-case letters for reciprocal lattice vectors and lower-case letters for vectors inside the first BZ.}, {\it i.e.}{}
\begin{equation}
	\vec{G} = \sum_{i=1}^{d} m_i \vec{b}_i, \qq{for} m_i \in \mathbb{Z},
\end{equation}
where $\vec{b}_i$ ($1 \leq i \leq d$) are the reciprocal lattice vectors, which obey
\begin{equation}
	\vec{a}_i \cdot \vec{b}_j = \delta_{ij}, \qq{with} 1 \leq i,j \leq d.
\end{equation}
Additionally, $N_0$ denotes the number of unit cells in the system, while $\Omega_0$ denotes the volume of one unit cell. The momentum space operators $\hat{a}^\dagger_{\vec{k},\vec{G},n}$ are defined for $\vec{k}$ in the first BZ
\begin{equation}
	\hat{a}^\dagger_{\vec{k},\vec{G},n} = \frac{1}{\sqrt{N_0 \Omega_0}} \int \dd[d]{r} \hat{\psi}^\dagger_{n} \left( \vec{r} \right) e^{i \left( \vec{k} - \vec{G} \right) \cdot \vec{r}}.
\end{equation}
For simplicity, we also \emph{define} the momentum-space operators $\hat{a}^\dagger_{\vec{k},\vec{G},n}$ for $\vec{k}$ \emph{outside} the first BZ via the following embedding condition~\cite{BER21,SON21,BER21a,LIA21,BER21b,XIE21}{}
\begin{equation}
	\hat{a}^\dagger_{\vec{k} + \vec{G}_0,\vec{G},n} = \hat{a}^\dagger_{\vec{k},\vec{G} - \vec{G}_0,n}, \qq{for any reciprocal vector $\vec{G}_0$.} 
\end{equation}

Using the newly defined plane-wave basis from \cref{app:eqn:ft_psi}, the Lagrangian of the system can be recast as 
\begin{align}
	\label{app:eqn:generic_lagrangian_k}
	\mathcal{L} \left[ \hat{\psi}^\dagger_{}, \hat{\psi}_{} \right] &= \frac{i}{2} \sum_{n} \sum_{\vec{G}} \sum_{\vec{k}} \left( \hat{a}^\dagger_{\vec{k},\vec{G},n} \dot{\hat{a}}_{\vec{k},\vec{G},n} - \dot{\hat{a}}^\dagger_{\vec{k},\vec{G},n} \hat{a}_{\vec{k},\vec{G},n} \right) - \mathcal{H} \left[ \hat{\psi}^\dagger_{}, \hat{\psi}_{} \right]
\end{align}
with the momentum-space Hamiltonian reading as
\begin{align}
	\mathcal{H} \left[ \hat{\psi}^\dagger_{}, \hat{\psi}_{} \right] &= \sum_{n,m} \sum_{\vec{G}_1,\vec{G}_2} \sum_{\vec{k}} \hat{a}^\dagger_{\vec{k},\vec{G}_1,n} \left( T_{nm} \left( \vec{k} - \vec{G}_2 \right) \delta_{\vec{G}_1, \vec{G}_2} + U_{nm} \left( \vec{G}_2 - \vec{G}_1  \right)  \right) \hat{a}_{\vec{k},\vec{G}_2,m} \nonumber \\
	&+ \frac{1}{2} \frac{1}{N_0 \Omega_0} \sum_{n,m} \sum_{\vec{G}_1,\vec{G}_2,\vec{Q}} \sum_{\vec{k}_1,\vec{k}_2,\vec{q}} V \left( \vec{q} + \vec{G} \right) \hat{a}^\dagger_{\vec{k}_2+\vec{q}, \vec{G}_2 - \vec{Q}, n} \hat{a}_{\vec{k}_2, \vec{G}_2, n}  \hat{a}^\dagger_{\vec{k}_1-\vec{q}, \vec{G}_1 + \vec{Q}, m} \hat{a}_{\vec{k}_1, \vec{G}_1, m}.
	\label{app:eqn:generic_hamiltonian_k}
\end{align}

\subsection{Derivation of the particle current}\label{app:sec:cur_deriv_cont:particle}

To derive the particle current, we will employ Noether's theorem~\cite{PAU03,PES95}. We consider infinitesimal, space-time dependent phase variations of the fermion field 
\begin{equation}
	\label{app:eqn:phase_rotation}
	\hat{\psi}_{n} \left( \vec{r}, t \right) \to \hat{\psi}_{n} \left( \vec{r}, t \right) e^{i \phi \left( \vec{r}, t \right)} = \hat{\psi}_{n} \left( \vec{r}, t \right) + i\phi \left( \vec{r}, t \right) \hat{\psi}_{n} \left( \vec{r}, t \right) + \mathcal{O} \left( \phi^2 \left( \vec{r}, t \right) \right),
\end{equation} 
where the real scalar $\phi \left( \vec{r}, t \right)$ denotes the phase rotating the fermionic fields at position $\vec{r}$ and time $t$. The particle current $\vec{j} \left( \vec{r}, t \right)$ and the particle density $\rho \left( \vec{r}, t \right) $ can be found by taking the following functional derivative~\cite{MOR96} 
\begin{equation}
	\label{app:eqn:noether_particle_current}
	\eval{ \fdv{S \left[ \hat{\psi}^\dagger_{} + \delta \hat{\psi}^\dagger_{}, \hat{\psi}_{} + \delta \hat{\psi}_{} \right] }{\phi \left( \vec{r}, t \right)}}_{\phi \left( \vec{r}, t \right) = 0} = \pdv{\rho \left( \vec{r}, t \right) }{t} + \nabla \cdot \vec{j} \left( \vec{r}, t \right),
\end{equation}
with the variation of the fermion field being defined as 
\begin{equation}
	\delta \hat{\psi}_{n} \left( \vec{r}, t \right) \equiv i \phi \left( \vec{r}, t \right) \hat{\psi}_{n} \left( \vec{r}, t \right).
\end{equation}
Working to first order in the phase $\phi \left( \vec{r}, t \right)$, the variation of the Lagrangian is given by 
\begin{align}
	\var{\mathcal{L}} \left[ \hat{\psi}^\dagger_{}, \hat{\psi}_{} \right] &= \mathcal{L} \left[ \hat{\psi}^\dagger_{} + \delta \hat{\psi}^\dagger_{}, \hat{\psi}_{} + \delta \hat{\psi}_{} \right] - \mathcal{L} \left[ \hat{\psi}^\dagger_{}, \hat{\psi}_{} \right] \nonumber \\
	&=  - \sum_{n} \int \dd[d]{r} \dot{\phi} \left( \vec{r}, t \right)  \hat{\psi}^\dagger_{n} \left( \vec{r}, t \right) \hat{\psi}_{n} \left( \vec{r}, t \right) \nonumber \\
	&+\sum_{n,m} \int \dd[d]{r} \left[ i \phi \left( \vec{r}, t \right) \hat{\psi}^\dagger_{n} \left( \vec{r}, t \right) T_{nm} \left( -i \nabla \right) \hat{\psi}_{m}\left( \vec{r}, t \right) 
	- i \hat{\psi}^\dagger_{n} \left( \vec{r}, t \right) T_{nm} \left( -i \nabla \right) \left( \hat{\psi}_{m} \left( \vec{r}, t \right) \phi \left( \vec{r}, t \right) \right) \right], \label{app:eqn:1st_order_var_particle}
\end{align}
where we have employed the expression of the Hamiltonian from \cref{app:eqn:generic_hamiltonian_non_herm}. Note that the interaction and crystalline potential parts of the Lagrangian do not produce any variations. To simplify the expression in \cref{app:eqn:1st_order_var_particle}, we now take the simpler case in which $T_{nm} \left( - i\nabla \right)$ contains a fixed number $N$ of derivative operators, {\it i.e.}{}
\begin{equation}
	\label{app:eqn:kinetic_term_one_derivative}
	\begin{split}
		T_{nm} \left(-i \nabla \right) &= t^{\mu_1 \mu_2 \dots \mu_N}_{nm} \partial^{\mu_1} \partial^{\mu_2} \dots \partial^{\mu_N} \\
		T_{nm} \left(\vec{k} \right) &=  i^N t^{\mu_1 \mu_2 \dots \mu_N}_{nm} k^{\mu_1} k^{\mu_2} \dots k^{\mu_N}
	\end{split}, 
	\qq{with $n \in \mathbb{N}$}. 
\end{equation}
In \cref{app:eqn:kinetic_term_one_derivative}, the tensor $t^{\mu_1 \mu_2 \dots \mu_N}_{nm}$ is fully symmetric in the spatial indices $\mu_i$ ($1 \leq i \leq N$), over which we have employed the Einstein summation convention (such that repeated spatial indices are summed over). Using the assumption from \cref{app:eqn:kinetic_term_one_derivative}, as well as the shorthand notation 
\begin{equation}
	\label{app:eqn:currents_deriv_shorthand}
	\nabla^{i \dots j} = \begin{cases}
		\partial^{\mu_i} \partial^{\mu_{i+1}} \dots \partial^{\mu_j} & \text{for $1 \leq i \leq j \leq N$}, \\
		1 & \text{otherwise}
	\end{cases},
\end{equation} 
we will now focus on the third row of \cref{app:eqn:1st_order_var_particle}, which we first simplify using integration by parts 
\begin{align}
	&\int \dd[d]{r} \left[ i \phi \left( \vec{r}, t \right) \hat{\psi}^\dagger_{n} \left( \vec{r}, t \right) T_{nm} \left( -i \nabla \right) \hat{\psi}_{m} \left( \vec{r}, t \right) 
	- i \hat{\psi}^\dagger_{n} \left( \vec{r}, t \right) T_{nm} \left( -i \nabla \right) \left( \hat{\psi}_{m} \left( \vec{r}, t \right) \phi \left( \vec{r}, t \right) \right) \right] \nonumber \\
	=& i \int \dd[d]{r} \phi \left( \vec{r}, t \right) t^{\mu_1 \mu_2 \dots \mu_N}_{nm} \left[  \hat{\psi}^\dagger_{n} \left( \vec{r}, t \right) \nabla^{1 \dots N} \hat{\psi}_{m} \left( \vec{r}, t \right) - (-1)^N \left( \nabla^{1 \dots N} \hat{\psi}^\dagger_{n} \left( \vec{r}, t \right) \right)  \hat{\psi}_{m} \left( \vec{r}, t \right) \right].
\end{align}
We will now repeatedly use the identity $f\left( \vec{r}, t \right) \partial^{\mu} g\left( \vec{r}, t \right) = \partial^{\mu} \left( f\left( \vec{r}, t \right) g\left( \vec{r}, t \right) \right) - \left( \partial^{\mu} f\left( \vec{r}, t \right) \right) g\left( \vec{r}, t \right) $, as well as the symmetry of the $t^{\mu_1 \mu_2 \dots \mu_N}_{nm}$ tensor in the spatial indices to derive
\begin{align}
	&t^{\mu_1 \mu_2 \dots \mu_N}_{nm} \hat{\psi}^\dagger_{n} \nabla^{1 \dots N} \hat{\psi}_{m}  = 
	t^{\mu_1 \mu_2 \dots \mu_N}_{nm} \left[ \partial^{\mu_1} \left( \hat{\psi}^\dagger_{n} \nabla^{2 \dots N} \hat{\psi}_{m} \right) - \left( \partial^{\mu_1} \hat{\psi}^\dagger_{n} \right) \nabla^{2 \dots N} \hat{\psi}_{m} \right] \nonumber \\
	= & t^{\mu_1 \mu_2 \dots \mu_N}_{nm} \left\lbrace \partial^{\mu_1} \left( \hat{\psi}^\dagger_{n} \nabla^{2 \dots N} \hat{\psi}_{m} \right) - \partial^{\mu_2}\left[ \left( \partial^{\mu_1} \hat{\psi}^\dagger_{n} \right) \nabla^{3 \dots N} \hat{\psi}_{m} \right] +  \left( \nabla^{1 \dots 2} \hat{\psi}^\dagger_{n} \right) \nabla^{3 \dots N} \hat{\psi}_{m}  \right\rbrace \nonumber \\
	= & t^{\mu_1 \mu_2 \dots \mu_N}_{nm} \left\lbrace  \partial^{\mu_1}\left[ \hat{\psi}^\dagger_{n} \nabla^{2 \dots N} \hat{\psi}_{m} - \left( \nabla^{2 \dots 2} \hat{\psi}^\dagger_{n} \right) \nabla^{3 \dots N} \hat{\psi}_{m} \right] +  \left( \nabla^{1 \dots 2} \hat{\psi}^\dagger_{n} \right) \nabla^{3 \dots N} \hat{\psi}_{m}  \right\rbrace \nonumber \\
	= & t^{\mu_1 \mu_2 \dots \mu_N}_{nm} \left\lbrace  \partial^{\mu_1}\left[ \hat{\psi}^\dagger_{n} \nabla^{2 \dots N} \hat{\psi}_{m} - \left( \nabla^{2 \dots 2} \hat{\psi}^\dagger_{n} \right) \nabla^{3 \dots N} \hat{\psi}_{m} + \left( \nabla^{2 \dots 3} \hat{\psi}^\dagger_{n} \right) \nabla^{4 \dots N} \hat{\psi}_{m} \right] -  \left( \nabla^{1 \dots 3} \hat{\psi}^\dagger_{n} \right) \nabla^{4 \dots N} \hat{\psi}_{m} \right\rbrace \nonumber \\
	\dots \nonumber \\
	= & t^{\mu_1 \mu_2 \dots \mu_N}_{nm}  \left\lbrace   \partial^{\mu_1}\left[ \sum_{j=1}^{N} (-1)^{j+1} \left( \nabla^{2 \dots j} \hat{\psi}^\dagger_{n} \right) \nabla^{(j+1) \dots N} \hat{\psi}_{m} \right] + (-1)^N \left( \nabla^{1 \dots N} \hat{\psi}^\dagger_{n} \right) \hat{\psi}_{m} \right\rbrace, \label{app:eqn:long_int_by_parts_expression}
\end{align}
where we have temporarily suppressed the space-time dependence of the field operators, for the sake of brevity. As a result, the third row of \cref{app:eqn:1st_order_var_particle} becomes simply
\begin{align}
	&\int \dd[d]{r} \left[ i \phi \left( \vec{r}, t \right) \hat{\psi}^\dagger_{n} \left( \vec{r}, t \right) T_{nm} \left( -i \nabla \right) \hat{\psi}_{m} \left( \vec{r}, t \right) 
	- i \hat{\psi}^\dagger_{n} \left( \vec{r}, t \right) T_{nm} \left( -i \nabla \right) \left( \hat{\psi}_{m} \left( \vec{r}, t \right) \phi \left( \vec{r}, t \right) \right) \right] \nonumber \\
	=& i \int \dd[d]{r} \phi \left( \vec{r}, t \right) t^{\mu_1 \mu_2 \dots \mu_N}_{nm} \partial^{\mu_1}\left[ \sum_{j=1}^{N} (-1)^{j+1} \left( \nabla^{2 \dots j} \hat{\psi}^\dagger_{n} \left( \vec{r}, t \right) \right) \nabla^{(j+1) \dots N} \hat{\psi}_{m} \left( \vec{r}, t \right) \right]. \label{app:eqn:second_row_simplification}
\end{align}
Taken together \cref{app:eqn:1st_order_var_particle,app:eqn:second_row_simplification} allow us to evaluate the functional derivative and determine the local particle current and density, respectively,
\begin{align}
	\rho \left( \vec{r}, t \right) &= \sum_{n} \hat{\psi}^\dagger_{n} \left( \vec{r}, t \right) \hat{\psi}_{n} \left( \vec{r}, t \right) \label{app:eqn:cont_part_density}, \\
	j^{\mu_1} \left( \vec{r}, t \right) &=  i \sum_{n,m} t^{\mu_1 \mu_2 \dots \mu_N}_{nm} \left[ \sum_{j=1}^{N} (-1)^{j+1} \left( \nabla^{2 \dots j} \hat{\psi}^\dagger_{n} \left( \vec{r}, t \right) \right) \nabla^{(j+1) \dots N} \hat{\psi}_{m} \left( \vec{r}, t \right) \right], \label{app:eqn:cont_part_current_complicated}
\end{align}
which obey the following continuity equation 
\begin{equation}
	\label{app:eqn:continuity_particle}
	\pdv{\rho \left( \vec{r}, t \right) }{t} + \nabla \cdot \vec{j} \left( \vec{r}, t \right) = 0.
\end{equation}
\Cref{app:eqn:continuity_particle} follows straight-forwardly from \cref{app:eqn:noether_particle_current} using the stationary action principle: for those field configurations that satisfy the equations of motion, the action will be stationary with respect to perturbations of the fields. \Cref{app:eqn:continuity_particle} simply implies that the particle density is conserved for those field configurations that satisfy the equations of motion.

\Cref{app:eqn:cont_part_density,app:eqn:cont_part_current_complicated} denote the expressions for the local particle current and density operators in the Heisenberg picture. The get the corresponding expressions in the Schro\"{o}dinger picture, one can simply set $t = 0$, as implied by \cref{app:eqn:rel_heis_schro}. In what follows, it is also useful to define the Fourier transformation of the density operator
\begin{equation}
	\label{app:eqn:ft_rho_op}
	\rho \left( \vec{q} + \vec{G} \right) = \int \dd[d]{r} \rho \left( \vec{r} \right) e^{i \left( \vec{q} + \vec{G} \right) \cdot \vec{r}} = \sum_{n} \sum_{\vec{G}'} \sum_{\vec{k}} \hat{a}^\dagger_{\vec{k}+\vec{q},\vec{G}'-\vec{G},n} \hat{a}_{\vec{k},\vec{G}',n}.
\end{equation}
The total charge $Q$ is obtained by integrating the $\rho \left( \vec{r} \right)$ over the entire space 
\begin{equation}
	Q = \eval{\rho \left( \vec{q} + \vec{G} \right)}_{\vec{q}+\vec{G}=\vec{0}}= \sum_{n} \int \dd[d]{r} \hat{\psi}^\dagger_{n} \left( \vec{r} \right) \hat{\psi}_{n} \left( \vec{r} \right) = \sum_{n} \sum_{\vec{G}} \sum_{\vec{k}} \hat{a}^\dagger_{\vec{k},\vec{G},n} \hat{a}_{\vec{k},\vec{G},n}. \label{app:eqn:part_charge_simple_real_particular}
\end{equation}
Similarly, we integrate the local current $j^{\mu_1} \left( \vec{r} \right)$ over the entire space to find the total current operator $\vec{J}$
\begin{align}
	J^{\mu_1} &= \int \dd[d]{r} j^{\mu_1} \left( \vec{r} \right) \nonumber \\
	&= i \sum_{n,m} \int \dd[d]{r} t^{\mu_1 \mu_2 \dots \mu_N}_{nm} \left[ \sum_{j=1}^{N} (-1)^{j+1} \left( \nabla^{2 \dots j} \hat{\psi}^\dagger_{n} \left( \vec{r} \right) \right) \nabla^{(j+1) \dots N} \hat{\psi}_{m} \left( \vec{r} \right) \right] \nonumber \\
	&= i N  \sum_{n,m} \int \dd[d]{r} t^{\mu_1 \mu_2 \dots \mu_N}_{nm} \hat{\psi}^\dagger_{n} \left( \vec{r} \right)  \nabla^{2 \dots N} \hat{\psi}_{m} \left( \vec{r} \right) \nonumber \\
	&= i^N N  \sum_{n,m} \sum_{\vec{G}} \sum_{\vec{k}} t^{\mu_1 \mu_2 \dots \mu_N}_{nm} \hat{a}^\dagger_{\vec{k},\vec{G},n} \left(\vec{k} - \vec{G}\right)^{\mu_2} \dots \left(\vec{k} - \vec{G}\right)^{\mu_N} \hat{a}_{\vec{k},\vec{G},m} \nonumber \\
	&= \sum_{n,m} \sum_{\vec{G}} \sum_{\vec{k}} \pdv{T_{nm} \left(\vec{k} -\vec{G} \right)}{k^{\mu_1}} \hat{a}^\dagger_{\vec{k},\vec{G},n} \hat{a}_{\vec{k},\vec{G},m}, 
	\label{app:eqn:part_current_simple_real_particular}
\end{align}
where, in going from the second to the third line, we have employed integration by parts. We remind the reader that \cref{app:eqn:part_charge_simple_real_particular,app:eqn:part_current_simple_real_particular} were derived assuming the form of the kinetic energy matrix is given by \cref{app:eqn:kinetic_term_one_derivative}. These expressions can be readily generalized by noting that: (1) the expression of the total charge does \emph{not} depend on the form of the kinetic energy matrix $T_{nm} \left( - i \nabla \right)$, and (2) the total current operator is linear in kinetic energy matrix. As such, one can consider the case where the kinetic energy matrix $T_{nm} \left( - i \nabla \right)$ is a sum of derivative terms of the form in \cref{app:eqn:kinetic_term_one_derivative} having different degrees $N$ and for which \cref{app:eqn:part_current_simple_real_particular} holds. We conclude that the \emph{generally-valid} expressions for the total particle charge and current are given by
\begin{align}
	Q &= \sum_{n} \int \dd[d]{r} \hat{\psi}^\dagger_{n} \left( \vec{r} \right) \hat{\psi}_{n} \left( \vec{r} \right) = \sum_{n} \sum_{\vec{G}} \sum_{\vec{k}} \hat{a}^\dagger_{\vec{k},\vec{G},n} \hat{a}_{\vec{k},\vec{G},n}, \label{app:eqn:part_charge_simple_real_general} \\
	\vec{J} &= \sum_{n,m} \sum_{\vec{G}} \sum_{\vec{k}} \pdv{T_{nm} \left(\vec{k} -\vec{G} \right)}{\vec{k}} \hat{a}^\dagger_{\vec{k},\vec{G},n} \hat{a}_{\vec{k},\vec{G},m}. 
	\label{app:eqn:part_current_simple_real_general}
\end{align}
Finally, we mention that \cref{app:eqn:part_charge_simple_real_general,app:eqn:part_current_simple_real_general} can equivalently be derived by minimally-coupling the fermionic field to a magnetic vector potential.

\subsection{Derivation of the energy current}\label{app:sec:cur_deriv_cont:energy}
With the particle charge and currents derived, we now move to obtain the energy current of the model. The strategy is similar to \cref{app:sec:cur_deriv_cont:particle}, but instead of phase variations, we now consider space-time dependent time translations of the fermion fields. Under this transformation, the time parameter changes as $t \to t - \phi \left( \vec{r}, t \right)$, resulting in the following variation of the fermion field\footnote{Here, we consider active, rather than passive transformations. In an active transformation, the field configuration itself is changed, while the  coordinate system is kept invariant. In contrast, in a passive transformation, the field is kept the same, while the coordinate system is changed.}
\begin{equation}
	\label{app:eqn:time_translation}
	\hat{\psi}_{n} \left( \vec{r}, t \right) \to \hat{\psi}_{n} \left( \vec{r}, t \right) - \phi \left( \vec{r}, t \right) \dot{\hat{\psi}}_{n} \left( \vec{r}, t \right) + \mathcal{O} \left( \phi^2 \left( \vec{r}, t \right) \right).
\end{equation}
In \cref{app:eqn:time_translation}, the time translation perturbation $\phi\left( \vec{r}, t \right)$ is a real space-time dependent scalar field. As in \cref{app:sec:cur_deriv_cont:particle}, the energy current $\vec{j}_E \left( \vec{r}, t \right)$ and the energy density $\rho_E \left( \vec{r}, t \right)$ can be found from Noether's theorem~\cite{MOR96} 
\begin{equation}
	\label{app:eqn:noether_energy_current}
	\eval{ \fdv{S \left[ \hat{\psi}^\dagger_{} + \delta \hat{\psi}^\dagger_{}, \hat{\psi}_{} + \delta \hat{\psi}_{} \right] }{\phi \left( \vec{r}, t \right)}}_{\phi \left( \vec{r}, t \right) = 0} =  \pdv{\rho_E \left( \vec{r}, t \right)}{t} + \nabla \cdot \vec{j}_E \left( \vec{r}, t \right) ,
\end{equation}
where the variation of the fermion field is given by 
\begin{equation}
	\delta \hat{\psi}_{n} \left( \vec{r}, t \right) \equiv - \phi \left( \vec{r}, t \right) \dot{\hat{\psi}}_{n} \left( \vec{r}, t \right).
\end{equation}

It is well known that the current and density derived through Noether's theorem are not unique~\cite{PES95}: for a general Noether four-current $J^{\mu}$ satisfying the conservation law $\partial_{\mu} J^{\mu} = 0$, the current $J^{\prime \mu} = J^{\mu} + \partial_{\nu} A^{\mu \nu}$, where $A^{\mu \nu} = - A^{\nu \mu}$ is also conserved. To ensure that the energy current and density we derived are hermitian operators, we will employ the formula of the Hamiltonian from \cref{app:eqn:generic_hamiltonian}, in which the Hamiltonian density is explicitly Hermitian.

To first order in the time translation $\phi \left( \vec{r} \right)$, the variation of the action is given by
\begin{align}
	\label{app:eqn:var_s_energy}
	\var{S} \left[ \hat{\psi}^\dagger_{}, \hat{\psi}_{} \right] &= S \left[ \hat{\psi}^\dagger_{} + \delta \hat{\psi}^\dagger_{}, \hat{\psi}_{} + \delta \hat{\psi}_{} \right] - S \left[ \hat{\psi}^\dagger_{}, \hat{\psi}_{} \right] \nonumber \\
	& = \var{S_1} \left[ \hat{\psi}^\dagger_{}, \hat{\psi}_{} \right] + \var{S_2} \left[ \hat{\psi}^\dagger_{}, \hat{\psi}_{} \right] + \var{S_3} \left[ \hat{\psi}^\dagger_{}, \hat{\psi}_{} \right] + \var{S_4} \left[ \hat{\psi}^\dagger_{}, \hat{\psi}_{} \right],
\end{align} 
where
\begin{align}
	\var{S_1} \left[ \hat{\psi}^\dagger_{}, \hat{\psi}_{} \right] &= -\frac{i}{2} \sum_{n} \int \dd{t} \int \dd[d]{r} \left( \hat{\psi}^\dagger_{n}  \dot{\hat{\psi}}_{n} \dot{\phi} + \hat{\psi}^\dagger_{n}  \ddot{\hat{\psi}}_{n} \phi + \dot{\hat{\psi}}^\dagger_{n}  \dot{\hat{\psi}}_{n} \phi - \ddot{\hat{\psi}}^\dagger_{n} \hat{\psi}_{n} \phi - \dot{\hat{\psi}}^\dagger_{n} \hat{\psi}_{n} \dot{\phi} - \dot{\hat{\psi}}^\dagger_{n}  \dot{\hat{\psi}}_{n} \phi \right), \label{app:eqn:1st_order_var_energy_1} \\ 
	\var{S_2} \left[ \hat{\psi}^\dagger_{}, \hat{\psi}_{} \right] &= \frac{1}{2} \sum_{n,m} \int \dd{t} \int \dd[d]{r} \left[ \phi \dot{\hat{\psi}}^\dagger_{n} T_{nm} \left( -i \nabla \right) \hat{\psi}_{m} + \hat{\psi}^\dagger_{n} T_{nm} \left( -i \nabla \right) \left( \dot{\hat{\psi}}_{m}  \phi  \right) + \text{h.c.} \right], \label{app:eqn:1st_order_var_energy_2} \\
	\var{S_3} \left[ \hat{\psi}^\dagger_{}, \hat{\psi}_{} \right] &= \sum_{n,m} \int \dd{t} \int \dd[d]{r} U_{nm} \phi \left( \dot{\hat{\psi}}^\dagger_{n} \hat{\psi}_{m} + \hat{\psi}^\dagger_{n} \dot{\hat{\psi}}_{m} \right), \label{app:eqn:1st_order_var_energy_3} \\
	\var{S_4} \left[ \hat{\psi}^\dagger_{}, \hat{\psi}_{} \right] &= \int \dd{t} \int \dd[d]{r_1} \dd[d]{r_2} V \left( \vec{r}_2 - \vec{r}_1 \right) \phi \left( \vec{r}_1 \right) \dot{\rho} \left( \vec{r}_{1}, t \right) \rho \left( \vec{r}_{2}, t \right). \label{app:eqn:1st_order_var_energy_4}
\end{align}
In \cref{app:eqn:1st_order_var_energy_1,app:eqn:1st_order_var_energy_2,app:eqn:1st_order_var_energy_3}, we have suppressed the space-time dependence of the time-translation $\phi \left( \vec{r}, t \right)$, the lattice potential $U_{nm} \left( \vec{r} \right)$, and of the field operators (in contrast, we have not done so in \cref{app:eqn:1st_order_var_energy_4}, where there exist two spatial integration variables, $\vec{r}_1$ and $\vec{r}_2$). Moreover, we have written \cref{app:eqn:1st_order_var_energy_4} in terms of the total fermion density operators defined in \cref{app:eqn:cont_part_density}. We will now simplify the expression for each of $\var{S_{1-4}} \left[ \hat{\psi}^\dagger_{}, \hat{\psi}_{} \right]$ such as to write the variation of the action in terms of space and time derivatives of the time-translation $\phi \left( \vec{r}, t \right)$. In doing so, the functional derivative of the action with respect to the time-translation $\phi \left( \vec{r}, t \right)$ will result in the continuity equation from \cref{app:eqn:noether_energy_current}, from which the energy current and energy density can be determined. Finally, we note that \cref{app:eqn:1st_order_var_energy_4} would be different for generic four-fermion interaction. However, the expression is valid for the Bistritzer-MacDonald models employed in this work and introduced in \cref{app:sec:BM_review}.

We start by noting that the Lagrangian in $\var{S_1} \left[ \hat{\psi}^\dagger_{}, \hat{\psi}_{} \right]$ is a total time derivative which implies that it vanishes, {\it i.e.}{},
\begin{equation}
	\label{app:eqn:1st_order_var_energy_1_simple_1}
	\var{S_1} \left[ \hat{\psi}^\dagger_{}, \hat{\psi}_{} \right] = -\frac{i}{2} \sum_{n} \int \dd{t} \int \dd[d]{r} \pdv{t} \left( \hat{\psi}^\dagger_{n}  \dot{\hat{\psi}}_{n} \phi - \dot{\hat{\psi}}^\dagger_{n} \hat{\psi}_{n} \phi \right) = 0.
\end{equation}
In simplifying the expression of $\var{S_2} \left[ \hat{\psi}^\dagger_{}, \hat{\psi}_{} \right]$, we will again assume that the kinetic energy matrix contains a fixed number of derivative operators (and generalize at the end), such that \cref{app:eqn:kinetic_term_one_derivative} holds, and use the shorthand notation from \cref{app:eqn:currents_deriv_shorthand}. Integrating by parts with respect to the time variable, we find that 
\begin{align}
	\var{S_2} \left[ \hat{\psi}^\dagger_{}, \hat{\psi}_{} \right] &= - \frac{1}{2} \sum_{n,m} \int \dd{t} \int \dd[d]{r} \left( \dot{\phi} \hat{\psi}^\dagger_{n} T_{nm} \left( -i \nabla \right) \hat{\psi}_{m} + \text{h.c.} \right) \nonumber \\
	&- \frac{1}{2} \sum_{n,m} \int \dd{t} \int \dd[d]{r} \left[ \phi \hat{\psi}^\dagger_{n} T_{nm} \left( -i \nabla \right) \dot{\hat{\psi}}_{m} - \hat{\psi}^\dagger_{n} T_{nm} \left( -i \nabla \right) \left( \dot{\hat{\psi}}_{m}  \phi  \right) + \text{h.c.} \right]. \label{app:eqn:1st_order_var_energy_2_simple_1}
\end{align}
Notice that the second row of \cref{app:eqn:1st_order_var_energy_2_simple_1} is identical to the first row of \cref{app:eqn:second_row_simplification}, with the exception of a time derivative appearing on the $\hat{\psi}_{m} \left( \vec{r}, t \right)$ field, an overall factor of $-i$, and an integral over time. As such, one can aply a similar type of manipulation as in \cref{app:eqn:second_row_simplification} to obtain  
\begin{align}
	\var{S_2} \left[ \hat{\psi}^\dagger_{}, \hat{\psi}_{} \right] &= - \frac{1}{2} \sum_{n,m} \int \dd{t} \int \dd[d]{r} \left( \dot{\phi} \hat{\psi}^\dagger_{n} T_{nm} \left( -i \nabla \right) \hat{\psi}_{m} + \text{h.c.} \right) \nonumber \\
	& -\frac{1}{2} \sum_{n,m} \int \dd{t} \int \dd[d]{r} \phi t^{\mu_1 \mu_2 \dots \mu_N}_{nm} \partial^{\mu_1}\left[ \sum_{j=1}^{N} (-1)^{j+1} \left( \nabla^{2 \dots j} \hat{\psi}^\dagger_{n} \right) \nabla^{(j+1) \dots N} \dot{\hat{\psi}}_{m} + \text{h.c.} \right]. \label{app:eqn:1st_order_var_energy_2_simple_2}
\end{align}
The expression of $\var{S_3} \left[ \hat{\psi}^\dagger_{}, \hat{\psi}_{} \right]$ can also be simplified using integration by parts to give
\begin{equation}
	\label{app:eqn:1st_order_var_energy_3_simple_1}
	\var{S_3} \left[ \hat{\psi}^\dagger_{}, \hat{\psi}_{} \right] = -\sum_{n,m} \int \dd{t} \int \dd[d]{r} U_{nm} \dot{\phi} \hat{\psi}^\dagger_{n} \hat{\psi}_{m}
\end{equation}

We now turn to the fourth term in the variation of the action, $\var{S_4} \left[ \hat{\psi}^\dagger_{}, \hat{\psi}_{} \right]$, which requires a more careful treatment. Before moving forward, it is worth noting that unlike the case of the particle current from \cref{app:sec:cur_deriv_cont:particle}, the local time translation does produce a variation of the interaction term ({\it i.e.}{}, $\delta S_4 \left[ \hat{\psi}^\dagger_{}, \hat{\psi}_{} \right]$). This is a direct consequence of the long-range nature of the interaction potential $V \left(\vec{r}\right)$.

We start by making the change of variables $\vec{r}_1 \to \vec{r}$ and $\vec{r}_2 \to \vec{r} + \Delta \vec{r}$, followed by a Taylor series expansion of the density operator
\begin{align}
	\var{S_4} \left[ \hat{\psi}^\dagger_{}, \hat{\psi}_{} \right] &= \int \dd{t} \int \dd[d]{r} \dd[d]{\Delta r} V \left( \Delta \vec{r} \right) \phi \left( \vec{r}, t \right) \dot{\rho} \left( \vec{r}, t \right)  \rho \left( \vec{r} + \Delta \vec{r}, t \right) \nonumber \\ 
	&= \int \dd{t} \int \dd[d]{r} \dd[d]{\Delta r} V \left( \Delta \vec{r} \right) \phi \left( \vec{r}, t \right) \dot{\rho} \left( \vec{r}, t \right) \sum_{n=0}^{\infty} \frac{\Delta r^{\mu_1}\dots \Delta r^{\mu_n}}{n!} \nabla^{1 \dots n}\rho \left( \vec{r}, t \right) \nonumber \\
	&= \sum_{n=0}^{\infty} \int \dd{t} \left( \int \dd[d]{\Delta r} V \left( \Delta \vec{r} \right) \frac{\Delta r^{\mu_1}\dots \Delta r^{\mu_n}}{n!} \right) \int \dd[d]{r} \phi \left( \vec{r}, t \right) \dot{\rho} \left( \vec{r}, t \right) \nabla^{1 \dots n}\rho \left( \vec{r}, t \right) \label{app:eqn:1st_order_var_energy_4_2}
\end{align}
where we have again made use of the shorthand notation defined in \cref{app:eqn:currents_deriv_shorthand}. To simplify \cref{app:eqn:1st_order_var_energy_4_2}, we now introduce the moments of the interaction potential
\begin{equation}
	\label{app:eqn:moments_interaction_potential}
	\mathcal{K}_n^{\mu_1 \dots \mu_n} = \int \dd[d]{\Delta r} V \left( \Delta \vec{r} \right) \frac{\Delta r^{\mu_1}\dots \Delta r^{\mu_n}}{n!}.
\end{equation}
In general, the integral in \cref{app:eqn:moments_interaction_potential} is not guaranteed to converge for arbitrary interaction potentials, which might not decay sufficiently quickly at infinity (in fact, it does \emph{not} converge for the Coulomb potential $V \left( \vec{r} \right) \propto r^{-1}$). To deal with these divergences, we regularize the potential according to $V \left( \vec{r} \right) \to V \left( \vec{r} \right) e^{-\epsilon r}$ for an infinitesimal $\epsilon > 0$. As we will show, the final answer does not depend on the moments from \cref{app:eqn:moments_interaction_potential}, but rather on the actual interaction potential, thus justifying the arbitrariness of the regularization scheme.  Due to the evenness of the interaction potential $V \left( \vec{r} \right)$, all odd moments vanish, allowing us to write
\begin{align}
	\var{S_4} \left[ \hat{\psi}^\dagger_{}, \hat{\psi}_{} \right] &= \sum_{n=0}^{\infty} \int \dd{t} \int \dd[d]{r} \mathcal{K}_{2n}^{\mu_1 \dots \mu_{2n}} \phi  \dot{\rho} \nabla^{1 \dots (2n)}\rho \nonumber \\
	&= \frac{1}{2}\sum_{n=0}^{\infty} \int \dd{t} \int \dd[d]{r} \mathcal{K}_{2n}^{\mu_1 \dots \mu_{2n}} \left( \phi  \dot{\rho} \nabla^{1 \dots (2n)}\rho - \phi  \rho \nabla^{1 \dots (2n)} \dot{\rho} - \dot{\phi}  \rho \nabla^{1 \dots (2n)}\rho \right),
	\label{app:eqn:1st_order_var_energy_4_3}
\end{align}
where we have also integrated by parts with respect to the time variable to rewrite the integrand in a more symmetrical form. In analogy with \cref{app:eqn:long_int_by_parts_expression}, and using the symmetry of the interaction moments tensor from \cref{app:eqn:moments_interaction_potential}, we can rewrite the first term in the integrand from \cref{app:eqn:1st_order_var_energy_4_3} as
\begin{equation}
	\mathcal{K}_{2n}^{\mu_1 \dots \mu_{2n}} \dot{\rho} \nabla^{1 \dots (2n)}\rho = \mathcal{K}_{2n}^{\mu_1 \dots \mu_{2n}}  \left\lbrace   \partial^{\mu_1}\left[ \sum_{j=1}^{2n} (-1)^{j+1} \left( \nabla^{2 \dots j} \dot{\rho} \right) \nabla^{(j+1) \dots (2n)} \rho \right] + (-1)^{2n} \left( \nabla^{1 \dots (2n)} \dot{\rho} \right) \rho \right\rbrace. \label{app:eqn:long_int_by_parts_expression_rho}
\end{equation}
Substituting \cref{app:eqn:long_int_by_parts_expression_rho} into \cref{app:eqn:1st_order_var_energy_4_3} and performing an integral by parts in the space variable, we obtain the final form of $\delta S_4$ which only depends on space and time derivatives of the time-translation $\phi \left(\vec{r} \right)$
\begin{align}
	\var{S_4} \left[ \hat{\psi}^\dagger_{}, \hat{\psi}_{} \right] &= -\frac{1}{2}\sum_{n=0}^{\infty} \int \dd{t} \int \dd[d]{r} \mathcal{K}_{2n}^{\mu_1 \dots \mu_{2n}} \left\lbrace \left( \partial^{\mu_1} \phi \right) \left[ \sum_{j=1}^{2n} (-1)^{j+1} \left( \nabla^{2 \dots j} \dot{\rho} \right) \nabla^{(j+1) \dots (2n)} \rho \right] + \dot{\phi}  \rho \nabla^{1 \dots (2n)}\rho \right\rbrace.
	\label{app:eqn:1st_order_var_energy_4_4}
\end{align}
By combining \cref{app:eqn:noether_energy_current,app:eqn:var_s_energy,app:eqn:1st_order_var_energy_1_simple_1,app:eqn:1st_order_var_energy_2_simple_2,app:eqn:1st_order_var_energy_3_simple_1,app:eqn:1st_order_var_energy_4_4}, we can now find the energy density and current of the theory to be, respectively,
\begin{align}
	\rho_E \left( \vec{r}, t \right) &= \frac{1}{2} \sum_{n,m} \left[ \hat{\psi}^\dagger_{n} \left( \vec{r}, t \right) \left( T_{nm} \left( -i \nabla \right) + U_{nm} \left( \vec{r}, t \right) \right) \hat{\psi}_{m} \left( \vec{r}, t \right) + \text{h.c.} \right] + \frac{1}{2}\sum_{n=0}^{\infty} \mathcal{K}_{2n}^{\mu_1 \dots \mu_{2n}}  \rho \left( \vec{r}, t \right) \nabla^{1 \dots (2n)} \rho \left( \vec{r}, t \right), \label{app:eqn:energy_density_1} \\
	j_E^{\mu_1} \left( \vec{r}, t \right) &= -\frac{1}{2} \sum_{n,m} \left[ t^{\mu_1 \mu_2 \dots \mu_N}_{nm} \sum_{j=1}^{N} (-1)^{j+1} \left( \nabla^{2 \dots j} \hat{\psi}^\dagger_{n} \left( \vec{r}, t \right) \right) \nabla^{(j+1) \dots N} \dot{\hat{\psi}}_{m} \left( \vec{r}, t \right) + \text{h.c.} \right] \nonumber \\
	&+ \frac{1}{2} \sum_{n=0}^{\infty} \mathcal{K}_{2n}^{\mu_1 \dots \mu_{2n}} \sum_{j=1}^{2n} (-1)^{j+1} \left( \nabla^{2 \dots j} \dot{\rho} \left( \vec{r}, t \right) \right) \nabla^{(j+1) \dots (2n)} \rho \left( \vec{r}, t \right) \label{app:eqn:energy_current_1},
\end{align}
where we remind the reader that the expressions were derived under the assumption of \cref{app:eqn:kinetic_term_one_derivative} that the kinetic energy tensor $T_{nm} \left( -i \nabla \right)$ contains a fixed number $N$ of spatial derivatives. Similarly to \cref{app:eqn:continuity_particle}, the energy density and current operators also obey the continuity equation 
\begin{equation}
	\label{app:eqn:continuity_energy}
	\pdv{\rho_E \left( \vec{r}, t \right) }{t} + \nabla \cdot \vec{j}_E \left( \vec{r}, t \right) = 0
\end{equation}
for the field configurations that obey the equations of motion. 

The expression of the energy density from \cref{app:eqn:energy_density_1} can be further simplified by substituting the expression of the interaction potential moments from \cref{app:eqn:moments_interaction_potential}
\begin{equation}
	\label{app:eqn:energy_density_2}
	\rho_E \left( \vec{r} \right) = \frac{1}{2} \sum_{n,m} \left[ \hat{\psi}^\dagger_{n} \left( \vec{r} \right) \left( T_{nm} \left( -i \nabla \right) + U_{nm} \left( \vec{r} \right) \right) \hat{\psi}_{m} \left( \vec{r} \right) + \text{h.c.} \right] + \frac{1}{2} \int \dd[d]{r'} V \left( \vec{r}' - \vec{r} \right) \rho \left( \vec{r} \right) \rho \left( \vec{r}' \right),
\end{equation}
implying that the total energy $Q_E$ (obtained by integrating $\rho_E \left( \vec{r} \right)$ over the entire space) is equal to the Hamiltonian of the system 
\begin{equation}
	Q_E = \mathcal{H}.\label{app:eqn:en_charge_simple_real_particular}
\end{equation}
It is worth noting that if we had derived the energy density and energy current from the Lagrangian \cref{app:eqn:generic_lagrangian}, but using \cref{app:eqn:generic_hamiltonian_non_herm} as the Hamiltonian instead of \cref{app:eqn:generic_hamiltonian}, we would have obtained the same expression for the total energy $Q_E$, but the energy density and current operators would not have been Hermitian [they would differ from their Hermitian counterparts from \cref{app:eqn:energy_density_1,app:eqn:energy_density_2} by a total divergence of an asymmetric tensor, as explained above \cref{app:eqn:var_s_energy}]. 

The total energy current $J_E^{\mu_1}$ is obtained by integrating the current $j_E^{\mu_1} \left( \vec{r} \right)$ over the entire space. Upon integrating by parts in the spatial variable, we obtain
\begin{align}
	J_E^{\mu_1} &= \int \dd[d]{r} j_E^{\mu_1} \left( \vec{r} \right) \nonumber \\
	&= -\frac{1}{2} \left[ \sum_{n,m} \int \dd[d]{r} t^{\mu_1 \mu_2 \dots \mu_N}_{nm} \sum_{j=1}^{N} (-1)^{j+1} \left( \nabla^{2 \dots j} \hat{\psi}^\dagger_{n} \left( \vec{r} \right) \right) \nabla^{(j+1) \dots N} \dot{\hat{\psi}}_{m} \left( \vec{r} \right) + \text{h.c.} \right] \nonumber \\
	&+ \frac{1}{2} \int \dd[d]{r} \sum_{n=0}^{\infty} \mathcal{K}_{2n}^{\mu_1 \dots \mu_{2n}} \sum_{j=1}^{2n} (-1)^{j+1} \left( \nabla^{2 \dots j} \dot{\rho} \left( \vec{r} \right) \right) \nabla^{(j+1) \dots (2n)} \rho \left( \vec{r} \right) \nonumber \\
	&=  -\frac{N}{2} \sum_{n,m} \left[ \int \dd[d]{r} t^{\mu_1 \mu_2 \dots \mu_N}_{nm} \hat{\psi}^\dagger_{n} \left( \vec{r} \right) \nabla^{2 \dots N} \dot{\hat{\psi}}_{m} \left( \vec{r} \right) + \text{h.c.} \right] + \frac{1}{2} \int \dd[d]{r} \sum_{n=0}^{\infty} (2n) \mathcal{K}_{2n}^{\mu_1 \dots \mu_{2n}} \dot{\rho} \left( \vec{r} \right) \nabla^{2 \dots (2n)} \rho \left( \vec{r} \right) \nonumber \\
	&=  -\frac{N}{2} \sum_{n,m} \left[ \int \dd[d]{r} t^{\mu_1 \mu_2 \dots \mu_N}_{nm} \hat{\psi}^\dagger_{n} \left( \vec{r} \right) \nabla^{2 \dots N} \dot{\hat{\psi}}_{m} \left( \vec{r} \right) + \text{h.c.} \right] + \frac{1}{2} \int \dd[d]{r_1} \int \dd[d]{r_2} \left( \vec{r}_2 - \vec{r}_1 \right)^{\mu_1} V \left( \vec{r}_2 - \vec{r}_1 \right) \dot{\rho} \left( \vec{r}_1 \right) \rho \left( \vec{r}_2 \right) \nonumber \\
	&=  -\frac{N}{2} \sum_{n,m} \left[ \int \dd[d]{r} t^{\mu_1 \mu_2 \dots \mu_N}_{nm} \hat{\psi}^\dagger_{n} \left( \vec{r} \right) \nabla^{2 \dots N} \dot{\hat{\psi}}_{m} \left( \vec{r} \right) + \text{h.c.} \right] \nonumber \\
	&+ \frac{1}{4} \int \dd[d]{r_1} \dd[d]{r_2} \left( \vec{r}_2 - \vec{r}_1 \right)^{\mu_1} V \left( \vec{r}_2 - \vec{r}_1 \right) \left( \dot{\rho} \left( \vec{r}_1 \right) \rho \left( \vec{r}_2 \right) - \dot{\rho} \left( \vec{r}_2 \right) \rho \left( \vec{r}_1 \right) \right) \nonumber \\
	&= -\frac{N}{2}  \sum_{n,m} \sum_{\vec{G}} \sum_{\vec{k}} \left[ i^{N-1} t^{\mu_1 \mu_2 \dots \mu_N}_{nm} \hat{a}^\dagger_{\vec{k},\vec{G},n} \left(\vec{k} - \vec{G}\right)^{\mu_2} \dots \left(\vec{k} - \vec{G}\right)^{\mu_N} \dot{\hat{a}}_{\vec{k},\vec{G},m} + \text{h.c.} \right] \nonumber \\
	&+ \frac{1}{4} \int \dd[d]{r_1} \dd[d]{r_2} \left( \vec{r}_2 - \vec{r}_1 \right)^{\mu_1} V \left( \vec{r}_2 - \vec{r}_1 \right) \left( \dot{\rho} \left( \vec{r}_1 \right) \rho \left( \vec{r}_2 \right) - \dot{\rho} \left( \vec{r}_2 \right) \rho \left( \vec{r}_1 \right) \right) \nonumber \\
	&= \frac{1}{2} \sum_{n,m} \sum_{\vec{G}} \sum_{\vec{k}} \left( i \pdv{T_{nm} \left(\vec{k} -\vec{G} \right)}{k^{\mu_1}} \hat{a}^\dagger_{\vec{k},\vec{G},n} \dot{\hat{a}}_{\vec{k},\vec{G},m} + \text{h.c.} \right) \nonumber \\
	&+ \frac{1}{4} \int \dd[d]{r_1} \dd[d]{r_2} \left( \vec{r}_2 - \vec{r}_1 \right)^{\mu_1} V \left( \vec{r}_2 - \vec{r}_1 \right) \left( \dot{\rho} \left( \vec{r}_1 \right) \rho \left( \vec{r}_2 \right) - \dot{\rho} \left( \vec{r}_2 \right) \rho \left( \vec{r}_1 \right) \right), \label{app:eqn:en_current_simple_real_particular}
\end{align}
where we remind the reader that all time derivatives are defined according to \cref{app:eqn:field_ops_time_deriv}.

Similarly to \cref{app:eqn:part_current_simple_real_particular}, \cref{app:eqn:en_current_simple_real_particular} was derived under the assumption that the kinetic energy matrix is given by \cref{app:eqn:kinetic_term_one_derivative}. Since the final expressions for the energy charge and current are also linear in the kinetic energy matrix, \cref{app:eqn:en_charge_simple_real_particular,app:eqn:en_current_simple_real_particular} will also hold for the general case for which the kinetic energy matrix $T_{nm} \left( - i \nabla \right)$ is a sum of derivative terms of the form in \cref{app:eqn:kinetic_term_one_derivative} having different degrees $N$. As a result the \emph{generally-valid} expressions of the energy charge and current are given by 
\begin{align}
	Q_E &= \mathcal{H}, \label{app:eqn:en_charge_simple_real_general} \\
	\vec{J}_E &= \vec{J}^{T}_{E} + 	\vec{J}^{V}_{E},
	\label{app:eqn:en_current_simple_real_general}
\end{align}
where the energy current has been separated into two contributions stemming from the kinetic ($\vec{J}^{T}_E$) and interaction ($\vec{J}^{V}_E$) energy part of the Hamiltonian, which are given respectively by
\begin{align}
	\vec{J}^{T}_{E} &= \frac{1}{2} \sum_{n,m} \sum_{\vec{G}} \sum_{\vec{k}} \left( i \pdv{T_{nm} \left(\vec{k} -\vec{G} \right)}{\vec{k}} \hat{a}^\dagger_{\vec{k},\vec{G},n} \dot{\hat{a}}_{\vec{k},\vec{G},m} \label{app:eqn:en_current_simple_real_general_sin} + \text{h.c.} \right) \\
	\vec{J}^{V}_{E} &= \frac{1}{4} \int \dd[d]{r_1} \dd[d]{r_2} \left( \vec{r}_2 - \vec{r}_1 \right) V \left( \vec{r}_2 - \vec{r}_1 \right) \left( \dot{\rho} \left( \vec{r}_1 \right) \rho \left( \vec{r}_2 \right) - \dot{\rho} \left( \vec{r}_2 \right) \rho \left( \vec{r}_1 \right) \right) \nonumber \\
	&=- \frac{i}{4} \frac{1}{N_0 \Omega_0} \sum_{\vec{G}} \sum_{\vec{k}} \pdv{V \left( \vec{k} + \vec{G} \right)}{\vec{k}} \left( \dot{\rho} \left( \vec{k}+\vec{G} \right) \rho \left( -\vec{k} - \vec{G} \right) - \dot{\rho} \left( -\vec{k} - \vec{G} \right) \rho \left( \vec{k} + \vec{G} \right) \right) \nonumber \\
	&=- \frac{i}{2} \frac{1}{N_0 \Omega_0} \sum_{\vec{G}} \sum_{\vec{k}} \pdv{V \left( \vec{k} + \vec{G} \right)}{\vec{k}} \dot{\rho} \left( \vec{k}+\vec{G} \right) \rho \left( -\vec{k} - \vec{G} \right) .\label{app:eqn:en_current_simple_real_general_interaction}
\end{align}
In \cref{app:eqn:en_current_simple_real_general_interaction}, the Fourier transformation of the local density operator $\rho \left(\vec{k} + \vec{G} \right)$ was defined in \cref{app:eqn:ft_rho_op}, while the time derivatives are defined according to \cref{app:eqn:field_ops_time_deriv}.

A number of points are worth mentioning:
\begin{itemize}
	\item The expressions in \cref{app:eqn:en_charge_simple_real_general,app:eqn:en_current_simple_real_general} agree and generalize the formulae quoted by Ref.~\cite{PAU03}.
	
	\item Note that unlike prototypical field theories featuring \emph{local} interactions (such as {\it e.g.}{} the $\phi^4$ theory), the energy current $\vec{J}_E$ features the formally quartic contribution $\vec{J}^{V}_{E}$. In this context, we consider both the fermion creation and annihilation operators $\hat{a}^\dagger_{}$ and $\hat{a}_{}$, as well as their time derivatives $\dot{\hat{a}}^\dagger_{}$ and $\dot{\hat{a}}_{}$ to be ``linear'' in the fermion operators (despite the fact that {\it e.g.}{} $\dot{\hat{a}}^\dagger_{} = i \commutator{ \mathcal{H} \left[ \hat{\psi}^\dagger_{}, \hat{\psi}_{}  \right]}{\hat{a}^\dagger_{}}= \hat{a}^\dagger_{} + \hat{a}^\dagger_{} \hat{a}^\dagger_{} \hat{a}_{}$ within an interacting theory). This is because correlation functions involving a fermion operator with a time derivative can be directly evaluated from the corresponding correlation function (containing the same number of fermion operators) without any time derivatives For example, a correlation function of the form $\left\langle \hat{a}^\dagger_{} \dot{\hat{a}}_{} \right\rangle$ can be directly evaluated as the time derivative of a two-point Green's function, as will be done in \cref{app:sec:thermoelectric_response:evaluating_the_transport_coefficients:dyn_lin_response}, rather than as a sum between a two-point function and a four-point function. 
	
	\item Thirdly, often times (as is in the case of twisted graphene materials), it is useful to measure the local particle density from charge neutrality (thus making explicit the particle-hole symmetry of the system). This is done by rewriting the Coulomb interaction in terms of the offset density matrix, such that the Hamiltonian is given by
	\begin{equation}
		\mathcal{H}' = \sum_{n,m} \int \dd[d]{r} \hat{\psi}^\dagger_{n} \left( \vec{r} \right) \left( T_{nm} \left( -i \nabla \right) + U_{nm} \left( \vec{r} \right)  \right) \hat{\psi}_{m} \left( \vec{r} \right) + \frac{1}{2} \int \dd[d]{r_1} \dd[d]{r_2} V \left( \vec{r}_2 - \vec{r}_1 \right) \delta\rho \left( \vec{r}_1 \right) \delta\rho \left( \vec{r}_2 \right),
		\label{app:eqn:generic_hamiltonian_ph_sym}
	\end{equation}
	where
	\begin{equation}
		\delta \rho \left( \vec{r} \right) = \frac{1}{2} \sum_{n} \left( \hat{\psi}^\dagger_{n} \left( \vec{r} \right) \hat{\psi}_{n} \left( \vec{r} \right) - \hat{\psi}_{n} \left( \vec{r} \right) \hat{\psi}^\dagger_{n} \left( \vec{r} \right) \right) = \rho \left( \vec{r} \right) - \frac{N_{\psi}}{2} \delta \left( \vec{r} \right),
	\end{equation}
	with $N_{\psi}$ being the number of fermion flavors.
	
	In this case, we note that $\mathcal{H}$ and $\mathcal{H}'$ are identical up to a redefinition of the chemical potential, while all the formulae defined in this \siSection{} remain valid as long as $\rho \left( \vec{r} \right)$ is replaced by $\delta \rho \left( \vec{r} \right)$ (the energy current of $\mathcal{H}$ and $\mathcal{H}'$ are in fact identical).	
\end{itemize}   

Finally, we note that, from the energy and particle charge and current, one can also \emph{define} the thermal charge and current whose expressions are given by~\cite{MAH00}
\begin{alignat}{4}
	\quad \rho_{Q} \left( \vec{r} \right) & &&= \rho_E \left( \vec{r} \right) - \mu \rho \left( \vec{r} \right), \quad &
	Q_{Q} & &&= Q_E - \mu Q, \label{app:eqn:generic_thermal_charge_density}\\
	\quad \vec{j}_{Q} \left( \vec{r} \right)& && = \vec{j}_E \left( \vec{r} \right) - \mu \vec{j} \left( \vec{r} \right), \quad & 
	\vec{J}_{Q} & &&= \vec{J}_E - \mu \vec{J}, \label{app:eqn:generic_thermal_current}	
\end{alignat}
where $\rho_{Q} \left( \vec{r} \right)$, $Q_Q$, $\vec{j}_{Q} \left( \vec{r} \right)$, and $\vec{J}_{Q}$ denote the thermal charge density, total thermal charge, thermal current density and total thermal current, respectively, whereas $\mu$ represents the chemical potential of the system. As a result of \cref{app:eqn:continuity_particle,app:eqn:continuity_energy}, the thermal density and current also obey a continuity equation 
\begin{equation}
	\label{app:eqn:continuity_thermal}
	\pdv{\rho_Q \left( \vec{r}, t \right) }{t} + \nabla \cdot \vec{j}_Q \left( \vec{r}, t \right) = 0.
\end{equation}
The thermal density and current will be useful when discussing thermal transport of electronic systems in \cref{app:sec:thermoelectric_response:thermoelectric}.

\subsection{Many-body charge-conjugation symmetry}\label{app:sec:cur_deriv_cont:ph_symmetry}

Having derived the particle, energy, and thermal currents, we now consider the effects of a many-body charge-conjugation symmetry on the total current operators. For concreteness, we assume that the system is described by the Hamiltonian in \cref{app:eqn:generic_hamiltonian_ph_sym} in which the interaction is normal ordered with respect to the charge neutrality point.

We start by defining a unitary charge-conjugation operator $\mathcal{P}$, whose action on the fermion operators is given by
\begin{equation}
	\label{app:eqn:general_ph_symmetry}
	\mathcal{P} \hat{\psi}^\dagger_{n} \left( \vec{r} \right) \mathcal{P}^{-1} = \sum_{m} \left[ D \left( \mathcal{P} \right) \right]_{mn} \hat{\psi}_{m} \left( \mathcal{P}_{\text{cart}} \vec{r} \right).
\end{equation} 
In \cref{app:eqn:general_ph_symmetry}, $D \left( \mathcal{P} \right)$ is the unitary matrix specifying the action of $\mathcal{P}$ on the fermion operators, while $\mathcal{P}_{\text{cart}} \vec{r}$ denotes the action of the charge conjugation symmetry $\mathcal{P}$ on a Cartesian vector $\vec{r}$. For the usual particle-hole symmetry (such as the many-body particle-hole symmetry of twisted bilayer graphene without complicated relaxation effects~\cite{SON19,BER21a}), $\mathcal{P}_{\text{cart}} \vec{r} = \vec{r}$. In contrast, twisted symmetric trilayer graphene features a spatial charge-conjugation symmetry for which $\mathcal{P}_{\text{cart}} \vec{r} = C_{2x} \vec{r}$~\cite{CAL21}. In what follows, we will work in the general case and leave $\mathcal{P}_{\text{cart}} \vec{r}$ unspecified. Fourier transforming according to \cref{app:eqn:ft_psi}, we can also cast \cref{app:eqn:general_ph_symmetry} in momentum space
\begin{equation}
	\label{app:eqn:general_ph_symmetry_k_space}
	\mathcal{P} \hat{a}^\dagger_{\vec{k},\vec{G},n} \mathcal{P}^{-1} = \sum_{m} \left[ D \left( \mathcal{P} \right) \right]_{mn} \hat{a}_{-\mathcal{P}_{\text{cart}} \vec{k},-\mathcal{P}_{\text{cart}} \vec{G}, m}.
\end{equation}

Requiring that $\mathcal{P}$ is a symmetry of the Hamiltonian $\mathcal{H}'$ ({\it i.e.}{}, $\commutator{\mathcal{P}}{\mathcal{H}'} =0$), we find that 
\begin{align}
	\sum_{m_1,m_2} \left[ D \left( \mathcal{P} \right) \right]_{m_1 n_1} T_{m_1 m_2} \left( -i \nabla \right) \left[ D \left( \mathcal{P} \right) \right]^{*}_{m_2 n_2} &= - T_{n_2 n_1} \left( i \mathcal{P}_{\text{cart}} \nabla \right), \label{app:eqn:ph_T} \\
	\sum_{m_1,m_2} \left[ D \left( \mathcal{P} \right) \right]_{m_1 n_1} U_{m_1 m_2} \left(  \vec{r} \right) \left[ D \left( \mathcal{P} \right) \right]^{*}_{m_2 n_2} &=  - U_{n_2 n_1} \left( \mathcal{P}_{\text{cart}} \vec{r} \right), \label{app:eqn:ph_U} \\
	V \left( \vec{r} \right) &= V \left( \mathcal{P}_{\text{cart}} \vec{r} \right), \label{app:eqn:ph_V}
\end{align}
together with 
\begin{equation}
	\label{app:eqn:ph_traceless}
	\sum_{n} T_{nn} \left( - i \nabla \right) = 0 \qq{and}
	\int \dd[d]{r} \sum_{n} U_{nn} \left( \vec{r} \right) = 0.
\end{equation}
\Crefrange{app:eqn:ph_T}{app:eqn:ph_traceless} can also be rewritten in momentum space as 
\begin{align}
	\sum_{m_1,m_2} \left[ D \left( \mathcal{P} \right) \right]_{m_1 n_1} T_{m_1 m_2} \left( \vec{q} \right) \left[ D \left( \mathcal{P} \right) \right]^{*}_{m_2 n_2} &= - T_{n_2 n_1} \left( - \mathcal{P}_{\text{cart}} \vec{q} \right), \label{app:eqn:ph_T_k} \\
	\sum_{m_1,m_2} \left[ D \left( \mathcal{P} \right) \right]_{m_1 n_1} U_{m_1 m_2} \left( \vec{G} \right) \left[ D \left( \mathcal{P} \right) \right]^{*}_{m_2 n_2} &=  - U_{n_2 n_1} \left( \mathcal{P}_{\text{cart}} \vec{G} \right), \label{app:eqn:ph_U_k}\\
	V \left( \vec{q} \right) &= V \left( \mathcal{P}_{\text{cart}} \vec{q} \right), \label{app:eqn:ph_V_k}
\end{align}
as well as
\begin{equation}
	\label{app:eqn:ph_traceless_k}
	\sum_{n} T_{nn} \left( \vec{k} \right) = 0 \qq{and}
	\sum_{n} U_{nn} \left( \vec{G} \right) \eval_{\vec{G} = \vec{0}} = 0.
\end{equation}
Using \cref{app:eqn:ph_T_k,app:eqn:ph_U_k,app:eqn:ph_V_k}, we can find the action of the charge-conjugation operator $\mathcal{P}$ on the total particle and energy current operators introduced in \cref{app:eqn:part_current_simple_real_general,app:eqn:en_current_simple_real_general}, respectively,
\begin{align}
	\mathcal{P} \vec{J} \mathcal{P}^{-1} = - \mathcal{P}_{\text{cart}} \vec{J}, \label{app:eqn:ph_symmetry_par_cur} \\
	\mathcal{P} \vec{J}_E \mathcal{P}^{-1} = \mathcal{P}_{\text{cart}} \vec{J}_E. \label{app:eqn:ph_symmetry_en_cur}
\end{align}
We will use \cref{app:eqn:ph_symmetry_par_cur,app:eqn:ph_symmetry_en_cur} in \cref{app:sec:thermoelectric_response:evaluating_the_transport_coefficients:PH_symmetry} to discuss the consequences of charge-conjugation symmetry on the thermoelectric transport coefficients of moir\'e graphene systems.

\section{Review of the Bistritzer-MacDonald models for twisted graphene heterostructures}\label{app:sec:BM_review}

In this \siSection{}, we briefly review the Bistritzer-MacDonald (BM) models~\cite{BIS11} of twisted bilayer graphene (TBG) and twisted symmetric trilayer graphene (TSTG). Throughout this work, we will employ the notation and conventions of Refs.~\cite{BER21,SON21,BER21a,LIA21,BER21b,XIE21} for TBG and Refs.~\cite{CAL21,XIE21b} for TSTG, to which the reader is referred for more details. We start by summarizing the TBG single-particle and interaction BM Hamiltonians. We then show how the TSTG Hamiltonian expressed in the appropriate mirror-symmetric basis can be obtained by the addition of a high-velocity Dirac fermion to the TBG Hamiltonian~\cite{LI19,KHA19,CAR20,CAL21}. 

\subsection{Twisted Bilayer Graphene}\label{app:sec:BM_review:TBG}
\subsubsection{Single-particle Hamiltonian}\label{app:sec:BM_review:TBG:single_particle}

TBG consists of two graphene layers indexed by $l = \pm$ (where $l=+$ corresponds to the top layer) that are rotated by an angle $\mp \frac{\theta}{2}$ relative to a reference coordinate system. Within layer $l$, $\hat{a}^\dagger_{\vec{p},\alpha,s,l}$ creates an electron with crystalline momentum $\vec{p}$, spin $s=\uparrow,\downarrow$, in the graphene sublattice $\alpha = A,B$. We define $\vec{K}_l$ to be the $K$ point of the single-layer graphene (SLG) BZ of layer $l$. For concreteness, $\vec{K}_l$ is taken to be along the direction with an angle $-l \frac{\theta}{2}$ to the $\hat{\vec{x}}$ axis. Each graphene layer has two valleys corresponding to the $K$ and $K'$ points of the corresponding BZ, labeled by $\eta = \pm$ and located at $\eta \vec{K}_{\pm}$. We also define three auxiliary vectors 
\begin{equation}
	\vec{q}_{1}=\left(\vec{K}_{+}-\vec{K}_{-}\right)=k_\theta \left( 0,-1 \right)^T, \qquad 
	\vec{q}_{2}=C_{3z}\vec{q}_{1}=k_\theta \left(\frac{\sqrt{3}}{2},\frac{1}{2} \right)^T,\qquad
	\vec{q}_{3}=C_{3z}^2\vec{q}_{1}=k_\theta \left(-\frac{\sqrt{3}}{2},\frac{1}{2} \right)^T,
\end{equation}
where $k_{\theta} = \abs{\vec{K}_+ - \vec{K}_-} = 2 \abs{\vec{K}_+}\sin \frac{\theta}{2}$. For small twist angles an approximate translation symmetry emerges, with the corresponding reciprocal moir\'e lattice being given by $\mathcal{Q}_0 = \mathbb{Z} \vec{b}_{M1} + \mathbb{Z} \vec{b}_{M2}$, where
\begin{equation}
	\vec{b}_{M1}=\vec{q}_3-\vec{q}_1\ ,\qquad  \vec{b}_{M2}=\vec{q}_3-\vec{q}_2,
\end{equation}
are the reciprocal lattice vectors. In real space, the moir\'e lattice vectors $\vec{a}_{Mi}$ (for $i=1,2$) are defined such that they obey $\vec{a}_{Mi} \cdot \vec{b}_{Mj} = 2 \pi \delta_{ij}$. For concreteness, $\vec{a}_{M1} = \frac{2\pi}{3 k_\theta} \left( \sqrt{3}, 1 \right)$ and $\vec{a}_{M2} = \frac{2\pi}{3 k_\theta} \left( - \sqrt{3}, 1 \right)$. Finally, we also define two auxiliary lattices $\mathcal{Q}_{\pm} = \mathcal{Q} \pm \vec{q}_1$, which together form a honeycomb lattice. 

The low-energy fermions governing the physics of TBG within the BM model are given by 
\begin{equation}
	\label{app:eqn:BM_low_energy_ops_TBG}
	\hat{c}^\dagger_{\vec{k},\vec{Q},\alpha,\eta,s} \equiv \hat{a}^\dagger_{\eta \vec{K}_l + \vec{k} - \vec{Q},\alpha,s,l}, \qq{for} \vec{Q} \in \mathcal{Q}_{\eta l },
\end{equation}
where $\hat{a}^\dagger_{\vec{p},\alpha,s,l}$ are the SLG fermions defined at the beginning of this section, $\vec{k}$ is in the moir\'e BZ (with $\vec{k}=\vec{0}$ corresponding to the $\Gamma_{M}$ point, {\it i.e.}{} the $\Gamma$ point of the moir\'e BZ). In this low-energy basis, the single-particle TBG Hamiltonian reads as~\cite{BER21,SON21,BER21a} 
\begin{equation}
	\label{app:eqn:spHamiltonian_BM_TBG}
	\hat{H}^{\text{TBG}}_{0} = \sum_{\vec{k}} \sum_{\eta, \alpha, \beta, s} \sum_{\vec{Q},\vec{Q}' \in \mathcal{Q}_{\pm}} \left[h^{\text{TBG},\eta}_{\vec{Q},\vec{Q}'} \left( \vec{k} \right) \right]_{\alpha \beta} \hat{c}^\dagger_{\vec{k},\vec{Q},\eta,\alpha,s} \hat{c}_{\vec{k},\vec{Q}',\eta,\beta,s},
\end{equation}
where the first-quantized TBG Hamiltonian in valley $\eta$, 
\begin{equation}
	\label{app:eqn:BM_TBG_ham}
	h^{\text{TBG},\eta}_{\vec{Q},\vec{Q}'} \left( \vec{k} \right) = h^{\text{D},\eta}_{\vec{Q}} \left( \vec{k} \right) \delta_{\vec{Q},\vec{Q}'} + h^{\text{I},\eta}_{\vec{Q},\vec{Q}'},
\end{equation}
comprises two contributions, the intra-layer Dirac cone Hamiltonian and the inter-layer tunneling matrix, whose expressions in the $\eta = +$ valley are given, respectively, by
\begin{align}
	h^{\text{D},+}_{\vec{Q}} \left( \vec{k} \right) &= v_F \left( \vec{k} - \vec{Q} \right) \cdot \boldsymbol{\sigma}, \label{app:eqn:BM_dirac_ham}\\
	h^{\text{I},+}_{\vec{Q},\vec{Q}'} &= \sum_{j=1}^{3} T_{j} \delta_{\vec{Q},\vec{Q}' \pm \vec{q}_j} \label{app:eqn:BM_interlayer_ham}.
\end{align}
In \cref{app:eqn:BM_dirac_ham,app:eqn:BM_interlayer_ham}, $v_F$ is the SLG Fermi velocity, $\boldsymbol{\sigma} = \left( \sigma_x,\sigma_y \right)$ is the two-dimensional Pauli vector, and $T_j$ represent the interlayer tunneling matrices defined according to 
\begin{equation}
	\label{app:eqn:BM_interlayerT}
	T_j = w_0 \sigma_0 + w_1 \left[ \sigma_x \cos \frac{2 \pi \left( j-1 \right)}{3} + \sigma_y \sin \frac{2 \pi \left( j-1 \right)}{3} \right], \quad \text{for} \quad j=1,2,3.
\end{equation}
The tunneling matrices $T_j$ ($1 \leq j \leq 3$) depend on two parameters, $w_0 \geq 0$ and $w_1 \geq 0$, which are the interlayer hoppings at the AA and AB/BA stacking centers of the two graphene sheets, respectively. Generically, in realistic systems $w_0 < w_1$ due to lattice relaxation and corrugation effects~\cite{UCH14,WIJ15,DAI16,JAI16,SON21}. Unless mentioned otherwise, throughout this work we will consider $\theta = 1.05 \degree$, $w_1 = \SI{110}{\milli\electronvolt}$, $v_F = \SI{5.944}{\electronvolt \angstrom}$, $\abs{\vec{K}_+} = \SI{1.703}{\angstrom^{-1}}$, and $w_0/w_1 = 0.8$. In valley $\eta = -$, the single-particle TBG Hamiltonian is obtained from time-reversal symmetry
\begin{equation}
	h^{\text{TBG},-}_{\vec{Q},\vec{Q}'} \left( \vec{k} \right) = \sigma_x h^{\text{TBG},+}_{-\vec{Q},-\vec{Q}'} \left( -\vec{k} \right) \sigma_x .
\end{equation} 

For later use, we also introduce the real-space fermion basis of the TBG BM model~\cite{SON22}
\begin{align}
	\hat{c}^\dagger_{l,\alpha,\eta,s} \left( \vec{r} \right) &= \frac{1}{\sqrt{N_0 \Omega_0}} \sum_{\vec{k}} \sum_{\vec{Q} \in \mathcal{Q}_{l\eta}} \hat{c}^\dagger_{\vec{k},\vec{Q},\alpha,\eta,s} e^{ - i \left( \vec{k} - \vec{Q} \right) \cdot \vec{r}} \\
	\hat{c}^\dagger_{\vec{k},\vec{Q},\alpha,\eta,s} &= \frac{1}{\sqrt{N_0 \Omega_0}} \int \dd[2]{r} \hat{c}^\dagger_{l,\alpha,\eta,s} \left( \vec{r} \right) e^{i \left( \vec{k} - \vec{Q} \right) \cdot \vec{r}}, \qq{for} \vec{Q} \in \mathcal{Q}_{\eta l}.
\end{align} 

\subsubsection{Many-body Hamiltonian}\label{app:sec:BM_review:TBG:many_body}
We now turn to the interaction Hamiltonian in TBG, which was derived and discussed at length in Refs.~\cite{BER21a,LIA21}. Similarly to \cref{app:sec:cur_deriv_cont:model}, we let $V \left( \vec{r} \right)$ denote the interaction potential between two electrons within the TBG sample, and define its Fourier transform according to \cref{app:eqn:ft_V}. Throughout this work, we will consider a double-gated experimental setup, for which the interaction potential and its Fourier transformation are given, respectively, by 
\begin{equation}
	\label{app:eqn:double_gate_interaction}
	V \left( \vec{r} \right) = U_{\xi} \sum_{n=-\infty}^{\infty} \frac{(-1)^n}{\sqrt{\left( \abs{\vec{r}}/\xi \right)^2 + n^2}} \qq{and}
	V \left( \vec{q} \right) = \left( \pi U_{\xi} \xi^2 \right) \frac{\tanh \left( \abs{\vec{q}} \xi/2 \right)}{\abs{\vec{q}}\xi},
\end{equation}
with $U_{\xi} = \frac{e^2}{4 \pi \epsilon_0 \epsilon \xi}$ denoting the interaction energy scale (in SI units), where $\epsilon$ is the dielectric constant, while $\xi$ is the distance between the two screening gates. Throughout this work, we consider $\epsilon = 6$, $\xi = \SI{10}{\nano\meter}$, and $U_{\xi} = \SI{24}{\milli\electronvolt}$. 

The interaction Hamiltonian of TBG can be written as~\cite{KAN19,BUL20a,BER21a} 
\begin{equation}
	\label{app:eqn:BM_interaction_ham_TBG}
	\hat{H}^{\text{TBG}}_I = \frac{1}{2} \frac{1}{N_0 \Omega_0} \sum_{\vec{q}} \sum_{\vec{G} \in \mathcal{Q}_0} V \left( \vec{q} + \vec{G} \right) \delta \rho \left( - \vec{q} - \vec{G} \right) \delta \rho \left( \vec{q} + \vec{G} \right),
\end{equation}
where the density operator $\delta \rho \left( \vec{q} + \vec{G} \right)$ has been shifted so as to vanish for symmetric states at charge neutrality, {\it i.e.}{},
\begin{equation}
	\label{app:eqn:BM_offset_density_op}
	\delta \rho \left( \vec{q} + \vec{G} \right) = \rho \left( \vec{q} + \vec{G} \right) - \frac{1}{2} \delta_{\vec{q},\vec{0}} \delta_{\vec{G},\vec{0}} = \sum_{\eta,\alpha,s} \sum_{\vec{k}} \sum_{\vec{Q} \in \mathcal{Q}_\pm} \left( \hat{c}^\dagger_{\vec{k} + \vec{q}, \vec{Q} - \vec{G}, \alpha, \eta, s} \hat{c}_{\vec{k}, \vec{Q}, \alpha, \eta, s} - \frac{1}{2} \delta_{\vec{q},\vec{0}} \delta_{\vec{G},\vec{0}} \right).
\end{equation}
Finally, the full many-body TBG Hamiltonian can be obtained by summing together the single-particle and interaction contributions $\hat{H}^{\text{TBG}} = \hat{H}^{\text{TBG}}_0 + \hat{H}^{\text{TBG}}_I$.

\subsection{Twisted Symmetric Trilayer Graphene}\label{app:sec:BM_review:TSTG}

\subsubsection{Single-particle Hamiltonian}\label{app:sec:BM_review:TSTG:single_particle}

The main focus of this work concerns TBG. Nevertheless, many of the techniques developed here can be readily applied to TSTG, as the former can be thought of as consisting a ``TBG-like'' contribution and a high-velocity Dirac fermion~\cite{LI19,KHA19,CAR20,CAL21}. To be more specific, TSTG consists of three graphene layers indexed by $l=1,2,3$ (corresponding to the bottom, middle, and top layers, respectively). The top and bottom layers ($l=1,3$) are located directly above one another and are rotated by an angle $-\theta/2$ such the K point of the SLG BZ corresponding to these two layers is given by $\vec{K}_+$. The middle layer is rotated by an angle $\theta/2$ such that the K point of its SLG BZ is given by $\vec{K}_-$. Similarly to \cref{app:sec:BM_review:TBG}, we let $\hat{a}^\dagger_{\vec{p},\alpha,s,l}$ creates an electron with crystalline momentum $\vec{p}$, spin $s=\uparrow,\downarrow$, in the graphene sublattice $\alpha = A,B$, and within layer $l$.

Keeping the same moir\'e lattice notation as in \cref{app:sec:BM_review:TBG}, we define the low-energy fermion operators as 
\begin{equation}
	\hat{a}^\dagger_{\vec{k},\vec{Q},\eta,\alpha,s,l} = 
	\begin{cases}
		\hat{a}^\dagger_{\eta \vec{K}_+ + \vec{k} - \vec{Q}, \alpha, s, l}, & \quad \text{for } \vec{Q} \in \mathcal{Q}_{\eta} \text{ and } l=1,3 \\
		\hat{a}^\dagger_{\eta \vec{K}_- + \vec{k} - \vec{Q}, \alpha, s, l}, & \quad \text{for } \vec{Q} \in \mathcal{Q}_{-\eta} \text{ and } l=2 \\
	\end{cases},
\end{equation}

Due to the mirror symmetry of the system with respect to reflections in the plane of the $l=2$ SLG (in the absence of any perpendicular displacement field), we can recombine the fermions from the top and bottom layer into symmetric and antisymmetric combinations, and construct a mirror-symmetric basis
\begin{align}
	\hat{c}^\dagger_{\vec{k},\vec{Q},\eta,\alpha,s} &\equiv \begin{cases}
		\frac{1}{\sqrt{2}} \left(
			\hat{a}^\dagger_{\vec{k},\vec{Q},\eta,\alpha,s,3} + 	\hat{a}^\dagger_{\vec{k},\vec{Q},\eta,\alpha,s,1}
		\right), & \vec{Q} \in \mathcal{Q}_{\eta} \\ 
		\hat{a}^\dagger_{\vec{k},\vec{Q},\eta,\alpha,s,2}, & \vec{Q} \in \mathcal{Q}_{-\eta}
	\end{cases}, \label{app:eqn:BM_low_energy_ops_TSTG_TBGlike}\\
	\hat{b}^\dagger_{\vec{k},\vec{Q},\eta,\alpha,s} &\equiv	\frac{1}{\sqrt{2}} \left(
		\hat{a}^\dagger_{\vec{k},\vec{Q},\eta,\alpha,s,3} - 	\hat{a}^\dagger_{\vec{k},\vec{Q},\eta,\alpha,s,1}
	\right),  
	\vec{Q} \in \mathcal{Q}_{\eta}.  \label{app:eqn:BM_low_energy_ops_TSTG_Dirac}
\end{align}
Note that the mirror-even operators $\hat{c}^\dagger_{\vec{k},\vec{Q},\eta,\alpha,s}$ are entirely equivalent to the TBG low-energy fermions introduced in \cref{app:sec:BM_review:TBG}.

Within the mirror-symmetric basis, the single-particle Hamiltonian of TSTG splits into three terms 
\begin{equation}
	\hat{H}^{\text{TSTG}}_{0} = \hat{H}^{\text{TBG}}_{0} + \hat{H}^{\text{D}}_{0}  + \hat{H}^{\mathcal{E}}_{0},
\end{equation}
which we will now discuss individually. First, the mirror-even fermions give rise to a ``TBG-like'' contribution 
\begin{equation}
	\hat{H}^{\text{TBG}}_{0} = \sum_{\vec{k}} \sum_{\eta, \alpha, \beta, s} \sum_{\vec{Q},\vec{Q}' \in \mathcal{Q}_{\pm}} \left[h^{\text{TBG},\eta}_{\vec{Q},\vec{Q}'} \left( \vec{k} \right) \right]_{\alpha \beta} \hat{c}^\dagger_{\vec{k},\vec{Q},\eta,\alpha,s} \hat{c}_{\vec{k},\vec{Q}',\eta,\beta,s},
\end{equation}
where the first-quantized TBG Hamiltonian $h^{\text{TBG},\eta}_{\vec{Q},\vec{Q}'} \left( \vec{k} \right)$ is identical to the one in \cref{app:eqn:BM_TBG_ham}, up to a $\sqrt{2}$ prefactor of the inter-layer coupling
\begin{equation}
	\label{app:eqn:BM_TSTG_TBG_ham}
	h^{\text{TBG},\eta}_{\vec{Q},\vec{Q}'} \left( \vec{k} \right) = h^{\text{D},\eta}_{\vec{Q}} \left( \vec{k} \right) \delta_{\vec{Q},\vec{Q}'} + \sqrt{2} h^{\text{I},\eta}_{\vec{Q},\vec{Q}'}.
\end{equation}
The two terms in \cref{app:eqn:BM_TSTG_TBG_ham} have been defined in \cref{app:eqn:BM_dirac_ham,app:eqn:BM_interlayer_ham}. For TSTG, we employ the same single-particle parameters as in \cref{app:sec:BM_review:TBG:single_particle}, with the exception of the twist angle, for which we choose $\theta = 1.4703 \degree$. On the other hand, the mirror-odd operators $\hat{b}^\dagger_{\vec{k},\vec{Q},\eta,\alpha,s}$ form a highly-dispersing Dirac cone Hamiltonian, whose Fermi velocity equals that of SLG
\begin{equation}
	\hat{H}^{\text{D}}_{0} = \sum_{\vec{k}} \sum_{\eta, \alpha, \beta, s} \sum_{\vec{Q} \in \mathcal{Q}_{\eta}} \left[h^{\text{D},\eta}_{\vec{Q}} \left( \vec{k} \right) \right]_{\alpha \beta} \hat{b}^\dagger_{\vec{k},\vec{Q},\eta,\alpha,s} \hat{b}_{\vec{k},\vec{Q},\eta,\beta,s}.
\end{equation}
Finally, the mirror symmetry of TSTG is often broken experimentally by the addition of a perpendicular displacement field. In the presence of displacement field the $l$-th graphene layer of TSTG acquires a potential $\frac{\mathcal{E}}{2} \left( l - 2 \right)$, thus breaking the mirror symmetry. Written in the mirror-symmetric basis, the displacement field contribution $\hat{H}^{U}_{0}$ is given by
\begin{equation}
	\label{app:eqn:BM_TSTG_displacement_field_ham}
	\hat{H}^{\mathcal{E}}_0 = \sum_{\vec{k}} \sum_{\eta,\alpha,s} \sum_{\vec{Q} \in \mathcal{Q}_{\eta}} \frac{\mathcal{E}}{2} \left(
		\hat{b}^\dagger_{\vec{k},\vec{Q},\eta,\alpha,s} \hat{c}_{\vec{k},\vec{Q},\eta,\alpha,s} +
		\hat{c}^\dagger_{\vec{k},\vec{Q},\eta,\alpha,s} \hat{b}_{\vec{k},\vec{Q},\eta,\alpha,s}
	\right).
\end{equation}
In this work, we will only consider the effects of small displacement fields $\mathcal{E} \leq \SI{25}{\milli\electronvolt}$.

\subsubsection{Many-body Hamiltonian}\label{app:sec:BM_review:TSTG:many_body}

For TSTG, we consider the same double-gated setup, with the electron-electron interaction potential being given by \cref{app:eqn:double_gate_interaction}. The TSTG interaction Hamiltonian reads as 
\begin{equation}
	\hat{H}^{\text{TSTG}}_I = \frac{1}{2} \frac{1}{N_0 \Omega_0} \sum_{\vec{q}} \sum_{\vec{G} \in \mathcal{Q}_0} V \left( \vec{q} + \vec{G} \right) \delta \rho \left( - \vec{q} - \vec{G} \right) \delta \rho \left( \vec{q} + \vec{G} \right),
\end{equation}
where, the offset density operators decouple into a mirror-odd and a mirror-even contribution $\delta \rho \left( \vec{q} + \vec{G} \right) = \delta \rho^{\hat{b}} \left( \vec{q} + \vec{G} \right) + \delta \rho^{\hat{c}} \left( \vec{q} + \vec{G} \right)$, with the two contributions being given, respectively, by
\begin{align}
	\delta \hat{\rho}^{\hat{b}} \left( \vec{q} + \vec{G} \right) &= \hat{\rho}^{\hat{b}} \left( \vec{q} + \vec{G} \right) - \frac{1}{2} \delta_{\vec{q},\vec{0}} \delta_{\vec{G},\vec{0}} = \sum_{\eta,\alpha,s} \sum_{\vec{k}} \sum_{\vec{Q} \in \mathcal{Q}_\eta} \left( \hat{b}^\dagger_{\vec{k} + \vec{q}, \vec{Q} - \vec{G}, \alpha, \eta, s} \hat{b}_{\vec{k}, \vec{Q}, \alpha, \eta, s} - \frac{1}{2} \delta_{\vec{q},\vec{0}} \delta_{\vec{G},\vec{0}} \right), \label{app:eqn:BM_offset_density_op_TSTG_TBG}\\
	\delta \hat{\rho}^{\hat{c}} \left( \vec{q} + \vec{G} \right) &=  \hat{\rho}^{\hat{c}} \left( \vec{q} + \vec{G} \right) - \frac{1}{2} \delta_{\vec{q},\vec{0}} \delta_{\vec{G},\vec{0}} = \sum_{\eta,\alpha,s} \sum_{\vec{k}} \sum_{\vec{Q} \in \mathcal{Q}_{\pm}} \left( \hat{c}^\dagger_{\vec{k} + \vec{q}, \vec{Q} - \vec{G}, \alpha, \eta, s} \hat{c}_{\vec{k}, \vec{Q}, \alpha, \eta, s} - \frac{1}{2} \delta_{\vec{q},\vec{0}} \delta_{\vec{G},\vec{0}} \right). \label{app:eqn:BM_offset_density_op_TSTG_D}
\end{align}
The many-body TSTG Hamiltonian is thus given by $\hat{H}^{\text{TSTG}}=\hat{H}_0^{\text{TSTG}}+\hat{H}_I^{\text{TSTG}}$. Note that whereas the fermions belonging to the two mirror symmetry sectors are only coupled at the single-particle level in the presence of a nonzero displacement field ({\it i.e.}{}, whenever the mirror symmetry is broken), the former are \emph{always} coupled by the interaction Hamiltonian~\cite{CAL21}.

\section{Review of the topological heavy fermion models for twisted graphene heterostructures}\label{app:sec:HF_review}

This \siSection{} presents a concise overview of the topological heavy fermion (THF) models for TBG~\cite{SON22} and TSTG~\cite{YU23a}. We begin by reviewing the fermionic species within both THF models, along with analytical approximations of their wave functions. An upper bound on the nonabelian Berry connection of the heavy fermions' wave functions is then derived. The former will be employed in \cref{app:sec:deriv_moire_cur} for obtaining the energy and particle currents of TBG and TSTG in the THF basis. Subsequently, we move to the single-particle THF Hamiltonians and outline how they are obtained by projecting the corresponding BM model Hamiltonians onto the THF basis. Finally, we give a brief overview of the interaction THF Hamiltonian both in terms of the THF interaction parameters, as was originally derived in Ref.~\cite{SON22}, and also in terms of form-factors which can be analytically approximated~\cite{CAL23}. The form-factor expression of the interaction Hamiltonian will be employed in \cref{app:sec:deriv_moire_cur:energy_interaction:comparison} to analyze the interaction contribution to the energy current within the THF models.

\subsection{The THF model fermions}
\label{app:sec:HF_review:fermions}
In this section, we briefly review the fermion species of the THF models for TBG and TSTG, first introduced in Refs.~\cite{SON22,YU23a}, respectively, and employed in Refs.~\cite{RAI23a,LI23a,HU23i,CHO23,CAL23,LAU23,ZHO24,HU23}. Both models feature two types of fermions corresponding to the ``heavy'' ($f$) and ``conduction'' ($c$) electrons. The $f$-fermions are localized at the moir\'e AA-sites and transform as a pair of $p_x \pm i p_y$ orbitals under the corresponding symmetry groups, whereas the $c$-fermions form semimetallic, highly dispersive, and anomalous electronic bands. The introduction of the two fermionic species resolves the stable topological obstruction of the entire BM model~\cite{SON21,SON22,SON19,BUL20a,AHN19,PO18c}. The THF model for TSTG also features additional Dirac ($d$) fermions, which correspond to the mirror-odd fermions of TSTG, as discussed in \cref{app:sec:BM_review:TSTG:single_particle}~\cite{YU23a}. We start by discussing first the $f$- and then the $c$-electron states. We then briefly explain how the THF model of TSTG extends the one for TBG by the addition of $d$-fermions. 
 
\subsubsection{The local \texorpdfstring{$f$}{f}-electron states}\label{app:sec:HF_review:fermions:f_fermions}

By construction, the local $f$-electron states transform as $p_x \pm ip_y$ orbitals within the symmetry group of TBG~\cite{SON22}. In terms of the low-energy operators of the BM model introduced in \cref{app:eqn:BM_low_energy_ops_TBG,app:eqn:BM_low_energy_ops_TSTG_TBGlike}, the $f$-fermions are given in momentum space by 
\begin{equation}
	\label{app:eqn:f_fermions_mom_def}
	\hat{f}^\dagger_{\vec{k},\alpha,\eta,s} = \sum_{\vec{Q},\beta} v^{\eta}_{\vec{Q}\beta;\alpha} \left( \vec{k} \right) \hat{c}^\dagger_{\vec{k},\vec{Q},\beta,\eta,s}.
\end{equation} 
In \cref{app:eqn:f_fermions_mom_def}, $v^{\eta}_{\vec{Q}\beta,\alpha} \left( \vec{k} \right)$ is the $f$-fermion momentum-space wave function in valley $\eta$, where $\alpha = 1,2$ denotes the orbital quantum number, with $\alpha = 1$ ($\alpha = 2$) corresponding to the orbital transforming as the $p_x+ip_y$ ($p_x-ip_y$). The $f$-fermion wave function can be obtained numerically [but also analytically, as we will see around \cref{app:eqn:wannier_states_I,app:eqn:wannier_states_II}] through a combined Wannierization and disentanglement procedure~\cite{SON22}. In real space, the $f$-fermion operators are given by 
\begin{equation}
	\label{app:eqn:f_fermions_real_def}
	\hat{f}^\dagger_{\vec{R},\alpha,\eta,s} = \frac{1}{\sqrt{N_0}}\sum_{\vec{k}} \hat{f}^\dagger_{\vec{k},\alpha,\eta,s} e^{-i \vec{k} \cdot \vec{R}} = \sum_{l,\beta} e^{-i \eta l \vec{q}_1 \cdot \vec{R}} \int \dd[2]{r} w^{\eta}_{l \beta;\alpha} \left( \vec{r} - \vec{R} \right) \hat{c}^\dagger_{l,\beta,\eta,s} \left( \vec{r} \right),
\end{equation} 
where $w^{\eta}_{l \beta;\alpha} \left( \vec{r} - \vec{R} \right)$ is the amplitude at position $\vec{r}$ within sublattice $\beta$ in the graphene layer $l$ of the $\alpha$-th Wannier orbital located in unit cell $\vec{R}$\footnote{Note that unlike \cref{app:eqn:f_fermions_mom_def}, which is valid for both TBG and TSTG, \cref{app:eqn:f_fermions_real_def} is only valid for TBG. For TSTG, the expression of the $f$-fermion operators in the terms of the SLG ones can be found by substituting the definition of the mirror-symmetric operators from \cref{app:eqn:BM_low_energy_ops_TSTG_TBGlike}, and the Fourier-transformation of the $f$-fermion wave function from \cref{app:eqn:fourier_transform_f_fermions_to_r} into \cref{app:eqn:f_fermions_mom_def}. The resulting expression is quite cumbersome and will not be repeated here.}. The real and momentum space $f$-fermion wave functions are related by the Fourier transformations
\begin{align}
	v^{\eta}_{\vec{Q}\beta;\alpha} \left( \vec{k} \right) &= \frac{1}{\sqrt{\Omega_0}} \int \dd[2]{r} w^{\eta}_{l\beta;\alpha} \left(\vec{r} \right) e^{-i\left( \vec{k} - \vec{Q} \right) \cdot \vec{r}}, \qq{for} \vec{Q} \in \mathcal{Q}_{\eta l}, \label{app:eqn:fourier_transform_f_fermions_to_k}\\
	w^{\eta}_{l\beta;\alpha} \left(\vec{r}\right) &= \frac{1}{N_0 \sqrt{\Omega_0}} \sum_{\vec{k}} \sum_{\vec{Q} \in \mathcal{Q}_{\eta l},\beta} v^{\eta}_{\vec{Q}\beta;\alpha} \left(\vec{k} \right) e^{i \left( \vec{k} - \vec{Q} \right) \cdot \vec{r}}.  \label{app:eqn:fourier_transform_f_fermions_to_r}
\end{align}
It is worth noting that although the $f$-fermion wave functions are obtained numerically, they can also be well-approximated analytically in terms of Gaussian profiles~\cite{SON22},
\begin{alignat}{4}
	& w_{l1;1}^{\eta}(\vec{r}) & =& \frac{\alpha_1}{\sqrt{2}}\frac{1}{\sqrt{\pi\lambda_1^2}}e^{i\frac{\pi}{4}l\eta - r^2/(2\lambda_1^2)}, \qquad &&
	w_{l2;1}^{\eta}(\vec{r}) &= &-l\frac{\alpha_2}{\sqrt{2}}\frac{r_x + i\eta r_y}{\lambda_2^2\sqrt{\pi}}e^{i\frac{\pi}{4}l\eta - r^2/(2\lambda_2^2)},
	\label{app:eqn:wannier_states_I} \\
	& w_{l1;2}^{\eta}(\vec{r}) & =& l\frac{\alpha_2}{\sqrt{2}}\frac{r_x - i\eta r_y}{\lambda_2^2\sqrt{\pi}}e^{-i\frac{\pi}{4}l\eta - r^2/(2\lambda_2^2)},
	\qquad && 
	w_{l2;2}^{\eta}(\vec{r}) & =& \frac{\alpha_1}{\sqrt{2}}\frac{1}{\sqrt{\pi\lambda_1^2}}e^{-i\frac{\pi}{4}l\eta - r^2/(2\lambda_1^2)},
	\label{app:eqn:wannier_states_II}
\end{alignat} 
where the amplitudes $\alpha_1$ and $\alpha_2$, as well as the spreads $\lambda_1$ and $\lambda_2$ have been obtained both numerically and analytically across a large parameter regime~\cite{SON22,CAL23}. We note that the Wannier wave functions are very well localized at the AA sites, with the spread parameters $\lambda_1$ and $\lambda_2$ being about $20\%$ of the moir\'e lattice constant near the magic angle~\cite{SON22,CAL23}.

For the upcoming discussion on the particle and energy currents of moir\'e systems in \cref{app:sec:deriv_moire_cur}, it is useful to compute the nonabelian Berry connection of the $f$-orbital wave function. Specifically, for the $f$-electron states in valley $\eta$, the nonabelian Berry connection is defined as 
\begin{equation}
	\label{app:eqn:f_nonabelian_berry}
	\mathbfcal{A}^{\eta,ff}_{\alpha \alpha'} \left( \vec{k} \right) = 
	\sum_{\vec{Q},\beta} v^{* \eta}_{\vec{Q}\beta;\alpha} \left( \vec{k} \right) \partial_{\vec{k}} v^{\eta}_{\vec{Q}\beta;\alpha'} \left( \vec{k} \right)
	=-\sum_{\vec{Q},\beta} \left( \partial_{\vec{k}} v^{* \eta}_{\vec{Q}\beta;\alpha} \left( \vec{k} \right) \right) v^{\eta}_{\vec{Q}\beta;\alpha'} \left( \vec{k} \right),
\end{equation}
and can be recast in terms of the real-space wave functions according to
\begin{align}
	\mathbfcal{A}^{\eta,ff}_{\alpha \alpha'} \left( \vec{k} \right) &= \frac{1}{\Omega_0} \int \dd[2]{r_1} \dd[2]{r_2} \sum_{l,\beta} w^{*\eta}_{l\beta;\alpha} \left( \vec{r}_1 \right) \left( -i \vec{r}_2 \right) w^{\eta}_{l\beta;\alpha'} \left( \vec{r}_2 \right) e^{i \left( \vec{k} - \eta l \vec{q}_1 \right) \cdot \left( \vec{r}_1 - \vec{r}_2 \right)} \sum_{\vec{G} \in \mathcal{Q}_0} e^{-i \vec{G} \cdot \left( \vec{r}_1 - \vec{r}_2 \right)} \nonumber \\
	&= \int \dd[2]{r_1} \dd[2]{r_2} \sum_{l,\beta} w^{*\eta}_{l\beta;\alpha} \left( \vec{r}_1 \right) \left( -i \vec{r}_2 \right) w^{\eta}_{l\beta;\alpha'} \left( \vec{r}_2 \right) e^{i \left( \vec{k} - \eta l \vec{q}_1 \right) \cdot \left( \vec{r}_1 - \vec{r}_2 \right)} \sum_{\vec{R}} \delta \left( \vec{r}_1 - \vec{r}_2 - \vec{R} \right) \nonumber \\
	&=  -i \sum_{\vec{R}} \sum_{l} e^{i \left( \vec{k} - \eta l \vec{q}_1 \right) \cdot \vec{R}} \mathbfcal{I}^{\eta}_{l; \alpha \alpha'} \left( \vec{R} \right),
	\label{app:eqn:berry_connection_in_real_space}
\end{align}
where we have introduced the following overlap integral
\begin{equation}
	\mathbfcal{I}^{\eta}_{l; \alpha \alpha'} \left( \vec{R} \right) = \sum_{\beta} \int \dd[2]{r} \vec{r}  w^{*\eta}_{l\beta;\alpha} \left( \vec{r} + \vec{R} \right) w^{\eta}_{l\beta;\alpha'} \left( \vec{r} \right),
\end{equation}
which can, in principle, be computed exactly using \cref{app:eqn:wannier_states_I,app:eqn:wannier_states_II}. As discussed above, the wave functions of the $f$-electrons are exponentially localized near the AA site with a spread that is much smaller than the moir\'e lattice constant. As a result, we expect $\mathbfcal{I}^{\eta}_{l; \alpha \alpha'} \left( \vec{R} \right) \approx \vec{0}$ for any $\vec{R} \neq \vec{0}$. For $\vec{R} = \vec{0}$, one finds by plugging in the expressions from \cref{app:eqn:wannier_states_I,app:eqn:wannier_states_II} that $\mathbfcal{I}^{\eta}_{l; \alpha \alpha'} \left( \vec{0} \right) = \vec{0}$, exactly. In fact, the same result can be deduced from the $C_{2z}T$ and $C_{3z}$ symmetries of TBG, as a result of which the $f$-electron wave function and the overlap integral obey~\cite{SON22}
\begin{alignat}{5}
	C_{2z}T &:& 
	\quad w^{\eta}_{l\beta;\alpha} \left( \vec{r} \right) &&=& w^{*\eta}_{l\bar{\beta};\bar{\alpha}} \left( -\vec{r} \right) & 
	\quad \mathbfcal{I}^{\eta}_{l; \alpha \alpha'} \left( \vec{0} \right) &&=& - \mathbfcal{I}^{*\eta}_{l; \bar{\alpha} \bar{\alpha}'} \left( \vec{0} \right), \label{app:eqn:c2zt_cond_overlap_wann}\\
	C_{3z} &:& 
	\quad 
	e^{i \frac{2\pi}{3} \left(2 \beta - 3 \right)}
	w^{\eta}_{l\beta;\alpha} \left( \vec{r} \right) &&=& 
	e^{i \frac{2\pi}{3} \left(2 \alpha - 3 \right)}
	w^{\eta}_{l\beta;\alpha} \left( C_{3z} \vec{r} \right) & 
	\quad \mathbfcal{I}^{\eta}_{l; \alpha \alpha} \left( \vec{0} \right) &&=& C_{3z} \mathbfcal{I}^{\eta}_{l; \alpha \alpha} \left( \vec{0} \right), \label{app:eqn:c3z_cond_overlap_wann}
\end{alignat}
where $\bar{\alpha} = 3 - \alpha$ and $C_{3z} \mathbfcal{I}^{\eta}_{l; \alpha \alpha} \left( \vec{0} \right)$ denotes the action of the $C_{3z}$ symmetry on the Cartesian vector $\mathbfcal{I}^{\eta}_{l; \alpha \alpha} \left( \vec{0} \right)$. Together with the Hermiticity condition, $\mathbfcal{I}^{\eta}_{l; \alpha \alpha'} \left( \vec{0} \right) = \mathbfcal{I}^{*\eta}_{l;\alpha' \alpha} \left( \vec{0} \right)$, \cref{app:eqn:c2zt_cond_overlap_wann,app:eqn:c3z_cond_overlap_wann} impose $\mathbfcal{I}^{\eta}_{l; \alpha \alpha'} \left( \vec{0} \right) = \vec{0}$. One can conclude that the Berry connection of the $f$-orbitals is approximately zero, {\it i.e.}{} $\mathbfcal{A}^{\eta,ff}_{\alpha \alpha'} \left( \vec{k} \right) \approx \vec{0}$ (as expected since the $f$-electrons form well-localized atomic orbitals). However, to see how large $\mathbfcal{A}^{\eta,ff}_{\alpha \alpha'} \left( \vec{k} \right)$ is in comparison with the other THF model parameters, we also need to perform an order of magnitude estimation, which will be done below. 

In \cref{app:sec:deriv_moire_cur:particle}, we show that, upon projecting the particle current operator in the THF fermion basis, a Berry connection contribution arises. In order to estimate this contribution, we will now obtain an order of magnitude estimation of the Berry connection. To do so, we can employ the analytical expressions of the $f$-electron wave functions from \cref{app:eqn:wannier_states_I,app:eqn:wannier_states_II} and approximate the sum over lattice vectors in \cref{app:eqn:berry_connection_in_real_space} by an integral over the entire space
\begin{equation}
	\mathbfcal{A}^{\eta,ff}_{\alpha \alpha'} \left( \vec{k} \right) \approx -\frac{i}{\Omega_0} \sum_{l} \int \dd[2]{r} e^{i \left( \vec{k} - \eta l \vec{q}_1 \right) \cdot \vec{r}} \mathbfcal{I}^{\eta}_{l; \alpha \alpha'} \left( \vec{r} \right).
\end{equation}
We then perform an order of magnitude estimation, since from \cref{app:eqn:wannier_states_I,app:eqn:wannier_states_II}, we have that
\begin{equation}
	\abs{\mathbfcal{I}^{\eta}_{l; \alpha \alpha'} \left( \vec{r} \right)} = \left[ \sum_{n=1}^{3}  \sum_{i=1}^{2} \mathcal{O}\left( \frac{r^{n}}{\lambda_i^{n-1}} \right) \right] \exp\left[- \sum_{i=1}^{2} \mathcal{O} \left( \frac{r^2}{\lambda_i^2} \right) \right], 
\end{equation}
As a result, we find that the Berry connection of the $f$-orbitals is given by
\begin{equation}
	\label{app:eqn:f_berry_connection_bound}
	\abs{\mathbfcal{A}^{\eta,ff}_{\alpha \alpha'} \left( \vec{k} \right)} = \frac{1}{\Omega_0}  \left[ \sum_{n=0}^{4} \sum_{i=1}^{2} \mathcal{O}\left( k^{n}\lambda_i^{n+3} \right) \right] \exp\left[ - \sum_{i=1}^{2} \mathcal{O} \left( k^2 \lambda_i^2 \right) \right] = \sum_{i=1}^{2} \mathcal{O} \left( \frac{\lambda_i^3}{\Omega_0} \right),
\end{equation}
which is indeed very small as both $\lambda_1$ and $\lambda_2$ are only about $20\%$ of the moir\'e lattice constant.

\subsubsection{The conduction \texorpdfstring{$c$}{c}-electron states}\label{app:sec:HF_review:fermions:c_fermions}

For TBG around the magic angle, the six bands closest to charge neutrality form the $\Gamma_1 \oplus \Gamma_2 \oplus 2 \Gamma_3$ representation at the $\Gamma_M$ point (for each spin and valley)~\cite{SON22}. The $f$-electrons contribute one $\Gamma_3$ representation per spin and per valley. In order to correctly capture the strong topology of the system, Ref.~\cite{SON22} also introduces four anomalous $c$-electrons around the $\Gamma_M$ point. The $c$-fermion operators are defined according to 
\begin{equation}
	\label{app:eqn:def:c_electrons}
	\hat{c}^\dagger_{\vec{k},a,\eta,s} = \sum_{\vec{Q},\beta} \tilde{u}^{\eta}_{\vec{Q}\beta;a} \left( \vec{k} \right) \hat{c}^\dagger_{\vec{k},\vec{Q},\beta,\eta,s}, \qq{for} 1 \leq a \leq 4,
\end{equation}
where $a$ indexes the four $c$-electron states (for each spin and valley), while $\tilde{u}^{\eta}_{\vec{Q}\beta;a} \left( \vec{k} \right)$ denotes their wave function. The latter is determined numerically, as explained in Ref.~\cite{SON22}. 

By construction, the $\hat{c}^\dagger_{\vec{k},a,\eta,s}$ fermions for $a=1,2$ ($a=3,4$) form the $\Gamma_3$ ($\Gamma_1 \oplus \Gamma_2$) representation at the $\Gamma_M$ point, thus correctly reproducing the topology and symmetry of the six TBG bands closest to charge neutrality therein~\cite{SON22}. Moreover, the $\hat{c}^\dagger_{\vec{k},a,\eta,s}$ fermions have a large kinetic energy away from the $\Gamma_M$ point, and as a result, they are only defined for momenta below a certain cutoff $\abs{\vec{k}} < \Lambda_c$. Since they are only relevant in the proximity of the $\Gamma_M$ point, we will follow Ref.~\cite{SON22} and approximate their wave function according to 
\begin{equation}
	\label{app:eqn:approx_of_c_wavf}
	\tilde{u}^{\eta}_{\vec{Q}\beta;a} \left( \vec{k} \right) \approx \tilde{u}^{\eta}_{\vec{Q}\beta;a} \left( \vec{0} \right).
\end{equation}  

Finally, we note that \textit{ad-hoc} analytical expressions for the $c$-electron wave functions have been obtained~\cite{CAL23}. Although not derived from any analytically-tractable approximations of the BM model~\cite{BER21,CAL23}, but rather by inspecting the numerical solutions of the BM single-particle Hamiltonian, these expressions match the numerical solutions very well. They also fully satisfy the symmetries of the system. The analytical approximations to the $c$-electron wave functions at $\vec{k} = \vec{0}$ are~\cite{CAL23}
{\footnotesize\begin{alignat}{4}
	&\tilde{u}_{\vec{Q} 1, 1}^{(\eta)}(\vec{0})
	&&= - \alpha_{c1} \sqrt{ \frac{2\pi \lambda_{c1}^2}{ \Omega_0 \mathcal{N}_{c1} } } e^{- i \frac{\pi}4 \zeta_\vec{Q} - \frac12 \vec{Q}^2 \lambda_{c1}^2 } ,\qquad &&
	\tilde{u}_{\vec{Q} 2, 1}^{(\eta)}(\vec{0})
	&&= \alpha_{c2} \sqrt{ \frac{\pi \lambda_{c2}^6 }{ \Omega_0 \mathcal{N}_{c1}} } (i\eta Q_x + Q_y)^2 e^{- i \frac{\pi}4 \zeta_\vec{Q} - \frac12 \vec{Q}^2 \lambda_{c2}^2 }, \label{app:eqn:anal_func_c_1} \\
	&\tilde{u}_{\vec{Q} 1, 2}^{(\eta)}(\vec{0})
	&&= \alpha_{c2} \sqrt{ \frac{\pi \lambda_{c2}^6 }{ \Omega_0 \mathcal{N}_{c2}} } (-i\eta Q_x + Q_y)^2 e^{i \frac{\pi}4 \zeta_\vec{Q} - \frac12 \vec{Q}^2 \lambda_{c2}^2 } ,\qquad &&
	\tilde{u}_{\vec{Q} 2, 2}^{(\eta)}(\vec{0})
	&&= - \alpha_{c1} \sqrt{ \frac{2\pi \lambda_{c1}^2}{ \Omega_0 \mathcal{N}_{c2}} } e^{i \frac{\pi}4 \zeta_\vec{Q} - \frac12 \vec{Q}^2 \lambda_{c1}^2 }, \label{app:eqn:anal_func_c_2} \\
	&\tilde{u}_{\vec{Q} 1, 3}^{(\eta)}(\vec{0})
	&&= \alpha_{c3} \sqrt{ \frac{2\pi \lambda_{c3}^4 }{ \Omega_0 \mathcal{N}_{c3}} } \zeta_\vec{Q} (-i \eta Q_x + Q_y) e^{ - i \frac{\pi}4 \zeta_\vec{Q} - \frac12 \vec{Q}^2 \lambda_{c3}^2 },\qquad &&
	\tilde{u}_{\vec{Q} 2, 3}^{(\eta)}(\vec{0})
	&&= \alpha_{c4} \sqrt{ \frac{\pi \lambda_{c4}^6 }{ \Omega_0 \mathcal{N}_{c3}} } (-i\eta Q_x +  Q_y)^2 e^{ - i \frac{\pi}4 \zeta_\vec{Q} - \frac12 \vec{Q}^2 \lambda_{c4}^2 }, \label{app:eqn:anal_func_c_3}\\
	&\tilde{u}_{\vec{Q} 1, 4}^{(\eta)}(\vec{0}) &&= \alpha_{c4} \sqrt{ \frac{\pi \lambda_{c4}^6 }{ \Omega_0 \mathcal{N}_{c4}} } (i\eta Q_x +  Q_y)^2 e^{ i \frac{\pi}4 \zeta_\vec{Q} - \frac12 \vec{Q}^2 \lambda_{c4}^2 },\qquad &&
	\tilde{u}_{\vec{Q} 2, 4}^{(\eta)}(\vec{0})
	&&= \alpha_{c3} \sqrt{ \frac{2\pi \lambda_{c3}^4 }{ \Omega_0 \mathcal{N}_{c4}} } \zeta_\vec{Q} (i \eta Q_x + Q_y) e^{ i \frac{\pi}4 \zeta_\vec{Q} - \frac12 \vec{Q}^2 \lambda_{c3}^2 }, \label{app:eqn:anal_func_c_4}
\end{alignat}}where we have used the sublattice factor $\zeta_\vec{Q} = + 1$ ($\zeta_\vec{Q} = - 1$) for $\vec{Q} \in \mathcal{Q}_{+}$ ($\vec{Q} \in \mathcal{Q}_{+}$). These expressions are also extended for small momenta around the $\Gamma_M$ point using \cref{app:eqn:approx_of_c_wavf}~\cite{SON22}. For the twist angle and tunneling amplitude ratio quoted below \cref{app:eqn:BM_interlayerT}, the parameters appearing in \cref{app:eqn:anal_func_c_1,app:eqn:anal_func_c_2,app:eqn:anal_func_c_3,app:eqn:anal_func_c_4} have been fitted to the numerical solution and are given by~\cite{CAL23}
\begin{alignat}{5}
	& \lambda_{c1} = 0.2194 \abs{\vec{a}_{M1}},\qquad 
	&& \lambda_{c2} = 0.3299 \abs{\vec{a}_{M1}},\qquad 
	&& \alpha_{c1} = 0.3958,\qquad 
	&& \alpha_{c2} = 0.9183, \qquad 
	&&\mathcal{N}_{c1} = \mathcal{N}_{c2} = 1.2905, \\ 
	&\lambda_{c3} = 0.2430 \abs{\vec{a}_{M1}},\qquad 
	&& \lambda_{c4} = 0.2241 \abs{\vec{a}_{M1}},\qquad 
	&& \alpha_{c3} = 0.9257,\qquad 
	&& \alpha_{c4} = 0.3783, \qquad 
	&&\mathcal{N}_{c3} = \mathcal{N}_{c4} = 1.1102.
\end{alignat}
These expressions were used in Ref.~\cite{CAL23} to analytically obtain the density form factors of TBG in the THF basis. We will employ the latter in \cref{app:sec:deriv_moire_cur:energy_interaction} to estimate the interaction contribution to the energy current of TBG.   

\subsubsection{The Dirac \texorpdfstring{$d$}{d}-electron states}\label{app:sec:HF_review:fermions:d_fermions}

As briefly reviewed in \cref{app:sec:BM_review:TSTG}, TSTG can be understood as consisting of a TBG-like contribution and a high-velocity Dirac fermion. In order to build a THF model of TSTG, Ref.~\cite{YU23a} introduces another fermion species besides the $f$- and $c$-electrons which captures the low-energy Dirac modes. These new $d$-fermions are simply the mirror-odd fermions defined in \cref{app:eqn:BM_low_energy_ops_TSTG_Dirac}, up to a redefinition in the momentum origin, namely
\begin{equation}
	\label{app:eqn:def:d_electrons}
	\hat{d}^\dagger_{\vec{p},\alpha,\eta,s} \equiv \hat{b}^\dagger_{\vec{p} + \eta \vec{q}_1,\eta \vec{q}_1,\eta,\alpha,s}.
\end{equation}
Here and in what follows, we will use $\vec{p}$ for the momentum of the $d$-fermions, as it is measured from the $\eta \vec{q}_1$ point ({\it i.e.}{}, the $\mathrm{K}_M$ and $\mathrm{K}'_M$ points), as opposed to the $\Gamma_M$ point. Similarly to the $c$-electrons defined in \cref{app:eqn:def:c_electrons}, the $d$-fermions are only relevant for small momenta near the $\mathrm{K}_M$ point, $\abs{\vec{k}} < \Lambda_d$.

\subsection{The single-particle THF model}
\label{app:sec:HF_review:single_particle}
Having reviewed the basis of the THF models of TBG and TSTG, we now turn to the corresponding single-particle Hamiltonians. The latter are obtained by simply projecting the BM model Hamiltonians reviewed in \cref{app:sec:BM_review} into the fermion basis from \cref{app:sec:HF_review:fermions}~\cite{SON22}. To be specific, the projection of the TBG fermions (as well as the mirror-even fermions in TSTG) is implemented via the approximation 
\begin{equation}
	\label{app:eqn:proj_to_THF_basis}
	\hat{c}^\dagger_{\vec{k},\vec{Q},\beta,\eta,s} \approx \begin{cases}
		 \sum_{\alpha} v^{*\eta}_{\vec{Q}\beta;\alpha} \left( \vec{k} \right) \hat{f}^\dagger_{\vec{k},\alpha,\eta,s} + \sum_{a} \tilde{u}^{*\eta}_{\vec{Q}\beta;a}\left( \vec{k} \right) \hat{c}^\dagger_{\vec{k},a,\eta,s} & \qq{for} \abs{\vec{k}} \leq \Lambda_c. \\
		 \sum_{\alpha} v^{*\eta}_{\vec{Q}\beta;\alpha} \left( \vec{k} \right) \hat{f}^\dagger_{\vec{k},\alpha,\eta,s} & \qq{for} \abs{\vec{k}} > \Lambda_c.
	\end{cases}
\end{equation}
In this section, we briefly summarize the THF model Hamiltonian for TBG and explain how the latter is extended~\cite{YU23a} in the case of TSTG by the addition of the $d$-fermions.

\subsubsection{The single-particle THF model of TBG}
\label{app:sec:HF_review:single_particle:TBG}

The single-particle THF model Hamiltonian for TBG is given by 
\begin{align}
	H^{\text{TBG}}_{0} =& \sum_{\substack{\abs{\vec{k}} \leq \Lambda_c\\ \eta, s}} \left[ \sum_{a,a'} h^{cc,\eta}_{a a'} \left( \vec{k} \right) \hat{c}^\dagger_{\vec{k},a,\eta,s} \hat{c}_{\vec{k},a,\eta,s} + \left( \sum_{a,\alpha} h^{cf,\eta}_{a \alpha} \left( \vec{k} \right) \hat{c}^\dagger_{\vec{k},a,\eta,s} \hat{f}_{\vec{k},\alpha,\eta,s}  + \text{h.c.} \right) \right] \nonumber \\
	+&  \sum_{\substack{\vec{k},\alpha,\alpha' \\ \eta,s}} h^{ff,\eta}_{\alpha \alpha'} \left( \vec{k} \right) \hat{f}^\dagger_{\vec{k},\alpha,\eta,s} \hat{f}_{\vec{k},\alpha',\eta,s} ,
	\label{app:eqn:single_part_THF_TBG}
\end{align}
where ``$+\text{h.c.}$'' denotes the addition of the Hermitian conjugate. In order to differentiate the BM and THF model Hamiltonians, the former are always denoted with a hat. The matrix blocks appearing in \cref{app:eqn:single_part_THF_TBG} are given by~\cite{SON22}
\begin{align}
	h^{cc,\eta} \left( \vec{k} \right) =& \begin{pmatrix}
		\mathbb{0} & v_{\star} \left(\eta k_x \sigma_0 + i k_y \sigma_z \right) \\ 
		v_{*} \left(\eta k_x \sigma_0 - i k_y \sigma_z \right) & M \sigma_z
	\end{pmatrix} \label{app:eqn:cc_thf_block},\\
	h^{cf,\eta} \left( \vec{k} \right) =& \begin{pmatrix}
		\gamma \sigma_0 + v_{*}' \left( \eta k_x \sigma_x + k_y \sigma_y \right) \\ 
		v_{*}'' \left( \eta k_x \sigma_x - k_y \sigma_y \right)
	\end{pmatrix} e^{-\frac{\abs{\vec{k}}^2 \lambda^2}{2}} \label{app:eqn:cf_thf_block}, \\
	h^{ff,\eta} \left( \vec{k} \right) =& \mathcal{O} \left( t \right) \approx \mathbb{0} \label{app:eqn:f_thf_block},
\end{align}
with $\mathbb{0}$ denoting the zero matrix. The single-particle parameters appearing in \cref{app:eqn:cc_thf_block,app:eqn:cf_thf_block} depend on the parameters of the BM model of TBG and have been obtained both numerically and analytically across a vast range of twist angles $\theta$ and tunneling ratios $w_0/w_1$~\cite{SON22,CAL23}. Unless mentioned otherwise, throughout this work we consider $\lambda = 0.3375 \abs{\vec{a}_{M1}}$, $\gamma = \SI{-24.75}{\milli\electronvolt}$, $M = \SI{3.697}{\milli\electronvolt}$, $v_{\star} = \SI{-4.303}{\electronvolt\angstrom}$, $v'_{\star} = \SI{1.623}{\electronvolt\angstrom}$, and $v''_{\star} = \SI{-0.0332}{\electronvolt\angstrom}$, which correspond to the BM model parameters chosen around \cref{app:eqn:BM_TBG_ham}. Near the magic angle, which will be the focus of this work, the $f$-electrons are dispersionless, which implies that their nearest neighbor hopping amplitude $t$ vanishes. As a result, we will approximate the matrix block from \cref{app:eqn:f_thf_block} to zero. 

Finally, we note that \cref{app:eqn:cc_thf_block,app:eqn:cf_thf_block,app:eqn:f_thf_block} have been obtained by projecting the BM model single-particle Hamiltonian of TBG into the THF basis according to \cref{app:eqn:proj_to_THF_basis}. As a result, the former obey~\cite{SON22}
\begin{align}
	h_{a a'}^{cc,\eta} \left( \vec{k} \right) &\approx \sum_{\beta, \beta' } \sum_{\vec{Q},\vec{Q}' \in \mathcal{Q}_{\pm}}  \left[h^{\text{TBG},\eta}_{\vec{Q},\vec{Q}'} \left( \vec{k} \right) \right]_{\beta \beta'} \tilde{u}^{*\eta}_{\vec{Q}\beta;a} \left( \vec{k} \right) \tilde{u}^{\eta}_{\vec{Q}'\beta';a'} \left( \vec{k} \right), \label{app:eqn:proj_in_sp_TBG_THF_c} \\
	h_{a \alpha'}^{cf,\eta} \left( \vec{k} \right) &\approx \sum_{\beta, \beta' } \sum_{\vec{Q},\vec{Q}' \in \mathcal{Q}_{\pm}}  \left[h^{\text{TBG},\eta}_{\vec{Q},\vec{Q}'} \left( \vec{k} \right) \right]_{\beta \beta'} \tilde{u}^{*\eta}_{\vec{Q}\beta;a} \left( \vec{k} \right) v^{\eta}_{\vec{Q}'\beta';\alpha'} \left( \vec{k} \right), \label{app:eqn:proj_in_sp_TBG_THF_cf} \\
	h^{ff,\eta}_{\alpha \alpha'} \left( \vec{k} \right) & \approx \sum_{\beta, \beta' } \sum_{\vec{Q},\vec{Q}' \in \mathcal{Q}_{\pm}}  \left[h^{\text{TBG},\eta}_{\vec{Q},\vec{Q}'} \left( \vec{k} \right) \right]_{\beta \beta'} v^{*\eta}_{\vec{Q}\beta;\alpha} \left( \vec{k} \right) v^{\eta}_{\vec{Q}'\beta';\alpha'} \left( \vec{k} \right) \approx 0. \label{app:eqn:proj_in_sp_TBG_THF_f}
\end{align} 
\subsubsection{The single-particle THF model of TSTG}
\label{app:sec:HF_review:single_particle:TSTG}
Starting from the single-particle THF model of TBG from \cref{app:eqn:single_part_THF_TBG}, Ref.~\cite{YU23a} constructed the single-particle THF model of TSTG $H_0^{\text{TSTG}}$ by the addition of the high-velocity Dirac fermion ({\it i.e.}{} the mirror-even fermions from the BM TSTG model)
\begin{equation}
	\label{app:eqn:single_part_THF_TSTG}
	H_0^{\text{TSTG}} = H_0^{\text{TBG}} + H_0^{\text{D}} + H_0^{\mathcal{E}}. 
\end{equation}
In \cref{app:eqn:single_part_THF_TSTG}, $H_0^{\text{TBG}}$ has the same form as in \cref{app:eqn:single_part_THF_TBG}, with the only change being that the $\gamma$, $v''_{*}$, and $M$ parameters now depend on the displacement field $\mathcal{E}$ according to~\cite{YU23a}
\begin{equation}
	\gamma \to \gamma + B_{\gamma} \mathcal{E}^2, \quad
	v''_{*} \to v''_{*} + B_{v''_{*}} \mathcal{E}^2,
	\quad
	M \to M + B_{M} \mathcal{E}^2,
\end{equation}
where the constants $B_{\gamma}$, $B_{v''_{*}}$, and $B_{M}$ have been determined through second-order perturbation theory~\cite{YU23a}. For TSTG, we employ slightly different zero-$\mathcal{E}$ parameters: $\lambda = 0.3359 \abs{\vec{a}_{M1}}$, $\gamma = \SI{-33.16}{\milli\electronvolt}$, $M = \SI{7.01}{\milli\electronvolt}$, $v_{\star} = \SI{-4.301}{\electronvolt\angstrom}$, $v'_{\star} = \SI{1.625}{\electronvolt\angstrom}$, and $v''_{\star} = \SI{-0.0346}{\electronvolt\angstrom}$, which correspond to the BM TSTG model parameters chosen around \cref{app:eqn:BM_TSTG_TBG_ham}. Additionally, we use $B_{\gamma} = \SI{-3.75e-4}{\milli\electronvolt^{-1}}$, $B_{v''_{*}} = \SI{7.65e3}{\electronvolt^{-1}\angstrom}$, and $B_{M} = \SI{3.28e-4}{\milli\electronvolt^{-1}}$~\cite{YU23a}.

The other two terms in \cref{app:eqn:single_part_THF_TSTG} are given by~\cite{YU23a}
\begin{equation}
	H_0^{\text{D}} = \sum_{\substack{\vec{p} \leq \Lambda_d \\ \alpha,\alpha',\eta,s}} h^{dd,\eta}_{\alpha \alpha'} \left( \vec{p} \right) \hat{d}^\dagger_{\vec{p},\alpha,\eta,s} \hat{d}_{\vec{p},\alpha',\eta,s} \qq{and}
	H_0^{\mathcal{E}} = \sum_{\substack{\vec{p} \leq \Lambda_d \\ \alpha,\eta,s}} h^{df,\eta}_{\alpha \alpha'} \left( \vec{p} \right)  \hat{d}^\dagger_{\vec{p},\alpha',\eta,s} \hat{f}_{\vec{p}+\eta \vec{q}_1,\alpha,\eta,s} + \text{h.c.},
\end{equation}
where the two matrix blocks read as
\begin{align}
	h^{dd,\eta} \left( \vec{p} \right) = & h^{\text{D},\eta}_{\eta \vec{q}_1} \left( \vec{p} + \eta \vec{q}_1 \right) = v_F \left( \eta p_x \sigma_x + p_y \sigma_y \right) \label{app:eqn:dd_thf_block},\\
	h^{df,\eta} \left( \vec{p} \right) =& M_1 \mathcal{E} \left( \sigma_0 - i \eta \sigma_z \right) e^{-\frac{\abs{\vec{p}}^2 \lambda^2}{2}} \label{app:eqn:df_thf_block}.
\end{align}
The Fermi velocity $v_F = \SI{5.944}{\electronvolt \angstrom}$ is given around \cref{app:eqn:BM_TBG_ham}, while the $d$-$f$ coupling constant is $M_1=-0.1394$.

\subsection{The interaction THF Hamiltonian}
\label{app:sec:HF_review:interaction}

In the case of TBG, the THF interaction Hamiltonian is obtained by projecting the density operators from \cref{app:eqn:BM_offset_density_op} into the THF model basis according to \cref{app:eqn:proj_to_THF_basis}. In this section, we will outline the main results and refer the readers to Refs.~\cite{SON22,CAL23} for the full details. For TSTG, the projection of the mirror-even density matrix from \cref{app:eqn:BM_offset_density_op_TSTG_TBG} proceeds analogously with TBG. The density matrix corresponding to the mirror-odd fermions from  \cref{app:eqn:BM_offset_density_op_TSTG_D} is projected by simply imposing a momentum cutoff for the latter~\cite{YU23a}. 

The projected TBG density matrix can be written in the $f$- and $c$-electron basis using the so-called \emph{form-factors}, which depend on the actual wave functions of the two fermionic species. Such an expression is however cumbersome to employ. Ref.~\cite{SON22} has shown that the THF interaction Hamiltonian can be approximated to a form that only depends on a handful of parameters (overlap integrals). Nevertheless, we will also review form-factors and their analytical expressions~\cite{CAL23}, which will be useful in estimating the interaction contribution to the energy current in TBG in \cref{app:sec:deriv_moire_cur:energy_interaction}. 

\subsubsection{Analytical form-factor expression for the TBG density matrix}
\label{app:sec:HF_review:interaction:analytical_with_ff}

In what follows, we will employ a similar convention to the one introduced in \cref{app:sec:HF_review:single_particle} for the Hamiltonians within the BM and THF models: the projected (unprojected) density operator will be denoted without (with) a hat. The BM model TBG density operator $\delta \hat{\rho}^{\hat{c}} \left( \vec{q}+\vec{G} \right)$ from \cref{app:eqn:BM_offset_density_op} can be projected into the $f$- and $c$-fermion states by using \cref{app:eqn:proj_to_THF_basis} to afford~\cite{CAL23}
\begin{align}
	\delta \rho^{\hat{c}} \left( \vec{q}+\vec{G} \right) =& \sum_{\vec{k},\eta,s,\alpha,\alpha'} \mathcal{M}_{\alpha\alpha'}^{ff,\eta} \left(\vec{k},\vec{q}+\vec{G} \right) \left( \hat{f}^\dagger_{\vec{p}_{\vec{k}+\vec{q}},\alpha,\eta,s} \hat{f}_{\vec{k},\alpha',\eta,s} - \frac{1}{2} \delta_{\vec{q},\vec{0}} \delta_{\vec{G},\vec{0}} \delta_{\alpha \alpha'} \right) \nonumber \\
	+& \sum_{\vec{k},\eta,s,a,a'} \Theta \left( \Lambda_c - \abs{\vec{k}} \right) \Theta \left( \Lambda_c - \abs{\vec{p}_{\vec{k}+\vec{q}}} \right) \mathcal{M}_{aa'}^{c,\eta} \left(\vec{k},\vec{q}+\vec{G} \right) \left( \hat{c}^\dagger_{\vec{p}_{\vec{k}+\vec{q}},a,\eta,s} \hat{c}_{\vec{k},a',\eta,s} - \frac{1}{2} \delta_{\vec{q},\vec{0}} \delta_{\vec{G},\vec{0}} \delta_{a a'} \right)  \nonumber \\
	+& \sum_{\vec{k},\eta,s,\alpha,a'} \left( \Theta \left( \Lambda_c - \abs{\vec{k}} \right) \mathcal{M}_{\alpha a'}^{fc,\eta} \left(\vec{k},\vec{q}+\vec{G} \right) \hat{f}^\dagger_{\vec{p}_{\vec{k}+\vec{q}},\alpha,\eta,s} \hat{c}_{\vec{k},a',\eta,s} + \text{h.c.} \right), \label{app:eqn:projection_of_TBG_rho_to_THF}
\end{align}
where $\Theta(x)$ denotes the Heaviside step function (but can also denote a soft truncation function~\cite{CAL23}) and $\vec{p}_{\vec{k} + \vec{q}}$ denotes the image of $\vec{k} + \vec{q}$ in the first moir\'e BZ, {\it i.e.}{}
\begin{equation}
	\vec{p}_{\vec{k} + \vec{q}} = \vec{k} + \vec{q} + \vec{G}, \qq{such that} \vec{k} + \vec{q} + \vec{G} \in \mathrm{MBZ}, 
\end{equation} 
with $\vec{G}$ being a reciprocal lattice vector. Additionally, in \cref{app:eqn:projection_of_TBG_rho_to_THF} we have introduced the following form factors, which depend on the wave functions of the $c$- and $f$-electrons
\begin{align}
	\mathcal{M}^{ff,\eta}_{\alpha \alpha'}\left(\vec{k},\vec{q}+\vec{G} \right) &= \sum_{\vec{Q}, \beta} v^{*\eta}_{\vec{Q}-\vec{G} \beta; \alpha}\left( \vec{k}+\vec{q} \right) v^{\eta}_{\vec{Q} \beta;\alpha'} \left(\vec{k} \right), \label{app:eqn:def_ffactors_ff}\\
	\mathcal{M}^{cc,\eta}_{a a'}\left(\vec{k},\vec{q}+\vec{G} \right) &= \sum_{\vec{Q}, \beta} \tilde{u}^{*\eta}_{\vec{Q}-\vec{G} \beta; a}\left( \vec{k}+\vec{q} \right) \tilde{u}^{\eta}_{\vec{Q} \beta;a'} \left(\vec{k} \right), \label{app:eqn:def_ffactors_cc}\\ 
	\mathcal{M}^{fc,\eta}_{\alpha a'}\left(\vec{k},\vec{q}+\vec{G} \right)  &= \sum_{\vec{Q},\beta} v^{*\eta}_{\vec{Q}-\vec{G} \beta;\alpha} \left( \vec{k}+\vec{q} \right) \tilde{u}^{\eta}_{\vec{Q}\beta;a'}\left(\vec{k} \right) \label{app:eqn:def_ffactors_fc}. 
\end{align}
By using the analytical expressions for the $f$- and $c$-electron wave functions from \cref{app:eqn:wannier_states_I,app:eqn:wannier_states_II} and \cref{app:eqn:anal_func_c_1,app:eqn:anal_func_c_2,app:eqn:anal_func_c_3,app:eqn:anal_func_c_4}, respectively, Ref.~\cite{CAL23} has obtained analytical expressions for the form factors: the matrices $\mathcal{M}^{ff,\eta}\left(\vec{k}, \vec{q} \right)$, $\mathcal{M}^{cc,\eta}\left(\vec{k}, \vec{q} \right)$, and $\mathcal{M}^{fc,\eta}\left(\vec{k}, \vec{q} \right)$ have been shown to be $\vec{k}$-independent and take the form of polynomials in $q_x$ and $q_y$ of degree no larger than four multiplied by a Gaussian in the momentum transfer magnitude $\abs{\vec{q}}$. For example, within this approximation, the $f$-electron form factor is given by 
\begin{equation}
	\mathcal{M}^{ff,\eta}_{\alpha\alpha'} \left( \vec{k}, \vec{q} \right) = \left[ \alpha_1^2 \exp \left( - \frac{1}{2} \abs{\vec{q}}^2 \lambda_1^2 \right) + \alpha_2^2 \left(1 - \frac{1}{4} \abs{\vec{q}}^2 \lambda_2 \right) \exp \left( -\frac{1}{4} \abs{\vec{q}}^2 \lambda_2^2 \right) \right] \delta_{\alpha \alpha'}.
\end{equation}
By virtue of the $f$-electron wave-function being exponentially-localized ({\it i.e.}{}, an atomic limit), the $f$-electron form factor does not depend on $\vec{k}$. For $\mathcal{M}^{cc,\eta}\left(\vec{k}, \vec{q} \right)$, and $\mathcal{M}^{fc,\eta}\left(\vec{k}, \vec{q} \right)$, the independence on $\vec{k}$ relies additionally on the approximation from \cref{app:eqn:approx_of_c_wavf}. 

The full analytical approximations of the form factors can be found in Appendix F of Ref.~\cite{CAL23}. We will employ them in \cref{app:sec:deriv_moire_cur:energy_interaction} to assess the magnitude of the interaction contribution to the energy current. 

\subsubsection{The interaction THF Hamiltonian for TBG}
\label{app:sec:HF_review:interaction:TBG}

The THF interaction Hamiltonian was derived by directly substituting the expression of the projected density operator from \cref{app:eqn:projection_of_TBG_rho_to_THF} into the BM interaction Hamiltonian from \cref{app:eqn:BM_interaction_ham_TBG}. Through a series of justified approximations, Ref.~\cite{SON22} has shown that the resulting expression can be simplified to a sum of just seven terms which depend on only six parameters and the Coulomb interaction potential from \cref{app:eqn:double_gate_interaction}. The THF interaction Hamiltonian is thus given by~\cite{SON22}
\begin{equation}
	\label{app:eqn:THF_interaction_TBG}
	H^{\text{TBG}}_I = H_{U_1} + H_{U_2} + H_V + H_W + H_J + H_{\tilde{J}} + H_K,
\end{equation}
where
\begin{align}
	H_{U_1} &= \frac{U_1}{2} \sum_{\vec{R}} \sum_{\substack{\alpha,\eta,s \\ \alpha',\eta',s'}} :\mathrel{\hat{f}^\dagger_{\vec{R},\alpha,\eta,s} \hat{f}_{\vec{R},\alpha,\eta,s}}: :\mathrel{\hat{f}^\dagger_{\vec{R},\alpha',\eta',s'} \hat{f}_{\vec{R},\alpha',\eta',s'}}:, \label{app:eqn:THF_int:U1} \\
H_{U_2} &= \frac{U_2}{2} \sum_{\left\langle \vec{R}, \vec{R}' \right\rangle} \sum_{\substack{\alpha,\eta,s \\ \alpha',\eta',s'}} :\mathrel{\hat{f}^\dagger_{\vec{R}',\alpha,\eta,s} \hat{f}_{\vec{R}',\alpha,\eta,s}}: :\mathrel{\hat{f}^\dagger_{\vec{R},\alpha',\eta',s'} \hat{f}_{\vec{R},\alpha',\eta',s'}}:, \label{app:eqn:THF_int:U2} \\
H_V &= \frac{1}{2 \Omega_0 N_0} \sum_{\abs{\vec{k}_1},\abs{\vec{k}_1} \leq \Lambda_c} \sum_{\substack{\vec{q} \\ \abs{\vec{k}_1+\vec{q}},\abs{\vec{k}_2+\vec{q}} \leq \Lambda_c}} \sum_{\substack{a,\eta,s \\ a',\eta',s'}} V \left( \vec{q} \right):\mathrel{\hat{c}^\dagger_{\vec{k}_1 + \vec{q} ,a,\eta,s} \hat{c}_{\vec{k}_1,a,\eta,s}}: :\mathrel{\hat{c}^\dagger_{\vec{k}_2 - \vec{q},a',\eta',s'} \hat{c}_{\vec{k}_2,a',\eta',s'}}:, \label{app:eqn:THF_int:V} \\
H_W &= \frac{1}{N_0} \sum_{\substack{\vec{k}_1 \\ \abs{\vec{k}_2} \leq \Lambda_c}} \sum_{\substack{\vec{q} \\ \abs{\vec{k}_2-\vec{q}} \leq \Lambda_c}} \sum_{\substack{\alpha,\eta,s \\ a',\eta',s'}} W_{a'} :\mathrel{\hat{f}^\dagger_{\vec{k}_1 + \vec{q},\alpha,\eta,s} \hat{f}_{\vec{k}_1,\alpha,\eta,s}}: :\mathrel{\hat{c}^\dagger_{\vec{k}_2 - \vec{q},a',\eta',s'} \hat{c}_{\vec{k}_2,a',\eta',s'}}:, \label{app:eqn:THF_int:W} \\
H_J &= -\frac{J}{2N_0} \sum_{\substack{\vec{k}_1 \\ \abs{\vec{k}_2} \leq \Lambda_c}} \sum_{\substack{\vec{q} \\ \abs{\vec{k}_2+\vec{q}} \leq \Lambda_c}} \sum_{\substack{\alpha,\eta,s \\ a',\eta',s'}} \left[ \eta \eta' + (-1)^{\alpha + \alpha'} \right] :\mathrel{\hat{f}^\dagger_{\vec{k}_1 + \vec{q},\alpha,\eta,s} \hat{f}_{\vec{k}_1,\alpha',\eta',s'}}: :\mathrel{\hat{c}^\dagger_{\vec{k}_2 - \vec{q},\alpha'+2,\eta',s'} \hat{c}_{\vec{k}_2,\alpha+2,\eta,s}}:, \label{app:eqn:THF_int:J} \\
H_{\tilde{J}} &= -\frac{J}{4N_0} \sum_{\abs{\vec{k}_1}, \abs{\vec{k}_2} \leq \Lambda_c} \sum_{\vec{q}} \sum_{\substack{\alpha,\eta,s \\ a',\eta',s'}} \left[ \eta \eta' - (-1)^{\alpha + \alpha'} \right] \hat{f}^\dagger_{\vec{k}_1 + \vec{q},\alpha,\eta,s} \hat{f}^\dagger_{\vec{k}_2 - \vec{q},\alpha',\eta',s'} \hat{c}_{\vec{k}_2,\alpha'+2,\eta',s'} \hat{c}_{\vec{k}_1,\alpha+2,\eta,s} + \text{h.c.} \label{app:eqn:THF_int:Jtilde} \\	
H_{K} &= \frac{K}{N_0 \Omega_0} \sum_{\abs{\vec{k}_1},\abs{\vec{k}_2},\abs{\vec{k}_3} \leq \Lambda_c} \sum_{\substack{\alpha,\eta,s \\ \eta',s'}} \eta \eta' \left(
	\hat{c}^\dagger_{\vec{k}_1,\bar{\alpha},\eta,s} \hat{c}_{\vec{k}_3,\alpha+2,\eta,s} \hat{f}^\dagger_{\vec{k}_1 - \vec{k}_2 - \vec{k}_3,\alpha,\eta',s'} \hat{c}_{\vec{k}_2,\alpha+2,\eta',s'} 
	\right. \nonumber \\
	&\left. -\hat{f}^\dagger_{\vec{k}_2 + \vec{k}_3 - \vec{k}_1, \alpha, \eta', s'} \hat{c}_{\vec{k}_2,\alpha+2,\eta',s'} \hat{c}^\dagger_{\vec{k}_1,\bar{\alpha}+2,\eta,s} \hat{c}_{\vec{k}_3,\alpha,\eta,s} 
	\right)
	+ \text{h.c.}, \label{app:eqn:THF_int:K}
 \end{align}
which correspond, respectively, to the onsite ($H_{U_1}$) and nearest-neighbor ($H_{U_2}$) $f$-electron repulsions, the $c$-electron Coulomb interaction ($H_V$), the $f$-$c$ density-density interaction ($H_W$), the $f$-$c$ exchange interaction ($H_J$), the double-hybridzation interaction ($H_{\tilde{J}}$), and the density-hybridization interaction ($H_{K}$). In \crefrange{app:eqn:THF_int:U1}{app:eqn:THF_int:J}, we have also defined the normal-ordered form of an operator $\mathcal{O}$ to be $:\mathrel{\mathcal{O}}: = \mathcal{O} - \ev{\mathcal{O}}{\mathrm{G}_0}$, where $\ket{\mathrm{G}_0}$ denotes a state at the charge neutrality point. We take this state to be such that 
\begin{align}
	\ev{\hat{c}^\dagger_{\vec{k},a,\eta,s} \hat{c}_{\vec{k}',a',\eta',s'}}{\mathrm{G}_0} &= \frac{1}{2} \delta_{\vec{k}, \vec{k}'} \delta_{a a'} \delta_{\eta \eta'} \delta_{s s'}, \nonumber \\
	\ev{\hat{f}^\dagger_{\vec{k},\alpha,\eta,s} \hat{f}_{\vec{k}',\alpha',\eta',s'}}{\mathrm{G}_0} &= \frac{1}{2} \delta_{\vec{k}, \vec{k}'} \delta_{\alpha \alpha'} \delta_{\eta \eta'} \delta_{s s'}, \nonumber \\
	\ev{\hat{f}^\dagger_{\vec{k},\alpha,\eta,s} \hat{c}_{\vec{k}',a',\eta',s'}}{\mathrm{G}_0} &= \ev{\hat{c}^\dagger_{\vec{k},a,\eta,s} \hat{f}_{\vec{k}',\alpha',\eta',s'}}{\mathrm{G}_0} = 0.
\end{align}
Additionally, in \cref{app:eqn:THF_int:U2}, $\left\langle \vec{R}, \vec{R}' \right\rangle$ denotes nearest-neighbor lattice sites. For the angle and tunneling ratio chosen around \cref{app:eqn:BM_TBG_ham}, as well as the relative permittivity and screening length chosen around \cref{app:eqn:double_gate_interaction}, the six interaction parameters ($W_1 = W_2$ and $W_3 = W_4$ follows by symmetry~\cite{SON22}) are given by~\cite{SON22}
\begin{align}
	U_1 &= \SI{57.95}{\milli\electronvolt}, &
	U_2 &= \SI{2.239}{\milli\electronvolt}, &
	W_1 &= W_2 = \SI{44.03}{\milli\electronvolt}, \nonumber \\ 
	J &= \SI{16.38}{\milli\electronvolt}, &
	K &= \SI{4.887}{\milli\electronvolt}, &
	W_3 &= W_4 = \SI{50.2}{\milli\electronvolt}, \label{app:eqn:THF_interaction_params_TBG}
\end{align}
while the $c$-$c$ interaction potential is the same as in \cref{app:eqn:double_gate_interaction}.

\subsubsection{The interaction THF Hamiltonian for TSTG}
\label{app:sec:HF_review:interaction:TSTG}
As pointed out in \cref{app:sec:BM_review:TSTG:many_body}, the mirror-even and mirror-odd fermions of TSTG are always coupled through density-density interactions, even in the absence of a perpendicularly applied displacement field. As a result the TSTG interaction Hamiltonian of TSTG ($H^{\text{TSTG}}_{I}$) can be constructed from the one of TBG by simply adding three terms to the latter,
\begin{equation}
	\label{app:eqn:THF_interaction_TSTG}
	H^{\text{TSTG}}_{I} = H^{\text{TBG}}_{I} + H_V^{d} + H_V^{cd} + H_W^{fd},
\end{equation}
which correspond to the Coulomb repulsion terms between the $d$-electrons ($H_V^{d}$), the $c$- and $d$-electrons ($H_V^{cd}$), as well as the $f$- and $d$-electrons ($H_W^{fd}$)~\cite{YU23a}. These three additional terms are given respectively by~\cite{YU23a}
\begin{align}
	H_V^{d} &= \frac{1}{2 \Omega_0 N_0} \sum_{\abs{\vec{p}_1},\abs{\vec{p}_1} \leq \Lambda_d} \sum_{\substack{\vec{q} \\ \abs{\vec{p}_1+\vec{q}},\abs{\vec{p}_2+\vec{q}} \leq \Lambda_d}} \sum_{\substack{\alpha,\eta,s \\ \alpha',\eta',s'}} V \left( \vec{q} \right):\mathrel{\hat{d}^\dagger_{\vec{p}_1 + \vec{q} ,\alpha ,\eta,s} \hat{d}_{\vec{p}_1,\alpha,\eta,s}}: :\mathrel{\hat{d}^\dagger_{\vec{p}_2 - \vec{q},\alpha',\eta',s'} \hat{d}_{\vec{p}_2,\alpha',\eta',s'}}:, \label{app:eqn:TSTG_int:Vd} \\
H_V^{cd} &= \frac{1}{\Omega_0 N_0} \sum_{\substack{\abs{\vec{k}_1} \leq \Lambda_c \\ \abs{\vec{p}_2}< \Lambda_d}} \sum_{\substack{\vec{q} \\ \abs{\vec{k}_1+\vec{q}} \leq \Lambda_c \\ \abs{\vec{p}_2+\vec{q}} \leq \Lambda_d}} \sum_{\substack{a,\eta,s \\ \alpha',\eta',s'}} V \left( \vec{q} \right):\mathrel{\hat{c}^\dagger_{\vec{k}_1 + \vec{q} ,a,\eta,s} \hat{c}_{\vec{k}_1,a,\eta,s}}: :\mathrel{\hat{d}^\dagger_{\vec{p}_2 - \vec{q},\alpha',\eta',s'} \hat{d}_{\vec{p}_2,\alpha',\eta',s'}}:, \label{app:eqn:TSTG_int:Vcd} \\
H_W^{fd} &= \frac{1}{N_0} \sum_{\substack{\vec{k}_1 \\ \abs{\vec{p}_2} \leq \Lambda_d}} \sum_{\substack{\vec{q} \\ \abs{\vec{p}_2-\vec{q}} \leq \Lambda_d}} \sum_{\substack{\alpha,\eta,s \\ \alpha',\eta',s'}} W_{fd} :\mathrel{\hat{f}^\dagger_{\vec{k}_1 + \vec{q},\alpha,\eta,s} \hat{f}_{\vec{k}_1,\alpha,\eta,s}}: :\mathrel{\hat{d}^\dagger_{\vec{p}_2 - \vec{q},\alpha',\eta',s'} \hat{d}_{\vec{p}_2,\alpha',\eta',s'}}:, \label{app:eqn:TSTG_int:Vfd}
 \end{align} 
where we note that the state $\ket{\mathrm{G}_0}$ at charge neutrality relative to which the normal ordering is defined obeys
\begin{align}
	\ev{\hat{d}^\dagger_{\vec{p},\alpha,\eta,s} \hat{d}_{\vec{p}',\alpha',\eta',s'}}{\mathrm{G}_0} &= \frac{1}{2} \delta_{\vec{p}, \vec{p}'} \delta_{\alpha \alpha'} \delta_{\eta \eta'} \delta_{s s'}, \nonumber \\
	\ev{\hat{d}^\dagger_{\vec{p},\alpha,\eta,s} \hat{c}_{\vec{k}',a',\eta',s'}}{\mathrm{G}_0} &= \ev{\hat{d}^\dagger_{\vec{p},\alpha,\eta,s} \hat{f}_{\vec{k}',\alpha',\eta',s'}}{\mathrm{G}_0} = 0, \nonumber \\
	\ev{\hat{c}^\dagger_{\vec{k},a,\eta,s} \hat{d}_{\vec{p}',\alpha',\eta',s'}}{\mathrm{G}_0} &= \ev{\hat{f}^\dagger_{\vec{k},\alpha,\eta,s} \hat{d}_{\vec{p}',\alpha',\eta',s'}}{\mathrm{G}_0} = 0.
\end{align}
Finally, for the single-particle parameters chosen around \cref{app:eqn:BM_TSTG_TBG_ham} and the Coulomb repulsion assumed in \cref{app:sec:BM_review:TSTG:many_body}, the THF interaction parameters for TSTG are given by~\cite{YU23a}
\begin{align}
	U_1 &= \SI{91.5}{\milli\electronvolt}, &
	U_2 &= \SI{6.203}{\milli\electronvolt}, &
	W_1 &= W_2 = \SI{88.54}{\milli\electronvolt}, \nonumber \\ 
	J &= \SI{24.25}{\milli\electronvolt}, &
	K &= \SI{7.061}{\milli\electronvolt}, &
	W_3 &= W_4 = \SI{97.67}{\milli\electronvolt}, &
	W_{fd}& = \SI{94.71}{\milli\electronvolt}. \label{app:eqn:THF_interaction_params_TSTG}
\end{align}
The approximate increase by a factor of $\sqrt{2}$ in the interaction parameters of TSTG relative to the ones of TBG from \cref{app:eqn:THF_interaction_params_TBG} can be traced back to the magic twist angle of TSTG, which is itself larger by a factor of $\sqrt{2}$ than the one of TBG~\cite{YU23a}. In TSTG at the magic angle, the moir\'e lattice constant is smaller by a factor of $\sqrt{2}$ compared to the one of TBG. If the gate distance $\xi$ had also been rescaled by a factor of $\frac{1}{\sqrt{2}}$ in TSTG compared to that of TBG, then the TSTG interaction parameters would have been exactly $\sqrt{2}$ larger than those of TBG. In this work, we consider the same gate distance for both TBG and TSTG. As a result, the $\sqrt{2}$ rescaling of the THF interactions parameters in TSTG relative to the ones of TBG is only \emph{approximate}.

\section{Review of the many-body theory of the THF model}\label{app:sec:many_body_rev}

In this \siSection{}, we provide a brief review of the many-body theory for the THF model from Ref.~\cite{CAL23b}. The main goal of this \siSection{} is the formalize the notation that will be used subsequent sections. We start by reviewing the notation of Ref.~\cite{CAL23b}, which treats all fermionic species on equal footing and define the Green's function for the system. Next, we briefly discuss the self-energy corrections to the THF model arising both at the static (Hartree-Fock) and dynamic levels, assuming a generic ground state. Finally, we list the correlated symmetry-breaking ground state candidates considered by this work and Ref.~\cite{CAL23b} and briefly discuss the many-body charge-conjugation symmetries of TBG and TSTG in the THF model. 

\subsection{Notation}\label{app:sec:many_body_rev:generic_not}

We start by reviewing the notation introduced in \cref*{DMFT:app:sec:hartree_fock} of Ref.~\cite{CAL23b} that treats all the fermionic species on equal footing. For TBG, we define the $\hat{\gamma}^\dagger_{\vec{k},i, \eta, s}$ fermions (for $1 \leq i \leq 6$) together with their wave functions $\phi_{\vec{Q} \beta; i} \left( \vec{k} \right)$
\begin{equation}
	\label{app:eqn:shorthand_gamma_not}
	\hat{\gamma}^\dagger_{\vec{k},i, \eta, s} \equiv \begin{cases}
		\hat{c}^\dagger_{\vec{k},i, \eta, s}, & \qq{for} 1 \leq i \leq 4 \\
		\hat{f}^\dagger_{\vec{k},i-4,\eta,s}, & \qq{for} 5 \leq i \leq 6
	\end{cases}, \quad
	\phi_{\vec{Q} \beta; i} \left( \vec{k} \right) \equiv \begin{cases}
		\tilde{u}_{\vec{Q} \beta; i} \left( \vec{k} \right), & \qq{for} 1 \leq i \leq 4 \\
		v_{\vec{Q} \beta; i-4} \left( \vec{k} \right), & \qq{for} 5 \leq i \leq 6
	\end{cases}.
\end{equation}
For TSTG, we introduce two additional fermions
\begin{equation}
	\label{app:eqn:shorthand_gamma_not_TSTG}
	\hat{\gamma}^\dagger_{\vec{k},i, \eta, s} \equiv \hat{d}^\dagger_{\vec{P}_{\vec{k}},i-6,\eta,s} \qq{for} 7 \leq  i \leq 8,
\end{equation}
where $\vec{P}_{\vec{k}}$ denotes the image of $\vec{k} - \eta \vec{q}_1$ in the first BZ. As in Ref.~\cite{CAL23b}, the momentum cutoffs $\Lambda_c$ and $\Lambda_d$ have been extended such that the $\hat{c}^\dagger_{\vec{k},a,\eta,s}$ ($\hat{d}^\dagger_{\vec{p},\alpha,\eta,s}$) fermion covers exactly one BZ around the $\Gamma_M$ ($\eta \vec{q}_1$) point. Written in the $\gamma$-fermion basis, the single-particle THF Hamiltonian matrices of TBG and TSTG are, respectively,
\begin{equation}
	\label{app:eqn:full_THF_Hamiltonian_gamma_bas}
	h^{\text{TBG},\eta} \left( \vec{k} \right) = \begin{pNiceMatrix}[last-col=3,first-row]
		4 & 2 \\
		h^{cc,\eta} \left( \vec{k} \right) & h^{cf,\eta} \left( \vec{k} \right) & 4 \\
		h^{\dagger cf,\eta} \left( \vec{k} \right) & h^{ff,\eta} \left( \vec{k} \right)  & 2 \\
	\end{pNiceMatrix},
	\quad
	h^{\text{TSTG},\eta} \left( \vec{k} \right) = \begin{pNiceMatrix}[last-col=4,first-row]
		4 & 2 & 2\\
		h^{cc,\eta} \left( \vec{k} \right) & h^{cf,\eta} \left( \vec{k} \right) & \mathbb{0} & 4 \\
		h^{\dagger cf,\eta} \left( \vec{k} \right) & h^{ff,\eta} \left( \vec{k} \right) & h^{df,\eta} \left( \vec{P}_{\vec{k}} \right) & 2 \\
		\mathbb{0} & h^{\dagger df,\eta} \left( \vec{P}_{\vec{k}} \right) & h^{dd,\eta} \left( \vec{P}_{\vec{k}} \right) & 2 \\
	\end{pNiceMatrix},
\end{equation}
where the matrix blocks are given by \cref{app:eqn:cc_thf_block,app:eqn:cf_thf_block,app:eqn:f_thf_block,app:eqn:dd_thf_block,app:eqn:df_thf_block}, and their dimensions have been explicitly indicated outside each matrix.

We define the grand canonical Hamiltonian for the THF model as $K^{\text{TBG}} = H^{\text{TBG}} - \mu \hat{N}$ for TBG and $K^{\text{TSTG}} = H^{\text{TSTG}} - \mu \hat{N}$ for TSTG. The many-body THF Hamiltonians for TBG and TSTG are given, respectively by 
\begin{equation}
	H^{\text{TBG}} = H^{\text{TBG}}_0 + H^{\text{TBG}}_I, \quad
	H^{\text{TSTG}} = H^{\text{TSTG}}_0 + H^{\text{TSTG}}_I,
\end{equation} 
while the total number operator is $\hat{N} = \sum_{\vec{k},\eta,s} \sum_{i=1}^{6} \hat{\gamma}^\dagger_{\vec{k},i,\eta,s} \hat{\gamma}_{\vec{k},i,\eta,s}$ for TBG and $\hat{N} = \sum_{\vec{k},\eta,s} \sum_{i=1}^{8} \hat{\gamma}^\dagger_{\vec{k},i,\eta,s} \hat{\gamma}_{\vec{k},i,\eta,s}$ for TSTG. As in Ref.~\cite{CAL23b}, we consider only states that maintain moiré translation symmetry. The interacting Matsubara Green's function is defined as
\begin{equation}
	\label{app:eqn:matsubara_gf_THF_tau}
	-\left\langle \mathcal{T}_{\tau} \hat{\gamma}_{\vec{k},i,\eta,s} \left( \tau \right) \hat{\gamma}^\dagger_{\vec{k}',i',\eta',s'} \left( 0 \right)  \right\rangle  = \delta_{\vec{k},\vec{k}'} \mathcal{G}_{i \eta s; i' \eta' s'} \left(\tau, \vec{k} \right),
\end{equation}
with $\tau$ being the imaginary time, $\mathcal{T}_{\tau}$ ordering the operators that follow it by imaginary time, and $\left\langle \hat{\mathcal{O}} \right\rangle$ denoting the expectation value of an operator $\hat{\mathcal{O}}$ in the grand canonical ensemble
\begin{equation}
	\left\langle \hat{\mathcal{O}} \right\rangle = \frac{\Tr\left[e^{-\beta K} \hat{\mathcal{O}} \right]}{\Tr\left[e^{-\beta K}\right]}.
\end{equation}
The fermion operators also evolve by the grand canonical ensemble Hamiltonian $K$
\begin{equation}
	\label{app:eqn:gr_can_evolved_gamma_ops}
	\hat{\gamma}^\dagger_{\vec{k},i,\eta,s} \left( \tau \right) = e^{K \tau} \hat{\gamma}^\dagger_{\vec{k},i,\eta,s} \left( 0 \right) e^{-K\tau}.
\end{equation}
The Fourier transformation of the Green's function is given by 
\begin{equation}
	\label{app:eqn:matsubara_gf_THF_ft}
	\mathcal{G}_{i \eta s; i' \eta' s'} \left(i \omega_n, \vec{k} \right) = \int_{0}^{\beta} \dd{\tau} e^{i \omega_n \tau} 	\mathcal{G}_{i \eta s; i' \eta' s'} \left(\tau, \vec{k} \right),
\end{equation}
with $\beta = 1/T$ being the inverse temperature and $\omega_n = \frac{(2 n + 1)\pi}{\beta}$, the fermionic Matsubara frequencies. We also define the Fermion spectral function of the system, $A_{i \eta s; i' \eta' s'} \left( \omega, \vec{k} \right)$, which is can be obtained from the retarded Green's function as\footnote{The spectral function $A_{i \eta s; i' \eta' s'} \left( \omega, \vec{k} \right)$ is denoted with a capital ``A'' and should not be confused with the nonabelian Berry connection introduced in \cref{app:eqn:f_nonabelian_berry}, which is denoted with a calligraphic ``$\mathbfcal{A}$''.}
\begin{equation}
	\label{app:eqn:spectral_function}
	A_{i \eta s; i' \eta' s'} \left( \omega, \vec{k} \right) = \frac{-1}{2 \pi i} \left( \mathcal{G}_{i \eta s; i' \eta' s'} \left(\omega + i 0^{+}, \vec{k} \right) - \mathcal{G}^{*}_{i' \eta' s',i \eta s} \left(\omega + i 0^{+}, \vec{k} \right) \right).
\end{equation}
The Green's function for generic complex frequencies $z$ can then be recovered from the spectral function
\begin{equation}
	\label{app:eqn:spectral_rep_of_GF}
	\mathcal{G}_{i \eta s; i' \eta' s'} \left(z, \vec{k} \right) = \int_{-\infty}^{\infty} \frac{\dd{\omega}}{z-\omega} A_{i \eta s; i' \eta' s'} \left( \omega, \vec{k} \right).
\end{equation} 
\Cref{app:eqn:spectral_rep_of_GF} can be proved using the Lehman representation of the Green's function~\cite{MAH00}, as shown in \cref*{DMFT:app:sec:hartree_fock:generic_not} of Ref.~\cite{CAL23b}. Finally, the density matrix of the system is given by
\begin{equation}
	\label{app:eqn:def_rho_HF}
	\varrho_{i \eta s; i' \eta' s'} \left(\vec{k} \right) = \left\langle  :\mathrel{\hat{\gamma}^\dagger_{\vec{k}, i, \eta, s} \hat{\gamma}_{\vec{k}, i' \eta' s'}}: \right\rangle = \int_{-\infty}^{\infty} \dd{\omega} n_{\mathrm{F}} \left( \omega \right) A_{i' \eta' s'; i \eta s} \left( \omega, \vec{k} \right) - \frac{1}{2} \delta_{ii'}\delta_{\eta \eta'}\delta_{s s'},
\end{equation}
where $n_{\mathrm{F}} \left( \omega \right)$ denotes the Fermi-Dirac distribution function 
\begin{equation}
	\label{app:eqn:fd_distribution}
	n_{\mathrm{F}} \left( \omega \right) = \frac{1}{e^{\beta \omega} + 1}.
\end{equation}
Additionally, we define the total fillings of the $c$-, $f$-, and $d$-electrons to be 
\begin{equation}
	\label{app:eqn:flavor_filing}
	\nu_c = \frac{1}{N_0} \sum_{i=1}^{4} \sum_{\vec{k},\eta,s} \varrho_{i \eta s; i \eta s} \left(\vec{k} \right), \quad
	\nu_f = \frac{1}{N_0} \sum_{i=5}^{6} \sum_{\vec{k},\eta,s} \varrho_{i \eta s; i \eta s} \left(\vec{k} \right), \quad
	\nu_d = \frac{1}{N_0} \sum_{i=7}^{8} \sum_{\vec{k},\eta,s} \varrho_{i \eta s; i \eta s} \left(\vec{k} \right),
\end{equation} 
as well as the total electron filling $\nu = \nu_c + \nu_f + \nu_d$.

It is also useful to define the \emph{non-interacting} Matsubara Green's function
\begin{equation}
	\label{app:eqn:matsubara_gf_THF_tau_noninteracting}
	-\left\langle \mathcal{T}_{\tau} \hat{\gamma}_{\vec{k},i,\eta,s} \left( \tau \right) \hat{\gamma}^\dagger_{\vec{k}',i',\eta',s'} \left( 0 \right)  \right\rangle_0  = \delta_{\vec{k},\vec{k}'} \mathcal{G}^{0}_{i \eta s; i' \eta' s'} \left(\tau, \vec{k} \right),
\end{equation}
whose definition is similar to \cref{app:eqn:matsubara_gf_THF_tau}, with the only exception being that the imaginary time-evolution and the averaging $\left\langle \dots \right\rangle_0$ in \cref{app:eqn:matsubara_gf_THF_tau_noninteracting} is performed within the \emph{non-interacting} grand canonical ensemble Hamiltonian $K_0 = H_0 - \mu \hat{N}$. The non-interacting Green's function can readily be expressed in terms of the single-particle Hamiltonian
from \cref{app:eqn:full_THF_Hamiltonian_gamma_bas}, 
\begin{equation}
	\label{app:eqn:gf_within_non_int}
	\mathcal{G}^{0} \left(i\omega_n,  \vec{k} \right) = \left[\left( i\omega_n + \mu \right) \mathbb{1} - h \left( \vec{k} \right)  \right]^{-1}, \qq{where} h \left( \vec{k} \right) = h^{\text{TBG}} \left( \vec{k} \right), h^{\text{TSTG}} \left( \vec{k} \right).
\end{equation}
Similarly, to \cref{app:eqn:spectral_function}, we can also define the non-interacting spectral function of the system by analytically continuing the non-interacting Green's function
\begin{equation}
	\label{app:eqn:spectral_function_non_int}
	A^{0}_{i \eta s; i' \eta' s'} \left( \omega, \vec{k} \right) = \frac{-1}{2 \pi i} \left( \mathcal{G}^{0}_{i \eta s; i' \eta' s'} \left(\omega + i 0^{+}, \vec{k} \right) - \mathcal{G}^{*0}_{i' \eta' s',i \eta s} \left(\omega + i 0^{+}, \vec{k} \right) \right).
\end{equation}

In the analysis of the spectral function of the system depicted in \crefrange{fig:seebeck_sym_br:e}{fig:seebeck_sym_br:h}, we adhere to a convention similar to that introduced in \cref*{DMFT:app:sec:results_corr_ins} of Ref.~\cite{CAL23b}. There, the color intensity corresponds to the total spectral weight at a specific $\vec{k}$ point, defined by the trace of the spectral function, $\Tr A \left( \omega, \vec{k} \right)$. Likewise, in \cref{fig:seebeck_sym:c}, the color intensity signifies the trace of the spectral function \emph{summed} over the entire moiré BZ, $\sum_{\vec{k}} \Tr A \left( \omega, \vec{k} \right)$. Furthermore, in both \cref{fig:seebeck_sym_br,fig:seebeck_sym}, the hue of the colormap indicates the $f$-character of the excitation, defined according to
\begin{equation}
	\%_{\hat{f}} = \frac{ \sum_{\alpha, \eta, s} A_{(\alpha + 4) \eta s; (\alpha + 4) \eta s} \left(\omega, \vec{k} \right)}{\Tr A \left( \omega, \vec{k} \right)	}, \qquad
	\%_{\hat{f}} = \frac{ \sum_{\vec{k},\alpha, \eta, s} A_{(\alpha + 4) \eta s; (\alpha + 4) \eta s} \left(\omega, \vec{k} \right)}{\sum_{\vec{k}} \Tr A \left( \omega, \vec{k} \right)	},
\end{equation} 
respectively. Excitations with $\%_{\hat{f}} = 1$ ($\%_{\hat{f}} = 0$) are shown in blue (orange), as indicated by the colormaps of \cref{fig:seebeck_sym_br,fig:seebeck_sym}.

\subsection{Self-energy corrections}\label{app:sec:many_body_rev:self-energy}

The fully-interacting Green's function and the non-interacting one are connected via the Dyson equation~\cite{MAH00}
\begin{equation}
	\label{app:eqn:dyson_equation}
	\mathcal{G} \left( i \omega_n, \vec{k} \right) = \left[ \left( \mathcal{G}^{0} \left( i \omega_n, \vec{k} \right) \right)^{-1} - \Sigma \left(i \omega_n, \vec{k} \right) \right]^{-1}, 
\end{equation}
where $\Sigma \left(i \omega_n, \vec{k} \right)$ is the self-energy matrix of the system. In general, the self-energy correction can be split into the static (or Hartree-Fock) and the dynamic contributions, which are, respectively, frequency-independent and frequency-dependent. As in Ref.~\cite{CAL23b}, we approximate the dynamic contribution to the self-energy to be $\vec{k}$-independent (or site diagonal in real space) such that 
\begin{equation}
	\label{app:eqn:dmft_approximation}
	\Sigma \left( i \omega_n, \vec{k} \right) \approx h^{I,\text{MF}} \left( \vec{k} \right) + \Sigma \left( i \omega_n \right).
\end{equation}
In \cref{app:eqn:dmft_approximation}, $h^{I,\text{MF}} \left( \vec{k} \right)$ denotes the Hartree-Fock interaction Hamiltonian of the system and $\Sigma \left( i \omega_n \right)$ is the $\vec{k}$-independent approximation of the dynamical self-energy correction. Following Ref.~\cite{CAL23b}, we take another approximation and only consider the dynamical self energy-correction for the $f$-electrons, as they are dispersionless, feature a strong onsite Hubbard interaction, and are thus expected to be strongly correlated
\begin{equation}
	\label{app:eqn:full_second_order_sigma}
	\Sigma_{i \eta s; i' \eta' s'} \left( i \omega_n \right) = \begin{cases}
		\Sigma^{f}_{(i-4) \eta s; (i'-4) \eta' s'} \left(i \omega_n \right) & \qq{if} 5 \leq i,i' \leq 6 \\
		- i \Gamma_{c} \delta_{i i'} \delta_{\eta \eta'} \delta_{s s'} & \qq{otherwise}
	\end{cases}.
\end{equation}
The $f$-electron dynamical self-energy correction $\Sigma^{f} \left(i \omega_n \right)$ was computed in Ref.~\cite{CAL23b} in both the symmetric and symmetry-broken phases of the THF model. For the other electron species ({\it i.e.}{}, the $c$-electrons in the case of TBG and the $c$- and $d$-electrons in the case of TSTG), the dynamical self-energy is approximated by a small imaginary ``broadening factor''. As in Ref.~\cite{CAL23b}, we employ $\Gamma_{c} = \SI{1}{\milli\electronvolt}$ ($\Gamma_{c} = \SI{1.5}{\milli\electronvolt}$) for TBG (TSTG). This specific choice was explained in \cref*{DMFT:app:sec:se_correction_beyond_HF:sc_problem_and_numerics} of Ref.~\cite{CAL23b}. Here, we only mention that the increase of the broadening factor for TSTG by a factor of $\frac{3}{2}$ mimics the relative increase of all the energy scales of TSTG relative to TBG by a factor of approximately $\sqrt{2} \approx \frac{3}{2}$, as discussed at the end of \cref{app:sec:HF_review:interaction:TSTG}.

Writing the THF interaction Hamiltonian for TBG in the generic form~\cite{CAL23b},  
{
	\small
	\begin{equation}
		\label{app:eqn:general_TBG_int_forHF}
		H^{\text{TBG}}_{I} = \sum_{\substack{i,\eta_1,s_1 \\ j,\eta_2,s_2}}  \sum_{\substack{k,\eta_3,s_3 \\ l,\eta_4,s_4}}  \sum_{\substack{\vec{k}_1,\vec{k}_2,\vec{q} \\ \vec{k}'_1, \vec{k}'_2}} V_{i \eta_1 s_1; j \eta_2 s_2; k \eta_3 s_3; l \eta_4 s_4} \left( \vec{q} \right) \delta_{\vec{k}'_2,\vec{k}_2 - \vec{q}} \delta_{\vec{k}'_1,\vec{k}_1 + \vec{q}} 
		:\mathrel{\hat{\gamma}^\dagger_{\vec{k}'_1 ,i,\eta_1,s_1} \hat{\gamma}_{\vec{k}_1,j,\eta_2,s_2}}:
		:\mathrel{\hat{\gamma}^\dagger_{\vec{k}'_2,k,\eta_3,s_3} \hat{\gamma}_{\vec{k}_2,l,\eta_4,s_4}}:,
\end{equation}}the Hartree-Fock interaction Hamiltonian matrix can be found using the standard Hartree-Fock decoupling procedure. To be specific, for a given state of the system characterized by the density, $\varrho_{i \eta s;i' \eta' s'} \left(\vec{k} \right)$, the TBG Hartree-Fock matrix is given by
\begin{align}
	\label{app:eqn:genera_TBG_int_HF}
	h^{\text{TBG},I,\text{MF}}_{i \eta s; i' \eta' s'} \left( \vec{k} \right) 
	=& 2 \sum_{\vec{k}'} \sum_{\substack{m,\eta_1,s_1 \\ n,\eta_2,s_2}} \left( V_{n \eta_1 s_1; m \eta_2 s_2; i \eta s; i' \eta' s'} \left( \vec{0} \right) 	\varrho_{n \eta_1 s_1; m \eta_2 s_2} \left(\vec{k}' \right) \right. \nonumber\\
	-& \left. V_{i \eta s; m \eta_2 s_2; n \eta_1 s_1; i' \eta' s'} \left( \vec{k} - \vec{k}' \right) \varrho_{n \eta_1 s_1; m \eta_2 s_2} \left(\vec{k}' \right) \right).
\end{align}
The interaction tensor from \cref{app:eqn:general_TBG_int_forHF} can be directly obtained by casting \crefrange{app:eqn:THF_int:U1}{app:eqn:THF_int:K} into the notation introduced in \cref{app:eqn:shorthand_gamma_not}, and equating \cref{app:eqn:general_TBG_int_forHF} with \cref{app:eqn:THF_interaction_TBG}. For TSTG, a similar linear relation to \cref{app:eqn:genera_TBG_int_HF} between the Hartree-Fock interaction Hamiltonian and the density matrix was given in Ref.~\cite{CAL23b}. The \emph{total} mean-field Hamiltonian is obtained by summing the single-particle Hamiltonian and the Hartree-Fock interaction one, with the corresponding matrix reading as
\begin{equation}
	\label{app:eqn:HF_Hamiltonian}
	h^{\text{MF}}_{i \eta s; i' \eta' s'} \left( \vec{k} \right) = h^{\eta}_{i i'} \left(\vec{k} \right) \delta_{\eta \eta'} \delta_{s s'}+  h^{I,\text{MF}}_{i \eta s; i' \eta' s'} \left( \vec{k} \right).
\end{equation}
Using \cref{app:eqn:HF_Hamiltonian}, the second-quantized Hartree-Fock Hamiltonian can be defined as simply~\cite{CAL23b}
\begin{equation}
	H_{\text{MF}} \equiv \sum_{\substack{\vec{k},i, \eta, s \\ i', \eta', s'}} h^{\text{MF}}_{i \eta s;i' \eta' s'} \left( \vec{k} \right) \hat{\gamma}^\dagger_{\vec{k},i, \eta, s} \hat{\gamma}_{\vec{k},i', \eta', s'}.
\end{equation}
In terms of the total mean-field Hamiltonian and under the approximation in \cref{app:eqn:dmft_approximation}, the fully-interacting Green's function of the THF model is given by
\begin{equation}
	\label{app:eqn:general_gf_self_energy}
	\mathcal{G} \left( i \omega_n, \vec{k} \right) = \left[\left( i\omega_n + \mu \right) \mathbb{1} - h^{\text{MF}} \left( \vec{k} \right) - \Sigma \left( i \omega_n \right) \right]^{-1}.
\end{equation}

\subsection{Correlated ground state candidates}\label{app:sec:many_body_rev:ground_states}

\subsubsection{Model states}\label{app:sec:many_body_rev:ground_states:model}

\begin{table}[t]
	\centering
	\begin{tabular}{|c|l|r|}
		\hline
		$\nu$ & State & Parent state \\
		\hline\hline
		$-4$ & $\IfStrEqCase{1}{{1}{\ket{\nu={-}4} }
		{2}{\ket{\nu={-}3, \mathrm{IVC}}}
		{3}{\ket{\nu={-}3, \mathrm{VP}}}
		{4}{\ket{\nu={-}2, \mathrm{K-IVC}}}
		{5}{\ket{\nu={-}2, \mathrm{VP}}}
		{6}{\ket{\nu={-}1, (\mathrm{K-IVC}+\mathrm{VP})}}
		{7}{\ket{\nu={-}1, \mathrm{VP}}}
		{8}{\ket{\nu=0, \mathrm{K-IVC}}}
		{9}{\ket{\nu=0, \mathrm{VP}}}
	}
	[nada]
$ & $\ket{\mathrm{FS}}$ \\
		\hline
		\multirow[b]{2}{*}{$-3$} & $\IfStrEqCase{2}{{1}{\ket{\nu={-}4} }
		{2}{\ket{\nu={-}3, \mathrm{IVC}}}
		{3}{\ket{\nu={-}3, \mathrm{VP}}}
		{4}{\ket{\nu={-}2, \mathrm{K-IVC}}}
		{5}{\ket{\nu={-}2, \mathrm{VP}}}
		{6}{\ket{\nu={-}1, (\mathrm{K-IVC}+\mathrm{VP})}}
		{7}{\ket{\nu={-}1, \mathrm{VP}}}
		{8}{\ket{\nu=0, \mathrm{K-IVC}}}
		{9}{\ket{\nu=0, \mathrm{VP}}}
	}
	[nada]
$ & $\displaystyle \prod_{\vec{k}} \frac{1}{\sqrt{2}} \left( \hat{f}^\dagger_{\vec{k},1,+,\uparrow} - i\hat{f}^\dagger_{\vec{k},2,-,\uparrow} \right)\ket{\mathrm{FS}}$ \\
		& $\IfStrEqCase{3}{{1}{\ket{\nu={-}4} }
		{2}{\ket{\nu={-}3, \mathrm{IVC}}}
		{3}{\ket{\nu={-}3, \mathrm{VP}}}
		{4}{\ket{\nu={-}2, \mathrm{K-IVC}}}
		{5}{\ket{\nu={-}2, \mathrm{VP}}}
		{6}{\ket{\nu={-}1, (\mathrm{K-IVC}+\mathrm{VP})}}
		{7}{\ket{\nu={-}1, \mathrm{VP}}}
		{8}{\ket{\nu=0, \mathrm{K-IVC}}}
		{9}{\ket{\nu=0, \mathrm{VP}}}
	}
	[nada]
$ & $\displaystyle \prod_{\vec{k}} \hat{f}^\dagger_{\vec{k},1,+,\uparrow} \ket{\mathrm{FS}}$ \\
		\hline
		\multirow[b]{2}{*}{$-2$} & $\IfStrEqCase{4}{{1}{\ket{\nu={-}4} }
		{2}{\ket{\nu={-}3, \mathrm{IVC}}}
		{3}{\ket{\nu={-}3, \mathrm{VP}}}
		{4}{\ket{\nu={-}2, \mathrm{K-IVC}}}
		{5}{\ket{\nu={-}2, \mathrm{VP}}}
		{6}{\ket{\nu={-}1, (\mathrm{K-IVC}+\mathrm{VP})}}
		{7}{\ket{\nu={-}1, \mathrm{VP}}}
		{8}{\ket{\nu=0, \mathrm{K-IVC}}}
		{9}{\ket{\nu=0, \mathrm{VP}}}
	}
	[nada]
$ & $\displaystyle \prod_{\vec{k}} \frac{1}{2} \left( \hat{f}^\dagger_{\vec{k},1,+,\uparrow} - i\hat{f}^\dagger_{\vec{k},2,-,\uparrow} \right) \left( \hat{f}^\dagger_{\vec{k},2,+,\uparrow} + i\hat{f}^\dagger_{\vec{k},1,-,\uparrow} \right) \ket{\mathrm{FS}}$ \\
		& $\IfStrEqCase{5}{{1}{\ket{\nu={-}4} }
		{2}{\ket{\nu={-}3, \mathrm{IVC}}}
		{3}{\ket{\nu={-}3, \mathrm{VP}}}
		{4}{\ket{\nu={-}2, \mathrm{K-IVC}}}
		{5}{\ket{\nu={-}2, \mathrm{VP}}}
		{6}{\ket{\nu={-}1, (\mathrm{K-IVC}+\mathrm{VP})}}
		{7}{\ket{\nu={-}1, \mathrm{VP}}}
		{8}{\ket{\nu=0, \mathrm{K-IVC}}}
		{9}{\ket{\nu=0, \mathrm{VP}}}
	}
	[nada]
$ & $\displaystyle \prod_{\vec{k}} \hat{f}^\dagger_{\vec{k},1,+,\uparrow} \hat{f}^\dagger_{\vec{k},2,+,\uparrow} \ket{\mathrm{FS}}$ \\
		\hline
		\multirow[b]{2}{*}{$-1$} & $\IfStrEqCase{6}{{1}{\ket{\nu={-}4} }
		{2}{\ket{\nu={-}3, \mathrm{IVC}}}
		{3}{\ket{\nu={-}3, \mathrm{VP}}}
		{4}{\ket{\nu={-}2, \mathrm{K-IVC}}}
		{5}{\ket{\nu={-}2, \mathrm{VP}}}
		{6}{\ket{\nu={-}1, (\mathrm{K-IVC}+\mathrm{VP})}}
		{7}{\ket{\nu={-}1, \mathrm{VP}}}
		{8}{\ket{\nu=0, \mathrm{K-IVC}}}
		{9}{\ket{\nu=0, \mathrm{VP}}}
	}
	[nada]
$ & $\displaystyle \prod_{\vec{k}} \frac{1}{2} \hat{f}^\dagger_{\vec{k},1,+,\downarrow} \left( \hat{f}^\dagger_{\vec{k},1,+,\uparrow} - i\hat{f}^\dagger_{\vec{k},2,-,\uparrow} \right) \left( \hat{f}^\dagger_{\vec{k},2,+,\uparrow} + i\hat{f}^\dagger_{\vec{k},1,-,\uparrow} \right) \ket{\mathrm{FS}}$ \\
		& $\IfStrEqCase{7}{{1}{\ket{\nu={-}4} }
		{2}{\ket{\nu={-}3, \mathrm{IVC}}}
		{3}{\ket{\nu={-}3, \mathrm{VP}}}
		{4}{\ket{\nu={-}2, \mathrm{K-IVC}}}
		{5}{\ket{\nu={-}2, \mathrm{VP}}}
		{6}{\ket{\nu={-}1, (\mathrm{K-IVC}+\mathrm{VP})}}
		{7}{\ket{\nu={-}1, \mathrm{VP}}}
		{8}{\ket{\nu=0, \mathrm{K-IVC}}}
		{9}{\ket{\nu=0, \mathrm{VP}}}
	}
	[nada]
$ & $\displaystyle \prod_{\vec{k}} \hat{f}^\dagger_{\vec{k},1,+,\downarrow} \hat{f}^\dagger_{\vec{k},1,+,\uparrow} \hat{f}^\dagger_{\vec{k},2,+,\uparrow} \ket{\mathrm{FS}}$ \\
		\hline
		\multirow[b]{2}{*}{$0$} & $\IfStrEqCase{8}{{1}{\ket{\nu={-}4} }
		{2}{\ket{\nu={-}3, \mathrm{IVC}}}
		{3}{\ket{\nu={-}3, \mathrm{VP}}}
		{4}{\ket{\nu={-}2, \mathrm{K-IVC}}}
		{5}{\ket{\nu={-}2, \mathrm{VP}}}
		{6}{\ket{\nu={-}1, (\mathrm{K-IVC}+\mathrm{VP})}}
		{7}{\ket{\nu={-}1, \mathrm{VP}}}
		{8}{\ket{\nu=0, \mathrm{K-IVC}}}
		{9}{\ket{\nu=0, \mathrm{VP}}}
	}
	[nada]
$ & $\displaystyle \prod_{\vec{k}} \prod_{s} \frac{1}{4} \left( \hat{f}^\dagger_{\vec{k},1,+,s} - i\hat{f}^\dagger_{\vec{k},2,-,s} \right) \left( \hat{f}^\dagger_{\vec{k},2,+,s} + i\hat{f}^\dagger_{\vec{k},1,-,s} \right) \ket{\mathrm{FS}}$ \\
		& $\IfStrEqCase{9}{{1}{\ket{\nu={-}4} }
		{2}{\ket{\nu={-}3, \mathrm{IVC}}}
		{3}{\ket{\nu={-}3, \mathrm{VP}}}
		{4}{\ket{\nu={-}2, \mathrm{K-IVC}}}
		{5}{\ket{\nu={-}2, \mathrm{VP}}}
		{6}{\ket{\nu={-}1, (\mathrm{K-IVC}+\mathrm{VP})}}
		{7}{\ket{\nu={-}1, \mathrm{VP}}}
		{8}{\ket{\nu=0, \mathrm{K-IVC}}}
		{9}{\ket{\nu=0, \mathrm{VP}}}
	}
	[nada]
$ & $\displaystyle \prod_{\vec{k}} \prod_{\alpha,s}\hat{f}^\dagger_{\vec{k},\alpha,+,s} \ket{\mathrm{FS}}$ \\
		\hline
		$\nu' > 0$ & $\ket{\nu = \nu',\text{Type}}$ & $\mathcal{P}^{(\prime)} \ket{\nu = -\nu',\text{Type}}$ \\
		\hline
	\end{tabular}
 	\caption{Correlated ground state candidates considered in this work (table adapted from Ref.~\cite{CAL23b}). For each filling $\nu$ we list the correlated ground state candidate, as well as the wave functions of the corresponding parent states~\cite{SON22,YU23a}. Starting from the corresponding parent state, the wave function of each correlated state is obtained using Hartree-Fock~\cite{CAL23b}. In turn, the parent states are obtained by occupying $(\nu+4)$ $f$-electron bands on top of a half-filled Fermi sea of $c$- (for TBG) or $c$- and $d$- (for TSTG) electrons (denoted as $\ket{\mathrm{FS}}$). The positive integer-filled states are obtained from the ones at negative filling using the many-body charge conjugation symmetry $\mathcal{P}$ in the case of TBG, or the many-body spatial charge conjugation symmetry $\mathcal{P}'$ in the case of TSTG, as detailed in \cref{app:sec:many_body_rev:ground_states:ph_symmetry}.}
	\label{app:tab:model_states}
\end{table}

This work and Ref.~\cite{CAL23b} consider nine different correlated ground state candidates at different integer fillings (together with their charge conjugated counterparts) for TBG and TSTG, as summarized in \cref{app:tab:model_states}. Within the Hartree-Fock approximation, the wave function of each state (or equivalently its self-consistent density matrix) is obtained at zero temperature by starting from the indicated parent states~\cite{SON22}, as was explained in Ref.~\cite{CAL23b}. In the absence of strain or relaxation effects, the states we consider here have been shown to be ground states or low-energy states in TBG~\cite{SON22} or TSTG at low values of the displacement field $\mathcal{E}$~\cite{YU23a}. For both TBG~\cite{YUA18,OCH18,THO18,DOD18,XU18b,KOS18,PO18a,PAD18,VEN18,KEN18,RAD18,HUA19,LIU19,WU19,CLA19,KAN19,SEO19,DA19,ANG19,WU20,XIE20b,BUL20b,CHA20,REP20,CEA20,ZHA20,KAN20a,BUL20a,CHI20b,SOE20,CHR20,EUG20,VAF20,LIU21,DA21,LIU21a,THO21,KWA21a,LIA21,XIE21,ZHA21,PAR21a,VAF21,KWA21b,CHE21,POT21,XIE21a,LED21,CHA21,KAN21,KWA21,HOF22,WAG22,SON22,BRI22,CAL22d,HON22,ZHA23a,BLA22,XIE23a,KWA23} and TSTG~\cite{XIE21b,LED21,CHR22,YU23a,WAN23c}, a myriad of correlated ground states have been proposed. Because the different states at a given integer filling are close in energy (with the precise hierarchy being determined by effects such as strain, relaxation, {\it etc.}{}), we will consider multiple correlated ground state candidates and assess the differences in their transport properties. As we explain in \cref{app:sec:seebeck_sym_br}, however, we find that the phenomenology reported by experiments (the negative Seebeck coefficient at low-temperature and positive fillings)~\cite{MER24,BAT24} can be reproduced by any correlated state with heavy hole and light electron excitations, such as, for example, the Intervalley Kekul\'e Spiral state~\cite{KWA21,WAG22,WAN23c}.

Starting from the zero-temperature integer-filled solutions obtained using Hartree-Fock, we employ second-order self-consistent perturbation theory as explained in detail in \cref*{DMFT:app:sec:se_correction_beyond_HF} of Ref.~\cite{CAL23b} to obtain the spectral function of the THF model in the corresponding symmetry-broken phases at finite temperature and doping. Specifically, we first obtain the self-consistent second-order solution at integer filling and finite temperature using the zero-temperature Hartree-Fock solution as initial condition. We then converge to the symmetry-broken phases at small doping \emph{around} the integer filling by starting from the integer-filled solution and increasing or decreasing the filling in small increments. In practice, this entails the self-consistent determination (at each temperature and filling) of the density matrix $\varrho \left( \vec{k} \right)$ and dynamical $f$-electron self-energy correction $\Sigma^{f} \left( i \omega_n \right)$. The latter contains the onsite second-order contributions arising from the $H_{U_1}$ and $H_{U_2}$ terms of the interaction Hamiltonian. At the same time, the density matrix determines through \cref{app:eqn:genera_TBG_int_HF}, the Hartree-Fock Hamiltonian of the system (or the static self-energy correction) $h^{\text{MF}} \left( \vec{k} \right)$. In turn, this allows us to compute the Green's function of the system $\mathcal{G} \left( i \omega_n, \vec{k} \right)$ from \cref{app:eqn:general_gf_self_energy}, and then the spectral function $A \left( \omega, \vec{k} \right)$, using \cref{app:eqn:spectral_function}. It is the spectral function at different fillings and temperatures that is finally employed to obtain the transport properties of the THF model, as we will show in \cref{app:sec:thermoelectric_response}. 

In addition to the symmetry-broken correlated phases from \cref{app:tab:model_states}, we also consider the \emph{symmetric} phase of both TSTG and TBG using iterative perturbation theory. The details on how this solution is obtained is presented din detail in \cref*{DMFT:app:sec:se_symmetric} of Ref.~\cite{CAL23b}  The self-consistent density matrix and dynamical $f$-electron self-energy (or equivalently the spectral function) of this phase are obtained starting from charge neutrality ($\nu = 0$) and then doping towards the band insulator filling in small increments. Note that in the symmetric state, the solution of the system obeys all the symmetries of the THF Hamiltonian. In particular, the $f$-electron density matrix in the symmetric phase is proportional to identity~\cite{CAL23b}
\begin{equation}
	\varrho_{(\alpha+4) \eta s; (\alpha'+4) \eta' s'} \left( \vec{k} \right) = \nu_f \delta_{\alpha \alpha'} \delta_{\eta \eta'} \delta_{s s'}.
\end{equation}

\subsubsection{Many-body charge conjugation}\label{app:sec:many_body_rev:ground_states:ph_symmetry}

Given that the THF models for both TBG and TSTG possess many-body (unitary) charge conjugation symmetry~\cite{SON19,SON21,BER21a,CAL21,SON22,YU23a}, we can restrict, without loss of generality, to positive fillings with $\nu \geq 0$~\cite{CAL23b}. For concreteness, the many-body charge conjugation operator $\mathcal{P}$ is defined for TBG as the single-particle transformation $C_{2z}TP$ (where $P$, $T$, and $C_{2z}$ are the unitary particle-hole, the antiunitary time-reversal, and two-fold $z$-directed unitary rotation symmetry operations) followed by an exchange of the creation and annihilation operators~\cite{BER21a,SON22}. TSTG at a finite displacement field does not maintain this symmetry. Instead, it exhibits a \emph{spatial} many-body charge conjugation symmetry, represented by $\mathcal{P}'$~\cite{CAL21}. The latter is defined as the combined antiunitary single-particle transformation $m_z C_{2x} C_{2z} T P$ (where $C_{2x}$ denotes a two-folder rotation around the $x$ axis and $m_z$ is the mirror symmetry operator perpendicular to the $z$ axis) followed by an interchange between the creation and annihilation operators~\cite{CAL21}. The action of the two many-body charge conjugation symmetries on the TBG and TSTG fermions within the THF model is given, respectively, by
\begin{alignat}{6}
	\mathcal{P} \hat{\gamma}^\dagger_{\vec{k},i,\eta,s} \mathcal{P}^{-1} &=&& \sum_{i',\eta'} \left[ D \left( \mathcal{P} \right) \right]_{i' \eta'; i \eta} \hat{\gamma}_{-\vec{k},i',\eta',s}, \quad&\text{with}&\quad & \commutator{H^{\text{TBG}}}{\mathcal{P}} &=& 0, \label{app:eqn:ph_sym_act_TBG} \\
	\mathcal{P}' \hat{\gamma}^\dagger_{\vec{k},i,\eta,s} \mathcal{P}^{\prime-1} &=&& \sum_{i',\eta'} \left[ D \left( \mathcal{P}' \right) \right]_{i' \eta'; i \eta} \hat{\gamma}_{-C_{2x} \vec{k},i',\eta',s}, \quad&\text{with}&\quad & \commutator{H^{\text{TSTG}}}{\mathcal{P}'} &=& 0. \label{app:eqn:ph_sym_act_TSTG}
\end{alignat}
In \cref{app:eqn:ph_sym_act_TBG,app:eqn:ph_sym_act_TSTG}, the representation matrices of $\mathcal{P}$ and $\mathcal{P}'$ are defined as
\begin{align}
	D \left( \mathcal{P} \right) &= \left[ \sigma_y \oplus \sigma_y \oplus \left( - \sigma_y \right) \right] \tau_z, \label{app:eqn:ph_sym_rep_TBG} \\
	D \left( \mathcal{P}' \right) &= \left[ i\sigma_z \oplus i\sigma_z \oplus \left( - i\sigma_z \right) \oplus \left( - \sigma_0 \right) \right] \tau_z, \label{app:eqn:ph_sym_rep_TSTG}
\end{align}
where $\sigma_{\mu}$ ($\mu=0,x,y,z$) represent the Pauli and identity matrices acting on the $i=1,2$, $i=3,4$, and $i=5,6$, orbital subspaces (and additionally in the $i=7,8$ orbital subspace in the case of TSTG), while $\tau_{\mu}$ ($\mu=0,x,y,z$) are the Pauli and identity matrices operating on the valley subspace. 

The two charge-conjugation operators $\mathcal{P}$ and $\mathcal{P}'$ map states at filling $\nu$ to states at filling $-\nu$. Alternatively, in the grand canonical ensemble, the charge conjugation operators map phases at chemical potential $\mu = \mu_0$ to ones at chemical potential $\mu = -\mu_0$. Denoting by $A^{\pm \mu_0} \left( \omega, \vec{k} \right)$ the spectral function of the system at chemical potential $\pm \mu_0$, it was shown in \cref*{DMFT:app:sec:hartree_fock:ground_states:ph_symmetry} of Ref.~\cite{CAL23b} that 
\begin{align}
	A^{\mu_0}_{i \eta s; i' \eta' s'} \left( \omega, \vec{k} \right) &= \sum_{\substack{i_1, \eta_1 \\ i_2, \eta_2}} \left[ D \left( \mathcal{P} \right) \right]^{*}_{i_1 \eta_1; i \eta} \left[ D \left( \mathcal{P} \right) \right]_{i_2 \eta_2; i' \eta'} A^{-\mu_0}_{i_2 \eta_2 s'; i_1 \eta_1 s} \left( - \omega, - \vec{k} \right), \label{app:eqn:TBG_sp_func_ph_sym} \\
	A^{\mu_0}_{i \eta s; i' \eta' s'} \left( \omega, \vec{k} \right) &= \sum_{\substack{i_1, \eta_1 \\ i_2, \eta_2}} \left[ D \left( \mathcal{P}' \right) \right]^{*}_{i_1 \eta_1; i \eta} \left[ D \left( \mathcal{P}' \right) \right]_{i_2 \eta_2; i' \eta'} A^{-\mu_0}_{i_2 \eta_2 s'; i_1 \eta_1 s} \left( - \omega, - C_{2x} \vec{k} \right), \label{app:eqn:TSTG_sp_func_ph_sym}
\end{align} 
for TBG and TSTG, respectively. As a result and without loss of generality, we will only consider the positive integer-filled insulators from \cref{app:tab:model_states}, as well as positive fillings in the symmetric phase. The spectral functions at negative fillings can be readily obtained through \cref{app:eqn:TBG_sp_func_ph_sym,app:eqn:TSTG_sp_func_ph_sym}

\section{Derivation of the particle and energy currents in twisted graphene heterostructures}\label{app:sec:deriv_moire_cur}

The THF model Hamiltonian for TBG or TSTG is obtained by projecting the corresponding BM model Hamiltonian in an appropriately constructed basis of heavy and conduction fermions, as reviewed in 
\cref{app:sec:HF_review}. The goal of this \siSection{} is to obtain also obtain the energy and particle currents of TBG and TSTG within their respective THF model bases. To do so, we start from the formulae of the currents derived in \cref{app:sec:cur_deriv_cont}. Being valid for generic fermionic systems in the continuum with density-density interactions, the former can be directly applied to the BM model Hamiltonians to obtain the corresponding currents. By projecting the resulting current operators into the THF model basis, we then obtain the low-energy particle and energy current operators, which characterize the physics of TBG and TSTG around charge neutrality. 

For the particle current $\vec{J}$ and the single-particle contribution to the energy current $\vec{J}^{T}_E$, we find that, to an excellent approximation, the corresponding operator only depends on the single-particle parameters of the THF model Hamiltonian, but not on the $f$- or $c$-electron wave function. Furthermore, we find that the quartic interaction contribution to the energy current $\vec{J}^{V}_E$ contributes at most $4 \%$ of the single-particle one, and as a result can be completely ignored. Physically, this is a consequence of the large Fermi velocity of the $c$-electrons (which enhances $\vec{J}^{T}_E$) and the short-range nature of the interaction (which diminishes $\vec{J}^{V}_E$).

\subsection{Derivation of the particle current}\label{app:sec:deriv_moire_cur:particle}

To avoid confusion, in what follows, we will employ a similar convention to \cref{app:sec:BM_review,app:sec:HF_review}, and label the currents within the BM (THF) model with (without) a hat. We start by deriving the particle current within the BM model using \cref{app:eqn:part_current_simple_real_general}. The resulting operator is then projected into the THF model basis by using \cref{app:eqn:proj_to_THF_basis}. 

\subsubsection{The particle current within the BM model}\label{app:sec:deriv_moire_cur:particle:BM}
From the single-particle BM Hamiltonian for TBG in \cref{app:eqn:spHamiltonian_BM_TBG}, we find using \cref{app:eqn:part_current_simple_real_general} that the TBG particle current within the BM basis is given by 
\begin{equation}
	\label{app:eqn:BM_curent_part_TBG}
	\hat{\vec{J}}^{\text{TBG}} =  \sum_{\vec{k}} \sum_{\eta, \alpha, \beta, s} \sum_{\vec{Q},\vec{Q}' \in \mathcal{Q}_{\pm}} \left[\pdv{h^{\text{TBG},\eta}_{\vec{Q},\vec{Q}'} \left( \vec{k} \right)}{\vec{k}} \right]_{\alpha \beta} \hat{c}^\dagger_{\vec{k},\vec{Q},\eta,\alpha,s} \hat{c}_{\vec{k},\vec{Q}',\eta,\beta,s},
\end{equation}
where the single-particle Hamiltonian matrix $h^{\text{TBG},\eta}_{\vec{Q},\vec{Q}'}$ was defined in \cref{app:eqn:BM_TBG_ham}. For TSTG, there is an additional contribution stemming from the mirror-odd fermions
\begin{equation}
	\label{app:eqn:BM_curent_part_TSTG}
	\hat{\vec{J}}^{\text{TSTG}} = \hat{\vec{J}}^{\text{TBG}} + \sum_{\vec{k}} \sum_{\eta, \alpha, \beta, s} \sum_{\vec{Q} \in \mathcal{Q}_{\eta}} \left[\pdv{h^{\text{D},\eta}_{\vec{Q}} \left( \vec{k} \right)}{\vec{k}} \right]_{\alpha \beta} \hat{b}^\dagger_{\vec{k},\vec{Q},\eta,\alpha,s} \hat{b}_{\vec{k},\vec{Q},\eta,\beta,s},
\end{equation}
with the Dirac cone Hamiltonian $h^{\text{D},\eta}_{\vec{Q}} \left( \vec{k} \right)$ being given in \cref{app:eqn:BM_dirac_ham}. Note that the displacement field contribution from \cref{app:eqn:BM_TSTG_displacement_field_ham} is momentum independent and, as a result, does not generate any additional current terms. For completeness, we note that the expressions for the particle current can be simplified by explicitly evaluating the partial derivative to afford 
\begin{align}
	\hat{\vec{J}}^{\text{TBG}} &=  \sum_{\vec{k}} \sum_{\eta, \alpha, \beta, s} \sum_{\vec{Q} \in \mathcal{Q}_{\pm}} \left[ v_F \boldsymbol{\sigma} \right]_{\alpha \beta} \hat{c}^\dagger_{\vec{k},\vec{Q},\eta,\alpha,s} \hat{c}_{\vec{k},\vec{Q},\eta,\beta,s}, \\
	\hat{\vec{J}}^{\text{TSTG}} &= \hat{\vec{J}}^{\text{TBG}} +  \sum_{\vec{k}} \sum_{\eta, \alpha, \beta, s} \sum_{\vec{Q} \in \mathcal{Q}_{\eta}} \left[ v_F \boldsymbol{\sigma} \right]_{\alpha \beta} \hat{b}^\dagger_{\vec{k},\vec{Q},\eta,\alpha,s} \hat{b}_{\vec{k},\vec{Q},\eta,\beta,s}.
\end{align}
Because they are written in terms of the BM model Hamiltonian, in \cref{app:sec:deriv_moire_cur:particle:THF}, we will employ the formulae from \cref{app:eqn:BM_curent_part_TBG,app:eqn:BM_curent_part_TSTG} to project the particle current operators in the THF basis. 

\subsubsection{The particle current within the THF model}\label{app:sec:deriv_moire_cur:particle:THF}

With the particle currents in the BM basis at hand, we now derive their low-energy counterparts by projecting them into the THF basis. Starting with the TBG current, we plug \cref{app:eqn:proj_to_THF_basis} into \cref{app:eqn:BM_curent_part_TBG} to obtain
\begin{equation}
	\label{app:eqn:BM_curent_part_TBG_projected_complicated}
	\vec{J}^{\text{TBG}} =  \sum_{1 \leq i,j \leq 6} \sum_{\abs{\vec{k}} \leq \Lambda_{ij}} \sum_{\substack{\eta,s \\ \beta, \beta'}} \sum_{\vec{Q},\vec{Q}' \in \mathcal{Q}_{\pm}} \left[\pdv{h^{\text{TBG},\eta}_{\vec{Q},\vec{Q}'} \left( \vec{k} \right)}{\vec{k}} \right]_{\beta \beta'} \phi^{*\eta}_{\vec{Q}\beta;i} \left( \vec{k} \right) \phi^{\eta}_{\vec{Q}'\beta';j} \left( \vec{k} \right) \hat{\gamma}^\dagger_{\vec{k},i, \eta, s} \hat{\gamma}_{\vec{k},j,\eta,s}, 
\end{equation} 
where we have employed the notation from \cref{app:eqn:shorthand_gamma_not} for the wave functions $\phi^{*\eta}_{\vec{Q}\beta;i} \left( \vec{k} \right)$ and operators $\hat{\gamma}^\dagger_{\vec{k},i, \eta, s}$. The momentum cutoff $\Lambda_{ij}$ is such that the summation over $\vec{k}$ runs over the entire moir\'e Brillouin zone when \emph{both} the $i$ and $j$ indices correspond to $f$-fermions ({\it i.e.}{}, $5 \leq i,j \leq 6$), but is limited for momenta smaller than $\Lambda_c$ whenever \emph{either} one of $i$ or $j$ corresponds to $c$-electrons ({\it i.e.}{}, $1 \leq i \leq 4$ or $1 \leq j \leq 4$). The expression \cref{app:eqn:BM_curent_part_TBG_projected_complicated} is quite cumbersome, as it also depends on the wave functions of the $c$- and $f$-electrons, rather than only on the THF single-particle parameters from \cref{app:sec:HF_review:single_particle:TBG} ({\it i.e.}{}, $\gamma$, $M$, $v_{\star}$ etc.). In fact, \textit{a priori}, it is not evident that the low-energy current can be described \emph{exclusively} using these parameters -- it depends on the momentum derivative of the BM model Hamiltonian and the wave functions of the $f$- and $c$-electrons expressed in \emph{the BM model basis}. The \emph{exact} form of the latter in terms of just the THF single-particle parameters is not known. However, in what follows, we will demonstrate that, with remarkable accuracy, the low-energy particle current \emph{can} indeed be expressed solely in terms of the THF single-particle parameters ({\it i.e.}{} without the $f$- and $c$-electron wave functions expressed in the BM model basis).

To this end, we first extend the definition of the $f$-electron nonabelian Berry connection \cref{app:eqn:f_nonabelian_berry} by introducing the following nonabelian Berry connections involving the $c$-electron wave functions
\begin{alignat}{3}
	\mathbfcal{A}^{\eta,cc}_{a a'} \left( \vec{k} \right) &=& 
	\sum_{\vec{Q},\beta} \tilde{u}^{* \eta}_{\vec{Q}\beta;a} \left( \vec{k} \right) \partial_{\vec{k}} \tilde{u}^{\eta}_{\vec{Q}\beta;a'} \left( \vec{k} \right)
	&=&-\sum_{\vec{Q},\beta} \left(\partial_{\vec{k}} \tilde{u}^{* \eta}_{\vec{Q}\beta;a} \left( \vec{k} \right) \right) \tilde{u}^{\eta}_{\vec{Q}\beta;a'} \left( \vec{k} \right), \label{app:eqn:c_nonabelian_berry} \\
	\mathbfcal{A}^{\eta,fc}_{\alpha a'} \left( \vec{k} \right) &=& \sum_{\vec{Q},\beta} \tilde{u}^{* \eta}_{\vec{Q}\beta;\alpha} \left( \vec{k} \right) \partial_{\vec{k}} v^{\eta}_{\vec{Q}\beta;a'} \left( \vec{k} \right)
	&=&-\sum_{\vec{Q},\beta} \left(\partial_{\vec{k}} \tilde{u}^{* \eta}_{\vec{Q}\beta;\alpha} \left( \vec{k} \right) \right) v^{\eta}_{\vec{Q}\beta;a'} \left( \vec{k} \right). \label{app:eqn:fc_nonabelian_berry}
\end{alignat}
The Berry connection for both the $c$- and the $f$-electron species can also be defined in analogy with \cref{app:eqn:full_THF_Hamiltonian_gamma_bas}
\begin{equation}
	\label{app:eqn:full_Berry_Hamiltonian_gamma_bas}
	\mathbfcal{A}_{ij}^{\eta} \left( \vec{k} \right) = \sum_{\vec{Q},\beta} \phi^{* \eta}_{\vec{Q}\beta;i} \left( \vec{k} \right) \partial_{\vec{k}} \phi^{\eta}_{\vec{Q}\beta;j} \left( \vec{k} \right), \quad
	\mathbfcal{A}^{\eta} \left( \vec{k} \right) = \begin{pNiceMatrix}[last-col=3,first-row]
		4 & 2 \\
		\mathbfcal{A}^{\eta,cc} \left( \vec{k} \right) & \mathbfcal{A}^{\eta,cf} \left( \vec{k} \right) & 4 \\
		-\mathbfcal{A}^{\dagger \eta,cf} \left( \vec{k} \right) & \mathbfcal{A}^{\eta,ff} \left( \vec{k} \right)  & 2 \\
	\end{pNiceMatrix}.
\end{equation}
We also note that when restricting its action on the $f$- and $c$-electron wave functions, the single-particle TBG Hamiltonian matrix can be rewritten as 
\begin{equation}
	\label{app:eqn:bm_ham_using_completeness}
	\left[h^{\text{TBG},\eta}_{\vec{Q},\vec{Q}'} \left( \vec{k} \right) \right]_{\beta \beta'} \approx \sum_{1 \leq i,j \leq 6} h^{\text{TBG},\eta}_{ij} \left( \vec{k} \right) \phi^{\eta}_{\vec{Q}\beta;i} \left( \vec{k} \right) \phi^{*\eta}_{\vec{Q}'\beta';j} \left( \vec{k} \right),
\end{equation}
which follows directly from \cref{app:eqn:proj_in_sp_TBG_THF_c,app:eqn:proj_in_sp_TBG_THF_cf,app:eqn:proj_in_sp_TBG_THF_f}. We note that \cref{app:eqn:bm_ham_using_completeness} is just the inverse of \cref{app:eqn:proj_in_sp_TBG_THF_c,app:eqn:proj_in_sp_TBG_THF_cf,app:eqn:proj_in_sp_TBG_THF_f}. Using the product rule, the orthonormality of the $f$- and $c$-electron wave functions, as well as \cref{app:eqn:full_Berry_Hamiltonian_gamma_bas,app:eqn:bm_ham_using_completeness}, one can express the matrix elements in \cref{app:eqn:BM_curent_part_TBG_projected_complicated} in terms of derivatives of the single-particle THF Hamiltonian together with Berry connection contributions
\begin{align}
	&\sum_{\substack{\beta, \beta' \\ \vec{Q},\vec{Q}' \in \mathcal{Q}_{\pm}}} \left[\partial_{\vec{k}} h^{\text{TBG},\eta}_{\vec{Q},\vec{Q}'} \left( \vec{k} \right) \right]_{\beta \beta'} \phi^{*\eta}_{\vec{Q}\beta;i} \left( \vec{k} \right) \phi^{\eta}_{\vec{Q}'\beta';j} \left( \vec{k} \right) = \nonumber \\
	=&\sum_{\substack{\beta, \beta',i',j' \\ \vec{Q},\vec{Q}' \in \mathcal{Q}_{\pm}}} \left[
		\left( \partial_{\vec{k}} h^{\text{TBG},\eta}_{i'j'} \left( \vec{k} \right) \right) \phi^{\eta}_{\vec{Q}\beta;i'} \left( \vec{k} \right) \phi^{*\eta}_{\vec{Q}'\beta';j'} \left( \vec{k} \right) 
		+ h^{\text{TBG},\eta}_{i'j'} \left( \vec{k} \right) \left( \partial_{\vec{k}} \phi^{\eta}_{\vec{Q}\beta;i'} \left( \vec{k} \right) \right) \phi^{*\eta}_{\vec{Q}'\beta';j'} \left( \vec{k} \right) \right. \nonumber \\ 
	&\left.+h^{\text{TBG},\eta}_{i'j'} \left( \vec{k} \right) \phi^{\eta}_{\vec{Q}\beta;i'} \left( \vec{k} \right) \left( \partial_{\vec{k}}  \phi^{*\eta}_{\vec{Q}'\beta';j'} \left( \vec{k} \right) \right)
	\right] \phi^{*\eta}_{\vec{Q}\beta;i} \left( \vec{k} \right) \phi^{\eta}_{\vec{Q}'\beta';j} \left( \vec{k} \right) \nonumber \\
	=&\partial_{\vec{k}} h^{\text{TBG},\eta}_{ij} \left( \vec{k} \right)
		+ \sum_{i'} \left( \mathbfcal{A}^{\eta}_{i i'} \left( \vec{k} \right) h^{\text{TBG},\eta}_{i'j} \left( \vec{k} \right) 
		- h^{\text{TBG},\eta}_{ii'} \left( \vec{k} \right) \mathbfcal{A}^{\eta}_{i' j} \left( \vec{k} \right) \right). 
\end{align}
As a result, the single-particle current $\vec{J}^{\text{TBG}}$ can be written as 
\begin{equation}
	\label{app:eqn:BM_curent_part_TBG_projected_complicated_2}
	\vec{J}^{\text{TBG}} =  \sum_{\eta,s} \sum_{1 \leq i,j \leq 6} \sum_{\abs{\vec{k}} \leq \Lambda_{ij}} \left[ 
		\partial_{\vec{k}} h^{\text{TBG},\eta}_{ij} \left( \vec{k} \right)
		+ \sum_{i'} \left( \mathbfcal{A}^{\eta}_{i i'} \left( \vec{k} \right) h^{\text{TBG},\eta}_{i'j} \left( \vec{k} \right) 
		- h^{\text{TBG},\eta}_{ii'} \left( \vec{k} \right) \mathbfcal{A}^{\eta}_{i' j} \left( \vec{k} \right) \right) 
	\right] \hat{\gamma}^\dagger_{\vec{k},i, \eta, s} \hat{\gamma}_{\vec{k},j,\eta,s}. 
\end{equation}

In order to further simplify the expression in \cref{app:eqn:BM_curent_part_TBG_projected_complicated_2}, we note that Ref.~\cite{SON22} approximates the wave function of the $c$-electron near $\Gamma_M$ by its value \emph{at} the $\Gamma_M$ point, as reviewed in \cref{app:eqn:approx_of_c_wavf}. By substituting \cref{app:eqn:approx_of_c_wavf} into \cref{app:eqn:c_nonabelian_berry}, we find that, in the same spirit, the Berry connection of the $c$-electron wave functions in the BM model basis is negligible\footnote{The THF model Hamiltonian is obtained by projecting the BM model Hamiltonian in the THF model basis. What we are ignoring here is the nonabelian Berry connection \emph{of the projection basis}. The Berry connection of the $c$-electrons is not ignored, as shown in \cref{app:eqn:non_ignoring_berry}.}
\begin{equation}
	\mathbfcal{A}^{\eta,cc}_{a a'} \left( \vec{k} \right) \approx 0 \qq{and} \mathbfcal{A}^{\eta,fc}_{\alpha a'} \left( \vec{k} \right) \approx 0 \qq{for} \vec{k} \leq \Lambda_c,
\end{equation} 
which leads to the following expression for the particle current
\begin{align}
	\vec{J}^{\text{TBG}} &= \sum_{\vec{k} \leq \Lambda_c} \left[ \sum_{\substack{a,a' \\ \eta, s}} \partial_{\vec{k}} h_{a a'}^{cc,\eta} \left( \vec{k} \right)	\hat{c}^\dagger_{\vec{k},a,\eta,s} \hat{c}_{\vec{k},a',\eta,s}  
	+ \sum_{\substack{a, \alpha' \\ \eta, s}} 
	\left( \partial_{\vec{k}} h_{a \alpha'}^{cf,\eta} \left( \vec{k} \right) \hat{c}^\dagger_{\vec{k},a,\eta,s} \hat{f}_{\vec{k},\alpha',\eta,s} + \text{h.c.} \right) \right] \nonumber \\
	&- \sum_{\vec{k} \leq \Lambda_c} \sum_{\substack{a,\alpha ',\alpha'' \\ \eta, s}} \left( h_{a \alpha''}^{cf,\eta} \left( \vec{k} \right) \mathbfcal{A}^{\eta,ff}_{\alpha '' \alpha'} \left( \vec{k} \right)	\hat{c}^\dagger_{\vec{k},a,\eta,s} \hat{f}_{\vec{k},\alpha',\eta,s} + \text{h.c.}  \right) \nonumber \\
	&+ \sum_{\vec{k}} \sum_{\substack{\alpha,\alpha '\\ \eta, s}} \partial_{\vec{k}} h_{\alpha \alpha'}^{ff,\eta} \left( \vec{k} \right) \hat{f}^\dagger_{\vec{k},\alpha,\eta,s} \hat{f}_{\vec{k},\alpha',\eta,s} \nonumber \\
	&+ \sum_{\vec{k}} \sum_{\substack{\alpha,\alpha ',\alpha'' \\ \eta, s}} \left( \mathbfcal{A}^{\eta,ff}_{\alpha \alpha''} \left( \vec{k} \right) h_{\alpha'' \alpha'}^{ff,\eta} \left( \vec{k} \right)  - h_{\alpha \alpha''}^{ff,\eta} \left( \vec{k} \right) \mathbfcal{A}^{\eta,ff}_{\alpha '' \alpha'} \left( \vec{k} \right)  \right) \hat{f}^\dagger_{\vec{k},\alpha,\eta,s} \hat{f}_{\vec{k},\alpha',\eta,s}. \label{app:eqn:BM_curent_part_TBG_projected_partial_approx} 
\end{align}
We will now discuss each term in \cref{app:eqn:BM_curent_part_TBG_projected_partial_approx} individually. For starters, the leading contributions in the first row of \cref{app:eqn:BM_curent_part_TBG_projected_partial_approx} are proportional to $v_{*}$ and $v'_{*}$, respectively. The contribution from the second row is of order $\mathcal{O} \left[ \left( \gamma + \frac{v'_{*}}{\sqrt{\Omega_0}} \right) \sum_{i=1}^{2} \frac{\lambda_i^3}{\Omega_0} \right]$, where we have used the bound of the $f$-electron Berry connection derived in \cref{app:eqn:f_berry_connection_bound} and $\lambda_i$ are the spread parameters of the $f$-fermions in real space (with $\lambda_i \sim 1/5 \sqrt{\Omega_0}$ for $i=1,2$). For the parameters chosen in \cref{app:sec:HF_review:single_particle}, we note that $\gamma \sqrt{\Omega_0} \sim v'_{*} $, implying that $\left( \gamma + \frac{v'_{*}}{\sqrt{\Omega_0}} \right) \sum_{i=1}^{2} \frac{\lambda_i^3}{\Omega_0} \sim \sum_{i=1}^{2} \left(\frac{\lambda_i}{\sqrt{\Omega_0}} \right)^3 v'_{*} \sim 10^{-2} v'_{*}$. Therefore, the second row in \cref{app:eqn:BM_curent_part_TBG_projected_partial_approx} can be neglected with respect to the first one. 

Both the third and fourth rows of \cref{app:eqn:BM_curent_part_TBG_projected_partial_approx} are proportional to the nearest neighbor hopping of the $f$-fermions $t$, but scale differently with the spread of the $f$-electron wave function, being of order $\mathcal{O} \left(t\sqrt{\Omega_0} \right)$ and  $\mathcal{O} \left( \sum_{i=1}^{2} \frac{t \lambda_i^3}{\Omega_0} \right)$, respectively. At the magic angle, where $t$ vanishes~\cite{SON22,CAL23}, we can completely neglect the last two rows of \cref{app:eqn:BM_curent_part_TBG_projected_partial_approx}. Although not relevant for this work, we also note that \emph{away} from the magic-angle, where the kinetic Hamiltonian of the $f$-electrons (and hence $t$) is non-negligible and the third row cannot be ignored, the contribution in the fourth row can still be discarded as it is suppressed relative to the former by a factor of $\left(\frac{\lambda_i}{\sqrt{\Omega_0}} \right)^3 \sim 10^{-2}$ (for $i=1,2$), due to the localized nature of the $f$-electrons across a wide range of parameters around the magic angle~\cite{CAL23}.

With all these approximations taken into account, the final expression of the particle current in TBG valid around the magic angle is given simply by 
\begin{equation}
	\vec{J}^{\text{TBG}} = \sum_{\vec{k} \leq \Lambda_c} \left[ \sum_{\substack{a,a' \\ \eta, s}} \partial_{\vec{k}} h_{a a'}^{cc,\eta} \left( \vec{k} \right)	\hat{c}^\dagger_{\vec{k},a,\eta,s} \hat{c}_{\vec{k},a',\eta,s}  
	+ \sum_{\substack{a, \alpha' \\ \eta, s}} 
	\left( \partial_{\vec{k}} h_{a \alpha'}^{cf,\eta} \left( \vec{k} \right) \hat{c}^\dagger_{\vec{k},a,\eta,s} \hat{f}_{\vec{k},\alpha',\eta,s} + \text{h.c.} \right) \right], \label{app:eqn:THF_curent_part_TBG_simple}
\end{equation}
which can indeed be expressed only in terms of the single-particle parameters of the corresponding THF Hamiltonian.

Finally, we note that the nonabelian Berry connection of the $c$-electrons is not ignored in \cref{app:eqn:THF_curent_part_TBG_simple}. To see this, we define the band basis of the $c$-electron Hamiltonian 
\begin{equation}
	\sum_{a'=1}^{4} h^{cc,\eta}_{aa'} \left( \vec{k} \right) \psi_{n,\eta,a'} \left( \vec{k} \right) = \epsilon^{\eta}_{n} \left( \vec{k} \right) \psi_{n,\eta,a} \left( \vec{k} \right), \qq{for} 1\leq n \leq 4,
\end{equation}
where $\epsilon^{\eta}_{n} \left( \vec{k} \right)$ and $\psi_{n,\eta,a} \left( \vec{k} \right)$ are, respectively, the energy and wave function of the $n$-th $c$-electron band at $\vec{k}$. We then take the corresponding band basis operators to be
\begin{equation}
	\hat{e}^\dagger_{\vec{k},n,\eta,s} \equiv \sum_{a=1}^{4} \psi_{n,\eta,a} \left( \vec{k} \right) \hat{c}^\dagger_{\vec{k},a,\eta,s}, \qq{for} 1 \leq n \leq 4,
\end{equation} 
such that the first term of the current operator can be rewritten as 
\begin{align}
	\sum_{\substack{a,a' \\ \eta, s}} \partial_{\vec{k}} h_{a a'}^{cc,\eta} \left( \vec{k} \right)	\hat{c}^\dagger_{\vec{k},a,\eta,s} \hat{c}_{\vec{k},a',\eta,s} = & \sum_{\eta,s} \sum_{n=1}^{4}  \partial_{\vec{k}} \epsilon^{\eta}_{n} \left( \vec{k} \right)	\hat{e}^\dagger_{\vec{k},n,\eta,s} \hat{e}_{\vec{k},n,\eta,s} \nonumber \\ 
	+ & \sum_{\eta,s} \sum_{n,n'=1}^{4} \left( \epsilon^{\eta}_{n} \left( \vec{k} \right) - \epsilon^{\eta}_{n'} \left( \vec{k} \right) \right) \mathbfcal{A}^{\eta,e}_{n n'} \left( \vec{k} \right) \hat{e}^\dagger_{\vec{k},n',\eta,s}  \hat{e}_{\vec{k},n,\eta,s},
	\label{app:eqn:non_ignoring_berry}
\end{align}
where 
\begin{equation}
	\mathbfcal{A}^{\eta,e}_{n n'} \left( \vec{k} \right) = \sum_{a=1}^{4} \psi^{*}_{n,\eta,a} \left(\vec{k} \right) \partial_{\vec{k}} \psi_{n',\eta,a} \left(\vec{k} \right)
\end{equation}
is the nonabelian Berry connection of the $c$-electrons.

To extend \cref{app:eqn:THF_curent_part_TBG_simple} to TSTG, we note that the latter only includes an additional contribution stemming from the mirror-odd fermions and given by the second term of \cref{app:eqn:BM_curent_part_TSTG}. The projection of the mirror-odd fermions is done by simply imposing a cutoff in their momentum. As a result, we can readily write the expression of the TSTG current
\begin{equation}
	\vec{J}^{\text{TSTG}} = \vec{J}^{\text{TBG}} + \sum_{\substack{\vec{p} \leq \Lambda_d \\ \alpha,\alpha',\eta,s}} \partial_{\vec{p}} h^{dd,\eta}_{\alpha \alpha'} \left( \vec{p} \right) \hat{d}^\dagger_{\vec{p},\alpha,\eta,s} \hat{d}_{\vec{p},\alpha',\eta,s}. \label{app:eqn:THF_curent_part_TSTG_simple}
\end{equation}

\subsection{Derivation of the single-particle contribution to the energy current \texorpdfstring{$\vec{J}^{T}_{E}$}{JTE}}\label{app:sec:deriv_moire_cur:energy_sp_part}

The energy current derived in \cref{app:eqn:en_current_simple_real_general} has two contributions: one stemming from the single-particle part of the Hamiltonian and one, from the interaction part. In this section, we discuss the single-particle contribution, whose derivation closely follows the one of the particle current from \cref{app:sec:deriv_moire_cur:particle}. The interaction part of the energy current will be derived in \cref{app:sec:deriv_moire_cur:energy_interaction}.

We note that the single-particle contribution to the energy current $\vec{J}^{T}_{E}$ from \cref{app:eqn:en_current_simple_real_general_sin} only differs from the particle current by an overall factor of $(-i)$ and one time derivative. As a result, we can employ all the derivations and approximations of \cref{app:sec:deriv_moire_cur:particle} and directly write
\begin{align}
	\vec{J}^{T,\text{TBG}}_{E} =& \frac{1}{2} \sum_{\vec{k} \leq \Lambda_c} \sum_{\substack{a,a' \\ \eta, s}} \left( i \partial_{\vec{k}} h_{a a'}^{cc,\eta} \left( \vec{k} \right)	\hat{c}^\dagger_{\vec{k},a,\eta,s} \dot{\hat{c}}_{\vec{k},a',\eta,s} + \text{h.c.} \right)  \nonumber \\
	+& \frac{1}{2} \sum_{\vec{k} \leq \Lambda_c} \sum_{\substack{a, \alpha' \\ \eta, s}} 
	\left[ 
		\left( i \partial_{\vec{k}} h_{a \alpha'}^{cf,\eta} \left( \vec{k} \right) \hat{c}^\dagger_{\vec{k},a,\eta,s} \dot{\hat{f}}_{\vec{k},\alpha',\eta,s} +i  \partial_{\vec{k}} h_{a \alpha'}^{*cf,\eta} \left( \vec{k} \right) \hat{f}^\dagger_{\vec{k},\alpha',\eta,s} \dot{\hat{c}}_{\vec{k},a,\eta,s} \right) 
		+ \text{h.c.}
	\right], \label{app:eqn:THF_en_current_sp_part_TBG_simple}\\
	\vec{J}^{T,\text{TSTG}}_{E} =& \vec{J}^{T,\text{TBG}}_{E} + \frac{1}{2} \sum_{\substack{\vec{p} \leq \Lambda_d \\ \alpha,\alpha',\eta,s}} \left( i\partial_{\vec{p}} h^{dd,\eta}_{\alpha \alpha'} \left( \vec{p} \right) \hat{d}^\dagger_{\vec{p},\alpha,\eta,s} \dot{\hat{d}}_{\vec{p},\alpha',\eta,s} + \text{h.c.} \right). \label{app:eqn:THF_en_current_sp_part_TSTG_simple}
\end{align}
One additional approximation is that in \cref{app:eqn:THF_en_current_sp_part_TBG_simple,app:eqn:THF_en_current_sp_part_TSTG_simple} the time-derivatives will be evaluated within the corresponding low-energy THF models, {\it i.e.}{} for any operator $\hat{\mathcal{O}}$
\begin{equation}
	\dot{\hat{\mathcal{O}}} = i \commutator{H}{\hat{\mathcal{O}}},
\end{equation} 
where $H$ is the THF model interacting Hamiltonian.

\subsection{Derivation of the interaction contribution to the energy current \texorpdfstring{$\vec{J}^{V}_{E}$}{JVE}}\label{app:sec:deriv_moire_cur:energy_interaction}

Having discussed the single-particle contribution to the energy current in \cref{app:sec:deriv_moire_cur:energy_sp_part}, we now turn to the interaction one. The interaction contribution to the energy current is a formally quartic operator. After deriving its formula for TBG and TSTG, we employ a Hartree-Fock decoupling procedure to obtain its mean-field approximation. We then show numerically that for the correlated ground state candidates considered in \cref{app:tab:model_states}, the interaction contribution to the energy current is at most $4\%$ of the single-particle one.

\subsubsection{The interaction contribution to the energy current for TBG and TSTG}\label{app:sec:deriv_moire_cur:energy_interaction:deriv}

The interaction contribution to the energy current within the BM model can be readily obtained by substituting the expression of the density operators from \cref{app:eqn:BM_offset_density_op} for TBG and from \cref{app:eqn:BM_offset_density_op_TSTG_TBG,app:eqn:BM_offset_density_op_TSTG_D} for TSTG into \cref{app:eqn:en_current_simple_real_general_interaction}. In what follows, we will project the resulting expressions in the THF basis.

Starting with the case of TBG, we note that the corresponding density operator can be projected into the THF basis using the form-factor expressions from \cref{app:eqn:projection_of_TBG_rho_to_THF}. With the notation from \cref{app:eqn:shorthand_gamma_not}, the interaction contribution to the energy current is then given by
\begin{align}
	\label{app:eqn:BM_curent_en_int_TBG_projected_complicated}
	\vec{J}^{V,\text{TBG}}_{E} =& \frac{-i}{2N_0 \Omega_0} \sum_{\substack{\eta,s,\eta',s' \\ 1 \leq i,j,k,l \leq 6}} \sum_{\vec{q}, \vec{G}} \sum_{\substack{\abs{\vec{k}'_1} \leq{\Lambda_i} \abs{\vec{k}_1} \leq \Lambda_{j} \\ \abs{\vec{k}'_2} \leq \Lambda_{l},\abs{\vec{k}_2} \leq \Lambda_{k}}} \delta_{\vec{k}'_2,\vec{k}_2 - \vec{q}} \delta_{\vec{k}'_1,\vec{k}_1 + \vec{q}} \pdv{V \left( \vec{q} + \vec{G} \right)}{\vec{q}} 
	\mathcal{M}^{\eta}_{ij}\left(\vec{k}_1,\vec{q}+\vec{G} \right) \mathcal{M}^{\eta'}_{kl}\left(\vec{k}_2, -\vec{q} - \vec{G} \right)
	\nonumber \\
	&\times   
	\left[ \dv{t}\left( \hat{\gamma}^\dagger_{\vec{k}'_1 ,i,\eta,s} \hat{\gamma}_{\vec{k}_1,j,\eta,s} - \frac{1}{2}\delta_{\vec{G},\vec{0}} \delta_{\vec{q},\vec{0}} \delta_{ij} \right) \right] 
	\left( \hat{\gamma}^\dagger_{\vec{k}'_2,k,\eta',s'} \hat{\gamma}_{\vec{k}_2,l,\eta',s'} - \frac{1}{2}\delta_{\vec{G},\vec{0}} \delta_{\vec{q},\vec{0}} \delta_{kl} \right),
\end{align}
where the index-dependent cutoff $\Lambda_{i}$ is such that $\Lambda_{i} = \Lambda_c$, for $1 \leq i \leq 4$, and $\sum_{\abs{\vec{k}} \leq \Lambda_{i}}$ sums over the entire BZ if $5 \leq i \leq 6$. In \cref{app:eqn:BM_curent_en_int_TBG_projected_complicated}, we have introduced the form-factors for the $\gamma$-fermions to be
\begin{equation}
	\mathcal{M}^{\eta}\left(\vec{k}, \vec{q} + \vec{G} \right) = \begin{pNiceMatrix}[last-col=3,first-row]
		4 & 2 \\
		\mathcal{M}^{cc,\eta}\left(\vec{k}, \vec{q} + \vec{G} \right) & \mathcal{M}^{\dagger,fc,\eta}\left(\vec{k} + \vec{q}, -\vec{q} - \vec{G} \right) & 4 \\
		\mathcal{M}^{fc,\eta}\left(\vec{k}, \vec{q} + \vec{G} \right) & \mathcal{M}^{ff,\eta}\left(\vec{k}, \vec{q} + \vec{G} \right)  & 2 \\
	\end{pNiceMatrix},
\end{equation}
where the $f$- and $c$-electron form factors are given in \cref{app:eqn:def_ffactors_ff,app:eqn:def_ffactors_cc,app:eqn:def_ffactors_fc}.

In the case of TSTG, the density operator for the mirror-odd fermions can be projected without a form factor by simply imposing a the cutoff $\Lambda_d$ for their momentum
\begin{equation}
	\label{app:eqn:projected_b_density_in_THF}
	\delta \rho^{\hat{b}} \left( \vec{q} \right) = \sum_{\eta,\alpha,s} \sum_{\substack{ \vec{p} \\ \abs{\vec{p}},\abs{\vec{p} + \vec{q}} \leq \Lambda_d }} \left( \hat{d}^\dagger_{ \vec{p} + \vec{q}, \alpha, \eta, s} \hat{d}_{\vec{p}, \alpha, \eta, s} - \frac{1}{2} \delta_{\vec{q},\vec{0}} \right),
\end{equation}
where the momentum transfer $\vec{q}$ is much smaller than a reciprocal vector, as a result of the small cutoff for the $d$-fermions' momenta. Using the projected density expression for the TSTG fermions obtained by summing \cref{app:eqn:projection_of_TBG_rho_to_THF,app:eqn:projected_b_density_in_THF}, we obtain the following expression for the interaction contribution to the energy current of TSTG
{	
	\small
	\begin{align}
		\label{app:eqn:BM_curent_en_int_TSTG_projected_complicated}
		\vec{J}^{V,\text{TSTG}}_{E} =& \frac{-i}{2N_0 \Omega_0} \sum_{\substack{\alpha, \eta,s \\ \alpha' \eta',s'}} \sum_{\abs{\vec{p}_1}, \abs{\vec{p}_2} \leq \Lambda_d} 
		\sum_{\substack{\vec{q} \\ \abs{\vec{p}_1 + \vec{q}} \leq \Lambda_d \\ \abs{\vec{p}_2 - \vec{q}} \leq \Lambda_d}}\pdv{V \left( \vec{q} \right)}{\vec{q}} 
		\left( \dv{t} :\mathrel{\hat{d}^\dagger_{ \vec{p}_1 + \vec{q}, \alpha, \eta, s} \hat{d}_{\vec{p}_1, \alpha, \eta, s}}: \right)
		:\mathrel{\hat{d}^\dagger_{ \vec{p}_2 - \vec{q}, \alpha', \eta', s'} \hat{d}_{\vec{p}_2, \alpha', \eta', s'}}:
		\nonumber \\
& - \frac{i}{2N_0 \Omega_0} \sum_{\substack{\eta,s,\eta',s',\alpha' \\ 1 \leq i,j \leq 6}} \sum_{\substack{\abs{\vec{k}_1} \leq \Lambda_{j} \\ \abs{\vec{p}_2} \leq \Lambda_d} } \sum_{\substack{\vec{q} \\ \abs{\vec{k}_1 + \vec{q}} \leq \Lambda_{i} \\ \abs{\vec{p}_2 - \vec{q}} \leq \Lambda_d}} \pdv{V \left( \vec{q} \right)}{\vec{q}} 
		\mathcal{M}^{\eta}_{ij}\left(\vec{k}_1,\vec{q} \right) \left( \dv{t} :\mathrel{ \hat{\gamma}^\dagger_{\vec{k}_1 + \vec{q} ,i,\eta,s} \hat{\gamma}_{\vec{k}_1,j,\eta,s}}: \right) 
		:\mathrel{\hat{d}^\dagger_{ \vec{p}_2 - \vec{q}, \alpha', \eta', s'} \hat{d}_{\vec{p}_2, \alpha', \eta', s'}}: \nonumber \\
& - \frac{i}{2N_0 \Omega_0} \sum_{\substack{\eta,s,\alpha,\eta',s' \\ 1 \leq k,l \leq 6}} 
		\sum_{\substack{\abs{\vec{k}_2} \leq \Lambda_{l} \\ \abs{\vec{p}_1} \leq \Lambda_d} } \sum_{\substack{\vec{q} \\ \abs{\vec{k}_2 - \vec{q}} \leq \Lambda_{k} \\ \abs{\vec{p}_1 + \vec{q}} \leq \Lambda_d}} \pdv{V \left( \vec{q} \right)}{\vec{q}}
		\mathcal{M}^{\eta'}_{kl}\left(\vec{k}_2, -\vec{q} \right)
		\nonumber \\
		&\quad\times   
		\left( \dv{t} :\mathrel{ \hat{d}^\dagger_{ \vec{p}_1 + \vec{q}, \alpha, \eta, s} \hat{d}_{\vec{p}_1, \alpha, \eta, s}}: \right)
		:\mathrel{\hat{\gamma}^\dagger_{\vec{k}_2-\vec{q},k,\eta',s'} \hat{\gamma}_{\vec{k}_2,l,\eta',s'}}: + \vec{J}^{V,\text{TBG}}_{E}.
	\end{align}}

\subsubsection{Hartree-Fock approximation to the TBG interaction energy current}\label{app:sec:deriv_moire_cur:energy_interaction:hf}

Unlike the single-particle contribution to the energy current, which is formally quadratic in the fermion operators, the interaction parts are quartic operators, making their manipulation rather cumbersome. In order to simplify the expression of $\vec{J}^{V,\text{TBG}}_{E}$, we will first approximate the form factors using their analytical expressions derived by Ref.~\cite{CAL23}, and briefly reviewed in \cref{app:sec:HF_review:interaction:analytical_with_ff} and then perform a Hartree-Fock decoupling procedure to render them quadratic. 

In the Gaussian approximation, the form factors become $\vec{k}$-independent~\cite{CAL23}, 
\begin{equation}
	\mathcal{M}^{\eta}\left(\vec{k}, \vec{q} + \vec{G} \right) \approx \mathcal{M}^{\eta}\left(\vec{q} + \vec{G} \right),
\end{equation} 
allowing us to rewrite $\vec{J}^{V,\text{TBG}}_{E}$ in a form similar to \cref{app:eqn:general_TBG_int_forHF},
{
	\small
	\begin{equation}
		\vec{J}^{V,\text{TBG}}_{E} \approx \sum_{\substack{i,\eta_1,s_1 \\ j,\eta_2,s_2 \\ k,\eta_3,s_3 \\ l,\eta_4,s_4}}   \sum_{\substack{\vec{k}_1,\vec{k}_2,\vec{q} \\ \vec{k}'_1, \vec{k}'_2}} \vec{W}_{i \eta_1 s_1; j \eta_2 s_2; k \eta_3 s_3; l \eta_4 s_4} \left( \vec{q} \right) \delta_{\vec{k}'_2,\vec{k}_2 - \vec{q}} \delta_{\vec{k}'_1,\vec{k}_1 + \vec{q}} 
	\pdv{t}\left( :\mathrel{\hat{\gamma}^\dagger_{\vec{k}'_1 ,i,\eta_1,s_1} \hat{\gamma}_{\vec{k}_1,j,\eta_2,s_2}}: \right)
		:\mathrel{\hat{\gamma}^\dagger_{\vec{k}'_2,k,\eta_3,s_3} \hat{\gamma}_{\vec{k}_2,l,\eta_4,s_4}}:,
	\end{equation}}where, similarly to \cref{app:sec:many_body_rev:generic_not}, we have extended the cutoff for the $c$-electrons to encompass the first moir\'e BZ (such that the sum over $\vec{k}_1$, $\vec{k}_2$, $\vec{k}'_1$ and $\vec{k}'_2$ runs over the entire BZ). The rank-four tensor describing the current operator has the following form
\begin{equation}
	\label{app:eqn:def_W_tensor_for_current}
	\vec{W}_{i \eta_1 s_1; j \eta_2 s_2; k \eta_3 s_3; l \eta_4 s_4} \left( \vec{q} \right) = - \frac{i}{2N_0 \Omega_0} \sum_{\vec{G}} \pdv{V \left( \vec{q} + \vec{G} \right)}{\vec{q}} 
	\mathcal{M}^{\eta_1}_{ij}\left(\vec{q}+\vec{G} \right) \mathcal{M}^{\eta_3}_{kl}\left(-\vec{q} - \vec{G} \right) \delta_{\eta_1 \eta_2} \delta_{s_1 s_2} \delta_{\eta_3 \eta_4} \delta_{s_3 s_4}.
\end{equation}
In principle, we can perform a similar analysis to Ref.~\cite{SON22} and further simplify the $\vec{W}$ tensor by approximating it to its leading contributions. We will see, however, in \cref{app:sec:deriv_moire_cur:energy_interaction:comparison} that the interaction makes a negligible contribution to the energy current and could be ignored completely. A quick calculation already reveals that for the single-particle parameters chosen around \cref{app:eqn:spHamiltonian_BM_TBG}, the largest element of $N_0\vec{W}_{i \eta_1 s_1; j \eta_2 s_2; k \eta_3 s_3; l \eta_4 s_4} \left( \vec{q} \right)$ is no more than $4\%$ of $v_{*}$, which quantifies the leading contribution to the single-particle energy current $\vec{J}^{T,\text{TBG}}_{E}$. 

For now, however, we proceed and perform a Hartree-Fock decoupling of $\vec{J}^{V,\text{TBG}}_{E}$, under which the latter becomes
\begin{align}
	\vec{J}^{V,\text{TBG},\text{MF}}_{E} =& \sum_{\substack{i,\eta_1,s_1 \\ j,\eta_2,s_2 \\ k,\eta_3,s_3 \\ l,\eta_4,s_4}} \sum_{\vec{k}_1,\vec{k}_2} \left\lbrace
	\vec{W}_{i \eta_1 s_1; j \eta_2 s_2; k \eta_3 s_3; l \eta_4 s_4} \left( \vec{0} \right)  \left[ 
	\left\langle \pdv{t}\left( :\mathrel{\hat{\gamma}^\dagger_{\vec{k}_1 ,i,\eta_1,s_1} \hat{\gamma}_{\vec{k}_1,j,\eta_2,s_2}}: \right) \right\rangle
	:\mathrel{\hat{\gamma}^\dagger_{\vec{k}_2,k,\eta_3,s_3} \hat{\gamma}_{\vec{k}_2,l,\eta_4,s_4}}: \right. \right. \nonumber \\
	&+\left. \left\langle :\mathrel{\hat{\gamma}^\dagger_{\vec{k}_2,k,\eta_3,s_3} \hat{\gamma}_{\vec{k}_2,l,\eta_4,s_4}}: \right\rangle
	\pdv{t}\left( :\mathrel{\hat{\gamma}^\dagger_{\vec{k}_1 ,i,\eta_1,s_1} \hat{\gamma}_{\vec{k}_1,j,\eta_2,s_2}}: \right) \right] \nonumber \\
	&-\vec{W}_{i \eta_1 s_1; j \eta_2 s_2; k \eta_3 s_3; l \eta_4 s_4} \left( \vec{k}_2 - \vec{k}_1 \right) \left[
	\left\langle :\mathrel{\dot{\hat{\gamma}}^\dagger_{\vec{k}_2 ,i,\eta_1,s_1} \hat{\gamma}_{\vec{k}_2,l,\eta_4,s_4}}: \right\rangle \hat{\gamma}^\dagger_{\vec{k}_1,k,\eta_3,s_3} \hat{\gamma}_{\vec{k}_1,j,\eta_2,s_2}
	\right. \nonumber \\
	&\quad + \left\langle :\mathrel{\hat{\gamma}^\dagger_{\vec{k}_1 ,k,\eta_3,s_3} \hat{\gamma}_{\vec{k}_1,j,\eta_2,s_2}}: \right\rangle \dot{\hat{\gamma}}^\dagger_{\vec{k}_2,i,\eta_1,s_1} \hat{\gamma}_{\vec{k}_2,l,\eta_4,s_4}
	+ \left\langle :\mathrel{\hat{\gamma}^\dagger_{\vec{k}_2 ,i,\eta_1,s_1} \hat{\gamma}_{\vec{k}_2,l,\eta_4,s_4}}: \right\rangle \hat{\gamma}^\dagger_{\vec{k}_1,k,\eta_3,s_3} \dot{\hat{\gamma}}_{\vec{k}_1,j,\eta_2,s_2} \nonumber \\
	&\left. \left. \quad + \left\langle :\mathrel{\hat{\gamma}^\dagger_{\vec{k}_1 ,k,\eta_3,s_3} \dot{\hat{\gamma}}_{\vec{k}_1,j,\eta_2,s_2}}: \right\rangle \hat{\gamma}^\dagger_{\vec{k}_2,i,\eta_1,s_1} \hat{\gamma}_{\vec{k}_2,l,\eta_4,s_4} \right]\right\rbrace.
	\label{app:eqn:BM_curent_en_int_TSTG_projected_complicated:HF_1}
\end{align}
In order to simplify the expression further, we now note that in equilibrium, the density matrix of the system is constant in time 
\begin{align}
	\pdv{t} \varrho_{i \eta s;i' \eta' s'} \left( \vec{k} \right) &= \left\langle \pdv{t} :\mathrel{\hat{\gamma}^\dagger_{\vec{k},i, \eta, s} \hat{\gamma}_{\vec{k},i', \eta', s'} }: \right\rangle = \left\langle i \commutator{H^{\text{TBG}}}{\hat{\gamma}^\dagger_{\vec{k},i, \eta, s} \hat{\gamma}_{\vec{k},i', \eta', s'}}  \right\rangle \nonumber \\
	&=i \frac{\Tr \left( e^{-\beta K^{\text{TBG}}} \commutator{H^{\text{TBG}}}{\hat{\gamma}^\dagger_{\vec{k},i, \eta, s} \hat{\gamma}_{\vec{k},i', \eta', s'}} \right) }{\Tr \left( e^{-\beta K^{\text{TBG}}} \right)} \nonumber \\
	&=i \frac{\Tr \left( e^{-\beta K^{\text{TBG}}} H^{\text{TBG}} \hat{\gamma}^\dagger_{\vec{k},i, \eta, s} \hat{\gamma}_{\vec{k},i', \eta', s'} \right) - \Tr \left( e^{-\beta K^{\text{TBG}}} \hat{\gamma}^\dagger_{\vec{k},i, \eta, s} \hat{\gamma}_{\vec{k},i', \eta', s'} H^{\text{TBG}} \right) }{\Tr \left( e^{-\beta K^{\text{TBG}}} \right)} = 0, \label{app:eqn:constant_of_density_matrix}
\end{align}
where we have used the cyclic property of the trace and the fact that the grand canonical Hamiltonian $K^{\text{TBG}}$ and the (canonical) Hamiltonian $H^{\text{TBG}}$ commute with one another. Employing \cref{app:eqn:constant_of_density_matrix}, as well as the following symmetry property of the $\vec{W}$ tensor, which follows straight-forwardly from its definition in \cref{app:eqn:def_W_tensor_for_current}, 
\begin{equation}
	\vec{W}_{i \eta_1 s_1; j \eta_2 s_2; k \eta_3 s_3; l \eta_4 s_4} \left( \vec{q} \right) = -\vec{W}_{k \eta_3 s_3; l \eta_4 s_4;i \eta_1 s_1; j \eta_2 s_2;} \left( -\vec{q} \right),
\end{equation}
we can rewrite \cref{app:eqn:BM_curent_en_int_TSTG_projected_complicated:HF_1} as 
\begin{align}
	\vec{J}^{V,\text{TBG},\text{MF}}_{E} =& \sum_{\substack{i,\eta_1,s_1 \\ j,\eta_2,s_2 \\ k,\eta_3,s_3 \\ l,\eta_4,s_4}} \sum_{\vec{k}_1,\vec{k}_2} \left\lbrace
	\vec{W}_{i \eta_1 s_1; j \eta_2 s_2; k \eta_3 s_3; l \eta_4 s_4} \left( \vec{0} \right) 
	\varrho_{k \eta_3 s_3; l \eta_4 s_4} \left( \vec{k}_2 \right) 
	\pdv{t}\left( \hat{\gamma}^\dagger_{\vec{k}_1 ,i,\eta_1,s_1} \hat{\gamma}_{\vec{k}_1,j,\eta_2,s_2} \right) \right. \nonumber \\
	&-\vec{W}_{i \eta_1 s_1; j \eta_2 s_2; k \eta_3 s_3; l \eta_4 s_4} \left( \vec{k}_2 - \vec{k}_1 \right) \left[
	2 \left\langle :\mathrel{\dot{\hat{\gamma}}^\dagger_{\vec{k}_2 ,i,\eta_1,s_1} \hat{\gamma}_{\vec{k}_2,l,\eta_4,s_4}}: \right\rangle \hat{\gamma}^\dagger_{\vec{k}_1,k,\eta_3,s_3} \hat{\gamma}_{\vec{k}_1,j,\eta_2,s_2}
	\right. \nonumber \\
	&\quad \left.\left.+ \varrho_{i \eta_1 s_1; l \eta_4 s_4} \left( \vec{k}_2 \right) \left( \hat{\gamma}^\dagger_{\vec{k}_1,k,\eta_3,s_3} \dot{\hat{\gamma}}_{\vec{k}_1,j,\eta_2,s_2} - \dot{\hat{\gamma}}^\dagger_{\vec{k}_1,k,\eta_3,s_3} \hat{\gamma}_{\vec{k}_1,j,\eta_2,s_2} \right) \right] \right\rbrace. \label{app:eqn:BM_curent_en_int_TSTG_projected_complicated:HF_2}
\end{align}

To move forward, we now note that within the Hartree-Fock approximation, the fermion operators evolve as
\begin{align}
	\dot{\hat{\gamma}}^\dagger_{\vec{k},i,\eta,s} &= i \comm{H^{\text{TBG}}_{\text{MF}}}{\hat{\gamma}^\dagger_{\vec{k},i,\eta,s}} = i \sum_{i', \eta', s'} h^{\text{TBG},\text{MF}}_{i' \eta' s'; i \eta s} \left( \vec{k} \right) \hat{\gamma}^\dagger_{\vec{k},i',\eta',s'}, \label{app:eqn:time_evolve_HF_cre}\\
	\dot{\hat{\gamma}}_{\vec{k},i,\eta,s} &= i \comm{H^{\text{TBG}}_{\text{MF}}}{\hat{\gamma}_{\vec{k},i,\eta,s}} = - i \sum_{i', \eta', s'} h^{\text{TBG},\text{MF}}_{i \eta s; i' \eta' s'} \left( \vec{k} \right) \hat{\gamma}_{\vec{k},i',\eta',s'}, \label{app:eqn:time_evolve_HF_des}
\end{align} 
which allow us to evaluate the expectation value in the second row of \cref{app:eqn:BM_curent_en_int_TSTG_projected_complicated:HF_2},
\begin{equation}
	\left\langle :\mathrel{\dot{\hat{\gamma}}^\dagger_{\vec{k}_2 ,i,\eta_1,s_1} \hat{\gamma}_{\vec{k}_2,l,\eta_4,s_4}}: \right\rangle = i \sum_{i', \eta', s'} h^{\text{TBG},\text{MF}}_{i' \eta' s'; i \eta_1 s_1} \left( \vec{k}_2 \right) \varrho_{i' \eta' s'; l \eta_4 s_4} \left( \vec{k}_2 \right),
\end{equation}
and obtain the final expression for the Hartree-Fock approximation to the interaction energy current
\begin{align}
	\vec{J}^{V,\text{TBG},\text{MF}}_{E} =& \sum_{\substack{i,\eta_1,s_1 \\ j,\eta_2,s_2 \\ k,\eta_3,s_3 \\ l,\eta_4,s_4}} \sum_{\vec{k}_1,\vec{k}_2} \left\lbrace
	\vec{W}_{i \eta_1 s_1; j \eta_2 s_2; k \eta_3 s_3; l \eta_4 s_4} \left( \vec{0} \right) 
	\varrho_{k \eta_3 s_3; l \eta_4 s_4} \left( \vec{k}_2 \right) 
	\pdv{t}\left( \hat{\gamma}^\dagger_{\vec{k}_1 ,i,\eta_1,s_1} \hat{\gamma}_{\vec{k}_1,j,\eta_2,s_2} \right) \right. \nonumber \\
	&-\vec{W}_{i \eta_1 s_1; j \eta_2 s_2; k \eta_3 s_3; l \eta_4 s_4} \left( \vec{k}_2 - \vec{k}_1 \right) \left[
	2 i \sum_{i', \eta', s'} h^{\text{TBG},\text{MF}}_{i' \eta' s'; i \eta_1 s_1} \left( \vec{k}_2 \right) \varrho_{i' \eta' s'; l \eta_4 s_4} \left( \vec{k}_2 \right) \hat{\gamma}^\dagger_{\vec{k}_1,k,\eta_3,s_3} \hat{\gamma}_{\vec{k}_1,j,\eta_2,s_2}
	\right. \nonumber \\
	&\quad \left.\left.+ \varrho_{i \eta_1 s_1; l \eta_4 s_4} \left( \vec{k}_2 \right) \left( \hat{\gamma}^\dagger_{\vec{k}_1,k,\eta_3,s_3} \dot{\hat{\gamma}}_{\vec{k}_1,j,\eta_2,s_2} - \dot{\hat{\gamma}}^\dagger_{\vec{k}_1,k,\eta_3,s_3} \hat{\gamma}_{\vec{k}_1,j,\eta_2,s_2} \right) \right] \right\rbrace. \label{app:eqn:BM_curent_en_int_TSTG_projected_complicated:HF_3}
\end{align}
We note that  similar expressions can be derived for the other three terms appearing in $\vec{J}^{V,\text{TSTG},\text{MF}}_{E}$ affording the Hartree-Fock approximation of the corresponding interaction energy current.

\subsubsection{Comparing the single-particle and interaction contributions to the energy current}\label{app:sec:deriv_moire_cur:energy_interaction:comparison}

Despite being formally quadratic in the fermionic operators, the expression in \cref{app:eqn:BM_curent_en_int_TSTG_projected_complicated:HF_3} is still complicated to manipulate, since it depends not on the Heavy-Fermion parameters, but on the $f$- and $c$-electron \emph{wave functions}, through the corresponding form-factors, as implied by \cref{app:eqn:def_W_tensor_for_current}. We will now estimate the interaction contribution to the energy current and show that it is much smaller than the single-particle one. As a result, the former can be ignored in our further calculations.

We start by noting that both $\vec{J}^{T,\text{TBG}}_{E}$, the single-particle contribution to the energy current defined in \cref{app:eqn:THF_en_current_sp_part_TBG_simple}, and $\vec{J}^{V,\text{TBG},\text{MF}}_{E}$, the mean-field decoupled interaction one derived in \cref{app:eqn:BM_curent_en_int_TSTG_projected_complicated:HF_3}, are fermion bilinear operators. In order to compare two fermion bilinear operators, one first obtains their matrix representation in the same orthonormal fermion basis, and then compares {\it e.g.}{} the Frobenius norm of the two matrices to assess the relative magnitude of the two operators. In our case, the single-particle energy current contribution from \cref{app:eqn:THF_en_current_sp_part_TBG_simple} contains terms in which one of the fermion operators is differentiated with respect to time, whereas in the Hartree-Fock-decoupled interaction one, there are also terms in which the fermion operators appear without any time derivative. To correctly compare the two contributions, we will now replace all fermion operators with a time derivative using the Hartree-Fock time-evolution from \cref{app:eqn:time_evolve_HF_cre,app:eqn:time_evolve_HF_des}, thereby expressing both operators in the $\hat{\gamma}^\dagger_{\vec{k},i, \eta, s}$ basis ({\it i.e.}{} without any time derivatives). Evaluating the time-derivatives in \cref{app:eqn:THF_en_current_sp_part_TBG_simple}, we obtain 
\begin{equation}
	\label{app:eqn:quadratic_j_T_noTDeriv}
	\vec{J}^{T,\text{TBG}}_E = \sum_{\vec{k}} \sum_{\substack{i, \eta, s \\ i', \eta', s'}} \vec{j}^T_{E,i \eta s; i' \eta' s'} \left( \vec{k} \right) \hat{\gamma}^\dagger_{\vec{k},i,\eta,s} \hat{\gamma}_{\vec{k},i',\eta',s'},
\end{equation} 
where the matrix $\vec{j}^T_{E,i \eta s; i' \eta' s'} \left( \vec{k} \right)$ is given by
\begin{equation}
	\label{app:eqn:j_T_matrix_HF}
	\vec{j}^T_{E,i \eta s; i' \eta' s'} \left( \vec{k} \right) = \frac{1}{2} \left[ 
		\sum_{j} \left(\partial_{\vec{k}} h^{\text{TBG},\eta}_{i j} \left( \vec{k} \right) \right) h^{\text{TBG},\text{MF}}_{j \eta s; i' \eta' s'} \left( \vec{k} \right) +
		\sum_{j'} \left(\partial_{\vec{k}} h^{\text{TBG},\eta'}_{j' i'} \left( \vec{k} \right) \right) h^{\text{TBG},\text{MF}}_{i \eta s; j' \eta' s'} \left( \vec{k} \right)  
	\right].
\end{equation}
Similarly, we can evaluate the time derivatives in the Hartree-Fock approximation to the energy current and obtain
\begin{equation}
	\label{app:eqn:quadratic_j_V_noTDeriv}
	\vec{J}^{V,\text{TBG}}_E = \sum_{\vec{k}} \sum_{\substack{i, \eta, s \\ i', \eta', s'}} \vec{j}^V_{E,i \eta s; i' \eta' s'} \left( \vec{k} \right) \hat{\gamma}^\dagger_{\vec{k},i,\eta,s} \hat{\gamma}_{\vec{k},i',\eta',s'},
\end{equation}
with the matrix $\vec{j}^V_{E,i \eta s; i' \eta' s'} \left( \vec{k} \right)$ being
{\small
	\begin{align}
	\vec{j}^V_{E,i \eta s; i' \eta' s'} \left( \vec{k} \right) =&i\sum_{\substack{j,\eta_2,s_2 \\ k,\eta_3,s_3 \\ l,\eta_4,s_4}} 	
	\left( \vec{W}_{j \eta_2 s_2; i' \eta' s'; k \eta_3 s_3; l \eta_4 s_4} \left( \vec{0} \right) 
		h^{\text{TBG},\text{MF}}_{i \eta s; j \eta_2 s_2} \left( \vec{k} \right)
		-\vec{W}_{i \eta s; j \eta_2 s_2; k \eta_3 s_3; l \eta_4 s_4} \left( \vec{0} \right) 
		h^{\text{TBG},\text{MF}}_{j \eta_2 s_2; i' \eta' s'} \left( \vec{k} \right)
	\right)
	\sum_{\vec{k}'} \varrho_{k \eta_3 s_3; l \eta_4 s_4} \left( \vec{k}' \right) \nonumber \\
-&2i \sum_{\substack{j,\eta_2,s_2 \\ k,\eta_3,s_3 \\ l,\eta_4,s_4}} \sum_{\vec{k}'}  
	\vec{W}_{k \eta_3 s_3; i' \eta' s';i \eta s ; l \eta_4 s_4} \left( \vec{k}' - \vec{k} \right)
	\varrho_{j \eta_2 s_2; l \eta_4 s_4} \left( \vec{k}' \right)
	h^{\text{TBG},\text{MF}}_{j \eta_2 s_2; k \eta_3 s_3} \left( \vec{k}' \right)  \nonumber \\
+&i \sum_{\substack{j,\eta_2,s_2 \\ k,\eta_3,s_3 \\ l,\eta_4,s_4}} \sum_{\vec{k}'} 
	\vec{W}_{k \eta_3 s_3; j \eta_2 s_2; i \eta s; l \eta_4 s_4} \left( \vec{k}' - \vec{k} \right)
	\varrho_{k \eta_3 s_3; l \eta_4 s_4} \left( \vec{k}' \right)
	h^{\text{TBG},\text{MF}}_{j \eta_2 s_2; i' \eta' s} \left( \vec{k}' \right) \nonumber \\
+&i \sum_{\substack{j,\eta_2,s_2 \\ k,\eta_3,s_3 \\ l,\eta_4,s_4}} \sum_{\vec{k}'} 
	\vec{W}_{j \eta_2 s_2; i' \eta' s'; k \eta_3 s_3; l \eta_4 s_4} \left( \vec{k}' - \vec{k} \right)
	\varrho_{j \eta_2 s_2; l \eta_4 s_4} \left( \vec{k}' \right)
	h^{\text{TBG},\text{MF}}_{i \eta s; k \eta_3 s_3} \left( \vec{k}' \right).
	\label{app:eqn:j_V_matrix_HF}
\end{align}}

We can now assess the relative contribution of the interaction to the energy current by computing the ratio $\alpha_{\abs{\vec{J}_E^V}/\abs{\vec{J}_E^T}}$ between the Frobenius norms of the two matrices from \cref{app:eqn:j_V_matrix_HF,app:eqn:j_T_matrix_HF},
\begin{equation}
	\label{app:eqn:ratio_interaction_kinetic}
	\alpha_{\abs{\vec{J}_E^V}/\abs{\vec{J}_E^T}} = \sqrt{\frac{
			\sum_{\vec{k},\mu} \sum_{\substack{i,\eta,s \\ i',\eta',s' }} \abs{j^{V,\mu}_{E,i \eta s; i' \eta' s'} \left( \vec{k} \right)}^{2} 
		}{
			\sum_{\vec{k},\mu} \sum_{\substack{i,\eta,s \\ i',\eta',s' }} \abs{j^{T,\mu}_{E,i \eta s; i' \eta' s'} \left( \vec{k} \right)}^{2} 
		}}.
\end{equation}
In \cref{app:tab:ratio_interaction_sp_en_cur}, we have computed the ratio from \cref{app:eqn:ratio_interaction_kinetic} for the nine correlated ground state candidates of \cref{app:tab:model_states} that we consider in this work. We find that the interaction contribution to the energy current is no larger than approximately $
4\%$ of the single-particle one. There are two reasons for this:
\begin{itemize}
	\item The large group velocity of the $c$-electrons enhances the single-particle contribution.
	\item Upon projecting into the THF model basis, Ref.~\cite{SON22} finds that the \emph{long-range} interaction Coulomb Hamiltonian of TBG can be excellently described by \emph{short-range} interactions terms ({\it i.e.}{} for which the scattering amplitude is independent on the momentum transfer). This is a consequence of the localized nature of the $f$-electrons. Such interaction terms do not give rise to contributions to the energy current~\cite{PAU03}. By analogy, projecting $\vec{J}^{V,\text{TBG}}_{E}$ from the BM model basis to THF one will also lead to a negligible contribution. 
\end{itemize}
As a result, in what follows, we will completely neglect the interaction contribution to the energy current of TBG, and consider solely the single-particle one. For TSTG the interaction contribution to the energy current can also be neglected for the same reasons as in TBG. As a result, the energy currents for TBG and TSTG will by given by
\begin{equation}
	\label{app:eqn:energy_currents_THF}
	\vec{J}^{\text{TBG}}_{E} \approx \vec{J}^{T,\text{TBG}}_{E} \qq{and}
	\vec{J}^{\text{TSTG}}_{E} \approx \vec{J}^{T,\text{TSTG}}_{E},
\end{equation}
where $	\vec{J}^{T,\text{TBG}}_{E}$ and $\vec{J}^{T,\text{TSTG}}_{E}$ have been obtained in \cref{app:eqn:THF_en_current_sp_part_TBG_simple,app:eqn:THF_en_current_sp_part_TSTG_simple}, respectively.
\begin{table}[t]
	\centering
	\begin{tabular}{|r|l|r|c|}
		\hline
		$\nu$ & State & $\alpha_{\abs{\vec{J}_E^V}/\abs{\vec{J}_E^T}}$ \\
		\hline\hline
		$4$ & $ \IfStrEqCase{1}{{1}{\ket{\nu={}4} }
		{2}{\ket{\nu={}3, \mathrm{IVC}}}
		{3}{\ket{\nu={}3, \mathrm{VP}}}
		{4}{\ket{\nu={}2, \mathrm{K-IVC}}}
		{5}{\ket{\nu={}2, \mathrm{VP}}}
		{6}{\ket{\nu={}1, (\mathrm{K-IVC}+\mathrm{VP})}}
		{7}{\ket{\nu={}1, \mathrm{VP}}}
		{8}{\ket{\nu=0, \mathrm{K-IVC}}}
		{9}{\ket{\nu=0, \mathrm{VP}}}
	}
	[nada]
 $ & $3.14\%$ \\
		\hline
		\multirow[b]{2}{*}{$3$} & $ \IfStrEqCase{2}{{1}{\ket{\nu={}4} }
		{2}{\ket{\nu={}3, \mathrm{IVC}}}
		{3}{\ket{\nu={}3, \mathrm{VP}}}
		{4}{\ket{\nu={}2, \mathrm{K-IVC}}}
		{5}{\ket{\nu={}2, \mathrm{VP}}}
		{6}{\ket{\nu={}1, (\mathrm{K-IVC}+\mathrm{VP})}}
		{7}{\ket{\nu={}1, \mathrm{VP}}}
		{8}{\ket{\nu=0, \mathrm{K-IVC}}}
		{9}{\ket{\nu=0, \mathrm{VP}}}
	}
	[nada]
 $ & $3.40\%$ \\
		& $ \IfStrEqCase{3}{{1}{\ket{\nu={}4} }
		{2}{\ket{\nu={}3, \mathrm{IVC}}}
		{3}{\ket{\nu={}3, \mathrm{VP}}}
		{4}{\ket{\nu={}2, \mathrm{K-IVC}}}
		{5}{\ket{\nu={}2, \mathrm{VP}}}
		{6}{\ket{\nu={}1, (\mathrm{K-IVC}+\mathrm{VP})}}
		{7}{\ket{\nu={}1, \mathrm{VP}}}
		{8}{\ket{\nu=0, \mathrm{K-IVC}}}
		{9}{\ket{\nu=0, \mathrm{VP}}}
	}
	[nada]
 $ & $3.42\%$ \\
		\hline
		\multirow[b]{2}{*}{$2$} & $ \IfStrEqCase{4}{{1}{\ket{\nu={}4} }
		{2}{\ket{\nu={}3, \mathrm{IVC}}}
		{3}{\ket{\nu={}3, \mathrm{VP}}}
		{4}{\ket{\nu={}2, \mathrm{K-IVC}}}
		{5}{\ket{\nu={}2, \mathrm{VP}}}
		{6}{\ket{\nu={}1, (\mathrm{K-IVC}+\mathrm{VP})}}
		{7}{\ket{\nu={}1, \mathrm{VP}}}
		{8}{\ket{\nu=0, \mathrm{K-IVC}}}
		{9}{\ket{\nu=0, \mathrm{VP}}}
	}
	[nada]
 $ & $3.78\%$ \\
		& $ \IfStrEqCase{5}{{1}{\ket{\nu={}4} }
		{2}{\ket{\nu={}3, \mathrm{IVC}}}
		{3}{\ket{\nu={}3, \mathrm{VP}}}
		{4}{\ket{\nu={}2, \mathrm{K-IVC}}}
		{5}{\ket{\nu={}2, \mathrm{VP}}}
		{6}{\ket{\nu={}1, (\mathrm{K-IVC}+\mathrm{VP})}}
		{7}{\ket{\nu={}1, \mathrm{VP}}}
		{8}{\ket{\nu=0, \mathrm{K-IVC}}}
		{9}{\ket{\nu=0, \mathrm{VP}}}
	}
	[nada]
 $ & $3.69\%$ \\
		\hline
		\multirow[b]{2}{*}{$1$} & $ \IfStrEqCase{6}{{1}{\ket{\nu={}4} }
		{2}{\ket{\nu={}3, \mathrm{IVC}}}
		{3}{\ket{\nu={}3, \mathrm{VP}}}
		{4}{\ket{\nu={}2, \mathrm{K-IVC}}}
		{5}{\ket{\nu={}2, \mathrm{VP}}}
		{6}{\ket{\nu={}1, (\mathrm{K-IVC}+\mathrm{VP})}}
		{7}{\ket{\nu={}1, \mathrm{VP}}}
		{8}{\ket{\nu=0, \mathrm{K-IVC}}}
		{9}{\ket{\nu=0, \mathrm{VP}}}
	}
	[nada]
 $ & $3.63\%$ \\
		& $ \IfStrEqCase{7}{{1}{\ket{\nu={}4} }
		{2}{\ket{\nu={}3, \mathrm{IVC}}}
		{3}{\ket{\nu={}3, \mathrm{VP}}}
		{4}{\ket{\nu={}2, \mathrm{K-IVC}}}
		{5}{\ket{\nu={}2, \mathrm{VP}}}
		{6}{\ket{\nu={}1, (\mathrm{K-IVC}+\mathrm{VP})}}
		{7}{\ket{\nu={}1, \mathrm{VP}}}
		{8}{\ket{\nu=0, \mathrm{K-IVC}}}
		{9}{\ket{\nu=0, \mathrm{VP}}}
	}
	[nada]
 $ & $3.59\%$ \\
		\hline
		\multirow[b]{2}{*}{$0$} & $ \IfStrEqCase{8}{{1}{\ket{\nu={}4} }
		{2}{\ket{\nu={}3, \mathrm{IVC}}}
		{3}{\ket{\nu={}3, \mathrm{VP}}}
		{4}{\ket{\nu={}2, \mathrm{K-IVC}}}
		{5}{\ket{\nu={}2, \mathrm{VP}}}
		{6}{\ket{\nu={}1, (\mathrm{K-IVC}+\mathrm{VP})}}
		{7}{\ket{\nu={}1, \mathrm{VP}}}
		{8}{\ket{\nu=0, \mathrm{K-IVC}}}
		{9}{\ket{\nu=0, \mathrm{VP}}}
	}
	[nada]
 $ & $3.26\%$ \\
		& $ \IfStrEqCase{9}{{1}{\ket{\nu={}4} }
		{2}{\ket{\nu={}3, \mathrm{IVC}}}
		{3}{\ket{\nu={}3, \mathrm{VP}}}
		{4}{\ket{\nu={}2, \mathrm{K-IVC}}}
		{5}{\ket{\nu={}2, \mathrm{VP}}}
		{6}{\ket{\nu={}1, (\mathrm{K-IVC}+\mathrm{VP})}}
		{7}{\ket{\nu={}1, \mathrm{VP}}}
		{8}{\ket{\nu=0, \mathrm{K-IVC}}}
		{9}{\ket{\nu=0, \mathrm{VP}}}
	}
	[nada]
 $ & $3.30\%$ \\
		\hline
	\end{tabular}
	\caption{Relative contribution of the interaction to the energy current in TBG. We consider the nine correlated ground state candidates from \cref{app:tab:model_states} and compute the ratio between the interaction and single-particle contribution to the energy current, as defined in \cref{app:eqn:ratio_interaction_kinetic}. Our calculations are performed with the self-consistent Hartree-Fock solution at zero temperature. We find that the interaction contribution to the energy current is no bigger than $4\%$ of the single-particle one.}
	\label{app:tab:ratio_interaction_sp_en_cur}
\end{table}

\section{Thermoelectric response in graphene moir\'e heterostructures}\label{app:sec:thermoelectric_response}

In this \siSection{}, we provide a comprehensive derivation of the thermoelectric response for TBG and TSTG, within the THF model. We begin with a general overview of linear response theory and illustrate how the linear response coefficients connect to current-current correlation functions. Subsequently, we focus on thermoelectric transport and derive the expression of the Seebeck coefficient. Using the particle and energy current operators from \cref{app:sec:deriv_moire_cur}, we obtain the Seebeck coefficient for both TBG and TSTG. Lastly, we explain the numerical computation of the Seebeck coefficient of TBG and TSTG in both the fully-interacting model (covering both symmetry-broken and fully symmetric phases) and the non-interacting limit of the model.

\subsection{Linear response theory}\label{app:sec:thermoelectric_response:lin_response}

In this section, we consider the response of a quantum mechanical system subjected to a set of generic static ``forces''. Such a ``force'' can be, for example, the electric field, to which the response of the system will be an electric current. To obtain the corresponding linear response coefficients, we will follow the derivation put forward by Luttinger in Ref.~\cite{LUT64}, and reviewed by Ref.~\cite{MAH00}. 

\subsubsection{Response to a static perturbation}
\label{app:sec:thermoelectric_response:lin_response:static_perturbation}

To this end, we let $H$ and $\hat{N}$ denote the unperturbed (but fully-interacting) Hamiltonian and the number operator of the system. We then consider adding the following perturbation to the Hamiltonian, which contains $N_{\phi}$ general ``forces''
\begin{equation}
	\label{app:eqn:general_perturbation_lin_theo}
	\delta H = \int \dd[d]{r} \sum_{n=1}^{N_{\phi}} \rho_n \left( \vec{r} \right) \phi_n \left( \vec{r} \right).
\end{equation}
In \cref{app:eqn:general_perturbation_lin_theo}, $\rho_n \left( \vec{r} \right)$ ($1 \leq n \leq N_{\phi}$) is the quantum mechanical density operator corresponding to the generalized ``charge'' of some conserved quantity and $\phi_n \left( \vec{r} \right)$ is the potential that couples with the said ``charge'' density. For example, $\rho_n \left( \vec{r} \right)$ can denote the electrical charge, while $\phi_n \left( \vec{r} \right)$ can be the electrostatic potential of an externally applied electric field. Note that the potentials $\phi_{n} \left( \vec{r} \right)$ do not depend on time and we will assume that the corresponding ``forces'' $\nabla \phi_n \left( \vec{r} \right)$ are spatially uniform (or $\vec{r}$-independent).

We now consider the scenario where the perturbation $\delta H$ is turned on gradually, such that the Hamiltonian of the \emph{perturbed} system is given by 
\begin{equation}
	H' \left( t \right) \equiv H  + \delta H e^{s t}.
\end{equation}
The inclusion of the exponential prefactor $e^{s t}$ with the \emph{small} $s > 0$ damping factor ensures that the perturbation vanishes in the infinite past $t \to -\infty$. Next, we assume that, in the absence of the perturbation, the system is in thermal equilibrium having the following density matrix \emph{operator}\footnote{We distinguish the density matrix introduced in \cref{app:eqn:def_rho_HF} from the density matrix operator from \cref{app:eqn:def_density_matrix_op} by adding a hat to the latter.}
\begin{equation}
	\label{app:eqn:def_density_matrix_op}
	\hat{\varrho}_0 = e^{\beta \left( \Omega - K \right)},
\end{equation}
where $K \equiv  H - \mu \hat{N}$ is the grand canonical Hamiltonian of the unperturbed system (with $\mu$ being the chemical potential) and $\Omega$ is related to the partition function $Z$ of the system
\begin{equation}
	\Omega \equiv - \frac{1}{\beta} \log  Z,  \quad
	Z \equiv \Tr \left( e^{-\beta K } \right).
\end{equation}
.

The density matrix describing the perturbed system takes the form
\begin{equation}
	\hat{\varrho} \left( t \right) \equiv \hat{\varrho}_0 + \delta \hat{\varrho} \left( t \right),
\end{equation}
and is equal to the equilibrium density matrix $\hat{\varrho}_0$ to which a small time-dependent contribution $\delta \hat{\varrho} \left( t\right)$ is added to account for the effects of the time-dependent perturbation in \cref{app:eqn:general_perturbation_lin_theo}.
In general, $\hat{\varrho} \left( t \right) $ is not simply proportional to $e^{-\beta K'\left(t \right)}$ with $K' \left( t \right) \equiv H' \left( t \right) - \mu \hat{N}$~\cite{MAH00}. Instead, its time-evolution is controlled by the von Neumann equation 
\begin{align}
	i \pdv{\hat{\varrho} \left( t \right)}{t} &= \commutator{K' \left( t \right)}{\hat{\varrho} \left( t \right)} \nonumber \\
	&= \commutator{K}{\hat{\varrho}_0}
	+ e^{s t} \commutator{\delta H}{\hat{\varrho}_0} 
	+ \commutator{K}{\delta \hat{\varrho} \left( t \right)} 
	+ e^{s t} \commutator{\delta H}{\delta \hat{\varrho} \left( t \right)} \nonumber \\
	&\approx  e^{s t} \commutator{\delta H}{\hat{\varrho}_0} 
	+ \commutator{K}{\delta \hat{\varrho} \left( t \right)}. \label{app:eqn:time_evolution_denMat}
\end{align}
In the last line of \cref{app:eqn:time_evolution_denMat} we have employed the definition of the equilibrium density matrix from \cref{app:eqn:def_density_matrix_op} which implies that $\commutator{K}{\hat{\varrho}_0} = 0$. Moreover, since we are interested in the \emph{linear} response of the system to the perturbation $\delta H$, we can assume that $\delta \hat{\varrho} \left( t \right)$ is small, implying that the $e^{s t} \commutator{\delta H}{\delta \hat{\varrho} \left( t \right)}$ is much smaller than the other two nonvanishing commutators and can be neglected. \Cref{app:eqn:time_evolution_denMat} can be rearranged into
\begin{align}
	i \left( \pdv{\delta \hat{\varrho} \left( t \right)}{t} + i K \delta \hat{\varrho} \left(t \right)  - i \delta \hat{\varrho} \left(t \right) K \right) &= e^{s t} \commutator{\delta H}{\hat{\varrho}_0}  \nonumber \\
	i e^{- i K t} \left[ \pdv{t} \left( e^{i K t} \delta \hat{\varrho} \left( t \right)  e^{-i K t} \right) \right] e^{i K t} &= e^{s t} \commutator{\delta H}{\hat{\varrho}_0} 
	\nonumber \\
	i \pdv{t} \left( e^{i K t} \delta \hat{\varrho} \left( t \right)  e^{-i K t} \right)  &= e^{s t} e^{i K t} \commutator{\delta H}{\hat{\varrho}_0}  e^{-i K t} \nonumber \\
	i \pdv{t} \left( e^{i K t} \delta \hat{\varrho} \left( t \right)  e^{-i K t} \right)  &= e^{s t} \commutator{\delta H \left( t \right)}{\hat{\varrho}_0}, \label{app:eqn:time_evolution_denMat_rearanged}
\end{align}
where the time-evolved perturbation $\delta H \left( t \right)$ is defined with respect to the grand canonical Hamiltonian of the unperturbed system
\begin{equation}
	\delta H \left( t \right) = e^{i K t} \delta H e^{-i K t}.
\end{equation}

\Cref{app:eqn:time_evolution_denMat_rearanged} can be readily integrated and, using the fact that $\lim_{t \to -\infty} \delta \hat{\varrho} \left( t \right) = 0$ ({\it i.e.}{}, in the absence of any perturbation, the system is in equilibrium), we have that  
\begin{align}
	e^{i K t}  \delta \hat{\varrho} \left( t \right) e^{-i K t} &= -i \int_{-\infty}^{t} \dd{t'} e^{s t'} \commutator{\delta H \left( t' \right)}{\hat{\varrho}_0} \nonumber \\
	\delta \hat{\varrho} \left( t \right) &= -i \int_{-\infty}^{t} \dd{t'} e^{s t'} e^{-i K t}  \commutator{\delta H \left( t' \right)}{\hat{\varrho}_0} e^{i K t}  \nonumber \\
	\delta \hat{\varrho} \left( t \right) &= -i e^{s t} \int_{-\infty}^{t} \dd{t'} e^{s \left( t' - t \right)}  \commutator{\delta H \left( t' - t \right)}{\hat{\varrho}_0},
\end{align}
which, after a change of integration variables, implies that $\delta \hat{\varrho} \left( t \right) = \delta \hat{\varrho}  e^{st} $, where we have defined 
\begin{equation}
	\label{app:eqn:perturbation_unsimplified}
	\delta \hat{\varrho} \equiv - i \int_{0}^{\infty} \dd{t} e^{-s t}  \commutator{\delta H \left(- t \right)}{\hat{\varrho}_0}.
\end{equation}
The expression of the commutator can be further simplified by means of the following trick~\cite{LUT64}
\begin{align}
	\commutator{\delta H \left(- t \right)}{\hat{\varrho}_0} &= \delta H \left(- t \right) e^{\beta \left( \Omega - K \right)} - e^{\beta \left( \Omega - K \right)} \delta H \left(- t \right) \nonumber \\
	 &= e^{\beta \left( \Omega - K \right)} \left[ e^{-\beta \left( \Omega - K \right)} \delta H \left(- t \right) e^{\beta \left( \Omega - K \right)} - \delta H \left(- t \right) \right] \nonumber \\
	 &= \hat{\varrho}_0  \left( \delta H \left(- t - i \beta \right) - \delta H \left(- t \right) \right) \nonumber \\
	 &=- i \hat{\varrho}_0  \int_{0}^{\beta} \dd{\tau} \pdv{\delta H \left(- t - i \tau \right)}{t}, \label{app:eqn:commutator_trick_lin_resp}
\end{align}
that follows directly from the definition in \cref{app:eqn:def_density_matrix_op}. Upon substituting \cref{app:eqn:commutator_trick_lin_resp} into \cref{app:eqn:perturbation_unsimplified}, we find that
\begin{equation}
	\label{app:eqn:perturbation_final}
	\delta \hat{\varrho} = - \int_{0}^{\infty} \dd{t} e^{-s t}  \int_{0}^{\beta} \dd{\tau}\hat{\varrho}_0  \pdv{\delta H \left(- t - i \tau \right)}{t} .
\end{equation}

As a result of applying the perturbation from \cref{app:eqn:general_perturbation_lin_theo}, the system will develop currents corresponding to the generalized ``charges'' to which the potentials $\phi_n \left( \vec{r} \right)$ couple. For instance, an applied electric field will induce a current flow in the system (in \cref{app:sec:thermoelectric_response:thermoelectric} we will also discuss the case where the potential correspond to a temperature gradient). The value of the current when the perturbation has been turned on fully ({\it i.e.}{}, at $t = 0$) can be computed from the density matrix of the system
\begin{equation}
	\label{app:eqn:expec_local_current_op}
	\left\langle j^{\alpha}_n \left( \vec{r} \right) \right\rangle = \Tr \left( \hat{\varrho}\left( t \right) j^{\alpha}_n \left( \vec{r} \right)  \right) \eval_{t = 0}  = \Tr \left[ \left( \hat{\varrho}_0 + \delta \hat{\varrho} \right) j^{\alpha}_n \left( \vec{r} \right)  \right]  = \Tr \left( \delta \hat{\varrho} j^{\alpha}_n \left( \vec{r} \right) \right) +  \Tr \left( \hat{\varrho}_0 j^{\alpha}_n \left( \vec{r} \right) \right) ,
\end{equation}
where $j^{\alpha}_n \left( \vec{r} \right)$ is the $\alpha$-th component of the current operator $\vec{j}_{n} \left( \vec{r} \right)$ corresponding to the $n$-th conserved charge. The current $\vec{j}_n \left( \vec{r} \right)$ and the corresponding charge density operator $\rho_{n} \left( \vec{r} \right)$ are related via the continuity equation 
\begin{equation}
	\label{app:eqn:lin_resp_cont_eqn}
	\pdv{\rho_n \left( \vec{r}, t \right)}{t} + \nabla \cdot \vec{j}_n \left( \vec{r}, t \right) = 0,
\end{equation} 
where the time evolution of the current and density operators is given by the unperturbed grand canonical Hamiltonian
\begin{equation}
	\rho_n \left( \vec{r}, t \right) = e^{i K t} \rho_n \left( \vec{r} \right) e^{- i K t}, \qquad
	\vec{j}_n \left( \vec{r}, t \right) = e^{i K t} \vec{j}_n \left( \vec{r} \right) e^{- i K t}.
\end{equation}

Since we are interested in uniform generalized ``forces'' $\nabla \phi_{n} \left( \vec{r} \right)$, we can average over the spatial coordinate in \cref{app:eqn:expec_local_current_op} to obtain 
\begin{equation}
	\label{app:eqn:expec_current_op}
	\left\langle j^{\alpha}_n \right\rangle \equiv \frac{1}{N_0 \Omega_{0}} \int \dd[d]{r} 	\left\langle j^{\alpha}_n \left( \vec{r} \right) \right\rangle = \frac{1}{N_0 \Omega_{0}} \Tr \left( \delta \hat{\varrho} J^{\alpha}_n \right),
\end{equation}
where we have used a notation analogous to \cref{app:eqn:part_current_simple_real_general,app:eqn:en_current_simple_real_general_interaction}, in which $N_0 \Omega_{0}$ is the total volume of the system while the \emph{total} current operator is defined as\footnote{Note that the average current and the expectation value of the total current operator are related by $\left\langle j^{\alpha}_n \right\rangle = \frac{1}{N_0 \Omega_0} \left\langle J^{\alpha}_n \right\rangle$.} 
\begin{equation}
	\label{app:eqn:general_total_current_definition}
	\vec{J}_{n} \equiv \int \dd[d]{r} \vec{j}_n \left( \vec{r} \right), \quad
	\vec{J}_{n} \left( t \right) \equiv e^{i K t} \vec{J}_{n} e^{- i K t}.
\end{equation}
In deriving \cref{app:eqn:general_total_current_definition}, we have also employed the fact that no currents flow in equilibrium, which is equivalent to $\Tr \left( \hat{\varrho}_0 J^{\alpha}_{n} \right) = 0$. We note that all the THF phases considered in this work have $C_{3z}$ symmetry (implying that $C_{3z} \hat{\varrho}_0 C^{-1}_{3z} = \hat{\varrho}_0$). Under $C_{3z}$, the local and total current operators transform as 
\begin{align}
	C_{3z}^{-1} j^{\alpha}_n \left( C_{3z} \vec{r} \right) C_{3z} &= \sum_{\beta} \mathcal{R}^{\alpha \beta}_{C_{3z}} j^{\beta}_n \left( \vec{r} \right), \\
	C_{3z}^{-1} J^{\alpha}_n C_{3z} &= \sum_{\beta} \mathcal{R}^{\alpha \beta}_{C_{3z}} J^{\beta}_n, \label{app:eqn:trafo_c3_J}
\end{align}
with the Cartesian rotation matrix for the $C_{3z}$ transformation being given by
\begin{equation}
	\label{app:eqn:matrix_c3}
	\mathcal{R}_{C_{3z}} = \frac{1}{2} \begin{pmatrix}
		-1 & - \sqrt{3} \\
		\sqrt{3} & -1
	\end{pmatrix}.
\end{equation}
In turn, this implies that there are no currents flowing in equilibrium, since $\Tr \left( \hat{\varrho}_0 J^{\alpha}_{n} \right) = \sum_{\beta} \mathcal{R}^{\alpha \beta}_{C_{3z}} \Tr \left( \hat{\varrho}_0 J^{\beta}_{n} \right)$, or, equivalently, $\left\langle \vec{J}_n \right\rangle$ is an eigenvector of $\mathcal{R}_{C_{3z}}$, $\sum_{\beta} \mathcal{R}^{\alpha \beta}_{C_{3z}} \left\langle J^{\beta}_n \right\rangle = \left\langle J^{\alpha}_n \right\rangle$. However, $\mathcal{R}_{C_{3z}}$ cannot have a nonzero eigenvector with eigenvalue one, meaning that $\left\langle J^{\alpha}_n \right\rangle = 0$.

By inserting \cref{app:eqn:commutator_trick_lin_resp} into \cref{app:eqn:expec_current_op}, we find that the current response to the externally applied fields $\phi_{n} \left( \vec{r} \right)$ is given by 
\begin{equation}
	\label{app:eqn:expec_current_op_interm}
	\left\langle j^{\alpha}_n \right\rangle = - \frac{1}{N_0 \Omega_0} \int_{0}^{\infty} \dd{t} e^{-s t}  \int_{0}^{\beta} \dd{\tau} \Tr \left(\hat{\varrho}_0  \pdv{\delta H \left(- t - i \tau \right)}{t} J^{\alpha}_n \left( 0 \right) \right).
\end{equation}
We now turn our attention to the time derivative of the perturbation. The latter can be evaluated using
\begin{align}
	\pdv{\delta H \left(- t - i \tau \right)}{t} &= \int \dd[d]{r} \sum_{n=1}^{N_{\phi}} \pdv{\rho_n \left( \vec{r}, - t - i \tau \right)}{t} \phi_n \left( \vec{r} \right) \nonumber \\
	&= - \int \dd[d]{r} \sum_{n=1}^{N_{\phi}} \nabla \cdot \vec{j}_{n} \left( \vec{r},  - t - i \tau \right) \phi_n \left( \vec{r} \right) \nonumber \\
	&= \int \dd[d]{r} \sum_{n=1}^{N_{\phi}} \vec{j}_{n} \left( \vec{r},  - t - i \tau \right) \cdot \left( \nabla \phi_n \left( \vec{r} \right) \right) \nonumber \\
	&= \sum_{n=1}^{N_{\phi}} \nabla \phi_n \cdot  \vec{J}_{n} \left( - t - i \tau \right), \label{app:eqn:perturbation_deriv_expression}
\end{align}
where we have used the continuity equation, \cref{app:eqn:lin_resp_cont_eqn}, as well as the fact that, by assumption, $\nabla \phi_{n} \left( \vec{r} \right) \equiv \nabla \phi_{n}$ is $\vec{r}$-independent (corresponding, for instance, to a uniform electric field). Using \cref{app:eqn:expec_current_op_interm,app:eqn:perturbation_deriv_expression}, we find that the induced currents $\left\langle j^{\alpha}_n \right\rangle$ are linearly related to the applied forces according to 
\begin{equation}
	\left\langle j^{\alpha}_n \right\rangle = \sum_{m=1}^{N_{\phi}} \sum_{\beta=1}^{d} \chi^{\alpha \beta}_{nm} \partial^{\beta} \phi_m,
\end{equation}
with the static bulk linear response coefficients $\chi^{\alpha \beta}_{nm}$ being given by~\cite{LUT64,MAH00}
\begin{equation}
	\label{app:eqn:lin_resp_coeff_expression_luttinger}
	\chi^{\alpha \beta}_{nm} \equiv - \frac{1}{N_0 \Omega_0} \int_{0}^{\infty} \dd{t} e^{-s t}  \int_{0}^{\beta} \dd{\tau} \Tr \left( \hat{\varrho}_0  J^{\beta}_m \left(- t - i \tau \right) J^{\alpha}_n \left( 0 \right) \right).
\end{equation}
In \cref{app:eqn:lin_resp_coeff_expression_luttinger} the limit $s \to 0^{+}$ should be taken at the end (corresponding to the case when the perturbation is turned on infinitely slowly). 

\subsubsection{Properties of the linear response coefficient}\label{app:sec:thermoelectric_response:lin_response:properties}

Due to the presence of two integrals, the expression of the linear response coefficient from \cref{app:eqn:lin_resp_coeff_expression_luttinger} is somewhat cumbersome to employ directly. To obtain a simpler form for $\chi^{\alpha \beta}_{nm}$, we start by introducing the complete set of eigenstates of the grand canonical Hamiltonian $\left \lbrace \ket{a} \right \rbrace$, such that 
\begin{equation}
	\label{app:eqn:complete_set_lin_resp_theory}
	K \ket{a} = E_a \ket{a}.
\end{equation}
Inserting resolutions of the identity in terms of the complete basis $\left \lbrace \ket{a} \right \rbrace$ into \cref{app:eqn:lin_resp_coeff_expression_luttinger}, we can rewrite it as 
\begin{align}
	\chi^{\alpha \beta}_{nm} =& - \frac{1}{N_0 \Omega_0} \frac{1}{Z} \sum_{a,b} \int_{0}^{\infty} \dd{t} e^{-st} \int_{0}^{\beta} \dd{\tau} e^{-\beta E_a} \bra{a} e^{-i E_a \left( t + i \tau \right)} J^{\beta}_{m} e^{i E_b \left( t + i \tau \right)}\ket{b} \bra{b} J^{\alpha}_{n} \ket{a} \nonumber \\
	=& - \frac{1}{N_0 \Omega_0} \frac{1}{Z} \sum_{a,b} e^{-\beta E_a} \bra{a} {J}^{\beta}_{m} \ket{b} \bra{b} {J}^{\alpha}_{n} \ket{a} \int_{0}^{\infty} \dd{t} e^{i t \left(E_b - E_a + i s \right)}  \int_{0}^{\beta} \dd{\tau} e^{ \tau \left( E_a - E_b \right)} \nonumber \\
	=& \frac{1}{N_0 \Omega_0} \frac{1}{Z} \sum_{a,b} \frac{e^{-\beta E_b} - e^{-\beta E_a}}{E_b - E_a} \bra{a} {J}^{\beta}_{m} \ket{b} \bra{b} {J}^{\alpha}_{n} \ket{a} \frac{-i\left[ \lim_{t \to \infty} e^{it \left( E_b - E_a + i s \right)} - 1 \right]}{E_b - E_a + i s}\nonumber\\
 =& \frac{1}{N_0 \Omega_0} \frac{1}{Z} \sum_{a,b} \frac{e^{-\beta E_b} - e^{-\beta E_a}}{E_b - E_a} \bra{a} {J}^{\beta}_{m} \ket{b} \bra{b} {J}^{\alpha}_{n} \ket{a} \frac{i}{E_b - E_a + i s}.
\label{app:eqn:spectral_decomposition_chi}
\end{align}

One key property that can be extracted from \cref{app:eqn:spectral_decomposition_chi} is that $\chi^{\alpha \beta}_{nm}$ is real. Indeed one can directly compute
\begin{align}
	\chi^{*\alpha \beta}_{nm} =& \frac{1}{N_0 \Omega_0} \frac{1}{Z} \sum_{a,b} \frac{e^{-\beta E_b} - e^{-\beta E_a}}{E_b - E_a} \bra{a} {J}^{\beta}_{m} \ket{b} \bra{b} {J}^{\alpha}_{n} \ket{a} \frac{i}{E_a - E_b + i s} \nonumber \\
	 =& \frac{1}{N_0 \Omega_0} \frac{1}{Z} \sum_{a,b} \frac{e^{-\beta E_a} - e^{-\beta E_b}}{E_a - E_b} \bra{a} {J}^{\beta}_{m} \ket{b} \bra{b} {J}^{\alpha}_{n} \ket{a} \frac{i}{E_b - E_a + i s} = \chi^{\alpha \beta}_{nm},
\end{align}
where, in going from the first to the second row, we have exchanged the dummy summations indices $a$ and $b$ and used the Hermiticity of the current operators. 

Because $\chi^{\alpha \beta}_{nm}$ is real, one trivially has $\chi^{\alpha \beta}_{nm} =  \Re \left( \chi^{\alpha \beta}_{nm} \right)$, which allows us to write 
\begin{equation}
	\chi^{\alpha \beta}_{nm} = \Re \left( \chi^{\alpha \beta}_{nm} \right) = \left( \chi^{\alpha \beta}_{nm} \right)_1 + \left( \chi^{\alpha \beta}_{nm} \right)_2,
\end{equation}
where we have separated $\chi^{\alpha \beta}_{nm}$ into two contributions
\begin{align}
	\left( \chi^{\alpha \beta}_{nm} \right)_1 &= \frac{1}{N_0 \Omega_0} \frac{1}{Z} \sum_{a,b} \frac{e^{-\beta E_b} - e^{-\beta E_a}}{E_b - E_a} \Re \left( \bra{a} {J}^{\beta}_{m} \ket{b} \bra{b} {J}^{\alpha}_{n} \ket{a} \right)  \Re \left( \frac{i}{E_b - E_a + i s} \right) \nonumber \\
	&= \frac{\pi}{N_0 \Omega_0} \frac{1}{Z} \sum_{a,b}  \frac{e^{-\beta E_b} - e^{-\beta E_a}}{E_b - E_a}  \Re \left( \bra{a} {J}^{\beta}_{m} \ket{b} \bra{b} {J}^{\alpha}_{n} \ket{a} \right) \delta \left( E_b - E_a \right) \nonumber \\
	&= -\frac{\pi \beta}{N_0 \Omega_0} \frac{1}{Z} \sum_{a,b}  e^{-\beta E_a}  \Re \left( \bra{a} {J}^{\beta}_{m} \ket{b} \bra{b} {J}^{\alpha}_{n} \ket{a} \right) \delta \left( E_b - E_a \right), \label{app:eqn:chi_original_1_final} \\
\left( \chi^{\alpha \beta}_{nm} \right)_2 &= - \frac{1}{N_0 \Omega_0} \frac{1}{Z} \sum_{a,b} \frac{e^{-\beta E_b} - e^{-\beta E_a}}{E_b - E_a} \Im \left( \bra{a} {J}^{\beta}_{m} \ket{b} \bra{b} {J}^{\alpha}_{n} \ket{a} \right)  \Im \left( \frac{i}{E_b - E_a + i s} \right) \nonumber \\
	&= - \frac{1}{N_0 \Omega_0} \frac{1}{Z} \sum_{a,b}  \frac{e^{-\beta E_b} - e^{-\beta E_a}}{E_b - E_a}  \Im \left( \bra{a} {J}^{\beta}_{m} \ket{b} \bra{b} {J}^{\alpha}_{n} \ket{a} \right) \mathcal{PV} \frac{1}{E_b - E_a}. \label{app:eqn:chi_original_2_final}
\end{align}
In both \cref{app:eqn:chi_original_1_final,app:eqn:chi_original_2_final} the limit $s \to 0^{+}$ was taken at the end, with $\mathcal{PV}$ denoting the Cauchy principal value.

\subsubsection{Alternative expression for the linear response coefficients}\label{app:sec:thermoelectric_response:lin_response:matsubara}

The expression in \cref{app:eqn:spectral_decomposition_chi} is somewhat reminiscent of the Lehmann representation of correlation functions~\cite{MAH00}. To make this connection more apparent, we start by introducing the following correlation function~\cite{MAH00,CZY08}
\begin{equation}
	\label{app:eqn:current_corr_func_matsubara}
	\chi^{\alpha \beta}_{nm} \left( i \tilde{\Omega}_n \right) = \frac{1}{N_0 \Omega_0} \frac{i}{i \tilde{\Omega}_n} \int_{0}^{\beta} \dd{\tau} e^{i \tilde{\Omega}_n \tau} \left\langle \mathcal{T}_{\tau} J^{\beta}_{m} \left( -i\tau \right) J^{\alpha}_{n} \left( 0 \right) \right\rangle,
\end{equation}
where $i\tilde{\Omega}_n = \frac{ 2 \pi i n}{\beta}$ ($n \in \mathbb{Z}$) is a bosonic Matsubara frequency\footnote{To avoid confusing $i\tilde{\Omega}_n$ with the unit-cell volume $\Omega_0$, a tilde was added to the former.},
\begin{equation}
	J^{\alpha}_{m} \left( - i \tau \right) = e^{K \tau} J^{\alpha}_{m} e^{-K \tau}
\end{equation}
is the imaginary-time-evolved current operator\footnote{Just for the current operators, we will stick to the convention of \cref{app:eqn:general_total_current_definition}, wherein the imaginary time argument is explicitly indicated by the imaginary unit $i$.}, $\left\langle \dots \right\rangle$ denotes the average in the grand canonical ensemble corresponding to $K$ and $\mathcal{T}_{\tau}$ enforces the time-ordering with respect to the imaginary time
\begin{equation}
	\mathcal{T}_{\tau} J^{\beta}_{m} \left( -i\tau \right) J^{\alpha}_{n} \left( 0 \right) = \begin{cases}
		J^{\beta}_{m} \left( -i\tau \right) J^{\alpha}_{n} \left( 0 \right), & \qq{if} \tau>0 \\
		J^{\alpha}_{n} \left( 0 \right) J^{\beta}_{m} \left( -i\tau \right), & \qq{if} \tau<0 \\
	\end{cases}.
\end{equation}
Up to prefactors, \cref{app:eqn:current_corr_func_matsubara} is nothing else but the imaginary-time, time-ordered Green's function of the current operators~\cite{MAH00}. Moreover, it can be shown that \cref{app:eqn:current_corr_func_matsubara} is the dynamical linear response coefficient~\cite{MAH00}. Note that in the expression of $\chi^{\alpha\beta}_{nm}$ from \cref{app:eqn:lin_resp_coeff_expression_luttinger}, no time-ordering is necessary. Our goal in what follows will be to relate the static linear response coefficients $\chi_{nm}^{\alpha\beta}$ to the correlation function $\chi^{\alpha \beta}_{nm} \left( i \tilde{\Omega}_n \right)$.

To relate \cref{app:eqn:current_corr_func_matsubara} to the static linear response coefficients $\chi^{\alpha \beta}_{nm}$, we construct the spectral representation of \cref{app:eqn:current_corr_func_matsubara} by inserting resolutions of the identity in terms of the eigenbasis of $K$
\begin{align}
	\chi^{\alpha \beta}_{nm} \left( i \tilde{\Omega}_n \right) &= \frac{1}{N_0 \Omega_0} \frac{i}{i \tilde{\Omega}_n} \frac{1}{Z} \sum_{a,b} \int_{0}^{\beta} \dd{\tau} e^{i \tilde{\Omega}_n \tau} e^{-\beta E_b} e^{E_b \tau} e^{-E_a \tau} \bra{b} J^{\alpha}_{n} \ket{a} \bra{a} J^{\beta}_{m} \ket{b} \nonumber \\
	&= \frac{1}{N_0 \Omega_0} \frac{i}{i \tilde{\Omega}_n} \frac{1}{Z} \sum_{a,b}  e^{-\beta E_b} \frac{e^{\beta \left(i \tilde{\Omega}_n - E_a + E_b \right)} - 1}{i \tilde{\Omega}_n - E_a + E_b} \bra{a} {J}^{\beta}_{m} \ket{b} \bra{b} {J}^{\alpha}_{n} \ket{a} \nonumber \\
	&= \frac{1}{N_0 \Omega_0} \frac{i}{i \tilde{\Omega}_n} \frac{1}{Z} \sum_{a,b}  \frac{e^{- \beta E_a} - e^{- \beta E_b}}{i \tilde{\Omega}_n - E_a + E_b} \bra{a} {J}^{\beta}_{m} \ket{b} \bra{b} {J}^{\alpha}_{n} \ket{a}. \label{app:eqn:current_corr_func_matsubara_anal_cont}
\end{align}
Next, we analytically continue the expression of $\chi^{\alpha \beta}_{nm} \left( i \Omega_n \right)$ along the real axis by substituting $i \tilde{\Omega}_n \to \omega + i 0^{+}$, while simultaneously taking the real part (as we will show later in this section, the real part is connected to the linear response coefficient $\chi_{nm}^{\alpha\beta}$, and we are not interested in the imaginary part) to obtain
\begin{equation}
	\label{app:eqn:real_part_linear_response_1}
	\Re \chi^{\alpha \beta}_{nm} \left( \omega + i 0^{+} \right) = \left( \Re \chi^{\alpha \beta}_{nm} \left( \omega + i 0^{+} \right) \right)_1 + \left( \Re \chi^{\alpha \beta}_{nm} \left( \omega + i 0^{+} \right) \right)_2,
\end{equation}
where similarly to \cref{app:eqn:chi_original_1_final,app:eqn:chi_original_2_final}, we have separated $\Re \chi^{\alpha \beta}_{nm} \left( \omega + i 0^{+} \right)$ into two contributions
\begin{align}
	\left( \Re \chi^{\alpha \beta}_{nm} \left( \omega + i 0^{+} \right) \right)_1 &= \frac{-1}{N_0 \Omega_0} \frac{1}{Z} \sum_{a,b} \frac{ e^{- \beta E_a} - e^{- \beta E_b}}{\omega} \Re \left( \bra{a} {J}^{\beta}_{m} \ket{b} \bra{b} {J}^{\alpha}_{n} \ket{a} \right) \Im \left( \frac{1}{\omega + i 0^{+} - E_a + E_b} \right), \label{app:eqn:chi_new_1_def}\\
	\left( \Re \chi^{\alpha \beta}_{nm} \left( \omega + i 0^{+} \right) \right)_2 &= \frac{-1}{N_0 \Omega_0} \frac{1}{Z} \sum_{a,b} \frac{ e^{- \beta E_a} - e^{- \beta E_b}}{\omega} \Im \left( \bra{a} {J}^{\beta}_{m} \ket{b} \bra{b} {J}^{\alpha}_{n} \ket{a} \right)\Re \left( \frac{1}{\omega + i 0^{+} - E_a + E_b}  \right). \label{app:eqn:chi_new_2_def}
\end{align}
In \cref{app:eqn:chi_new_1_def,app:eqn:chi_new_2_def} we have also dropped the small imaginary offset ``$+i 0^{+}$'' in the $\frac{1}{\omega + i 0^{+}}$ factor. In practice, this small imaginary offset will generate a $- i \pi \delta \left( \omega \right)$ term. However, since we are interested in taking the limit $\omega \to 0$, rather than in evaluating $\Re \chi^{\alpha \beta}_{nm} \left( \omega + i 0^{+} \right)$ directly \emph{at} $\omega = 0$, this additional contribution will vanish (since $\delta \left( \omega \right) = 0$ for $\omega \neq 0$). We now take the zero-frequency \emph{limit} ($\omega \to 0$). The first contribution becomes equal to the first term of the static linear response coefficient $\chi^{\alpha \beta}_{nm}$ from \cref{app:eqn:chi_original_1_final}
\begin{align}
	\lim_{\omega \to 0} \left( \Re \chi^{\alpha \beta}_{nm} \left( \omega + i 0^{+} \right) \right)_1 &= \lim_{\omega \to 0}\frac{\pi}{N_0 \Omega_0} \frac{1}{Z} \sum_{a,b} \Re \left( \bra{a} {J}^{\beta}_{m} \ket{b} \bra{b} {J}^{\alpha}_{n} \ket{a} \right) \frac{e^{- \beta E_a} - e^{- \beta E_b}}{\omega} \delta \left( \omega - E_a + E_b \right) \nonumber \\
	&= \lim_{\omega \to 0}\frac{\pi}{N_0 \Omega_0} \frac{1}{Z} \sum_{a,b} \Re \left( \bra{a} {J}^{\beta}_{m} \ket{b} \bra{b} {J}^{\alpha}_{n} \ket{a} \right) \frac{e^{- \beta \left( \omega + E_b \right)} - e^{- \beta E_b} }{\omega} \delta \left( \omega - E_a + E_b \right) \nonumber \\
	&= -\frac{\pi \beta}{N_0 \Omega_0} \frac{1}{Z} \sum_{a,b} \Re \left( \bra{a} {J}^{\beta}_{m} \ket{b} \bra{b} {J}^{\alpha}_{n} \ket{a} \right) e^{- \beta E_b} \delta \left(E_a - E_b \right) = \left( \chi^{\alpha \beta}_{nm} \right)_1. \label{app:eqn:chi_new_1_final}
\end{align}

Evaluating the second contribution in \cref{app:eqn:chi_new_2_def} requires a more careful treatment. For starters, we replace the $+ i 0^{+}$ term by $+is$ and defer taking the limit $s \to 0^{+}$ to the end. We then expand in powers of $\omega$ up to and including the zeroth order
\begin{align}
	\left( \Re \chi^{\alpha \beta}_{nm} \left( \omega + i 0^{+} \right) \right)_2 &= \frac{-1}{N_0 \Omega_0} \frac{1}{Z} \sum_{a,b} \left( e^{- \beta E_a} - e^{- \beta E_b} \right) \Im \left( \bra{a} {J}^{\beta}_{m} \ket{b} \bra{b} {J}^{\alpha}_{n} \ket{a} \right) \nonumber \\
	&\times \left\lbrace - \frac{E_a - E_b}{\left( E_a - E_b \right)^2 + s^2} \frac{1}{\omega} - \frac{\left( E_a - E_b \right)^2 - s^2}{ \left[ \left( E_a - E_b \right)^2 + s^2 \right]^2} + \mathcal{O} \left( \omega \right) \right\rbrace . \label{app:eqn:chi_new_2_inter_1} 
\end{align}
It can be seen that the first term in the braces vanishes upon interchanging the dummy summation variables $a \leftrightarrow b$, because it changes sign under the exchange, while
\begin{align}
	\left( e^{-\beta E_a} - e^{-\beta E_b} \right) \Im \left( \bra{a} {J}^{\beta}_{m} \ket{b} \bra{b} {J}^{\alpha}_{n} \ket{a} \right) &= \left( e^{-\beta E_b} - e^{-\beta E_a}  \right) \Im \left[ \left( \bra{a} {J}^{\beta}_{m} \ket{b} \right)^* \left( \bra{b} {J}^{\alpha}_{n} \ket{a} \right)^* \right] \nonumber \\
	&= \left( e^{-\beta E_b} - e^{-\beta E_a}  \right) \Im \left( \bra{b} {J}^{\beta}_{m} \ket{a}  \bra{a} {J}^{\alpha}_{n} \ket{b}  \right). 
\end{align}
Taking the limit $\omega \to 0$, we obtain
{\small
	\begin{equation}
	\label{app:eqn:chi_new_2_inter_2}
	\lim_{\omega \to 0} \left( \Re \chi^{\alpha \beta}_{nm} \left( \omega + i 0^{+} \right) \right)_2 = \frac{1}{N_0 \Omega_0} \frac{1}{Z} \sum_{a,b} \frac{e^{- \beta E_a} - e^{- \beta E_b}}{E_a -E_b} \Im \left( \bra{a} {J}^{\beta}_{m} \ket{b} \bra{b} {J}^{\alpha}_{n} \ket{a} \right) \frac{\left[ \left( E_a - E_b \right)^2 - s^2\right] \left(E_a - E_b \right)}{ \left[ \left( E_a - E_b \right)^2 + s^2 \right]^2},
\end{equation}}where the limit $s \to 0^{+}$ is still left to be imposed. In addition, we have kept a $ \left( E_a-E_b \right)$ in both denominator and numerator in \cref{app:eqn:chi_new_2_inter_2}, which will simplify our subsequent calculations as we discuss now. Letting $x = E_a - E_b$ and replacing $s \to \sqrt{s}$ (which is allowed, since we will take the limit $s \to 0^{+}$ at the end), the final factor in \cref{app:eqn:chi_new_2_inter_2} becomes
\begin{equation}
	\label{app:eqn:fancy_distribution}
	\frac{\left(x^2 - s^2\right) x}{\left( x^2 + s^2 \right)^2} = \frac{x}{x^2 + s^2} - \frac{2 s^2 x}{\left( x^2 + s^2 \right)^2} = \Re \frac{1}{x+ i s} - \frac{2 s^2 x}{\left( x^2 + s^2 \right)^2}.
\end{equation}
Upon taking the limit $s \to 0^{+}$, the first term in \cref{app:eqn:fancy_distribution} will approach $\mathcal{PV} \frac{1}{x}$. The second term, on the other hand, will tend to zero for $x \neq 0$, but can still give to nonzero \emph{distribution} (such as a Dirac $\delta$-function) for $x = 0$ (which would correspond to $E_a = E_b$). To understand the behavior of $\frac{2 s^2 x}{\left( x^2 + s^2 \right)^2}$ near the origin, we consider the general analytic function 
\begin{equation}
	g (x) = \sum_{n=0}^{\infty} a_n x^{n},
\end{equation}
which is defined in terms of its Taylor series for a small interval around the origin. We now integrate $g (x)$ against the second term of \cref{app:eqn:fancy_distribution} around the origin
\begin{equation}
	\mathcal{I}_{\epsilon} = \int_{-\epsilon}^{\epsilon} \dd{x} g (x) \frac{2 s^2 x}{\left( x^2 + s^2 \right)^2} = \sum_{n=0}^{\infty} \int_{-\frac{\epsilon}{s}}^{\frac{\epsilon}{s}} \dd{\lambda} \frac{2 s^4 \lambda}{ s^4 \left(1 + \lambda^2 \right)^2} a_n \left( \lambda s \right)^n = \sum_{n=0}^{\infty} a_{2 n + 1} s^{2 n + 1} \int_{-\frac{\epsilon}{s}}^{\frac{\epsilon}{s}} \dd{\lambda} \frac{2 \lambda^{2n + 2}}{ \left(1 + \lambda^2 \right)^2}.
\end{equation}
The integration interval $\left[-\epsilon, \epsilon \right]$ is chosen to lie within the convergence radius of $g (x)$. As $s$ approaches zero, the main contributions to the integral $\mathcal{I}_{\epsilon}$ will stem from $\lambda \gg 1$, such that
\begin{equation}
	\mathcal{I}_{\epsilon} 
	= \sum_{n=0}^{\infty} a_{2 n + 1} s^{2 n + 1} \left( \int_{-\frac{\epsilon}{s}}^{\frac{\epsilon}{s}} \dd{\lambda} 2 \lambda^{2n - 2} + \mathcal{O} \left( 1 \right) \right)
	= \sum_{n=0}^{\infty} a_{2 n + 1} s^{2 n + 1} \left( \frac{4}{2n-1}  \left( \frac{\epsilon}{s} \right)^{2n-1} + \mathcal{O} \left( 1 \right) \right) = \mathcal{O} \left( s \right),
\end{equation} 
which tends to zero as $s \to 0^{+}$. As such, we conclude that 
\begin{equation}
	\label{app:eqn:fancy_distribution_solved}
	\lim_{s \to 0^{+}} \frac{\left(x^2 - s^2\right) x}{\left( x^2 + s^2 \right)^2} = \mathcal{PV} \frac{1}{x}.
\end{equation}
Using \cref{app:eqn:fancy_distribution_solved} in \cref{app:eqn:chi_new_2_inter_2}, we obtain
\begin{equation}
	\label{app:eqn:chi_new_2_final}
	\lim_{\omega \to 0} \left( \Re \chi^{\alpha \beta}_{nm} \left( \omega + i 0^{+} \right) \right)_2 =- \frac{1}{N_0 \Omega_0} \frac{1}{Z} \sum_{a,b} \frac{e^{- \beta E_b} - e^{- \beta E_a}}{E_b -E_a} \Im \left( \bra{a} {J}^{\beta}_{m} \ket{b} \bra{b} {J}^{\alpha}_{n} \ket{a} \right) \mathcal{PV} \frac{1}{E_b - E_a} = \left( \chi^{\alpha \beta}_{nm} \right)_2.
\end{equation}
Taken together, \cref{app:eqn:chi_new_1_final,app:eqn:chi_new_2_final} imply that 
\begin{equation}
	\label{app:eqn:linear-response_with_anal_cont}
	\chi^{\alpha \beta}_{nm} = \lim_{\omega \to 0} \left[ \Re\left( \eval{\chi^{\alpha \beta}_{n m} \left( i \tilde{\Omega}_n \right) }_{i \tilde{\Omega}_n \to \omega + i 0^{+}} \right)\right].
\end{equation}
Because the correlation function $\chi^{\alpha \beta}_{n m} \left( i \tilde{\Omega}_n \right)$ can be computed using standard diagramatic techniques, \cref{app:eqn:linear-response_with_anal_cont} provides a crucial simplification for obtaining the linear response coefficients, and will be employed instead of \cref{app:eqn:lin_resp_coeff_expression_luttinger} in our upcoming derivations. 

Finally, we mention that one intermediate result derived and employed in this section (which will also be used in subsequent calculations) is that for vanishing (but nonzero) $\omega$ 
\begin{align}
	\frac{E}{\omega} \frac{1}{\omega + i 0^{+} - E} &= \lim_{s \to 0^+} \frac{E}{\omega} \frac{1}{\omega - E + i s } \nonumber \\
	&= \lim_{s \to 0^+} \frac{E}{\omega} \left[ \frac{\omega - E}{\left( \omega - E \right)^2 + s^2 } - i \frac{s}{\left( \omega - E \right)^2 + s^2 } \right] \nonumber \\
	&= \lim_{s \to 0^+} \left[ \frac{1}{\omega} \frac{- E^2}{E^2 + s^2 } - \frac{\left(E^2 - s^2 \right) E}{\left( E^2 + s^2 \right)^2} - i \frac{E}{\omega} \frac{s}{\left( \omega - E \right)^2 + s^2 } + \mathcal{O} \left( \omega \right) \right] \nonumber \\
	&= -\frac{1}{\omega} - \mathcal{PV} \frac{1}{E} - i \pi \frac{E}{\omega} \delta \left( \omega - E \right) + \mathcal{O} \left( \omega \right), \label{app:eqn:one_over_omega_plemelj}
\end{align}
where we have used \cref{app:eqn:fancy_distribution_solved}.

\subsubsection{Some remarks on time-reversal symmetry}\label{app:sec:thermoelectric_response:lin_response:trev_sym}

In \cref{app:sec:thermoelectric_response:lin_response:matsubara}, we have shown that using \cref{app:eqn:linear-response_with_anal_cont}, the static susceptibility can be computed from the current-current correlation function introduced in \cref{app:eqn:current_corr_func_matsubara}. Here, we would like to point out that our proof of \cref{app:eqn:linear-response_with_anal_cont} can be simplified in the presence of \emph{spinless} time-reversal symmetry $\mathcal{T}$, with $\mathcal{T}^2=1$.

To this end, we assume that the many-body unperturbed grand canonical Hamiltonian $K$ has spinless time-reversal symmetry ({\it i.e.}{}, $\commutator{\mathcal{T}}{H} = 0$, where $\mathcal{T}^2 = 1$, such is the case of the THF model). As a result of this, the eigenbasis of $K$ can be chosen to be time-reversal symmetric, {\it i.e.}{}
\begin{equation}
	\mathcal{T} \ket{a} = \ket{\mathcal{T} a} = \ket{a}.
\end{equation}
Next, we also assume that the generalized charge density operators to which the external fields couple in the perturbation from \cref{app:eqn:general_perturbation_lin_theo} are \emph{also} time-reversal symmetric, namely
\begin{equation}
	\label{app:eqn:time_reversal_action_dens_ops}
	\mathcal{T} \rho_{n} \left( \vec{r} \right) \mathcal{T}^{-1} = \rho_{n} \left( \vec{r} \right), \quad 
	\mathcal{T} \rho_{n} \left( \vec{r}, t \right) \mathcal{T}^{-1} = \rho_{n} \left( \vec{r}, -t \right),
	\qq{for} 1 \leq n \leq N_{\phi}.
\end{equation}

In passing, we mention that for the particle and energy density operators introduced in \cref{app:eqn:cont_part_density,app:eqn:energy_density_2}, respectively, this is indeed the case, whenever the Hamiltonian of the system is itself time-reversal symmetric. To see why, we start by denoting the action of the time-reversal symmetry operator on the fermions introduced in \cref{app:sec:cur_deriv_cont:model} by 
\begin{equation}
	\mathcal{T} \hat{\psi}^\dagger_{n} \left( \vec{r} \right) \mathcal{T}^{-1} = \sum_{m} \left[ D \left( \mathcal{T} \right) \right]_{m n} \hat{\psi}^\dagger_{m} \left( \vec{r} \right),
\end{equation}
where the matrix $D \left( \mathcal{T} \right)$ is unitary. Next, we note that the particle density operator from \cref{app:eqn:cont_part_density} is time-reversal symmetric
\begin{equation}
	\label{app:eqn:t_rev_rho_charge}
	\mathcal{T} \rho \left( \vec{r} \right) \mathcal{T}^{-1} = \sum_{n,m_1,m_2} \left[ D \left( \mathcal{T} \right) \right]_{m_1 n} \left[ D \left( \mathcal{T} \right) \right]^{*}_{m_2 n} \hat{\psi}^\dagger_{m_1} \left( \vec{r} \right) \hat{\psi}_{m_2} \left( \vec{r} \right) = \sum_{m_1,m_2} \delta_{m_1 m_2} \hat{\psi}^\dagger_{m_1} \left( \vec{r} \right) \hat{\psi}_{m_2} \left( \vec{r} \right) = \rho \left( \vec{r} \right),
\end{equation}
which follows straight-forwardly from the unitarity of $D \left( \mathcal{T} \right)$. At the same time, requiring that the Hamiltonian from \cref{app:eqn:generic_hamiltonian} is time-reversal symmetric places the following constraints on the kinetic energy and crystalline potential matrices 
\begin{align}
	\sum_{m_1,m_2} \left[ D \left( \mathcal{T} \right) \right]_{m_1 n_1} T^{*}_{m_1 m_2} \left( -i \nabla \right) \left[ D \left( \mathcal{T} \right) \right]^{*}_{m_2 n_2} &=  T_{n_1 n_2} \left( -i \nabla \right), \label{app:eqn:t_rev_T} \\
	\sum_{m_1,m_2} \left[ D \left( \mathcal{T} \right) \right]_{m_1 n_1} U^{*}_{m_1 m_2} \left( \vec{r} \right) \left[ D \left( \mathcal{T} \right) \right]^{*}_{m_2 n_2} &=  U_{n_1 n_2} \left( \vec{r} \right). \label{app:eqn:t_rev_U}
\end{align}
In turn, \cref{app:eqn:t_rev_rho_charge,app:eqn:t_rev_T,app:eqn:t_rev_U} imply that the energy density $\rho_{E} \left( \vec{r} \right)$ from \cref{app:eqn:energy_density_2} is also time-reversal symmetric. This is a consequence of the locality of the time-reversal symmetry operator: as $\mathcal{T}$ is both local \emph{and} a symmetry of the entire Hamiltonian, it will also be a symmetry of the energy density operator. 

From \cref{app:eqn:time_reversal_action_dens_ops} and the continuity equation in \cref{app:eqn:lin_resp_cont_eqn}, we find that 
\begin{equation}
	\label{app:eqn:time_reversal_action_local_cur}
	\mathcal{T} \vec{j}_{n} \left( \vec{r} \right) \mathcal{T}^{-1} = -\vec{j}_{n} \left( \vec{r} \right), \quad 
	\mathcal{T} \vec{j}_{n} \left( \vec{r}, t \right) \mathcal{T}^{-1} = -\vec{j}_{n} \left( \vec{r}, -t \right),
	\qq{for} 1 \leq n \leq N_{\phi},
\end{equation}
which, in turn, implies that total currents defined in \cref{app:eqn:general_total_current_definition} also anticommute with the time-reversal operator 
\begin{equation}
	\label{app:eqn:time_reversal_action_total_cur}
	\mathcal{T} \vec{J}_{n} \mathcal{T}^{-1} = -\vec{J}_{n}, \quad 
	\mathcal{T} \vec{J}_{n} \left( t \right) \mathcal{T}^{-1} = -\vec{J}_{n} \left( -t \right),
	\qq{for} 1 \leq n \leq N_{\phi}.
\end{equation}

As a consequence of time-reversal symmetry, the current matrix elements $\bra{a} {J}^{\beta}_{m} \ket{b} \bra{b} {J}^{\alpha}_{n} \ket{a}$ can be shown to be real. To avoid any ambiguities arising when combining the antiunitarity of the time-reversal symmetry operator with the Dirac bra-ket notation, we will temporarily denote the inner product as $\left\langle a,b \right\rangle \equiv \bra{a} \ket{b}$. In this convention, we find 
\begin{align}
	& \bra{a}  J^{\beta}_{m} \ket{b} \bra{b} J^{\alpha}_{n} \ket{a} 
	= \left\langle a,J^{\beta}_{m} b \right\rangle \left\langle b,J^{\alpha}_{n} a \right\rangle \nonumber \\
	=& \left\langle \mathcal{T} a, \mathcal{T} J^{\beta}_{m} b \right\rangle^{*} \left\langle \mathcal{T} b, \mathcal{T} J^{\alpha}_{n} a \right\rangle^{*}
	= \left\langle \mathcal{T} a, \left( -J^{\beta}_{m} \right) \mathcal{T} b \right\rangle^{*} \left\langle \mathcal{T} b, \left( -J^{\alpha}_{n} \right) \mathcal{T} a \right\rangle^{*} \nonumber \\
	=& \left\langle a, \left( - J^{\beta}_{m} \right) b \right\rangle^{*} \left\langle b, \left( -J^{\alpha}_{n} \right) a \right\rangle^{*}
	= \bra{a} \left( - J^{\beta}_{m} \right) \ket{b}^{*} \bra{b} \left( -J^{\alpha}_{n} \right) \ket{a}^{*}
	= \left( \bra{a} J^{\beta}_{m} \ket{b} \bra{b} J^{\alpha}_{n} \ket{a} \right)^{*}. \label{app:eqn:reality_cur_mat_elem}
\end{align}
In turn, \cref{app:eqn:reality_cur_mat_elem} implies that the $\left( \chi^{\alpha \beta}_{nm} \right)_2$ and $\left( \Re \chi^{\alpha \beta}_{nm} \left( \omega + i 0^{+} \right) \right)_2$ contributions from \cref{app:eqn:chi_original_2_final,app:eqn:chi_new_2_def} vanish. This simplifies the proof of \cref{app:eqn:linear-response_with_anal_cont} by avoiding the rather technical manipulation of expressions in \cref{app:eqn:chi_new_2_inter_1} involving Cauchy's principal value. 

When proving \cref{app:eqn:linear-response_with_anal_cont}, Ref.~\cite{MAH00} implicitly assumes the reality of the current matrix elements, without justifying it. Even more confusingly, in the current-current correlator expressions analogous to \cref{app:eqn:current_corr_func_matsubara}, Ref.~\cite{MAH00} also incorrectly swaps the indices $n \leftrightarrow m$ and  $\alpha \leftrightarrow \beta$. This mistake, which can be confirmed by comparing with similar expressions from Ref.~\cite{CZY08}, is inconsequential whenever the current matrix elements are real, but leads to incorrect results generally. We hereby have corrected it.

In the present work, the odd-filling states break time-reversal symmetry, by picking a positive or a negative Chern number~\cite{LIA21}. To reflect this spontaneous symmetry breaking, one can include a small magnetic field in the Hamiltonian, as a result of which the latter will no longer be time-reversal symmetric. Therefore, for our purposes, the simpler proof of \cref{app:eqn:linear-response_with_anal_cont} shown in Ref.~\cite{MAH00} does not hold and instead had to be re-derived in \cref{app:sec:thermoelectric_response:lin_response:matsubara}.

\subsection{Thermoelectric transport} \label{app:sec:thermoelectric_response:thermoelectric}

Having briefly reviewed linear response theory more generally in \cref{app:sec:thermoelectric_response:lin_response:static_perturbation}, in this section we particularize the formulae derived therein in the case of thermoelectric transport. 

\subsubsection{The thermoelectric linear response coefficients \texorpdfstring{$L^{\alpha \beta}_{ij}$}{Labij}} \label{app:sec:thermoelectric_response:thermoelectric:Lij}

Consider the situation in which a uniform electric field and a constant temperature gradient is applied to our system. As a result of these two ``forces'', the electrons will flow and give rise to both energy and particle currents. In what follows, it is conventional to work in terms of the \emph{thermal} current introduced in \cref{app:eqn:generic_thermal_current}, as opposed to working with the energy current directly. The two currents differ by a $\mu \vec{j}\left( \vec{r} \right)$ term as described in \cref{app:eqn:generic_thermal_current}. As we show later, the correlators of involving the particle and thermal (as opposed to energy) currents are directly connected to the Seebeck coefficient. 

To investigate the current generated by temperature and electric potential, we also introduce the \emph{electrochemical} potential\footnote{The electrochemical potential $\bar{\mu} \left( \vec{r} \right)$ (denoted with a ``bar'') should not be confused with the fictitious chemical potential $\tilde{\mu}$ (denoted with a ``tilde'') that was introduced in Ref.~\cite{CAL23b} for the modified iterative perturbation theory.}
\begin{equation}
	\bar{\mu} \left( \vec{r} \right) \equiv \mu + e \Phi \left( \vec{r} \right),
 \label{app:eqn:electrochemical potential}
\end{equation}
where $e$ is the electron's charge and $\Phi \left( \vec{r} \right)$ is the electrostatic potential which is in general $\vec{r}$ dependent. The temperature gradients and the electrochemical potential give rise to both particle and thermal currents. In the linear-response regime, both currents are linearly dependent on the gradients 
$\nabla \tilde{\mu}$ and $ \nabla \left( 1/T \right)$. This linear relation allows us to define the thermal transport coefficients $L^{\alpha \beta}_{ij}$ (with $1\leq i,j \leq 2$) via the following phenomenological equations~\cite{MAH00}
\begin{align}
	\left\langle j^{\alpha} \right\rangle &= \sum_{\beta} \left[ - \frac{1}{T} L^{\alpha \beta}_{11} \partial^{\beta} \bar{\mu} + L^{\alpha \beta}_{12} \partial^{\beta} \left( \frac{1}{T} \right) \right], \label{app:eqn:thermo_response_part} \\
	\left\langle j_{Q}^{\alpha} \right\rangle &= \sum_{\beta} \left[- \frac{1}{T} L^{\alpha \beta}_{21} \partial^{\beta} \bar{\mu} + L^{\alpha \beta}_{22} \partial^{\beta} \left( \frac{1}{T} \right) \right], \label{app:eqn:thermo_response_energy}
\end{align}
where the transport coefficients are in general $T$-dependent. 
  
We now provide the microscopic expressions of the thermal transport coefficients via linear response theory. Concretely, we assume that throughout the system, the (spatially-dependent) electrochemical potential and temperature field are given by 
\begin{equation}
	\bar{\mu} \left( \vec{r} \right) = \mu + \delta \bar{\mu} \left( \vec{r} \right) \qq{and}
	\beta \left( \vec{r} \right) = \beta + \delta \beta \left( \vec{r} \right).
\end{equation}
where $\mu$ and $\beta$ are the unperturbed values of the chemical potential and inverse temperature (equivalently speaking, the inverse temperature in equilibrium), to which small spatially-dependent perturbations $\delta \bar{\mu} \left( \vec{r} \right)$ and $\delta \beta \left( \vec{r} \right)$ are added (where $\abs{\delta \bar{\mu} \left( \vec{r} \right) / \mu} \ll 1$ and $\abs{\delta \beta \left( \vec{r} \right) / \beta} \ll 1$). 
Here, $\delta \bar{\mu} \left( \vec{r} \right) $ corresponds to the $e\Phi \left( \vec{r} \right)$ from \cref{app:eqn:electrochemical potential}. 
As in \cref{app:sec:thermoelectric_response:lin_response:static_perturbation}, we take the gradients of the electrochemical potential and inverse temperature fields to be $\vec{r}$-independent
\begin{equation}
	\nabla \delta \bar{\mu} \left( \vec{r} \right) = \nabla \delta \bar{\mu} \qq{and}
	\nabla \delta \beta \left( \vec{r} \right) = \nabla \delta \beta.
\end{equation}

Modifying the uniform chemical potential and temperature is equivalent to adding the following perturbation to the grand canonical Hamiltonian of the system\footnote{Second order terms such as $\delta \beta \left( \vec{r} \right) \delta \bar{\mu} \left( \vec{r} \right) $ are ignored.}~\cite{LUT64}
\begin{equation}
	\label{app:eqn:perturbation_thermoelectric}
	\delta H = \int \dd[d]{r} \left(  \frac{\delta \beta \left( \vec{r} \right)}{\beta} \rho_{Q} \left( \vec{r} \right) - \delta \bar{\mu} \left( \vec{r} \right) \rho \left( \vec{r} \right) \right), \qq{such that}
	\pdv{\delta H \left( t \right)}{t} = \frac{1}{\beta} \nabla \delta \beta \cdot \vec{J}_{Q} \left(t \right) - \nabla \delta \bar{\mu} \cdot \vec{J} \left(t \right).
\end{equation} 
We now discuss the meaning of each term. For a system without temperature gradients, the grand canonical Hamiltonian is given by $H-\mu \hat{N}$, with $H$ being the Hamiltonian and $\hat{N}$ the particle number. Therefore, in the presence of a temperature perturbation $\delta \beta \left( \vec{r} \right)$, the local grand canonical Hamiltonian density $\rho_Q \left( \vec{r} \right)$, equal with the thermal density defined in \cref{app:eqn:generic_thermal_current}, couples to $\delta \beta(\vec{r}) $ (which locally modifies the grand canonical Hamiltonian). In analogy to the $ H-\mu N$ term, $ \rho_{Q} \left( \vec{r} \right) = \rho_E \left( \vec{r} \right) - \mu \rho \left( \vec{r} \right)$ consists of two terms. The first term denotes the energy density $\rho_E \left( \vec{r} \right)$ defined in \cref{app:eqn:energy_density_1}, which corresponds to the local Hamiltonian density, and the second term is the charge density $\rho \left( \vec{r} \right)$ defined in \cref{app:eqn:cont_part_density}, which corresponds to the local particle density. Moreover, the electrochemical potential $\delta \bar{\mu} \left( \vec{r} \right)$ directly couples to the particle density $\rho \left( \vec{r} \right)$. Combining the two contributions, we can write down the perturbation to the grand canonical Hamiltonian, $\delta H$, as given in \cref{app:eqn:perturbation_thermoelectric}. As in \cref{app:eqn:perturbation_deriv_expression}, taking the time-derivative of $\delta H$, introducing the total thermal $\vec{J}_{Q}$ and particle $\vec{J}$ current operators from \cref{app:eqn:generic_thermal_current,app:eqn:part_current_simple_real_particular}, and using the continuity equations from \cref{app:eqn:continuity_particle,app:eqn:continuity_thermal}, we can also derive the second equality of \cref{app:eqn:perturbation_thermoelectric}.

The perturbation in \cref{app:eqn:perturbation_thermoelectric} is of the form of \cref{app:eqn:general_perturbation_lin_theo}, and as a result, one can express the thermoelectric linear response coefficients from \cref{app:eqn:thermo_response_part,app:eqn:thermo_response_energy} using \cref{app:eqn:lin_resp_coeff_expression_luttinger,app:eqn:current_corr_func_matsubara,app:eqn:linear-response_with_anal_cont}. For concreteness, if we define 
\begin{align}
	L^{\alpha \beta}_{11} \left( i \tilde{\Omega}_n \right) &= \frac{1}{N_0 \Omega_0} \frac{1}{\beta \tilde{\Omega}_n} \int_{0}^{\beta} \dd{\tau} e^{i \tilde{\Omega}_n \tau} \left\langle \mathcal{T}_{\tau} J^{\beta} \left( -i\tau \right) J^{\alpha} \left( 0 \right) \right\rangle, \label{app:eqn:L11_mats} \\
	L^{\alpha \beta}_{12} \left( i \tilde{\Omega}_n \right) &= \frac{1}{N_0 \Omega_0} \frac{1}{\beta \tilde{\Omega}_n} \int_{0}^{\beta} \dd{\tau} e^{i \tilde{\Omega}_n \tau} \left\langle \mathcal{T}_{\tau} J_{Q}^{\beta} \left( -i\tau \right) J^{\alpha} \left( 0 \right) \right\rangle, \label{app:eqn:L12_mats} \\
	L^{\alpha \beta}_{21} \left( i \tilde{\Omega}_n \right) &= \frac{1}{N_0 \Omega_0} \frac{1}{\beta \tilde{\Omega}_n} \int_{0}^{\beta} \dd{\tau} e^{i \tilde{\Omega}_n \tau} \left\langle \mathcal{T}_{\tau} J^{\beta} \left( -i\tau \right) J_{Q}^{\alpha} \left( 0 \right) \right\rangle, \label{app:eqn:L21_mats} \\
	L^{\alpha \beta}_{22} \left( i \tilde{\Omega}_n \right) &= \frac{1}{N_0 \Omega_0} \frac{1}{\beta \tilde{\Omega}_n} \int_{0}^{\beta} \dd{\tau} e^{i \tilde{\Omega}_n \tau} \left\langle \mathcal{T}_{\tau} J_{Q}^{\beta} \left( -i\tau \right) J_{Q}^{\alpha} \left( 0 \right) \right\rangle, \label{app:eqn:L22_mats}
\end{align}
then the static thermoelectric response coefficients can be determined from 
\begin{equation}
	\label{app:eqn:Ls_anal_cont}
	L^{\alpha \beta}_{ij} = \lim_{\omega \to 0} \left[ \Re\left( \eval{L^{\alpha \beta}_{ij} \left( i \tilde{\Omega}_n \right) }_{i \tilde{\Omega}_n \to \omega + i 0^{+}} \right)\right], \qq{for} 1 \leq i,j \leq 2.
\end{equation}
When performing the analytical continuation in \cref{app:eqn:Ls_anal_cont}, it is important that one follows the same sequence of steps as in the derivation from \cref{app:sec:thermoelectric_response:lin_response:matsubara}, namely:
\begin{enumerate}
	\item Compute the correlation functions from \cref{app:eqn:L11_mats,app:eqn:L12_mats,app:eqn:L22_mats} in Matsubara frequency using standard diagramatic techniques.
	\item Perform the analytical continuation $i \tilde{\Omega}_n \to \omega + i 0^{+}$.
	\item Evaluate $ \Re L^{\alpha \beta}_{ij} \left( \omega + i 0^{+} \right) $ for $\omega \neq 0$. 
	\item Take the limit $\omega \to 0$ of the previous result to obtain $L^{\alpha \beta}_{ij}$. 
\end{enumerate}

Finally, we note that throughout this work, all the phases of the THF model we consider have $C_{3z}$ symmetry. This implies that the thermoelectric linear response coefficients obey 
\begin{equation}
	\label{app:eqn:c3_on_lij}
	\mathcal{R}_{C_{3z}} L_{ij} \mathcal{R}_{C_{3z}}^{-1} = L_{ij}, 
\end{equation}
where $\mathcal{R}_{C_{3z}}$ is the cartesian representation matrix of the $C_{3z}$ symmetry defined in \cref{app:eqn:matrix_c3}. \Cref{app:eqn:c3_on_lij} follows straightforwardly from \crefrange{app:eqn:L11_mats}{app:eqn:L22_mats} using the transformation laws of the current operators from \cref{app:eqn:trafo_c3_J}.

As a result, the thermoelectric coefficients have only two independent components, 
\begin{equation}
	\label{app:eqn:indep_comp_of_L_tensors}
	L^{xx}_{ij} = L^{yy}_{ij} \qq{and}
	L^{xy}_{ij} = -L^{yx}_{ij}.
\end{equation} 
Note that with $C_{3z}$ symmetry the thermoelectric linear response coefficients have just as many independent components as in the fully rotationally-symmetric case ({\it i.e.}{}, two independent components each). 
 
\subsubsection{Relating the thermoelectric coefficients to transport measurements} \label{app:sec:thermoelectric_response:thermoelectric:Lij_to_transport}

The thermoelectric linear response coefficients defined in \cref{app:sec:thermoelectric_response:thermoelectric} are rather abstract, so it is useful to relate them to commonly measured physical quantities~\cite{MAH00}. For example, the conductivity $\sigma^{\alpha \beta}$ is measured in the absence of any temperature ($\nabla \beta = \vec{0}$) or chemical potential ($\nabla \mu = \vec{0}$) gradients. The particle current is related to the gradient of the electrostatic potential $\Phi$ by 
\begin{equation}
	-e \left\langle j^{\alpha} \right\rangle = \sum_{\beta} \sigma^{\alpha \beta} \left( - \partial^{\beta} \Phi \right).
\end{equation}
Comparing the definition of the conductivity with \cref{app:eqn:thermo_response_part}, we find that 
\begin{equation}
	\label{app:eqn:relation_between_L11_and_sigma}
	\sigma^{\gamma \delta}=-\beta e^{2} L^{\gamma \delta}_{11}.
\end{equation}

Another experimentally relevant scenario concerns the thermoelectric effect. In this case, a temperature gradient is applied along the {\it e.g.}{} $x$ direction, in the absence of any chemical potential gradients $\nabla \mu = \vec{0}$, or particle currents along the $x$ direction $\left\langle j^{x} \right\rangle=0$. \Cref{app:eqn:thermo_response_part} then implies that
\begin{align}
	0 = - \frac{e}{T} L_{11}^{xx} \partial^{x} \Phi + L_{12}^{xx} \partial^{x} \left(\frac{1}{T} \right), 
\end{align}
or, otherwise stated, a potential difference along the $x$ direction develops, which is given by
\begin{equation}
	\label{app:eqn:seebeck_voltage}
	\partial^{x} \Phi = - \frac{\beta}{e} \frac{L_{12}^{xx}}{L_{11}^{xx}} \partial^{x} T.
\end{equation}
\Cref{app:eqn:seebeck_voltage} is known as the Seebeck effect and enables one to define an associated Seebeck coefficient
\begin{equation}
	\label{app:eqn:seebeck_definition}
	S = \frac{\beta}{e} \frac{L_{12}^{xx}}{L_{11}^{xx}} = -\beta^2e \frac{L_{12}^{xx}}{\sigma^{xx}}.
\end{equation}

For the case of the THF model, as a result of \cref{app:eqn:indep_comp_of_L_tensors}, the same answer can be obtained irrespective of the direction along which the temperature gradient is applied. 

In what follows, we will be exclusively concerned with obtaining the Seebeck coefficient within the THF model.

\subsubsection{Consequences of many-body charge conjugation symmetry} \label{app:sec:thermoelectric_response:evaluating_the_transport_coefficients:PH_symmetry}

As mentioned in \cref{app:sec:many_body_rev:ground_states:ph_symmetry}, the THF models for both TBG and TSTG (without the more complicated lattice relaxation effects) are symmetric with respect to the charge-conjugation symmetry operators $\mathcal{P}$ and $\mathcal{P}'$, which were defined in \cref{app:eqn:ph_sym_act_TBG,app:eqn:ph_sym_act_TSTG}, respectively. In this section, we derive the consequences of these symmetries on the thermal linear response coefficients. In doing so, we will work with the current operators of TBG and TSTG in the THF basis, introduced in \cref{app:sec:deriv_moire_cur}, and employ the results of \cref{app:sec:cur_deriv_cont:ph_symmetry} on the action of the many-body charge conjugation symmetry on the total particle and energy current operators.

The action of the many-body charge conjugation symmetries of TBG and TSTG (denoted by $\mathcal{P}$ and $\mathcal{P}'$, respectively) on a spatial variable $\vec{r}$ is given by 
\begin{equation}
	\mathcal{P} \vec{r} \mathcal{P}^{-1} = \vec{r} \qq{and}
	\mathcal{P}' \vec{r} \mathcal{P}^{\prime -1} = C_{2x} \vec{r}.
\end{equation} 
As a result of this and \cref{app:eqn:ph_symmetry_par_cur,app:eqn:ph_symmetry_en_cur}, the action of the many-body charge conjugation symmetry on the total particle and energy current operators is given by
\begin{alignat}{4}
	\mathcal{P} J^{\alpha,\text{TBG}} (t) \mathcal{P}^{-1} &=& - \xi^{\alpha}_{\text{TBG}} J^{\alpha,\text{TBG}} (t) & \quad \mathcal{P}' J^{\alpha,\text{TSTG}} (t) \mathcal{P}^{\prime-1} &=& - \xi^{\alpha}_{\text{TSTG}} J^{\alpha,\text{TSTG}} (t), \label{app:eqn:ph_part_cur} \\
	\mathcal{P} J_E^{\alpha,\text{TBG}} (t) \mathcal{P}^{-1} &=& - \xi^{\alpha}_{\text{TBG}} J_E^{\alpha,\text{TBG}} (t) & \quad \mathcal{P}' J_E^{\alpha,\text{TSTG}} (t) \mathcal{P}^{\prime-1} &=& - \xi^{\alpha}_{\text{TSTG}} J_E^{\alpha,\text{TSTG}} (t), \label{app:eqn:ph_en_cur}
\end{alignat}
where, for simplicity, we have introduced the factor $\xi^{\alpha}$, which is given by $\xi_{\text{TBG}}^{\alpha} = 1$ (for $\alpha = x,y$) in the case of TBG and by 
\begin{equation}
	\xi_{\text{TSTG}}^{\alpha} = \begin{cases}
		+1 \qq{if} \alpha = x \\
		-1 \qq{if} \alpha = y 
	\end{cases}
\end{equation}
in the case of TSTG.

As the many-body charge-conjugation operator maps phases at chemical potential $\mu$ to phases at chemical potential $-\mu$, we will find it useful to add an index $\mu$ to the thermal current $\vec{J}_Q$, the grand canonical Hamiltonian $K$, and the thermal expectation $\left\langle \dots \right\rangle$ operation to indicate the chemical potential at which the operator is evaluated or at which the operation is performed. Under the many-body charge conjugation symmetry we have 
\begin{align}
	\mathcal{P}^{(\prime)} K_{\mu} \mathcal{P}^{(\prime)-1} &=\mathcal{P}^{(\prime)} \left( H-\mu N \right) \mathcal{P}^{(\prime)-1} =H-\mu \left( N_\gamma-N \right)= K_{-\mu} - \mu N_{\gamma} \nonumber\\ 
	\mathcal{P}^{(\prime)} J^{\alpha}_{Q,\mu} \mathcal{P}^{(\prime)-1} &= \xi^{\alpha} J^{\alpha}_{Q,-\mu}, \label{app:eqn:particle_hole_action_gc_and_curr}
\end{align}
where $N_{\gamma}$ is the total number of fermionic species in the THF model ($N_{\gamma} = 24 N_0$ for TBG and $N_{\gamma} = 32 N_0$ for TSTG), $\mu$ is a parameter which will not be affected by the transformation, and $\mathcal{P}^{(\prime)} = \mathcal{P}$ for TBG and $\mathcal{P}^{(\prime)} = \mathcal{P}'$ for TSTG. Moreover, we remind the reader that since $\mathcal{P}^{(\prime)}$ is a symmetry of the Hamiltonian, one must have that $\mathcal{P}^{(\prime)} H \mathcal{P}^{(\prime)-1} = H$. In addition to \cref{app:eqn:particle_hole_action_gc_and_curr}, we also find that
\begin{equation}
	\label{app:eqn:particle_hole_action_averaging}
	\left\langle \mathcal{O} \right\rangle_{\mu} = \frac{\Tr \left( e^{-\beta K_{\mu}} \mathcal{O} \right)}{\Tr \left( e^{-\beta K_{\mu}} \right)}  = \frac{\Tr \left( e^{-\beta K_{\mu}} \mathcal{P}^{(\prime)-1} \mathcal{P}^{(\prime)} \mathcal{O} \right)}{\Tr \left( e^{-\beta K_{\mu}} \mathcal{P}^{(\prime)} \mathcal{P}^{(\prime)-1} \right)}  = \frac{e^{\beta \mu N_{\gamma}} \Tr \left( e^{-\beta K_{-\mu}} \mathcal{P}^{(\prime)} \mathcal{O} \mathcal{P}^{(\prime)-1} \right)}{e^{\beta \mu N_{\gamma}} \Tr \left( e^{-\beta K_{-\mu}} \mathcal{P}^{(\prime)-1} \mathcal{P}^{(\prime)} \right)} = \left\langle \mathcal{P}^{(\prime)} \mathcal{O} \mathcal{P}^{(\prime)-1} \right\rangle_{-\mu}.
\end{equation}

\Cref{app:eqn:particle_hole_action_gc_and_curr,app:eqn:particle_hole_action_averaging} enable us to relate the current-current correlation functions at opposite chemical potentials. For example, 
\begin{align}
	\left\langle \mathcal{T}_{\tau} J_{Q,\mu}^{\beta} \left( -i\tau \right) J^{\alpha} \left( 0 \right) \right\rangle_{\mu} &= \left\langle \mathcal{T}_{\tau} \mathcal{P}^{(\prime)} J_{Q,\mu}^{\beta} \left( -i\tau \right) \mathcal{P}^{(\prime)-1} \mathcal{P}^{(\prime)} J^{\alpha} \left( 0 \right) \mathcal{P}^{(\prime)-1} \right\rangle_{-\mu} \nonumber \\
	&= -\xi^{\alpha} \xi^{\beta} \left\langle \mathcal{T}_{\tau} J_{Q,-\mu}^{\beta} \left( -i\tau \right) J^{\alpha} \left( 0 \right) \right\rangle_{-\mu}, \label{app:eqn:ph_effect_expec_JJ_12}
\end{align}
and similarly
\begin{align}
	\left\langle \mathcal{T}_{\tau} J^{\beta} \left( -i\tau \right) J^{\alpha} \left( 0 \right) \right\rangle_{\mu} &= \xi^{\alpha} \xi^{\beta} \left\langle \mathcal{T}_{\tau} J^{\beta} \left( -i\tau \right) J^{\alpha} \left( 0 \right) \right\rangle_{-\mu}, \label{app:eqn:ph_effect_expec_JJ_11} \\
	\left\langle \mathcal{T}_{\tau} J_{Q,\mu}^{\beta} \left( -i\tau \right) J_{Q,\mu}^{\alpha} \left( 0 \right) \right\rangle_{\mu} &= \xi^{\alpha} \xi^{\beta} \left\langle \mathcal{T}_{\tau} J_{Q,-\mu}^{\beta} \left( -i\tau \right) J_{Q,-\mu}^{\alpha} \left( 0 \right) \right\rangle_{-\mu}. \label{app:eqn:ph_effect_expec_JJ_22}
\end{align}
In turn, \cref{app:eqn:ph_effect_expec_JJ_12,app:eqn:ph_effect_expec_JJ_11,app:eqn:ph_effect_expec_JJ_22} imply that for TBG and TSTG the thermoelectric response coefficients at opposite chemical potentials can be related by 
\begin{equation}
	\label{app:eqn:relation_of_L_correlators}
	L^{\alpha \beta}_{11,\mu} = \xi^{\alpha} \xi^{\beta} L^{\alpha \beta}_{11,-\mu}, \quad
	L^{\alpha \beta}_{12,\mu} = -\xi^{\alpha} \xi^{\beta} L^{\alpha \beta}_{12,-\mu}, \quad
	L^{\alpha \beta}_{22,\mu} = \xi^{\alpha} \xi^{\beta} L^{\alpha \beta}_{22,-\mu}
\end{equation}
where the subscript indicates the chemical potential at which the correlator is evaluated. In particular \cref{app:eqn:relation_of_L_correlators} implies that the conductivity is the same for phases which are charge-conjugated to one another, while the Seebeck coefficient is antisymmetric with respect to the filling. As a result, we can focus exclusively on positive fillings in what follows.  

\subsection{Computing the thermoelectric transport coefficients in the THF model} \label{app:sec:thermoelectric_response:evaluating_the_transport_coefficients}

{\renewcommand{\arraystretch}{1.5}
	\begin{table}[t]
		\centering
		\begin{tabular}{|c|c|c|}
			\hline
			Operator/Quantity & Expression & Introduced \\
			\hline
			$\hat{\gamma}^\dagger_{\vec{k},i, \eta, s} \left( t \right)$ & $\hat{\gamma}^\dagger_{\vec{k},i, \eta, s} \left( t \right) = e^{i H t} \hat{\gamma}^\dagger_{\vec{k},i, \eta, s} e^{- i H t}$ &  \cref{app:eqn:shorthand_gamma_not,app:eqn:rel_heis_schro} \\
			\hline
			$\dot{\hat{\gamma}}^\dagger_{\vec{k},i, \eta, s} \left( t \right) \equiv \pdv{t} \hat{\gamma}^\dagger_{\vec{k},i, \eta, s} \left( t \right)$ & $\dot{\hat{\gamma}}^\dagger_{\vec{k},i, \eta, s} \left( t \right) = i\commutator{H}{\hat{\gamma}^\dagger_{\vec{k},i, \eta, s} \left( t \right)}$ &  \cref{app:eqn:field_ops_time_deriv} \\
			\hline
			$\hat{\gamma}^\dagger_{\vec{k},i, \eta, s} \left( \tau \right)$ & $\hat{\gamma}^\dagger_{\vec{k},i, \eta, s} \left( \tau \right) = e^{K \tau} \hat{\gamma}^\dagger_{\vec{k},i, \eta, s} e^{- K \tau}$ &  \cref{app:eqn:shorthand_gamma_not,app:eqn:gr_can_evolved_gamma_ops} \\
			\hline
			$\pdv{\tau} \hat{\gamma}^\dagger_{\vec{k},i, \eta, s} \left( \tau \right)$ & $\pdv{\tau} \hat{\gamma}^\dagger_{\vec{k},i, \eta, s} \left( \tau \right) = \commutator{K}{\hat{\gamma}^\dagger_{\vec{k},i, \eta, s} \left( t \right)}$ &  \cref{app:eqn:time_deriv_imaginary_gamma} \\
			\hline
			$\mathcal{G} \left(\tau, \vec{k} \right)$ & $\mathcal{G}_{i \eta s; i' \eta' s'} \left(\tau, \vec{k} \right) = -\left\langle \mathcal{T}_{\tau} \hat{\gamma}_{\vec{k},i,\eta,s} \left( \tau \right) \hat{\gamma}^\dagger_{\vec{k}',i',\eta',s'} \left( 0 \right)  \right\rangle$ &  \cref{app:eqn:matsubara_gf_THF_tau} \\
			\hline
			$\vec{J}_{n} \left(- t - i \tau \right)$ & $\vec{J}_{n} \left( - t - i \tau \right) = e^{-i K t + K \tau} \vec{J}_{n} e^{i K t - K \tau}$ & \cref{app:eqn:general_total_current_definition} \\
			\hline
		\end{tabular}
		\caption{Imaginary and real time dependence of various operators and quantities used in this \siSection{}. For reference, we list here the definition employed for the real and imaginary time-evolved fermionic operators and their time derivatives, the Matsubara Green's function, and the total current operators evolved for a general complex time $-t - i \tau$.}
		\label{app:tab:notation_recap}
	\end{table}
}
This section outlines the main calculation performed in this work: obtaining the Seebeck coefficient of the THF model. We start from the expressions of the particle and energy currents derived in \cref{app:sec:deriv_moire_cur}. These are then substituted into \cref{app:eqn:L11_mats,app:eqn:L12_mats} allowing us to compute the corresponding current-current correlation function using the interacting Green's function of the THF model introduced in \cref{app:eqn:matsubara_gf_THF_tau}. The calculation of the interacting Green's function has been described in \cref*{DMFT:app:sec:se_correction_beyond_HF,DMFT:app:sec:se_symmetric} of Ref.~\cite{CAL23b} for the symmetry-broken and symmetric phases, respectively. After performing the analytical continuation $i \tilde{\Omega}_n \to \omega + i 0^{+}$ and taking the limit $\omega \to 0$, the static thermoelectric linear response coefficients $L^{xx}_{12}$ and $L^{xx}_{11}$ can be computed, which allows us to obtain the Seebeck coefficient using \cref{app:eqn:seebeck_definition}. 

\subsubsection{The current operators in imaginary time} \label{app:sec:thermoelectric_response:evaluating_the_transport_coefficients:curr_imag_time}

As shown in \cref{app:sec:deriv_moire_cur} the particle [\cref{app:eqn:THF_curent_part_TBG_simple,app:eqn:THF_curent_part_TSTG_simple}] and energy [\cref{app:eqn:energy_currents_THF}] currents in TBG and TSTG are given by 
\begin{align}
	J^{\alpha} &= \sum_{\vec{k}} \sum_{\substack{i, \eta, s \\ i', \eta', s'}} T^{\alpha}_{i \eta s;i' \eta' s'} \left( \vec{k} \right) \hat{\gamma}^\dagger_{\vec{k},i, \eta, s} \hat{\gamma}_{\vec{k},i', \eta', s'}, \label{app:eqn:generic_particle_current} \\
	J_E^{\alpha} &= \frac{1}{2} \sum_{\vec{k}} \sum_{\substack{i, \eta, s \\ i', \eta', s'}} T^{\alpha}_{i \eta s;i' \eta' s'} \left( \vec{k} \right) \left( i \hat{\gamma}^\dagger_{\vec{k},i, \eta, s} \dot{\hat{\gamma}}_{\vec{k},i', \eta', s'} - i \dot{\hat{\gamma}}^\dagger_{\vec{k},i, \eta, s} \hat{\gamma}_{\vec{k},i', \eta', s'} \right), \label{app:eqn:generic_energy_current}
\end{align}
where we have extended the cutoff for the $c$- and $d$-electrons to encompass one moir\'e BZ and introduced the following tensor in the notation of \cref{app:sec:many_body_rev:generic_not}  
\begin{equation}
	\vec{T}_{i \eta s;i' \eta' s'} \left( \vec{k} \right) = \begin{cases}
		\partial_{\vec{k}} h^{cc,\eta}_{i i'} \left( \vec{k} \right) \delta_{\eta \eta'} \delta_{s s'} & \qq{if} 1 \leq i,i' \leq 4 \\
		\partial_{\vec{k}} h^{cf,\eta}_{i (i'-4)} \left( \vec{k} \right) \delta_{\eta \eta'} \delta_{s s'} & \qq{if} 1 \leq i \leq 4 \qq{and} 5 \leq i' \leq 6 \\
		\partial_{\vec{k}} h^{*cf,\eta}_{(i-4) i'} \left( \vec{k} \right) \delta_{\eta \eta'} \delta_{s s'} & \qq{if} 5 \leq i \leq 6 \qq{and} 1 \leq i' \leq 4 \\
		\partial_{\vec{k}} h^{dd,\eta}_{(i-6) (i'-6)} \left( \vec{P}_{\vec{k}} \right) \delta_{\eta \eta'} \delta_{s s'} & \qq{for TSTG and if} 7 \leq i,i' \leq 8 
	\end{cases},
\end{equation}
where, as defined around \cref{app:eqn:full_THF_Hamiltonian_gamma_bas}, $\vec{P}_{\vec{k}}$ denotes the image of $\vec{k} - \eta \vec{q}_1$ in the first moir\'e BZ. Both the particle and energy currents are quadratic operators since the quartic interaction contribution to the energy current only accounts for less than $4 \%$ of the kinetic one, as derived in \cref{app:sec:deriv_moire_cur:energy_interaction:comparison}. Finally, it is worth noting that the time derivative appearing in \cref{app:eqn:generic_energy_current} is computed by assuming that the fermion operators evolve with the many-body THF Hamiltonian $H$ ({\it i.e.}{}, \emph{not} with the grand canonical Hamiltonian $K$). This follows from our derivation of the total energy current operator in \cref{app:sec:cur_deriv_cont:energy}: the time derivative of a fermion operator in \cref{app:eqn:en_current_simple_real_general_sin} corresponds to the commutator between that operator and Hamiltonian, according to \cref{app:eqn:field_ops_time_deriv}. Therefore, the same notation should be applied here, which leads us to 
\begin{equation}
	\dot{\hat{\gamma}}^\dagger_{\vec{k},i, \eta, s} = i \commutator{H}{\hat{\gamma}^\dagger_{\vec{k},i, \eta, s}} = i \commutator{K}{\hat{\gamma}^\dagger_{\vec{k},i, \eta, s}} + i \mu \hat{\gamma}^\dagger_{\vec{k},i, \eta, s}.
\end{equation}
For reference, the real and imaginary time evolution of the fermion creation and annihilation operators is summarized in \cref{app:tab:notation_recap}. 

Using the time-evolved fermion operators from \cref{app:eqn:gr_can_evolved_gamma_ops}, we rewrite the imaginary-time current operators as
\begin{align}
	J^{\alpha} \left( -i \tau \right) &= \sum_{\vec{k}} \sum_{\substack{i, \eta, s \\ i', \eta', s'}} T^{\alpha}_{i \eta s;i' \eta' s'} \left( \vec{k} \right) \hat{\gamma}^\dagger_{\vec{k},i, \eta, s} (\tau) \hat{\gamma}_{\vec{k},i', \eta', s'} (\tau), \label{app:eqn:mats_particle_current} \\
	J_E^{\alpha} \left( -i \tau \right) &= \frac{1}{2} \sum_{\vec{k}} \sum_{\substack{i, \eta, s \\ i', \eta', s'}} T^{\alpha}_{i \eta s;i' \eta' s'} \left( \vec{k} \right) \left[ - \hat{\gamma}^\dagger_{\vec{k},i, \eta, s} (\tau) \pdv{\tau} \hat{\gamma}_{\vec{k},i', \eta', s'} (\tau) + \mu \hat{\gamma}^\dagger_{\vec{k},i, \eta, s} (\tau) \hat{\gamma}_{\vec{k},i', \eta', s'} (\tau) \right. \nonumber \\
	& \left. +  \left( \pdv{\tau} \hat{\gamma}^\dagger_{\vec{k},i, \eta, s} (\tau) \right) \hat{\gamma}_{\vec{k},i', \eta', s'} (\tau) + \mu \hat{\gamma}^\dagger_{\vec{k},i, \eta, s} (\tau) \hat{\gamma}_{\vec{k},i', \eta', s'} (\tau) \right], \label{app:eqn:mats_energy_current}
\end{align}
from which the thermal current operator $\vec{J}_Q = \vec{J}_E - \mu \vec{J}$ introduced in \cref{app:eqn:generic_thermal_current} follows straightforwardly
\begin{equation}
	J_Q^{\alpha} \left( -i \tau \right) = \frac{1}{2} \sum_{\vec{k}} \sum_{\substack{i, \eta, s \\ i', \eta', s'}} T^{\alpha}_{i \eta s;i' \eta' s'} \left( \vec{k} \right) \left[ \left( \pdv{\tau} \hat{\gamma}^\dagger_{\vec{k},i, \eta, s} (\tau) \right) \hat{\gamma}_{\vec{k},i', \eta', s'} (\tau) - \hat{\gamma}^\dagger_{\vec{k},i, \eta, s} (\tau) \pdv{\tau} \hat{\gamma}_{\vec{k},i', \eta', s'} (\tau)  \right]. \label{app:eqn:mats_thermal_current}
\end{equation}
We remind the reader that the imaginary-time-evolved fermion operators $\hat{\gamma}^\dagger_{\vec{k},i, \eta, s} (\tau)$ are defined according to \cref{app:eqn:gr_can_evolved_gamma_ops}, which implies that 
\begin{equation}
	\label{app:eqn:time_deriv_imaginary_gamma}
	\pdv{\tau} \hat{\gamma}^\dagger_{\vec{k},i, \eta, s} (\tau)  = \commutator{K}{ \hat{\gamma}^\dagger_{\vec{k},i, \eta, s} (\tau)}.
\end{equation}

\subsubsection{General remarks on four-fermion correlation functions} \label{app:sec:thermoelectric_response:evaluating_the_transport_coefficients:four_fermi_corr}

\begin{figure}[!t]
	\centering
	\begin{tikzpicture}[baseline=(current bounding box.north)]
		\begin{feynman}
			\vertex (a) {{$\substack{ \vec{k}_{1} \\ i_{1}, \eta_{1}, s_{1}}$}};
			\vertex [right=3.5 cm of a] (b) {{$\substack{ \vec{k}_{1} \\ i_{2}, \eta_{2}, s_{2}}$}};
			\vertex [below=3.5 cm of a] (c) {{$\substack{ \vec{k}_{2} \\ i_{4}, \eta_{4}, s_{4}}$}};
			\vertex [right=3.5 cm of c] (d) {{$\substack{ \vec{k}_{2} \\ i_{3}, \eta_{3}, s_{3}}$}};
			\vertex[blob] (cnt) at (1.75,-1.75) {};
			\diagram*{
				(a) -- [fermion] (cnt) -- [fermion] (b),
				(d) -- [fermion] (cnt) -- [fermion] (c)
			};
		\end{feynman}
		\path (a) ++(0,0.5cm) node{a)};
	\end{tikzpicture}
	\qquad
	\begin{tikzpicture}[baseline=(current bounding box.north)]
		\begin{feynman}
			\vertex (a) {{$\substack{ \vec{k}_{1} \\ i_{1}, \eta_{1}, s_{1}}$}};
			\vertex [right=3.5 cm of a] (b) {{$\substack{ \vec{k}_{1} \\ i_{2}, \eta_{2}, s_{2}}$}};
			\vertex [below=3.5 cm of a] (c) {{$\substack{ \vec{k}_{2} \\ i_{4}, \eta_{4}, s_{4}}$}};
			\vertex [right=3.5 cm of c] (d) {{$\substack{ \vec{k}_{2} \\ i_{3}, \eta_{3}, s_{3}}$}};
			\vertex[blob] (cnt1) at (1.75,0) {};
			\vertex[blob] (cnt2) at (1.75,-3.5) {};
			\diagram*{
				(a) -- [fermion] (cnt1) -- [fermion] (b),
				(d) -- [fermion] (cnt2) -- [fermion] (c)
			};
		\end{feynman}
		\path (a) ++(0,0.5cm) node{b)};
	\end{tikzpicture}
	\qquad
	\begin{tikzpicture}[baseline=(current bounding box.north)]
		\begin{feynman}
			\vertex (a) {{$\substack{ \vec{k}_{1} \\ i_{1}, \eta_{1}, s_{1}}$}};
			\vertex [right=3.5 cm of a] (b) {{$\substack{ \vec{k}_{1} \\ i_{2}, \eta_{2}, s_{2}}$}};
			\vertex [below=3.5 cm of a] (c) {{$\substack{ \vec{k}_{2} \\ i_{4}, \eta_{4}, s_{4}}$}};
			\vertex [right=3.5 cm of c] (d) {{$\substack{ \vec{k}_{2} \\ i_{3}, \eta_{3}, s_{3}}$}};
			\vertex[blob] (cnt1) at (0,-1.75) {};
			\vertex[blob] (cnt2) at (3.5,-1.75) {};
			\diagram*{
				(a) -- [fermion] (cnt1) -- [fermion] (c),
				(d) -- [fermion] (cnt2) -- [fermion] (b)
			};
		\end{feynman}
		\path (a) ++(0,0.5cm) node{c)};
	\end{tikzpicture}
	\subfloat{\label{app:fig:cur_diags:a}}
	\subfloat{\label{app:fig:cur_diags:b}}
	\subfloat{\label{app:fig:cur_diags:c}}
	\caption{Feynman diagrams that contribute to the four-fermion correlation function from \cref{app:eqn:four_fermion_general}. The edges correspond to fermion propagators for which the $\mathrm{U} \left(1\right)$ charge flow is indicated by the arrow. For clarity, we have also labeled the external edges of the diagrams with indices of the four-fermion correlation function $\mathcal{I}_{i_{1} \eta_{1} s_{1};i_{2} \eta_{2} s_{2};i_{3} \eta_{3} s_{3};i_{4} \eta_{4} s_{4}} \left( \vec{k}_1, \vec{k}_2; \tau_1, \tau_2 \right)$. The hatched circles denote fully-connected Feynman graphs. The contribution in (a) is fully-connected and is known as the vertex correction to the current-current correlation function, while the contributions from (b) and (c) are disconnected. Each of the diagrams in (b) and (c) corresponds to a product of two fully-interacting Green's functions. When evaluating the current-current correlation functions, we will ignore the vertex corrections.}
	\label{app:fig:cur_diags}
\end{figure}
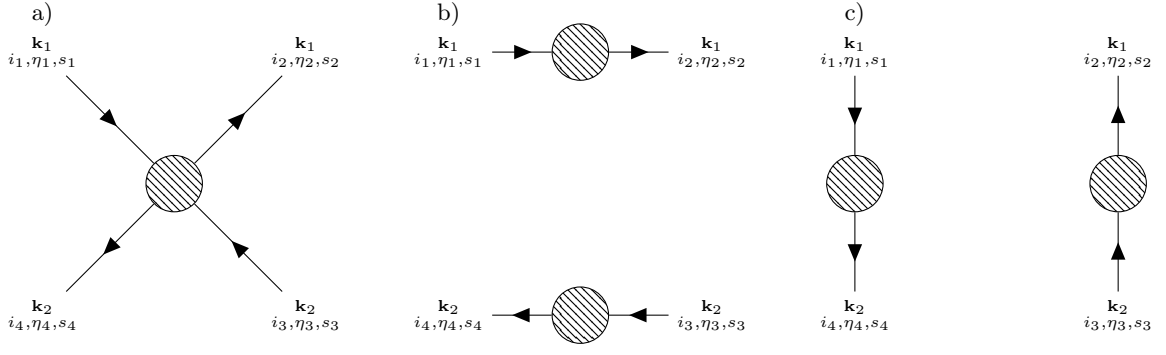

Because both the thermal and particle currents are quadratic operators, computing the current-current correlation functions from \crefrange{app:eqn:L11_mats}{app:eqn:L22_mats} will involve computing four-fermion expectation values of the form 
\begin{equation}
	\label{app:eqn:four_fermion_general}
	\mathcal{I}_{i_{1} \eta_{1} s_{1};i_{2} \eta_{2} s_{2};i_{3} \eta_{3} s_{3};i_{4} \eta_{4} s_{4}} \left( \vec{k}_1, \vec{k}_2; \tau_1, \tau_2 \right) = \left\langle \mathcal{T}_{\tau} \hat{\gamma}^\dagger_{\vec{k}_1,i_{1}, \eta_{1}, s_{1}} \left( \tau_1 \right) \hat{\gamma}_{\vec{k}_1,i_{2}, \eta_{2}, s_{2}} \left( \tau_1 \right)  \hat{\gamma}^\dagger_{\vec{k}_2,i_{3}, \eta_{3}, s_{3}} \left( \tau_2 \right)  \hat{\gamma}_{\vec{k}_2,i_{4}, \eta_{4}, s_{4}} \left( \tau_2 \right) \right\rangle,
\end{equation}
within an interacting theory governed by the grand canonical Hamiltonian $K$ (with additional time derivatives appearing in the case of the thermal current). If $\mathcal{I}_{i_{1} \eta_{1} s_{1};i_{2} \eta_{2} s_{2};i_{3} \eta_{3} s_{3};i_{4} \eta_{4} s_{4}} \left( \vec{k}_1, \vec{k}_2; \tau_1, \tau_2 \right)$ had been computed within a non-interacting theory, then Wick's theorem would have been applicable, allowing us to decouple the four-fermion correlation function into products of two-particle Green's functions, {\it i.e.}{}
\begin{align}
	\mathcal{I}^{\text{non-interacting}}_{i_{1} \eta_{1} s_{1};i_{2} \eta_{2} s_{2};i_{3} \eta_{3} s_{3};i_{4} \eta_{4} s_{4}} \left( \vec{k}_1, \vec{k}_2; \tau_1, \tau_2 \right) =& \left\langle \mathcal{T}_{\tau} \hat{\gamma}^\dagger_{\vec{k}_1,i_{1}, \eta_{1}, s_{1}} \left( \tau_1 \right) \hat{\gamma}_{\vec{k}_1,i_{2}, \eta_{2}, s_{2}} \left( \tau_1 \right)  \hat{\gamma}^\dagger_{\vec{k}_2,i_{3}, \eta_{3}, s_{3}} \left( \tau_2 \right)  \hat{\gamma}_{\vec{k}_2,i_{4}, \eta_{4}, s_{4}} \left( \tau_2 \right) \right\rangle_0 \nonumber \\
	=&\left\langle \hat{\gamma}^\dagger_{\vec{k}_1,i_{1}, \eta_{1}, s_{1}} \left( \tau_1 \right) \hat{\gamma}_{\vec{k}_1,i_{2}, \eta_{2}, s_{2}} \left( \tau_1 \right) \right\rangle_0 \left\langle \hat{\gamma}^\dagger_{\vec{k}_2,i_{3}, \eta_{3}, s_{3}} \left( \tau_2 \right)  \hat{\gamma}_{\vec{k}_2,i_{4}, \eta_{4}, s_{4}} \left( \tau_2 \right) \right\rangle_0 \nonumber \\
	&- \left\langle \mathcal{T}_{\tau} \hat{\gamma}^\dagger_{\vec{k}_1,i_{1}, \eta_{1}, s_{1}} \left( \tau_1 \right)  \hat{\gamma}_{\vec{k}_2,i_{4}, \eta_{4}, s_{4}} \left( \tau_2 \right)  \right\rangle_0 \left\langle  \mathcal{T}_{\tau} \hat{\gamma}^\dagger_{\vec{k}_2,i_{3}, \eta_{3}, s_{3}} \left( \tau_2 \right) \hat{\gamma}_{\vec{k}_1,i_{2}, \eta_{2}, s_{2}} \left( \tau_1 \right) \right\rangle_0 \nonumber \\
	=& \mathcal{G}^{0}_{i_{2} \eta_{2} s_{2};i_{1} \eta_{1} s_{1}} \left(0^{-}, \vec{k}_1 \right) \mathcal{G}^{0}_{i_{4} \eta_{4} s_{4};i_{3} \eta_{3} s_{3}} \left(0^{-}, \vec{k}_2 \right) \nonumber \\
	&- \mathcal{G}^{0}_{i_{4} \eta_{4} s_{4};i_{1} \eta_{1} s_{1}} \left(\tau_2 - \tau_1, \vec{k}_1 \right) \mathcal{G}^{0}_{i_{2} \eta_{2} s_{2};i_{3} \eta_{3} s_{3}} \left(\tau_1 - \tau_2, \vec{k}_2 \right) \delta_{\vec{k}_1, \vec{k}_2}. \label{app:eqn:four_fermion_general_eval_nonint}
\end{align}

Within the fully interacting theory, a diagramatic expansion is necessary to evaluate $\mathcal{I}_{i_{1} \eta_{1} s_{1};i_{2} \eta_{2} s_{2};i_{3} \eta_{3} s_{3};i_{4} \eta_{4} s_{4}}$. In \cref{app:fig:cur_diags}, we illustrate schematically the three types of diagrams that contribute to the four-fermion correlation function. Depending on the connectivity of the graphs, the diagrams can either be fully-connected, such as the ones in \cref{app:fig:cur_diags:a}, or disconnected, such as the ones in \cref{app:fig:cur_diags:b,app:fig:cur_diags:c}. In the context of current-current correlation functions, the fully-connected diagrams are known as \emph{vertex corrections}\footnote{A detailed discussion of the types of vertex corrections and their relative contributions is beyond the scope of this work.}. In our calculations, we employ a standard approximation, wherein the vertex corrections are omitted. We also comment that the vertex corrections will vanish in infinite dimensions, as proved in Ref.~\cite{PAU03}. 
In \cref{app:fig:cur_diags:b,app:fig:cur_diags:c}, each of the disconnected components corresponds to a fully-dressed (or fully interacting) Green's function. As a result, evaluating the four-fermion correlation function from \cref{app:eqn:four_fermion_general} within the interacting theory, but without vertex corrections, simply amounts to replacing the non-interacting Green's function in \cref{app:eqn:four_fermion_general_eval_nonint} with the fully-interacting ones
\begin{align}
	\mathcal{I}_{i_{1} \eta_{1} s_{1};i_{2} \eta_{2} s_{2};i_{3} \eta_{3} s_{3};i_{4} \eta_{4} s_{4}} \left( \vec{k}_1, \vec{k}_2; \tau_1, \tau_2 \right) \approx& \left\langle \hat{\gamma}^\dagger_{\vec{k}_1,i_{1}, \eta_{1}, s_{1}} \left( \tau_1 \right) \hat{\gamma}_{\vec{k}_1,i_{2}, \eta_{2}, s_{2}} \left( \tau_1 \right) \right\rangle \left\langle \hat{\gamma}^\dagger_{\vec{k}_2,i_{3}, \eta_{3}, s_{3}} \left( \tau_2 \right)  \hat{\gamma}_{\vec{k}_2,i_{4}, \eta_{4}, s_{4}} \left( \tau_2 \right) \right\rangle \nonumber \\
	&- \left\langle \mathcal{T}_{\tau} \hat{\gamma}^\dagger_{\vec{k}_1,i_{1}, \eta_{1}, s_{1}} \left( \tau_1 \right)  \hat{\gamma}_{\vec{k}_2,i_{4}, \eta_{4}, s_{4}} \left( \tau_2 \right)  \right\rangle \left\langle  \mathcal{T}_{\tau} \hat{\gamma}^\dagger_{\vec{k}_2,i_{3}, \eta_{3}, s_{3}} \left( \tau_2 \right) \hat{\gamma}_{\vec{k}_1,i_{2}, \eta_{2}, s_{2}} \left( \tau_1 \right) \right\rangle \nonumber \\
	=& \mathcal{G}_{i_{2} \eta_{2} s_{2};i_{1} \eta_{1} s_{1}} \left(0^{-}, \vec{k}_1 \right) \mathcal{G}_{i_{4} \eta_{4} s_{4};i_{3} \eta_{3} s_{3}} \left(0^{-}, \vec{k}_2 \right) \nonumber \\
	&- \mathcal{G}_{i_{4} \eta_{4} s_{4};i_{1} \eta_{1} s_{1}} \left(\tau_2 - \tau_1, \vec{k}_1 \right) \mathcal{G}_{i_{2} \eta_{2} s_{2};i_{3} \eta_{3} s_{3}} \left(\tau_1 - \tau_2, \vec{k}_2 \right) \delta_{\vec{k}_1, \vec{k}_2}. \label{app:eqn:four_fermion_general_eval_int}
\end{align}

In what follows, we will approximate the four-fermion correlation functions appearing in \cref{app:eqn:L11_mats,app:eqn:L12_mats} by dropping the vertex corrections and using the decoupling prescription from \cref{app:eqn:four_fermion_general_eval_int}. 

\subsubsection{Computing the dynamic linear response thermal coefficients} \label{app:sec:thermoelectric_response:evaluating_the_transport_coefficients:dyn_lin_response}

We start by evaluating current-current correlation function from \cref{app:eqn:L11_mats}, which, after dropping the vertex corrections becomes 
\begin{align}
	&\left\langle \mathcal{T}_{\tau} J^{\beta} \left(- i \tau \right) J^{\alpha} \left( 0 \right) \right\rangle \approx  \left\langle J^{\beta} \left(- i \tau \right) \right\rangle \left\langle J^{\alpha} \left( 0 \right) \right\rangle- \sum_{\vec{k}_1, \vec{k}_2} \sum_{\substack{i_{1}, \eta_{1}, s_{1} \\ i_{2}, \eta_{2}, s_{2}}} \sum_{\substack{i_{3}, \eta_{3}, s_{3} \\ i_{4}, \eta_{4}, s_{4}}} T^{\beta}_{i_{1} \eta_{1} s_{1};i_{2} \eta_{2} s_{2}} \left( \vec{k}_1 \right) T^{\alpha}_{i_{3} \eta_{3} s_{3};i_{4} \eta_{4} s_{4}} \left( \vec{k}_2 \right)  \nonumber \\
	&\times \mathcal{G}_{i_{4} \eta_{4} s_{4};i_{1} \eta_{1} s_{1}} \left(- \tau, \vec{k}_1 \right) \mathcal{G}_{i_{2} \eta_{2} s_{2};i_{3} \eta_{3} s_{3}} \left(\tau, \vec{k}_2 \right) \delta_{\vec{k}_1, \vec{k}_2} \nonumber \\
	= &-\sum_{\substack{\vec{k}, i_{1}, \eta_{1}, s_{1} \\ i_{2}, \eta_{2}, s_{2}}} \sum_{\substack{i_{3}, \eta_{3}, s_{3} \\ i_{4}, \eta_{4}, s_{4}}} T^{\beta}_{i_{1} \eta_{1} s_{1};i_{2} \eta_{2} s_{2}} \left( \vec{k} \right) T^{\alpha}_{i_{3} \eta_{3} s_{3};i_{4} \eta_{4} s_{4}} \left( \vec{k} \right) \mathcal{G}_{i_{4} \eta_{4} s_{4};i_{1} \eta_{1} s_{1}} \left(- \tau, \vec{k} \right) \mathcal{G}_{i_{2} \eta_{2} s_{2};i_{3} \eta_{3} s_{3}} \left(\tau, \vec{k} \right) \nonumber \\
	= &\frac{-1}{\beta^2} \sum_{\substack{\vec{k}, i_{1}, \eta_{1}, s_{1} \\ i_{2}, \eta_{2}, s_{2} \\ i_{3}, \eta_{3}, s_{3} \\ i_{4}, \eta_{4}, s_{4}}} T^{\beta}_{i_{1} \eta_{1} s_{1};i_{2} \eta_{2} s_{2}} \left( \vec{k} \right) T^{\alpha}_{i_{3} \eta_{3} s_{3};i_{4} \eta_{4} s_{4}} \left( \vec{k} \right) \sum_{i \omega_x, i \omega_y} \mathcal{G}_{i_{4} \eta_{4} s_{4};i_{1} \eta_{1} s_{1}} \left(i \omega_x, \vec{k} \right) \mathcal{G}_{i_{2} \eta_{2} s_{2};i_{3} \eta_{3} s_{3}} \left(i \omega_y, \vec{k} \right) e^{\left(i\omega_x - i\omega_y \right) \tau} \label{app:eqn:L11_corr_curr_final}
\end{align}
where we have used the fact that no currents flow in equilibrium ({\it i.e.}{}, $\left\langle J^{\alpha} \left(-i \tau \right) \right\rangle = 0$) and have ignored the vertex corrections, as mentioned before. To avoid confusion, we also comment on the notation we used, which is also summarized in \cref{app:tab:notation_recap}. The current operators have been evaluated at general complex times ({\it i.e.}{} containing both a real and imaginary part), as for example in \cref{app:eqn:lin_resp_coeff_expression_luttinger}, where the current operator is denoted as $J^{\alpha}( t - i \tau)$. As a result, we always explicitly include the imaginary unit $i$ when evaluating the current operator at an imaginary time, as mentioned already around \cref{app:eqn:current_corr_func_matsubara}. On the other hand, the Matsubara Green's function is only evaluated for an imaginary time. As a result we stick to the standard convention of \cref{app:eqn:matsubara_gf_THF_tau}, where the imaginary unit $i$ is not explicitly indicated in the argument of Matsubara Green's function. 

The expression from \cref{app:eqn:L11_corr_curr_final} can be substituted into \cref{app:eqn:L11_mats}, and, after evaluating the integral over $\tau$, the result reads as
\begin{equation}
	\label{app:eqn:L11_eval_interm_1}
	L^{\alpha \beta}_{11} \left( i \tilde{\Omega}_n \right) = \frac{-1}{N_0 \Omega_0} \frac{i}{\beta} \sum_{\substack{\vec{k}, i_{1}, \eta_{1}, s_{1} \\ i_{2}, \eta_{2}, s_{2}}} \sum_{\substack{i_{3}, \eta_{3}, s_{3} \\ i_{4}, \eta_{4}, s_{4}}} T^{\beta}_{i_{1} \eta_{1} s_{1};i_{2} \eta_{2} s_{2}} \left( \vec{k} \right) T^{\alpha}_{i_{3} \eta_{3} s_{3};i_{4} \eta_{4} s_{4}} \left( \vec{k} \right) \mathcal{S}_{i_{1} \eta_{1} s_{1};i_{2} \eta_{2} s_{2};i_{3} \eta_{3} s_{3};i_{4} \eta_{4} s_{4}} \left(i \tilde{\Omega}_n, \vec{k} \right).
\end{equation}
For simplicity, in \cref{app:eqn:L11_eval_interm_1}, we have introduced
\begin{align}
	& \mathcal{S}_{i_{1} \eta_{1} s_{1};i_{2} \eta_{2} s_{2};i_{3} \eta_{3} s_{3};i_{4} \eta_{4} s_{4}} \left(i \tilde{\Omega}_n, \vec{k} \right) = \frac{1}{i \tilde{\Omega}_n \beta} \sum_{i \omega_x} \mathcal{G}_{i_{4} \eta_{4} s_{4};i_{1} \eta_{1} s_{1}} \left(i \omega_x, \vec{k} \right) \mathcal{G}_{i_{2} \eta_{2} s_{2};i_{3} \eta_{3} s_{3}} \left(i \omega_x + i\tilde{\Omega}_n, \vec{k} \right) \nonumber \\
	=& \frac{1}{i \tilde{\Omega}_n \beta} \int_{-\infty}^{\infty} \dd{\omega_1} \int_{-\infty}^{\infty} \dd{\omega_2}  A_{i_{4} \eta_{4} s_{4};i_{1} \eta_{1} s_{1}} \left(\omega_1, \vec{k} \right) A_{i_{2} \eta_{2} s_{2};i_{3} \eta_{3} s_{3}} \left(\omega_2, \vec{k} \right) \sum_{i \omega_x} \frac{1}{i \omega_x - \omega_1} \frac{1}{i \omega_x + i \tilde{\Omega}_n - \omega_2} \nonumber \\
	=& \frac{1}{i \tilde{\Omega}_n} \int_{-\infty}^{\infty} \dd{\omega_1} \int_{-\infty}^{\infty} \dd{\omega_2}  A_{i_{4} \eta_{4} s_{4};i_{1} \eta_{1} s_{1}} \left(\omega_1, \vec{k} \right) A_{i_{2} \eta_{2} s_{2};i_{3} \eta_{3} s_{3}} \left(\omega_2,\vec{k} \right) \frac{n_{\mathrm{F}} \left( \omega_1 \right) - n_{\mathrm{F}} \left( \omega_2 \right)}{\omega_1 - \omega_2 + i\tilde{\Omega}_n}, \label{app:eqn:L11_eval_interm_2}
\end{align}
with the Matsubara summation being readily evaluated via the spectral function representation of the Green's function from \cref{app:eqn:spectral_rep_of_GF}. In \cref{app:eqn:L11_eval_interm_2}, $\omega_x \in  \left( 2 \mathbb{Z} + 1 \right) \pi/\beta$ is a fermioninc Matsubara frequency. The spectral function of the Green's function $A \left( \omega, \vec{k} \right)$ is determined for interacting systems, as discussed in \cref*{DMFT:app:sec:se_correction_beyond_HF,DMFT:app:sec:se_symmetric} of Ref.~\cite{CAL23b}, or as will be discussed in \cref{app:sec:thermoelectric_response:details_on_numerics:non_interactions} in the absence of interactions. Upon substituting \cref{app:eqn:L11_eval_interm_2} back into \cref{app:eqn:L11_eval_interm_1} we obtain 
\begin{equation}
	\label{app:eqn:L11_eval_stage_1}
	L^{\alpha \beta}_{11} \left( i \tilde{\Omega}_n \right) = \frac{-1}{N_0 \Omega_0} \frac{i}{\beta} \sum_{\vec{k}} \int_{-\infty}^{\infty} \dd{\omega_1} \int_{-\infty}^{\infty} \dd{\omega_2} \Tr \left( T^{\beta} \left( \vec{k} \right) A \left( \omega_2, \vec{k} \right) T^{\alpha} \left( \vec{k} \right) A \left( \omega_1, \vec{k} \right) \right) \frac{1}{i\tilde{\Omega}_n} \frac{n_{\mathrm{F}} \left( \omega_1 \right) - n_{\mathrm{F}} \left( \omega_2 \right)}{i \tilde{\Omega}_n - \omega_2 + \omega_1 }. 
\end{equation}

For the time being, we leave \cref{app:eqn:L11_eval_stage_1} in its current form and consider the current-current correlator from \cref{app:eqn:L12_mats}, which, using \cref{app:eqn:mats_thermal_current} as well as the same approximation that neglects the vertex corrections, can be decoupled as 
\begin{align}
	&\left\langle \mathcal{T}_{\tau} J_Q^{\beta} \left(- i \tau \right) J^{\alpha} \left( 0 \right) \right\rangle \approx  \left\langle J_Q^{\beta} \left(- i \tau \right) \right\rangle \left\langle J^{\alpha} \left( 0 \right) \right\rangle- \sum_{\vec{k}_1, \vec{k}_2} \sum_{\substack{i_{1}, \eta_{1}, s_{1} \\ i_{2}, \eta_{2}, s_{2}}} \sum_{\substack{i_{3}, \eta_{3}, s_{3} \\ i_{4}, \eta_{4}, s_{4}}} T^{\beta}_{i_{1} \eta_{1} s_{1};i_{2} \eta_{2} s_{2}} \left( \vec{k}_1 \right) T^{\alpha}_{i_{3} \eta_{3} s_{3};i_{4} \eta_{4} s_{4}} \left( \vec{k}_2 \right)  \nonumber \\
	& \times \frac{1}{2} \left[ \left\langle \mathcal{T}_{\tau} \left( \pdv{\tau} \hat{\gamma}^\dagger_{\vec{k}_1,i_{1}, \eta_{1}, s_{1}} \left( \tau \right) \right)  \hat{\gamma}_{\vec{k}_2,i_{4}, \eta_{4}, s_{4}} \left( 0 \right)  \right\rangle \left\langle  \mathcal{T}_{\tau} \hat{\gamma}^\dagger_{\vec{k}_2,i_{3}, \eta_{3}, s_{3}} \left( 0 \right) \hat{\gamma}_{\vec{k}_1,i_{2}, \eta_{2}, s_{2}} \left( \tau \right) \right\rangle \right. \nonumber \\
	&  \quad \left. - \left\langle \mathcal{T}_{\tau} \hat{\gamma}^\dagger_{\vec{k}_1,i_{1}, \eta_{1}, s_{1}} \left( \tau \right)  \hat{\gamma}_{\vec{k}_2,i_{4}, \eta_{4}, s_{4}} \left( 0 \right)  \right\rangle \left\langle  \mathcal{T}_{\tau} \hat{\gamma}^\dagger_{\vec{k}_2,i_{3}, \eta_{3}, s_{3}} \left( 0 \right) \left( \pdv{\tau} \hat{\gamma}_{\vec{k}_1,i_{2}, \eta_{2}, s_{2}} \left( \tau \right) \right) \right\rangle
	\right]. \label{app:eqn:L12_corr_curr_interm}
\end{align} 
The time derivatives inside the two-fermion correlation functions can be evaluated using the definition of the Green's function from \cref{app:eqn:matsubara_gf_THF_tau}
\begin{align}
	\pdv{\tau }\mathcal{G}_{i \eta s;i' \eta' s'} \left(\tau, \vec{k} \right) &= -\pdv{\tau} \left\langle \mathcal{T}_{\tau} \hat{\gamma}_{\vec{k},i, \eta, s} \left( \tau \right) \hat{\gamma}^\dagger_{\vec{k},i', \eta', s'} \left( 0 \right)  \right\rangle \nonumber \\
	&= -\pdv{\tau} \left( \Theta \left( \tau \right) \left\langle \hat{\gamma}_{\vec{k},i, \eta, s} \left( \tau \right) \hat{\gamma}^\dagger_{\vec{k},i', \eta', s'} \left( 0 \right)  \right\rangle - \Theta \left( - \tau \right) \left\langle \hat{\gamma}^\dagger_{\vec{k},i', \eta', s'} \left( 0 \right) \hat{\gamma}_{\vec{k},i, \eta, s} \left( \tau \right) \right\rangle \right) \nonumber \\
	&= - \delta \left( \tau \right) \left\langle \anticommutator{\hat{\gamma}_{\vec{k},i, \eta, s} \left( \tau \right)}{\hat{\gamma}^\dagger_{\vec{k},i', \eta', s'} \left( 0 \right)} \right\rangle - \left\langle \mathcal{T}_{\tau} \left( \pdv{\tau} \hat{\gamma}_{\vec{k},i, \eta, s} \left( \tau \right)  \right) \hat{\gamma}^\dagger_{\vec{k},i', \eta', s'} \left( 0 \right)  \right\rangle,
\end{align}
which holds for $-\beta < \tau < \beta$, and from which we obtain that 
\begin{align}
	\label{app:eqn:correlator_with_derivative_1}
	\left\langle \mathcal{T}_{\tau} \left( \pdv{\tau} \hat{\gamma}_{\vec{k},i, \eta, s} \left( \tau \right)  \right) \hat{\gamma}^\dagger_{\vec{k},i', \eta', s'} \left( 0 \right)  \right\rangle &= - \pdv{\tau }\mathcal{G}_{i \eta s;i' \eta' s'} \left(\tau, \vec{k} \right) - \delta \left( \tau \right) \delta_{i i'} \delta_{\eta \eta'} \delta_{s s'}, \\
	\label{app:eqn:correlator_with_derivative_2}
	\left\langle \mathcal{T}_{\tau} \hat{\gamma}_{\vec{k},i, \eta, s} \left( 0 \right) \left( \pdv{\tau}  \hat{\gamma}^\dagger_{\vec{k},i', \eta', s'} \left( \tau \right) \right)  \right\rangle &= - \pdv{\tau }\mathcal{G}_{i \eta s;i' \eta' s'} \left(-\tau, \vec{k} \right) + \delta \left( \tau \right) \delta_{i i'} \delta_{\eta \eta'} \delta_{s s'}.
\end{align}
Upon substituting \cref{app:eqn:correlator_with_derivative_1,app:eqn:correlator_with_derivative_2} into \cref{app:eqn:L12_corr_curr_interm}, we find that
\begin{align}
	&\left\langle \mathcal{T}_{\tau} J_Q^{\beta} \left(- i \tau \right) J^{\alpha} \left( 0 \right) \right\rangle \approx - \sum_{\vec{k}} \sum_{\substack{i_{1}, \eta_{1}, s_{1} \\ i_{2}, \eta_{2}, s_{2}}} \sum_{\substack{i_{3}, \eta_{3}, s_{3} \\ i_{4}, \eta_{4}, s_{4}}} T^{\beta}_{i_{1} \eta_{1} s_{1};i_{2} \eta_{2} s_{2}} \left( \vec{k} \right) T^{\alpha}_{i_{3} \eta_{3} s_{3};i_{4} \eta_{4} s_{4}} \left( \vec{k} \right)  \nonumber \\
	\times & \frac{1}{2} \left[ \left( \delta \left( \tau \right) \delta_{i_1 i_4} \delta_{\eta_1 \eta_4} \delta_{s_1 s_4} - \pdv{\tau} \mathcal{G}_{i_{4} \eta_{4} s_{4};i_{1} \eta_{1} s_{1}} \left( - \tau, \vec{k} \right)  \right) \mathcal{G}_{i_{2} \eta_{2} s_{2};i_{3} \eta_{3} s_{3}} \left( \tau, \vec{k} \right) \right. \nonumber \\
	&  \quad \left. - \mathcal{G}_{i_{4} \eta_{4} s_{4};i_{1} \eta_{1} s_{1}} \left( - \tau, \vec{k} \right) \left(  - \delta \left( \tau \right) \delta_{i_2 i_3} \delta_{\eta_2 \eta_3} \delta_{s_2 s_3} - \pdv{\tau}\mathcal{G}_{i_{2} \eta_{2} s_{2};i_{3} \eta_{3} s_{3}} \left(\tau, \vec{k} \right) \right)
	\right] \nonumber \\
	\approx & \mathcal{C}^{\beta \alpha} (\tau) \delta (\tau) + \frac{1}{2} \sum_{\vec{k}} \Tr \left[T^{\beta} \left( \vec{k} \right) \mathcal{G}\left( \tau, \vec{k} \right) T^{\alpha} \left( \vec{k} \right) \left( \pdv{\tau} \mathcal{G} \left( -\tau, \vec{k} \right) \right) \right] -\frac{1}{2} \sum_{\vec{k}} \Tr \left[T^{\beta} \left( \vec{k} \right) \left( \pdv{\tau} \mathcal{G}\left( \tau, \vec{k} \right) \right) T^{\alpha} \left( \vec{k} \right)  \mathcal{G} \left( -\tau, \vec{k} \right) \right] \nonumber \\
	\approx &  \mathcal{C}^{\beta \alpha} (\tau) \delta (\tau) + \frac{1}{2 \beta^2} \sum_{\substack{\vec{k} \\ i\omega_x, i\omega_y }} \Tr \left( T^{\beta} \left( \vec{k} \right) \mathcal{G}\left( i \omega_y, \vec{k} \right) T^{\alpha} \left( \vec{k} \right)  \mathcal{G} \left(i \omega_x, \vec{k} \right) \right) \left( i\omega_x + i \omega_y \right) e^{\left(i\omega_x - i\omega_y \right) \tau}, \qq{for} \abs{\tau} < \beta, \label{app:eqn:L12_corr_curr_final}
\end{align}
where, for brevity, we have also introduced the following function
\begin{equation}
	\mathcal{C}^{\beta \alpha} \left( \tau \right) =- \frac{1}{2} \sum_{\vec{k}} \left( \Tr \left( T^{\beta} \left( \vec{k} \right) \mathcal{G} \left( \tau, \vec{k} \right) T^{\alpha} \left( \vec{k} \right) \right) + \Tr \left( \mathcal{G} \left( -\tau, \vec{k} \right) T^{\beta} \left( \vec{k} \right) T^{\alpha} \left( \vec{k} \right) \right) \right)
\end{equation}

Note that the current-current correlator from \cref{app:eqn:L12_corr_curr_final} is integrated in the interval $0 < \tau < \beta$ in \cref{app:eqn:L12_mats}. However, \cref{app:eqn:L12_corr_curr_final} contains a Dirac $\delta$-function singularity at $\tau=0$. It is therefore useful to change the integration interval in \cref{app:eqn:L12_mats}. This is possible because of the periodicity of the integrand. Specifically, for $0 < \tau < \beta$, we find that 
\begin{align}
	\left\langle \mathcal{T}_{\tau} J_Q^{\beta} \left(- i \left( \tau - \beta \right) \right) J^{\alpha} \left( 0 \right) \right\rangle &= \left\langle J^{\alpha} \left( 0 \right) e^{K \left( \tau - \beta \right)} J_Q^{\beta} \left(0 \right) e^{-K \left( \tau - \beta \right)}  \right\rangle \nonumber \\
	&= \frac{1}{Z} \Tr \left[ e^{-\beta K} J^{\alpha} \left( 0 \right) e^{-\beta K} e^{K \tau}  J_Q^{\beta} \left(0 \right) e^{-K \tau} e^{\beta K} \right] \nonumber \\
	&= \frac{1}{Z} \Tr \left[ e^{-\beta K} e^{K \tau}  J_Q^{\beta} \left(0 \right) e^{-K \tau} J^{\alpha} \left( 0 \right) \right] = \left\langle \mathcal{T}_{\tau} J_Q^{\beta} \left(- i \tau \right) J^{\alpha} \left( 0 \right) \right\rangle,
\end{align}
which, combined with the fact that $e^{i \beta \Omega_n } = 1$, allows us to change the integration interval in \cref{app:eqn:L12_mats} to
\begin{align}
	L^{\alpha \beta}_{12} \left( i \tilde{\Omega}_n \right) &= \frac{1}{N_0 \Omega_0} \frac{1}{\beta \tilde{\Omega}_n} \int_{0}^{\beta} \dd{\tau} e^{i \tilde{\Omega}_n \tau} \left\langle \mathcal{T}_{\tau} J_{Q}^{\beta} \left( -i\tau \right) J^{\alpha} \left( 0 \right) \right\rangle \nonumber \\
	&= \frac{1}{N_0 \Omega_0} \frac{1}{\beta \tilde{\Omega}_n} \int_{-\frac{\beta}{2}}^{\frac{\beta}{2}} \dd{\tau} e^{i \tilde{\Omega}_n \tau} \left\langle \mathcal{T}_{\tau} J_{Q}^{\beta} \left( -i\tau \right) J^{\alpha} \left( 0 \right) \right\rangle. \label{app:eqn:L12_mats_change_interval}
\end{align}

Even after the change of the interval, integrating the first term of \cref{app:eqn:L12_corr_curr_final} necessitates a more careful treatment, because the Matsubara Green's function $\mathcal{G} \left( \tau, \vec{k} \right)$ on which it depends is not continuous at the origin
\begin{equation}
	\label{app:eqn:gf_origin}
	\lim_{\tau \to 0^{\pm}} \mathcal{G}_{i \eta s;i' \eta' s'} \left(\tau, \vec{k} \right) = \varrho_{i' \eta' s';i \eta s} \left( \vec{k} \right) \mp \frac{1}{2} \delta_{i i'} \delta_{\eta \eta'} \delta_{s s'}, 
\end{equation}
with the density matrix being defined in \cref{app:eqn:def_rho_HF}. Taking the integral of just the first term in \cref{app:eqn:L12_corr_curr_final} and performing a change of variables $\tau \to -\tau$, we find that
\begin{align}
	&\int_{-\frac{\beta}{2}}^{\frac{\beta}{2}} \dd{\tau} e^{i\tilde{\Omega}_n \tau} \mathcal{C}^{\beta \alpha} (\tau) \delta (\tau) =- \frac{1}{2} \int_{-\frac{\beta}{2}}^{\frac{\beta}{2}} \dd{\tau} \sum_{\vec{k}} \left( \Tr \left( T^{\beta} \left( \vec{k} \right) \mathcal{G} \left( \tau, \vec{k} \right) T^{\alpha} \left( \vec{k} \right) + T^{\alpha} \left( \vec{k} \right) \mathcal{G} \left( -\tau, \vec{k} \right) T^{\beta} \left( \vec{k} \right) \right) \right) \delta \left( \tau \right) \nonumber \\
	=&- \frac{1}{4} \int_{-\frac{\beta}{2}}^{\frac{\beta}{2}} \dd{\tau} \sum_{\vec{k}} \left[ \Tr \left( T^{\beta} \left( \vec{k} \right) \left( \mathcal{G} \left( \tau, \vec{k} \right) + \mathcal{G} \left( - \tau, \vec{k} \right) \right) T^{\alpha} \left( \vec{k} \right) + T^{\alpha} \left( \vec{k} \right) \left( \mathcal{G} \left( -\tau, \vec{k} \right) + \mathcal{G} \left( \tau, \vec{k} \right) \right) T^{\beta} \left( \vec{k} \right) \right) \right] \delta \left( \tau \right) \nonumber \\
	=& - \frac{1}{2} \sum_{\vec{k}} \left( \Tr \left( T^{\beta} \left( \vec{k} \right) \varrho^{T} \left( \vec{k} \right) T^{\alpha} \left( \vec{k} \right) \right) + \Tr \left( \varrho^{T} \left( \vec{k} \right) T^{\beta} \left( \vec{k} \right) T^{\alpha} \left( \vec{k} \right) \right) \right) \equiv \mathcal{C}^{\beta \alpha}, \label{app:eqn:def_L12_constant_cal_C}
\end{align}
where we have used the fact that $\mathcal{G} \left( -\tau, \vec{k} \right) + \mathcal{G} \left( \tau, \vec{k} \right)$ is continuous at $\tau = 0$, with 
\begin{equation}
	\lim_{\tau \to 0} \left( \mathcal{G} \left( -\tau, \vec{k} \right) + \mathcal{G} \left( \tau, \vec{k} \right) \right) = \varrho^{T} \left( \vec{k} \right).
\end{equation}
Because both $T^{\alpha} \left( \vec{k} \right)$ and $\varrho \left( \vec{k} \right)$ are Hermitian matrices, the constant $\mathcal{C}^{\beta \alpha} $ defined in \cref{app:eqn:def_L12_constant_cal_C} is real.

After substituting \cref{app:eqn:L12_corr_curr_final} into \cref{app:eqn:L12_mats}, using \cref{app:eqn:def_L12_constant_cal_C}, and evaluating the integral over $\tau$ for the second term, we find that 
\begin{align}
	L^{\alpha \beta}_{12} \left( i \tilde{\Omega}_n \right) &= \frac{1}{N_0 \Omega_0} \frac{1}{\beta \tilde{\Omega}_n} \left[ \mathcal{C}^{\beta \alpha} + \frac{1}{\beta} \sum_{\vec{k}, i\omega_x} \Tr \left( T^{\beta} \left( \vec{k} \right) \mathcal{G}\left( i \omega_x  + i\tilde{\Omega}_n, \vec{k} \right) T^{\alpha} \left( \vec{k} \right)  \mathcal{G} \left(i \omega_x, \vec{k} \right) \right) \left( i\omega_x + \frac{i \tilde{\Omega}_n}{2} \right) \right] \nonumber \\
	&= \frac{1}{N_0 \Omega_0} \frac{1}{\beta \tilde{\Omega}_n} \left[ \mathcal{C}^{\beta \alpha} + \sum_{\vec{k}} \int_{-\infty}^{\infty} \dd{\omega_1} \int_{-\infty}^{\infty} \dd{\omega_2} \Tr \left( T^{\beta} \left( \vec{k} \right) A\left( \omega_2, \vec{k} \right) T^{\alpha} \left( \vec{k} \right)  A\left(\omega_1, \vec{k} \right) \right) \right. \nonumber \\
	&\times \frac{1}{\beta} \sum_{i \omega_x} \left. \frac{1}{i\omega_x - \omega_1} \frac{1}{i\omega_x + i \tilde{\Omega}_n - \omega_2} \left( i\omega_x + \frac{i \tilde{\Omega}_n}{2} \right) \right]. \label{app:eqn:L12_eval_stage_1_interm_1}
\end{align}
The Matsubara frequency summation can be computed using standard techniques~\cite{WIK23}
\begin{align}
	&\frac{1}{\beta} \sum_{i \omega_x} \frac{1}{i\omega_x - \omega_1} \frac{1}{i\omega_x + i \tilde{\Omega}_n - \omega_2} \left( i\omega_x + \frac{i \tilde{\Omega}_n}{2} \right) = \frac{1}{2 \beta} \sum_{i \omega_x} \frac{i \omega_x - \omega_1 + i\omega_x + i\tilde{\Omega}_n - \omega_2 + \left( \omega_1 + \omega_2 \right)}{\left( i \omega_x - \omega_1 \right) \left( i \omega_x + i \tilde{\Omega}_n - \omega_2 \right)} \nonumber \\
	&=\frac{1}{2 \beta} \sum_{i \omega_x} \left( \frac{1}{i\omega_x - \omega_1} +  \frac{1}{i\omega_x + i \tilde{\Omega}_n - \omega_2} + \frac{\omega_1 + \omega_2}{\left( i \omega_x - \omega_1 \right) \left( i \omega_x + i \tilde{\Omega}_n - \omega_2 \right)} \right) \nonumber \\
	&=\frac{1}{2} \left[ n_{\mathrm{F}} \left( \omega_1 \right) + n_{\mathrm{F}} \left( \omega_2 \right) +  \left( \omega_1 + \omega_2 \right)\frac{n_{\mathrm{F}} \left( \omega_1 \right) - n_{\mathrm{F}} \left( \omega_2 \right)}{\omega_1 - \omega_2 + i\tilde{\Omega}_n} \right]. \label{app:eqn:mats_sumation_for_L12}
\end{align}

Upon substituting \cref{app:eqn:mats_sumation_for_L12} into \cref{app:eqn:L12_eval_stage_1_interm_1}, the integrals involving the first two terms of the former can be evaluated by using the expression of the density matrix $\rho \left( \vec{k} \right)$ in terms of the spectral function, {\it i.e.}{}
\begin{align}
	&\sum_{\vec{k}} \int_{-\infty}^{\infty} \dd{\omega_1} \int_{-\infty}^{\infty} \dd{\omega_2}  \Tr \left( T^{\beta} \left( \vec{k} \right) A\left( \omega_2, \vec{k} \right) T^{\alpha} \left( \vec{k} \right)  A\left(\omega_1, \vec{k} \right) \right) \frac{n_{\mathrm{F}} \left( \omega_1 \right) + n_{\mathrm{F}} \left( \omega_2 \right)}{2} \nonumber \\
	=& \frac{1}{2} \sum_{\vec{k}} \left( \Tr \left( T^{\beta} \left( \vec{k} \right) T^{\alpha} \left( \vec{k} \right)  \varrho^{T} \left( \vec{k} \right) \right) + \Tr \left( T^{\beta} \left( \vec{k} \right) \varrho^{T} \left( \vec{k} \right) T^{\alpha} \left( \vec{k} \right) \right) \right) \nonumber \\
	=& \frac{1}{2} \sum_{\vec{k}} \left( \Tr \left( T^{\beta} \left( \vec{k} \right) \varrho^{T} \left( \vec{k} \right) T^{\alpha} \left( \vec{k} \right) \right) + \Tr \left( \varrho^{T} \left( \vec{k} \right) T^{\beta} \left( \vec{k} \right) T^{\alpha} \left( \vec{k} \right) \right) \right) = - \mathcal{C}^{\beta \alpha}. \label{app:eqn:first_summation_in_mats_sumation_for_L12}
\end{align}
Using \cref{app:eqn:first_summation_in_mats_sumation_for_L12}, we find from \cref{app:eqn:L12_eval_stage_1_interm_1,app:eqn:mats_sumation_for_L12} that 
\begin{align}
	L^{\alpha \beta}_{12} \left( i \tilde{\Omega}_n \right) &= \frac{1}{N_0 \Omega_0} \frac{1}{\beta \tilde{\Omega}_n} \left[ \mathcal{C}^{\beta \alpha} - \mathcal{C}^{\beta \alpha}+ \sum_{\vec{k}} \Tr \left( T^{\beta} \left( \vec{k} \right) T^{\alpha} \left( \vec{k} \right) \right)  \right. \nonumber \\
	& \left. + \sum_{\vec{k}} \int_{-\infty}^{\infty} \dd{\omega_1} \int_{-\infty}^{\infty} \dd{\omega_2} \Tr \left( T^{\beta} \left( \vec{k} \right) A \left( \omega_2, \vec{k} \right) T^{\alpha} \left( \vec{k} \right) A \left( \omega_1, \vec{k} \right) \right)  \frac{\omega_1 + \omega_2}{2} \frac{n_{\mathrm{F}} \left( \omega_1 \right) - n_{\mathrm{F}} \left( \omega_2 \right)}{\omega_1 - \omega_2 + i\tilde{\Omega}_n} \right]. \nonumber \\
	&= \frac{1}{N_0 \Omega_0} \frac{i}{\beta } \left[ \sum_{\vec{k}} \int_{-\infty}^{\infty} \dd{\omega_1} \int_{-\infty}^{\infty} \dd{\omega_2} \Tr \left( T^{\beta} \left( \vec{k} \right) A \left( \omega_2, \vec{k} \right) T^{\alpha} \left( \vec{k} \right) A \left( \omega_1, \vec{k} \right) \right)  \frac{\omega_1 + \omega_2}{2 i \tilde{\Omega}_n} \frac{n_{\mathrm{F}} \left( \omega_1 \right) - n_{\mathrm{F}} \left( \omega_2 \right)}{\omega_1 - \omega_2 + i\tilde{\Omega}_n}   \right. \nonumber \\
	&+ \frac{1}{i \tilde{\Omega}_n} \left. \sum_{\vec{k}} \Tr \left( T^{\beta} \left( \vec{k} \right) T^{\alpha} \left( \vec{k} \right) \right)  \right]. \label{app:eqn:L12_eval_stage_1}
\end{align}
\Cref{app:eqn:L11_eval_stage_1,app:eqn:L12_eval_stage_1} are the final expressions for the dynamic correlation functions from \cref{app:eqn:L11_mats,app:eqn:L12_mats} in terms of the THF spectral function. They are valid for any bosonic Matsubara frequency $i \tilde{\Omega}_n$ and will be used to compute the static linear response thermal coefficients in \cref{app:sec:thermoelectric_response:evaluating_the_transport_coefficients:static_lin_response}. It is worth mentioning the only two approximations employed in deriving them were neglecting the vertex corrections to the current-current correlation functions and discarding the interaction ({\it i.e.}{} quartic) contribution to the energy current, which was quantitatively justified in \cref{app:sec:deriv_moire_cur:energy_interaction:comparison}. Note that discarding the quartic interaction contribution to the energy current does not mean that we are performing a non-interacting calculations: the Green's function that we employ (and, consequently, the corresponding spectral function) is still computed within the interacting theory. Discarding the interaction contribution to the energy current simply means that we drop the terms in the energy current that arise from the interaction Hamiltonian, which were shown to be small in \cref{app:sec:deriv_moire_cur:energy_interaction:comparison}. 

\subsubsection{Computing the static linear response thermal coefficients} \label{app:sec:thermoelectric_response:evaluating_the_transport_coefficients:static_lin_response}

With the expressions of the dynamic thermoelectric response coefficients in terms of the THF spectral function from \cref{app:eqn:L11_mats,app:eqn:L12_mats} at hand, we are now ready to obtain the static ones. To do so, we employ the algorithmic prescription from below \cref{app:eqn:Ls_anal_cont}.

To begin with, we define 
\begin{equation}
	\label{app:eqn:def_integrand_for_L11_L12}
	\mathcal{S}^{\alpha \beta} \left( \omega_1, \omega_2 \right) = \frac{1}{N_0 \Omega_0} \sum_{\vec{k}} \Tr \left( T^{\beta} \left( \vec{k} \right) A \left( \omega_2, \vec{k} \right) T^{\alpha} \left( \vec{k} \right) A \left( \omega_1, \vec{k} \right) \right),
\end{equation}
and perform the analytical continuation $i \tilde{\Omega}_n \to \omega  + i 0^{+}$ in \cref{app:eqn:L11_mats,app:eqn:L12_mats} 
\begin{align}
	L^{\alpha \beta}_{11} \left( \omega + i 0^{+} \right) =&  \frac{-i}{\beta} \int_{-\infty}^{\infty} \dd{\omega_1} \int_{-\infty}^{\infty} \dd{\omega_2} \mathcal{S}^{\alpha \beta} \left( \omega_1, \omega_2 \right) \frac{1}{\omega} \frac{n_{\mathrm{F}} \left( \omega_1 \right) - n_{\mathrm{F}} \left( \omega_2 \right)}{\omega - \omega_2 + \omega_1 + i 0^{+}  }, \label{app:eqn:L11_anal_cont_interm_1} \\
	L^{\alpha \beta}_{12} \left( \omega + i 0^{+} \right) =&  \frac{i}{\beta} \int_{-\infty}^{\infty} \dd{\omega_1} \int_{-\infty}^{\infty} \dd{\omega_2} \mathcal{S}^{\alpha \beta} \left( \omega_1, \omega_2 \right) \frac{\omega_1 + \omega_2}{2\omega} \frac{n_{\mathrm{F}} \left( \omega_1 \right) - n_{\mathrm{F}} \left( \omega_2 \right)}{\omega - \omega_2 + \omega_1 + i 0^{+} } \nonumber \\
	&+ \frac{i}{\beta \omega} \frac{1}{N_0 \Omega_0} \sum_{\vec{k}} \Tr\left( T^{\beta} \left( \vec{k} \right) T^{\alpha} \left( \vec{k} \right) \right) .  \label{app:eqn:L12_anal_cont_interm_1}
\end{align}
Note that, as we have discussed below Eq.~\cref{app:eqn:chi_new_2_def}, the real-frequency correlators are evaluated in the $\omega \to 0$ limit, but not \emph{at} $\omega = 0$. As such, the small imaginary offset can be and has been discarded in the $\frac{1}{\omega + i 0^{+}}$ prefactors. Additionally, note that the last term in \cref{app:eqn:L12_anal_cont_interm_1} is purely imaginary, as the corresponding traces are real
\begin{equation}
	 \Tr\left( T^{\beta} \left( \vec{k} \right) T^{\alpha} \left( \vec{k} \right) \right)^{*} = \Tr\left( T^{\dagger \beta} \left( \vec{k} \right) T^{\dagger \alpha} \left( \vec{k} \right) \right) = \Tr\left( T^{\alpha} \left( \vec{k} \right) T^{\beta} \left( \vec{k} \right) \right) =  \Tr\left( T^{\beta} \left( \vec{k} \right) T^{\alpha} \left( \vec{k} \right) \right),
\end{equation}
a property which follows from the Hermiticity of the current matrices $T^{\alpha} \left( \vec{k} \right)$. Taking the real part, as well as the limit $\omega \to 0$ using \cref{app:eqn:one_over_omega_plemelj}, we obtain the following expressions for the static thermoelectric coefficients 
\begin{align}
	L^{\alpha \beta}_{11} =&  - \frac{\pi}{\beta} \lim_{\omega \to 0} \int_{-\infty}^{\infty} \dd{\omega_1} \int_{-\infty}^{\infty} \dd{\omega_2} \Re \left( \mathcal{S}^{\alpha \beta} \left( \omega_1, \omega_2 \right) \right) \delta\left(\omega - \omega_2 + \omega_1 \right) \frac{n_{\mathrm{F}} \left( \omega_1 \right) - n_{\mathrm{F}} \left( \omega_2 \right)}{\omega }   \nonumber \\
	&- \frac{1}{\beta} \lim_{\omega \to 0} \int_{-\infty}^{\infty} \dd{\omega_1} \int_{-\infty}^{\infty} \dd{\omega_2} \Im \left( \mathcal{S}^{\alpha \beta} \left( \omega_1, \omega_2 \right) \right) \left( n_{\mathrm{F}} \left( \omega_1 \right) - n_{\mathrm{F}} \left( \omega_2 \right) \right) \left( \frac{1}{\omega} + \frac{1}{\omega_2 - \omega_1} \right) \mathcal{PV} \frac{1}{\omega_2 - \omega_1}, \label{app:eqn:L11_anal_cont_interm_2} \\
L^{\alpha \beta}_{12} =&  \frac{\pi}{\beta} \lim_{\omega \to 0} \int_{-\infty}^{\infty} \dd{\omega_1} \int_{-\infty}^{\infty} \dd{\omega_2} \Re \left( \mathcal{S}^{\alpha \beta} \left( \omega_1, \omega_2 \right) \right) \delta\left(\omega - \omega_2 + \omega_1 \right) \frac{n_{\mathrm{F}} \left( \omega_1 \right) - n_{\mathrm{F}} \left( \omega_2 \right)}{\omega } \frac{ \omega_1 + \omega_2 }{2} \nonumber \\
	&+\frac{1}{\beta} \lim_{\omega \to 0} \int_{-\infty}^{\infty} \dd{\omega_1} \int_{-\infty}^{\infty} \dd{\omega_2} \Im \left( \mathcal{S}^{\alpha \beta} \left( \omega_1, \omega_2 \right) \right)  \left( n_{\mathrm{F}} \left( \omega_1 \right) - n_{\mathrm{F}} \left( \omega_2 \right) \right) \left( \frac{1}{\omega} + \frac{1}{\omega_2 - \omega_1} \right) \mathcal{PV} \frac{\left( \omega_1 + \omega_2 \right) / 2}{\omega_2 - \omega_1}. \label{app:eqn:L12_anal_cont_interm_2}
\end{align}

To further simplify the expressions in \cref{app:eqn:L11_anal_cont_interm_2}, we note that due to cyclic property of the trace, as well as the Hermiticity of the current tensors $T^{\alpha} \left( \vec{k} \right)$ and the spectral function $A \left( \omega, \vec{k} \right)$, the function $\mathcal{S}^{\alpha \beta} \left( \omega_1, \omega_2 \right)$ obeys the following property 
\begin{align}
	\mathcal{S}^{* \alpha \beta} \left( \omega_1, \omega_2 \right) 
	=& \frac{1}{N_0 \Omega_0} \sum_{\vec{k}} \Tr \left( A^{\dagger} \left( \omega_1, \vec{k} \right) T^{\dagger \alpha} \left( \vec{k} \right) A^{\dagger} \left( \omega_2, \vec{k} \right) T^{\dagger \beta} \left( \vec{k} \right) \right) \nonumber \\  
	=& \frac{1}{N_0 \Omega_0} \sum_{\vec{k}} \Tr \left( T^{\beta} \left( \vec{k} \right) A \left( \omega_1, \vec{k} \right) T^{\alpha} \left( \vec{k} \right) A \left( \omega_2, \vec{k} \right) \right) = \mathcal{S}^{\alpha \beta}  \left( \omega_2, \omega_1 \right),
\end{align}
which implies that 
\begin{equation}
	\label{app:eqn:symmetries_of_S_in_omega}
	\Re \left( \mathcal{S}^{\alpha \beta} \left( \omega_1, \omega_2 \right) \right) = \Re \left( \mathcal{S}^{\alpha \beta} \left( \omega_2, \omega_1 \right) \right) \qq{and}
	\Im \left( \mathcal{S}^{\alpha \beta} \left( \omega_1, \omega_2 \right) \right) = -\Im \left( \mathcal{S}^{\alpha \beta} \left( \omega_2, \omega_1 \right) \right).
\end{equation}
\Cref{app:eqn:symmetries_of_S_in_omega} allows us to evaluate the limit $\omega \to 0$ in \cref{app:eqn:L11_anal_cont_interm_2,app:eqn:L12_anal_cont_interm_2} by dropping the terms proportional to $\frac{1}{\omega}$ in the second row of each equation (which vanish under the antisymmetry of the $\omega_1 \leftrightarrow \omega_2$ integration variable exchange). 
\begin{align}
	L^{\alpha \beta}_{11} =&  \frac{\pi}{\beta} \int_{-\infty}^{\infty} \dd{\omega} \pdv{n_{\mathrm{F}} \left( \omega \right)}{\omega} \Re \left( \mathcal{S}^{\alpha \beta} \left( \omega, \omega \right) \right) \nonumber \\
	&+ \frac{1}{\beta} \int_{-\infty}^{\infty} \dd{\omega_1} \int_{-\infty}^{\infty} \dd{\omega_2} \Im \left( \mathcal{S}^{\alpha \beta} \left( \omega_1, \omega_2 \right) \right) \frac{ n_{\mathrm{F}} \left( \omega_2 \right) - n_{\mathrm{F}} \left( \omega_1 \right)}{\omega_2 - \omega_1} \mathcal{PV} \frac{1}{\omega_2 - \omega_1}, \label{app:eqn:L11_static_general_final} \\
L^{\alpha \beta}_{12} =& - \frac{\pi}{\beta} \int_{-\infty}^{\infty} \dd{\omega} \omega \pdv{n_{\mathrm{F}} \left( \omega \right)}{\omega} \Re \left( \mathcal{S}^{\alpha \beta} \left( \omega, \omega \right) \right) \nonumber \\
	&-\frac{1}{\beta} \lim_{\omega \to 0} \int_{-\infty}^{\infty} \dd{\omega_1} \int_{-\infty}^{\infty} \dd{\omega_2} \Im \left( \mathcal{S}^{\alpha \beta} \left( \omega_1, \omega_2 \right) \right)  \frac{ n_{\mathrm{F}} \left( \omega_2 \right) - n_{\mathrm{F}} \left( \omega_1 \right)}{\omega_2 - \omega_1} \mathcal{PV} \frac{\left( \omega_1 + \omega_2 \right) / 2}{\omega_2 - \omega_1}. \label{app:eqn:L12_static_general_final}
\end{align}

\Cref{app:eqn:L11_static_general_final,app:eqn:L12_static_general_final} are the final expressions for the static thermoelectric transport coefficients. Throughout this work, we are exclusively concerned with the Seebeck coefficient defined in \cref{app:eqn:seebeck_definition}, which only depends on the diagonal parts of the linear response coefficients from \cref{app:eqn:L11_static_general_final,app:eqn:L12_static_general_final}. When $\alpha = \beta$, one can again employ the cyclic property of the trace and additionally derive
\begin{align}
	\mathcal{S}^{\alpha \alpha} \left( \omega_1, \omega_2 \right) &= \frac{1}{N_0 \Omega_0} \sum_{\vec{k}} \Tr \left( T^{\alpha} \left( \vec{k} \right) A \left( \omega_2, \vec{k} \right) T^{\alpha} \left( \vec{k} \right) A \left( \omega_1, \vec{k} \right) \right) \nonumber \\
	&= \frac{1}{N_0 \Omega_0} \sum_{\vec{k}} \Tr \left( T^{\alpha} \left( \vec{k} \right) A \left( \omega_1, \vec{k} \right) T^{\alpha} \left( \vec{k} \right) A \left( \omega_2, \vec{k} \right) \right) = \mathcal{S}^{\alpha \alpha} \left( \omega_2, \omega_1 \right), \label{app:eqn:symmetry_of_diagonal_of_S}
\end{align}
which combined with the $\omega_1 \leftrightarrow \omega_2$ exchange symmetry properties from \cref{app:eqn:symmetries_of_S_in_omega} implies that $\mathcal{S}^{\alpha \alpha} \left( \omega_1, \omega_2 \right)$ is real. This property enables us to compute the diagonal parts of the $L_{11}$ and $L_{12}$ matrices in a much simpler form which involves only a single integral over $\omega$
\begin{align}
	L^{\alpha \alpha}_{11} =&  \frac{\pi}{\beta} \int_{-\infty}^{\infty} \dd{\omega} \pdv{n_{\mathrm{F}} \left( \omega \right)}{\omega} \sigma^{\alpha} \left( \omega \right), \label{app:eqn:L11_static_diagonal_final} \\
L^{\alpha \alpha}_{12} =& - \frac{\pi}{\beta} \int_{-\infty}^{\infty} \dd{\omega} \omega \pdv{n_{\mathrm{F}} \left( \omega \right)}{\omega} \sigma^{\alpha} \left( \omega \right), \label{app:eqn:L12_static_diagonal_final}
\end{align}
where we have defined the \emph{real} conductivity function\footnote{While confusing, the term ``conductivity function'' is conventional. One should not confuse $\sigma^{\alpha} \left( \omega \right)$ with the dynamical conductivity tensor, nor relate it to the static conductivity as $\sigma^{\alpha \alpha} = \eval{\sigma^{\alpha} \left( \omega \right)}_{\omega = 0}$. $\sigma^{\alpha} \left( \omega \right)$ is simply the integrand of \cref{app:eqn:L11_static_diagonal_final,app:eqn:L12_static_diagonal_final} and can only be related to the static conductivity via \cref{app:eqn:relation_between_L11_and_sigma,app:eqn:L11_static_diagonal_final}. In the zero temperature limit, one can employ a Sommerfeld expansion, as will be discussed in \cref{app:eqn:sommerfeld_exp} and show that the static conductivity can be \emph{approximated} as $\sigma^{\alpha \alpha} = - \pi e^2 \eval{\sigma^{\alpha} \left( \omega \right)}_{\omega = 0} + \mathcal{O} \left( T^2 \right)$.}, which can be expressed in terms of the spectral function of the system as 
\begin{equation}
	\label{app:eqn:def_conduc_func}
	\sigma^{\alpha} \left( \omega \right) =  \frac{1}{N_0 \Omega_0} \sum_{\vec{k}} \Tr \left( T^{\alpha} \left( \vec{k} \right) A \left( \omega, \vec{k} \right) T^{\alpha} \left( \vec{k} \right) A \left( \omega, \vec{k} \right) \right).
\end{equation}
Because of the $C_{3z}$ rotational symmetry of all the phases of the THF model considered in this work, $\sigma^{\alpha} \left( \omega \right)$ is independent on the direction along which we consider the Seebeck coefficient. Once the conductivity function for a given phase is computed, the Seebeck coefficient can be determined by substituting \cref{app:eqn:L11_static_diagonal_final,app:eqn:L12_static_diagonal_final} into its definition from \cref{app:eqn:seebeck_definition}. 

\subsection{Numerical computation of the Seebeck coefficient} \label{app:sec:thermoelectric_response:details_on_numerics}

\begin{table}[t]
	\centering
	\begin{tabular}{|l|l|l|l|}
		\hline
		Type & Interactions included & Computation of $A \left( \omega, \vec{k} \right)$ & Seebeck coefficient results \\
		\hline
		Non-interacting & No & \cref{app:sec:thermoelectric_response:details_on_numerics:non_interactions} & \cref{app:sec:seebeck_non_int} \\
		\hline
		Symmetry-broken states & Yes & \cref*{DMFT:app:sec:se_correction_beyond_HF} of Ref.~\cite{CAL23b} & \cref{app:sec:seebeck_sym_br} \\
		\hline
		Symmetric phase & Yes & \cref*{DMFT:app:sec:se_symmetric} of Ref.~\cite{CAL23b} & \cref{app:sec:seebeck_sym} \\		
		\hline
	\end{tabular}
	\caption{The three different phases of the THF model for which we compute the Seebeck coefficient in this work. For each phase of the THF model, we list whether interactions are considered, where the spectral function $A \left( \omega, \vec{k} \right)$ was obtained, and the \siSection{} where the corresponding results are shown.}
	\label{app:table:types_of_seebecks}
\end{table}

As shown in \cref{app:eqn:L11_static_diagonal_final,app:eqn:L12_static_diagonal_final}, the main hurdle in computing the Seebeck coefficient is evaluating the conductivity function $\sigma^{\alpha} \left( \omega \right)$, which is defined as a sum over all the momenta in the moir\'e BZ, for different values of $\omega$. In this section, we briefly summarize the procedure we employed to obtain it numerically. 

Throughout this work, we consider the Seebeck coefficient for three different phases of the THF model, which are summarized in \cref{app:table:types_of_seebecks}. In the interacting case, we consider the Hartree-Fock corrections arising from all terms of the interaction Hamiltonian. Additionally, we also include the dynamical correlation effects stemming from the onsite Hubbard interaction of the $f$-electrons $H_{U_1}$ (in both the symmetric and symmetry-broken phases), as well as the dynamical correlation effects arising from the nearest neighbor $f$-electron interaction $H_{U_2}$ (in the symmetry-broken case). In the non-interacting case, we drop all interaction terms. The procedure we employ to obtain the Seebeck coefficient is different depending on whether interactions are included or not. In what follows we will discuss both of these cases individually.

\subsubsection{Numerical computation of the Seebeck coefficient without interactions} \label{app:sec:thermoelectric_response:details_on_numerics:non_interactions}

To obtain a better understanding of the Seebeck coefficient in the symmetric state of the THF model, it is useful to also obtain it within the \emph{non-interacting} THF model. In the absence of interactions, the lifetimes of \emph{both} the $f$- and $c$-electrons (together with the $d$-electrons in the case of TSTG) need to be introduced by hand (unlike the interacting case, which will be discussed in \cref{app:sec:thermoelectric_response:details_on_numerics:interactions} and where only the broadening of the $c$- and $d$-electrons is introduced by hand). The means that the expression of the non-interacting Green's function from \cref{app:eqn:gf_within_non_int} is modified to 
\begin{equation}
	\label{app:eqn:gf_non_int_for_Seebeck}
	\mathcal{G}^{0} \left( \omega + i 0^{+}, \vec{k} \right) = \left[\left( \omega + \mu \right) \mathbb{1} - h \left( \vec{k} \right) + i \Gamma \right]^{-1},
\end{equation}
where $h \left( \vec{k} \right)$ is the non-interacting THF Hamiltonian matrix from \cref{app:eqn:full_THF_Hamiltonian_gamma_bas}, and the lifetime matrix is defined as 
\begin{equation}
	\label{app:eqn:broadening_in_noninteracting}
	\Gamma_{i \eta s;i' \eta' s'} = \begin{cases}
		\Gamma_f \delta_{i i'} \delta_{\eta \eta'} \delta_{s s'}, & \qq{if} 5 \leq i,i' \leq 6 \\
		\Gamma_c \delta_{i i'} \delta_{\eta \eta'} \delta_{s s'}, & \qq{otherwise} \\
	\end{cases},
\end{equation}
with the $c$- and $f$-electron broadening factors being given by $\Gamma_c$ and $\Gamma_f$ (for TSTG, the $d$-electrons have the same broadening factor as the $c$-electrons). As in the interacting simulations from \cref*{DMFT:app:sec:se_correction_beyond_HF,DMFT:app:sec:se_symmetric} of Ref.~\cite{CAL23b}, we employ $\Gamma_c = \SI{1}{\milli\electronvolt}$ for TBG and $\Gamma_c = \SI{1.5}{\milli\electronvolt}$ for TSTG\footnote{The fact that $\Gamma^{\text{TSTG}}_{c}/ \Gamma^{\text{TBG}}_{c} =  3/2  \approx{\sqrt{2}}$ reflects the scaling of the TSTG band structure by a factor of approximately $\sqrt{2}$ relative to the TBG one~\cite{YU23a}.}, while $\Gamma_f$ is varied as will be shown in \cref{app:sec:seebeck_non_int}. Note that the small imaginary offset in \cref{app:eqn:gf_non_int_for_Seebeck} can be discarded as a non-vanishing imaginary offset is already introduced by the broadening matrix. Finally, we also mention that because we introduce a cutoff for the $c$- and/or $d$-electrons' momenta, as explained in the beginning of \cref{app:sec:many_body_rev:generic_not}, the spectrum of $h \left( \vec{k} \right)$ is bounded. The exact value of the cutoff is irrelevant for the Seebeck coefficient around charge neutrality ({\it i.e.}{} for fillings $\abs{\nu} \lesssim 4$), for which the chemical potential lies far from the spectrum bounds of $h \left( \vec{k} \right)$.

From \cref{app:eqn:gf_non_int_for_Seebeck}, it is easy to see that the chemical potential $\mu$ can be absorbed by a shift in the frequency $\omega$. To see this explicitly, we introduce the non-interacting Green's function at charge neutrality
\begin{equation}
	\label{app:eqn:gf_non_int_for_Seebeck_CN}
	\mathcal{G}_{0}^{0} \left( \omega + i 0^{+}, \vec{k} \right) = \left( \omega \mathbb{1} - h \left( \vec{k} \right) + i \Gamma \right)^{-1},
\end{equation}
in terms of which the non-interacting Green's function at chemical potential $\mu$ is given by 
\begin{equation}
	\mathcal{G}^{0} \left( \omega + i 0^{+}, \vec{k} \right) = \mathcal{G}_{0}^{0} \left( \omega + \mu + i 0^{+}, \vec{k} \right). 
\end{equation}
Note that because $h \left( \vec{k} \right)$ is block-diagonal containing $4 = 2\text{(spin)} \times 2\text{(valley)}$ six-dimensional matrices (eight-dimensional in the case of TSTG), the charge-neutral Green's function from \cref{app:eqn:gf_non_int_for_Seebeck_CN} can be computed explicitly as a function of $\vec{k}$ and $\omega$. The resulting analytical expression is too complicated to reproduce here, but from a computational standpoint, it is much faster to evaluate than the numerical inversion of $\omega \mathbb{1} - h \left( \vec{k} \right) + i \Gamma$. 

From its definition in \cref{app:eqn:spectral_function}, we can obtain the non-interacting spectral function of the system $A^{0} \left( \omega, \vec{k} \right)$, which, once again, can be expressed in terms of the charge neutral spectral function $A^{0}_{0} \left( \omega, \vec{k} \right)$,
\begin{align}
	A^{0}_{0,i \eta s;i' \eta' s'} \left(\omega, \vec{k} \right) =& \frac{-1}{2 \pi i} \left[\mathcal{G}^{0}_{0,i \eta s; i' \eta' s'} \left(\omega + i 0^{+}, \vec{k} \right) - 
	\left(\mathcal{G}^{0}_{0,i' \eta' s'; i \eta s} \left(\omega + i 0^{+}, \vec{k} \right)\right)^* \right], \\
	A^{0} \left(\omega, \vec{k} \right) =& A^{0}_{0} \left( \omega + \mu, \vec{k} \right). \label{app:eqn:spectral_function_seebeck_shift}
\end{align}
We note that, for multi-flavor (or multi-orbital) systems, where the Green's function is a matrix, the spectral function should be calculated via \cref{app:eqn:spectral_function_seebeck_shift}, instead of directly taking the imaginary part of the Green's function. In turn, \cref{app:eqn:spectral_function_seebeck_shift} implies that the conductivity function at arbitrary chemical potential is related to the one at charge neutrality by a shift in frequency
\begin{align}
	\sigma^{\alpha}_{0} \left( \omega \right) &= \frac{1}{N_0 \Omega_0} \sum_{\vec{k}} \Tr \left( T^{\alpha} \left( \vec{k} \right) A^{0}_{0} \left( \omega, \vec{k} \right) T^{\alpha} \left( \vec{k} \right) A^{0}_{0} \left( \omega, \vec{k} \right) \right), \label{app:eqn:def_non_int_conductivity}\\
	\sigma^{\alpha} \left( \omega \right) &= \sigma^{\alpha}_{0} \left( \omega + \mu \right).
\end{align}
In the numerical evaluation of the conductivity function, we use a relatively dense mesh of $144 \times 144$ $\vec{k}$-points and $2^{12}$ $\omega$-points. Such a dense momentum-space discretization is necessary to ensure a smooth result for the conductivity function. Additionally, the dense discretization in frequency space ensures that all the features of $\sigma^{\alpha} \left( \omega \right)$ (which changes relatively quickly on frequency scale comparable to the bandwidth of the TBG or TSTG active bands) are correctly captured. 

Once the conductivity function is computed, one can obtain the thermoelectric linear response coefficients and electron filling as a function of the chemical potential $\mu$ using
\begin{align}
	L^{\alpha \alpha}_{11} (\mu) =&  \frac{\pi}{\beta} \int_{-\infty}^{\infty} \dd{\omega} \pdv{n_{\mathrm{F}} \left( \omega - \mu \right)}{\omega} \sigma_{0}^{\alpha} \left( \omega \right), \label{app:eqn:L11_static_diagonal_nonint} \\
L^{\alpha \alpha}_{12} (\mu) =& - \frac{\pi}{\beta} \int_{-\infty}^{\infty} \dd{\omega} \left( \omega - \mu \right) \pdv{n_{\mathrm{F}} \left( \omega - \mu \right)}{\omega} \sigma_{0}^{\alpha} \left( \omega \right), \label{app:eqn:L12_static_diagonal_nonint} \\
	\nu \left( \mu \right) =& \int_{-\infty}^{\infty} \dd{\omega} \left( n_{\mathrm{F}} \left( \omega - \mu \right) - \frac{1}{2} \right) \sum_{\vec{k}} \Tr \left( A^{0}_{0} \left( \omega, \vec{k} \right) \right), \label{app:eqn:nu_nonint}
\end{align}
from which the Seebeck coefficient as a function of $\nu$ is obtained using \cref{app:eqn:seebeck_definition}. Note that the conductivity function only needs to be computed once: the dependence on the chemical potential (filling) and temperature only comes into play through the Fermi-Dirac distribution when the thermoelectric linear response coefficients are computed. In contrast, in the interacting case, which will be discussed in \cref{app:sec:thermoelectric_response:details_on_numerics:interactions}, the conductivity function \emph{itself} depends on the chemical potential and the temperature, making the computation of the Seebeck coefficient much more numerically intensive.

\subsubsection{Numerical computation of the Seebeck coefficient with interactions} \label{app:sec:thermoelectric_response:details_on_numerics:interactions}

For each interacting phase of the THF model (at a given temperature $T$ and filling $\nu$), a self-consistent density matrix $\varrho \left( \vec{k} \right)$ and a $\vec{k}$-independent self-consistent dynamical self-energy $\Sigma \left( \omega + i 0^{+} \right)$ are obtained through the methods outlined in \cref*{DMFT:app:sec:se_correction_beyond_HF,DMFT:app:sec:se_symmetric} of Ref.~\cite{CAL23b}, depending on whether or not the corresponding phase breaks any symmetries of the model. The Seebeck coefficient is then obtained using the following algorithmic prescription:
\begin{enumerate}
	\item From the density matrix $\varrho \left( \vec{k} \right)$, which is computed on a $48 \times 48$ $\vec{k}$-point mesh, the corresponding Hartree-Fock Hamiltonian $h^{\text{MF}} \left( \vec{k} \right)$ is obtained through \cref{app:eqn:genera_TBG_int_HF,app:eqn:HF_Hamiltonian} for a $144 \times 144$ $\vec{k}$-point mesh, for TBG. For TSTG, we use \cref*{DMFT:app:eqn:TSTG_HF_Hamiltonian} of Ref.~\cite{CAL23b} and \cref{app:eqn:HF_Hamiltonian}
	
	\item We then compute the Green's function of the system via \cref{app:eqn:general_gf_self_energy}, which, upon analytical continuation, becomes
	\begin{equation}
		\label{app:eqn:GF_for_interacting_seebeck}
		\mathcal{G} \left( \omega + i 0^{+}, \vec{k} \right) = \left[\left( \omega + i 0^{+} + \mu \right) \mathbb{1} - h^{\text{MF}} \left( \vec{k} \right) - \Sigma \left( \omega + i 0^{+} \right) \right]^{-1},
	\end{equation}
	which is therefore defined on a relatively dense mesh of $144 \times 144$ $\vec{k}$-points and $2^{12}$ $\omega$-points ({\it i.e.}{}, the same frequency discretization as the self-energy). As in \cref{app:sec:thermoelectric_response:details_on_numerics:non_interactions}, we employ a dense momentum-space mesh in order to obtain a smooth result for the conductivity function $\sigma^{\alpha} \left( \omega \right)$. The range of $\omega$ that is sampled (which matches the range of $\omega$ for which the dynamical self-energy $\Sigma \left( \omega \right)$ is computed) is chosen to be approximately twice as large as the bandwidth of the Hartree-Fock Hammiltonian $h^{\text{MF}} \left( \vec{k} \right)$, thus ensuring that the tail of the spectral function of the system is correctly captured.
	
	\item The spectral function $A \left( \omega, \vec{k}  \right)$ can then be obtained from the retarded Green's function via \cref{app:eqn:spectral_function}. 
	
	\item Finally, from the spectral function $A \left( \omega, \vec{k}  \right)$, we employ \cref{app:eqn:def_conduc_func} to obtain the conductivity function. Once $\sigma^{x} \left( \omega \right)$ is obtained, we compute the $L^{xx}_{11}$ and $L^{xx}_{12}$ coefficients using \cref{app:eqn:L11_static_diagonal_final,app:eqn:L12_static_diagonal_final}, which then allow us to compute Seebeck coefficient using \cref{app:eqn:seebeck_definition}.
\end{enumerate}

As seen above, computing the conductivity function \cref{app:eqn:def_conduc_func} requires inverting $144 \times 144 \times 2^{12} \approx 8.5 \times 10^7$ small ($24 \times 24$ for TBG or $32 \times 32$ for TSTG) matrices, and three times as many matrix-matrix multiplications. Moreover, unlike the non-interacting case from \cref{app:sec:thermoelectric_response:details_on_numerics:non_interactions}, the conductivity function needs to be computed for \emph{each} filling $\nu$, temperature $T$, and type of correlated phase. In order to speed up the computation, we implement a parallel version of the algorithm on an NVIDIA Ampere GPU.

\section{Analytic expressions of the Seebeck coefficient }\label{app:sec:asymptotes}

In this \siSection{}, we provide an analytical intuition on the Seebeck coefficient. To do so, we start by introducing a family of quadratic band Hamiltonians, featuring non-hybridized bands with different effective masses and lifetimes. For these models, we compute the thermoelectric response coefficients and show that they can be expressed as integral of elementary functions convolved with the derivative of the Fermi-Dirac distribution function. We then obtain analytical asymptotic expressions of the thermoelectric coefficients in the limit of large electron lifetime and in the low-temperature limit. Working in the experimentally-relevant~\cite{MER24} low-temperature limit, we consider a simple heavy-light model featuring a quadratic band of heavy holes and a quadratic band of light electrons, akin to the charge-one excitation spectrum of the TBG correlated insulators (see for example \cref*{DMFT:app:sec:results_corr_ins} of Ref.~\cite{CAL23b} or Refs.~\cite{BER21b,KAN21,SON22}). Using the low-temperature asymptotic expressions of the thermoelectric coefficients, we discuss the effects of mass and lifetime asymmetry on the Seebeck coefficient. Finally, we investigate the effects of wave function topology on the Seebeck coefficient and show that unlike the idealized non-interacting flat-band limit~\cite{KRU23}, the quantum-geometric contribution to the thermoelectric effect are negligible in realistic models of TBG \emph{with} interactions. For the latter, the charge-one excitation dispersion is non-negligible, and the kinetic contribution to the thermoelectric obscure the ones arising from quantum geometry.

\subsection{Quadratic band Hamiltonians}\label{app:sec:asymptotes:model}

\subsubsection{Model}\label{app:sec:asymptotes:model:defitinion}

For the purposes of building an intuition on the dependence of the Seebeck coefficient on the energetics of band structures, in this section we consider simple quadratic band Hamiltonians in $d=2$ dimensions
\begin{equation}
	\label{app:eqn:simple_ham_op}
	K = \sum_{\abs{\vec{k}} \leq \Lambda} \sum_{1 \leq n,m \leq N_b} \left( h_{nm} \left( \vec{k} \right) - \mu \delta_{nm} \right) \hat{a}^\dagger_{\vec{k},n} \hat{a}_{\vec{k},m} ,
\end{equation}
in which $\mu$ is the chemical potential and the Hamiltonian matrix is given by 
\begin{equation}
	\label{app:eqn:simple_ham_matrix}
	h_{nm} \left( \vec{k} \right) = \left( \frac{\abs{\vec{k}}^2}{2 m_n} + \Delta_{n} \right) \delta_{nm}.
\end{equation}
The grand canonical Hamiltonian from \cref{app:eqn:simple_ham_op} corresponds to $N_b$ non-hybridized quadratic bands with different fermion effective masses $m_n$ (which can be either positive or negative), having different offsets $\Delta_{n}$ from zero energy (with $1 \leq n \leq N_b$). In \cref{app:eqn:simple_ham_op}, we have also introduced a momentum cutoff $\Lambda$ equal to the extent of the BZ. 

We take the Green's function of the system to be
\begin{equation}
	\label{app:eqn:quadratic_gf}
	-\left\langle \mathcal{T}_{\tau} \hat{a}_{\vec{k},n} \left( \tau \right) \hat{a}^\dagger_{\vec{k}',m} \left( 0 \right)  \right\rangle  = \delta_{\vec{k},\vec{k}'} \delta_{nm} \mathcal{G}_{n} \left(\tau, \vec{k} \right),
\end{equation}
where the imaginary time-evolution is performed with respect to the grand canonical Hamiltonian $K$. The Green's function's expression in Matsubara frequency is given by
\begin{equation}
	\label{app:eqn:quadratic_gf_mats}
	\mathcal{G}_{n} \left( i \omega_m, \vec{k} \right) = \frac{1}{i \omega_m + \mu - \frac{\abs{\vec{k}}^2}{2 m_n} - \Delta_{n} + i \Gamma_n }.
\end{equation}
In \cref{app:eqn:quadratic_gf_mats}, we have introduced a phenomenological band-dependent broadening term $+i \Gamma_n$, where $\Gamma_n$ is positive and inversely proportional to the lifetime of the $\hat{a}^\dagger_{\vec{k},n}$ fermions. The simple quadratic band Hamiltonian can be thought of as the $\vec{k} \cdot \vec{p}$ expansion of the Hartree-Fock Hamiltonian of the THF model near integer fillings. The broadening terms can therefore be understood as arising from interactions beyond the Hartree-Fock level. The corresponding spectral function is a Lorentzian given by 
\begin{equation}
	\label{app:eqn:quadratic_sp_func}
	A_{n} \left( \omega, \vec{k} \right) = - \frac{1}{\pi} \Im \left[\mathcal{G}_{n} \left( \omega + i 0^{+}, \vec{k} \right) \right] =  \frac{1}{\pi} \frac{\Gamma_n}{\left( \omega + \mu- \frac{\abs{\vec{k}}^2}{2 m_n} - \Delta_n  \right)^2 + \Gamma_n^2}.
\end{equation}

\subsubsection{Thermoelectric response}\label{app:sec:asymptotes:model:thermo}

Given that the Hamiltonian from \cref{app:eqn:simple_ham_op} is non-interacting\footnote{Alternatively, it can be considered interacting, but with a localized interaction term that leads to no contribution to the energy current, as is the case with the THF model.}, the current operators can be found from \cref{app:eqn:part_current_simple_real_general,app:eqn:en_current_simple_real_general_sin,app:eqn:generic_thermal_current} in analogy with \cref{app:eqn:generic_particle_current,app:eqn:mats_thermal_current},
\begin{align}
	J^{\alpha} \left( -i \tau \right) &= \sum_{\abs{\vec{k}} \leq \Lambda} \sum_{1 \leq n,m \leq N_b} T^{\alpha}_{nm} \left( \vec{k} \right) \hat{a}^\dagger_{\vec{k},n} (\tau) \hat{a}_{\vec{k},m} (\tau), \label{app:eqn:quad_particle_current} \\
	J_Q^{\alpha} \left( -i \tau \right) &= \frac{1}{2} \sum_{\abs{\vec{k}} \leq \Lambda} \sum_{1 \leq n,m \leq N_b} T^{\alpha}_{nm} \left( \vec{k} \right) \left[ \left( \pdv{\tau} \hat{a}^\dagger_{\vec{k},n} (\tau) \right) \hat{a}_{\vec{k},m} (\tau) - \hat{a}^\dagger_{\vec{k},n} (\tau) \pdv{\tau} \hat{a}_{\vec{k},m} (\tau)  \right], \label{app:eqn:quad_thermal_current}
\end{align}
where the matrix $T^{\alpha}_{n m} \left( \vec{k} \right)$ has a simple expression, 
\begin{equation}
	\label{app:eqn:cur_mat_elems_quadratic}
	T^{\alpha}_{nm} \left( \vec{k} \right) = \pdv{h_{nm} \left( \vec{k} \right)}{k^{\alpha}} = \frac{k^{\alpha}}{m_n} \delta_{nm},
\end{equation} 
with no off-diagonal terms. 

Because the current operators for the quadratic band Hamiltonian in \cref{app:eqn:simple_ham_op} have similar forms to the ones of the THF model from \cref{app:eqn:generic_particle_current,app:eqn:mats_thermal_current}, we can immediately conclude that diagonal parts of the thermal linear coefficients $L_{11}$ and $L_{12}$ are given by \cref{app:eqn:L11_static_diagonal_final,app:eqn:L12_static_diagonal_final}, with the conductivity function being 
\begin{equation}
	\label{app:eqn:cond_func_quadratic}
	\sigma^{\alpha} \left( \omega \right) = \sum_{n=1}^{N_b} \sigma_n^{\alpha} \left( \omega \right),
\end{equation}
where we have introduced the individual band contributions to the conductivity function
\begin{equation}
	\label{app:eqn:cond_function_nth_band_def}
	\sigma_n^{\alpha} \left( \omega \right) \equiv \frac{1}{N_0 \Omega_0} \sum_{\abs{\vec{k}} \leq \Lambda} T^{\alpha}_{nn} \left( \vec{k} \right) A_{n} \left( \omega, \vec{k} \right) T^{\alpha}_{nn} \left( \vec{k} \right) A_{n} \left( \omega, \vec{k} \right), \qq{for} 1 \leq n \leq N_b.
\end{equation}
In deriving \cref{app:eqn:cond_function_nth_band_def}, we have used the fact that both the spectral function of this simple model from \cref{app:eqn:quadratic_sp_func}, and the current matrix elements from \cref{app:eqn:cur_mat_elems_quadratic} are diagonal in the band index. As a result we find that the diagonal parts of the $L_{11}$ and $L_{12}$ coefficients can also be expressed as a sum of independent contributions from each band
\begin{equation}
	L_{11}^{\alpha \alpha} = \sum_{n=1}^{N_b} L_{11,n}^{\alpha \alpha} \qq{and} 
	L_{12}^{\alpha \alpha} = \sum_{n=1}^{N_b} L_{12,n}^{\alpha \alpha}, 
\end{equation}
with the individual band contributions to the thermal response coefficients being given by
\begin{align}
	L^{\alpha \alpha}_{11,n} =&  \frac{\pi}{\beta} \int_{-\infty}^{\infty} \dd{\omega} \pdv{n_{\mathrm{F}} \left( \omega \right)}{\omega} \sigma_n^{\alpha} \left( \omega \right), \label{app:eqn:L11_static_band_final} \\
L^{\alpha \alpha}_{12,n} =& - \frac{\pi}{\beta} \int_{-\infty}^{\infty} \dd{\omega} \omega \pdv{n_{\mathrm{F}} \left( \omega \right)}{\omega} \sigma_n^{\alpha} \left( \omega \right). \label{app:eqn:L12_static_band_final}
\end{align}
The thermal linear response coefficients stemming from the $n$-th band are written in terms of the $n$-th band conductivity function, which can be evaluated analytically from \cref{app:eqn:cond_function_nth_band_def}
\begin{align}
	 \sigma_n^{\alpha} \left( \omega \right) &= \frac{1}{N_0 \Omega_0} \frac{N_0 \Omega_0}{(2 \pi)^2} \int_{\abs{\vec{k}} \leq \Lambda} \dd[2]{k} \left(\frac{k^{\alpha}}{m_n}\right)^2 \left[ \frac{\Gamma_n}{\left( \omega + \mu- \frac{\abs{\vec{k}}^2}{2 m_n} - \Delta_n  \right)^2 + \Gamma_n^2} \right]^2 \nonumber \\
	 &= \frac{1}{4 \pi^4} \frac{1}{2} \int_{\abs{\vec{k}} \leq \Lambda} \dd[2]{k} \left(\frac{\abs{\vec{k}}}{m_n}\right)^2 \left[ \frac{\Gamma_n}{\left( \omega + \mu- \frac{\abs{\vec{k}}^2}{2 m_n} - \Delta_n  \right)^2 + \Gamma_n^2} \right]^2 \nonumber \\
	 &= \frac{1}{4 \pi^3} \int_{0}^{\Lambda} k \dd{k} \left(\frac{k}{m_n}\right)^2 \left[ \frac{\Gamma_n}{\left( \omega + \mu- \frac{k^2}{2 m_n} - \Delta_n  \right)^2 + \Gamma_n^2} \right]^2 \nonumber \\
	 &= \frac{1}{4 \pi^3} \int_{0}^{\frac{\Lambda^2}{2 m_n}} m_n \dd{u} \left(\frac{2 u}{m_n} \right) \left[ \frac{\Gamma_n}{\left( \omega + \mu- u - \Delta_n  \right)^2 + \Gamma_n^2} \right]^2 \nonumber \\
	 &= \frac{1}{2 \pi^3} \int_{0}^{\frac{\Lambda^2}{2 m_n}} \frac{u \dd{u}}{\Gamma_n^2} \left[ \frac{1}{\left(\frac{ \omega + \mu- u - \Delta_n}{\Gamma_n}  \right)^2 + 1 } \right]^2 \nonumber \\
	 &= \frac{1}{2 \pi^3} f \left(\frac{\Lambda^2}{2 \Gamma_n m_n} , \frac{\omega + \mu - \Delta_n}{\Gamma_n} \right) \label{app:eqn:cond_function_nth_band}.
\end{align}
In \cref{app:eqn:cond_function_nth_band}, we have introduced the following analytical function
\begin{equation}
	\label{app:eqn:simple_model_f}
	f \left(x, y\right) = \int_{0}^{x} \frac{w \dd{w}}{\left[ 1 + \left( w - y \right)^2 \right]^2} = \frac{1}{2} \left[\frac{x(x-y)}{(x-y)^2+1}+ y \left(\arctan(x-y)+\arctan{y} \right)\right].
\end{equation}

{\renewcommand{\arraystretch}{1.5}
	\begin{table}[t]
		\centering
		\begin{tabular}{|p{2.5cm}|l|p{2.5cm}|l|p{5cm}|}
			\hline
			Main approximation & Formulae & Additional simplifications & Formulae & Validity of the additional simplification \\
			\hline
			\multirow{2}{*}{$\Gamma_n \to 0$} & \multirow{2}{*}{\cref{app:eqn:correlator_L11_gamma_approx,app:eqn:correlator_L12_gamma_approx,app:eqn:fill_gamma_approx}} & $\beta \to \infty$ & \multirow{1}{*}{\cref{app:eqn:correlator_L11_gamma_approx_hb,app:eqn:correlator_L12_gamma_approx_hb,app:eqn:fill_gamma_approx_hb}} & $\Gamma_n \ll \frac{1}{\beta}$, but $\frac{1}{\beta}$ is much smaller than the bandwidths and energy gaps of the system, and $\mu$ is more than $\mathcal{O} \left( \frac{1}{\beta} \right)$ from the band edge. \\
			\cline{3-5}
			& & $\beta \to 0$ & \multirow{1}{*}{\cref{app:eqn:correlator_L11_quad_band_gamma_approx_lb,app:eqn:correlator_L12_quad_band_gamma_approx_lb,app:eqn:correlator_fill_quad_band_gamma_approx_lb}} & $\frac{1}{\beta}$ is the largest energy scale of the system, while $\Gamma_n$ is the smallest one. \\
			\hline
\multirow{2}{*}{$\beta \to \infty$} & \multirow{2}{*}{\cref{app:eqn:correlator_L11_quad_band_high_beta,app:eqn:correlator_L12_quad_band_high_beta,app:eqn:fill_quad_band_high_beta}} & $\Gamma_n \to \infty$ & \multirow{1}{*}{\cref{app:eqn:correlator_L11_quad_band_high_beta_lg,app:eqn:correlator_L12_quad_band_high_beta_lg}} & $\frac{1}{\beta}$ is the smallest energy scale of the system, $\Gamma_n \gg \kappa_n$, and the chemical potential is within $\Gamma_n$ of the $n$-th band. \\
			\cline{3-5}
			& & $\Gamma_n \to 0$ & \multirow{1}{*}{\cref{app:eqn:correlator_L11_quad_band_high_beta_hg,app:eqn:correlator_L12_quad_band_high_beta_hg}} & $\frac{1}{\beta} \ll \Gamma_n$, but $\Gamma_n$ is much smaller than the bandwidths and energy gaps of the system, and $\mu$ is more than $\mathcal{O} \left( \Gamma_n \right)$ from the band edge. \\
			\hline
		\end{tabular}
		\caption{Summary of analytical approximations for thermoelectric linear response coefficients obtained for the quadratic Hamiltonian from \cref{app:eqn:simple_ham_op}. We consider two main approximations which correspond respectively to the zero-lifetime ($\Gamma_n \to 0$) and zero-temperature ($\beta \to \infty$) limits. Each of these approximations is further simplified under two different additional assumptions. For every additional simplification, the conditions of validity are listed in the last column of the table. In particular, the order in which the $\Gamma_n \to 0$ and $\beta \to \infty$ limits are taken leads to different results, which are valid under different regimes.}
		\label{app:tab:two_band_summary_approx}
	\end{table}
}

Defining the (signed) bandwidth of the $n$-th quadratic band to be
\begin{equation}
	\label{app:eqn:definition_kapp}
	\kappa_{n} = \frac{\Lambda^2}{2 m_n},
\end{equation}
which can be positive or negative depending on the sign of the mass $m_n$, the $n$-th band contribution to the thermal linear response coefficients is given by 
\begin{align}
	L_{11,n} &= \frac{1}{2 \pi^2 \beta} \int_{-\infty}^{\infty} \dd{\omega} \pdv{n_{\mathrm{F}} \left( \omega \right)}{\omega} f \left(\frac{\kappa_n}{\Gamma_n} , \frac{\omega + \mu_n}{\Gamma_n} \right), \label{app:eqn:correlator_L11_quad_band}\\
	L_{12,n} &= - \frac{1}{2 \pi^2 \beta} \int_{-\infty}^{\infty} \dd{\omega} \omega \pdv{n_{\mathrm{F}} \left( \omega \right)}{\omega} f \left(\frac{\kappa_n}{\Gamma_n} , \frac{\omega + \mu_n}{\Gamma_n} \right), \label{app:eqn:correlator_L12_quad_band}
\end{align} 
where for simplicity we have dropped the component superscripts $^{\alpha \alpha}$, since the diagonal part of the conductivity tensors is independent of direction for the Hamiltonian from \cref{app:eqn:simple_ham_op}. Additionally, for brevity of notation, we have defined 
\begin{equation}
	\mu_n \equiv \mu - \Delta_n, \qq{for}  1 \leq n \leq N_b. 
\end{equation} 

The thermoelectric coefficients in \cref{app:eqn:correlator_L11_quad_band,app:eqn:correlator_L12_quad_band} are given in terms of the chemical potential $\mu$. The experimental measurements are often performed as a function of the total filling of the system 
\begin{equation}
	N^e = \sum_{n=1}^{N_b} N^e_n,
\end{equation}
such that $N^e = 0$ ($N^e = N_b$) corresponds to the fully-filled (fully empty) case and where we have introduced the filling of the $n$-th band to be 
\begin{align}
	N^e_n &= \int_{-\infty}^{\infty} \dd{\omega} n_{\mathrm{F}} \left( \omega \right) \left( \frac{\pi \Lambda^2 N_0 \Omega_0}{(2 \pi)^2} \right)^{-1} \sum_{\abs{\vec{k}} \leq \Lambda} A_n \left( \omega, \vec{k} \right) \nonumber \\
	&= \int_{-\infty}^{\infty} \dd{\omega} n_{\mathrm{F}} \left( \omega \right) \left( \frac{\pi \Lambda^2 N_0 \Omega_0}{(2 \pi)^2} \right)^{-1} \frac{N_0 \Omega_0}{(2 \pi)^2} \int_{\abs{\vec{k}} \leq \Lambda} \dd[2]{k}  \frac{1}{\pi} \frac{\Gamma_n}{\left( \omega + \mu_n - \frac{\abs{\vec{k}}^2}{2 m_n} \right)^2 + \Gamma_n^2} \nonumber \\
	&= \frac{1}{\pi} \int_{-\infty}^{\infty} \dd{\omega} n_{\mathrm{F}} \left( \omega \right) \frac{2 \pi}{\pi \Lambda^2} \int_{0}^{\Lambda} k \dd{k}  \frac{\Gamma_n}{\left( \omega + \mu_n - \frac{\abs{\vec{k}}^2}{2 m_n} \right)^2 + \Gamma_n^2}\nonumber \\
	&= \frac{1}{\pi} \int_{-\infty}^{\infty} \dd{\omega} n_{\mathrm{F}} \left( \omega \right) \frac{2}{\Lambda^2} \frac{1}{\Gamma_n} \int_{0}^{\kappa_n} m_n \dd{u}  \frac{1}{\left( \frac{ \omega + \mu_n - u}{\Gamma_n} \right)^2 + 1} \nonumber \\
	&= \frac{1}{\pi}\int_{-\infty}^{\infty} \dd{\omega} n_{\mathrm{F}} \left( \omega \right) \frac{1}{\kappa_n} g \left(\frac{\kappa_n}{\Gamma_n} , \frac{\omega + \mu_n}{\Gamma_n} \right). \label{app:eqn:fill_quad_band}
\end{align}
In \cref{app:eqn:fill_quad_band}, we have defined the following integral, which can be evaluated analytically
\begin{equation}
	\label{app:eqn:simple_model_g}
	g \left(x, y\right) = \int_{0}^{x} \frac{\dd{w}}{1 + \left( w - y \right)^2} = \arctan \left( x - y \right) + \arctan \left( y \right).
\end{equation}

Owing to the presence of the Fermi function, the formulae in \cref{app:eqn:correlator_L11_quad_band,app:eqn:correlator_L12_quad_band,app:eqn:fill_quad_band} cannot be evaluated analytically. Nevertheless, asymptotic expressions can still be derived in the limit of large fermion lifetimes $\Gamma_n \to 0$ or in the low temperature limit $\beta \to \infty$, as we will show in what follows. A summary of all the analytical approximations and their validity conditions is provided in \cref{app:tab:two_band_summary_approx}.

\subsubsection{Asymptotic expressions thermoelectric coefficients \texorpdfstring{$L_{11,n}$}{L11n} and \texorpdfstring{$L_{12,n}$}{L12n} in the \texorpdfstring{$\Gamma_n \to 0$}{long lifetime} limit}\label{app:sec:asymptotes:model:formula_gamma}

To obtain approximations of the thermoelectric response coefficients in the limit $\Gamma_{n} \to 0$, we first note that the functions defined in \cref{app:eqn:simple_model_f,app:eqn:simple_model_g} admit the following asymptotes
\begin{alignat}{3}
	\label{app:eqn:asymptote_f}
	f \left( \frac{x}{\Gamma}, \frac{y}{\Gamma} \right) & &= & \frac{\pi}{4 \Gamma} \left( \frac{y \abs{x-y}}{x-y}+ \abs{y} \right) + \mathcal{O}\left(\Gamma^2\right) &= & \begin{cases}
		-\frac{\pi y}{2 \Gamma} + \mathcal{O}\left(\Gamma ^2\right) & x < y < 0 \\
		\frac{\pi y}{2 \Gamma}  + \mathcal{O}\left(\Gamma ^2\right) & x \geq y \geq 0 \\
		\mathcal{O}\left(\Gamma ^2\right) & \text{otherwise}
	\end{cases}, \\
	\label{app:eqn:asymptote_g}
	g \left( \frac{x}{\Gamma}, \frac{y}{\Gamma} \right) & &= & \frac{\pi}{2} \left( \frac{\abs{x-y}}{x-y}+ \frac{\abs{y}}{y} \right) + \mathcal{O}\left(\Gamma\right) &= & \begin{cases}
		- \pi + \mathcal{O}\left(\Gamma \right) & x < y < 0 \\
		\pi  + \mathcal{O}\left(\Gamma \right) & x \geq y \geq 0 \\
		\mathcal{O}\left(\Gamma \right) & \text{otherwise}
	\end{cases},
\end{alignat}
where the leading order contribution only depends on $x$ implicitly, via the piece-wise conditions. In turn, \cref{app:eqn:asymptote_f,app:eqn:asymptote_g} allow us to evaluate the integrals in \cref{app:eqn:correlator_L11_quad_band,app:eqn:correlator_L12_quad_band} analytically up to order $\mathcal{O} \left(\Gamma_{n} ^2\right)$, as well as the integral in \cref{app:eqn:fill_quad_band} analytically up to linear order in $\Gamma_n$
\begin{align}
	L_{11,n} &= \frac{1}{4 \pi \beta^2 \Gamma_{n}} \left[
	\frac{\beta \kappa _n e^{\beta  \mu_n }}{e^{\beta  \mu_n }+e^{\beta  \kappa _n}}
	- \log \left( 1 + e^{\beta  \mu_n} \right) 
	+ \log \left(1 + e^{\beta  \left(\mu_n - \kappa _n\right)}\right) \right] + \mathcal{O}\left(\Gamma_{n} ^2\right), \label{app:eqn:correlator_L11_gamma_approx} \\
	L_{12,n} &= \frac{1}{4 \pi \beta^2 \Gamma_{n}} \left[
	\frac{\beta  \kappa _n e^{\beta  \kappa _n} \left(\kappa _n - \mu_n \right)}{e^{\beta  \mu_n }+e^{\beta  \kappa_n }}
	+\left(\mu_n - 2 \kappa _n\right) \log \left(1 + e^{-\beta  \left(\mu_n-\kappa _n\right)} \right)
	\right. \nonumber \\
	&\left.
	-\mu_n \log \left(1 + e^{-\beta \mu_n } \right)
	-\frac{2}{\beta} \Li_2\left(-e^{-\beta  \left(\mu_n-\kappa _n\right)}\right) + \frac{2}{\beta} \Li_2\left(-e^{-\beta  \mu_n }\right)\right] + \mathcal{O}\left(\Gamma_{n}^2\right) \label{app:eqn:correlator_L12_gamma_approx}, \\
	N^e_n &= 1 + \frac{1}{\beta \kappa_n} \log \left( \frac{1 + e^{\beta \mu_n}}{e^{\beta \kappa_n} + e^{\beta \mu_n}} \right) + \mathcal{O}\left(\Gamma_{n}\right) = \frac{1}{\beta \kappa_n} \left[ \log \left( 1 + e^{\beta \mu_n} \right) - \log \left( 1 + e^{\beta \left( \mu_n - \kappa_n \right)} \right) \right] + \mathcal{O}\left(\Gamma_{n}\right), \label{app:eqn:fill_gamma_approx}
\end{align} 
where $\Li_s \left( x \right)$ is the polylogarithm function of order $s$. While exact in the limit of vanishing $\Gamma_{n}$, the expressions in \cref{app:eqn:correlator_L11_gamma_approx,app:eqn:correlator_L12_gamma_approx,app:eqn:fill_gamma_approx} are not particularly illuminating. As a result, we will now proceed to simplify them by assuming two different temperature regimes:
\begin{enumerate}
	\item \textbf{The low-temperature limit $\beta \to \infty$}. In this limit, the temperature is much smaller than the bandwidth and energy gaps of the system, but still larger than $\Gamma_n$, as summarized in \cref{app:tab:two_band_summary_approx}. In order to obtain an approximation in this limit, we note that the polylogarithm function admits the following asymptotic expansions
	\begin{equation}
		\label{app:eqn:polylog_approx}
		\begin{split}	
			\Li_{1} \left( - e^{\beta x} \right) &= - \log \left(1 + e^{\beta x} \right)= -\beta x \Theta(x) - e^{-\beta \abs{x}} + \mathcal{O} \left( e^{-2 \beta \abs{x}} \right), \\
			\Li_{2} \left( - e^{\beta x} \right) &= -\left( \frac{\pi^2}{6} + \frac{\beta^2 x^2}{2} \right) \Theta(x) +\sgn(x) e^{-\beta \abs{x}} + \mathcal{O} \left( e^{-2 \beta \abs{x}} \right),
		\end{split}
	\end{equation}
	for $\beta > 0$ and any $x \in \mathbb{R}$. As such, one can obtain a remarkably (exponentially) accurate approximation of $\Li_{s} \left( - e^{\beta x} \right)$ (for $s=1,2$), which is valid for $\abs{x} \gg 1 / \beta$. Inspecting  \cref{app:eqn:correlator_L11_gamma_approx,app:eqn:correlator_L12_gamma_approx,app:eqn:fill_gamma_approx}, $x$ can be either $-\mu_n$ or $-\left( \mu_n - \kappa_n \right)$, and we find that there are only four cases \textup{\uppercase\expandafter{\romannumeral1}}-\textup{\uppercase\expandafter{\romannumeral4}} that we need to consider, depending on the signs of $\mu_n$ and $\mu_n - \kappa_n$, which we denote by 
	\begin{equation}
		\textup{\uppercase\expandafter{\romannumeral1}}: \mu_n<0, \mu_n < \kappa _n,
		\quad 
		\textup{\uppercase\expandafter{\romannumeral2}}: \mu_n<0, \mu_n > \kappa _n,
		\quad
		\textup{\uppercase\expandafter{\romannumeral3}}: \mu_n>0, \mu_n < \kappa _n,
		\quad
		\textup{\uppercase\expandafter{\romannumeral4}}: \mu_n>0, \mu_n > \kappa _n.
	\end{equation}
	In each of the four cases, we will employ \cref{app:eqn:polylog_approx} to find approximations of the thermoelectric coefficients and the individual filling of each band to $\mathcal{O} \left( e^{-2 \beta \abs{\mu_n}}, e^{-2 \beta \abs{\mu_n - \kappa_n}}  \right)$. These approximations will therefore be valid whenever the chemical potential $\mu$ is $\mathcal{O} \left( \frac{1}{\beta} \right)$ from the band edges (which are located at $\Delta_n$ and $\Delta_n + \kappa_n$ and thus correspond to $\mu_n = 0$ or $\mu_n = \kappa_n$, where we remind the reader that $\kappa_n$ can be both positive and negative). We find the following approximations
	\begin{align}
		L_{11,n} &\approx \frac{1}{4 \pi \beta^2 \Gamma_n}\begin{cases}
			\beta  \kappa _n e^{\beta  \left(\mu _n-\kappa _n\right)}-e^{\beta  \mu _n} & \textup{\uppercase\expandafter{\romannumeral1}} \\
			\beta  \kappa _n \left(-e^{\beta  \left(\kappa _n-\mu _n\right)}\right)-e^{\beta  \mu _n}+\beta  \mu _n & \textup{\uppercase\expandafter{\romannumeral2}} \\
			\beta  \kappa _n e^{\beta  \left(\mu _n-\kappa _n\right)}-e^{-\beta  \mu _n}-\beta  \mu _n & \textup{\uppercase\expandafter{\romannumeral3}} \\
			\beta  \kappa _n \left(-e^{\beta  \left(\kappa _n-\mu _n\right)}\right)-e^{-\beta  \mu _n} & \textup{\uppercase\expandafter{\romannumeral4}} \\
		\end{cases}, \label{app:eqn:correlator_L11_gamma_approx_hb}\\
L_{12,n} &\approx \frac{1}{4 \pi \beta^2 \Gamma_n}\begin{cases}
			\beta  \kappa _n \left(\kappa _n-\mu _n\right) e^{\beta  \left(\mu _n-\kappa _n\right)}+\mu _n e^{\beta  \mu _n} & \textup{\uppercase\expandafter{\romannumeral1}} \\
			\frac{3 \beta  \left(\beta  \kappa _n \left(\mu _n-\kappa _n\right) e^{\beta  \left(\kappa _n-\mu _n\right)}+\mu _n e^{\beta  \mu _n}\right)+\pi ^2}{3 \beta } & \textup{\uppercase\expandafter{\romannumeral2}} \\
			-\frac{\pi ^2-3 \beta  \left(\beta  \kappa _n \left(\kappa _n-\mu _n\right) e^{\beta  \left(\mu _n-\kappa _n\right)}+\mu _n e^{-\beta  \mu _n}\right)}{3 \beta } & \textup{\uppercase\expandafter{\romannumeral3}} \\
			\beta  \kappa _n \left(\mu _n-\kappa _n\right) e^{\beta  \left(\kappa _n-\mu _n\right)}+\mu _n e^{-\beta  \mu _n} & \textup{\uppercase\expandafter{\romannumeral4}} \\
		\end{cases}, \label{app:eqn:correlator_L12_gamma_approx_hb}\\
N^e_n &\approx \frac{1}{\beta \kappa_n}\begin{cases}
		e^{\beta  \mu _n}-e^{\beta  \left(\mu _n-\kappa _n\right)} & \textup{\uppercase\expandafter{\romannumeral1}} \\
		-e^{\beta  \left(\kappa _n-\mu _n\right)}+\beta  \kappa _n+e^{\beta  \mu _n}-\beta  \mu _n & \textup{\uppercase\expandafter{\romannumeral2}} \\
		-e^{\beta  \left(\mu _n-\kappa _n\right)}+\beta  \mu _n+e^{-\beta  \mu _n} & \textup{\uppercase\expandafter{\romannumeral3}} \\
		-e^{\beta  \left(\kappa _n-\mu _n\right)}+\beta  \kappa _n+e^{-\beta  \mu _n} & \textup{\uppercase\expandafter{\romannumeral4}} \\
		\end{cases}.\label{app:eqn:fill_gamma_approx_hb}\\
	\end{align}
	In \cref{app:sec:asymptotes:heavylight}, we will be interested in investigating the effects of mass and lifetime asymmetry on the Seebeck coefficient. As the expressions in \cref{app:eqn:correlator_L11_gamma_approx_hb,app:eqn:correlator_L12_gamma_approx_hb,app:eqn:fill_gamma_approx_hb} are only valid in the infinite lifetime limit $\Gamma_n \to 0$, they are unsuitable for this purpose. As a result, they are only mentioned here for completeness and will not be discussed further.
	
	\item \textbf{The high-temperature limit $\beta \to 0$}. The high-temperature limit corresponds to the case where $\frac{1}{\beta}$ is much higher than all the bandwidths and energy gaps of the system \emph{and} $\abs{\mu_n}, \abs{\kappa_n} \ll \frac{1}{\beta}$. This limit is often not relevant for experiments (which for TBG and TSTG are carried out at temperatures of up to tens of kelvin~\cite{PAU22,GHA22,MER24,BAT24}, while the corresponding interacting charge-one excitation bands are characterized by energy scales of tens of millielectronvolts, which correspond to hundreds of kelvin~\cite{CAL23b}), but for the sake of completeness, we will present the results here. The formulae for the two thermoelectric linear response coefficients and the filling of each band in the $\beta \to 0$ limit can be obtained by Taylor-expanding \cref{app:eqn:correlator_L11_gamma_approx,app:eqn:correlator_L12_gamma_approx,app:eqn:fill_gamma_approx} and keeping only the leading contribution
\begin{align}
	L_{11,n} &= -\frac{\kappa_n^2}{32 \pi \Gamma_{n} }+\mathcal{O}\left(\beta^2\right), \label{app:eqn:correlator_L11_quad_band_gamma_approx_lb}\\
	L_{12,n} &= \frac{\left[2 \kappa _n^3 - 3 \kappa_n^2 \mu_n \right]}{96 \pi \Gamma_{n}} + \mathcal{O} \left(\beta ^2\right). \label{app:eqn:correlator_L12_quad_band_gamma_approx_lb}\\
	N^e_n &= \frac{1}{2} + \frac{2 \mu_n - \kappa_n}{8} \beta + \mathcal{O}\left(\beta ^3\right). \label{app:eqn:correlator_fill_quad_band_gamma_approx_lb}
\end{align} 
	The fact that $N^e_n = \frac{1}{2} + \mathcal{O} \left( \beta \right)$ can be understood as follows: at infinite temperature, all grand canonical system configurations have equal probability (all Boltzmann factors are one); as a result all electronic states can be occupied or empty with an equal probability, leading to an effective filling of $\frac{1}{2}$ (and small corrections) for every band.

\end{enumerate}

\subsubsection{Asymptotic expressions thermoelectric coefficients \texorpdfstring{$L_{11,n}$}{L11n} and \texorpdfstring{$L_{12,n}$}{L12n} in the \texorpdfstring{$\beta \to \infty$}{low temperature} limit}\label{app:sec:asymptotes:model:formula_beta}

Another limit in which one can evaluate the thermoelectric response coefficients from \cref{app:eqn:correlator_L11_quad_band,app:eqn:correlator_L12_quad_band}, as well as the individual band filling from \cref{app:eqn:fill_quad_band} is the low temperature limit $\beta \to \infty$ (or rather the limit in which the temperature is the smallest energy scale of the system). As summarized in \cref{app:tab:two_band_summary_approx}, this is different from the approximations considered in \cref{app:eqn:correlator_L11_gamma_approx_hb,app:eqn:correlator_L12_gamma_approx_hb,app:eqn:fill_gamma_approx_hb}, which take $\Gamma_n$ to be the smallest energy scale of the system and \emph{then} take the low temperature approximation (thus effectively working in the case where the temperature is low, but is not the lowest energy scale of the system). To do so, we employ the Sommerfeld expansion~\cite{SOM28}, which states that for a sufficiently well-behaved function $h( \omega )$, we have
\begin{align}
	\int_{-\infty}^{\infty} \dd{\omega} \left(-\dv{n_{\mathrm{F}} \left( \omega \right)}{\omega} \right) h (\omega) &= \left( h(\omega) + \frac{\pi ^2}{6 \beta ^2} \dv[2]{h(\omega)}{\omega} +\frac{7 \pi ^4}{360 \beta ^4} \dv[4]{h(\omega)}{\omega} \right) \biggr\rvert_{\omega = 0} + \mathcal{O} \left( \beta^{-6} \right), \label{app:eqn:sommerfeld_exp} \\
	\int_{-\infty}^{\infty} \dd{\omega} n_{\mathrm{F}} ( \omega ) h (\omega) &= \int_{-\infty}^{0} h(\omega) \dd{\omega} + \left(\frac{\pi ^2}{6 \beta ^2} h(\omega) +\frac{7 \pi ^4}{120 \beta ^4} \dv[2]{h(\omega)}{\omega} \right) \biggr\rvert_{\omega = 0} + \mathcal{O} \left( \beta^{-6} \right), \label{app:eqn:sommerfeld_exp_distr}
\end{align}
Applying the Sommerfeld expansion for the computation of the integrals in \cref{app:eqn:correlator_L11_quad_band,app:eqn:correlator_L12_quad_band,app:eqn:fill_quad_band} and keeping only the leading contribution, we obtain
{
\small
\begin{align}
	L_{11,n} &= \frac{-1}{8 \pi ^2 \beta} \left[ \frac{2 \kappa _n \left(\kappa _n-\mu _n\right)}{\Gamma _n^2+\left(\kappa _n-\mu _n\right){}^2}+\frac{2 \mu _n \left(\arctan\left(\frac{\kappa _n-\mu _n}{\Gamma _n}\right)+\arctan\left(\frac{\mu _n}{\Gamma _n}\right)\right)}{\Gamma _n} \right]  + \mathcal{O} \left( \beta^{-3} \right), \label{app:eqn:correlator_L11_quad_band_high_beta} \\
	L_{12,n} &= \frac{1}{12 \beta ^3} \left\lbrace \frac{\kappa _n \left[ \Gamma _n^2 \left(\kappa _n-2 \mu _n\right) \left(\kappa _n+\mu _n\right)-\Gamma _n^4+\mu _n \left( \kappa _n-\mu _n\right)^3 \right]}{\left( \Gamma _n^2+\mu _n^2 \right) \left[ \Gamma _n^2 + \left(\kappa _n-\mu _n\right)^2 \right]^2}+\frac{\arctan\left(\frac{\kappa _n-\mu _n}{\Gamma _n}\right)+\arctan\left(\frac{\mu _n}{\Gamma _n}\right)}{\Gamma _n} \right\rbrace  + \mathcal{O} \left( \beta^{-5} \right), \label{app:eqn:correlator_L12_quad_band_high_beta} \\
	N^e_n &= \frac{1}{2 \pi  \kappa _n} \left[ \Gamma _n \log \left( \frac{\Gamma _n^2+\left(\kappa _n-\mu _n\right)^2}{\Gamma _n^2+\mu _n^2} \right)+2 \left(\mu _n-\kappa _n\right) \arctan\left(\frac{\kappa _n-\mu _n}{\Gamma _n}\right) +2 \mu _n \arctan\left(\frac{\mu _n}{\Gamma _n}\right) \right] + \frac{1}{2} + \mathcal{O} \left( \beta \right). \label{app:eqn:fill_quad_band_high_beta}
\end{align}}

We note that the expressions of the thermal response coefficient can be further simplified \cref{app:eqn:correlator_L11_quad_band_high_beta,app:eqn:correlator_L12_quad_band_high_beta} in the limit of large or small lifetime. We note that in the large lifetime limit $\Gamma_n \to 0$, we cannot simply Taylor expand in $\Gamma_n$, as that would generate singular terms proportional to $\frac{1}{\mu_n}$ or $\frac{1}{\kappa_n - \mu_n}$, which would diverge near the band edge ({\it i.e.}{}, when $\mu_n=0$ or $\mu_n = \kappa_n$). To obtain more well-behaved asymptotes, we will use the Euler series for the arctangent function\cite{CHI05},
\begin{equation}
	\label{app:eqn:arctan_euler_series}
	\arctan x = \sum_{n=0}^{\infty} \frac{2^{2n} \left( n! \right)^2}{\left( 2 n + 1 \right)!} \frac{x^{2 n + 1}}{\left( 1 + x^2 \right)^{n+1}},
\end{equation}
which has the advantage of converging faster than the corresponding Taylor series. To approximate the arctangent function in our expressions, we will truncate \cref{app:eqn:arctan_euler_series} to the first term and approximate
\begin{align}
	\arctan \left( \frac{x}{\Gamma} \right) &\approx \frac{\Gamma x}{\Gamma^2 + x^2}, \qq{for} \Gamma \to \infty \label{app:eqn:approx_arctan_large} \\
	\arctan \left( \frac{x}{\Gamma} \right) = \frac{\pi}{2} \sgn \left( \frac{x}{\Gamma} \right) - \arctan \left( \frac{\Gamma}{x} \right) &\approx \frac{\pi}{2} \sgn (x) -  \frac{\Gamma x}{\Gamma^2 + x^2}, \qq{for} \Gamma \to 0 \qq{and} \Gamma > 0. \label{app:eqn:approx_arctan_small}
\end{align}

Using \cref{app:eqn:approx_arctan_large,app:eqn:approx_arctan_small}, we can obtain relatively simple expressions for the thermoelectric linear response coefficients in two different limits (and additionally in the $\beta \to \infty$ limit):
\begin{enumerate}
	\item \textbf{The $\Gamma_n \to \infty$ limit.} This limit corresponds to the case where the lifetime broadening of the $n$-th band fermions is much larger than its bandwidth $\kappa_n$. Equivalently speaking, this indicates the lifetime of electron goes to zero, and electrons will behave like local moments. We can use \cref{app:eqn:approx_arctan_large} in \cref{app:eqn:correlator_L11_quad_band_high_beta,app:eqn:correlator_L12_quad_band_high_beta} to obtain
	\begin{align}
		L_{11,n} &\approx -\frac{\kappa _n}{4 \pi ^2 \beta} \frac{ \Gamma _n^2 \kappa _n+2 \mu _n^2 \left(\kappa _n-\mu _n\right) }{ \left(\Gamma _n^2+\mu _n^2\right) \left[ \Gamma _n^2+\left(\kappa _n-\mu _n\right)^2 \right]}, \label{app:eqn:correlator_L11_quad_band_high_beta_lg}\\
		L_{12,n} &\approx \frac{\kappa _n}{6 \beta ^3} \frac{ \Gamma _n^2 \left(-\kappa _n \mu _n+\kappa _n^2-\mu _n^2\right)+\mu _n \left(\kappa _n-\mu _n\right)^3 }{ \left(\Gamma _n^2+\mu _n^2\right) \left[ \Gamma _n^2+\left( \kappa _n-\mu _n \right)^2 \right]^2}. \label{app:eqn:correlator_L12_quad_band_high_beta_lg}
	\end{align}
	These approximations assume that $\abs{ \frac{\kappa_n - \mu_n}{\Gamma_n} } \ll 1$ and $\abs{ \frac{\mu_n}{\Gamma_n} } \ll 1$. For the case when $\Gamma_n$ is large but not infinite, this implies that \cref{app:eqn:correlator_L11_quad_band_high_beta_lg,app:eqn:correlator_L12_quad_band_high_beta_lg} are only valid when the chemical potential is within $\mathcal{O} \left( \Gamma_n \right)$ of the band \emph{and} the bandwidth is much smaller than the lifetime broadening. For large enough lifetime broadening $\Gamma_n$, this assumption holds by definition. 
	
	\item \textbf{The $\Gamma_n \to 0$ limit.} This limit corresponds to case where the lifetime broadening of the $n$-th band excitations $\Gamma_n$ is much smaller than its bandwidth $\kappa_n$. In this limit, we can obtain approximations of the thermoelectric linear response coefficients using \cref{app:eqn:approx_arctan_small},
	\begin{align}
		L_{11,n} &\approx -\frac{1}{8 \pi ^2 \beta  \Gamma _n} \left\lbrace \pi  \mu_n \left(\sgn\left(\kappa _n-\mu _n\right)+\sgn\left(\mu _n\right)\right)  + \frac{2 \Gamma _n^3 \kappa _n \left(\kappa _n-2 \mu _n\right)}{\left(\Gamma _n^2+\mu _n^2\right) \left[\Gamma _n^2+\left(\kappa _n-\mu _n\right)^2 \right]} \right\rbrace, \label{app:eqn:correlator_L11_quad_band_high_beta_hg}\\
		L_{12,n} &\approx \frac{1}{24 \beta ^3 \Gamma _n} \left \lbrace \pi  \left(\sgn\left(\kappa _n-\mu _n\right)+\sgn\left(\mu _n\right)\right)-\frac{4 \Gamma _n^3 \kappa _n}{\left[ \Gamma _n^2+\left(\kappa _n-\mu _n\right)^2 \right]^2} \right \rbrace. \label{app:eqn:correlator_L12_quad_band_high_beta_hg}
	\end{align}
	The approximations in \cref{app:eqn:correlator_L11_quad_band_high_beta_hg,app:eqn:correlator_L12_quad_band_high_beta_hg} are valid whenever $\abs{ \frac{\kappa_n - \mu_n}{\Gamma_n} } \gg 1$ and $\abs{ \frac{\mu_n}{\Gamma_n} } \gg 1$, meaning that they are accurate everywhere \emph{except} within $\mathcal{O} \left( \Gamma_n \right)$ from the band edges. 
\end{enumerate}

\subsubsection{Mott formula for the Seebeck coefficient}\label{app:sec:asymptotes:model:mott_formula}

It is worth noting that in the low-temperature limit, a more general approximation can be derived whenever the thermoelectric coefficients $L^{\alpha \alpha}_{11}$ and $L^{\alpha \alpha}_{12}$ can be written as single integrals over an energy-dependent conductivity function, as is the case for the quadratic band Hamiltonians of \cref{app:eqn:simple_ham_op} and for the THF model (provided that the vertex corrections are ignored), as derived in \cref{app:eqn:L11_static_diagonal_final,app:eqn:L12_static_diagonal_final}. Starting from \cref{app:eqn:L11_static_diagonal_final,app:eqn:L12_static_diagonal_final}, performing the Sommerfeld expansion with the aid of \cref{app:eqn:sommerfeld_exp}, and keeping only the leading non-vanishing contribution (first-order term in $\beta^{-1}$ for $L_{11}^{\alpha\alpha}$ and third-order term in $\beta^{-1}$ for $L_{22}^{\alpha\alpha}$)
\begin{align}
	L^{\alpha \alpha}_{11} =& \frac{\pi}{\beta} \int_{-\infty}^{\infty} \dd{\omega} \pdv{n_{\mathrm{F}} \left( \omega \right)}{\omega} \sigma^{\alpha} \left( \omega \right) = \frac{\pi}{\beta}\left[ - \sigma^{\alpha} \left( \omega \right) \eval_{\omega=0} + \mathcal{O}\left( \beta^{-2} \right) \right] =-\frac{\pi}{\beta} \sigma^{\alpha} \left( 0 \right) + \mathcal{O} \left( \beta^{-3} \right), \label{app:eqn:L11_Mott} \\
L^{\alpha \alpha}_{12} =& -\frac{\pi}{\beta} \int_{-\infty}^{\infty} \dd{\omega} \omega \pdv{n_{\mathrm{F}} \left( \omega \right)}{\omega} \sigma^{\alpha} \left( \omega \right) =  \frac{\pi}{\beta}
 \left[ \left( \omega \sigma^\alpha(\omega) \right) \eval_{\omega=0} + \frac{\pi^2}{6\beta^2} \left( \pdv[2]{ \left( \omega\sigma^\alpha \left( \omega \right) \right)}{\omega} \right) \eval_{\omega=0} + \mathcal{O} \left( \beta^{-4} \right) \right] \nonumber\\ 
 =& \frac{\pi^3}{3 \beta^3} \pdv{\sigma^{\alpha} \left( \omega \right)}{\omega} \eval_{\omega = 0} + \mathcal{O} \left( \beta^{-5} \right). \label{app:eqn:L12_Mott}
\end{align}
Through \cref{app:eqn:seebeck_definition}, \cref{app:eqn:L11_Mott,app:eqn:L12_Mott} lead to the so-called Mott formula for Seebeck coefficient~\cite{JON80}
\begin{equation}
	\label{app:eqn:mott_formula}
	S = - \frac{\pi^2}{3 \beta e} \frac{1}{\sigma^{x} \left( \omega \right)} \pdv{\sigma^{x} \left( \omega \right)}{\omega} \eval_{\omega = 0} + \mathcal{O} \left( \beta^{-3} \right) = - \frac{\pi^2}{3 e \beta} \pdv{\log \left( \sigma^{x} \left( \omega \right) \right)}{\omega} \eval_{\omega = 0} + \mathcal{O} \left( \beta^{-3} \right) ,
\end{equation}
where we have used the fact that we work in units for which the conductivity function is dimensionless. In the same low-temperature limit, we can also obtain the static conductivity from \cref{app:eqn:relation_between_L11_and_sigma}
\begin{equation}
	\label{app:eqn:low_temperature_conductivity}
	\sigma^{\alpha \alpha}= \pi e^{2} \sigma^{\alpha} \left( 0 \right) + \mathcal{O} \left( \beta^{-2} \right),
\end{equation} 
which shows that the static conductivity is related to $\sigma^{\alpha} \left( 0 \right)$ only up to corrections of order $T^2$ ({\it i.e.}{}, in the low temperature limit). 

In the case of the quadratic band Hamiltonian from \cref{app:eqn:simple_ham_op}, the conductivity function can be computed directly from \cref{app:eqn:cond_func_quadratic,app:eqn:cond_function_nth_band} and, because the bands are rigid with respect to doping, it is given by
\begin{equation}
	\label{app:eqn:cond_function_rigid_doping}
	\sigma^{x} \left( \omega \right) = \sigma_0^{x} \left( \omega + \mu \right),
\end{equation}
where the conductivity at zero chemical potential is defined by 
\begin{equation}
	\sigma_0^{x} \left( \omega \right) \equiv \frac{1}{2 \pi^3} \sum_{n=1}^{N_b} f \left(\frac{\kappa_n}{\Gamma_n} , \frac{\omega - \Delta_n}{\Gamma_n} \right),
\end{equation}
with the function $f (x,y)$ being given by \cref{app:eqn:simple_model_f}. Similar to \cref{app:sec:asymptotes:model:formula_beta}, we can obtain further approximations of the conductivity function by employing the following asymptotic expressions for the $f(x,y)$ function, which are derived with the aid of \cref{app:eqn:approx_arctan_large,app:eqn:approx_arctan_small},
\begin{align}
	f \left( \frac{x}{\Gamma}, \frac{y}{\Gamma} \right) &\approx \frac{x \left[ \Gamma ^2 x+2 y^2 (x-y) \right]}{2 \left(\Gamma ^2+y^2\right) \left[ \Gamma ^2+(x-y)^2\right]}, \qq{for} \Gamma \to \infty \label{app:eqn:approx_f_large} \\
	f \left( \frac{x}{\Gamma}, \frac{y}{\Gamma} \right) &\approx  \frac{\pi  y}{4 \Gamma} \left( \sgn(x-y)+\sgn(y) \right) + \frac{\Gamma ^2 x (x-2 y)}{2\left(\Gamma ^2+y^2\right) \left[ \Gamma ^2+(x-y)^2\right]} , \qq{for} \Gamma \to 0 \qq{and} \Gamma > 0. \label{app:eqn:approx_f_small}
\end{align}

Finally, we mention that experiments in both TBG~\cite{MER24} and TSTG~\cite{BAT24} have uncovered Mott formula violations. In order to probe the validity of the Mott formula, both experiments estimate the zero-$\omega$ conductivity function at different fillings (or alternatively at different chemical potentials) from \cref{app:eqn:low_temperature_conductivity}, and then make the rigid doping assumption from \cref{app:eqn:cond_function_rigid_doping} to obtain the conductivity function $\sigma^{\alpha} \left( \omega \right)$ as a function of $\omega$. The Seebeck coefficient computed using \cref{app:eqn:mott_formula} has been shown not to agree with the measured Seebeck coefficient~\cite{MER24,BAT24}. This is an indication of strong correlation effects, as the rigid doping assumption from \cref{app:eqn:cond_function_rigid_doping} does \emph{not} hold, as seen also in our simulations from \cref*{DMFT:app:sec:results_corr_ins,DMFT:app:sec:results_symmetry} of Ref.~\cite{CAL23b}. 

\subsection{Simple heavy-light model}\label{app:sec:asymptotes:heavylight}

In this section, we apply the formulae derived in \cref{app:sec:asymptotes:model}, as well as their corresponding asymptotes to compute the Seebeck coefficient in the case of a simple two-band model with heavy and light fermions. The resulting heavy-light model is an effective toy model for the charge-one excitations of TBG around nonzero integer fillings. We will focus not only on the effective mass asymmetry between the electron and hole bands, but also on the effects that their markedly different lifetimes have on the Seebeck coefficient. 

\subsubsection{General phenomenology of the Seebeck coefficient}\label{app:sec:asymptotes:heavylight:phenomenology}
\begin{figure}[!t]\includegraphics[width=\textwidth]{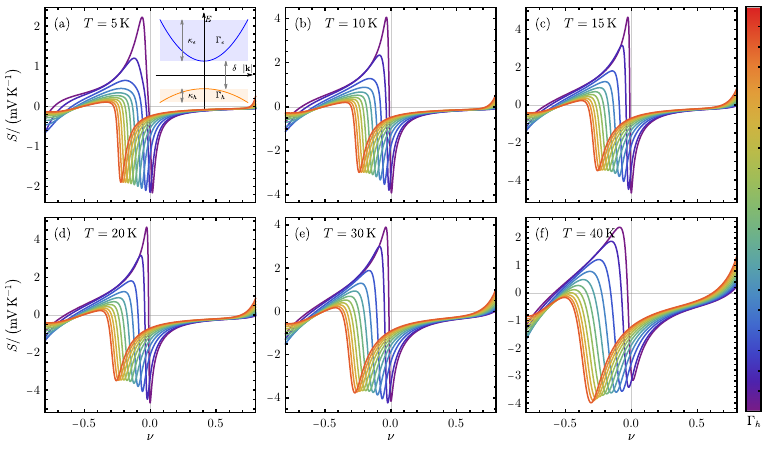}\subfloat{\label{app:fig:seebeck_two_band_numerical:a}}\subfloat{\label{app:fig:seebeck_two_band_numerical:b}}\subfloat{\label{app:fig:seebeck_two_band_numerical:c}}\subfloat{\label{app:fig:seebeck_two_band_numerical:d}}\subfloat{\label{app:fig:seebeck_two_band_numerical:e}}\subfloat{\label{app:fig:seebeck_two_band_numerical:f}}\caption{Numerical results for the Seebeck coefficient in the heavy-light model. We plot the Seebeck coefficient as a function of the doping $\nu$ for the quadratic band Hamiltonian from \cref{app:eqn:simple_ham_op} with $N_b = 2$ bands. The schematic dispersion of the resulting two-band model is shown in the inset of (a). The electron and hole quadratic bands are shown in blue and orange, respectively. The bandwidth of the two bands and the band gap are indicated by the gray arrows. The parameters $\kappa_h = \SI{-8}{\milli\electronvolt}$, $\kappa_e = \SI{30}{\milli\electronvolt}$, $\delta = \SI{16}{\milli\electronvolt}$ are chosen to qualitatively reproduce the charge-one excitation spectrum of the $\protect\IfStrEqCase{6}{{1}{\ket{\nu={}4} }
		{2}{\ket{\nu={}3, \mathrm{IVC}}}
		{3}{\ket{\nu={}3, \mathrm{VP}}}
		{4}{\ket{\nu={}2, \mathrm{K-IVC}}}
		{5}{\ket{\nu={}2, \mathrm{VP}}}
		{6}{\ket{\nu={}1, (\mathrm{K-IVC}+\mathrm{VP})}}
		{7}{\ket{\nu={}1, \mathrm{VP}}}
		{8}{\ket{\nu=0, \mathrm{K-IVC}}}
		{9}{\ket{\nu=0, \mathrm{VP}}}
	}
	[nada]
$ correlated ground state of TBG. Each panel in (a)-(f) shows numerical results at a different temperature, as indicated in the upper left corner of each panel. For each temperature, we compute the Seebeck coefficient for a fixed electron lifetime broadening $\Gamma_e = \SI{1}{\milli\electronvolt}$ and for $\Gamma_h / \si{\milli\electronvolt} = 1$, $2$, $4$, $6$, $8$, $10$, $12$, $14$, $16$, $18$, $20$. The colors of the plots indicate the value of $\Gamma_h$ with blue (red) corresponding to the lowest (highest) lifetime broadening.}\label{app:fig:seebeck_two_band_numerical}\end{figure}

For concreteness, we consider the quadratic band Hamiltonian from \cref{app:eqn:simple_ham_op} for $N_b = 2$ bands. Its band structure is depicted schematically in the inset of \cref{app:fig:seebeck_two_band_numerical:a}. The bandwidth and lifetime broadening of the hole band are given by $\kappa_1 = \kappa_{h}$ and $\Gamma_1 = \Gamma_h$, while those of the electron bands are given by $\kappa_2 = \kappa_{e}$ and $\Gamma_2 = \Gamma_e$, where $\kappa_h<0$, $\kappa_e>0$, $\Gamma_h$, and $\Gamma_e$ will be fixed in what follows. We also denote the gap between the hole and the electron band by $\delta$, implying that $\Delta_1 = - \delta/2$ and $\Delta_2 = + \delta/2$. 

In the numerical results we will present, we chose $\kappa_h = \SI{-8}{\milli\electronvolt}$, $\kappa_e = \SI{30}{\milli\electronvolt}$, $\delta = \SI{16}{\milli\electronvolt}$ so as to qualitatively reproduce the charge-one excitation dispersion of the $\IfStrEqCase{6}{{1}{\ket{\nu={}4} }
		{2}{\ket{\nu={}3, \mathrm{IVC}}}
		{3}{\ket{\nu={}3, \mathrm{VP}}}
		{4}{\ket{\nu={}2, \mathrm{K-IVC}}}
		{5}{\ket{\nu={}2, \mathrm{VP}}}
		{6}{\ket{\nu={}1, (\mathrm{K-IVC}+\mathrm{VP})}}
		{7}{\ket{\nu={}1, \mathrm{VP}}}
		{8}{\ket{\nu=0, \mathrm{K-IVC}}}
		{9}{\ket{\nu=0, \mathrm{VP}}}
	}
	[nada]
$ correlated ground state of TBG shown in \cref*{DMFT:app:fig:sym_br_bs_6_TBG_low} of Ref.~\cite{CAL23b}. Because the electron excitations consist mainly of $c$-electrons, we also choose $\Gamma_e = \SI{1}{\milli\electronvolt}$ to be consistent with the imaginary part of the $c$-electron self-energy, as fixed in \cref{app:eqn:full_second_order_sigma}. On the other hand, the lifetime broadening of the hole excitations $\Gamma_h$ is taken to be a free parameter which is varied in the interval $\SI{1}{\milli\electronvolt} \leq \Gamma_h \leq \SI{20}{\milli\electronvolt}$. 

The Seebeck coefficient obtained by numerically evaluating the integrals over frequency in \cref{app:eqn:correlator_L11_quad_band,app:eqn:correlator_L12_quad_band} are shown in \cref{app:fig:seebeck_two_band_numerical} for different values of the temperature $T$ and lifetime broadening of the hole excitation band $\Gamma_h$. The results are shown as a function of the filling $\nu$ which is measured with respect to charge neutrality
\begin{equation}
	\nu \equiv N^{e}_1 + N^{e}_2 - 1. 
\end{equation}
In what follows, we will find useful to summarize all the general trends of the Seebeck coefficient here and then analytically discuss the first three points in the following \cref{app:sec:asymptotes:heavylight:analytical}.
\begin{enumerate}
	\item Without any lifetime asymmetry ({\it i.e.}{} $\Gamma_h = \Gamma_e$), the Seebeck coefficient is approximately antisymmetric with respect to the charge neutrality point for all the temperatures we considered (they are not \emph{perfectly} antisymmetric as a result of the different effective masses of the electron and hole excitation bands). There is a positive (negative) peak in the Seebeck coefficient corresponding to small hole (electron) doping. The Seebeck coefficient also has a inflection point which is close to the charge neutrality point $\nu = 0$ and where the Seebeck coefficient is approximately zero. Due to the mass difference, the inflection point is not exactly at $\nu=0$. This is the conventional expectation for the Seebeck coefficient in a typical semiconductor.\label{app:enum:conventional_expectation}
	\item As lifetime asymmetry is introduced by increasing the lifetime broadening $\Gamma_h$ of the hole charge-one excitations three changes can be observed in the Seebeck coefficient:
	\begin{enumerate}
		\item The positive peak corresponding to the hole excitations shrinks. \label{app:enum:shrink_of_hole_peak}
		\item The negative peak corresponding to the electron excitations remains approximately unchanged in height ({\it i.e.}{} its height changes much less than the height of the positive peak). \label{app:enum:const_elec_peak}
		\item The negative peak corresponding to the electron excitations shifts its position towards the hole doping side. \label{app:enum:shift_elec_peak}
		\item The inflection point in between the hole and electron Seebeck peaks shifts towards the hole doping side. \label{app:enum:shift_inflection_point}
	\end{enumerate} 
	\item For the largest lifetime broadening of the hole excitations, the Seebeck coefficient is negative for most of the doping region we considered. At $T = \SI{40}{\kelvin}$, the case with the largest lifetime asymmetry ($\Gamma_h = \SI{20}{\milli\electronvolt}$) is fully negative for the entire doping region.  \label{app:enum:fully_negative_high_temperature}
	\item The Seebeck coefficient is seen to indicate the type of carriers that predominate at a given filling. Without lifetime asymmetry, the Seebeck coefficient is negative (positive) for electron (hole) doping indicating that electron (hole) excitations dominate transport. When the lifetime of the hole excitations is reduced (which in effect makes them ``bad'' carriers), the Seebeck coefficient becomes mostly negative, indicating that the dominant type of carriers for most of the doping range considered is electron excitations. \label{app:enum:type_of_carrier}  
	\item The qualitative behavior of the Seebeck coefficient is approximately temperature-independent. \label{app:enum:temperature_independence}
\end{enumerate}

\subsubsection{Analytical understanding of the effects of lifetime asymmetry on the Seebeck coefficient}\label{app:sec:asymptotes:heavylight:analytical}
\begin{figure}[!t]\includegraphics[width=\textwidth]{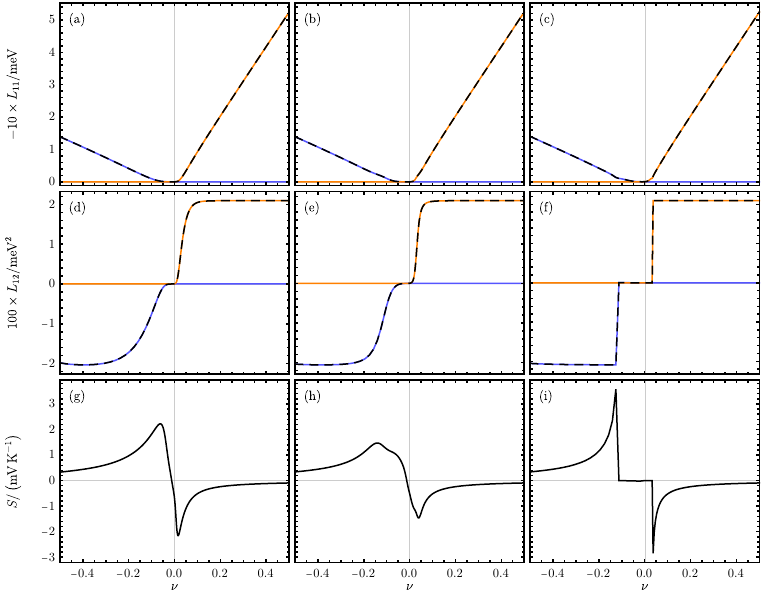}\subfloat{\label{app:fig:seebeck_correlators_1:a}}\subfloat{\label{app:fig:seebeck_correlators_1:b}}\subfloat{\label{app:fig:seebeck_correlators_1:c}}\subfloat{\label{app:fig:seebeck_correlators_1:d}}\subfloat{\label{app:fig:seebeck_correlators_1:e}}\subfloat{\label{app:fig:seebeck_correlators_1:f}}\subfloat{\label{app:fig:seebeck_correlators_1:g}}\subfloat{\label{app:fig:seebeck_correlators_1:h}}\subfloat{\label{app:fig:seebeck_correlators_1:i}}\caption{The Seebeck coefficient of the heavy-light model without lifetime asymmetry, or, alternatively, with lifetime symmetry ($\Gamma_e=\Gamma_h$). We use $T = \SI{5}{\kelvin}$, $\Gamma_e = \Gamma_h = \SI{1}{\milli\electronvolt}$, and the same band structure parameters as in \cref{app:fig:seebeck_two_band_numerical}. In (a)-(f) we plot the thermoelectric linear response coefficients in the heavy-light model together with their analytical approximations. The exact results obtained by numerical integration through \cref{app:eqn:correlator_L11_quad_band,app:eqn:correlator_L12_quad_band} are shown in (a) and (d); the low-temperature approximations from \cref{app:eqn:correlator_L11_quad_band_high_beta,app:eqn:correlator_L12_quad_band_high_beta} are plotted in (b) and (e); in (c) and (f), we plot the asymptotic forms from \cref{app:eqn:correlator_L11_quad_band_high_beta_hg,app:eqn:correlator_L12_quad_band_high_beta_hg} of the low-temperature thermoelectric coefficients obtained in the limit of small broadening. In (a)-(f), the electron (hole) contribution to the corresponding thermoelectric coefficient is shown in orange (blue), while the total contribution is denoted by the black dashed line. Each of (g)-(i) illustrates the Seebeck coefficient computed using the thermoelectric coefficients plotted in the two panels of the same column.}\label{app:fig:seebeck_correlators_1}\end{figure}\begin{figure}[!t]\includegraphics[width=\textwidth]{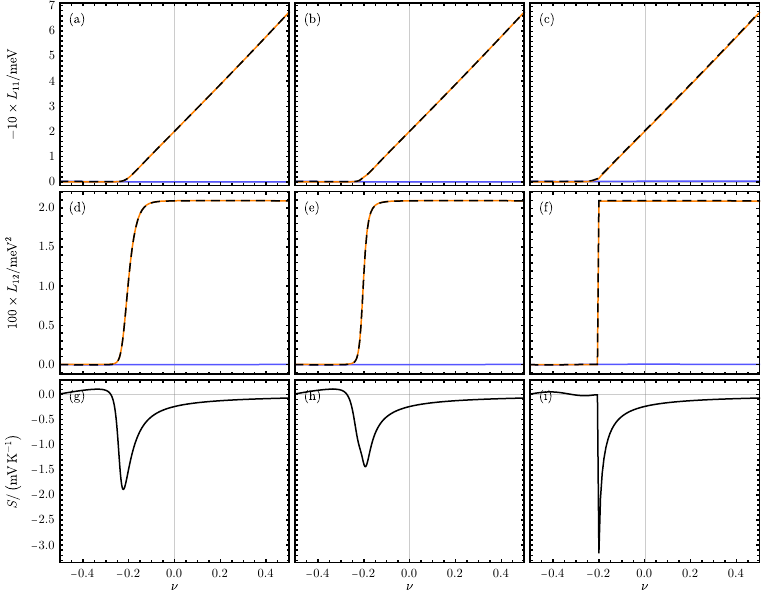}\subfloat{\label{app:fig:seebeck_correlators_2:a}}\subfloat{\label{app:fig:seebeck_correlators_2:b}}\subfloat{\label{app:fig:seebeck_correlators_2:c}}\subfloat{\label{app:fig:seebeck_correlators_2:d}}\subfloat{\label{app:fig:seebeck_correlators_2:e}}\subfloat{\label{app:fig:seebeck_correlators_2:f}}\subfloat{\label{app:fig:seebeck_correlators_2:g}}\subfloat{\label{app:fig:seebeck_correlators_2:h}}\subfloat{\label{app:fig:seebeck_correlators_2:i}}\caption{The Seebeck coefficient of the heavy-light model with lifetime asymmetry. We use $T = \SI{5}{\kelvin}$, $\Gamma_e = \SI{1}{\milli\electronvolt}$, $\Gamma_h = \SI{20}{\milli\electronvolt}$, and the same band structure parameters as in \cref{app:fig:seebeck_two_band_numerical}. In (a)-(f) we plot the thermoelectric linear response coefficients in the heavy-light model together with their analytical approximations. The exact results obtained by numerical integration through \cref{app:eqn:correlator_L11_quad_band,app:eqn:correlator_L12_quad_band} are shown in (a) and (d); the low-temperature approximations from \cref{app:eqn:correlator_L11_quad_band_high_beta,app:eqn:correlator_L12_quad_band_high_beta} are plotted in (b) and (e); in (c) and (f), we plot the asymptotic forms from \cref{app:sec:asymptotes:model:formula_beta} of the low-temperature thermoelectric coefficients obtained in the limit of small broadening for the electron bands and in the limit of large broadening for the hole band: we use \cref{app:eqn:correlator_L11_quad_band_high_beta_lg,app:eqn:correlator_L12_quad_band_high_beta_lg} to compute the thermoelectric coefficients for the hole band and \cref{app:eqn:correlator_L11_quad_band_high_beta_hg,app:eqn:correlator_L12_quad_band_high_beta_hg} to compute them for the electron band. In (a)-(f), the electron (hole) contribution to the corresponding thermoelectric coefficient is shown in orange (blue), while the total contribution is denoted by the black dashed line. Each of (g)-(i) illustrates the Seebeck coefficient computed using the thermoelectric coefficients plotted in the two panels of the same column.}\label{app:fig:seebeck_correlators_2}\end{figure}

In what follows, we will provide an analytical understanding of the anomalous behavior of the Seebeck coefficient in the presence of lifetime asymmetry as shown in \cref{app:fig:seebeck_two_band_numerical}. Given that the Seebeck coefficient's general trend remains consistent across the temperature range of interest, specifically from $\SI{5}{\kelvin}$ to $\SI{40}{\kelvin}$, we can focus exclusively on the low-temperature regime (taking $T = \SI{5}{\kelvin}$, which is equivalent to $k_B T \approx \SI{0.4}{\milli\electronvolt}$). Here, by low-temperature regime, we mean the parameter regime where the temperature $T$ is smaller than the other energy scales (the lifetime broadenings $\Gamma_n$, the bandwidths $\abs{\kappa_n}$, and the gap $\delta$) of the systems such that the analytical approximations of the thermoelectric linear response coefficients given in \cref{app:eqn:correlator_L11_quad_band_high_beta,app:eqn:correlator_L12_quad_band_high_beta} hold. 

To understand \cref{app:enum:conventional_expectation}, we first investigate the behaviors of the thermal response coefficients in the lifetime symmetric case ($\Gamma_e = \Gamma_h = \SI{1}{\milli\electronvolt}$). The thermal response coefficients and their approximations are plotted in \cref{app:fig:seebeck_correlators_1} for the same band structure parameters as in \cref{app:fig:seebeck_two_band_numerical}. We find that the exact Seebeck coefficient and the ones obtained through the low-temperature approximations from \cref{app:eqn:correlator_L11_quad_band_high_beta,app:eqn:correlator_L12_quad_band_high_beta} qualitatively agree. It can also be seen that for positive (negative) filling $\nu$, the thermoelectric coefficients are dominated by the electron (hole) contribution. We can therefore approximate
\begin{equation}
	\label{app:eqn:seebeck_separate_per_band}
	S \approx \begin{cases} 
		\frac{\beta}{e} \frac{L_{12,1}}{L_{11,1}}, & \qq{for electron doping} \\
		\frac{\beta}{e} \frac{L_{12,2}}{L_{11,2}}, & \qq{for hole doping} \\
	\end{cases}.
\end{equation}
We can then obtain a simpler expression for the Seebeck coefficient by employing the asymptotes of the linear response coefficients in the limit of small lifetime broadening obtained in \cref{app:eqn:correlator_L11_quad_band_high_beta_hg,app:eqn:correlator_L12_quad_band_high_beta_hg}
\begin{equation}
	\label{app:eqn:approx_seebeck_separate_band_eq_lifetime}
	S \approx -\frac{\pi^2}{3 \beta e} \frac{ \pi  \left(\sgn\left(\kappa _n-\mu _n\right)+\sgn\left(\mu _n\right)\right)-\frac{4 \Gamma _n^3 \kappa _n}{\left[ \Gamma _n^2+\left(\kappa _n-\mu _n\right)^2 \right]^2}}{\pi  \mu_n \left(\sgn\left(\kappa _n-\mu _n\right)+\sgn\left(\mu _n\right)\right)  + \frac{2 \Gamma _n^3 \kappa _n \left(\kappa _n-2 \mu _n\right)}{\left(\Gamma _n^2+\mu _n^2\right) \left[\Gamma _n^2+\left(\kappa _n-\mu _n\right)^2 \right]}}, 
\end{equation}
where $n = 1$ for electron doping and $n = 2$ for hole doping. As seen in \cref{app:fig:seebeck_correlators_1:i}, the expression of the Seebeck coefficient from \cref{app:eqn:approx_seebeck_separate_band_eq_lifetime}, which was computed through the asymptotes from \cref{app:eqn:correlator_L11_quad_band_high_beta_hg,app:eqn:correlator_L12_quad_band_high_beta_hg} is not accurate in the band gap. In this region, the thermoelectric coefficients are small, making the relative errors of these approximations non-negligible. This is precisely because the so-called leading contribution to the thermoelectric coefficients from \cref{app:eqn:correlator_L11_quad_band_high_beta_hg,app:eqn:correlator_L12_quad_band_high_beta_hg} vanish in the gap. However, when $\mu_n \approx 0$ and $0 < \abs{\mu_n} < \abs{\kappa_n}$ ({\it i.e.}{} when the chemical potential is near the band edge), \cref{app:eqn:approx_seebeck_separate_band_eq_lifetime} can qualitatively explain the origin of the peaks in the Seebeck coefficient. Using also the fact that $\Gamma_n \ll \abs{\kappa_n}$ (which is valid for the parameters we choose in \cref{app:fig:seebeck_correlators_1}), we find that
\begin{equation}
	\label{app:eqn:approx_seebeck_band_edge}
	S \approx -\frac{\pi^2}{3 \beta e} \frac{ \pi  \sgn\left( \kappa_n \right) }{\pi  \abs{\mu_n}  + 2 \Gamma_n}, \qq{for} \mu_n \approx 0 \qq{and} 0 < \abs{\mu_n} < \abs{\kappa_n}.
\end{equation}
\Cref{app:eqn:approx_seebeck_band_edge} implies that the Seebeck coefficient peaks at the band edges, with a negative peak for electron doping and a positive one for hole doping. Additionally, we find that within this approximation, near the band edge (with $\mu_n\approx 0$), the Seebeck coefficient is dependent \emph{only} on the excitation lifetimes and \emph{not} on the effective masses (which only influence the \emph{sign} of the Seebeck coefficient). This explains why in the absence of any lifetime asymmetry, the Seebeck coefficient is approximately ``particle-hole antisymmetric'', despite a significant effective mass asymmetry $\abs{\frac{\kappa_e}{\kappa_h}} \approx 4$, thereby explaining \cref{app:enum:conventional_expectation}.

In \cref{app:fig:seebeck_correlators_2}, we plot the thermoelectric coefficients, their approximations, and the Seebeck coefficient of the heavy-light model for $\Gamma_e = \SI{1}{\milli\electronvolt}$ and $\Gamma_h = \SI{20}{\milli\electronvolt}$. Similar to the lifetime symmetric case from \cref{app:fig:seebeck_correlators_2}, the low-temperature approximations of the thermoelectric coefficients from \cref{app:eqn:correlator_L11_quad_band_high_beta_hg,app:eqn:correlator_L12_quad_band_high_beta_hg} qualitatively reproduces the exact results. Also similar to the lifetime symmetric case, we find that the Seebeck coefficient is dominated by the electron (hole) band contribution for electron (hole) doping, such that \cref{app:eqn:seebeck_separate_per_band} still holds. To gain some analytical intuition on the behavior of the Seebeck coefficient in the two doping regimes, we can employ the asymptotic expressions of the thermoelectric linear response coefficients derived at the end of \cref{app:sec:asymptotes:model:formula_beta} in the limit of large or small lifetime broadening. 

Because \cref{app:eqn:seebeck_separate_per_band} holds even in the lifetime asymmetric case shown in \cref{app:fig:seebeck_correlators_2}, when the chemical potential is near the electron band edge, (which harbors long-lived excitations with small lifetime broadening $\Gamma_e = \SI{1}{\milli\electronvolt}$) the Seebeck coefficient is still given by \cref{app:eqn:approx_seebeck_band_edge}. This means that in the low-temperature limit defined at the beginning of this section and with the $\Gamma_n \to 0$ approximation from \cref{app:eqn:correlator_L11_quad_band_high_beta_hg,app:eqn:correlator_L12_quad_band_high_beta_hg}, the height of the electron peak in the Seebeck coefficient is unaffected by the enhanced broadening of the hole excitations, thereby explaining \cref{app:enum:const_elec_peak}. 

Near the hole band edge, we need to employ the small-lifetime asymptotes for the thermoelectric coefficients from \cref{app:eqn:correlator_L11_quad_band_high_beta_lg,app:eqn:correlator_L12_quad_band_high_beta_lg}. In this case, the Seebeck coefficient is therefore given by
\begin{equation}
	\label{app:eqn:approx_seebeck_band_edge_small_live}
	S \approx -\frac{2 \pi^2}{3 \beta e} \frac{\Gamma _1^2 \left(-\kappa _1 \mu _n+\kappa _1^2-\mu _1^2\right)+\mu _1 \left(\kappa _1-\mu _1\right)^3}{\Gamma _1^2+\left( \kappa _1-\mu _1 \right)^2} \frac{1}{\Gamma _1^2 \kappa _1+2 \mu _1^2 \left(\kappa _1-\mu _1\right)} \approx -\frac{2 \pi^2}{3 \beta e} \frac{\kappa_1}{ \Gamma_1^2 + \kappa_1^2}, 
\end{equation}
which is positive since $\kappa_1 = \kappa_h <1$. In the second approximation of \cref{app:eqn:approx_seebeck_band_edge_small_live}, we have additionally taken the chemical potential to be at the hole band edge such that $\abs{\mu_1} \approx 0$. Using \cref{app:eqn:approx_seebeck_band_edge,app:eqn:approx_seebeck_band_edge_small_live}, we can compare the peaks in the Seebeck coefficient at the band edge for the hole and the electron bands and find 
\begin{equation}
	\label{app:eqn:approx_seebeck_ratio_e_h_peak}
	\abs{\frac{S_{\text{hole}}}{S_{\text{electron}}}} \approx \frac{2 \abs{\kappa_h}}{\Gamma_h^2 + \kappa_h^2} \frac{2 \Gamma_e}{\pi} \sim 0.02,
\end{equation}
thus explaining why the positive hole peak in the Seebeck coefficient shrinks once the lifetime of the hole excitations is decreased ({\it i.e.}{} \cref{app:enum:shrink_of_hole_peak}). As the temperature is increased, the electron peak is broadened and, as a result, the hole peak shrinks even more leading to a fully negative Seebeck coefficient, as depicted in \cref{app:fig:seebeck_two_band_numerical:f}, thereby explaining \cref{app:enum:fully_negative_high_temperature}. There are two other points worth mentioning in this case:
\begin{itemize}
	\item In the limit of vanishing temperature and vanishing electron excitation lifetime broadening $\Gamma_e$, the height of the electron peak does not change at all. However, once finite temperature and finite $\Gamma_e$ effects are introduced, its height will change slightly as $\Gamma_h$ is changed, as seen in \cref{app:fig:seebeck_two_band_numerical:f}. This change, however, is a ``second-order'' effect, while the shrinking of the positive hole peak is already a leading order effect. 
	\item \Cref{app:eqn:approx_seebeck_band_edge_small_live} implies that in the limit of vanishing temperature and large hole excitation lifetime broadening $\Gamma_h$, the Seebeck coefficient at the hole band edge will always be positive. Finite temperature effects are needed in order to achieve a fully-negative Seebeck coefficient even in the lifetime asymmetric case, as shown in \cref{app:fig:seebeck_two_band_numerical:f}. 
\end{itemize}

We also note that in the lifetime asymmetric case, the electron peak of the Seebeck coefficient shifts towards the hole doping side. This is again an effect of the lifetime asymmetry between the electron and hole excitations. To be more quantitative, we note that the electron peak of the Seebeck coefficient occurs roughly whenever the chemical potential is at the electron band edge. Because the hole excitations are very broad ({\it i.e.}{} incoherent), at that point the hole band is not fully filled, meaning that the total filling $\nu$ is negative. To be precise, the peak in the Seebeck coefficient occurs roughly for $\mu = \delta/2$ (where $\delta$ is the gap between the hole and electron bands). Inserting this into the analytical expression for the filling of each individual band from \cref{app:eqn:fill_quad_band_high_beta}, we find that the electron peak in the Seebeck coefficient occurs roughly at $\nu \approx -0.2$, which is consistent with the exact results from \cref{app:fig:seebeck_correlators_2:g} and hereby explains \cref{app:enum:shift_elec_peak}. Because the inflection point in the Seebeck coefficient occurs in between the electron- and hole-doped peaks, the former also shifts towards the hole doping side, thus explaining \cref{app:enum:shift_inflection_point}.

Finally, we also mention that, as will be discussed in more detail in \cref{app:sec:seebeck_non_int}, the simple heavy-light model mimics the behavior of the Seebeck coefficient in the THF model in its symmetry-broken phases. In the THF model, near a positive integer filling, the electron charge-one excitations are light, have a pronounced $c$-electron character, and have a small lifetime broadening. Conversely, on the hole doping side, the charge-one excitations are heavy, have mostly $f$-electron character, and have a large lifetime broadening. Therefore, as we have analytically shown using the simple heavy-light model, the lifetime asymmetry will lead to an overall negative Seebeck coefficient in the THF model around integer fillings $\nu=1,2$, as observed in experiments~\cite{MER24,BAT24}. Other scenarios that have been proposed to explain the shape of the Seebeck coefficient as a function of filling in TBG, such as the Dirac-revival picture~\cite{PAU22}, cannot reproduce the negative offset observed by experiments~\cite{MER24}. 

\subsection{Quantum-geometric contributions to the Seebeck coefficient}\label{app:sec:asymptotes:quantum_geometric}

The aim of the first part of this \siSection{} was to provide an intuitive understanding of the effects governing the thermoelectric response of TBG, paying special attention to the interplay between the effective mass and the lifetime of the charge-one excitation. To do so, we have employed a simple two-band model as a proxy for the charge-one excitations of TBG. Within the two-band model, the Hamiltonian is diagonal in the band basis, meaning that the two bands, representing the electron and hole excitations, are topologically trivial. 

However, the charge-one excitation bands in TBG are \emph{not} topologically trivial. In fact, Ref.~\cite{KRU23} suggests that the origin of the large Seebeck coefficient ({\it i.e.}{} $\abs{S} \gg \frac{k_B}{e} \approx \SI{86}{\micro\volt\kelvin^{-1}}$) measured by Ref.~\cite{PAU22} in TBG can only be explained by invoking the quantum-geometric contributions. Therefore, a natural question that arises is whether the heavy-light model (which ignores the topology of the charge-one excitations of TBG) constitutes the correct framework for interpreting the thermoelectric response of TBG. In other words, does the topology of the charge-one excitation bands contribute significantly to the thermoelectric response of TBG around the integer-filled correlated insulators?

In this section, we investigate the topological contribution to the thermoelectric response of TBG. To do so, we follow Ref.~\cite{KRU23} and split the conductivity function into two contributions: a kinetic one and a quantum geometric one (both of which will be defined below). We then consider both the non-interacting and the interacting cases\footnote{More care is needed when separating the kinetic and quantum geometric contributions to the Seebeck coefficient in the interacting case. In the presence of interactions, the dispersion of the charge-one excitation bands arises from the non-trivial quantum geometry of the active TBG bands~\cite{BER21b,XIE21,KAN21}. In other words, ``turning off'' the quantum geometry of the TBG flat bands would modify the dispersion of the charge-one excitation bands. As such, in the interacting case, we define the kinetic contribution as arising from the actual interactions and quantum geometry-induced dispersion of the charge-one excitation band. In contrast the quantum-geometric contribution in this context is defined as the part of the conductivity function which arises from the \emph{wave functions} of the charge-one excitations.} 

In the (experimentally unrealistic, but theoretically relevant) non-interacting limit, we find that the topological contribution to the conductivity function is significant. This is to be expected because in the absence of interactions, the dispersion of the active TBG bands is very small, meaning that the quantum geometric contribution cannot be neglected. We will numerically investigate the thermoelectric response of both TBG and TSTG in the non-interacting limit in \cref{app:sec:seebeck_non_int}.

In the more realistic interacting case, the charge-one excitation bands of TBG disperse much more strongly, leading to a large kinetic contribution to the conductivity function. As a result, we find that the topological contribution to the conductivity function is negligible in the interacting case. Therefore, the heavy-light model is suitable for developing an analytical understanding of the Seebeck coefficient of TBG's correlated insulators. Full numerical results for the Seebeck coefficient of both TBG and TSTG in the symmetry-broken phases will be provided in \cref{app:sec:seebeck_sym_br}.

Finally, we mention that the toy model used by Ref.~\cite{KRU23} to explain the large Seebeck coefficient measured in TBG comprises two Chern-$\pm 1$ bands, with a small lifetime broadening of just $\Gamma = \SI{0.01}{\milli\electronvolt}$. Due to the exactly vanishing dispersion, the nonzero conductivity function in this model arises solely from quantum-geometric effects. However, this model is unrealistic for TBG because the TBG charge-one excitations are not flat, but have a significant dispersion. Moreover, as already shown using simple calculations within the heavy-light model from \cref{app:sec:asymptotes:heavylight}, quantum geometric contributions are not needed to obtain a large Seebeck coefficient.

\subsubsection{Kinetic and quantum-geometric contributions to the conductivity function in the non-interacting limit}\label{app:sec:asymptotes:quantum_geometric:cond_func_contrib_non_int}

To estimate the effects of quantum-geometry on the Seebeck coefficient, we start with the definition of the conductivity function in the THF model given by \cref{app:eqn:def_conduc_func}. The longitudinal thermoelectric coefficients both depend on the conductivity function through \cref{app:eqn:L11_static_diagonal_final,app:eqn:L12_static_diagonal_final}. Our goal will be to express separate the conductivity function into two terms: a \emph{kinetic} term which depends only on the dispersion of the charge-excitation bands and a \emph{quantum-geometric} terms which additionally depends on their wave functions.

Working in the non-interacting limit, we first diagonalize the single-particle TBG Hamiltonian in the THF model
\begin{equation}
	\label{app:eqn:diagonalizing_non_interaction_TBG}
	\sum_{i', \eta', s'} h^{\text{TBG}}_{i \eta s;i' \eta' s'} \left( \vec{k} \right) \varphi_{n;i' \eta' s'} \left( \vec{k} \right) = \epsilon_{n} \left( \vec{k} \right) \varphi_{n;i \eta s} \left( \vec{k} \right),
\end{equation} 
where $\varphi_{n;i \eta s} \left( \vec{k} \right)$ and $\epsilon_{n} \left( \vec{k} \right)$ are, respectively, the wave function and energy eigenvalue of $h^{\text{TBG}}_{i \eta s;i' \eta' s'} \left( \vec{k} \right)$ indexed in the order of increasing energy by $1 \leq n \leq 24$. Assuming an identical broadening factor for both the $f$- and $c$-electron, which amounts to setting $\Gamma_f = \Gamma_c = \Gamma_0$ in \cref{app:eqn:broadening_in_noninteracting}, we can employ \cref{app:eqn:spectral_function_non_int,app:eqn:gf_non_int_for_Seebeck} to express the non-interacting spectral function of TBG in terms of the wave functions and energies from \cref{app:eqn:diagonalizing_non_interaction_TBG}
\begin{equation}
	\label{app:eqn:lorentz_spectral_func_non_int}
	A^{0}_{i \eta s;i' \eta' s'} \left(\omega, \vec{k} \right) = \sum_{n} \frac{\Gamma_0 \left( \vec{k} \right)}{\pi} \frac{\varphi_{n;i \eta s} \left( \vec{k} \right) \varphi^*_{n;i' \eta' s'} \left( \vec{k} \right)}{ \left( \omega + \mu - \epsilon_n \left( \vec{k} \right) \right)^2 + \Gamma^2_0}.
\end{equation}
From \cref{app:eqn:lorentz_spectral_func_non_int}, we can define the spectral function of TBG \emph{projected} in the (non-interacting) band basis of TBG
\begin{equation}
	\label{app:eqn:band_basis_spectral_function_non_int}
	A^0_{n} \left( \omega, \vec{k} \right) =  \sum_{\substack{i, \eta, s \\ i', \eta', s'}} \varphi^*_{n;i \eta s} \left( \vec{k} \right) A^0_{i \eta s;i' \eta' s'} \left( \omega, \vec{k} \right) \varphi_{n;i' \eta' s'} \left( \vec{k} \right).
\end{equation}

Using \cref{app:eqn:lorentz_spectral_func_non_int}, we can compute the conductivity function of the TBG defined in \cref{app:eqn:def_conduc_func} to be 
\begin{equation}
	\label{app:eqn:conductivity_function_band_basis_non_int}
	\sigma^{\alpha} \left( \omega \right) = \frac{1}{N_0 \Omega_0} \sum_{\vec{k}} \sum_{n,m} v^{\alpha}_{mn} \left( \vec{k} \right) A^0_{n} \left( \omega, \vec{k} \right) v^{\alpha}_{nm} \left( \vec{k} \right) A^0_{m} \left( \omega, \vec{k} \right),
\end{equation}
where we have introduced the projected group velocity of TBG in the non-interacting limit
\begin{equation}
	\label{app:eqn:qp_group_velocity}
	v^{\alpha}_{nm} \left( \vec{k} \right) = \sum_{\substack{i, \eta, s \\ i', \eta', s'}} \varphi^*_{n;i \eta s} \left( \vec{k} \right) \left( \partial_{k^{\alpha}} h^{\text{TBG}}_{i \eta s;i' \eta' s'} \left( \vec{k} \right) \right) \varphi_{m;i' \eta' s'} \left( \vec{k} \right).
\end{equation}
Using the fact that $\varphi_{n;i \eta s} \left( \vec{k} \right)$ are the eigenstates of the single-particle TBG Hamiltonian with energies $\epsilon_n \left( \vec{k} \right)$, we can rewrite the projected group velocity as 
\begin{equation}
	\label{app:eqn:approx_qp_group_velocity_non_int}
	v^{\alpha}_{nm} \left( \vec{k} \right) = \sum_{\substack{i, \eta, s \\ i', \eta', s'}} \varphi^*_{n;i \eta s} \left( \vec{k} \right) \left( \partial_{k^{\alpha}} h^{\text{TBG}}_{i \eta s;i' \eta' s'} \left( \vec{k} \right) \right) \varphi_{m;i' \eta' s'} \left( \vec{k} \right) =  \partial_{k^{\alpha}} \epsilon_{n} \left( \vec{k} \right) \delta_{nm} + \left( \epsilon_{n} \left( \vec{k} \right) - \epsilon_{m} \left( \vec{k} \right) \right) \mathcal{A}_{nm}^{\alpha} \left( \vec{k} \right),
\end{equation}
with the nonabelian Berry connection of the TBG bands being defined as 
\begin{equation}
	\mathcal{A}_{nm}^{\alpha} \left( \vec{k} \right) \equiv \sum_{i, \eta, s} \varphi^{*}_{n;i \eta s} \left( \vec{k} \right) \partial_{k^{\alpha}} \varphi_{m;i \eta s} \left( \vec{k} \right).
\end{equation}

Plugging \cref{app:eqn:approx_qp_group_velocity_non_int} into \cref{app:eqn:conductivity_function_band_basis_non_int}, we find that in the non-interacting limit, we can split the conductivity function $\sigma^{\alpha} \left( \omega \right)$ into a ``kinetic'' and a ``quantum-geometric'' contribution
\begin{equation}
	\label{app:eqn:split_of_conductivity_function_non_int}
	\sigma^{\alpha} \left( \omega \right) = \sigma^{\alpha}_{\text{Kinetic}} \left( \omega \right) +  \sigma^{\alpha}_{\text{Geometric}} \left( \omega \right),
\end{equation}
which are given respectively by the $n=m$ and $n \neq m$ terms of in the summation from \cref{app:eqn:conductivity_function_band_basis_non_int}
\begin{align}
	\sigma^{\alpha}_{\text{Kinetic}} \left( \omega \right) &= \frac{1}{N_0 \Omega_0} \sum_{\vec{k}} \sum_{n} \left[ v^{\alpha}_{nn} \left( \vec{k} \right) A^0_{n} \left( \omega, \vec{k} \right)\right]^2 \nonumber \\
	&= \frac{1}{N_0 \Omega_0} \sum_{\vec{k}} \sum_{n} \left[ \frac{\Gamma_0 }{\pi} \frac{\partial_{k^{\alpha}} \epsilon_{n} \left( \vec{k} \right)}{ \left( \omega + \mu - \epsilon_n \left( \vec{k} \right) \right)^2 + \Gamma^2_0} \right]^2 \label{app:eqn:cond_func_kin_non_int} \\
	\sigma^{\alpha}_{\text{Geometric}} \left( \omega \right) &= \frac{1}{N_0 \Omega_0} \sum_{\vec{k}} \sum_{n \neq m} v^{\alpha}_{mn} \left( \vec{k} \right) A^0_{n} \left( \omega, \vec{k} \right) v^{\alpha}_{nm} \left( \vec{k} \right) A_{m} \left( \omega, \vec{k} \right) \nonumber \\
	&= \frac{1}{N_0 \Omega_0} \sum_{\vec{k}} \sum_{n \neq m} \frac{\Gamma^2_0}{\pi^2} \frac{\left[ \left( \epsilon_{n} \left( \vec{k} \right) - \epsilon_{m} \left( \vec{k} \right) \right) \mathcal{A}_{nm}^{\alpha,\text{HF}} \left( \vec{k} \right) \right]^{2}}{\left[ \left( \omega + \mu - \epsilon_n \left( \vec{k} \right) \right)^2 + \Gamma^2_0 \right] \left[ \left( \omega + \mu - \epsilon_m \left( \vec{k} \right) \right)^2 + \Gamma^2_0 \right] }. \label{app:eqn:cond_func_geo_non_int}
\end{align}

To assess the relative contribution of the quantum-geometric part of the conductivity function in the non-interacting limit, we compute both the kinetic and the quantum-geometric contribution to the conductivity function from \cref{app:eqn:cond_func_kin_non_int,app:eqn:cond_func_geo_non_int} at charge neutrality ($\mu = 0$) taking $\Gamma_0 = \SI{1}{\milli\electronvolt}$. We then compute the relative magnitude of the quantum-geometric contribution using 
\begin{equation}
	\epsilon_{\text{Geometric}} = \sqrt{ \frac{ \int_{-\omega_0}^{\omega_0} \dd{\omega} \abs{ \sigma_{\text{Geometric}}^{\alpha} \left( \omega \right)}^2}{\int_{-\omega_0}^{\omega_0} \dd{\omega} \abs{ \sigma^{\alpha} \left( \omega \right) }^2} },
\end{equation} 
where $\omega_0$ is an energy cutoff that we will specify below. In the non-interacting limit, we want focus exclusively on the active TBG bands. Given that the gap between the active and the remote bands of TBG is approximately equal to $\abs{\gamma}$~\cite{SON22}, which was defined in \cref{app:sec:HF_review:single_particle:TBG}, we take $\omega_0 = \gamma /2$. Because the bandwidth of the active TBG bands is set by $2 \abs{M}$, where $\abs{M} \ll \gamma/2$, as mentioned in \cref{app:sec:HF_review:single_particle:TBG}, we note that our choice for the cutoff in $\omega_0$ is large enough so that it includes the active TBG band contribution, but small enough such that it does not include the strongly-dispersive remote bands of TBG. Numerically, we find that $\epsilon_{\text{Geometric}} \approx 40\%$, which implies that there is a significant geometric contribution to the conductivity function. This is to be expected, since the kinetic contribution from \cref{app:eqn:cond_func_kin_non_int} is proportional to the group velocity of the TBG active bands, which is suppressed near the magic angle.

\subsubsection{Kinetic and quantum-geometric contributions to the conductivity function in the interacting limit}\label{app:sec:asymptotes:quantum_geometric:cond_func_contrib_int}

In the interacting case, the spectral function cannot be computed analytically. To make some analytical progress, we note that in our calculations in the symmetry-broken phase, the system still features well-defined quasi-particle excitations (at least in the limit of small temperature). For concreteness, we assume that at each momentum $\vec{k}$, there are quasi-particle excitations of energy $E_{n} \left( \vec{k} \right)$, wave function $\varphi_{n;i \eta s} \left( \vec{k} \right)$ having lifetime broadening $\Gamma_{n} \left( \vec{k}  \right)$. In this notation $n$ indexes the quasi-particle excitation in the order of increasing energy such that $1 \leq n \leq 24$ for TBG. The spectral function of the system can therefore be \emph{approximated} as 
\begin{equation}
	\label{app:eqn:lorentz_approx_spectral_func}
	A_{i \eta s;i' \eta' s'} \left(\omega, \vec{k} \right) \approx \sum_{n} \frac{\Gamma_n \left( \vec{k} \right)}{\pi} \frac{\varphi_{n;i \eta s} \left( \vec{k} \right) \varphi^*_{n;i' \eta' s'} \left( \vec{k} \right)}{ \left( \omega - E_n \left( \vec{k} \right) \right)^2 + \Gamma^2_n \left( \vec{k} \right)}.
\end{equation}
In principle, the quasi-particle wave functions, energies, and lifetimes can be found at each momentum point by fitting the approximation ansatz from \cref{app:eqn:lorentz_approx_spectral_func} to the computed spectral function. For our purposes, we fix the quasi-particle wave functions to be the eigenstates of the Hartree-Fock Hamiltonian corresponding to the state of interest. Once the quasi-particle wave functions are fixed, one can determine their energy and lifetime by fitting the spectral function to a Lorentzian profile. To this end, we define the spectral function \emph{projected} in the quasi-particle basis (which is the same as the band basis of the Hartree-Fock Hamiltonian) to be given by
\begin{equation}
	\label{app:eqn:band_basis_spectral_function}
	A_{n} \left( \omega, \vec{k} \right) =  \sum_{\substack{i, \eta, s \\ i', \eta', s'}} \varphi^*_{n;i \eta s} \left( \vec{k} \right) A_{i \eta s;i' \eta' s'} \left( \omega, \vec{k} \right) \varphi_{n;i' \eta' s'} \left( \vec{k} \right).
\end{equation}
The quasi-particle energies and lifetimes can be determined by fitting $A_{n} \left( \omega, \vec{k} \right)$ to a Lorentzian profile
\begin{equation}
	A_{n} \left( \omega, \vec{k} \right) \approx \frac{\Gamma_n \left( \vec{k} \right)}{\pi} \frac{1}{ \left( \omega - E_n \left( \vec{k} \right) \right)^2 + \Gamma^2_n \left( \vec{k} \right)}.
\end{equation}

Using \cref{app:eqn:lorentz_approx_spectral_func}, we can approximate the conductivity function of the THF model as 
\begin{equation}
	\label{app:eqn:conductivity_function_band_basis}
	\sigma^{\alpha} \left( \omega \right) \approx \frac{1}{N_0 \Omega_0} \sum_{\vec{k}} \sum_{n,m} v^{\alpha}_{mn} \left( \vec{k} \right) A_{n} \left( \omega, \vec{k} \right) v^{\alpha}_{nm} \left( \vec{k} \right) A_{m} \left( \omega, \vec{k} \right),
\end{equation}
where the quasi-particle projected group velocity $v^{\alpha}_{mn}$ is defined exactly as in \cref{app:eqn:qp_group_velocity} for the non-interacting case, with the only difference being that now $\varphi_{n;i' \eta' s'} \left( \vec{k} \right)$ are the quasi-paprticle wave functions. In \cref{app:eqn:qp_group_velocity}, $h \left( \vec{k} \right)$ is the single-particle THF Hamiltonian of TBG. We now note that the Hartree-Fock THF interaction Hamiltonian matrix $h^{\text{TBG},I,\text{MF}} \left( \vec{k} \right)$ defined in \cref{app:eqn:genera_TBG_int_HF} is \emph{approximately} $\vec{k}$-independent. This is because, in general, the Hartree contribution is always momentum-independent, while the Fock contribution is only momentum-dependent whenever the amplitude of the corresponding scattering process depends on the momentum transfer. In the THF model for TBG, the only interaction terms for which the scattering process depends on the momentum transfer (and for which the corresponding Fock terms will be $\vec{k}$-dependent) are $H_{U_2}$ and $H_{V}$ from \cref{app:eqn:THF_int:U2,app:eqn:THF_int:V}. As a result, $\partial_{\vec{k}} h^{\text{TBG},I,\text{MF}} \left( \vec{k} \right) \approx \vec{0}$, and so we can make the following approximation
\begin{equation}
	\partial_{\vec{k}} h^{\text{TBG}} \left( \vec{k} \right) \approx \partial_{\vec{k}} h^{\text{TBG},\text{MF}} \left( \vec{k} \right).
\end{equation}
Using the fact that $\varphi_{n;i \eta s} \left( \vec{k} \right)$ are the eigenstates of the Hartree-Fock Hamiltonian with energies $\epsilon_n \left( \vec{k} \right)$ 
\begin{equation}
	\label{app:eqn:diag_hf_ham_TBG}
	\sum_{i', \eta', s'} h^{\text{TBG},\text{MF}}_{i \eta s; i' \eta' s'} 
	\left( \vec{k} \right) \varphi_{n; i' \eta' s'} \left( \vec{k} \right) = \epsilon_n \left( \vec{k} \right) \varphi_{n;i \eta s} \left( \vec{k} \right),
\end{equation}
we can rewrite the quasi-particle group velocity as 
\begin{equation}
	\label{app:eqn:approx_qp_group_velocity}
	v^{\alpha}_{nm} \left( \vec{k} \right) \approx \sum_{\substack{i, \eta, s \\ i', \eta', s'}} \varphi^*_{n;i \eta s} \left( \vec{k} \right) \left( \partial_{k^{\alpha}} h^{\text{TBG},\text{MF}}_{i \eta s;i' \eta' s'} \left( \vec{k} \right) \right) \varphi_{m;i' \eta' s'} \left( \vec{k} \right) =  \partial_{k^{\alpha}} \epsilon_{n} \left( \vec{k} \right) \delta_{nm} + \left( \epsilon_{n} \left( \vec{k} \right) - \epsilon_{m} \left( \vec{k} \right) \right) \mathcal{A}_{nm}^{\alpha,\text{HF}} \left( \vec{k} \right),
\end{equation}
where we have introduced the nonabelian Berry connection of the Hartree-Fock bands 
\begin{equation}
	\mathcal{A}_{nm}^{\alpha,\text{HF}} \left( \vec{k} \right) \equiv \sum_{i, \eta, s} \varphi^{*}_{n;i \eta s} \left( \vec{k} \right) \partial_{k^{\alpha}} \varphi_{m;i \eta s} \left( \vec{k} \right).
\end{equation}

Plugging \cref{app:eqn:approx_qp_group_velocity} into \cref{app:eqn:conductivity_function_band_basis}, we find that in the limit of well-defined quasiparticles, we can approximately split the conductivity function $\sigma^{\alpha} \left( \omega \right)$ into a ``kinetic'' and a ``quantum-geometric'' contribution
\begin{equation}
	\label{app:eqn:split_of_conductivity_function}
	\sigma^{\alpha} \left( \omega \right) \approx \sigma^{\alpha}_{\text{Kinetic}} \left( \omega \right) +  \sigma^{\alpha}_{\text{Geometric}} \left( \omega \right),
\end{equation}
which are given respectively by the $n=m$ and $n \neq m$ terms of in the summation from \cref{app:eqn:conductivity_function_band_basis}
\begin{align}
	\sigma^{\alpha}_{\text{Kinetic}} \left( \omega \right) &= \frac{1}{N_0 \Omega_0} \sum_{\vec{k}} \sum_{n} \left[ v^{\alpha}_{nn} \left( \vec{k} \right) A_{n} \left( \omega, \vec{k} \right)\right]^2 \nonumber \\
	&\approx \frac{1}{N_0 \Omega_0} \sum_{\vec{k}} \sum_{n} \left[ \frac{\Gamma_n \left( \vec{k} \right)}{\pi} \frac{\partial_{k^{\alpha}} \epsilon_{n} \left( \vec{k} \right)}{ \left( \omega - E_n \left( \vec{k} \right) \right)^2 + \Gamma^2_n \left( \vec{k} \right)} \right]^2 \label{app:eqn:cond_func_kin} \\
	\sigma^{\alpha}_{\text{Geometric}} \left( \omega \right) &= \frac{1}{N_0 \Omega_0} \sum_{\vec{k}} \sum_{n \neq m} v^{\alpha}_{mn} \left( \vec{k} \right) A_{n} \left( \omega, \vec{k} \right) v^{\alpha}_{nm} \left( \vec{k} \right) A_{m} \left( \omega, \vec{k} \right) \nonumber \\
	&\approx \frac{1}{N_0 \Omega_0} \sum_{\vec{k}} \sum_{n \neq m} \frac{\Gamma_n \left( \vec{k} \right)}{\pi} \frac{\Gamma_m \left( \vec{k} \right)}{\pi} \frac{\left[ \left( \epsilon_{n} \left( \vec{k} \right) - \epsilon_{m} \left( \vec{k} \right) \right) \mathcal{A}_{nm}^{\alpha,\text{HF}} \left( \vec{k} \right) \right]^{2}}{\left[ \left( \omega - E_n \left( \vec{k}\underline{} \right) \right)^2 + \Gamma^2_n \left( \vec{k} \right) \right] \left[ \left( \omega - E_m \left( \vec{k} \right) \right)^2 + \Gamma^2_m \left( \vec{k} \right) \right] }. \label{app:eqn:cond_func_geo}
\end{align}
We remind the reader that in order to separate the conductivity function into kinetic and geometric part, we have employed the assumption that a quasi-particle basis exist in which the spectral function is diagonal. In deriving the second lines of both \cref{app:eqn:cond_func_kin,app:eqn:cond_func_geo} we have \emph{additionally} assumed that:
\begin{enumerate}
	\item The quasi-particle wave functions of the state we consider are the same as the eigenstates of the Hartree-Fock Hamiltonian of the same state.
	\item The Hartree-Fock interaction Hamiltonian has negligible $\vec{k}$-dependence.
	\item The diagonal part of the spectral function in the band basis takes the form of a Lorentzian.
\end{enumerate}

Approximating the quasi-particle excitation energies by the Hartree-Fock charge-one excitation ones $E_n \left( \vec{k} \right) \approx \epsilon_n \left( \vec{k} \right)$, we find that the kinetic contribution to the conductivity function is given by
\begin{equation}
	\label{app:eqn:cond_func_kin_approx}
	\sigma^{\alpha}_{\text{Kinetic}} \left( \omega \right) \approx \frac{1}{N_0 \Omega_0} \sum_{\vec{k}} \sum_{n} \left[ \frac{\Gamma_n \left( \vec{k} \right)}{\pi} \frac{\partial_{k^{\alpha}} E_{n} \left( \vec{k} \right)}{ \left( \omega - E_n \left( \vec{k} \right) \right)^2 + \Gamma^2_n \left( \vec{k} \right)} \right]^2.
\end{equation}
Upon performing the quadratic expansion of the quasi-particle energies $E_{n} \left( \vec{k} \right) \approx \frac{\abs{\vec{k}}^2}{2 m_n} + \Delta_n - \mu$ and also assuming that $\Gamma_{n} \left( \vec{k} \right)$ is momentum-independent, it is easy to see that the kinetic part of the conductivity is precisely what is modeled by \cref{app:eqn:cond_func_quadratic} -- the conductivity function computed for the family of quadratic band Hamiltonians from \cref{app:sec:asymptotes:model}. The second contribution from \cref{app:eqn:split_of_conductivity_function} depends on the nonabelian Berry connection and therefore on the wave function on the quasi-particles (not just on their energy). In our toy-model calculations from \cref{app:sec:asymptotes:model}, this topological contribution is ignored. In the following, we will show that this approximation is justified.

To assess the relative contribution of the quantum-geometric part of the conductivity function, we consider the nine correlated insulators of TBG from \cref{app:tab:model_states}. First, we determine the Hartree-Fock Hamiltonian of these states as outlined in \cref{app:sec:many_body_rev:ground_states}. Using the Hartree-Fock Hamiltonian, the Green's function of the system is
\begin{equation}
	\label{app:eqn:gf_for_geometric_contribution}
	\mathcal{G} \left( \omega + i 0^{+}, \vec{k} \right) = \left[ \left(\omega + \mu \right) \mathbb{1} - h^{\text{TBG},\text{MF}} \left( \vec{k} \right) + i \Gamma \right]^{-1},
\end{equation}
where we employ $\vec{k}$-independent lifetime matrix 
\begin{equation}
	\Gamma_{i \eta s;i' \eta' s'} = \begin{cases}
		\Gamma_f \delta_{i i'} \delta_{\eta \eta'} \delta_{s s'}, & \qq{if} 5 \leq i,i' \leq 6 \\
		\Gamma_c \delta_{i i'} \delta_{\eta \eta'} \delta_{s s'}, & \qq{otherwise} \\
	\end{cases},
\end{equation}
with $\Gamma_f$ and $\Gamma_c$ denoting the lifetime broadening factors for the $f$- and $c$-electrons. We set $\Gamma_c = \SI{1}{\milli\electronvolt}$ and vary the $f$-electron lifetime within th interval $\SI{1}{\milli\electronvolt} \leq \Gamma_f \leq \SI{10}{\milli\electronvolt}$. Using the Green's function from \cref{app:eqn:gf_for_geometric_contribution}, we obtain the spectral function of the system using \cref{app:eqn:spectral_function}. 

The \emph{exact} conductivity function of the system can be obtained through \cref{app:eqn:def_conduc_func}. By projecting the spectral function in the Hartree-Fock band basis using \cref{app:eqn:band_basis_spectral_function}, we then compute the band-diagonal spectral function $A_{n} \left(\omega, n \right)$. The group velocity $v^{\alpha}_{nm} \left( \vec{k} \right)$ of the quasi-particle excitations is then obtained from the Hartree-Fock Hamiltonian from \cref{app:eqn:approx_qp_group_velocity}. Using the group velocity $v^{\alpha}_{nm} \left( \vec{k} \right)$ and the band-diagonal spectral function $A_{n} \left(\omega, n \right)$, we then compute the kinetic part of the conductivity function from the first line of \cref{app:eqn:cond_func_kin}.

With both the exact and the kinetic part of the conductivity functions computed, we assess the relative quantum-geometric contribution using 
\begin{equation}
	\epsilon_{\text{Geometric}} = \sqrt{ \frac{ \int_{-\omega_0}^{\omega_0} \dd{\omega} \abs{ \sigma^{\alpha} \left( \omega \right) - \sigma_{\text{Kinetic}}^{\alpha} \left( \omega \right)}^2}{\int_{-\omega_0}^{\omega_0} \dd{\omega} \abs{ \sigma^{\alpha} \left( \omega \right) }^2} },
\end{equation} 
for $\omega_0 = \SI{100}{\milli\electronvolt}$. Since the thermoelectric coefficients are computed from the conductivity function, $\epsilon_{\text{Geometric}}$ offers a quantitative estimate for the quantum-geometric contribution to the Seebeck coefficient. For all nine insulators we consider and any lifetime of the $f$-electrons $\SI{1}{\milli\electronvolt} \leq \Gamma_f \leq \SI{10}{\milli\electronvolt}$, we find that $\epsilon_{\text{Geometric}} \approx 3\%$, implying that the quantum-geometric contribution to the conductivity function is negligible.

\section{Numerical results for the Seebeck coefficient in the non-interacting limit} \label{app:sec:seebeck_non_int}

This \siSection{} presents comprehensive numerical results for the Seebeck coefficient of the THF model for both TBG and TSTG in the non-interacting limit. The full results are plotted in \cref{app:sec:seebeck_non_int:results}, while a brief discussion on the main features of the results is provided in \cref{app:sec:seebeck_non_int:discussion}.

We employ the method outlined in \cref{app:sec:thermoelectric_response:details_on_numerics:non_interactions} in which we fix the lifetime broadening of the $c$-electrons in TBG to $\Gamma_c = \SI{1}{\milli\electronvolt}$, and the lifetimes of the $c$- and $d$-electrons in TSTG to $\Gamma_c = \SI{1.5}{\milli\electronvolt}$. We provide results for a range of positive fillings $0 \leq \nu \leq 5$, temperatures $T$, and $f$-electron broadening factors $\Gamma_f$. For TSTG, we also consider the effects of a finite perpendicular displacement field $\mathcal{E}$, showing results for multiple values of the latter in the range $0 \leq \mathcal{E}/\si{\milli\electronvolt} \leq 25$. As demonstrated in \cref{app:sec:thermoelectric_response:evaluating_the_transport_coefficients:PH_symmetry}, the Seebeck coefficient of TBG (TSTG) is antisymmetric with respect to filling, due to many-body (spatial) particle-hole symmetry. Consequently, we focus exclusively on positive fillings.

\subsection{Discussion} \label{app:sec:seebeck_non_int:discussion}

\subsubsection{Main features of the non-interacting Seebeck coefficient} \label{app:sec:seebeck_non_int:discussion:main_features}

\begin{figure}[!t]\includegraphics[width=\textwidth]{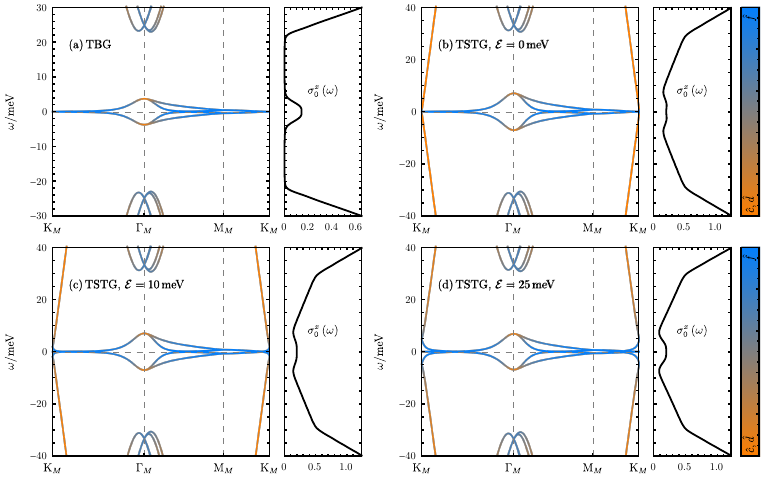}\subfloat{\label{app:fig:conductivity_func_nonint:a}}\subfloat{\label{app:fig:conductivity_func_nonint:b}}\subfloat{\label{app:fig:conductivity_func_nonint:c}}\subfloat{\label{app:fig:conductivity_func_nonint:d}}\caption{Non-interacting band structures and conductivity functions of the THF model. The results for TBG are shown in (a), while (b)-(c) show results for TSTG with $\mathcal{E}/\si{\milli\electronvolt} = 0,\,10,\,25$. In each panel, the band structure is shown on the left, while the conductivity function (corresponding to charge neutrality) is shown on the right. The band structures show the bands in \emph{both} valleys. The color of each band indicates its fermionic character: bluer shades signify a stronger $f$-electron character, while orange shades indicate a predominant $c$- and/or $d$-fermion character. Bands with an orange hue, due to the dispersive nature of the corresponding fermions, contribute more significantly to the conductivity function. Note that the conductivity function is dimensionless in units where $\hbar = 1$.}\label{app:fig:conductivity_func_nonint}\end{figure}

\begin{figure}[!h]\includegraphics[width=\textwidth]{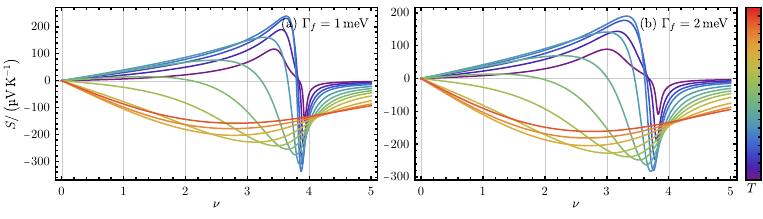}\subfloat{\label{app:fig:seebeck_non_int_1:a}}\subfloat{\label{app:fig:seebeck_non_int_1:b}}\caption{Seebeck coefficient in the non-interacting limit of TBG. The $f$-electron lifetime is shown in each panel, while the line colors denote the temperature (purple indicates the lowest temperature and red, the highest). In order of increasing temperatures, the thermopower curves correspond to $T/\si{\kelvin}=5$, $10$, $15$, $20$, $30$, $40$, $50$, $60$, $75$, $100$, $125$, $150$.}\label{app:fig:seebeck_non_int_1}\end{figure}

\begin{figure}[!h]\includegraphics[width=\textwidth]{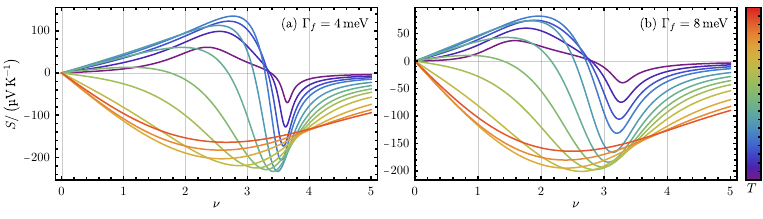}\subfloat{\label{app:fig:seebeck_non_int_2:a}}\subfloat{\label{app:fig:seebeck_non_int_2:b}}\caption{Seebeck coefficient in the non-interacting limit of TBG. The $f$-electron lifetime is shown in each panel, while the line colors denote the temperature (purple indicates the lowest temperature and red, the highest). In order of increasing temperatures, the thermopower curves correspond to $T/\si{\kelvin}=5$, $10$, $15$, $20$, $30$, $40$, $50$, $60$, $75$, $100$, $125$, $150$.}\label{app:fig:seebeck_non_int_2}\end{figure}

\begin{figure}[!h]\includegraphics[width=\textwidth]{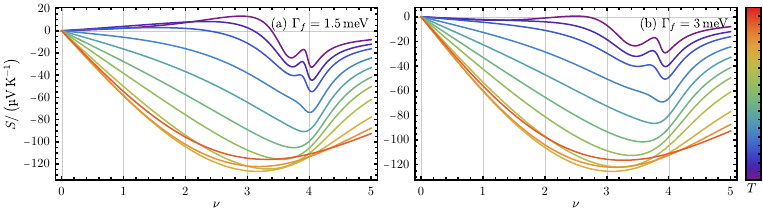}\subfloat{\label{app:fig:seebeck_non_int_3:a}}\subfloat{\label{app:fig:seebeck_non_int_3:b}}\caption{Seebeck coefficient in the non-interacting limit of TSTG at $\mathcal{E}=\SI{0}{\milli\electronvolt}$. The $f$-electron lifetime is shown in each panel, while the line colors denote the temperature (purple indicates the lowest temperature and red, the highest). In order of increasing temperatures, the thermopower curves correspond to $T/\si{\kelvin}=15$, $22.5$, $30$, $45$, $60$, $75$, $90$, $112.5$, $150$, $187.5$, $225$.}\label{app:fig:seebeck_non_int_3}\end{figure}

In order to provide an understanding of the Seebeck coefficient in the non-interacting limit, we plot the single-particle band structures together with the corresponding conductivity functions $\sigma^{x}_{0} \left( \omega \right)$ for both TBG and TSTG in \cref{app:fig:conductivity_func_nonint}. The latter is defined in the non-interacting limit by \cref{app:eqn:def_non_int_conductivity}. The main features of the thermoelectric response of TBG and TSTG can be summarized as follows:
\begin{itemize}
	\item For $\Gamma_f= \SI{1}{\milli\electronvolt}$ and low temperatures ($T \lesssim \SI{50}{\kelvin}$), the Seebeck coefficient of TBG shown in \cref{app:fig:seebeck_non_int_1:a} is positive for $0 \leq \nu \lesssim 4$, crosses zero for $\nu \approx 4$, then becomes negative. Focusing on the band insulator filling $\nu=4$, this behavior aligns with the conventional expectation for the Seebeck coefficient of an insulator (without lifetime asymmetry), as discussed in \cref{app:sec:asymptotes:heavylight:phenomenology}. The same conclusion can be obtained by inspecting the conductivity function from \cref{app:fig:conductivity_func_nonint:a}: using the Mott formula from \cref{app:eqn:mott_formula} (which should is valid for the lowest temperature considered here) and starting from charge neutrality, the conductivity function initially decreases (leading to a positive Seebeck coefficient) and then sharply increases at the remote band edge (which leads to a negative Seebeck coefficient).

	\item For higher temperatures ($T \gtrsim \SI{50}{\kelvin}$), the Seebeck coefficient of TBG from \cref{app:fig:seebeck_non_int_1:a} becomes fully negative for positive fillings. In \cref{app:sec:seebeck_non_int:discussion:analytic_cross}, we provide an analytical explanation for this crossover, and also compute an analytical estimate of the crossover temperature at which the Seebeck coefficient turns negative. For now, we only mention that for $0 \leq \nu \lesssim 4$, the Seebeck coefficient is dominated by the active band contribution at low temperatures (as the chemical potential is close to the active band edge). For larger temperatures, the electron excitation carriers from the remote band become thermally activated. Their significantly stronger dispersion makes the remote band's contribution to the Seebeck coefficient predominant at high temperatures, resulting in an overall negative Seebeck coefficient for $0 \leq \nu \lesssim 4$.
	
	\item In \cref{app:fig:seebeck_non_int_1:b,app:fig:seebeck_non_int_2:a,app:fig:seebeck_non_int_2:b}, we investigate the effects of a larger $f$-electron broadening factor $\Gamma_f$. The findings are as follows:
	\begin{itemize}
		\item For low-temperatures ($T \lesssim \SI{50}{\kelvin}$), the Seebeck coefficient for $\Gamma_f > \SI{1}{\milli\electronvolt}$ is qualitatively similar to the one computed for $\Gamma_f = \SI{1}{\milli\electronvolt}$. One key difference is that as $\Gamma_f$ is increased, the electron and hole peaks, as well as the point where the Seebeck coefficient crosses zero move to lower fillings than in the $\Gamma_f= \SI{1}{\milli\electronvolt}$ case. This shift occurs because the $f$-electron-dominated active TBG bands are not completely filled when the chemical potential is at the edge of the active (remote) band, which corresponds to the hole (electron) peak in the Seebeck coefficient.
		
		\item In the low-temperature regime, \emph{both} the hole and electron peaks decrease as $\Gamma_f$ is increased. This is because the $f$- and $c$-electrons hybridize with one another and an increase in $\Gamma_f$ leads to overall increased lifetime broadening for both the active and the remote TBG bands. However, since the active TBG bands exhibit a larger $f$-electron character, an increase in $\Gamma_f$ leads to a more pronounced decrease in the positive hole peak of the Seebeck coefficient.
		
		\item The crossover temperature, where the remote band contribution becomes dominant for $\nu > 0$, remains similar with increasing $f$-electron broadening (around $T \approx \SI{50}{\kelvin}$). Above this temperature, the influence of the remote band leads to a fully negative Seebeck coefficient for all positive fillings considered.
	\end{itemize}
	
	\item For TSTG without any displacement field, the conductivity function $\sigma_{0} \left( \omega \right)$ is shown in \cref{app:fig:conductivity_func_nonint:b}. Focusing on $\omega > 0$, we find that, compared to the conductivity function of TBG from \cref{app:fig:conductivity_func_nonint:a}, which vanishes in the gap between the active and remote bands, for TSTG, the conductivity function is nonzero in the gap between the active and remote bands of the TBG sector (due to the presence of the additional Dirac cone) and increases almost linearly towards the remote bands. At the remote bands, the slope in the conductivity function increases even more due to the strongly-dispersing remote TBG bands. As a result of the extra Dirac cone, the low-temperature Seebeck coefficient of TSTG at $\Gamma_f = \SI{1.5}{\milli\electronvolt}$, shown in \cref{app:fig:seebeck_non_int_3:a}, exhibits several qualitative changes compared to the TBG Seebeck coefficient in \cref{app:fig:seebeck_non_int_1:a}:
	\begin{itemize}
			
		\item For $\omega > 0$, the decreasing portion of the conductivity function at the active band edge is less pronounced in TSTG than in TBG, as observed in \cref{app:fig:conductivity_func_nonint:a} and \cref{app:fig:conductivity_func_nonint:b}, respectively. Using the Mott formula from \cref{app:eqn:mott_formula}, this leads to a much smaller hole peak in the Seebeck coefficient at the lowest temperatures. Qualitatively, we note that for $0 \leq \nu \lesssim 4$, the TBG sector of TSTG is hole-doped (which should lead to positive Seebeck coefficient), but the strongly-dispersive Dirac sector is \emph{electron}-doped (which should lead to a \emph{negative} Seebeck) coefficient. The overall effect is a smaller positive peak in the Seebeck coefficient at low temperatures for $0 \leq \nu \lesssim 4$.
		
		\item In TSTG at low temperatures, there exist \emph{two} (rather than one) negative electron peaks in the Seebeck coefficient. From the Mott formula in \cref{app:eqn:mott_formula}, the first peak corresponds to the case when the chemical potential is at the \emph{active} band edge, when the conductivity function in \cref{app:fig:conductivity_func_nonint:b} turns from decreasing to strongly increasing (as a result of the mirror-odd Dirac cone contribution). The second peak arises at the remote band edge, where the slope of the conductivity function sharply increases due to the strongly-dispersive remote TBG bands.
	\end{itemize}
	
	\item As the temperature increases in TSTG, the Seebeck coefficient turns entirely negative, influenced predominantly by the remote band contribution. The presence of the Dirac cone, which consistently favors a negative Seebeck coefficient across all temperatures, results in a lower crossover temperature for TSTG (where the Seebeck coefficient becomes completely negative) compared to TBG.
	
	\item In \cref{app:fig:seebeck_non_int_3:b,app:fig:seebeck_non_int_4:a,app:fig:seebeck_non_int_4:b}, we illustrate the effects of an increased $f$-electron broadening factor for TSTG at zero-displacement field. It can be seen that:
		\begin{itemize}
			
			\item The primary effect of a larger $f$-electron broadening factor is the reduction of the positive hole peak in the Seebeck coefficient at low temperatures and, to a lesser degree, the diminishment of the second negative electron peak (which is attributable to the remote bands).
			
			\item For $\Gamma_f = \SI{3}{\milli\electronvolt}$, the low-temperature positive and negative peaks in the Seebeck coefficient, shown in \cref{app:fig:seebeck_non_int_3:b}, shift towards charge neutrality. This happens for the same reason as in TBG: when the chemical potential is at the corresponding band edges, the active TBG bands are less filled when $\Gamma_f$ is increased. 
			
			\item With $\mathcal{E}=0$, the Dirac cone and TBG-like contributions remain uncoupled, leaving the Dirac cone unaffected by the increased $\Gamma_f$. Consequently, for the $\Gamma \geq \SI{6}{\milli\electronvolt}$ results shown in \cref{app:fig:seebeck_non_int_4}, the Seebeck coefficient becomes dominated by the Dirac cone and is negative for \emph{all} the temperature considered
			
			\item In all cases, as the temperature rises, the electron excitations from the strongly-dispersing remote band become thermally activated. These excitations dominate the thermoelectric coefficients for all positive fillings, leading to an overall \emph{negative} Seebeck coefficient.
		\end{itemize}
		
	\item In \crefrange{app:fig:seebeck_non_int_5}{app:fig:seebeck_non_int_14} we also illustrate the effects of a finite displacement field $5 \leq \mathcal{E}/\si{\milli\electronvolt} \leq 25$ on the Seebeck coefficient of TSTG. The primary qualitative effect of an increased displacement field is the enlargement of the positive peak of the Seebeck coefficient at low temperatures and low $f$-electron broadening factors $\Gamma_f$. This phenomenon can be understood by examining the band structures and non-interacting conductivity functions of TSTG, as shown in \cref{app:fig:conductivity_func_nonint:c,app:fig:conductivity_func_nonint:d} for $\mathcal{E} = \SI{10}{\milli\electronvolt}$ and $\mathcal{E} = \SI{25}{\milli\electronvolt}$: as the displacement field is increased, the Dirac cones formed by the $f$- and $d$-electrons at the $\mathrm{K}^{(')}_M$ point hybridize; as a result, the $d$-electrons move away from the active bands, which leads to a more pronounced decrease in the conductivity function near the active band edge. In turn, through the Mott formula from \cref{app:eqn:mott_formula}, this leads to an enhanced positive hole peak in the Seebeck coefficient at low temperatures and small $f$-electron broadening.
	
\end{itemize}

\subsubsection{Origin of the temperature crossover in TBG} \label{app:sec:seebeck_non_int:discussion:analytic_cross}

As shown in \cref{app:fig:seebeck_non_int_1:a}, the Seebeck coefficient of TBG for $0 \leq \nu \lesssim 4$ is positive at low temperatures, but turns negative at higher temperature. In this section, we provide an analytical explanation for this behavior, as well as an analytical estimate of the critical temperature at which the Seebeck coefficient switches sign.

As will be analytically shown below, the origin of this crossover can be found in the formula of the Seebeck coefficient from \cref{app:eqn:seebeck_definition}: 
\begin{itemize}
	\item The $L^{xx}_{11} (\mu)$ thermoelectric coefficient, given by \cref{app:eqn:L11_static_diagonal_nonint} in the non-interacting regime is strictly negative (the integrand \cref{app:eqn:L11_static_diagonal_nonint} is the product of a negative function -- the derivative of the Fermi-Dirac distribution function -- and a positive function -- the non-interacting conductivity function).
	\item As a result, the \emph{sign} of the Seebeck coefficient is determined only by the sign of the $L^{xx}_{12} (\mu)$ thermoelectric coefficient, whose expression is given in \cref{app:eqn:L12_static_diagonal_nonint}: a positive $L^{xx}_{12} (\mu)$ implies a negative sign of the Seebeck coefficient.
	\item From the numerical results of \cref{app:fig:seebeck_non_int_1:a}, we find that it is enough to consider the sign of the Seebeck coefficient near $\nu=0$ (which corresponds to $\mu=0$), since the sign of Seebeck coefficient does not change from small positive fillings to large positive fillings ($0<\nu\lesssim 3.5$ at low temperatures and $0<\nu\lesssim 5$ at high temperatures). At the crossover temperature, the Seebeck coefficient at small positive fillings turns from positive to negative.
\end{itemize} 

To obtain an analytical description of the crossover between the positive and negative Seebeck regimes, we start by splitting the conductivity function into two contributions, one from the active bands, and one from the remote bands
\begin{equation}
	\sigma_{0}^{x} \left( \omega \right) = \sigma_{0,\text{active}}^{x} \left( \omega \right) + \sigma_{0,\text{remote}}^{x} \left( \omega \right),
\end{equation}
where we have defined
\begin{equation}
	\label{app:eqn:separation_cond_function_def}
	\sigma^{x}_{0,\text{active}} \left( \omega \right) = \begin{cases}
		\sigma_{0}^{x} \left( \omega \right) & \qq{if} \abs{\omega} \leq \abs{\frac{\gamma}{2}} \\
		0 & \qq{otherwise}
	\end{cases} \qq{and}
	\sigma^{x}_{0,\text{remote}} \left( \omega \right) = \begin{cases}
		\sigma_{0}^{x} \left( \omega \right) & \qq{if} \abs{\omega} > \abs{\frac{\gamma}{2}} \\
		0 & \qq{otherwise}
	\end{cases}.
\end{equation}
In \cref{app:eqn:separation_cond_function_def}, the THF parameter $\abs{\gamma}$, which was defined in \cref{app:sec:HF_review:single_particle:TBG}, is approximately equal to the gap between the active and the remote bands. From \cref{app:eqn:L12_static_diagonal_nonint}, we can also split the $L^{xx}_{12} ( \mu )$ thermoelectric coefficient into an active and a remote contribution
\begin{equation}
	L^{xx}_{12} (\mu) = L^{xx}_{12,\text{active}} (\mu) + L^{xx}_{12,\text{remote}} (\mu),
\end{equation}
where 
\begin{align}
	L^{xx}_{12,\text{active}} (\mu) =& - \frac{\pi}{\beta} \int_{\abs{\omega} \leq \abs{\frac{\gamma}{2}}} \dd{\omega} \left( \omega - \mu \right) \pdv{n_{\mathrm{F}} \left( \omega - \mu \right)}{\omega} \sigma_{0,\text{active}}^{x} \left( \omega \right), \label{app:eqn:L12_active_contrib_def} \\
	L^{xx}_{12,\text{remote}} (\mu) =& - \frac{\pi}{\beta} \int_{\abs{\omega} > \abs{\frac{\gamma}{2}}} \dd{\omega} \left( \omega - \mu \right) \pdv{n_{\mathrm{F}} \left( \omega - \mu \right)}{\omega} \sigma_{0,\text{remote}}^{x} \left( \omega \right).
	\label{app:eqn:L12_remote_contrib_def}
\end{align}
To make some analytical progress, we now \emph{qualitatively} approximate the active and remote band contributions to the conductivity functions by linear expressions
\begin{align}
	\sigma_{0,\text{active}}^{x} \left( \omega \right) &= \begin{cases}
		\alpha_{\text{active}} (M - \abs{\omega}) &\qq{if} \abs{\omega} \leq M \\
		0 & \qq{otherwise}
	\end{cases}, \label{app:eqn:approx_non_int_cond_active}\\
	\sigma_{0,\text{remote}}^{x} \left( \omega \right) &= \begin{cases}
		\alpha_{\text{remote}} (\abs{\omega} - \abs{\gamma}) &\qq{if} \abs{\omega} \geq \abs{\gamma} \\
		0 & \qq{otherwise}
	\end{cases}, \label{app:eqn:approx_non_int_cond_remote}
\end{align}
where the slopes of the non-interacting conductivity function can be directly estimated from \cref{app:fig:conductivity_func_nonint:a},
\begin{equation}
	\alpha_{\text{active}} \approx \SI{0.05}{\milli\electronvolt^{-1}} \qq{and}
	\alpha_{\text{remote}} \approx \SI{0.075}{\milli\electronvolt^{-1}}.
\end{equation}

\begin{figure}[!t]\includegraphics[width=0.5\textwidth]{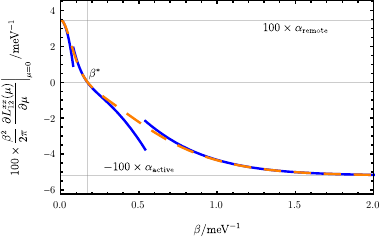}\caption{The temperature crossover of the Seebeck coefficient of TBG in the non-interacting limit. We plot the slope of the $L^{xx}_{12} \left( \mu \right)$ coefficient at charge neutrality as a function of the inverse temperature $\beta$. The slope computed from \cref{app:eqn:dL12_mu_active,app:eqn:dL12_mu_remote} is shown by the dashed orange line, while the blue line corresponds to the approximations from \cref{app:eqn:dL12_mu_active_approx,app:eqn:dL12_mu_remote_approx}.	The low- and high-temperature asymptotes for the slope of $L^{xx}_{12} \left( \mu \right)$ from \cref{app:eqn:low_high_temp_asymptote_slope_L12} are also shown by the two labeled gray horizontal hairlines. The critical inverse temperature $\beta^{*}$, where the slope changes sign, is shown by the vertical gray hairline.}\label{app:fig:temp_cross_non_int_seebeck}\end{figure}

The simple approximations for the conductivity functions from \cref{app:eqn:approx_non_int_cond_active,app:eqn:approx_non_int_cond_remote} allow us to analytically evaluate the integrals in \cref{app:eqn:L12_active_contrib_def,app:eqn:L12_remote_contrib_def} and obtain 
\begin{align}
	\frac{L^{xx}_{12,\text{active}} (\mu)}{\alpha_{\text{active}}} =& -\frac{\pi}{\beta^2} \left[ -2 \mu  \log \left(1 + e^{-\beta  \mu }\right)+(\mu -M) \log \left(1 + e^{\beta  (M-\mu )}\right)+(\mu +M) \log \left( 1 + e^{-\beta (\mu+M) } \right) \right] \nonumber \\
	+& \frac{2\pi}{\beta^3} \left[ \text{Li}_2\left(-e^{\beta  (M-\mu )}\right) + \Li_{2} \left(-e^{-\beta  (M+\mu )}\right) - 2 \Li_{2} \left(-e^{-\beta  \mu }\right) \right], \\
\frac{L^{xx}_{12,\text{remote}} (\mu)}{\alpha_{\text{remote}}} =& -\frac{\pi}{\beta^2}  \left[(\mu - \abs{\gamma} ) \log \left( 1 + e^{\beta  (\abs{\gamma} -\mu)} \right)+   (\abs{\gamma} +\mu ) \log \left( 1 + e^{\beta  (\abs{\gamma} +\mu)} \right) \right] \nonumber \\
	+& \frac{2\pi}{\beta^3} \left[ \Li_{2} \left(-e^{\beta  (\abs{\gamma} -\mu )}\right) - \Li_{2} \left(-e^{\beta  (\abs{\gamma} +\mu )}\right) \right],
\end{align}
where $\text{Li}_2(x)$ is the polylogarithm function of order $2$ employed also in \cref{app:sec:asymptotes:model:formula_gamma}. The sign of the $L^{xx}_{12} ( \mu )$ coefficient around charge neutrality can be determined through a Taylor expansion
\begin{equation}
	L^{xx}_{12} ( \mu ) = \eval{\pdv{L^{xx}_{12} ( \mu )}{\mu}}_{\mu = 0} \mu + \mathcal{O} \left( \mu^{3} \right) 
	= \left( \eval{\pdv{L^{xx}_{12,\text{active}} ( \mu )}{\mu}}_{\mu = 0} + \eval{\pdv{L^{xx}_{12,\text{remote}} ( \mu )}{\mu}}_{\mu = 0} \right) \mu + \mathcal{O} \left( \mu^{3} \right),
\end{equation}
where the fact that $L^{xx}_{12} ( \mu )$ is an odd function of $\mu$ follows from the many-body charge conjugation symmetry of TBG, as shown in \cref{app:eqn:relation_of_L_correlators}, while
\begin{align}
	\eval{\pdv{L^{xx}_{12,\text{active}} ( \mu )}{\mu}}_{\mu = 0} &= \frac{2\pi \alpha_{\text{active}}}{\beta^2} \left( \log \left( \cosh \frac{\beta M}{2} \right) - \frac{\beta M}{2} \tanh \frac{\beta M}{2} \right), \label{app:eqn:dL12_mu_active} \\
	\eval{\pdv{L^{xx}_{12,\text{remote}} ( \mu )}{\mu}}_{\mu = 0} &= \frac{2\pi \alpha_{\text{remote}}}{\beta^2} \left( \log \left(2 \cosh \frac{\beta \abs{\gamma}}{2} \right) - \frac{\beta \abs{\gamma}}{2}  \tanh \frac{\beta  \abs{\gamma} }{2} \right) \label{app:eqn:dL12_mu_remote}.
\end{align}
We can employ the following asymptotic series
\begin{align}
	\log \left( \cosh x \right) &= \begin{cases}
		\frac{x^2}{2} + \mathcal{O} \left( x^4 \right)& \qq{for} x \ll 1 \\
		x - \log (2) + e^{-2 x} + \mathcal{O} \left( e^{-4 x} \right)& \qq{for} x \gg 1
	\end{cases} \\
	\tanh x &= \begin{cases}
		x + \mathcal{O} \left( x^3 \right)& \qq{for} x \ll 1 \\
		1 -  2 e^{-2 x} + \mathcal{O} \left( e^{-4 x} \right)& \qq{for} x \gg 1
	\end{cases},
\end{align}
to obtain simpler expressions for \cref{app:eqn:dL12_mu_active,app:eqn:dL12_mu_remote}
\begin{align}
	\eval{\pdv{L^{xx}_{12,\text{active}} ( \mu )}{\mu}}_{\mu = 0} &= \frac{2\pi \alpha_{\text{active}}}{\beta^2}
	\begin{cases}
		e^{- \beta M} \left( 1 + \beta M \right) - \log(2) + \mathcal{O} \left(e^{- 2 \beta M} \right) & \qq{for} \beta M > 2 \\
		-\frac{1}{8} \beta^2 M^2  + \mathcal{O} \left( \beta^4 M^4 \right) & \qq{for} \beta M < 2\\
	\end{cases}, \label{app:eqn:dL12_mu_active_approx} \\
	\eval{\pdv{L^{xx}_{12,\text{remote}} ( \mu )}{\mu}}_{\mu = 0} &= \frac{2\pi \alpha_{\text{remote}}}{\beta^2}
	\begin{cases}
		e^{- \beta \abs{\gamma}} \left( 1 + \beta \abs{\gamma} \right) + \mathcal{O} \left(e^{- 2 \beta \abs{\gamma}} \right) & \qq{for} \beta \abs{\gamma} > 2 \\
		-\frac{1}{8} \beta^2 \abs{\gamma}^2 + \log (2) + \mathcal{O} \left( \beta^4 \abs{\gamma}^4 \right) & \qq{for} \beta \abs{\gamma} < 2 
	\end{cases}. \label{app:eqn:dL12_mu_remote_approx}
\end{align}

In \cref{app:fig:temp_cross_non_int_seebeck}, we plot the partial derivative of $L^{xx}_{12} ( \mu )$ with respect to the chemical potential at charge neutrality as a function of the inverse temperature $\beta$, as well as its asymptotic expressions from \cref{app:eqn:dL12_mu_active_approx,app:eqn:dL12_mu_remote_approx}. The fact that $	L^{xx}_{12} ( \mu )$ for $\mu>0$ changes sign as a function of temperature can be immediately inferred from the latter
\begin{equation}
	\label{app:eqn:low_high_temp_asymptote_slope_L12}
	\eval{\pdv{L^{xx}_{12} ( \mu )}{\mu}}_{\mu = 0} \approx \frac{2 \pi \log(2)}{\beta^2} \times \begin{cases}
		- \alpha_{\text{active}} & \qq{for} \beta M, \beta \abs{\gamma} \gg 1 \\
		\alpha_{\text{remote}} & \qq{for} \beta M, \beta \abs{\gamma} \ll 1 \\
	\end{cases}.
\end{equation}
From \cref{app:fig:temp_cross_non_int_seebeck}, we find that the inverse  crossover temperature is $\beta^{*} = \SI{0.173}{\milli\electronvolt^{-1}}$, which corresponds to a crossover temperature $T^{*} = \SI{67.2}{\kelvin}$. This estimated value for $T^{*}$ matches the numerical result in \cref{app:fig:seebeck_non_int_1:a} quite well: from the latter we find that $50 < T^{*} / \si{\milli\electronvolt} <60$.

An analytical expression for the crossover temperature can also be obtained \textit{a posteriori} by noting that $\beta^{*} \abs{\gamma} > 2 > \beta^{*} M$ (which follows from \cref{app:fig:temp_cross_non_int_seebeck} by inspection). In this interval, we can use the analytical expressions in \cref{app:eqn:dL12_mu_active_approx,app:eqn:dL12_mu_remote_approx} and set $\eval{\pdv{L^{xx}_{12} ( \mu )}{\mu}}_{\mu = 0, \beta = \beta^{*}} = 0$ to obtain
\begin{equation}
	\frac{1}{8} \beta^{*2} M^2 = \frac{\alpha_{\text{remote}}}{\alpha_{\text{active}}} \left( 1 + \beta^{*} \abs{\gamma} \right) e^{- \beta^{*} \abs{\gamma}},
\end{equation}
Making the further approximation $\beta^{*} \abs{\gamma} \gg 1$, we obtain
\begin{equation}
	\beta^{*} \abs{\gamma}  \approx \frac{8 \alpha_{\text{remote}}}{\alpha_{\text{active}}} \left(\frac{\abs{\gamma}}{M} \right)^2  e^{- \beta^{*} \abs{\gamma}},
\end{equation}
whose solution is given by the Lambert $W$ function~\cite{ABR65}
\begin{equation}
	\beta^{*} \approx \frac{1}{\abs{\gamma}} W_{0} \left( \frac{8 \alpha_{\text{remote}}}{\alpha_{\text{active}}} \left(\frac{\abs{\gamma}}{M} \right)^2  \right).
\end{equation}
For large arguments, the Lambert $W$ function admits the following asymptote~\cite{ABR65}
\begin{equation}
	W_{0} (x) = \log (x) - \log\log(x) + \mathcal{O} \left( 1 \right). 
\end{equation}
Since $\frac{8 \alpha_{\text{remote}}}{\alpha_{\text{active}}} \left(\frac{\abs{\gamma}}{M} \right)^2 \gg 1$, we find that 
\begin{equation}
	\beta^{*} \approx \frac{1}{\abs{\gamma}} \eval{\left( \log (x) - \log \log(x) \right)}_{x = \frac{8 \alpha_{\text{remote}}}{\alpha_{\text{active}}} \left(\frac{\abs{\gamma}}{M} \right)^2} = \SI{0.173}{\milli\electronvolt^{-1}}.
\end{equation}

\newcommand{\lastFigsNo}{2}
\clearpage
\subsection{Results} \label{app:sec:seebeck_non_int:results}

\ifthenelse{\equal{\lastFigsNo}{1}}{\FloatBarrier\clearpage}{}

\ifthenelse{\equal{\lastFigsNo}{3}}{\FloatBarrier\clearpage}{}

\ifthenelse{\equal{\lastFigsNo}{2}}{\FloatBarrier\clearpage}{}

\begin{figure}[!h]\includegraphics[width=\textwidth]{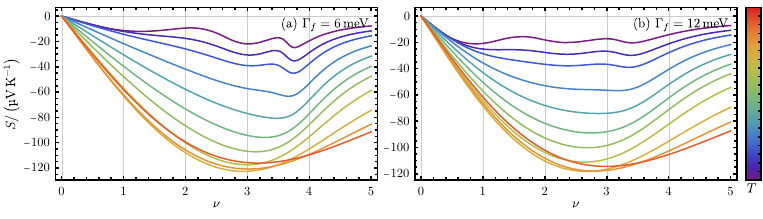}\subfloat{\label{app:fig:seebeck_non_int_4:a}}\subfloat{\label{app:fig:seebeck_non_int_4:b}}\caption{Seebeck coefficient in the non-interacting limit of TSTG at $\mathcal{E}=\SI{0}{\milli\electronvolt}$. The $f$-electron lifetime is shown in each panel, while the line colors denote the temperature (purple indicates the lowest temperature and red, the highest). In order of increasing temperatures, the thermopower curves correspond to $T/\si{\kelvin}=15$, $22.5$, $30$, $45$, $60$, $75$, $90$, $112.5$, $150$, $187.5$, $225$.}\label{app:fig:seebeck_non_int_4}\end{figure}

\ifthenelse{\equal{\lastFigsNo}{1}}{\FloatBarrier\clearpage}{}

\begin{figure}[!h]\includegraphics[width=\textwidth]{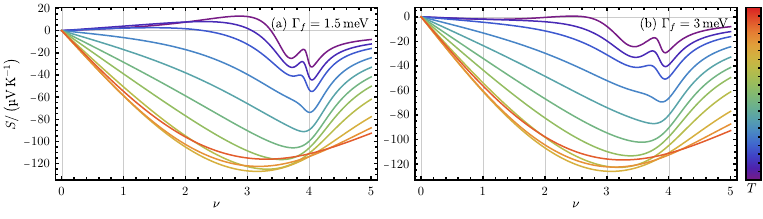}\subfloat{\label{app:fig:seebeck_non_int_5:a}}\subfloat{\label{app:fig:seebeck_non_int_5:b}}\caption{Seebeck coefficient in the non-interacting limit of TSTG at $\mathcal{E}=\SI{5}{\milli\electronvolt}$. The $f$-electron lifetime is shown in each panel, while the line colors denote the temperature (purple indicates the lowest temperature and red, the highest). In order of increasing temperatures, the thermopower curves correspond to $T/\si{\kelvin}=15$, $22.5$, $30$, $45$, $60$, $75$, $90$, $112.5$, $150$, $187.5$, $225$.}\label{app:fig:seebeck_non_int_5}\end{figure}

\ifthenelse{\equal{\lastFigsNo}{3}}{\FloatBarrier\clearpage}{}

\begin{figure}[!h]\includegraphics[width=\textwidth]{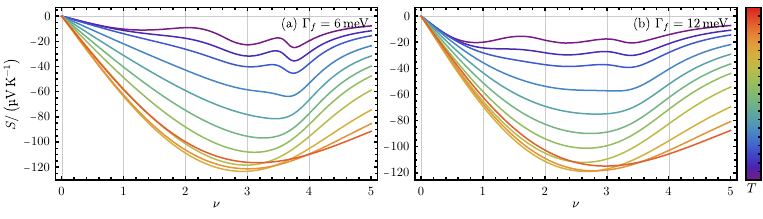}\subfloat{\label{app:fig:seebeck_non_int_6:a}}\subfloat{\label{app:fig:seebeck_non_int_6:b}}\caption{Seebeck coefficient in the non-interacting limit of TSTG at $\mathcal{E}=\SI{5}{\milli\electronvolt}$. The $f$-electron lifetime is shown in each panel, while the line colors denote the temperature (purple indicates the lowest temperature and red, the highest). In order of increasing temperatures, the thermopower curves correspond to $T/\si{\kelvin}=15$, $22.5$, $30$, $45$, $60$, $75$, $90$, $112.5$, $150$, $187.5$, $225$.}\label{app:fig:seebeck_non_int_6}\end{figure}

\ifthenelse{\equal{\lastFigsNo}{2}}{\FloatBarrier\clearpage}{}

\begin{figure}[!h]\includegraphics[width=\textwidth]{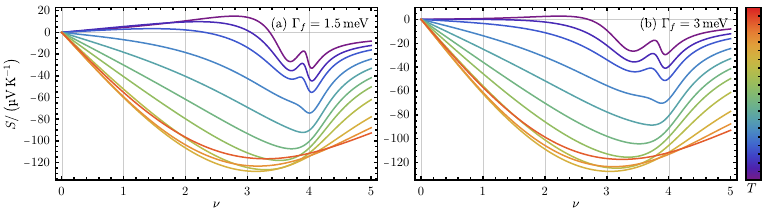}\subfloat{\label{app:fig:seebeck_non_int_7:a}}\subfloat{\label{app:fig:seebeck_non_int_7:b}}\caption{Seebeck coefficient in the non-interacting limit of TSTG at $\mathcal{E}=\SI{10}{\milli\electronvolt}$. The $f$-electron lifetime is shown in each panel, while the line colors denote the temperature (purple indicates the lowest temperature and red, the highest). In order of increasing temperatures, the thermopower curves correspond to $T/\si{\kelvin}=15$, $22.5$, $30$, $45$, $60$, $75$, $90$, $112.5$, $150$, $187.5$, $225$.}\label{app:fig:seebeck_non_int_7}\end{figure}

\ifthenelse{\equal{\lastFigsNo}{1}}{\FloatBarrier\clearpage}{}

\begin{figure}[!h]\includegraphics[width=\textwidth]{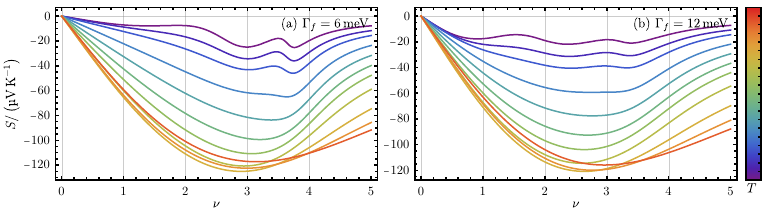}\subfloat{\label{app:fig:seebeck_non_int_8:a}}\subfloat{\label{app:fig:seebeck_non_int_8:b}}\caption{Seebeck coefficient in the non-interacting limit of TSTG at $\mathcal{E}=\SI{10}{\milli\electronvolt}$. The $f$-electron lifetime is shown in each panel, while the line colors denote the temperature (purple indicates the lowest temperature and red, the highest). In order of increasing temperatures, the thermopower curves correspond to $T/\si{\kelvin}=15$, $22.5$, $30$, $45$, $60$, $75$, $90$, $112.5$, $150$, $187.5$, $225$.}\label{app:fig:seebeck_non_int_8}\end{figure}

\ifthenelse{\equal{\lastFigsNo}{3}}{\FloatBarrier\clearpage}{}

\begin{figure}[!h]\includegraphics[width=\textwidth]{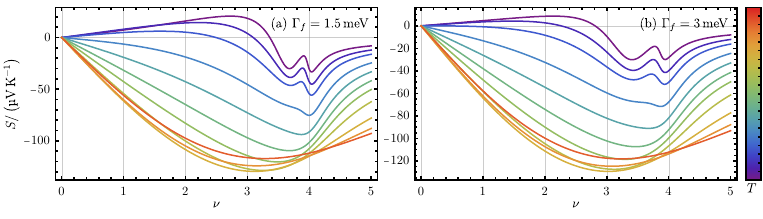}\subfloat{\label{app:fig:seebeck_non_int_9:a}}\subfloat{\label{app:fig:seebeck_non_int_9:b}}\caption{Seebeck coefficient in the non-interacting limit of TSTG at $\mathcal{E}=\SI{15}{\milli\electronvolt}$. The $f$-electron lifetime is shown in each panel, while the line colors denote the temperature (purple indicates the lowest temperature and red, the highest). In order of increasing temperatures, the thermopower curves correspond to $T/\si{\kelvin}=15$, $22.5$, $30$, $45$, $60$, $75$, $90$, $112.5$, $150$, $187.5$, $225$.}\label{app:fig:seebeck_non_int_9}\end{figure}

\ifthenelse{\equal{\lastFigsNo}{2}}{\FloatBarrier\clearpage}{}

\begin{figure}[!h]\includegraphics[width=\textwidth]{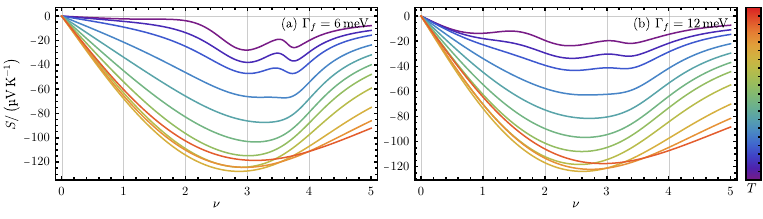}\subfloat{\label{app:fig:seebeck_non_int_10:a}}\subfloat{\label{app:fig:seebeck_non_int_10:b}}\caption{Seebeck coefficient in the non-interacting limit of TSTG at $\mathcal{E}=\SI{15}{\milli\electronvolt}$. The $f$-electron lifetime is shown in each panel, while the line colors denote the temperature (purple indicates the lowest temperature and red, the highest). In order of increasing temperatures, the thermopower curves correspond to $T/\si{\kelvin}=15$, $22.5$, $30$, $45$, $60$, $75$, $90$, $112.5$, $150$, $187.5$, $225$.}\label{app:fig:seebeck_non_int_10}\end{figure}

\ifthenelse{\equal{\lastFigsNo}{1}}{\FloatBarrier\clearpage}{}

\begin{figure}[!h]\includegraphics[width=\textwidth]{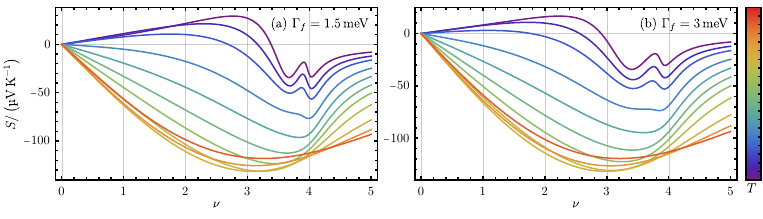}\subfloat{\label{app:fig:seebeck_non_int_11:a}}\subfloat{\label{app:fig:seebeck_non_int_11:b}}\caption{Seebeck coefficient in the non-interacting limit of TSTG at $\mathcal{E}=\SI{20}{\milli\electronvolt}$. The $f$-electron lifetime is shown in each panel, while the line colors denote the temperature (purple indicates the lowest temperature and red, the highest). In order of increasing temperatures, the thermopower curves correspond to $T/\si{\kelvin}=15$, $22.5$, $30$, $45$, $60$, $75$, $90$, $112.5$, $150$, $187.5$, $225$.}\label{app:fig:seebeck_non_int_11}\end{figure}

\ifthenelse{\equal{\lastFigsNo}{3}}{\FloatBarrier\clearpage}{}

\begin{figure}[!h]\includegraphics[width=\textwidth]{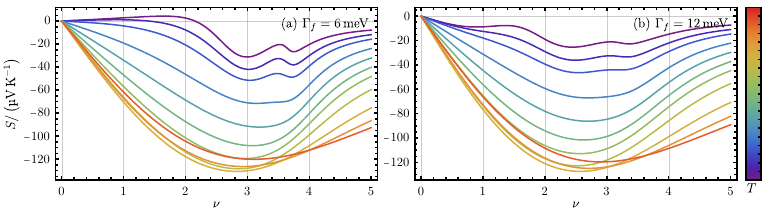}\subfloat{\label{app:fig:seebeck_non_int_12:a}}\subfloat{\label{app:fig:seebeck_non_int_12:b}}\caption{Seebeck coefficient in the non-interacting limit of TSTG at $\mathcal{E}=\SI{20}{\milli\electronvolt}$. The $f$-electron lifetime is shown in each panel, while the line colors denote the temperature (purple indicates the lowest temperature and red, the highest). In order of increasing temperatures, the thermopower curves correspond to $T/\si{\kelvin}=15$, $22.5$, $30$, $45$, $60$, $75$, $90$, $112.5$, $150$, $187.5$, $225$.}\label{app:fig:seebeck_non_int_12}\end{figure}

\ifthenelse{\equal{\lastFigsNo}{2}}{\FloatBarrier\clearpage}{}

\begin{figure}[!t]\includegraphics[width=\textwidth]{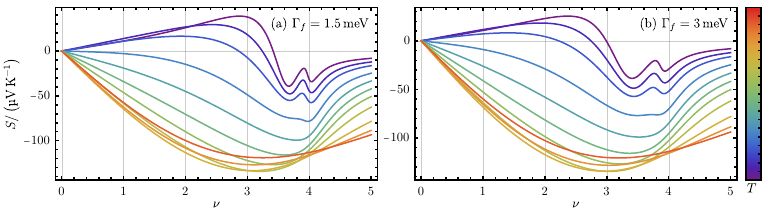}\subfloat{\label{app:fig:seebeck_non_int_13:a}}\subfloat{\label{app:fig:seebeck_non_int_13:b}}\caption{Seebeck coefficient in the non-interacting limit of TSTG at $\mathcal{E}=\SI{25}{\milli\electronvolt}$. The $f$-electron lifetime is shown in each panel, while the line colors denote the temperature (purple indicates the lowest temperature and red, the highest). In order of increasing temperatures, the thermopower curves correspond to $T/\si{\kelvin}=15$, $22.5$, $30$, $45$, $60$, $75$, $90$, $112.5$, $150$, $187.5$, $225$.}\label{app:fig:seebeck_non_int_13}\end{figure}

\ifthenelse{\equal{\lastFigsNo}{1}}{\FloatBarrier\clearpage}{}

\ifthenelse{\equal{\lastFigsNo}{1} \OR \equal{\lastFigsNo}{2} }{\begin{figure}[!t]}{\begin{figure}[!h]}\includegraphics[width=\textwidth]{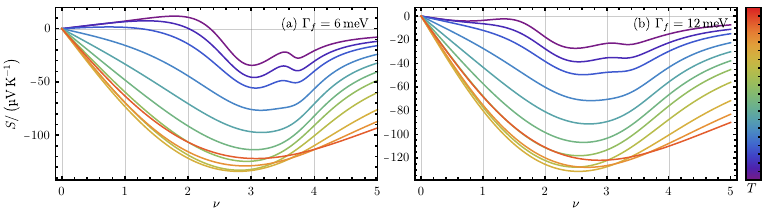}\subfloat{\label{app:fig:seebeck_non_int_14:a}}\subfloat{\label{app:fig:seebeck_non_int_14:b}}\caption{Seebeck coefficient in the non-interacting limit of TSTG at $\mathcal{E}=\SI{25}{\milli\electronvolt}$. The $f$-electron lifetime is shown in each panel, while the line colors denote the temperature (purple indicates the lowest temperature and red, the highest). In order of increasing temperatures, the thermopower curves correspond to $T/\si{\kelvin}=15$, $22.5$, $30$, $45$, $60$, $75$, $90$, $112.5$, $150$, $187.5$, $225$.}\label{app:fig:seebeck_non_int_14}\end{figure}

\ifthenelse{\equal{\lastFigsNo}{3}}{\FloatBarrier\clearpage}{}%

\section{Numerical results for the Seebeck coefficient in the symmetry-broken states} \label{app:sec:seebeck_sym_br}

In this \siSection{}, we detail numerical results for the Seebeck coefficient of the correlated ground state candidates of TBG and TSTG from \cref{app:tab:model_states}. These calculations follow the methods described in \cref{app:sec:thermoelectric_response:details_on_numerics:interactions}.

\subsection{Overview of the results}\label{app:sec:seebeck_sym_br:overview}

We have computed the Seebeck coefficient for each positively filled correlated ground state candidate listed in \cref{app:tab:model_states}, both at integer filling and for the correlated phases obtained by doping the latter up to a maximum of $\Delta \nu = \pm 0.5$. Our analysis includes TBG, TSTG with no displacement field ($\mathcal{E} = \SI{0}{\milli\electronvolt}$), and TSTG with a small value of displacement field ($\mathcal{E} = \SI{25}{\milli\electronvolt}$). We consider a range of temperatures $5 \lesssim T/\si{\kelvin} \lesssim 50$ (for TBG) and $5 \lesssim T/\si{\kelvin} \lesssim 70$ (for TSTG). The results are shown in \cref{app:sec:seebeck_sym_br:results} and discussed in \cref{app:sec:seebeck_sym_br:discussion}.

For each correlated ground state from \cref{app:tab:model_states}, the results are shown in a single figure ({\it e.g.}{} the Seebeck coefficient for the $\protect\IfStrEqCase{6}{{1}{\ket{\nu={}4} }
		{2}{\ket{\nu={}3, \mathrm{IVC}}}
		{3}{\ket{\nu={}3, \mathrm{VP}}}
		{4}{\ket{\nu={}2, \mathrm{K-IVC}}}
		{5}{\ket{\nu={}2, \mathrm{VP}}}
		{6}{\ket{\nu={}1, (\mathrm{K-IVC}+\mathrm{VP})}}
		{7}{\ket{\nu={}1, \mathrm{VP}}}
		{8}{\ket{\nu=0, \mathrm{K-IVC}}}
		{9}{\ket{\nu=0, \mathrm{VP}}}
	}
	[nada]
$ correlated ground state candidate is shown in \cref{app:fig:seebeck_ord_6}). To maintain clarity, all figures share a consistent format. The three panels (a)-(c) show the computed Seebeck coefficient for TBG, TSTG without displacement field ($\mathcal{E} = \SI{0}{\milli\electronvolt}$), and TSTG with displacement field ($\mathcal{E} = \SI{25}{\milli\electronvolt}$), respectively. In each panel, the Seebeck coefficient for different temperatures is color-coded, with purple corresponding to the lowest temperature and red corresponding to the highest one, as indicated in the figure caption and in the accompanying color bar. 

For the $0 \leq \nu \leq 3$ correlated ground state candidates, as we dope away from the integer filling, the system can transition to a disordered ({\it i.e.}{} symmetric) phase~\cite{RAI23a}. From a numerical standpoint, we find that the symmetry-broken solution cannot be stabilized for large values of doping, especially at high temperature. Consequently, the Seebeck coefficient is only plotted for the symmetry-broken phase. The degree of symmetry-breaking in the numerical solution is assessed by computing the onsite $f$-electron density matrix, which is defined by 
\begin{equation}
	\label{app:eqn:def_onsite_f_electron_den_mat}
	\mathcal{O}^f_{\alpha \eta s;\alpha' \eta' s'} \equiv \frac{1}{N_0} \sum_{\vec{k}} \varrho_{(\alpha + 4) \eta s; (\alpha' + 4) \eta' s'} \left( \vec{k} \right).
\end{equation} 
We only plot the results of which $\mathcal{O}^f$ is different from identity such that 
\begin{equation}
	\mathcal{O}^{f} - \frac{ \Tr \left( \mathcal{O}^{f} \right) }{N_f}  \mathbb{1} \neq \mathbb{0}, \qq{(up to numerical accuracy),}
\end{equation}
where $N_f = 8$ is the number of $f$-electron flavors. For instance, the Seebeck coefficient computed by doping the $\protect\IfStrEqCase{4}{{1}{\ket{\nu={}4} }
		{2}{\ket{\nu={}3, \mathrm{IVC}}}
		{3}{\ket{\nu={}3, \mathrm{VP}}}
		{4}{\ket{\nu={}2, \mathrm{K-IVC}}}
		{5}{\ket{\nu={}2, \mathrm{VP}}}
		{6}{\ket{\nu={}1, (\mathrm{K-IVC}+\mathrm{VP})}}
		{7}{\ket{\nu={}1, \mathrm{VP}}}
		{8}{\ket{\nu=0, \mathrm{K-IVC}}}
		{9}{\ket{\nu=0, \mathrm{VP}}}
	}
	[nada]
$ correlated ground state candidate for TBG at $T = \SI{26}{\kelvin}$ shown in \cref{app:fig:seebeck_ord_4:a} is only plotted in the interval $1.9 \lesssim \nu \lesssim 2.35$, beyond which the system transitions to a symmetric state (for which $\mathcal{O}^{f} \propto \mathbb{1}$).

Finally, we note that the results are shown only for the $\nu \geq 0$ correlated ground state candidates from \cref{app:tab:model_states}. The results for the $\nu < 0$ correlated phases can be inferred directly from the $\nu \geq 0$ ones, given that the Seebeck coefficient is an odd function of the filling, as shown in \cref{app:sec:thermoelectric_response:evaluating_the_transport_coefficients:PH_symmetry}. Despite our analysis covering only nine correlated ground state candidates, we anticipate our conclusions to be robust and not heavily dependent on the specific choice of ground state wave function.  This is supported by our analytical calculations from \cref{app:sec:asymptotes:heavylight}, where we have shown that the key ingredient necessary to reproduce the experimentally-observed negative Seebeck coefficient of TBG at low temperatures around $\nu = 1$ and $\nu = 2$~\cite{MER24} is the lifetime asymmetry between the heavy hole and light electron excitations. Because this dichotomy between the electron and hole excitation~\cite{KAN21} can be reproduced in other ground state candidates, such as the intervalley Kelul\'e spiral state~\cite{KWA21,WAG22,WAN23c,NUC23} or the time-reversal symmetric intervalley coherent state~\cite{BUL20a,SON22,CAL22d,NUC23}, we expect the phenomenology of the Seebeck coefficient to be largely insensitive to the exact ground state wave function under consideration. Our findings also indicate that for the correlated ground states at integer fillings $0 \leq \nu < 4$ that we consider, the Seebeck coefficient is virtually independent on the exact ground state candidate wave function.

\subsection{Discussion} \label{app:sec:seebeck_sym_br:discussion}

In this section, we provide a point-by-point discussion of our comprehensive numerical results on the Seebeck coefficient of the THF model in its symmetry-broken phases from \cref{app:tab:model_states}. We begin by discussing the results for TBG and then move on to consider TSTG without and with a perpendicularly applied displacement field.

\subsubsection{Discussion of the Seebeck coefficient in TBG} \label{app:sec:seebeck_sym_br:discussion:TBG}

For TBG, our numerical results, as presented in \cref{app:sec:seebeck_sym_br:results}, reveal several notable features. We summarize these starting from the charge neutrality point:
\begin{itemize}
	
	\item First, we note that the various correlated ground state candidates at the same integer filling exhibit qualitatively similar Seebeck coefficients. This similarity is attributed to the charge-one excitation spectrum of these insulators, which also displays qualitative resemblance, as seen in \cref*{DMFT:app:sec:results_corr_ins} of Ref.~\cite{CAL23b}
	
	\item For the $\nu = 0$ correlated states shown in \cref{app:fig:seebeck_ord_8:a,app:fig:seebeck_ord_9:a}, the Seebeck coefficient is odd with respect to the filling $\nu$. This is a consequence of the particle-hole symmetry of the model, as shown in \cref{app:sec:thermoelectric_response:evaluating_the_transport_coefficients:PH_symmetry}. The hole and electron peaks have the same absolute height and the Seebeck coefficient has the typical shape expected for a particle-hole symmetric semiconductor: at positive (negative) doping, thermoelectric transport is dominated by electron (hole) excitations leading to a negative (positive) Seebeck coefficient.

	\item The thermoelectric response of the $\nu = 1$ correlated insulators shown in \cref{app:fig:seebeck_ord_6:a,app:fig:seebeck_ord_7:a} differs significantly from the Seebeck coefficient obtained near charge neutrality. This can be traced to the different character of the charge-one excitations around the integer filling on the electron and hole doping sides, as discussed already in \cref*{DMFT:app:sec:results_corr_ins:general_remarks} of Ref.~\cite{CAL23b}. On the one hand, the electron excitations consist of strongly-dispersive, weakly-correlated, and, as a result, long-lived $c$-electrons. On the other hand, the hole excitations include both $c$-electrons (around the $\Gamma_M$ point) and $f$-electron (at the edges of the moir\'e BZ). The latter, which dominate the hole bands, are strongly-correlated and have a temperature-dependent scattering rate, with large temperature leading to larger scattering-rates, a more incoherent $f$-electron spectral function, and a shorter $f$-electron lifetime, as mentioned in \cref*{DMFT:app:sec:results_corr_ins:general_remarks} of Ref.~\cite{CAL23b}~\footnote{For example the high-temperature spectral functions at $\nu=0.8$ shown in \cref*{DMFT:app:fig:sym_br_bs_6_TBG_high:d,DMFT:app:fig:sym_br_bs_7_TBG_high:d} of Ref.~\cite{CAL23b} have a more incoherent hole band than the low-temperature ones shown in \cref*{DMFT:app:fig:sym_br_bs_6_TBG_low:d,DMFT:app:fig:sym_br_bs_7_TBG_low:d} of Ref.~\cite{CAL23b}.}. As a result, at a sufficiently high temperature, the hole excitations are expected to have a small lifetime. As discussed in \cref{app:sec:seebeck_sym_br:discussion}, this should lead to electron-dominated thermoelectric transport (yielding a negative Seebeck coefficient) at elevated temperatures. Our numerical findings indicate that:
	\begin{itemize}
		\item At low temperature, the lifetime asymmetry between the electron (mainly $c$-electrons) and hole excitations (mainly $f$-electrons) is not significant. The Seebeck coefficient therefore resembles the thermoelectric response of a particle-hole asymmetric semiconductor: while the Seebeck coefficient is \emph{not} odd around $\nu = 1$, the electron and hole peaks have approximately equal absolute heights, implying that the thermoelectric transport is electron (hole) dominated on the electron (hole) doping side.

		\item As the temperature is increased, the lifetime of the $f$-electrons decreases (while the lifetime of the $c$-electrons, which is added by hand, as explained in \cref{app:sec:thermoelectric_response:details_on_numerics:interactions} remains constant). This leads to a significantly larger broadening factor for the hole band, diminishing the hole contribution to the thermoelectric transport. As a result, the hole peak of the Seebeck coefficient shrinks, while the electron one remains approximately the same. At higher temperature, the Seebeck coefficient is negative for most if not all the doping range we consider (for example the Seebeck coefficient of the $\protect\IfStrEqCase{7}{{1}{\ket{\nu={}4} }
		{2}{\ket{\nu={}3, \mathrm{IVC}}}
		{3}{\ket{\nu={}3, \mathrm{VP}}}
		{4}{\ket{\nu={}2, \mathrm{K-IVC}}}
		{5}{\ket{\nu={}2, \mathrm{VP}}}
		{6}{\ket{\nu={}1, (\mathrm{K-IVC}+\mathrm{VP})}}
		{7}{\ket{\nu={}1, \mathrm{VP}}}
		{8}{\ket{\nu=0, \mathrm{K-IVC}}}
		{9}{\ket{\nu=0, \mathrm{VP}}}
	}
	[nada]
$ insulator at $T\geq \SI{27}{\kelvin}$). In this regime, transport is electron-dominated \emph{even} on the hole doping side. For large lifetime asymmetry, we also observe that the negative electron peak of the Seebeck coefficient moves towards charge neutrality, as expected from the heavy-light model discussed in \cref{app:sec:asymptotes:heavylight}.

		\item Finally, we note that the discontinuity in the Seebeck coefficient at $\nu \approx 0.6$ can be traced to the system transitioning between two different ordered states, as one more $f$-electron flavor becomes unoccupied, as seen in \cref*{DMFT:app:fig:sym_br_bs_6_TBG_low:a,DMFT:app:fig:sym_br_bs_7_TBG_low:a} of Ref.~\cite{CAL23b}.  
	\end{itemize}

	\item For the parameters we considered in this work and in Ref.~\cite{CAL23b}, the $\nu = 2$ correlated insulators do not feature an indirect gap, as shown in \cref*{DMFT:app:fig:sym_br_bs_4_TBG_low,DMFT:app:fig:sym_br_bs_5_TBG_low,DMFT:app:fig:sym_br_bs_4_TBG_high,DMFT:app:fig:sym_br_bs_5_TBG_high} of Ref.~\cite{CAL23b}. The Seebeck coefficient shown in \cref{app:fig:seebeck_ord_4:a,app:fig:seebeck_ord_5:a} exhibits the following characteristics:
	\begin{itemize}
		\item Starting from the lowest temperatures, due to the lack of an indirect gap, the negative electron peak of the Seebeck coefficient (which appears when the chemical potential is at the electron excitation band edge) occurs for $\nu < 2$. Because for $\nu < 2$ the chemical potential is also within the hole band (which by itself would give rise to a positive Seebeck coefficient), the magnitude of the electron peak is much smaller for the $\nu = 2$ correlated insulators compared to the $\nu=1$ ones.
		
		\item Similar to the $\nu = 1$ correlated insulators, there is a pronounced asymmetry between the electron (mainly $c$-electrons, strongly-dispersive, long-lived) and hole (mainly $f$-electrons, strongly-correlated with small dispersion, more incoherent) excitations. At higher temperatures, when the hole excitations experience increased lifetime broadening, the Seebeck coefficient becomes entirely negative.
		
		\item At low temperatures, the Seebeck coefficient displays two negative peaks: one at $\nu \lesssim 2$ and another one at $\nu \approx 2.4$. The second peak corresponds to the chemical potential reaching the remote band edge at higher dopings as indicated by the band structure calculations from \cref*{DMFT:app:sec:results_corr_ins:4,DMFT:app:sec:results_corr_ins:5} of Ref.~\cite{CAL23b}. Upon increasing the temperature, the Seebeck coefficient is smoothed-out and the two negative peaks merge together. Consequently, the negative peak at $\nu \lesssim 2$ \emph{appears} to be moving away from charge neutrality as the temperature is increased. This is different from the $\nu = 1$ case, where the negative electron peak moves \emph{towards} charge neutrality.
		
	\end{itemize}
	
	\item The $\nu = 3$ correlated ground state candidates considered in this work cannot be stabilized for large temperature and doping. We also mention that a moir\'e translation invariant solution is likely not the ground state of the system in at this filling~\cite{XIE23a}. For the sake of completeness however, we mention that:
	\begin{itemize}
		\item From the band structure calculations shown in \cref*{DMFT:app:sec:results_corr_ins:2,DMFT:app:sec:results_corr_ins:3} of Ref.~\cite{CAL23b}, both the hole and electron excitations of the $\protect\IfStrEqCase{2}{{1}{\ket{\nu={}4} }
		{2}{\ket{\nu={}3, \mathrm{IVC}}}
		{3}{\ket{\nu={}3, \mathrm{VP}}}
		{4}{\ket{\nu={}2, \mathrm{K-IVC}}}
		{5}{\ket{\nu={}2, \mathrm{VP}}}
		{6}{\ket{\nu={}1, (\mathrm{K-IVC}+\mathrm{VP})}}
		{7}{\ket{\nu={}1, \mathrm{VP}}}
		{8}{\ket{\nu=0, \mathrm{K-IVC}}}
		{9}{\ket{\nu=0, \mathrm{VP}}}
	}
	[nada]
$ and $\protect\IfStrEqCase{3}{{1}{\ket{\nu={}4} }
		{2}{\ket{\nu={}3, \mathrm{IVC}}}
		{3}{\ket{\nu={}3, \mathrm{VP}}}
		{4}{\ket{\nu={}2, \mathrm{K-IVC}}}
		{5}{\ket{\nu={}2, \mathrm{VP}}}
		{6}{\ket{\nu={}1, (\mathrm{K-IVC}+\mathrm{VP})}}
		{7}{\ket{\nu={}1, \mathrm{VP}}}
		{8}{\ket{\nu=0, \mathrm{K-IVC}}}
		{9}{\ket{\nu=0, \mathrm{VP}}}
	}
	[nada]
$ correlated ground state candidates are formed by strongly dispersing $c$-electrons. As a result, the Seebeck coefficient vanishes approximately near the integer filling.
		
		\item The band structure results from \cref*{DMFT:app:sec:results_corr_ins:2,DMFT:app:sec:results_corr_ins:3} of Ref.~\cite{CAL23b} show that the remote band is far away from the chemical potential for all the dopings for which the symmetry-broken solution is stable. On the one hand, under hole doping, the $c$-electron states of the seven filled active TBG bands are brought closer to the chemical potential leading also to a positive Seebeck coefficient. On the other hand, under electron doping, all the $c$-electron states above the fermi energy at $\nu = 3$ become occupied, leading to a hole-dominated \emph{positive} Seebeck coefficient. This qualitatively explains why the Seebeck coefficient increases under both hole and electron doping away from integer filling.
		
		\item As the temperature is increased, the $c$-electron states of the seven occupied active TBG bands become more thermally activated leading to an overall increase in the Seebeck coefficient.
	\end{itemize}
	
	\item The Seebeck coefficient behavior for the $\nu = 4$ band insulator, as shown in \cref{app:fig:seebeck_ord_1:a}, displays the following characteristics:
	\begin{itemize}
		\item The magnitude of the Seebeck coefficient is comparable to the $\nu=0$ and $\nu=1$ correlated insulators and much larger than that of the $\nu=2$ and $\nu=3$ correlated ground state candidates. This is attributed to the presence of an indirect gap, as evidenced in the band structure in \cref*{DMFT:app:sec:results_corr_ins:1} of Ref.~\cite{CAL23b}.
		
		\item Similar to the $\nu = 1$ and $\nu = 2$ correlated ground state candidates, there is a pronounced asymmetry between the electron (mainly $c$-electrons, strongly-dispersive, long-lived) and hole (mainly $f$-electrons, strongly-correlated, less dispersive, more incoherent) excitations. At low temperatures, when the $f$-electron lifetime broadening is relatively small, the system does not feature a pronounced lifetime asymmetry and the hole and electron peaks in the Seebeck coefficient are roughly equal in magnitude. The flatter nature of the hole bands and larger lifetime broadening shift the peaks and inflection point of the Seebeck coefficient towards hole doping, even at low temperatures.
		
		\item As the temperature is increased, the $f$-electron scattering rate increases accordingly, leading to a larger broadening of the hole band. In effect the hole peak of the Seebeck coefficient shrinks and the Seebeck coefficient becomes electron-dominated and, thus, negative. 
	\end{itemize}	
	
\end{itemize}

\subsubsection{Discussion of the Seebeck coefficient in TSTG without displacement field} \label{app:sec:seebeck_sym_br:discussion:TSTG_no_e}

Before delving into the results for the Seebeck coefficient of TSTG, it is useful to relate it to the Seebeck coefficient of TBG. As discussed in \cref{app:sec:BM_review:TSTG}, TSTG can be viewed as comprising a TBG-like contribution to which a high-velocity Dirac cone is added. In the presence of mirror symmetry, the TBG and Dirac fermions of TSTG remain uncoupled at the single-particle level. Numerically, we find that all self-consistent solutions obtained in the absence of a displacement field do not spontaneously break mirror symmetry. In other words, the spectral function of TSTG shows no off-diagonal matrix elements between the $d$-fermions and the $c$- and $f$-electrons
\begin{equation}
	A_{i \eta s;i' \eta' s'} \left( \omega, \vec{k} \right) = 0, \qq{for $1 \leq i \leq 6$ and $7 \leq i' \leq 8$ or for $1 \leq i' \leq 6$ and $7 \leq i \leq 8$}.   
\end{equation}
As the $\vec{T}(\vec{k})$ tensors introduced in \cref{app:sec:thermoelectric_response:evaluating_the_transport_coefficients:curr_imag_time} also have no off-diagonal matrix elements coupling the Dirac and TBG-like fermions of TSTG, we can separate the conductivity function for TSTG defined in \cref{app:eqn:def_conduc_func} in the absence of a displacement field
\begin{equation}
	\sigma^{\alpha} \left( \omega \right) = \sigma_{\text{TBG}}^{\alpha} \left( \omega \right) + \sigma_{\text{D}}^{\alpha} \left( \omega \right),
\end{equation}
where the TBG and Dirac part of the conductivity function are respectively defined by
\begin{equation}
	\label{app:eqn:separation_cond_func_TSTG}
	\sigma_{\text{TBG},\text{D}}^{\alpha} \left( \omega \right) =  \frac{1}{N_0 \Omega_0} \sum_{\vec{k}} \Tr \left( T_{\text{TBG},\text{D}}^{\alpha} \left( \vec{k} \right) A_{\text{TBG},\text{D}} \left( \omega, \vec{k} \right) T_{\text{TBG},\text{D}}^{\alpha} \left( \vec{k} \right) A_{\text{TBG},\text{D}} \left( \omega, \vec{k} \right) \right).
\end{equation} 
In \cref{app:eqn:separation_cond_func_TSTG} we have defined the matrix blocks for the $\vec{T} \left( \vec{k} \right)$ tensor and the spectral function $A \left( \omega, \vec{k} \right) $ corresponding to the TBG ($f$- and $c$-electrons) and Dirac ($d$-fermions) sectors of TSTG
\begin{alignat}{6}
	T^{\alpha}_{\text{TBG},i \eta s;i' \eta' s'} \left( \vec{k} \right) &&=& T^{\alpha}_{i \eta s;i' \eta' s'} \left( \vec{k} \right), &&\qq{for} 1 \leq i,i' \leq 6, \quad
	&T^{\alpha}_{\text{D},\alpha \eta s;\alpha' \eta' s'} \left( \vec{k} \right) &&= & T^{\alpha}_{(\alpha + 4)\eta s;(\alpha' + 4) \eta' s'} \left( \vec{k} \right), \\
	A_{\text{TBG},i \eta s;i' \eta' s'} \left( \omega, \vec{k} \right) &&=& A_{i \eta s;i' \eta' s'} \left( \omega, \vec{k} \right), &&\qq{for} 1 \leq i,i' \leq 6, \quad
	&A_{\text{D},\alpha \eta s;\alpha' \eta' s'} \left( \omega, \vec{k} \right) &&= & A_{(\alpha + 4)\eta s;(\alpha' + 4) \eta' s'} \left( \omega, \vec{k} \right),
\end{alignat}

Through \cref{app:eqn:L11_static_diagonal_final,app:eqn:L12_static_diagonal_final}, \cref{app:eqn:separation_cond_func_TSTG} also allows us to separate the thermoelectric response coefficients of TSTG into a TBG and a Dirac cone contribution 
\begin{equation}
	L^{\alpha\alpha}_{11} = L^{\alpha\alpha}_{\text{TBG},11} + L^{\alpha\alpha}_{\text{D},11}, \quad
	L^{\alpha\alpha}_{12} = L^{\alpha\alpha}_{\text{TBG},12} + L^{\alpha\alpha}_{\text{D},12},
\end{equation}
where $L^{\alpha\alpha}_{\text{TBG},11}$ and $L^{\alpha\alpha}_{\text{TBG},12}$ ($L^{\alpha\alpha}_{\text{D},11}$ and $L^{\alpha\alpha}_{\text{D},12}$) are derived from the TBG (Dirac cone) conductivity function as per \cref{app:eqn:separation_cond_func_TSTG}, using the integral formulae from \cref{app:eqn:L11_static_diagonal_final,app:eqn:L12_static_diagonal_final}. Considering the band structure results of the TSTG correlated ground state candidates from \cref*{DMFT:app:sec:results_corr_ins} of Ref.~\cite{CAL23b}, we see that Dirac cone sector of TSTG is more dispersive than the TBG one. Consequently, we anticipate that $\sigma_{\text{D}}^{\alpha} \left( \omega \right) \gg \sigma_{\text{TBG}}^{\alpha} \left( \omega \right)$, which qualitatively holds numerically. In turn this implies that $L^{\alpha\alpha}_{\text{D},11} \gg L^{\alpha\alpha}_{\text{TBG},11}$ and $L^{\alpha\alpha}_{\text{D},12} \gg L^{\alpha\alpha}_{\text{TBG},12}$. Defining 
\begin{equation}
	S_{\text{TBG}} = \frac{\beta}{e} \frac{L^{\alpha\alpha}_{\text{TBG},12}}{L^{\alpha\alpha}_{\text{TBG},11}} \qq{and}
	S_{\text{D}} = \frac{\beta}{e} \frac{L^{\alpha\alpha}_{\text{D},12}}{L^{\alpha\alpha}_{\text{D},11}},
\end{equation}
to be the Seebeck coefficients of the TBG and Dirac cone sectors by themselves, we can perform the following approximation for the Seebeck coefficient of TSTG
\begin{align}
	\frac{e S}{\beta} &=  \frac{L^{\alpha\alpha}_{\text{D},12} + L^{\alpha\alpha}_{\text{TBG},12}}{L^{\alpha\alpha}_{\text{D},11} + L^{\alpha\alpha}_{\text{TBG},11}} 
	=  \frac{L^{\alpha\alpha}_{\text{D},12} \left(1 + \frac{L^{\alpha\alpha}_{\text{TBG},12}}{L^{\alpha\alpha}_{\text{D},12}} \right)}{L^{\alpha\alpha}_{\text{D},11} \left(1 + \frac{L^{\alpha\alpha}_{\text{TBG},11}}{L^{\alpha\alpha}_{\text{D},11}} \right)}
	\approx  \frac{L^{\alpha\alpha}_{\text{D},12}}{L^{\alpha\alpha}_{\text{D},11}} \left(1 + \frac{L^{\alpha\alpha}_{\text{TBG},12}}{L^{\alpha\alpha}_{\text{D},12}} \right) \left(1 - \frac{L^{\alpha\alpha}_{\text{TBG},11}}{L^{\alpha\alpha}_{\text{D},11}} \right)  \nonumber \\ 
	&\approx  \frac{L^{\alpha\alpha}_{\text{D},12}}{L^{\alpha\alpha}_{\text{D},11}} \left(1 + \frac{L^{\alpha\alpha}_{\text{TBG},12}}{L^{\alpha\alpha}_{\text{D},12}} - \frac{L^{\alpha\alpha}_{\text{TBG},11}}{L^{\alpha\alpha}_{\text{D},11}} \right)  = \frac{e S_{\text{D}}}{\beta} \left(1 - \frac{L^{\alpha\alpha}_{\text{TBG},11}}{L^{\alpha\alpha}_{\text{D},11}} \right) + \frac{L^{\alpha\alpha}_{\text{TBG},11}}{L^{\alpha\alpha}_{\text{D},11}} \frac{e S_{\text{TBG}}}{\beta},
	\label{app:eqn:trilayer_seebeck_as_func_of_bilayer_interm}
\end{align}
where we have used the fact that $\frac{L^{\alpha\alpha}_{\text{TBG},11}}{L^{\alpha\alpha}_{\text{D},11}} \ll 1$ and $\frac{L^{\alpha\alpha}_{\text{TBG},12}}{L^{\alpha\alpha}_{\text{D},12}} \ll 1$, following from the greater conductivity of the Dirac sector compared to the TBG one. \Cref{app:eqn:trilayer_seebeck_as_func_of_bilayer_interm} can be equivalently cast as 
\begin{equation}
	\label{app:eqn:trilayer_seebeck_as_func_of_bilayer}
	S \approx  \left(1 - \frac{L^{\alpha\alpha}_{\text{TBG},11}}{L^{\alpha\alpha}_{\text{D},11}} \right) S_{\text{D}} + \frac{L^{\alpha\alpha}_{\text{TBG},11}}{L^{\alpha\alpha}_{\text{D},11}} S_{\text{TBG}} \approx S_{\text{D}} + \frac{L^{\alpha\alpha}_{\text{TBG},11}}{L^{\alpha\alpha}_{\text{D},11}} S_{\text{TBG}}.
\end{equation}

\Cref{app:eqn:trilayer_seebeck_as_func_of_bilayer} is an important qualitative result which warrants further discussion:
\begin{itemize}
	\item Because the mirror symmetry is not broken, in the correlated ground state candidates of TSTG at filling $\nu$ are the direct product of the corresponding correlated TBG insulator and an \emph{approximately} half-filled Dirac cone. For $\nu \geq 0$, the Dirac cone sector is only very lightly electron-doped (due to its large dispersion), while most of the doping happends in the TBG-like sector. As a result, we can approximate $S_{\text{TBG}}$ from \cref{app:eqn:trilayer_seebeck_as_func_of_bilayer} as the Seebeck coefficient of TBG at the corresponding filling.
	
	\item Since the Dirac cone sector is moderately electron-doped beyond half-filling, $S_{\text{D}}$ in \cref{app:eqn:trilayer_seebeck_as_func_of_bilayer} contributes negatively to the Seebeck coefficient (for $\nu > 0$). Notably, $S_{\text{D}}$ remains relatively constant with doping, and as a result most of the filling dependence of the Seebeck coefficient of TSTG for small doping around the integer filling will come from the filling dependence of $S_{\text{TBG}}$. At the same time, $S_{\text{D}}$ will become more negative for larger $\nu$ and so will have a more significant contribution for larger fillings. 
	
	\item The second term in \cref{app:eqn:trilayer_seebeck_as_func_of_bilayer} comes with a small prefactor $\frac{L^{\alpha\alpha}_{\text{TBG},11}}{L^{\alpha\alpha}_{\text{D},11}}$. Thus, we infer that the Seebeck coefficient of TSTG is \emph{qualitatively} akin to that of TBG at the corresponding filling, albeit by a small prefactor ($L^{\alpha\alpha}_{\text{TBG},11}/L^{\alpha\alpha}_{\text{D},11}$) to which a constant negative contribution ($S_{\text{D}}$) is added. 
\end{itemize}

With \cref{app:eqn:trilayer_seebeck_as_func_of_bilayer} at hand, we can qualitatively explain the main features of the Seebeck coefficient in TSTG:
\begin{itemize}
	\item Similar to TBG, different ground state candidates at identical filling levels exhibit qualitatively comparable Seebeck coefficients.
	
	\item Compared the TBG, the magnitude of the particle-hole antisymmetric Seebeck coefficient for the $\nu = 0$ correlated insulators shown in \cref{app:fig:seebeck_ord_8:b,app:fig:seebeck_ord_9:b} is smaller for TSTG. Furthermore, its magnitude increases significantly with temperature. This is attributable to the $S_{\text{D}}$ contribution from \cref{app:eqn:trilayer_seebeck_as_func_of_bilayer}, which significantly increases with temperature: for an electron (hole) doped Dirac cone, elevating the temperature substantially excites more electron (hole) carriers than hole (electron) ones, leading to a more pronounced negative (positive) Seebeck coefficient.
	
	\item For the $\nu = 1$ correlated ground state candidates, the magnitude of the Seebeck coefficient depicted in \cref{app:fig:seebeck_ord_6:b,app:fig:seebeck_ord_7:b} is again diminished by the presence of the strongly-dispersing Dirac cone compared to TBG. Moreover, given that $S_{\text{D}}$ is invariably negative for $\nu  > 0$ (with an increasing magnitude at higher temperatures, as previously discussed), the Seebeck coefficient predominantly remains negative across most considered temperatures, and the magnitude of the electron peak at $\nu \approx 1$ significantly decreases for higher temperatures.		
		
	\item At $\nu = 2$, the Seebeck coefficient shown in \cref{app:fig:seebeck_ord_4:b,app:fig:seebeck_ord_5:b} changes between TBG and TSTG in a similar manner to the $\nu = 1$ correlated ground state candidates. Specifically:
	\begin{itemize}
		\item The magnitude of the Seebeck coefficient is reduced in TSTG compared to TBG.
		\item The Seebeck coefficient remains negative throughout the doping range considered, due to the negative Dirac cone contribution.
		\item The increase in magnitude of the Seebeck coefficient is attributed to the increasing magnitude of $S_{\text{D}}$ as a function of temperature.
		\item The Dirac cone being slightly electron-doped implies that the TBG sector filling is less than $\nu$. Hence, the negative electron peak emerging from the TBG electron band edge, observed at $\nu \lesssim 2$ in TBG [as illustrated in \cref{app:fig:seebeck_ord_4:a,app:fig:seebeck_ord_5:a}], now manifests at marginally higher overall fillings. The second negative peak observed in TBG at $\nu \approx 2.4$ at low temperatures is now absent in the considered doping range for TSTG.
	\end{itemize}
	
	\item At $\nu = 4$, the active TBG bands are completely filled. The Dirac cone sector of TSTG is substantially more electron-doped than for the $\nu \leq 3$ ground state candidates, resulting in a larger density of states of Dirac electrons near the chemical potential, as evidenced in \cref*{DMFT:app:fig:sym_br_bs_1_TSTG_low,DMFT:app:fig:sym_br_bs_1_TSTG_high} of Ref.~\cite{CAL23b}. This leads to a heightened contribution of the Dirac cone sector to the conductivity function of TSTG, and via \cref{app:eqn:trilayer_seebeck_as_func_of_bilayer}, to a marked suppression of the Seebeck coefficient of TSTG in comparison to TBG: the Seebeck coefficient of TSTG at $\nu = 4$ shown in \cref{app:fig:seebeck_ord_1:b} is dominated by the Dirac cone contribution. Because the filling of the TBG sector is smaller that $\nu$, the negative electron peak of the Seebeck coefficient expected when the chemical potential is at the remote band edge becomes shifted towards larger overall fillings, beyond the doping interval we considered.  
\end{itemize}

\subsubsection{Discussion of the Seebeck coefficient in TSTG with displacement field} \label{app:sec:seebeck_sym_br:discussion:TSTG_e}

A nonzero perpendicular displacement field breaks the mirror symmetry of TSTG and couples Dirac cone and TBG-like fermions at the single-particle level. For small values of the displacement field $\mathcal{E}$ (within the range $0 \leq \mathcal{E} / \si{\milli\electronvolt} \leq 25$), Refs.\cite{XIE21b,YU23a} have demonstrated that the ground state of TSTG remains equivalent to that in the absence of a displacement field. Examining the $\vec{k}$-resolved spectral functions from \cref*{DMFT:app:sec:results_corr_ins} of Ref.~\cite{CAL23b}, we observe that the nonzero displacement field leaves the band structures of TSTG in the symmetry-broken phases largely similar to those at zero displacement field. The notable exception occurs near the $\mathrm{K}^{(\prime)}_M$ points, where the nonzero displacement field hybridizes the $f$- and $d$-electrons. Although the system retains its semimetallic nature for small displacement fields~\cite{XIE21b,YU23a}, this hybridization ``pushes away'' some of the $d$-electron spectral weight when the Dirac cone bands and $f$-electron flat charge-one excitation bands cross, as further discussed below.

Consequently, we find that:
\begin{itemize}
	\item The Seebeck coefficient of TSTG in the $\nu = 1$ and $\nu = 0$ correlated ground state candidates, depicted in \cref{app:fig:seebeck_ord_6,app:fig:seebeck_ord_7,app:fig:seebeck_ord_8,app:fig:seebeck_ord_9}, remains virtually unchanged from the zero-field result [panel (b)] when a nonzero displacement field is introduced [panel (c)].
	
	\item For the $\nu = 2$ correlated ground state candidates, the effects of a nonzero displacement field on the Seebeck coefficient, as shown in \cref{app:fig:seebeck_ord_4:c,app:fig:seebeck_ord_5:c}, varies depending on the doping level:
	\begin{itemize}
		\item Under hole doping, the magnitude of the negative Seebeck coefficient with displacement field increases markedly. This is attributed to the proximity of the flat $f$-electron hole bands to the chemical potential. The $f$-electrons, not contributing to the conductivity function, hybridize with the $d$-electrons near the $\mathrm{K}^{(\prime)}_M$ points, shifting the $d$-electron spectral weight away from the hole flat band, as seen in \cref*{DMFT:app:fig:sym_br_bs_4_TSTGu_low:d,DMFT:app:fig:sym_br_bs_5_TSTGu_low:d} of Ref.~\cite{CAL23b}. Specifically, this results in a diminished $d$-electron spectral weight on the hole side within a thermally activated window around the chemical potential. Compared to the zero displacement field case, this leads to a significant negative peak in the Seebeck coefficient at around $\nu = 1.8$, only observed with displacement field.
				
		\item For electron doping, the $f$-electrons are distant from the chemical potential, rendering the hybridization with the $d$-electrons negligible in terms of its effect on the Seebeck coefficient, which remains almost identical to the case without displacement field.
		
	\end{itemize} 
	
	\item The Seebeck coefficient of the $\nu = 3$ correlated ground state insulators shown in \cref{app:fig:seebeck_ord_2:c,app:fig:seebeck_ord_3:c} changes significantly in the presence of displacement field. For $\nu \approx 3.4$, the hole $f$-electron bands are far from the chemical potential and the Seebeck coefficient is roughly similar to the case without displacement field. As the filling is decreased, the hole bands are brought closer to the chemical potential. The hybridization between the $f$- and $d$-electrons leads to an enhancement of the $d$-electron spectral weight \emph{above} the hole flat bands. This explains the decrease in the Seebeck coefficient as the system becomes more hole-doped.
	
	\item Similar to the $\nu = 2$ and $\nu = 3$ cases, at low temperatures, the Seebeck coefficient around $\nu = 4$ illustrated in \cref{app:fig:seebeck_ord_1:c} diminishes (becomes more negative) with hole doping away from integer filling. This trend is again traced to the $f$-$d$ hybridization, which displaces the $d$-electron spectral weight from the filled $f$-electron bands. When the chemical potential is near the hole band edge, an increased $d$-electrons spectral weight above the chemical potential leads to a negative peak in the Seebeck coefficient.

\end{itemize}	

\enlargethispage{10\baselineskip}
\renewcommand{\lastFigsNo}{1}
\subsection{Results} \label{app:sec:seebeck_sym_br:results}
\begin{figure}[!h]\includegraphics[width=\textwidth]{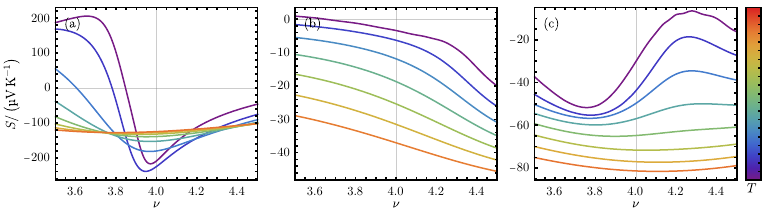}\subfloat{\label{app:fig:seebeck_ord_1:a}}\subfloat{\label{app:fig:seebeck_ord_1:b}}\subfloat{\label{app:fig:seebeck_ord_1:c}}\caption{Seebeck coefficient for the $\protect\IfStrEqCase{1}{{1}{\ket{\nu={}4} }
		{2}{\ket{\nu={}3, \mathrm{IVC}}}
		{3}{\ket{\nu={}3, \mathrm{VP}}}
		{4}{\ket{\nu={}2, \mathrm{K-IVC}}}
		{5}{\ket{\nu={}2, \mathrm{VP}}}
		{6}{\ket{\nu={}1, (\mathrm{K-IVC}+\mathrm{VP})}}
		{7}{\ket{\nu={}1, \mathrm{VP}}}
		{8}{\ket{\nu=0, \mathrm{K-IVC}}}
		{9}{\ket{\nu=0, \mathrm{VP}}}
	}
	[nada]
$ THF ground state candidate: (a) TBG, (b) TSTG at $\mathcal{E}=\SI{0}{\milli\electronvolt}$, (c) TSTG at $\mathcal{E}=\SI{25}{\milli\electronvolt}$. Colors indicate temperature: purple (lowest) to red (highest). (a) $T/\si{\kelvin}=5$, $10$, $15$, $20$, $25$, $30$, $35$, $40$, (b) $T/\si{\kelvin}=14$, $21$, $28$, $35$, $42$, $49$, $56$, and (c) $T/\si{\kelvin}=7$, $14$, $21$, $28$, $35$, $42$, $49$, $56$.}\label{app:fig:seebeck_ord_1}\end{figure}

\ifthenelse{\equal{\lastFigsNo}{2}}{\FloatBarrier}{}

\begin{figure}[!h]\includegraphics[width=\textwidth]{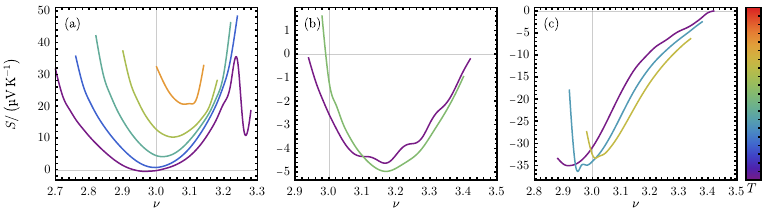}\subfloat{\label{app:fig:seebeck_ord_2:a}}\subfloat{\label{app:fig:seebeck_ord_2:b}}\subfloat{\label{app:fig:seebeck_ord_2:c}}\caption{Seebeck coefficient for the $\protect\IfStrEqCase{2}{{1}{\ket{\nu={}4} }
		{2}{\ket{\nu={}3, \mathrm{IVC}}}
		{3}{\ket{\nu={}3, \mathrm{VP}}}
		{4}{\ket{\nu={}2, \mathrm{K-IVC}}}
		{5}{\ket{\nu={}2, \mathrm{VP}}}
		{6}{\ket{\nu={}1, (\mathrm{K-IVC}+\mathrm{VP})}}
		{7}{\ket{\nu={}1, \mathrm{VP}}}
		{8}{\ket{\nu=0, \mathrm{K-IVC}}}
		{9}{\ket{\nu=0, \mathrm{VP}}}
	}
	[nada]
$ THF ground state candidate: (a) TBG, (b) TSTG at $\mathcal{E}=\SI{0}{\milli\electronvolt}$, (c) TSTG at $\mathcal{E}=\SI{25}{\milli\electronvolt}$. Colors indicate temperature: purple (lowest) to red (highest). (a) $T/\si{\kelvin}=4$, $6$, $8$, $10$, $12$, (b) $T/\si{\kelvin}=11.2$, $14$, and (c) $T/\si{\kelvin}=8.4$, $11.2$, $14$.}\label{app:fig:seebeck_ord_2}\end{figure}

\ifthenelse{\equal{\lastFigsNo}{1}}{\FloatBarrier}{}

\begin{figure}[!h]\includegraphics[width=\textwidth]{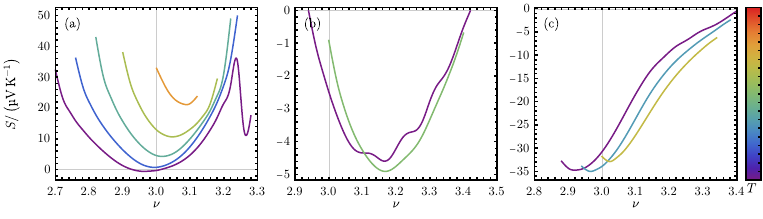}\subfloat{\label{app:fig:seebeck_ord_3:a}}\subfloat{\label{app:fig:seebeck_ord_3:b}}\subfloat{\label{app:fig:seebeck_ord_3:c}}\caption{Seebeck coefficient for the $\protect\IfStrEqCase{3}{{1}{\ket{\nu={}4} }
		{2}{\ket{\nu={}3, \mathrm{IVC}}}
		{3}{\ket{\nu={}3, \mathrm{VP}}}
		{4}{\ket{\nu={}2, \mathrm{K-IVC}}}
		{5}{\ket{\nu={}2, \mathrm{VP}}}
		{6}{\ket{\nu={}1, (\mathrm{K-IVC}+\mathrm{VP})}}
		{7}{\ket{\nu={}1, \mathrm{VP}}}
		{8}{\ket{\nu=0, \mathrm{K-IVC}}}
		{9}{\ket{\nu=0, \mathrm{VP}}}
	}
	[nada]
$ THF ground state candidate: (a) TBG, (b) TSTG at $\mathcal{E}=\SI{0}{\milli\electronvolt}$, (c) TSTG at $\mathcal{E}=\SI{25}{\milli\electronvolt}$. Colors indicate temperature: purple (lowest) to red (highest). (a) $T/\si{\kelvin}=4$, $6$, $8$, $10$, $12$, (b) $T/\si{\kelvin}=11.2$, $14$, and (c) $T/\si{\kelvin}=8.4$, $11.2$, $14$.}\label{app:fig:seebeck_ord_3}\end{figure}

\ifthenelse{\equal{\lastFigsNo}{3}}{\FloatBarrier}{}

\begin{figure}[!h]\includegraphics[width=\textwidth]{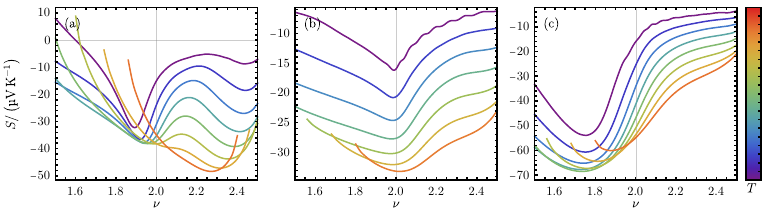}\subfloat{\label{app:fig:seebeck_ord_4:a}}\subfloat{\label{app:fig:seebeck_ord_4:b}}\subfloat{\label{app:fig:seebeck_ord_4:c}}\caption{Seebeck coefficient for the $\protect\IfStrEqCase{4}{{1}{\ket{\nu={}4} }
		{2}{\ket{\nu={}3, \mathrm{IVC}}}
		{3}{\ket{\nu={}3, \mathrm{VP}}}
		{4}{\ket{\nu={}2, \mathrm{K-IVC}}}
		{5}{\ket{\nu={}2, \mathrm{VP}}}
		{6}{\ket{\nu={}1, (\mathrm{K-IVC}+\mathrm{VP})}}
		{7}{\ket{\nu={}1, \mathrm{VP}}}
		{8}{\ket{\nu=0, \mathrm{K-IVC}}}
		{9}{\ket{\nu=0, \mathrm{VP}}}
	}
	[nada]
$ THF ground state candidate: (a) TBG, (b) TSTG at $\mathcal{E}=\SI{0}{\milli\electronvolt}$, (c) TSTG at $\mathcal{E}=\SI{25}{\milli\electronvolt}$. Colors indicate temperature: purple (lowest) to red (highest). (a) $T/\si{\kelvin}=5$, $8$, $11$, $14$, $17$, $20$, $23$, $26$, (b) $T/\si{\kelvin}=11.2$, $15.4$, $19.6$, $23.8$, $28$, $32.2$, $36.4$, and (c) $T/\si{\kelvin}=7$, $11.2$, $15.4$, $19.6$, $23.8$, $28$, $32.2$, $36.4$.}\label{app:fig:seebeck_ord_4}\end{figure}

\ifthenelse{\equal{\lastFigsNo}{2}}{\FloatBarrier}{}

\begin{figure}[!h]\includegraphics[width=\textwidth]{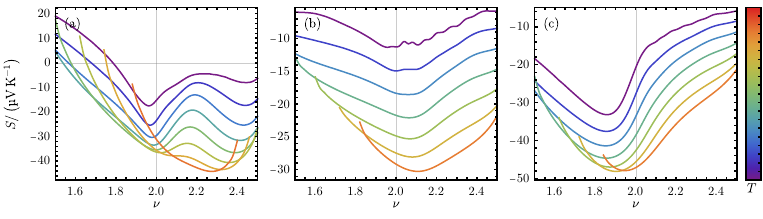}\subfloat{\label{app:fig:seebeck_ord_5:a}}\subfloat{\label{app:fig:seebeck_ord_5:b}}\subfloat{\label{app:fig:seebeck_ord_5:c}}\caption{Seebeck coefficient for the $\protect\IfStrEqCase{5}{{1}{\ket{\nu={}4} }
		{2}{\ket{\nu={}3, \mathrm{IVC}}}
		{3}{\ket{\nu={}3, \mathrm{VP}}}
		{4}{\ket{\nu={}2, \mathrm{K-IVC}}}
		{5}{\ket{\nu={}2, \mathrm{VP}}}
		{6}{\ket{\nu={}1, (\mathrm{K-IVC}+\mathrm{VP})}}
		{7}{\ket{\nu={}1, \mathrm{VP}}}
		{8}{\ket{\nu=0, \mathrm{K-IVC}}}
		{9}{\ket{\nu=0, \mathrm{VP}}}
	}
	[nada]
$ THF ground state candidate: (a) TBG, (b) TSTG at $\mathcal{E}=\SI{0}{\milli\electronvolt}$, (c) TSTG at $\mathcal{E}=\SI{25}{\milli\electronvolt}$. Colors indicate temperature: purple (lowest) to red (highest). (a) $T/\si{\kelvin}=5$, $8$, $11$, $14$, $17$, $20$, $23$, $26$, (b) $T/\si{\kelvin}=11.2$, $15.4$, $19.6$, $23.8$, $28$, $32.2$, $36.4$, and (c) $T/\si{\kelvin}=11.2$, $15.4$, $19.6$, $23.8$, $28$, $32.2$, $36.4$.}\label{app:fig:seebeck_ord_5}\end{figure}

\ifthenelse{\equal{\lastFigsNo}{1}}{\FloatBarrier}{}

\begin{figure}[!h]\includegraphics[width=\textwidth]{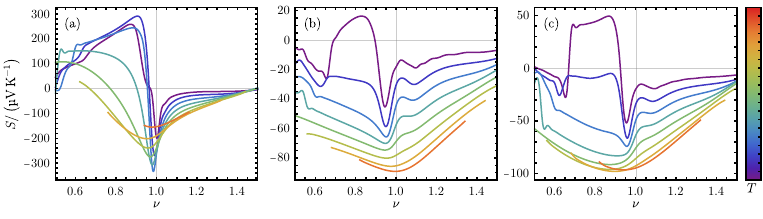}\subfloat{\label{app:fig:seebeck_ord_6:a}}\subfloat{\label{app:fig:seebeck_ord_6:b}}\subfloat{\label{app:fig:seebeck_ord_6:c}}\caption{Seebeck coefficient for the $\protect\IfStrEqCase{6}{{1}{\ket{\nu={}4} }
		{2}{\ket{\nu={}3, \mathrm{IVC}}}
		{3}{\ket{\nu={}3, \mathrm{VP}}}
		{4}{\ket{\nu={}2, \mathrm{K-IVC}}}
		{5}{\ket{\nu={}2, \mathrm{VP}}}
		{6}{\ket{\nu={}1, (\mathrm{K-IVC}+\mathrm{VP})}}
		{7}{\ket{\nu={}1, \mathrm{VP}}}
		{8}{\ket{\nu=0, \mathrm{K-IVC}}}
		{9}{\ket{\nu=0, \mathrm{VP}}}
	}
	[nada]
$ THF ground state candidate: (a) TBG, (b) TSTG at $\mathcal{E}=\SI{0}{\milli\electronvolt}$, (c) TSTG at $\mathcal{E}=\SI{25}{\milli\electronvolt}$. Colors indicate temperature: purple (lowest) to red (highest). (a) $T/\si{\kelvin}=7$, $12$, $17$, $22$, $27$, $32$, $37$, $42$, (b) $T/\si{\kelvin}=9.8$, $16.8$, $23.8$, $30.8$, $37.8$, $44.8$, $51.8$, $58.8$, and (c) $T/\si{\kelvin}=9.8$, $16.8$, $23.8$, $30.8$, $37.8$, $44.8$, $51.8$, $58.8$.}\label{app:fig:seebeck_ord_6}\end{figure}

\ifthenelse{\equal{\lastFigsNo}{3}}{\FloatBarrier}{}

\begin{figure}[!h]\includegraphics[width=\textwidth]{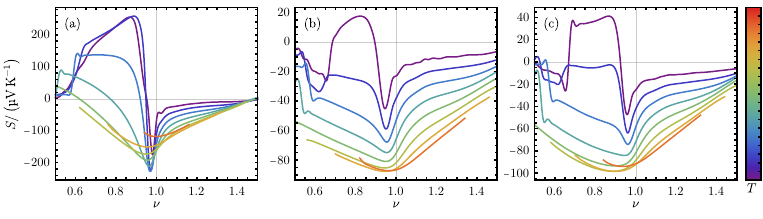}\subfloat{\label{app:fig:seebeck_ord_7:a}}\subfloat{\label{app:fig:seebeck_ord_7:b}}\subfloat{\label{app:fig:seebeck_ord_7:c}}\caption{Seebeck coefficient for the $\protect\IfStrEqCase{7}{{1}{\ket{\nu={}4} }
		{2}{\ket{\nu={}3, \mathrm{IVC}}}
		{3}{\ket{\nu={}3, \mathrm{VP}}}
		{4}{\ket{\nu={}2, \mathrm{K-IVC}}}
		{5}{\ket{\nu={}2, \mathrm{VP}}}
		{6}{\ket{\nu={}1, (\mathrm{K-IVC}+\mathrm{VP})}}
		{7}{\ket{\nu={}1, \mathrm{VP}}}
		{8}{\ket{\nu=0, \mathrm{K-IVC}}}
		{9}{\ket{\nu=0, \mathrm{VP}}}
	}
	[nada]
$ THF ground state candidate: (a) TBG, (b) TSTG at $\mathcal{E}=\SI{0}{\milli\electronvolt}$, (c) TSTG at $\mathcal{E}=\SI{25}{\milli\electronvolt}$. Colors indicate temperature: purple (lowest) to red (highest). (a) $T/\si{\kelvin}=7$, $12$, $17$, $22$, $27$, $32$, $37$, $42$, (b) $T/\si{\kelvin}=9.8$, $16.8$, $23.8$, $30.8$, $37.8$, $44.8$, $51.8$, $58.8$, and (c) $T/\si{\kelvin}=9.8$, $16.8$, $23.8$, $30.8$, $37.8$, $44.8$, $51.8$, $58.8$.}\label{app:fig:seebeck_ord_7}\end{figure}

\ifthenelse{\equal{\lastFigsNo}{2}}{\FloatBarrier}{}

\ifthenelse{\equal{\lastFigsNo}{2} }{\begin{figure}[!t]}{\begin{figure}[!h]}\includegraphics[width=\textwidth]{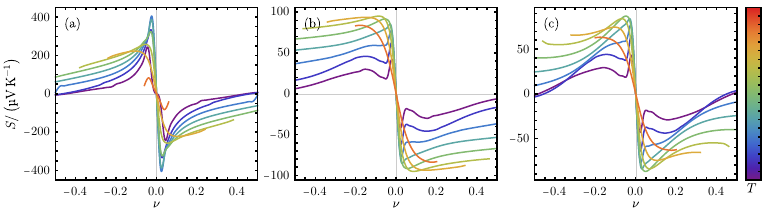}\subfloat{\label{app:fig:seebeck_ord_8:a}}\subfloat{\label{app:fig:seebeck_ord_8:b}}\subfloat{\label{app:fig:seebeck_ord_8:c}}\caption{Seebeck coefficient for the $\protect\IfStrEqCase{8}{{1}{\ket{\nu={}4} }
		{2}{\ket{\nu={}3, \mathrm{IVC}}}
		{3}{\ket{\nu={}3, \mathrm{VP}}}
		{4}{\ket{\nu={}2, \mathrm{K-IVC}}}
		{5}{\ket{\nu={}2, \mathrm{VP}}}
		{6}{\ket{\nu={}1, (\mathrm{K-IVC}+\mathrm{VP})}}
		{7}{\ket{\nu={}1, \mathrm{VP}}}
		{8}{\ket{\nu=0, \mathrm{K-IVC}}}
		{9}{\ket{\nu=0, \mathrm{VP}}}
	}
	[nada]
$ THF ground state candidate: (a) TBG, (b) TSTG at $\mathcal{E}=\SI{0}{\milli\electronvolt}$, (c) TSTG at $\mathcal{E}=\SI{25}{\milli\electronvolt}$. Colors indicate temperature: purple (lowest) to red (highest). (a) $T/\si{\kelvin}=8$, $14$, $20$, $26$, $32$, $38$, $44$, $50$, (b) $T/\si{\kelvin}=11.2$, $19.6$, $28$, $36.4$, $44.8$, $53.2$, $61.6$, $70$, and (c) $T/\si{\kelvin}=11.2$, $19.6$, $28$, $36.4$, $44.8$, $53.2$, $61.6$, $70$.}\label{app:fig:seebeck_ord_8}\end{figure}

\ifthenelse{\equal{\lastFigsNo}{1}}{\FloatBarrier}{}

\ifthenelse{\equal{\lastFigsNo}{1} \OR \equal{\lastFigsNo}{2} }{\begin{figure}[!t]}{\begin{figure}[!h]}\includegraphics[width=\textwidth]{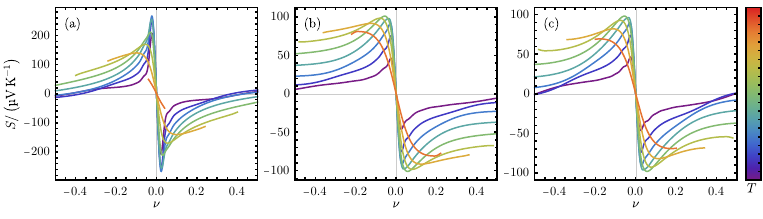}\subfloat{\label{app:fig:seebeck_ord_9:a}}\subfloat{\label{app:fig:seebeck_ord_9:b}}\subfloat{\label{app:fig:seebeck_ord_9:c}}\caption{Seebeck coefficient for the $\protect\IfStrEqCase{9}{{1}{\ket{\nu={}4} }
		{2}{\ket{\nu={}3, \mathrm{IVC}}}
		{3}{\ket{\nu={}3, \mathrm{VP}}}
		{4}{\ket{\nu={}2, \mathrm{K-IVC}}}
		{5}{\ket{\nu={}2, \mathrm{VP}}}
		{6}{\ket{\nu={}1, (\mathrm{K-IVC}+\mathrm{VP})}}
		{7}{\ket{\nu={}1, \mathrm{VP}}}
		{8}{\ket{\nu=0, \mathrm{K-IVC}}}
		{9}{\ket{\nu=0, \mathrm{VP}}}
	}
	[nada]
$ THF ground state candidate: (a) TBG, (b) TSTG at $\mathcal{E}=\SI{0}{\milli\electronvolt}$, (c) TSTG at $\mathcal{E}=\SI{25}{\milli\electronvolt}$. Colors indicate temperature: purple (lowest) to red (highest). (a) $T/\si{\kelvin}=8$, $14$, $20$, $26$, $32$, $38$, $44$, $50$, (b) $T/\si{\kelvin}=11.2$, $19.6$, $28$, $36.4$, $44.8$, $53.2$, $61.6$, $70$, and (c) $T/\si{\kelvin}=11.2$, $19.6$, $28$, $36.4$, $44.8$, $53.2$, $61.6$, $70$.}\label{app:fig:seebeck_ord_9}\end{figure}

\ifthenelse{\equal{\lastFigsNo}{3}}{\FloatBarrier}{}%

\section{Numerical results for the Seebeck coefficient in the symmetric state}\label{app:sec:seebeck_sym}

In this \siSection{}, we present results on the Seebeck coefficient in the symmetric state of the THF model. Similarly to the symmetry-broken case presented in \cref{app:sec:seebeck_sym_br}, these calculations follow the methods described in \cref{app:sec:thermoelectric_response:details_on_numerics:interactions}.

\subsection{Discussion} \label{app:sec:seebeck_sym:discussion}

The symmetric phase retains all symmetries inherent to the THF model, including moir\'e translation, $C_{6z}$ rotation, time-reversal, and $\mathrm{SU} \left(2\right) \times \mathrm{SU} \left(2\right)$ spin-valley rotation~\cite{BER21a,CAL21}. As proved in \cref*{DMFT:app:sec:se_symmetric} of Ref.~\cite{CAL23b}, in the symmetric phase, the onsite $f$-electron density matrix $\mathcal{O}^f$ introduced in \cref{app:eqn:def_onsite_f_electron_den_mat} is proportional to identity. The Seebeck coefficient in the symmetric phase is shown in \cref{app:fig:seebeck_sym_1} for TBG at $5 \leq T/\si{\kelvin} \leq 150$, while \crefrange{app:fig:seebeck_sym_2}{app:fig:seebeck_sym_7} illustrate the Seebeck coefficient of TSTG in the symmetric phase for $7.5 \leq T/\si{\kelvin} \leq 225$ and for different values of the displacement field $0 \leq \mathcal{E}/\si{\milli\electronvolt} \leq 25$. We consider $0 \leq \nu \leq 4.5$. 

The Seebeck coefficient of both TBG and TSTG in the symmetric phase features oscillations around integer fillings on top of a background that varies slowly with $\nu$. Notably, the baseline trend of the Seebeck coefficient in this symmetric phase mirrors that of the non-interacting phase shown in \cref{app:sec:seebeck_non_int:results}: the offset of the Seebeck coefficient decreases as temperature is increased. This can be understood as follows:

\begin{itemize}
	\item The $f$-electrons are dispersionless and, as a zeroth order approximation, do not contribute to thermoelectric transport. As a result, they can be ignored implying that from a transport standpoint, one can think of the TBG excitation spectrum within the THF model as being formed only of a few $c$-electron states around the $\Gamma_{M}$ belonging to the active TBG bands and numerous $c$-electron states forming the TBG remote bands.
	
	\item At positive fillings $\nu > 0$, in both the symmetric phase and the non-interacting limit, the chemical potential is closer to the $c$-electron states near the $\Gamma_{M}$ point (which form hole-like excitations), than to the $c$-electron states belonging to the remote TBG bands (which form high-energy electron-like excitations). As a result, for low temperatures, the Seebeck coefficient in both the symmetric phase and in the non-interacting limit will be hole-dominated and therefore largely positive. At higher temperatures, the larger number of $c$-electrons in the remote bands will become thermally activated and will dominate transport leading to an electron-dominated and therefore largely negative Seebeck coefficient. We conclude that the temperature dependence of the Seebeck's coefficient's offset in the symmetric phase qualitatively mirrors the Seebeck coefficient in the non-interacting limit.

	\item Around integer fillings in the symmetric phase, interaction-driven gaps form as a result of the onsite $f$-electron repulsion. This leads to the appearance of sawtooth oscillations in the Seebeck coefficient around the integer fillings. 
	
\end{itemize}

\renewcommand{\lastFigsNo}{1}
\subsection{Results} \label{app:sec:seebeck_sym:results}
\begin{figure}[!h]\includegraphics[width=0.6667\textwidth]{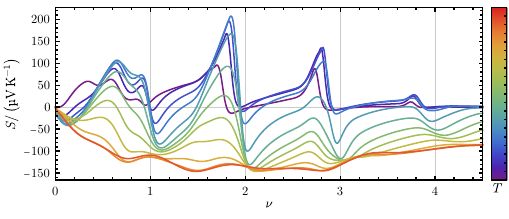}\caption{Seebeck coefficient observed in the symmetric phase of TBG. The varied line colors denote distinct temperatures utilized in the simulations. Purple signifies the lowest temperature, while red indicates the highest. In order of increasing temperatures, the thermopower curves correspond to $T/\si{\kelvin}=5$, $10$, $15$, $20$, $30$, $40$, $50$, $60$, $75$, $100$, $125$, $150$.}\label{app:fig:seebeck_sym_1}\end{figure}

\ifthenelse{\equal{\lastFigsNo}{3}}{\FloatBarrier\clearpage}{}

\begin{figure}[!h]\includegraphics[width=0.6667\textwidth]{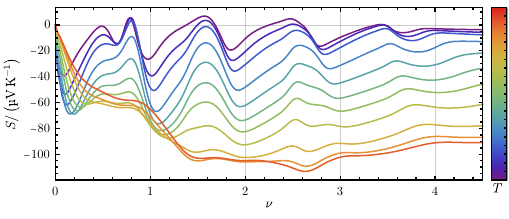}\caption{Seebeck coefficient observed in the symmetric phase of TSTG at $\mathcal{E}=\SI{0}{\milli\electronvolt}$. The varied line colors denote distinct temperatures utilized in the simulations. Purple signifies the lowest temperature, while red indicates the highest. In order of increasing temperatures, the thermopower curves correspond to $T/\si{\kelvin}=15$, $22.5$, $30$, $45$, $60$, $75$, $90$, $112.5$, $150$, $187.5$, $225$.}\label{app:fig:seebeck_sym_2}\end{figure}

\ifthenelse{\equal{\lastFigsNo}{2}}{\FloatBarrier\clearpage}{}

\begin{figure}[!h]\includegraphics[width=0.6667\textwidth]{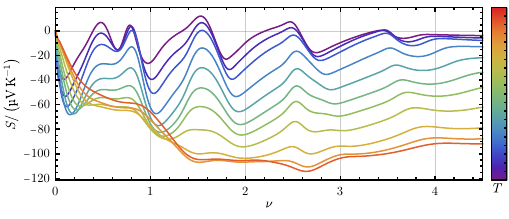}\caption{Seebeck coefficient observed in the symmetric phase of TSTG at $\mathcal{E}=\SI{5}{\milli\electronvolt}$. The varied line colors denote distinct temperatures utilized in the simulations. Purple signifies the lowest temperature, while red indicates the highest. In order of increasing temperatures, the thermopower curves correspond to $T/\si{\kelvin}=15$, $22.5$, $30$, $45$, $60$, $75$, $90$, $112.5$, $150$, $187.5$, $225$.}\label{app:fig:seebeck_sym_3}\end{figure}

\ifthenelse{\equal{\lastFigsNo}{1}}{\FloatBarrier\clearpage}{}

\begin{figure}[!h]\includegraphics[width=0.6667\textwidth]{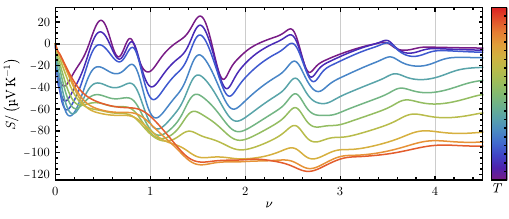}\caption{Seebeck coefficient observed in the symmetric phase of TSTG at $\mathcal{E}=\SI{10}{\milli\electronvolt}$. The varied line colors denote distinct temperatures utilized in the simulations. Purple signifies the lowest temperature, while red indicates the highest. In order of increasing temperatures, the thermopower curves correspond to $T/\si{\kelvin}=15$, $22.5$, $30$, $45$, $60$, $75$, $90$, $112.5$, $150$, $187.5$, $225$.}\label{app:fig:seebeck_sym_4}\end{figure}

\ifthenelse{\equal{\lastFigsNo}{3}}{\FloatBarrier\clearpage}{}

\begin{figure}[!h]\includegraphics[width=0.6667\textwidth]{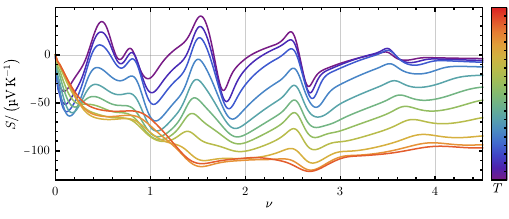}\caption{Seebeck coefficient observed in the symmetric phase of TSTG at $\mathcal{E}=\SI{15}{\milli\electronvolt}$. The varied line colors denote distinct temperatures utilized in the simulations. Purple signifies the lowest temperature, while red indicates the highest. In order of increasing temperatures, the thermopower curves correspond to $T/\si{\kelvin}=15$, $22.5$, $30$, $45$, $60$, $75$, $90$, $112.5$, $150$, $187.5$, $225$.}\label{app:fig:seebeck_sym_5}\end{figure}

\ifthenelse{\equal{\lastFigsNo}{2}}{\FloatBarrier\clearpage}{}

\ifthenelse{\equal{\lastFigsNo}{2} }{\begin{figure}[!t]}{\begin{figure}[!h]}\includegraphics[width=0.6667\textwidth]{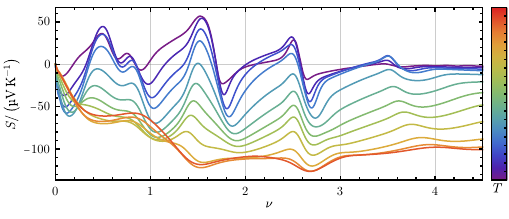}\caption{Seebeck coefficient observed in the symmetric phase of TSTG at $\mathcal{E}=\SI{20}{\milli\electronvolt}$. The varied line colors denote distinct temperatures utilized in the simulations. Purple signifies the lowest temperature, while red indicates the highest. In order of increasing temperatures, the thermopower curves correspond to $T/\si{\kelvin}=7.5$, $15$, $22.5$, $30$, $45$, $60$, $75$, $90$, $112.5$, $150$, $187.5$, $225$.}\label{app:fig:seebeck_sym_6}\end{figure}

\ifthenelse{\equal{\lastFigsNo}{1}}{\FloatBarrier\clearpage}{}

\ifthenelse{\equal{\lastFigsNo}{1} \OR \equal{\lastFigsNo}{2} }{\begin{figure}[!t]}{\begin{figure}[!h]}\includegraphics[width=0.6667\textwidth]{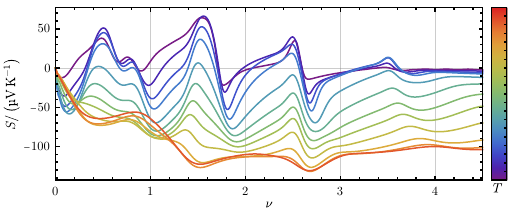}\caption{Seebeck coefficient observed in the symmetric phase of TSTG at $\mathcal{E}=\SI{25}{\milli\electronvolt}$. The varied line colors denote distinct temperatures utilized in the simulations. Purple signifies the lowest temperature, while red indicates the highest. In order of increasing temperatures, the thermopower curves correspond to $T/\si{\kelvin}=7.5$, $15$, $22.5$, $30$, $45$, $60$, $75$, $90$, $112.5$, $150$, $187.5$, $225$.}\label{app:fig:seebeck_sym_7}\end{figure}

\ifthenelse{\equal{\lastFigsNo}{3}}{\FloatBarrier\clearpage}{}%
\section{Comparison with experiments}\label{app:sec:comparison_with_exp}

In this \siSection{}, we provide a summary comparing our numerical results for the Seebeck coefficient of TBG and TSTG with the experimental measurements performed by the accompanying Refs.~\cite{MER24,BAT24}. A point-by-point discussion of the different results obtained in this work and the connection with the different regimes seen in experiments is provided in \cref{app:tab:experiment_summary}. In addition, the following points are worth noting:
\begin{enumerate}

	\item \label{app:enum:temp_mism_sym_br} For the numerical calculations of the Seebeck coefficient in the symmetry-broken case presented in \cref{app:sec:seebeck_sym_br}, there is a mismatch between the temperature used in the simulations and the actual temperature at which the experiments are carried out. This mismatch is likely due to an \emph{underestimation} of the $f$-electron interaction-induced fluctuations by the second-order self-consistent perturbation theory~\cite{CAL23b}. Indeed, using the latter method, the symmetry-broken phase at $\nu = 1$ of TBG can be stabilized up to a critical temperature of $T^{*} \approx \SI{50}{\kelvin}$, while more accurate calculations using dynamical mean field theory and a quantum Monte-Carlo impurity solver~\cite{RAI23a} find much smaller critical temperatures $T^{*} \sim \SIrange{10}{15}{\kelvin}$. In effect, the lifetime asymmetry between the hole (mainly $f$-electrons) and electron (mainly $c$-electrons) excitations is underestimated by our method, implying that higher temperatures are needed to replicate the negative Seebeck coefficient seen in experiments around $\nu = 1$~\cite{MER24}.
	
	\item \label{app:enum:not_thermo_3} Experimentally, Ref.~\cite{MER24} finds that often the response around $\nu = 3$ is superimposed with the very large response coming from the band insulator. This effect can also be observed in the longitudinal resistance of many TBG devices, including those studied in Ref.~\cite{MER24}. The proximity in doping to the band insulator implies that any twist angle inhomogeneity and/or finite temperature will give raise to a combination of the responses coming from $\nu = 3$ and $\nu = 4$. Importantly, the response around the band insulator is likely of different nature, possibly photovoltaic. The current picture for hot electron dynamics in TBG suggests that the photo-excited electron-hole pairs are generated in the dispersive bands and share their energy through electron-electron scattering until reaching the bottom (top) of the dispersive conduction (valence) band. For full occupation of the flat bands, it is natural to think of TBG (energetically) as a conventional semiconductor in which the hot carriers must directly recombine across the band gap. This stands in clear contrast to the case where many states are empty in the flat bands, where the hot carriers may relax all the way to the Fermi level before losing their energy through electron-phonon scattering.
	
	Thus, at the band insulator, we hypothesize that the response may be photovoltaic, occurring through the separation of the hot electron and hole distributions. This hypothesis is supported by the observation of increased cooling times for the electrons in TBG around the magic angle whenever the flat bands are fully occupied ($\SIrange{3}{30}{\pico\second}$)~\cite{MEH24}. Although cooling is aided by the phonon bath, this increased cooling time at the band insulator suggests that the hot carriers cannot thermalize down to the low-energy bands to then relax their energy through low-energy phonon modes. Further support for this picture comes from the dependence on optical power of the cooling time at the band insulator. Unlike the response for $-3.5 < \nu < 3.5$, the cooling time exhibits a clear dependence on optical power when the flat bands are occupied. The cooling time increases up to $\SI{70}{\pico\second}$, as the excitation power is increased (increasing the density of photo-excited electrons)~\cite{MEH24}. Even slower cooling times have been observed at the band insulator filling. Further experimental studies are required to shed light on the nature of the band insulator photoresponse. 
	
	As a result, we tentatively argue that the very large response from the band insulator, whose origin is possibly not thermoelectric, overwhelms the ''intrinsic`` thermoelectric response at $\nu = 3$, making it an unreliable observable in our experiment.
	
	\item \label{app:enum:temp_mism_sym} As discussed in \cref{app:sec:seebeck_sym:discussion}, the temperature dependence of the offset of the Seebeck coefficient in the symmetric phase qualitatively follows the non-interacting result from \cref{app:fig:seebeck_non_int_1,app:fig:seebeck_non_int_2}. In particular the offset decreases from positive at low temperature to negative at high temperature in both the interacting and non-interacting cases. Qualitatively, however, the crossover temperature strongly depends on the detailed non-interacting TBG band structure, as shown analytically in \cref{app:sec:seebeck_non_int:discussion:analytic_cross}. The TBG single-particle band structure is significantly affected by strain or relaxation effects, which are not included in our simulations. As such, we find that slightly lower temperatures are required for the simulations in the symmetric phase of TBG to reproduce the experimentally-observed Seebeck coefficient at higher temperature~\cite{MER24}.
	  
	\item \label{app:enum:mismatch_close_nu0} Ref.~\cite{BAT24} measures the Seebeck coefficient in a device that features a junction between TSTG and trilayer graphene (TG) twisted at a large unspecified angle. The filling of the two sides of the device (the TG and TSTG parts) cannot be independently controlled~\cite{BAT24}. Additionally, exactly at the junction between the TSTG and TG parts, the experiment measures $S_{\text{TSTG}} \left( \nu \right) - S_{\text{TG}} \left( \nu \right)$ (where $S_{\text{TSTG}} \left( \nu \right)$ and $S_{\text{TG}} \left( \nu \right)$ are the filling-dependent Seebeck coefficients of TSTG and TG, respectively), as opposed to purely $S_{\text{TSTG}} \left( \nu \right)$. Since $S_{\text{TG}} \left( \nu \right)$ has a strong filling dependence around charge neutrality (but then quickly turns to zero away from it), measurements at the TSTG and TG junction deviate from the pure $S_{\text{TSTG}} \left( \nu \right)$ response around $\nu = 0$. 
\end{enumerate}

\begin{longtable}{|p{4cm}|p{2cm}|p{4cm}|p{6cm}|}
		\hline
		Simulations of the Seebeck coefficient & Numerical results & Experimental regime with which we compare it & Comments \\
		\hline\hline
		TBG around $\nu=0$ in the symmetry-broken phase & \cref{app:fig:seebeck_ord_8:a,app:fig:seebeck_ord_9:a} & {\color{darkgreen} \textbf{Matched}} by the $T = \SI{10}{\kelvin}$ measurements of Ref.~\cite{MER24}. &  \\
		\hline
		TBG around $\nu=1$ in the symmetry-broken phase at $T < \SI{27}{\kelvin}$ & \cref{app:fig:seebeck_ord_6:a,app:fig:seebeck_ord_7:a} & {\color{orange} \textbf{Not observed}} in Ref.~\cite{MER24}. & Due to the temperature mismatch between the theoretical calculations in the symmetry-broken case and the experiment (see \cref{app:enum:temp_mism_sym_br}), we only expect this regime to be experimentally observed at very low temperatures.  \\
		\hline
		TBG around $\nu=1$ in the symmetry-broken phase at $T \geq \SI{27}{\kelvin}$ & \cref{app:fig:seebeck_ord_6:a,app:fig:seebeck_ord_7:a} & {\color{darkgreen} \textbf{Matched}} by the $T = \SI{10}{\kelvin}$ measurements of Ref.~\cite{MER24}. & Higher temperature are required in the theoretical calculations in order to reproduce the experimental results (see \cref{app:enum:temp_mism_sym_br}). The negative electron peak is slightly shifted towards charge neutrality in both theory and experiment.\\
		\hline
		TBG around $\nu=2$ in the symmetry-broken phase & \cref{app:fig:seebeck_ord_4:a,app:fig:seebeck_ord_5:a} & {\color{darkgreen} \textbf{Matched}} by the $T = \SI{10}{\kelvin}$ measurements of Ref.~\cite{MER24}. & The theoretical data shows a second negative peak at $\nu \approx 2.4$, which is not observed in the experiment. The shifting of the negative electron peak at $\nu \approx1.9$ towards charge neutrality is seen in both theory and experiment.\\
		\hline
		TBG around $\nu=3$ in the symmetry-broken phase & \cref{app:fig:seebeck_ord_2:a,app:fig:seebeck_ord_3:a} & {\color{red} \textbf{Not matched}} by the $T = \SI{10}{\kelvin}$ measurements of Ref.~\cite{MER24}. & The measured response close to the band gap is likely not exclusively thermoelectric (see \cref{app:enum:not_thermo_3}). The theoretically considered states are likely not the ground states of the system at $\nu = 3$.\\
		\hline
		TBG around $\nu=4$ & \cref{app:fig:seebeck_ord_1:a} & {\color{red} \textbf{Not matched}} by the $T = \SI{10}{\kelvin}$ measurements of Ref.~\cite{MER24}. & The measured response around the band gap is likely not thermoelectric (see \cref{app:enum:not_thermo_3}).\\
		\hline
		TBG for $\nu \geq 0$ at low temperatures $T \lesssim \SI{50}{\kelvin}$ in the non-interacting limit & \cref{app:fig:seebeck_non_int_1,app:fig:seebeck_non_int_2} & {\color{darkgreen} \textbf{Matched}} by the high-temperature measurements ($T = \SI{30}{\kelvin}$ or $T = \SI{50}{\kelvin}$) of Ref.~\cite{MER24} in offset. & Does not reproduce the experimentally-observed oscillations around the integer fillings since it is a non-interacting simulation.  \\
		\hline
		TBG for $\nu \geq 0$ at high temperatures $T \gtrsim \SI{50}{\kelvin}$ in the non-interacting limit & \cref{app:fig:seebeck_non_int_1,app:fig:seebeck_non_int_2} &  {\color{orange} \textbf{Not observed}} in Ref.~\cite{MER24}. &  \\
		\hline
		TBG for $\nu \geq 0$ at low temperatures $T \leq \SI{20}{\kelvin}$ in the symmetric phase & \cref{app:fig:seebeck_sym_1} & {\color{darkgreen} \textbf{Matched}} by the high-temperature measurements ($T = \SI{30}{\kelvin}$ or $T = \SI{50}{\kelvin}$) of Ref.~\cite{MER24} in both offset and oscillations. & Note that a temperature mismatch exists between the theory and experiment (see \cref{app:enum:temp_mism_sym}).  \\
		\hline
		TBG for $\nu \geq 0$ at high temperatures $T > \SI{20}{\kelvin} $ in the symmetric phase & \cref{app:fig:seebeck_sym_1} & {\color{orange} \textbf{Not observed}} in Ref.~\cite{MER24}. &	Due to the temperature mismatch between the theoretical calculations in the symmetric phase and the experiment (see \cref{app:enum:temp_mism_sym}), we only expect this regime to be experimentally observed at very high temperatures.  \\
		\hline\hline
		TSTG around $\nu = 0$ in the symmetry-broken phase & \cref{app:fig:seebeck_ord_8:b,app:fig:seebeck_ord_8:c,app:fig:seebeck_ord_9:b,app:fig:seebeck_ord_9:c} & {\color{darkgreen} \textbf{Matched}} by the low-temperature measurements ($T \lesssim \SI{40}{\kelvin}$) of Ref.~\cite{BAT24}. & Apparent mismatch for the local Seebeck coefficient measured close to the junction between the TSTG and the untwisted trilayer graphene parts of the device (see \cref{app:enum:mismatch_close_nu0}). \\
		\hline
		TSTG around $\nu = 1$ in the symmetry-broken phase & \cref{app:fig:seebeck_ord_6:b,app:fig:seebeck_ord_6:c,app:fig:seebeck_ord_7:b,app:fig:seebeck_ord_7:c} & {\color{darkgreen} \textbf{Matched}} by the low-temperature measurements ($T \lesssim \SI{40}{\kelvin}$) of Ref.~\cite{BAT24}. & Apparent mismatch for the local Seebeck coefficient measured close to the junction between the TSTG and the untwisted trilayer graphene parts of the device (see \cref{app:enum:mismatch_close_nu0}). \\
		\hline
		TSTG around $\nu = 2$ in the symmetry-broken phase & \cref{app:fig:seebeck_ord_4:b,app:fig:seebeck_ord_4:c,app:fig:seebeck_ord_5:b,app:fig:seebeck_ord_5:c} & {\color{darkgreen} \textbf{Matched}} by the low-temperature measurements ($T \lesssim \SI{50}{\kelvin}$) of Ref.~\cite{BAT24}. & Simulations reproduce the negative offset, the negative peak around integer filling, and the increase of the Seebeck coefficient in magnitude with increasing temperature. \\
		\hline
		TSTG around $\nu = 3$ in the symmetry-broken phase & \cref{app:fig:seebeck_ord_2:b,app:fig:seebeck_ord_2:c,app:fig:seebeck_ord_3:b,app:fig:seebeck_ord_3:c} & Apparently {\color{darkgreen} \textbf{matched}} by the low-temperature measurements ($T \lesssim \SI{50}{\kelvin}$) of Ref.~\cite{BAT24}. & Simulations reproduce the negative offset and the negative peak around integer filling, but the theoretically considered states are likely not the ground states of the system at $\nu = 3$. \\
		\hline
		TSTG around $\nu = 4$ & \cref{app:fig:seebeck_ord_1:b,app:fig:seebeck_ord_1:c} & Approximately {\color{darkgreen} \textbf{matched}} by the low-temperature measurements of Ref.~\cite{BAT24}. & Simulations reproduce the negative offset, but the negative peak is only captured at the lowest temperatures and in the presence of a displacement field. In contrast, the experiment features a much larger negative peak in the Seebeck coefficient at $\nu=4$, which is seen over a range of temperatures. \\
		\hline
		TSTG for $\nu \geq 0$ at low temperatures in the non-interacting limit with largely positive Seebeck coefficient & \crefrange{app:fig:seebeck_non_int_3}{app:fig:seebeck_non_int_14} & {\color{orange} \textbf{Not observed}} in Ref.~\cite{BAT24}. & The effects of electron-electron interactions are strong in TSTG and experiments cannot be captured by non-interacting simulations at low temperatures. \\
		\hline
		TSTG for $\nu \geq 0$ at high temperatures in the non-interacting limit with largely negative Seebeck coefficient & \crefrange{app:fig:seebeck_non_int_3}{app:fig:seebeck_non_int_14} &  {\color{darkgreen} \textbf{Matched}} by the high-temperature measurements ($T \gtrsim \SI{50}{\kelvin}$) of Ref.~\cite{BAT24} in offset. &  \\
		\hline
		TSTG for $\nu \geq 0$ at low temperatures in the symmetric phase with largely positive Seebeck coefficient & \crefrange{app:fig:seebeck_sym_2}{app:fig:seebeck_sym_7} & {\color{red} \textbf{Not matched}} by the low-temperature measurements of Ref.~\cite{BAT24}. & The negative offset observed in experiments is matched by the simulations in the symmetry-broken case, which is consistent with the tendency of TSTG to form symmetry-broken phases at low temperatures. \\
		\hline
		TSTG for $\nu \geq 0$ at high temperatures in the symmetric phase with largely negative Seebeck coefficient & \crefrange{app:fig:seebeck_sym_2}{app:fig:seebeck_sym_7} & {\color{darkgreen} \textbf{Matched}} by the high-temperature measurements of Ref.~\cite{BAT24} in both offset and oscillations. & \\
		\hline
	\caption{Point-by-point comparison between our numerical results of the Seebeck coefficient of TBG and TSTG from \cref{app:sec:seebeck_non_int,app:sec:seebeck_sym_br,app:sec:seebeck_sym} and the accompanying experimental measurements of Refs.~\cite{MER24,BAT24}.}
	\label{app:tab:experiment_summary}
\end{longtable}


\begin{thebibliography}{268}%
\makeatletter
\providecommand \@ifxundefined [1]{%
 \@ifx{#1\undefined}
}%
\providecommand \@ifnum [1]{%
 \ifnum #1\expandafter \@firstoftwo
 \else \expandafter \@secondoftwo
 \fi
}%
\providecommand \@ifx [1]{%
 \ifx #1\expandafter \@firstoftwo
 \else \expandafter \@secondoftwo
 \fi
}%
\providecommand \natexlab [1]{#1}%
\providecommand \enquote  [1]{``#1''}%
\providecommand \bibnamefont  [1]{#1}%
\providecommand \bibfnamefont [1]{#1}%
\providecommand \citenamefont [1]{#1}%
\providecommand \href@noop [0]{\@secondoftwo}%
\providecommand \href [0]{\begingroup \@sanitize@url \@href}%
\providecommand \@href[1]{\@@startlink{#1}\@@href}%
\providecommand \@@href[1]{\endgroup#1\@@endlink}%
\providecommand \@sanitize@url [0]{\catcode `\\12\catcode `\$12\catcode
  `\&12\catcode `\#12\catcode `\^12\catcode `\_12\catcode `\%12\relax}%
\providecommand \@@startlink[1]{}%
\providecommand \@@endlink[0]{}%
\providecommand \url  [0]{\begingroup\@sanitize@url \@url }%
\providecommand \@url [1]{\endgroup\@href {#1}{\urlprefix }}%
\providecommand \urlprefix  [0]{URL }%
\providecommand \Eprint [0]{\href }%
\providecommand \doibase [0]{https://doi.org/}%
\providecommand \selectlanguage [0]{\@gobble}%
\providecommand \bibinfo  [0]{\@secondoftwo}%
\providecommand \bibfield  [0]{\@secondoftwo}%
\providecommand \translation [1]{[#1]}%
\providecommand \BibitemOpen [0]{}%
\providecommand \bibitemStop [0]{}%
\providecommand \bibitemNoStop [0]{.\EOS\space}%
\providecommand \EOS [0]{\spacefactor3000\relax}%
\providecommand \BibitemShut  [1]{\csname bibitem#1\endcsname}%
\let\auto@bib@innerbib\@empty
\bibitem [{\citenamefont {Bistritzer}\ and\ \citenamefont
  {MacDonald}(2011)}]{BIS11}%
  \BibitemOpen
  \bibfield  {author} {\bibinfo {author} {\bibfnamefont {R.}~\bibnamefont
  {Bistritzer}}\ and\ \bibinfo {author} {\bibfnamefont {A.~H.}\ \bibnamefont
  {MacDonald}},\ }\href {https://doi.org/10.1073/pnas.1108174108} {\bibfield
  {journal} {\bibinfo  {journal} {PNAS}\ }\textbf {\bibinfo {volume} {108}},\
  \bibinfo {pages} {12233} (\bibinfo {year} {2011})}\BibitemShut {NoStop}%
\bibitem [{\citenamefont {Su{\'a}rez~Morell}\ \emph {et~al.}(2010)\citenamefont
  {Su{\'a}rez~Morell}, \citenamefont {Correa}, \citenamefont {Vargas},
  \citenamefont {Pacheco},\ and\ \citenamefont {Barticevic}}]{SUA10}%
  \BibitemOpen
  \bibfield  {author} {\bibinfo {author} {\bibfnamefont {E.}~\bibnamefont
  {Su{\'a}rez~Morell}}, \bibinfo {author} {\bibfnamefont {J.~D.}\ \bibnamefont
  {Correa}}, \bibinfo {author} {\bibfnamefont {P.}~\bibnamefont {Vargas}},
  \bibinfo {author} {\bibfnamefont {M.}~\bibnamefont {Pacheco}},\ and\ \bibinfo
  {author} {\bibfnamefont {Z.}~\bibnamefont {Barticevic}},\ }\href
  {https://doi.org/10.1103/PhysRevB.82.121407} {\bibfield  {journal} {\bibinfo
  {journal} {Phys. Rev. B}\ }\textbf {\bibinfo {volume} {82}},\ \bibinfo
  {pages} {121407} (\bibinfo {year} {2010})}\BibitemShut {NoStop}%
\bibitem [{\citenamefont {{Lopes~dos~Santos}}\ \emph
  {et~al.}(2007)\citenamefont {{Lopes~dos~Santos}}, \citenamefont {Peres},\
  and\ \citenamefont {Castro~Neto}}]{LOP07}%
  \BibitemOpen
  \bibfield  {author} {\bibinfo {author} {\bibfnamefont {J.~M.~B.}\
  \bibnamefont {{Lopes~dos~Santos}}}, \bibinfo {author} {\bibfnamefont
  {N.~M.~R.}\ \bibnamefont {Peres}},\ and\ \bibinfo {author} {\bibfnamefont
  {A.~H.}\ \bibnamefont {Castro~Neto}},\ }\href
  {https://doi.org/10.1103/PhysRevLett.99.256802} {\bibfield  {journal}
  {\bibinfo  {journal} {Phys. Rev. Lett.}\ }\textbf {\bibinfo {volume} {99}},\
  \bibinfo {pages} {256802} (\bibinfo {year} {2007})}\BibitemShut {NoStop}%
\bibitem [{\citenamefont {Cao}\ \emph {et~al.}(2018{\natexlab{a}})\citenamefont
  {Cao}, \citenamefont {Fatemi}, \citenamefont {Demir}, \citenamefont {Fang},
  \citenamefont {Tomarken}, \citenamefont {Luo}, \citenamefont
  {{Sanchez-Yamagishi}}, \citenamefont {Watanabe}, \citenamefont {Taniguchi},
  \citenamefont {Kaxiras}, \citenamefont {Ashoori},\ and\ \citenamefont
  {{Jarillo-Herrero}}}]{CAO18}%
  \BibitemOpen
  \bibfield  {author} {\bibinfo {author} {\bibfnamefont {Y.}~\bibnamefont
  {Cao}}, \bibinfo {author} {\bibfnamefont {V.}~\bibnamefont {Fatemi}},
  \bibinfo {author} {\bibfnamefont {A.}~\bibnamefont {Demir}}, \bibinfo
  {author} {\bibfnamefont {S.}~\bibnamefont {Fang}}, \bibinfo {author}
  {\bibfnamefont {S.~L.}\ \bibnamefont {Tomarken}}, \bibinfo {author}
  {\bibfnamefont {J.~Y.}\ \bibnamefont {Luo}}, \bibinfo {author} {\bibfnamefont
  {J.~D.}\ \bibnamefont {{Sanchez-Yamagishi}}}, \bibinfo {author}
  {\bibfnamefont {K.}~\bibnamefont {Watanabe}}, \bibinfo {author}
  {\bibfnamefont {T.}~\bibnamefont {Taniguchi}}, \bibinfo {author}
  {\bibfnamefont {E.}~\bibnamefont {Kaxiras}}, \bibinfo {author} {\bibfnamefont
  {R.~C.}\ \bibnamefont {Ashoori}},\ and\ \bibinfo {author} {\bibfnamefont
  {P.}~\bibnamefont {{Jarillo-Herrero}}},\ }\href
  {https://doi.org/10.1038/nature26154} {\bibfield  {journal} {\bibinfo
  {journal} {Nature}\ }\textbf {\bibinfo {volume} {556}},\ \bibinfo {pages}
  {80} (\bibinfo {year} {2018}{\natexlab{a}})}\BibitemShut {NoStop}%
\bibitem [{\citenamefont {Kerelsky}\ \emph {et~al.}(2019)\citenamefont
  {Kerelsky}, \citenamefont {McGilly}, \citenamefont {Kennes}, \citenamefont
  {Xian}, \citenamefont {Yankowitz}, \citenamefont {Chen}, \citenamefont
  {Watanabe}, \citenamefont {Taniguchi}, \citenamefont {Hone}, \citenamefont
  {Dean}, \citenamefont {Rubio},\ and\ \citenamefont {Pasupathy}}]{KER19}%
  \BibitemOpen
  \bibfield  {author} {\bibinfo {author} {\bibfnamefont {A.}~\bibnamefont
  {Kerelsky}}, \bibinfo {author} {\bibfnamefont {L.~J.}\ \bibnamefont
  {McGilly}}, \bibinfo {author} {\bibfnamefont {D.~M.}\ \bibnamefont {Kennes}},
  \bibinfo {author} {\bibfnamefont {L.}~\bibnamefont {Xian}}, \bibinfo {author}
  {\bibfnamefont {M.}~\bibnamefont {Yankowitz}}, \bibinfo {author}
  {\bibfnamefont {S.}~\bibnamefont {Chen}}, \bibinfo {author} {\bibfnamefont
  {K.}~\bibnamefont {Watanabe}}, \bibinfo {author} {\bibfnamefont
  {T.}~\bibnamefont {Taniguchi}}, \bibinfo {author} {\bibfnamefont
  {J.}~\bibnamefont {Hone}}, \bibinfo {author} {\bibfnamefont {C.}~\bibnamefont
  {Dean}}, \bibinfo {author} {\bibfnamefont {A.}~\bibnamefont {Rubio}},\ and\
  \bibinfo {author} {\bibfnamefont {A.~N.}\ \bibnamefont {Pasupathy}},\ }\href
  {https://doi.org/10.1038/s41586-019-1431-9} {\bibfield  {journal} {\bibinfo
  {journal} {Nature}\ }\textbf {\bibinfo {volume} {572}},\ \bibinfo {pages}
  {95} (\bibinfo {year} {2019})}\BibitemShut {NoStop}%
\bibitem [{\citenamefont {Xie}\ \emph {et~al.}(2019)\citenamefont {Xie},
  \citenamefont {Lian}, \citenamefont {J{\"a}ck}, \citenamefont {Liu},
  \citenamefont {Chiu}, \citenamefont {Watanabe}, \citenamefont {Taniguchi},
  \citenamefont {Bernevig},\ and\ \citenamefont {Yazdani}}]{XIE19}%
  \BibitemOpen
  \bibfield  {author} {\bibinfo {author} {\bibfnamefont {Y.}~\bibnamefont
  {Xie}}, \bibinfo {author} {\bibfnamefont {B.}~\bibnamefont {Lian}}, \bibinfo
  {author} {\bibfnamefont {B.}~\bibnamefont {J{\"a}ck}}, \bibinfo {author}
  {\bibfnamefont {X.}~\bibnamefont {Liu}}, \bibinfo {author} {\bibfnamefont
  {C.-L.}\ \bibnamefont {Chiu}}, \bibinfo {author} {\bibfnamefont
  {K.}~\bibnamefont {Watanabe}}, \bibinfo {author} {\bibfnamefont
  {T.}~\bibnamefont {Taniguchi}}, \bibinfo {author} {\bibfnamefont {B.~A.}\
  \bibnamefont {Bernevig}},\ and\ \bibinfo {author} {\bibfnamefont
  {A.}~\bibnamefont {Yazdani}},\ }\href
  {https://doi.org/10.1038/s41586-019-1422-x} {\bibfield  {journal} {\bibinfo
  {journal} {Nature}\ }\textbf {\bibinfo {volume} {572}},\ \bibinfo {pages}
  {101} (\bibinfo {year} {2019})}\BibitemShut {NoStop}%
\bibitem [{\citenamefont {Sharpe}\ \emph {et~al.}(2019)\citenamefont {Sharpe},
  \citenamefont {Fox}, \citenamefont {Barnard}, \citenamefont {Finney},
  \citenamefont {Watanabe}, \citenamefont {Taniguchi}, \citenamefont
  {Kastner},\ and\ \citenamefont {{Goldhaber-Gordon}}}]{SHA19}%
  \BibitemOpen
  \bibfield  {author} {\bibinfo {author} {\bibfnamefont {A.~L.}\ \bibnamefont
  {Sharpe}}, \bibinfo {author} {\bibfnamefont {E.~J.}\ \bibnamefont {Fox}},
  \bibinfo {author} {\bibfnamefont {A.~W.}\ \bibnamefont {Barnard}}, \bibinfo
  {author} {\bibfnamefont {J.}~\bibnamefont {Finney}}, \bibinfo {author}
  {\bibfnamefont {K.}~\bibnamefont {Watanabe}}, \bibinfo {author}
  {\bibfnamefont {T.}~\bibnamefont {Taniguchi}}, \bibinfo {author}
  {\bibfnamefont {M.~A.}\ \bibnamefont {Kastner}},\ and\ \bibinfo {author}
  {\bibfnamefont {D.}~\bibnamefont {{Goldhaber-Gordon}}},\ }\href
  {https://doi.org/10.1126/science.aaw3780} {\bibfield  {journal} {\bibinfo
  {journal} {Science}\ }\textbf {\bibinfo {volume} {365}},\ \bibinfo {pages}
  {605} (\bibinfo {year} {2019})}\BibitemShut {NoStop}%
\bibitem [{\citenamefont {Jiang}\ \emph {et~al.}(2019)\citenamefont {Jiang},
  \citenamefont {Lai}, \citenamefont {Watanabe}, \citenamefont {Taniguchi},
  \citenamefont {Haule}, \citenamefont {Mao},\ and\ \citenamefont
  {Andrei}}]{JIA19}%
  \BibitemOpen
  \bibfield  {author} {\bibinfo {author} {\bibfnamefont {Y.}~\bibnamefont
  {Jiang}}, \bibinfo {author} {\bibfnamefont {X.}~\bibnamefont {Lai}}, \bibinfo
  {author} {\bibfnamefont {K.}~\bibnamefont {Watanabe}}, \bibinfo {author}
  {\bibfnamefont {T.}~\bibnamefont {Taniguchi}}, \bibinfo {author}
  {\bibfnamefont {K.}~\bibnamefont {Haule}}, \bibinfo {author} {\bibfnamefont
  {J.}~\bibnamefont {Mao}},\ and\ \bibinfo {author} {\bibfnamefont {E.~Y.}\
  \bibnamefont {Andrei}},\ }\href {https://doi.org/10.1038/s41586-019-1460-4}
  {\bibfield  {journal} {\bibinfo  {journal} {Nature}\ }\textbf {\bibinfo
  {volume} {573}},\ \bibinfo {pages} {91} (\bibinfo {year} {2019})}\BibitemShut
  {NoStop}%
\bibitem [{\citenamefont {Choi}\ \emph {et~al.}(2019)\citenamefont {Choi},
  \citenamefont {Kemmer}, \citenamefont {Peng}, \citenamefont {Thomson},
  \citenamefont {Arora}, \citenamefont {Polski}, \citenamefont {Zhang},
  \citenamefont {Ren}, \citenamefont {Alicea}, \citenamefont {Refael},
  \citenamefont {{von Oppen}}, \citenamefont {Watanabe}, \citenamefont
  {Taniguchi},\ and\ \citenamefont {{Nadj-Perge}}}]{CHO19}%
  \BibitemOpen
  \bibfield  {author} {\bibinfo {author} {\bibfnamefont {Y.}~\bibnamefont
  {Choi}}, \bibinfo {author} {\bibfnamefont {J.}~\bibnamefont {Kemmer}},
  \bibinfo {author} {\bibfnamefont {Y.}~\bibnamefont {Peng}}, \bibinfo {author}
  {\bibfnamefont {A.}~\bibnamefont {Thomson}}, \bibinfo {author} {\bibfnamefont
  {H.}~\bibnamefont {Arora}}, \bibinfo {author} {\bibfnamefont
  {R.}~\bibnamefont {Polski}}, \bibinfo {author} {\bibfnamefont
  {Y.}~\bibnamefont {Zhang}}, \bibinfo {author} {\bibfnamefont
  {H.}~\bibnamefont {Ren}}, \bibinfo {author} {\bibfnamefont {J.}~\bibnamefont
  {Alicea}}, \bibinfo {author} {\bibfnamefont {G.}~\bibnamefont {Refael}},
  \bibinfo {author} {\bibfnamefont {F.}~\bibnamefont {{von Oppen}}}, \bibinfo
  {author} {\bibfnamefont {K.}~\bibnamefont {Watanabe}}, \bibinfo {author}
  {\bibfnamefont {T.}~\bibnamefont {Taniguchi}},\ and\ \bibinfo {author}
  {\bibfnamefont {S.}~\bibnamefont {{Nadj-Perge}}},\ }\href
  {https://doi.org/10.1038/s41567-019-0606-5} {\bibfield  {journal} {\bibinfo
  {journal} {Nat. Phys.}\ }\textbf {\bibinfo {volume} {15}},\ \bibinfo {pages}
  {1174} (\bibinfo {year} {2019})}\BibitemShut {NoStop}%
\bibitem [{\citenamefont {Polshyn}\ \emph {et~al.}(2019)\citenamefont
  {Polshyn}, \citenamefont {Yankowitz}, \citenamefont {Chen}, \citenamefont
  {Zhang}, \citenamefont {Watanabe}, \citenamefont {Taniguchi}, \citenamefont
  {Dean},\ and\ \citenamefont {Young}}]{POL19}%
  \BibitemOpen
  \bibfield  {author} {\bibinfo {author} {\bibfnamefont {H.}~\bibnamefont
  {Polshyn}}, \bibinfo {author} {\bibfnamefont {M.}~\bibnamefont {Yankowitz}},
  \bibinfo {author} {\bibfnamefont {S.}~\bibnamefont {Chen}}, \bibinfo {author}
  {\bibfnamefont {Y.}~\bibnamefont {Zhang}}, \bibinfo {author} {\bibfnamefont
  {K.}~\bibnamefont {Watanabe}}, \bibinfo {author} {\bibfnamefont
  {T.}~\bibnamefont {Taniguchi}}, \bibinfo {author} {\bibfnamefont {C.~R.}\
  \bibnamefont {Dean}},\ and\ \bibinfo {author} {\bibfnamefont {A.~F.}\
  \bibnamefont {Young}},\ }\href {https://doi.org/10.1038/s41567-019-0596-3}
  {\bibfield  {journal} {\bibinfo  {journal} {Nat. Phys.}\ }\textbf {\bibinfo
  {volume} {15}},\ \bibinfo {pages} {1011} (\bibinfo {year}
  {2019})}\BibitemShut {NoStop}%
\bibitem [{\citenamefont {Yankowitz}\ \emph {et~al.}(2019)\citenamefont
  {Yankowitz}, \citenamefont {Chen}, \citenamefont {Polshyn}, \citenamefont
  {Zhang}, \citenamefont {Watanabe}, \citenamefont {Taniguchi}, \citenamefont
  {Graf}, \citenamefont {Young},\ and\ \citenamefont {Dean}}]{YAN19}%
  \BibitemOpen
  \bibfield  {author} {\bibinfo {author} {\bibfnamefont {M.}~\bibnamefont
  {Yankowitz}}, \bibinfo {author} {\bibfnamefont {S.}~\bibnamefont {Chen}},
  \bibinfo {author} {\bibfnamefont {H.}~\bibnamefont {Polshyn}}, \bibinfo
  {author} {\bibfnamefont {Y.}~\bibnamefont {Zhang}}, \bibinfo {author}
  {\bibfnamefont {K.}~\bibnamefont {Watanabe}}, \bibinfo {author}
  {\bibfnamefont {T.}~\bibnamefont {Taniguchi}}, \bibinfo {author}
  {\bibfnamefont {D.}~\bibnamefont {Graf}}, \bibinfo {author} {\bibfnamefont
  {A.~F.}\ \bibnamefont {Young}},\ and\ \bibinfo {author} {\bibfnamefont
  {C.~R.}\ \bibnamefont {Dean}},\ }\href
  {https://doi.org/10.1126/science.aav1910} {\bibfield  {journal} {\bibinfo
  {journal} {Science}\ }\textbf {\bibinfo {volume} {363}},\ \bibinfo {pages}
  {1059} (\bibinfo {year} {2019})}\BibitemShut {NoStop}%
\bibitem [{\citenamefont {Lu}\ \emph {et~al.}(2019)\citenamefont {Lu},
  \citenamefont {Stepanov}, \citenamefont {Yang}, \citenamefont {Xie},
  \citenamefont {Aamir}, \citenamefont {Das}, \citenamefont {Urgell},
  \citenamefont {Watanabe}, \citenamefont {Taniguchi}, \citenamefont {Zhang},
  \citenamefont {Bachtold}, \citenamefont {MacDonald},\ and\ \citenamefont
  {Efetov}}]{LU19}%
  \BibitemOpen
  \bibfield  {author} {\bibinfo {author} {\bibfnamefont {X.}~\bibnamefont
  {Lu}}, \bibinfo {author} {\bibfnamefont {P.}~\bibnamefont {Stepanov}},
  \bibinfo {author} {\bibfnamefont {W.}~\bibnamefont {Yang}}, \bibinfo {author}
  {\bibfnamefont {M.}~\bibnamefont {Xie}}, \bibinfo {author} {\bibfnamefont
  {M.~A.}\ \bibnamefont {Aamir}}, \bibinfo {author} {\bibfnamefont
  {I.}~\bibnamefont {Das}}, \bibinfo {author} {\bibfnamefont {C.}~\bibnamefont
  {Urgell}}, \bibinfo {author} {\bibfnamefont {K.}~\bibnamefont {Watanabe}},
  \bibinfo {author} {\bibfnamefont {T.}~\bibnamefont {Taniguchi}}, \bibinfo
  {author} {\bibfnamefont {G.}~\bibnamefont {Zhang}}, \bibinfo {author}
  {\bibfnamefont {A.}~\bibnamefont {Bachtold}}, \bibinfo {author}
  {\bibfnamefont {A.~H.}\ \bibnamefont {MacDonald}},\ and\ \bibinfo {author}
  {\bibfnamefont {D.~K.}\ \bibnamefont {Efetov}},\ }\href
  {https://doi.org/10.1038/s41586-019-1695-0} {\bibfield  {journal} {\bibinfo
  {journal} {Nature}\ }\textbf {\bibinfo {volume} {574}},\ \bibinfo {pages}
  {653} (\bibinfo {year} {2019})}\BibitemShut {NoStop}%
\bibitem [{\citenamefont {Stepanov}\ \emph {et~al.}(2020)\citenamefont
  {Stepanov}, \citenamefont {Das}, \citenamefont {Lu}, \citenamefont
  {Fahimniya}, \citenamefont {Watanabe}, \citenamefont {Taniguchi},
  \citenamefont {Koppens}, \citenamefont {Lischner}, \citenamefont {Levitov},\
  and\ \citenamefont {Efetov}}]{STE20}%
  \BibitemOpen
  \bibfield  {author} {\bibinfo {author} {\bibfnamefont {P.}~\bibnamefont
  {Stepanov}}, \bibinfo {author} {\bibfnamefont {I.}~\bibnamefont {Das}},
  \bibinfo {author} {\bibfnamefont {X.}~\bibnamefont {Lu}}, \bibinfo {author}
  {\bibfnamefont {A.}~\bibnamefont {Fahimniya}}, \bibinfo {author}
  {\bibfnamefont {K.}~\bibnamefont {Watanabe}}, \bibinfo {author}
  {\bibfnamefont {T.}~\bibnamefont {Taniguchi}}, \bibinfo {author}
  {\bibfnamefont {F.~H.~L.}\ \bibnamefont {Koppens}}, \bibinfo {author}
  {\bibfnamefont {J.}~\bibnamefont {Lischner}}, \bibinfo {author}
  {\bibfnamefont {L.}~\bibnamefont {Levitov}},\ and\ \bibinfo {author}
  {\bibfnamefont {D.~K.}\ \bibnamefont {Efetov}},\ }\href
  {https://doi.org/10.1038/s41586-020-2459-6} {\bibfield  {journal} {\bibinfo
  {journal} {Nature}\ }\textbf {\bibinfo {volume} {583}},\ \bibinfo {pages}
  {375} (\bibinfo {year} {2020})}\BibitemShut {NoStop}%
\bibitem [{\citenamefont {Saito}\ \emph {et~al.}(2020)\citenamefont {Saito},
  \citenamefont {Ge}, \citenamefont {Watanabe}, \citenamefont {Taniguchi},\
  and\ \citenamefont {Young}}]{SAI20}%
  \BibitemOpen
  \bibfield  {author} {\bibinfo {author} {\bibfnamefont {Y.}~\bibnamefont
  {Saito}}, \bibinfo {author} {\bibfnamefont {J.}~\bibnamefont {Ge}}, \bibinfo
  {author} {\bibfnamefont {K.}~\bibnamefont {Watanabe}}, \bibinfo {author}
  {\bibfnamefont {T.}~\bibnamefont {Taniguchi}},\ and\ \bibinfo {author}
  {\bibfnamefont {A.~F.}\ \bibnamefont {Young}},\ }\href
  {https://doi.org/10.1038/s41567-020-0928-3} {\bibfield  {journal} {\bibinfo
  {journal} {Nat. Phys.}\ }\textbf {\bibinfo {volume} {16}},\ \bibinfo {pages}
  {926} (\bibinfo {year} {2020})}\BibitemShut {NoStop}%
\bibitem [{\citenamefont {Serlin}\ \emph {et~al.}(2020)\citenamefont {Serlin},
  \citenamefont {Tschirhart}, \citenamefont {Polshyn}, \citenamefont {Zhang},
  \citenamefont {Zhu}, \citenamefont {Watanabe}, \citenamefont {Taniguchi},
  \citenamefont {Balents},\ and\ \citenamefont {Young}}]{SER20}%
  \BibitemOpen
  \bibfield  {author} {\bibinfo {author} {\bibfnamefont {M.}~\bibnamefont
  {Serlin}}, \bibinfo {author} {\bibfnamefont {C.~L.}\ \bibnamefont
  {Tschirhart}}, \bibinfo {author} {\bibfnamefont {H.}~\bibnamefont {Polshyn}},
  \bibinfo {author} {\bibfnamefont {Y.}~\bibnamefont {Zhang}}, \bibinfo
  {author} {\bibfnamefont {J.}~\bibnamefont {Zhu}}, \bibinfo {author}
  {\bibfnamefont {K.}~\bibnamefont {Watanabe}}, \bibinfo {author}
  {\bibfnamefont {T.}~\bibnamefont {Taniguchi}}, \bibinfo {author}
  {\bibfnamefont {L.}~\bibnamefont {Balents}},\ and\ \bibinfo {author}
  {\bibfnamefont {A.~F.}\ \bibnamefont {Young}},\ }\href
  {https://doi.org/10.1126/science.aay5533} {\bibfield  {journal} {\bibinfo
  {journal} {Science}\ }\textbf {\bibinfo {volume} {367}},\ \bibinfo {pages}
  {900} (\bibinfo {year} {2020})}\BibitemShut {NoStop}%
\bibitem [{\citenamefont {Chen}\ \emph {et~al.}(2020)\citenamefont {Chen},
  \citenamefont {Sharpe}, \citenamefont {Fox}, \citenamefont {Zhang},
  \citenamefont {Wang}, \citenamefont {Jiang}, \citenamefont {Lyu},
  \citenamefont {Li}, \citenamefont {Watanabe}, \citenamefont {Taniguchi},
  \citenamefont {Shi}, \citenamefont {Senthil}, \citenamefont
  {{Goldhaber-Gordon}}, \citenamefont {Zhang},\ and\ \citenamefont
  {Wang}}]{CHE20b}%
  \BibitemOpen
  \bibfield  {author} {\bibinfo {author} {\bibfnamefont {G.}~\bibnamefont
  {Chen}}, \bibinfo {author} {\bibfnamefont {A.~L.}\ \bibnamefont {Sharpe}},
  \bibinfo {author} {\bibfnamefont {E.~J.}\ \bibnamefont {Fox}}, \bibinfo
  {author} {\bibfnamefont {Y.-H.}\ \bibnamefont {Zhang}}, \bibinfo {author}
  {\bibfnamefont {S.}~\bibnamefont {Wang}}, \bibinfo {author} {\bibfnamefont
  {L.}~\bibnamefont {Jiang}}, \bibinfo {author} {\bibfnamefont
  {B.}~\bibnamefont {Lyu}}, \bibinfo {author} {\bibfnamefont {H.}~\bibnamefont
  {Li}}, \bibinfo {author} {\bibfnamefont {K.}~\bibnamefont {Watanabe}},
  \bibinfo {author} {\bibfnamefont {T.}~\bibnamefont {Taniguchi}}, \bibinfo
  {author} {\bibfnamefont {Z.}~\bibnamefont {Shi}}, \bibinfo {author}
  {\bibfnamefont {T.}~\bibnamefont {Senthil}}, \bibinfo {author} {\bibfnamefont
  {D.}~\bibnamefont {{Goldhaber-Gordon}}}, \bibinfo {author} {\bibfnamefont
  {Y.}~\bibnamefont {Zhang}},\ and\ \bibinfo {author} {\bibfnamefont
  {F.}~\bibnamefont {Wang}},\ }\href
  {https://doi.org/10.1038/s41586-020-2049-7} {\bibfield  {journal} {\bibinfo
  {journal} {Nature}\ }\textbf {\bibinfo {volume} {579}},\ \bibinfo {pages}
  {56} (\bibinfo {year} {2020})}\BibitemShut {NoStop}%
\bibitem [{\citenamefont {Wong}\ \emph {et~al.}(2020)\citenamefont {Wong},
  \citenamefont {Nuckolls}, \citenamefont {Oh}, \citenamefont {Lian},
  \citenamefont {Xie}, \citenamefont {Jeon}, \citenamefont {Watanabe},
  \citenamefont {Taniguchi}, \citenamefont {Bernevig},\ and\ \citenamefont
  {Yazdani}}]{WON20}%
  \BibitemOpen
  \bibfield  {author} {\bibinfo {author} {\bibfnamefont {D.}~\bibnamefont
  {Wong}}, \bibinfo {author} {\bibfnamefont {K.~P.}\ \bibnamefont {Nuckolls}},
  \bibinfo {author} {\bibfnamefont {M.}~\bibnamefont {Oh}}, \bibinfo {author}
  {\bibfnamefont {B.}~\bibnamefont {Lian}}, \bibinfo {author} {\bibfnamefont
  {Y.}~\bibnamefont {Xie}}, \bibinfo {author} {\bibfnamefont {S.}~\bibnamefont
  {Jeon}}, \bibinfo {author} {\bibfnamefont {K.}~\bibnamefont {Watanabe}},
  \bibinfo {author} {\bibfnamefont {T.}~\bibnamefont {Taniguchi}}, \bibinfo
  {author} {\bibfnamefont {B.~A.}\ \bibnamefont {Bernevig}},\ and\ \bibinfo
  {author} {\bibfnamefont {A.}~\bibnamefont {Yazdani}},\ }\href
  {https://doi.org/10.1038/s41586-020-2339-0} {\bibfield  {journal} {\bibinfo
  {journal} {Nature}\ }\textbf {\bibinfo {volume} {582}},\ \bibinfo {pages}
  {198} (\bibinfo {year} {2020})}\BibitemShut {NoStop}%
\bibitem [{\citenamefont {Choi}\ \emph {et~al.}(2020)\citenamefont {Choi},
  \citenamefont {Kim}, \citenamefont {Peng}, \citenamefont {Thomson},
  \citenamefont {Lewandowski}, \citenamefont {Polski}, \citenamefont {Zhang},
  \citenamefont {Arora}, \citenamefont {Watanabe}, \citenamefont {Taniguchi},
  \citenamefont {Alicea},\ and\ \citenamefont {{Nadj-Perge}}}]{CHO20}%
  \BibitemOpen
  \bibfield  {author} {\bibinfo {author} {\bibfnamefont {Y.}~\bibnamefont
  {Choi}}, \bibinfo {author} {\bibfnamefont {H.}~\bibnamefont {Kim}}, \bibinfo
  {author} {\bibfnamefont {Y.}~\bibnamefont {Peng}}, \bibinfo {author}
  {\bibfnamefont {A.}~\bibnamefont {Thomson}}, \bibinfo {author} {\bibfnamefont
  {C.}~\bibnamefont {Lewandowski}}, \bibinfo {author} {\bibfnamefont
  {R.}~\bibnamefont {Polski}}, \bibinfo {author} {\bibfnamefont
  {Y.}~\bibnamefont {Zhang}}, \bibinfo {author} {\bibfnamefont {H.~S.}\
  \bibnamefont {Arora}}, \bibinfo {author} {\bibfnamefont {K.}~\bibnamefont
  {Watanabe}}, \bibinfo {author} {\bibfnamefont {T.}~\bibnamefont {Taniguchi}},
  \bibinfo {author} {\bibfnamefont {J.}~\bibnamefont {Alicea}},\ and\ \bibinfo
  {author} {\bibfnamefont {S.}~\bibnamefont {{Nadj-Perge}}},\ }\href@noop {}
  {\bibfield  {journal} {\bibinfo  {journal} {arXiv:2008.11746 [cond-mat]}\ }
  (\bibinfo {year} {2020})},\ \Eprint {https://arxiv.org/abs/2008.11746}
  {arxiv:2008.11746 [cond-mat]} \BibitemShut {NoStop}%
\bibitem [{\citenamefont {Nuckolls}\ \emph {et~al.}(2020)\citenamefont
  {Nuckolls}, \citenamefont {Oh}, \citenamefont {Wong}, \citenamefont {Lian},
  \citenamefont {Watanabe}, \citenamefont {Taniguchi}, \citenamefont
  {Bernevig},\ and\ \citenamefont {Yazdani}}]{NUC20}%
  \BibitemOpen
  \bibfield  {author} {\bibinfo {author} {\bibfnamefont {K.~P.}\ \bibnamefont
  {Nuckolls}}, \bibinfo {author} {\bibfnamefont {M.}~\bibnamefont {Oh}},
  \bibinfo {author} {\bibfnamefont {D.}~\bibnamefont {Wong}}, \bibinfo {author}
  {\bibfnamefont {B.}~\bibnamefont {Lian}}, \bibinfo {author} {\bibfnamefont
  {K.}~\bibnamefont {Watanabe}}, \bibinfo {author} {\bibfnamefont
  {T.}~\bibnamefont {Taniguchi}}, \bibinfo {author} {\bibfnamefont {B.~A.}\
  \bibnamefont {Bernevig}},\ and\ \bibinfo {author} {\bibfnamefont
  {A.}~\bibnamefont {Yazdani}},\ }\href
  {https://doi.org/10.1038/s41586-020-3028-8} {\bibfield  {journal} {\bibinfo
  {journal} {Nature}\ }\textbf {\bibinfo {volume} {588}},\ \bibinfo {pages}
  {610} (\bibinfo {year} {2020})}\BibitemShut {NoStop}%
\bibitem [{\citenamefont {Choi}\ \emph
  {et~al.}(2021{\natexlab{a}})\citenamefont {Choi}, \citenamefont {Kim},
  \citenamefont {Peng}, \citenamefont {Thomson}, \citenamefont {Lewandowski},
  \citenamefont {Polski}, \citenamefont {Zhang}, \citenamefont {Arora},
  \citenamefont {Watanabe}, \citenamefont {Taniguchi}, \citenamefont {Alicea},\
  and\ \citenamefont {{Nadj-Perge}}}]{CHO21}%
  \BibitemOpen
  \bibfield  {author} {\bibinfo {author} {\bibfnamefont {Y.}~\bibnamefont
  {Choi}}, \bibinfo {author} {\bibfnamefont {H.}~\bibnamefont {Kim}}, \bibinfo
  {author} {\bibfnamefont {Y.}~\bibnamefont {Peng}}, \bibinfo {author}
  {\bibfnamefont {A.}~\bibnamefont {Thomson}}, \bibinfo {author} {\bibfnamefont
  {C.}~\bibnamefont {Lewandowski}}, \bibinfo {author} {\bibfnamefont
  {R.}~\bibnamefont {Polski}}, \bibinfo {author} {\bibfnamefont
  {Y.}~\bibnamefont {Zhang}}, \bibinfo {author} {\bibfnamefont {H.~S.}\
  \bibnamefont {Arora}}, \bibinfo {author} {\bibfnamefont {K.}~\bibnamefont
  {Watanabe}}, \bibinfo {author} {\bibfnamefont {T.}~\bibnamefont {Taniguchi}},
  \bibinfo {author} {\bibfnamefont {J.}~\bibnamefont {Alicea}},\ and\ \bibinfo
  {author} {\bibfnamefont {S.}~\bibnamefont {{Nadj-Perge}}},\ }\href
  {https://doi.org/10.1038/s41586-020-03159-7} {\bibfield  {journal} {\bibinfo
  {journal} {Nature}\ }\textbf {\bibinfo {volume} {589}},\ \bibinfo {pages}
  {536} (\bibinfo {year} {2021}{\natexlab{a}})}\BibitemShut {NoStop}%
\bibitem [{\citenamefont {Saito}\ \emph
  {et~al.}(2021{\natexlab{a}})\citenamefont {Saito}, \citenamefont {Ge},
  \citenamefont {Rademaker}, \citenamefont {Watanabe}, \citenamefont
  {Taniguchi}, \citenamefont {Abanin},\ and\ \citenamefont {Young}}]{SAI21}%
  \BibitemOpen
  \bibfield  {author} {\bibinfo {author} {\bibfnamefont {Y.}~\bibnamefont
  {Saito}}, \bibinfo {author} {\bibfnamefont {J.}~\bibnamefont {Ge}}, \bibinfo
  {author} {\bibfnamefont {L.}~\bibnamefont {Rademaker}}, \bibinfo {author}
  {\bibfnamefont {K.}~\bibnamefont {Watanabe}}, \bibinfo {author}
  {\bibfnamefont {T.}~\bibnamefont {Taniguchi}}, \bibinfo {author}
  {\bibfnamefont {D.~A.}\ \bibnamefont {Abanin}},\ and\ \bibinfo {author}
  {\bibfnamefont {A.~F.}\ \bibnamefont {Young}},\ }\href
  {https://doi.org/10.1038/s41567-020-01129-4} {\bibfield  {journal} {\bibinfo
  {journal} {Nat. Phys.}\ }\textbf {\bibinfo {volume} {17}},\ \bibinfo {pages}
  {478} (\bibinfo {year} {2021}{\natexlab{a}})}\BibitemShut {NoStop}%
\bibitem [{\citenamefont {Liu}\ \emph {et~al.}(2021{\natexlab{a}})\citenamefont
  {Liu}, \citenamefont {Wang}, \citenamefont {Watanabe}, \citenamefont
  {Taniguchi}, \citenamefont {Vafek},\ and\ \citenamefont {Li}}]{LIU21c}%
  \BibitemOpen
  \bibfield  {author} {\bibinfo {author} {\bibfnamefont {X.}~\bibnamefont
  {Liu}}, \bibinfo {author} {\bibfnamefont {Z.}~\bibnamefont {Wang}}, \bibinfo
  {author} {\bibfnamefont {K.}~\bibnamefont {Watanabe}}, \bibinfo {author}
  {\bibfnamefont {T.}~\bibnamefont {Taniguchi}}, \bibinfo {author}
  {\bibfnamefont {O.}~\bibnamefont {Vafek}},\ and\ \bibinfo {author}
  {\bibfnamefont {J.~I.~A.}\ \bibnamefont {Li}},\ }\href
  {https://doi.org/10.1126/science.abb8754} {\bibfield  {journal} {\bibinfo
  {journal} {Science}\ }\textbf {\bibinfo {volume} {371}},\ \bibinfo {pages}
  {1261} (\bibinfo {year} {2021}{\natexlab{a}})}\BibitemShut {NoStop}%
\bibitem [{\citenamefont {Park}\ \emph
  {et~al.}(2021{\natexlab{a}})\citenamefont {Park}, \citenamefont {Cao},
  \citenamefont {Watanabe}, \citenamefont {Taniguchi},\ and\ \citenamefont
  {{Jarillo-Herrero}}}]{PAR21c}%
  \BibitemOpen
  \bibfield  {author} {\bibinfo {author} {\bibfnamefont {J.~M.}\ \bibnamefont
  {Park}}, \bibinfo {author} {\bibfnamefont {Y.}~\bibnamefont {Cao}}, \bibinfo
  {author} {\bibfnamefont {K.}~\bibnamefont {Watanabe}}, \bibinfo {author}
  {\bibfnamefont {T.}~\bibnamefont {Taniguchi}},\ and\ \bibinfo {author}
  {\bibfnamefont {P.}~\bibnamefont {{Jarillo-Herrero}}},\ }\href
  {https://doi.org/10.1038/s41586-021-03366-w} {\bibfield  {journal} {\bibinfo
  {journal} {Nature}\ }\textbf {\bibinfo {volume} {592}},\ \bibinfo {pages}
  {43} (\bibinfo {year} {2021}{\natexlab{a}})}\BibitemShut {NoStop}%
\bibitem [{\citenamefont {Wu}\ \emph {et~al.}(2021{\natexlab{a}})\citenamefont
  {Wu}, \citenamefont {Zhang}, \citenamefont {Watanabe}, \citenamefont
  {Taniguchi},\ and\ \citenamefont {Andrei}}]{WU21a}%
  \BibitemOpen
  \bibfield  {author} {\bibinfo {author} {\bibfnamefont {S.}~\bibnamefont
  {Wu}}, \bibinfo {author} {\bibfnamefont {Z.}~\bibnamefont {Zhang}}, \bibinfo
  {author} {\bibfnamefont {K.}~\bibnamefont {Watanabe}}, \bibinfo {author}
  {\bibfnamefont {T.}~\bibnamefont {Taniguchi}},\ and\ \bibinfo {author}
  {\bibfnamefont {E.~Y.}\ \bibnamefont {Andrei}},\ }\href
  {https://doi.org/10.1038/s41563-020-00911-2} {\bibfield  {journal} {\bibinfo
  {journal} {Nat. Mater.}\ }\textbf {\bibinfo {volume} {20}},\ \bibinfo {pages}
  {488} (\bibinfo {year} {2021}{\natexlab{a}})}\BibitemShut {NoStop}%
\bibitem [{\citenamefont {Cao}\ \emph {et~al.}(2021{\natexlab{a}})\citenamefont
  {Cao}, \citenamefont {{Rodan-Legrain}}, \citenamefont {Park}, \citenamefont
  {Yuan}, \citenamefont {Watanabe}, \citenamefont {Taniguchi}, \citenamefont
  {Fernandes}, \citenamefont {Fu},\ and\ \citenamefont
  {{Jarillo-Herrero}}}]{CAO21}%
  \BibitemOpen
  \bibfield  {author} {\bibinfo {author} {\bibfnamefont {Y.}~\bibnamefont
  {Cao}}, \bibinfo {author} {\bibfnamefont {D.}~\bibnamefont
  {{Rodan-Legrain}}}, \bibinfo {author} {\bibfnamefont {J.~M.}\ \bibnamefont
  {Park}}, \bibinfo {author} {\bibfnamefont {N.~F.~Q.}\ \bibnamefont {Yuan}},
  \bibinfo {author} {\bibfnamefont {K.}~\bibnamefont {Watanabe}}, \bibinfo
  {author} {\bibfnamefont {T.}~\bibnamefont {Taniguchi}}, \bibinfo {author}
  {\bibfnamefont {R.~M.}\ \bibnamefont {Fernandes}}, \bibinfo {author}
  {\bibfnamefont {L.}~\bibnamefont {Fu}},\ and\ \bibinfo {author}
  {\bibfnamefont {P.}~\bibnamefont {{Jarillo-Herrero}}},\ }\href
  {https://doi.org/10.1126/science.abc2836} {\bibfield  {journal} {\bibinfo
  {journal} {Science}\ }\textbf {\bibinfo {volume} {372}},\ \bibinfo {pages}
  {264} (\bibinfo {year} {2021}{\natexlab{a}})}\BibitemShut {NoStop}%
\bibitem [{\citenamefont {Das}\ \emph {et~al.}(2021)\citenamefont {Das},
  \citenamefont {Lu}, \citenamefont {{Herzog-Arbeitman}}, \citenamefont {Song},
  \citenamefont {Watanabe}, \citenamefont {Taniguchi}, \citenamefont
  {Bernevig},\ and\ \citenamefont {Efetov}}]{DAS21}%
  \BibitemOpen
  \bibfield  {author} {\bibinfo {author} {\bibfnamefont {I.}~\bibnamefont
  {Das}}, \bibinfo {author} {\bibfnamefont {X.}~\bibnamefont {Lu}}, \bibinfo
  {author} {\bibfnamefont {J.}~\bibnamefont {{Herzog-Arbeitman}}}, \bibinfo
  {author} {\bibfnamefont {Z.-D.}\ \bibnamefont {Song}}, \bibinfo {author}
  {\bibfnamefont {K.}~\bibnamefont {Watanabe}}, \bibinfo {author}
  {\bibfnamefont {T.}~\bibnamefont {Taniguchi}}, \bibinfo {author}
  {\bibfnamefont {B.~A.}\ \bibnamefont {Bernevig}},\ and\ \bibinfo {author}
  {\bibfnamefont {D.~K.}\ \bibnamefont {Efetov}},\ }\href
  {https://doi.org/10.1038/s41567-021-01186-3} {\bibfield  {journal} {\bibinfo
  {journal} {Nat. Phys.}\ }\textbf {\bibinfo {volume} {17}},\ \bibinfo {pages}
  {710} (\bibinfo {year} {2021})}\BibitemShut {NoStop}%
\bibitem [{\citenamefont {Tschirhart}\ \emph {et~al.}(2021)\citenamefont
  {Tschirhart}, \citenamefont {Serlin}, \citenamefont {Polshyn}, \citenamefont
  {Shragai}, \citenamefont {Xia}, \citenamefont {Zhu}, \citenamefont {Zhang},
  \citenamefont {Watanabe}, \citenamefont {Taniguchi}, \citenamefont {Huber},\
  and\ \citenamefont {Young}}]{TSC21}%
  \BibitemOpen
  \bibfield  {author} {\bibinfo {author} {\bibfnamefont {C.~L.}\ \bibnamefont
  {Tschirhart}}, \bibinfo {author} {\bibfnamefont {M.}~\bibnamefont {Serlin}},
  \bibinfo {author} {\bibfnamefont {H.}~\bibnamefont {Polshyn}}, \bibinfo
  {author} {\bibfnamefont {A.}~\bibnamefont {Shragai}}, \bibinfo {author}
  {\bibfnamefont {Z.}~\bibnamefont {Xia}}, \bibinfo {author} {\bibfnamefont
  {J.}~\bibnamefont {Zhu}}, \bibinfo {author} {\bibfnamefont {Y.}~\bibnamefont
  {Zhang}}, \bibinfo {author} {\bibfnamefont {K.}~\bibnamefont {Watanabe}},
  \bibinfo {author} {\bibfnamefont {T.}~\bibnamefont {Taniguchi}}, \bibinfo
  {author} {\bibfnamefont {M.~E.}\ \bibnamefont {Huber}},\ and\ \bibinfo
  {author} {\bibfnamefont {A.~F.}\ \bibnamefont {Young}},\ }\href
  {https://doi.org/10.1126/science.abd3190} {\bibfield  {journal} {\bibinfo
  {journal} {Science}\ }\textbf {\bibinfo {volume} {372}},\ \bibinfo {pages}
  {1323} (\bibinfo {year} {2021})}\BibitemShut {NoStop}%
\bibitem [{\citenamefont {Pierce}\ \emph {et~al.}(2021)\citenamefont {Pierce},
  \citenamefont {Xie}, \citenamefont {Park}, \citenamefont {Khalaf},
  \citenamefont {Lee}, \citenamefont {Cao}, \citenamefont {Parker},
  \citenamefont {Forrester}, \citenamefont {Chen}, \citenamefont {Watanabe},
  \citenamefont {Taniguchi}, \citenamefont {Vishwanath}, \citenamefont
  {{Jarillo-Herrero}},\ and\ \citenamefont {Yacoby}}]{PIE21}%
  \BibitemOpen
  \bibfield  {author} {\bibinfo {author} {\bibfnamefont {A.~T.}\ \bibnamefont
  {Pierce}}, \bibinfo {author} {\bibfnamefont {Y.}~\bibnamefont {Xie}},
  \bibinfo {author} {\bibfnamefont {J.~M.}\ \bibnamefont {Park}}, \bibinfo
  {author} {\bibfnamefont {E.}~\bibnamefont {Khalaf}}, \bibinfo {author}
  {\bibfnamefont {S.~H.}\ \bibnamefont {Lee}}, \bibinfo {author} {\bibfnamefont
  {Y.}~\bibnamefont {Cao}}, \bibinfo {author} {\bibfnamefont {D.~E.}\
  \bibnamefont {Parker}}, \bibinfo {author} {\bibfnamefont {P.~R.}\
  \bibnamefont {Forrester}}, \bibinfo {author} {\bibfnamefont {S.}~\bibnamefont
  {Chen}}, \bibinfo {author} {\bibfnamefont {K.}~\bibnamefont {Watanabe}},
  \bibinfo {author} {\bibfnamefont {T.}~\bibnamefont {Taniguchi}}, \bibinfo
  {author} {\bibfnamefont {A.}~\bibnamefont {Vishwanath}}, \bibinfo {author}
  {\bibfnamefont {P.}~\bibnamefont {{Jarillo-Herrero}}},\ and\ \bibinfo
  {author} {\bibfnamefont {A.}~\bibnamefont {Yacoby}},\ }\href
  {https://doi.org/10.1038/s41567-021-01347-4} {\bibfield  {journal} {\bibinfo
  {journal} {Nat. Phys.}\ }\textbf {\bibinfo {volume} {17}},\ \bibinfo {pages}
  {1210} (\bibinfo {year} {2021})}\BibitemShut {NoStop}%
\bibitem [{\citenamefont {Stepanov}\ \emph {et~al.}(2021)\citenamefont
  {Stepanov}, \citenamefont {Xie}, \citenamefont {Taniguchi}, \citenamefont
  {Watanabe}, \citenamefont {Lu}, \citenamefont {MacDonald}, \citenamefont
  {Bernevig},\ and\ \citenamefont {Efetov}}]{STE21}%
  \BibitemOpen
  \bibfield  {author} {\bibinfo {author} {\bibfnamefont {P.}~\bibnamefont
  {Stepanov}}, \bibinfo {author} {\bibfnamefont {M.}~\bibnamefont {Xie}},
  \bibinfo {author} {\bibfnamefont {T.}~\bibnamefont {Taniguchi}}, \bibinfo
  {author} {\bibfnamefont {K.}~\bibnamefont {Watanabe}}, \bibinfo {author}
  {\bibfnamefont {X.}~\bibnamefont {Lu}}, \bibinfo {author} {\bibfnamefont
  {A.~H.}\ \bibnamefont {MacDonald}}, \bibinfo {author} {\bibfnamefont {B.~A.}\
  \bibnamefont {Bernevig}},\ and\ \bibinfo {author} {\bibfnamefont {D.~K.}\
  \bibnamefont {Efetov}},\ }\href
  {https://doi.org/10.1103/PhysRevLett.127.197701} {\bibfield  {journal}
  {\bibinfo  {journal} {Phys. Rev. Lett.}\ }\textbf {\bibinfo {volume} {127}},\
  \bibinfo {pages} {197701} (\bibinfo {year} {2021})}\BibitemShut {NoStop}%
\bibitem [{\citenamefont {Choi}\ \emph
  {et~al.}(2021{\natexlab{b}})\citenamefont {Choi}, \citenamefont {Kim},
  \citenamefont {Lewandowski}, \citenamefont {Peng}, \citenamefont {Thomson},
  \citenamefont {Polski}, \citenamefont {Zhang}, \citenamefont {Watanabe},
  \citenamefont {Taniguchi}, \citenamefont {Alicea},\ and\ \citenamefont
  {{Nadj-Perge}}}]{CHO21a}%
  \BibitemOpen
  \bibfield  {author} {\bibinfo {author} {\bibfnamefont {Y.}~\bibnamefont
  {Choi}}, \bibinfo {author} {\bibfnamefont {H.}~\bibnamefont {Kim}}, \bibinfo
  {author} {\bibfnamefont {C.}~\bibnamefont {Lewandowski}}, \bibinfo {author}
  {\bibfnamefont {Y.}~\bibnamefont {Peng}}, \bibinfo {author} {\bibfnamefont
  {A.}~\bibnamefont {Thomson}}, \bibinfo {author} {\bibfnamefont
  {R.}~\bibnamefont {Polski}}, \bibinfo {author} {\bibfnamefont
  {Y.}~\bibnamefont {Zhang}}, \bibinfo {author} {\bibfnamefont
  {K.}~\bibnamefont {Watanabe}}, \bibinfo {author} {\bibfnamefont
  {T.}~\bibnamefont {Taniguchi}}, \bibinfo {author} {\bibfnamefont
  {J.}~\bibnamefont {Alicea}},\ and\ \bibinfo {author} {\bibfnamefont
  {S.}~\bibnamefont {{Nadj-Perge}}},\ }\href
  {https://doi.org/10.1038/s41567-021-01359-0} {\bibfield  {journal} {\bibinfo
  {journal} {Nat. Phys.}\ }\textbf {\bibinfo {volume} {17}},\ \bibinfo {pages}
  {1375} (\bibinfo {year} {2021}{\natexlab{b}})}\BibitemShut {NoStop}%
\bibitem [{\citenamefont {Xie}\ \emph {et~al.}(2021{\natexlab{a}})\citenamefont
  {Xie}, \citenamefont {Pierce}, \citenamefont {Park}, \citenamefont {Parker},
  \citenamefont {Khalaf}, \citenamefont {Ledwith}, \citenamefont {Cao},
  \citenamefont {Lee}, \citenamefont {Chen}, \citenamefont {Forrester},
  \citenamefont {Watanabe}, \citenamefont {Taniguchi}, \citenamefont
  {Vishwanath}, \citenamefont {{Jarillo-Herrero}},\ and\ \citenamefont
  {Yacoby}}]{XIE21d}%
  \BibitemOpen
  \bibfield  {author} {\bibinfo {author} {\bibfnamefont {Y.}~\bibnamefont
  {Xie}}, \bibinfo {author} {\bibfnamefont {A.~T.}\ \bibnamefont {Pierce}},
  \bibinfo {author} {\bibfnamefont {J.~M.}\ \bibnamefont {Park}}, \bibinfo
  {author} {\bibfnamefont {D.~E.}\ \bibnamefont {Parker}}, \bibinfo {author}
  {\bibfnamefont {E.}~\bibnamefont {Khalaf}}, \bibinfo {author} {\bibfnamefont
  {P.}~\bibnamefont {Ledwith}}, \bibinfo {author} {\bibfnamefont
  {Y.}~\bibnamefont {Cao}}, \bibinfo {author} {\bibfnamefont {S.~H.}\
  \bibnamefont {Lee}}, \bibinfo {author} {\bibfnamefont {S.}~\bibnamefont
  {Chen}}, \bibinfo {author} {\bibfnamefont {P.~R.}\ \bibnamefont {Forrester}},
  \bibinfo {author} {\bibfnamefont {K.}~\bibnamefont {Watanabe}}, \bibinfo
  {author} {\bibfnamefont {T.}~\bibnamefont {Taniguchi}}, \bibinfo {author}
  {\bibfnamefont {A.}~\bibnamefont {Vishwanath}}, \bibinfo {author}
  {\bibfnamefont {P.}~\bibnamefont {{Jarillo-Herrero}}},\ and\ \bibinfo
  {author} {\bibfnamefont {A.}~\bibnamefont {Yacoby}},\ }\href
  {https://doi.org/10.1038/s41586-021-04002-3} {\bibfield  {journal} {\bibinfo
  {journal} {Nature}\ }\textbf {\bibinfo {volume} {600}},\ \bibinfo {pages}
  {439} (\bibinfo {year} {2021}{\natexlab{a}})}\BibitemShut {NoStop}%
\bibitem [{\citenamefont {Das}\ \emph {et~al.}(2022)\citenamefont {Das},
  \citenamefont {Shen}, \citenamefont {Jaoui}, \citenamefont
  {{Herzog-Arbeitman}}, \citenamefont {Chew}, \citenamefont {Cho},
  \citenamefont {Watanabe}, \citenamefont {Taniguchi}, \citenamefont {Piot},
  \citenamefont {Bernevig},\ and\ \citenamefont {Efetov}}]{DAS22}%
  \BibitemOpen
  \bibfield  {author} {\bibinfo {author} {\bibfnamefont {I.}~\bibnamefont
  {Das}}, \bibinfo {author} {\bibfnamefont {C.}~\bibnamefont {Shen}}, \bibinfo
  {author} {\bibfnamefont {A.}~\bibnamefont {Jaoui}}, \bibinfo {author}
  {\bibfnamefont {J.}~\bibnamefont {{Herzog-Arbeitman}}}, \bibinfo {author}
  {\bibfnamefont {A.}~\bibnamefont {Chew}}, \bibinfo {author} {\bibfnamefont
  {C.-W.}\ \bibnamefont {Cho}}, \bibinfo {author} {\bibfnamefont
  {K.}~\bibnamefont {Watanabe}}, \bibinfo {author} {\bibfnamefont
  {T.}~\bibnamefont {Taniguchi}}, \bibinfo {author} {\bibfnamefont {B.~A.}\
  \bibnamefont {Piot}}, \bibinfo {author} {\bibfnamefont {B.~A.}\ \bibnamefont
  {Bernevig}},\ and\ \bibinfo {author} {\bibfnamefont {D.~K.}\ \bibnamefont
  {Efetov}},\ }\href {https://doi.org/10.1103/PhysRevLett.128.217701}
  {\bibfield  {journal} {\bibinfo  {journal} {Phys. Rev. Lett.}\ }\textbf
  {\bibinfo {volume} {128}},\ \bibinfo {pages} {217701} (\bibinfo {year}
  {2022})}\BibitemShut {NoStop}%
\bibitem [{\citenamefont {Nuckolls}\ \emph {et~al.}(2023)\citenamefont
  {Nuckolls}, \citenamefont {Lee}, \citenamefont {Oh}, \citenamefont {Wong},
  \citenamefont {Soejima}, \citenamefont {Hong}, \citenamefont {C{\u a}lug{\u
  a}ru}, \citenamefont {{Herzog-Arbeitman}}, \citenamefont {Bernevig},
  \citenamefont {Watanabe}, \citenamefont {Taniguchi}, \citenamefont
  {Regnault}, \citenamefont {Zaletel},\ and\ \citenamefont {Yazdani}}]{NUC23}%
  \BibitemOpen
  \bibfield  {author} {\bibinfo {author} {\bibfnamefont {K.~P.}\ \bibnamefont
  {Nuckolls}}, \bibinfo {author} {\bibfnamefont {R.~L.}\ \bibnamefont {Lee}},
  \bibinfo {author} {\bibfnamefont {M.}~\bibnamefont {Oh}}, \bibinfo {author}
  {\bibfnamefont {D.}~\bibnamefont {Wong}}, \bibinfo {author} {\bibfnamefont
  {T.}~\bibnamefont {Soejima}}, \bibinfo {author} {\bibfnamefont {J.~P.}\
  \bibnamefont {Hong}}, \bibinfo {author} {\bibfnamefont {D.}~\bibnamefont
  {C{\u a}lug{\u a}ru}}, \bibinfo {author} {\bibfnamefont {J.}~\bibnamefont
  {{Herzog-Arbeitman}}}, \bibinfo {author} {\bibfnamefont {B.~A.}\ \bibnamefont
  {Bernevig}}, \bibinfo {author} {\bibfnamefont {K.}~\bibnamefont {Watanabe}},
  \bibinfo {author} {\bibfnamefont {T.}~\bibnamefont {Taniguchi}}, \bibinfo
  {author} {\bibfnamefont {N.}~\bibnamefont {Regnault}}, \bibinfo {author}
  {\bibfnamefont {M.~P.}\ \bibnamefont {Zaletel}},\ and\ \bibinfo {author}
  {\bibfnamefont {A.}~\bibnamefont {Yazdani}},\ }\href
  {https://doi.org/10.1038/s41586-023-06226-x} {\bibfield  {journal} {\bibinfo
  {journal} {Nature}\ }\textbf {\bibinfo {volume} {620}},\ \bibinfo {pages}
  {525} (\bibinfo {year} {2023})}\BibitemShut {NoStop}%
\bibitem [{\citenamefont {Yu}\ \emph {et~al.}(2023{\natexlab{a}})\citenamefont
  {Yu}, \citenamefont {Foutty}, \citenamefont {Kwan}, \citenamefont {Barber},
  \citenamefont {Watanabe}, \citenamefont {Taniguchi}, \citenamefont {Shen},
  \citenamefont {Parameswaran},\ and\ \citenamefont {Feldman}}]{YU23c}%
  \BibitemOpen
  \bibfield  {author} {\bibinfo {author} {\bibfnamefont {J.}~\bibnamefont
  {Yu}}, \bibinfo {author} {\bibfnamefont {B.~A.}\ \bibnamefont {Foutty}},
  \bibinfo {author} {\bibfnamefont {Y.~H.}\ \bibnamefont {Kwan}}, \bibinfo
  {author} {\bibfnamefont {M.~E.}\ \bibnamefont {Barber}}, \bibinfo {author}
  {\bibfnamefont {K.}~\bibnamefont {Watanabe}}, \bibinfo {author}
  {\bibfnamefont {T.}~\bibnamefont {Taniguchi}}, \bibinfo {author}
  {\bibfnamefont {Z.-X.}\ \bibnamefont {Shen}}, \bibinfo {author}
  {\bibfnamefont {S.~A.}\ \bibnamefont {Parameswaran}},\ and\ \bibinfo {author}
  {\bibfnamefont {B.~E.}\ \bibnamefont {Feldman}},\ }\href
  {https://doi.org/10.1038/s41467-023-42275-6} {\bibfield  {journal} {\bibinfo
  {journal} {Nat. Commun.}\ }\textbf {\bibinfo {volume} {14}},\ \bibinfo
  {pages} {6679} (\bibinfo {year} {2023}{\natexlab{a}})},\ \Eprint
  {https://arxiv.org/abs/2206.11304} {arxiv:2206.11304 [cond-mat]} \BibitemShut
  {NoStop}%
\bibitem [{\citenamefont {Cao}\ \emph {et~al.}(2018{\natexlab{b}})\citenamefont
  {Cao}, \citenamefont {Fatemi}, \citenamefont {Fang}, \citenamefont
  {Watanabe}, \citenamefont {Taniguchi}, \citenamefont {Kaxiras},\ and\
  \citenamefont {{Jarillo-Herrero}}}]{CAO18a}%
  \BibitemOpen
  \bibfield  {author} {\bibinfo {author} {\bibfnamefont {Y.}~\bibnamefont
  {Cao}}, \bibinfo {author} {\bibfnamefont {V.}~\bibnamefont {Fatemi}},
  \bibinfo {author} {\bibfnamefont {S.}~\bibnamefont {Fang}}, \bibinfo {author}
  {\bibfnamefont {K.}~\bibnamefont {Watanabe}}, \bibinfo {author}
  {\bibfnamefont {T.}~\bibnamefont {Taniguchi}}, \bibinfo {author}
  {\bibfnamefont {E.}~\bibnamefont {Kaxiras}},\ and\ \bibinfo {author}
  {\bibfnamefont {P.}~\bibnamefont {{Jarillo-Herrero}}},\ }\href
  {https://doi.org/10.1038/nature26160} {\bibfield  {journal} {\bibinfo
  {journal} {Nature}\ }\textbf {\bibinfo {volume} {556}},\ \bibinfo {pages}
  {43} (\bibinfo {year} {2018}{\natexlab{b}})}\BibitemShut {NoStop}%
\bibitem [{\citenamefont {{de Vries}}\ \emph {et~al.}(2021)\citenamefont {{de
  Vries}}, \citenamefont {Portol{\'e}s}, \citenamefont {Zheng}, \citenamefont
  {Taniguchi}, \citenamefont {Watanabe}, \citenamefont {Ihn}, \citenamefont
  {Ensslin},\ and\ \citenamefont {Rickhaus}}]{DE21a}%
  \BibitemOpen
  \bibfield  {author} {\bibinfo {author} {\bibfnamefont {F.~K.}\ \bibnamefont
  {{de Vries}}}, \bibinfo {author} {\bibfnamefont {E.}~\bibnamefont
  {Portol{\'e}s}}, \bibinfo {author} {\bibfnamefont {G.}~\bibnamefont {Zheng}},
  \bibinfo {author} {\bibfnamefont {T.}~\bibnamefont {Taniguchi}}, \bibinfo
  {author} {\bibfnamefont {K.}~\bibnamefont {Watanabe}}, \bibinfo {author}
  {\bibfnamefont {T.}~\bibnamefont {Ihn}}, \bibinfo {author} {\bibfnamefont
  {K.}~\bibnamefont {Ensslin}},\ and\ \bibinfo {author} {\bibfnamefont
  {P.}~\bibnamefont {Rickhaus}},\ }\href
  {https://doi.org/10.1038/s41565-021-00896-2} {\bibfield  {journal} {\bibinfo
  {journal} {Nat. Nanotechnol.}\ }\textbf {\bibinfo {volume} {16}},\ \bibinfo
  {pages} {760} (\bibinfo {year} {2021})}\BibitemShut {NoStop}%
\bibitem [{\citenamefont {Oh}\ \emph {et~al.}(2021)\citenamefont {Oh},
  \citenamefont {Nuckolls}, \citenamefont {Wong}, \citenamefont {Lee},
  \citenamefont {Liu}, \citenamefont {Watanabe}, \citenamefont {Taniguchi},\
  and\ \citenamefont {Yazdani}}]{OH21}%
  \BibitemOpen
  \bibfield  {author} {\bibinfo {author} {\bibfnamefont {M.}~\bibnamefont
  {Oh}}, \bibinfo {author} {\bibfnamefont {K.~P.}\ \bibnamefont {Nuckolls}},
  \bibinfo {author} {\bibfnamefont {D.}~\bibnamefont {Wong}}, \bibinfo {author}
  {\bibfnamefont {R.~L.}\ \bibnamefont {Lee}}, \bibinfo {author} {\bibfnamefont
  {X.}~\bibnamefont {Liu}}, \bibinfo {author} {\bibfnamefont {K.}~\bibnamefont
  {Watanabe}}, \bibinfo {author} {\bibfnamefont {T.}~\bibnamefont
  {Taniguchi}},\ and\ \bibinfo {author} {\bibfnamefont {A.}~\bibnamefont
  {Yazdani}},\ }\href {https://doi.org/10.1038/s41586-021-04121-x} {\bibfield
  {journal} {\bibinfo  {journal} {Nature}\ }\textbf {\bibinfo {volume} {600}},\
  \bibinfo {pages} {240} (\bibinfo {year} {2021})}\BibitemShut {NoStop}%
\bibitem [{\citenamefont {Tian}\ \emph {et~al.}(2023)\citenamefont {Tian},
  \citenamefont {Gao}, \citenamefont {Zhang}, \citenamefont {Che},
  \citenamefont {Xu}, \citenamefont {Cheung}, \citenamefont {Watanabe},
  \citenamefont {Taniguchi}, \citenamefont {Randeria}, \citenamefont {Zhang},
  \citenamefont {Lau},\ and\ \citenamefont {Bockrath}}]{TIA23}%
  \BibitemOpen
  \bibfield  {author} {\bibinfo {author} {\bibfnamefont {H.}~\bibnamefont
  {Tian}}, \bibinfo {author} {\bibfnamefont {X.}~\bibnamefont {Gao}}, \bibinfo
  {author} {\bibfnamefont {Y.}~\bibnamefont {Zhang}}, \bibinfo {author}
  {\bibfnamefont {S.}~\bibnamefont {Che}}, \bibinfo {author} {\bibfnamefont
  {T.}~\bibnamefont {Xu}}, \bibinfo {author} {\bibfnamefont {P.}~\bibnamefont
  {Cheung}}, \bibinfo {author} {\bibfnamefont {K.}~\bibnamefont {Watanabe}},
  \bibinfo {author} {\bibfnamefont {T.}~\bibnamefont {Taniguchi}}, \bibinfo
  {author} {\bibfnamefont {M.}~\bibnamefont {Randeria}}, \bibinfo {author}
  {\bibfnamefont {F.}~\bibnamefont {Zhang}}, \bibinfo {author} {\bibfnamefont
  {C.~N.}\ \bibnamefont {Lau}},\ and\ \bibinfo {author} {\bibfnamefont {M.~W.}\
  \bibnamefont {Bockrath}},\ }\href
  {https://doi.org/10.1038/s41586-022-05576-2} {\bibfield  {journal} {\bibinfo
  {journal} {Nature}\ }\textbf {\bibinfo {volume} {614}},\ \bibinfo {pages}
  {440} (\bibinfo {year} {2023})}\BibitemShut {NoStop}%
\bibitem [{\citenamefont {Di~Battista}\ \emph {et~al.}(2022)\citenamefont
  {Di~Battista}, \citenamefont {Seifert}, \citenamefont {Watanabe},
  \citenamefont {Taniguchi}, \citenamefont {Fong}, \citenamefont {Principi},\
  and\ \citenamefont {Efetov}}]{DI22a}%
  \BibitemOpen
  \bibfield  {author} {\bibinfo {author} {\bibfnamefont {G.}~\bibnamefont
  {Di~Battista}}, \bibinfo {author} {\bibfnamefont {P.}~\bibnamefont
  {Seifert}}, \bibinfo {author} {\bibfnamefont {K.}~\bibnamefont {Watanabe}},
  \bibinfo {author} {\bibfnamefont {T.}~\bibnamefont {Taniguchi}}, \bibinfo
  {author} {\bibfnamefont {K.~C.}\ \bibnamefont {Fong}}, \bibinfo {author}
  {\bibfnamefont {A.}~\bibnamefont {Principi}},\ and\ \bibinfo {author}
  {\bibfnamefont {D.~K.}\ \bibnamefont {Efetov}},\ }\href
  {https://doi.org/10.1021/acs.nanolett.1c04512} {\bibfield  {journal}
  {\bibinfo  {journal} {Nano Lett.}\ }\textbf {\bibinfo {volume} {22}},\
  \bibinfo {pages} {6465} (\bibinfo {year} {2022})}\BibitemShut {NoStop}%
\bibitem [{\citenamefont {C{\u a}lug{\u a}ru}\ \emph
  {et~al.}(2022)\citenamefont {C{\u a}lug{\u a}ru}, \citenamefont {Regnault},
  \citenamefont {Oh}, \citenamefont {Nuckolls}, \citenamefont {Wong},
  \citenamefont {Lee}, \citenamefont {Yazdani}, \citenamefont {Vafek},\ and\
  \citenamefont {Bernevig}}]{CAL22d}%
  \BibitemOpen
  \bibfield  {author} {\bibinfo {author} {\bibfnamefont {D.}~\bibnamefont {C{\u
  a}lug{\u a}ru}}, \bibinfo {author} {\bibfnamefont {N.}~\bibnamefont
  {Regnault}}, \bibinfo {author} {\bibfnamefont {M.}~\bibnamefont {Oh}},
  \bibinfo {author} {\bibfnamefont {K.~P.}\ \bibnamefont {Nuckolls}}, \bibinfo
  {author} {\bibfnamefont {D.}~\bibnamefont {Wong}}, \bibinfo {author}
  {\bibfnamefont {R.~L.}\ \bibnamefont {Lee}}, \bibinfo {author} {\bibfnamefont
  {A.}~\bibnamefont {Yazdani}}, \bibinfo {author} {\bibfnamefont
  {O.}~\bibnamefont {Vafek}},\ and\ \bibinfo {author} {\bibfnamefont {B.~A.}\
  \bibnamefont {Bernevig}},\ }\href
  {https://doi.org/10.1103/PhysRevLett.129.117602} {\bibfield  {journal}
  {\bibinfo  {journal} {Phys. Rev. Lett.}\ }\textbf {\bibinfo {volume} {129}},\
  \bibinfo {pages} {117602} (\bibinfo {year} {2022})}\BibitemShut {NoStop}%
\bibitem [{\citenamefont {Tomarken}\ \emph {et~al.}(2019)\citenamefont
  {Tomarken}, \citenamefont {Cao}, \citenamefont {Demir}, \citenamefont
  {Watanabe}, \citenamefont {Taniguchi}, \citenamefont {{Jarillo-Herrero}},\
  and\ \citenamefont {Ashoori}}]{TOM19}%
  \BibitemOpen
  \bibfield  {author} {\bibinfo {author} {\bibfnamefont {S.~L.}\ \bibnamefont
  {Tomarken}}, \bibinfo {author} {\bibfnamefont {Y.}~\bibnamefont {Cao}},
  \bibinfo {author} {\bibfnamefont {A.}~\bibnamefont {Demir}}, \bibinfo
  {author} {\bibfnamefont {K.}~\bibnamefont {Watanabe}}, \bibinfo {author}
  {\bibfnamefont {T.}~\bibnamefont {Taniguchi}}, \bibinfo {author}
  {\bibfnamefont {P.}~\bibnamefont {{Jarillo-Herrero}}},\ and\ \bibinfo
  {author} {\bibfnamefont {R.~C.}\ \bibnamefont {Ashoori}},\ }\href
  {https://doi.org/10.1103/PhysRevLett.123.046601} {\bibfield  {journal}
  {\bibinfo  {journal} {Phys. Rev. Lett.}\ }\textbf {\bibinfo {volume} {123}},\
  \bibinfo {pages} {046601} (\bibinfo {year} {2019})}\BibitemShut {NoStop}%
\bibitem [{\citenamefont {Cao}\ \emph {et~al.}(2020)\citenamefont {Cao},
  \citenamefont {Chowdhury}, \citenamefont {{Rodan-Legrain}}, \citenamefont
  {{Rubies-Bigorda}}, \citenamefont {Watanabe}, \citenamefont {Taniguchi},
  \citenamefont {Senthil},\ and\ \citenamefont {{Jarillo-Herrero}}}]{CAO20}%
  \BibitemOpen
  \bibfield  {author} {\bibinfo {author} {\bibfnamefont {Y.}~\bibnamefont
  {Cao}}, \bibinfo {author} {\bibfnamefont {D.}~\bibnamefont {Chowdhury}},
  \bibinfo {author} {\bibfnamefont {D.}~\bibnamefont {{Rodan-Legrain}}},
  \bibinfo {author} {\bibfnamefont {O.}~\bibnamefont {{Rubies-Bigorda}}},
  \bibinfo {author} {\bibfnamefont {K.}~\bibnamefont {Watanabe}}, \bibinfo
  {author} {\bibfnamefont {T.}~\bibnamefont {Taniguchi}}, \bibinfo {author}
  {\bibfnamefont {T.}~\bibnamefont {Senthil}},\ and\ \bibinfo {author}
  {\bibfnamefont {P.}~\bibnamefont {{Jarillo-Herrero}}},\ }\href
  {https://doi.org/10.1103/PhysRevLett.124.076801} {\bibfield  {journal}
  {\bibinfo  {journal} {Phys. Rev. Lett.}\ }\textbf {\bibinfo {volume} {124}},\
  \bibinfo {pages} {076801} (\bibinfo {year} {2020})}\BibitemShut {NoStop}%
\bibitem [{\citenamefont {Zondiner}\ \emph {et~al.}(2020)\citenamefont
  {Zondiner}, \citenamefont {Rozen}, \citenamefont {{Rodan-Legrain}},
  \citenamefont {Cao}, \citenamefont {Queiroz}, \citenamefont {Taniguchi},
  \citenamefont {Watanabe}, \citenamefont {Oreg}, \citenamefont {{von Oppen}},
  \citenamefont {Stern}, \citenamefont {Berg}, \citenamefont
  {{Jarillo-Herrero}},\ and\ \citenamefont {Ilani}}]{ZON20}%
  \BibitemOpen
  \bibfield  {author} {\bibinfo {author} {\bibfnamefont {U.}~\bibnamefont
  {Zondiner}}, \bibinfo {author} {\bibfnamefont {A.}~\bibnamefont {Rozen}},
  \bibinfo {author} {\bibfnamefont {D.}~\bibnamefont {{Rodan-Legrain}}},
  \bibinfo {author} {\bibfnamefont {Y.}~\bibnamefont {Cao}}, \bibinfo {author}
  {\bibfnamefont {R.}~\bibnamefont {Queiroz}}, \bibinfo {author} {\bibfnamefont
  {T.}~\bibnamefont {Taniguchi}}, \bibinfo {author} {\bibfnamefont
  {K.}~\bibnamefont {Watanabe}}, \bibinfo {author} {\bibfnamefont
  {Y.}~\bibnamefont {Oreg}}, \bibinfo {author} {\bibfnamefont {F.}~\bibnamefont
  {{von Oppen}}}, \bibinfo {author} {\bibfnamefont {A.}~\bibnamefont {Stern}},
  \bibinfo {author} {\bibfnamefont {E.}~\bibnamefont {Berg}}, \bibinfo {author}
  {\bibfnamefont {P.}~\bibnamefont {{Jarillo-Herrero}}},\ and\ \bibinfo
  {author} {\bibfnamefont {S.}~\bibnamefont {Ilani}},\ }\href
  {https://doi.org/10.1038/s41586-020-2373-y} {\bibfield  {journal} {\bibinfo
  {journal} {Nature}\ }\textbf {\bibinfo {volume} {582}},\ \bibinfo {pages}
  {203} (\bibinfo {year} {2020})}\BibitemShut {NoStop}%
\bibitem [{\citenamefont {Lisi}\ \emph {et~al.}(2021)\citenamefont {Lisi},
  \citenamefont {Lu}, \citenamefont {Benschop}, \citenamefont {{de Jong}},
  \citenamefont {Stepanov}, \citenamefont {Duran}, \citenamefont {Margot},
  \citenamefont {Cucchi}, \citenamefont {Cappelli}, \citenamefont {Hunter},
  \citenamefont {Tamai}, \citenamefont {Kandyba}, \citenamefont {Giampietri},
  \citenamefont {Barinov}, \citenamefont {Jobst}, \citenamefont {Stalman},
  \citenamefont {Leeuwenhoek}, \citenamefont {Watanabe}, \citenamefont
  {Taniguchi}, \citenamefont {Rademaker}, \citenamefont {{van der Molen}},
  \citenamefont {Allan}, \citenamefont {Efetov},\ and\ \citenamefont
  {Baumberger}}]{LIS21}%
  \BibitemOpen
  \bibfield  {author} {\bibinfo {author} {\bibfnamefont {S.}~\bibnamefont
  {Lisi}}, \bibinfo {author} {\bibfnamefont {X.}~\bibnamefont {Lu}}, \bibinfo
  {author} {\bibfnamefont {T.}~\bibnamefont {Benschop}}, \bibinfo {author}
  {\bibfnamefont {T.~A.}\ \bibnamefont {{de Jong}}}, \bibinfo {author}
  {\bibfnamefont {P.}~\bibnamefont {Stepanov}}, \bibinfo {author}
  {\bibfnamefont {J.~R.}\ \bibnamefont {Duran}}, \bibinfo {author}
  {\bibfnamefont {F.}~\bibnamefont {Margot}}, \bibinfo {author} {\bibfnamefont
  {I.}~\bibnamefont {Cucchi}}, \bibinfo {author} {\bibfnamefont
  {E.}~\bibnamefont {Cappelli}}, \bibinfo {author} {\bibfnamefont
  {A.}~\bibnamefont {Hunter}}, \bibinfo {author} {\bibfnamefont
  {A.}~\bibnamefont {Tamai}}, \bibinfo {author} {\bibfnamefont
  {V.}~\bibnamefont {Kandyba}}, \bibinfo {author} {\bibfnamefont
  {A.}~\bibnamefont {Giampietri}}, \bibinfo {author} {\bibfnamefont
  {A.}~\bibnamefont {Barinov}}, \bibinfo {author} {\bibfnamefont
  {J.}~\bibnamefont {Jobst}}, \bibinfo {author} {\bibfnamefont
  {V.}~\bibnamefont {Stalman}}, \bibinfo {author} {\bibfnamefont
  {M.}~\bibnamefont {Leeuwenhoek}}, \bibinfo {author} {\bibfnamefont
  {K.}~\bibnamefont {Watanabe}}, \bibinfo {author} {\bibfnamefont
  {T.}~\bibnamefont {Taniguchi}}, \bibinfo {author} {\bibfnamefont
  {L.}~\bibnamefont {Rademaker}}, \bibinfo {author} {\bibfnamefont {S.~J.}\
  \bibnamefont {{van der Molen}}}, \bibinfo {author} {\bibfnamefont {M.~P.}\
  \bibnamefont {Allan}}, \bibinfo {author} {\bibfnamefont {D.~K.}\ \bibnamefont
  {Efetov}},\ and\ \bibinfo {author} {\bibfnamefont {F.}~\bibnamefont
  {Baumberger}},\ }\href {https://doi.org/10.1038/s41567-020-01041-x}
  {\bibfield  {journal} {\bibinfo  {journal} {Nat. Phys.}\ }\textbf {\bibinfo
  {volume} {17}},\ \bibinfo {pages} {189} (\bibinfo {year} {2021})}\BibitemShut
  {NoStop}%
\bibitem [{\citenamefont {Benschop}\ \emph {et~al.}(2021)\citenamefont
  {Benschop}, \citenamefont {{de Jong}}, \citenamefont {Stepanov},
  \citenamefont {Lu}, \citenamefont {Stalman}, \citenamefont {{van der Molen}},
  \citenamefont {Efetov},\ and\ \citenamefont {Allan}}]{BEN21}%
  \BibitemOpen
  \bibfield  {author} {\bibinfo {author} {\bibfnamefont {T.}~\bibnamefont
  {Benschop}}, \bibinfo {author} {\bibfnamefont {T.~A.}\ \bibnamefont {{de
  Jong}}}, \bibinfo {author} {\bibfnamefont {P.}~\bibnamefont {Stepanov}},
  \bibinfo {author} {\bibfnamefont {X.}~\bibnamefont {Lu}}, \bibinfo {author}
  {\bibfnamefont {V.}~\bibnamefont {Stalman}}, \bibinfo {author} {\bibfnamefont
  {S.~J.}\ \bibnamefont {{van der Molen}}}, \bibinfo {author} {\bibfnamefont
  {D.~K.}\ \bibnamefont {Efetov}},\ and\ \bibinfo {author} {\bibfnamefont
  {M.~P.}\ \bibnamefont {Allan}},\ }\href
  {https://doi.org/10.1103/PhysRevResearch.3.013153} {\bibfield  {journal}
  {\bibinfo  {journal} {Phys. Rev. Res.}\ }\textbf {\bibinfo {volume} {3}},\
  \bibinfo {pages} {013153} (\bibinfo {year} {2021})}\BibitemShut {NoStop}%
\bibitem [{\citenamefont {Lian}(2021)}]{LIA21c}%
  \BibitemOpen
  \bibfield  {author} {\bibinfo {author} {\bibfnamefont {B.}~\bibnamefont
  {Lian}},\ }\href {https://doi.org/10.1038/d41586-021-00843-0} {\bibfield
  {journal} {\bibinfo  {journal} {Nature}\ }\textbf {\bibinfo {volume} {592}},\
  \bibinfo {pages} {191} (\bibinfo {year} {2021})}\BibitemShut {NoStop}%
\bibitem [{\citenamefont {Rozen}\ \emph {et~al.}(2021)\citenamefont {Rozen},
  \citenamefont {Park}, \citenamefont {Zondiner}, \citenamefont {Cao},
  \citenamefont {{Rodan-Legrain}}, \citenamefont {Taniguchi}, \citenamefont
  {Watanabe}, \citenamefont {Oreg}, \citenamefont {Stern}, \citenamefont
  {Berg}, \citenamefont {{Jarillo-Herrero}},\ and\ \citenamefont
  {Ilani}}]{ROZ21}%
  \BibitemOpen
  \bibfield  {author} {\bibinfo {author} {\bibfnamefont {A.}~\bibnamefont
  {Rozen}}, \bibinfo {author} {\bibfnamefont {J.~M.}\ \bibnamefont {Park}},
  \bibinfo {author} {\bibfnamefont {U.}~\bibnamefont {Zondiner}}, \bibinfo
  {author} {\bibfnamefont {Y.}~\bibnamefont {Cao}}, \bibinfo {author}
  {\bibfnamefont {D.}~\bibnamefont {{Rodan-Legrain}}}, \bibinfo {author}
  {\bibfnamefont {T.}~\bibnamefont {Taniguchi}}, \bibinfo {author}
  {\bibfnamefont {K.}~\bibnamefont {Watanabe}}, \bibinfo {author}
  {\bibfnamefont {Y.}~\bibnamefont {Oreg}}, \bibinfo {author} {\bibfnamefont
  {A.}~\bibnamefont {Stern}}, \bibinfo {author} {\bibfnamefont
  {E.}~\bibnamefont {Berg}}, \bibinfo {author} {\bibfnamefont {P.}~\bibnamefont
  {{Jarillo-Herrero}}},\ and\ \bibinfo {author} {\bibfnamefont
  {S.}~\bibnamefont {Ilani}},\ }\href
  {https://doi.org/10.1038/s41586-021-03319-3} {\bibfield  {journal} {\bibinfo
  {journal} {Nature}\ }\textbf {\bibinfo {volume} {592}},\ \bibinfo {pages}
  {214} (\bibinfo {year} {2021})}\BibitemShut {NoStop}%
\bibitem [{\citenamefont {Saito}\ \emph
  {et~al.}(2021{\natexlab{b}})\citenamefont {Saito}, \citenamefont {Yang},
  \citenamefont {Ge}, \citenamefont {Liu}, \citenamefont {Taniguchi},
  \citenamefont {Watanabe}, \citenamefont {Li}, \citenamefont {Berg},\ and\
  \citenamefont {Young}}]{SAI21a}%
  \BibitemOpen
  \bibfield  {author} {\bibinfo {author} {\bibfnamefont {Y.}~\bibnamefont
  {Saito}}, \bibinfo {author} {\bibfnamefont {F.}~\bibnamefont {Yang}},
  \bibinfo {author} {\bibfnamefont {J.}~\bibnamefont {Ge}}, \bibinfo {author}
  {\bibfnamefont {X.}~\bibnamefont {Liu}}, \bibinfo {author} {\bibfnamefont
  {T.}~\bibnamefont {Taniguchi}}, \bibinfo {author} {\bibfnamefont
  {K.}~\bibnamefont {Watanabe}}, \bibinfo {author} {\bibfnamefont {J.~I.~A.}\
  \bibnamefont {Li}}, \bibinfo {author} {\bibfnamefont {E.}~\bibnamefont
  {Berg}},\ and\ \bibinfo {author} {\bibfnamefont {A.~F.}\ \bibnamefont
  {Young}},\ }\href {https://doi.org/10.1038/s41586-021-03409-2} {\bibfield
  {journal} {\bibinfo  {journal} {Nature}\ }\textbf {\bibinfo {volume} {592}},\
  \bibinfo {pages} {220} (\bibinfo {year} {2021}{\natexlab{b}})}\BibitemShut
  {NoStop}%
\bibitem [{\citenamefont {Lu}\ \emph {et~al.}(2021)\citenamefont {Lu},
  \citenamefont {Lian}, \citenamefont {Chaudhary}, \citenamefont {Piot},
  \citenamefont {Romagnoli}, \citenamefont {Watanabe}, \citenamefont
  {Taniguchi}, \citenamefont {Poggio}, \citenamefont {MacDonald}, \citenamefont
  {Bernevig},\ and\ \citenamefont {Efetov}}]{LU21}%
  \BibitemOpen
  \bibfield  {author} {\bibinfo {author} {\bibfnamefont {X.}~\bibnamefont
  {Lu}}, \bibinfo {author} {\bibfnamefont {B.}~\bibnamefont {Lian}}, \bibinfo
  {author} {\bibfnamefont {G.}~\bibnamefont {Chaudhary}}, \bibinfo {author}
  {\bibfnamefont {B.~A.}\ \bibnamefont {Piot}}, \bibinfo {author}
  {\bibfnamefont {G.}~\bibnamefont {Romagnoli}}, \bibinfo {author}
  {\bibfnamefont {K.}~\bibnamefont {Watanabe}}, \bibinfo {author}
  {\bibfnamefont {T.}~\bibnamefont {Taniguchi}}, \bibinfo {author}
  {\bibfnamefont {M.}~\bibnamefont {Poggio}}, \bibinfo {author} {\bibfnamefont
  {A.~H.}\ \bibnamefont {MacDonald}}, \bibinfo {author} {\bibfnamefont {B.~A.}\
  \bibnamefont {Bernevig}},\ and\ \bibinfo {author} {\bibfnamefont {D.~K.}\
  \bibnamefont {Efetov}},\ }\href {https://doi.org/10.1073/pnas.2100006118}
  {\bibfield  {journal} {\bibinfo  {journal} {PNAS}\ }\textbf {\bibinfo
  {volume} {118}},\ \bibinfo {pages} {e2100006118} (\bibinfo {year}
  {2021})}\BibitemShut {NoStop}%
\bibitem [{\citenamefont {Hesp}\ \emph {et~al.}(2021)\citenamefont {Hesp},
  \citenamefont {Torre}, \citenamefont {{Rodan-Legrain}}, \citenamefont
  {Novelli}, \citenamefont {Cao}, \citenamefont {Carr}, \citenamefont {Fang},
  \citenamefont {Stepanov}, \citenamefont {{Barcons-Ruiz}}, \citenamefont
  {Herzig~Sheinfux}, \citenamefont {Watanabe}, \citenamefont {Taniguchi},
  \citenamefont {Efetov}, \citenamefont {Kaxiras}, \citenamefont
  {{Jarillo-Herrero}}, \citenamefont {Polini},\ and\ \citenamefont
  {Koppens}}]{HES21}%
  \BibitemOpen
  \bibfield  {author} {\bibinfo {author} {\bibfnamefont {N.~C.~H.}\
  \bibnamefont {Hesp}}, \bibinfo {author} {\bibfnamefont {I.}~\bibnamefont
  {Torre}}, \bibinfo {author} {\bibfnamefont {D.}~\bibnamefont
  {{Rodan-Legrain}}}, \bibinfo {author} {\bibfnamefont {P.}~\bibnamefont
  {Novelli}}, \bibinfo {author} {\bibfnamefont {Y.}~\bibnamefont {Cao}},
  \bibinfo {author} {\bibfnamefont {S.}~\bibnamefont {Carr}}, \bibinfo {author}
  {\bibfnamefont {S.}~\bibnamefont {Fang}}, \bibinfo {author} {\bibfnamefont
  {P.}~\bibnamefont {Stepanov}}, \bibinfo {author} {\bibfnamefont
  {D.}~\bibnamefont {{Barcons-Ruiz}}}, \bibinfo {author} {\bibfnamefont
  {H.}~\bibnamefont {Herzig~Sheinfux}}, \bibinfo {author} {\bibfnamefont
  {K.}~\bibnamefont {Watanabe}}, \bibinfo {author} {\bibfnamefont
  {T.}~\bibnamefont {Taniguchi}}, \bibinfo {author} {\bibfnamefont {D.~K.}\
  \bibnamefont {Efetov}}, \bibinfo {author} {\bibfnamefont {E.}~\bibnamefont
  {Kaxiras}}, \bibinfo {author} {\bibfnamefont {P.}~\bibnamefont
  {{Jarillo-Herrero}}}, \bibinfo {author} {\bibfnamefont {M.}~\bibnamefont
  {Polini}},\ and\ \bibinfo {author} {\bibfnamefont {F.~H.~L.}\ \bibnamefont
  {Koppens}},\ }\href {https://doi.org/10.1038/s41567-021-01327-8} {\bibfield
  {journal} {\bibinfo  {journal} {Nat. Phys.}\ }\textbf {\bibinfo {volume}
  {17}},\ \bibinfo {pages} {1162} (\bibinfo {year} {2021})}\BibitemShut
  {NoStop}%
\bibitem [{\citenamefont {{D{\'i}ez-M{\'e}rida}}\ \emph
  {et~al.}(2023)\citenamefont {{D{\'i}ez-M{\'e}rida}}, \citenamefont
  {{D{\'i}ez-Carl{\'o}n}}, \citenamefont {Yang}, \citenamefont {Xie},
  \citenamefont {Gao}, \citenamefont {Senior}, \citenamefont {Watanabe},
  \citenamefont {Taniguchi}, \citenamefont {Lu}, \citenamefont {Higginbotham},
  \citenamefont {Law},\ and\ \citenamefont {Efetov}}]{DIE23}%
  \BibitemOpen
  \bibfield  {author} {\bibinfo {author} {\bibfnamefont {J.}~\bibnamefont
  {{D{\'i}ez-M{\'e}rida}}}, \bibinfo {author} {\bibfnamefont {A.}~\bibnamefont
  {{D{\'i}ez-Carl{\'o}n}}}, \bibinfo {author} {\bibfnamefont {S.~Y.}\
  \bibnamefont {Yang}}, \bibinfo {author} {\bibfnamefont {Y.-M.}\ \bibnamefont
  {Xie}}, \bibinfo {author} {\bibfnamefont {X.-J.}\ \bibnamefont {Gao}},
  \bibinfo {author} {\bibfnamefont {J.}~\bibnamefont {Senior}}, \bibinfo
  {author} {\bibfnamefont {K.}~\bibnamefont {Watanabe}}, \bibinfo {author}
  {\bibfnamefont {T.}~\bibnamefont {Taniguchi}}, \bibinfo {author}
  {\bibfnamefont {X.}~\bibnamefont {Lu}}, \bibinfo {author} {\bibfnamefont
  {A.~P.}\ \bibnamefont {Higginbotham}}, \bibinfo {author} {\bibfnamefont
  {K.~T.}\ \bibnamefont {Law}},\ and\ \bibinfo {author} {\bibfnamefont {D.~K.}\
  \bibnamefont {Efetov}},\ }\href {https://doi.org/10.1038/s41467-023-38005-7}
  {\bibfield  {journal} {\bibinfo  {journal} {Nat. Commun.}\ }\textbf {\bibinfo
  {volume} {14}},\ \bibinfo {pages} {2396} (\bibinfo {year}
  {2023})}\BibitemShut {NoStop}%
\bibitem [{\citenamefont {Hubmann}\ \emph {et~al.}(2022)\citenamefont
  {Hubmann}, \citenamefont {Soul}, \citenamefont {Di~Battista}, \citenamefont
  {Hild}, \citenamefont {Watanabe}, \citenamefont {Taniguchi}, \citenamefont
  {Efetov},\ and\ \citenamefont {Ganichev}}]{HUB22}%
  \BibitemOpen
  \bibfield  {author} {\bibinfo {author} {\bibfnamefont {S.}~\bibnamefont
  {Hubmann}}, \bibinfo {author} {\bibfnamefont {P.}~\bibnamefont {Soul}},
  \bibinfo {author} {\bibfnamefont {G.}~\bibnamefont {Di~Battista}}, \bibinfo
  {author} {\bibfnamefont {M.}~\bibnamefont {Hild}}, \bibinfo {author}
  {\bibfnamefont {K.}~\bibnamefont {Watanabe}}, \bibinfo {author}
  {\bibfnamefont {T.}~\bibnamefont {Taniguchi}}, \bibinfo {author}
  {\bibfnamefont {D.~K.}\ \bibnamefont {Efetov}},\ and\ \bibinfo {author}
  {\bibfnamefont {S.~D.}\ \bibnamefont {Ganichev}},\ }\href
  {https://doi.org/10.1103/PhysRevMaterials.6.024003} {\bibfield  {journal}
  {\bibinfo  {journal} {Phys. Rev. Mater.}\ }\textbf {\bibinfo {volume} {6}},\
  \bibinfo {pages} {024003} (\bibinfo {year} {2022})}\BibitemShut {NoStop}%
\bibitem [{\citenamefont {Ghawri}\ \emph {et~al.}(2022)\citenamefont {Ghawri},
  \citenamefont {Mahapatra}, \citenamefont {Garg}, \citenamefont {Mandal},
  \citenamefont {Bhowmik}, \citenamefont {Jayaraman}, \citenamefont {Soni},
  \citenamefont {Watanabe}, \citenamefont {Taniguchi}, \citenamefont
  {Krishnamurthy}, \citenamefont {Jain}, \citenamefont {Banerjee},
  \citenamefont {Chandni},\ and\ \citenamefont {Ghosh}}]{GHA22}%
  \BibitemOpen
  \bibfield  {author} {\bibinfo {author} {\bibfnamefont {B.}~\bibnamefont
  {Ghawri}}, \bibinfo {author} {\bibfnamefont {P.~S.}\ \bibnamefont
  {Mahapatra}}, \bibinfo {author} {\bibfnamefont {M.}~\bibnamefont {Garg}},
  \bibinfo {author} {\bibfnamefont {S.}~\bibnamefont {Mandal}}, \bibinfo
  {author} {\bibfnamefont {S.}~\bibnamefont {Bhowmik}}, \bibinfo {author}
  {\bibfnamefont {A.}~\bibnamefont {Jayaraman}}, \bibinfo {author}
  {\bibfnamefont {R.}~\bibnamefont {Soni}}, \bibinfo {author} {\bibfnamefont
  {K.}~\bibnamefont {Watanabe}}, \bibinfo {author} {\bibfnamefont
  {T.}~\bibnamefont {Taniguchi}}, \bibinfo {author} {\bibfnamefont {H.~R.}\
  \bibnamefont {Krishnamurthy}}, \bibinfo {author} {\bibfnamefont
  {M.}~\bibnamefont {Jain}}, \bibinfo {author} {\bibfnamefont {S.}~\bibnamefont
  {Banerjee}}, \bibinfo {author} {\bibfnamefont {U.}~\bibnamefont {Chandni}},\
  and\ \bibinfo {author} {\bibfnamefont {A.}~\bibnamefont {Ghosh}},\ }\href
  {https://doi.org/10.1038/s41467-022-29198-4} {\bibfield  {journal} {\bibinfo
  {journal} {Nat. Commun.}\ }\textbf {\bibinfo {volume} {13}},\ \bibinfo
  {pages} {1522} (\bibinfo {year} {2022})}\BibitemShut {NoStop}%
\bibitem [{\citenamefont {Jaoui}\ \emph {et~al.}(2022)\citenamefont {Jaoui},
  \citenamefont {Das}, \citenamefont {Di~Battista}, \citenamefont
  {{D{\'i}ez-M{\'e}rida}}, \citenamefont {Lu}, \citenamefont {Watanabe},
  \citenamefont {Taniguchi}, \citenamefont {Ishizuka}, \citenamefont
  {Levitov},\ and\ \citenamefont {Efetov}}]{JAO22}%
  \BibitemOpen
  \bibfield  {author} {\bibinfo {author} {\bibfnamefont {A.}~\bibnamefont
  {Jaoui}}, \bibinfo {author} {\bibfnamefont {I.}~\bibnamefont {Das}}, \bibinfo
  {author} {\bibfnamefont {G.}~\bibnamefont {Di~Battista}}, \bibinfo {author}
  {\bibfnamefont {J.}~\bibnamefont {{D{\'i}ez-M{\'e}rida}}}, \bibinfo {author}
  {\bibfnamefont {X.}~\bibnamefont {Lu}}, \bibinfo {author} {\bibfnamefont
  {K.}~\bibnamefont {Watanabe}}, \bibinfo {author} {\bibfnamefont
  {T.}~\bibnamefont {Taniguchi}}, \bibinfo {author} {\bibfnamefont
  {H.}~\bibnamefont {Ishizuka}}, \bibinfo {author} {\bibfnamefont
  {L.}~\bibnamefont {Levitov}},\ and\ \bibinfo {author} {\bibfnamefont {D.~K.}\
  \bibnamefont {Efetov}},\ }\href {https://doi.org/10.1038/s41567-022-01556-5}
  {\bibfield  {journal} {\bibinfo  {journal} {Nat. Phys.}\ }\textbf {\bibinfo
  {volume} {18}},\ \bibinfo {pages} {633} (\bibinfo {year} {2022})}\BibitemShut
  {NoStop}%
\bibitem [{\citenamefont {Paul}\ \emph {et~al.}(2022)\citenamefont {Paul},
  \citenamefont {Ghosh}, \citenamefont {Chakraborty}, \citenamefont {Roy},
  \citenamefont {Dutta}, \citenamefont {Watanabe}, \citenamefont {Taniguchi},
  \citenamefont {Panda}, \citenamefont {Agarwala}, \citenamefont {Mukerjee},
  \citenamefont {Banerjee},\ and\ \citenamefont {Das}}]{PAU22}%
  \BibitemOpen
  \bibfield  {author} {\bibinfo {author} {\bibfnamefont {A.~K.}\ \bibnamefont
  {Paul}}, \bibinfo {author} {\bibfnamefont {A.}~\bibnamefont {Ghosh}},
  \bibinfo {author} {\bibfnamefont {S.}~\bibnamefont {Chakraborty}}, \bibinfo
  {author} {\bibfnamefont {U.}~\bibnamefont {Roy}}, \bibinfo {author}
  {\bibfnamefont {R.}~\bibnamefont {Dutta}}, \bibinfo {author} {\bibfnamefont
  {K.}~\bibnamefont {Watanabe}}, \bibinfo {author} {\bibfnamefont
  {T.}~\bibnamefont {Taniguchi}}, \bibinfo {author} {\bibfnamefont
  {A.}~\bibnamefont {Panda}}, \bibinfo {author} {\bibfnamefont
  {A.}~\bibnamefont {Agarwala}}, \bibinfo {author} {\bibfnamefont
  {S.}~\bibnamefont {Mukerjee}}, \bibinfo {author} {\bibfnamefont
  {S.}~\bibnamefont {Banerjee}},\ and\ \bibinfo {author} {\bibfnamefont
  {A.}~\bibnamefont {Das}},\ }\href
  {https://doi.org/10.1038/s41567-022-01574-3} {\bibfield  {journal} {\bibinfo
  {journal} {Nat. Phys.}\ }\textbf {\bibinfo {volume} {18}},\ \bibinfo {pages}
  {691} (\bibinfo {year} {2022})}\BibitemShut {NoStop}%
\bibitem [{\citenamefont {Grover}\ \emph {et~al.}(2022)\citenamefont {Grover},
  \citenamefont {Bocarsly}, \citenamefont {Uri}, \citenamefont {Stepanov},
  \citenamefont {Di~Battista}, \citenamefont {Roy}, \citenamefont {Xiao},
  \citenamefont {Meltzer}, \citenamefont {Myasoedov}, \citenamefont {Pareek},
  \citenamefont {Watanabe}, \citenamefont {Taniguchi}, \citenamefont {Yan},
  \citenamefont {Stern}, \citenamefont {Berg}, \citenamefont {Efetov},\ and\
  \citenamefont {Zeldov}}]{GRO22}%
  \BibitemOpen
  \bibfield  {author} {\bibinfo {author} {\bibfnamefont {S.}~\bibnamefont
  {Grover}}, \bibinfo {author} {\bibfnamefont {M.}~\bibnamefont {Bocarsly}},
  \bibinfo {author} {\bibfnamefont {A.}~\bibnamefont {Uri}}, \bibinfo {author}
  {\bibfnamefont {P.}~\bibnamefont {Stepanov}}, \bibinfo {author}
  {\bibfnamefont {G.}~\bibnamefont {Di~Battista}}, \bibinfo {author}
  {\bibfnamefont {I.}~\bibnamefont {Roy}}, \bibinfo {author} {\bibfnamefont
  {J.}~\bibnamefont {Xiao}}, \bibinfo {author} {\bibfnamefont {A.~Y.}\
  \bibnamefont {Meltzer}}, \bibinfo {author} {\bibfnamefont {Y.}~\bibnamefont
  {Myasoedov}}, \bibinfo {author} {\bibfnamefont {K.}~\bibnamefont {Pareek}},
  \bibinfo {author} {\bibfnamefont {K.}~\bibnamefont {Watanabe}}, \bibinfo
  {author} {\bibfnamefont {T.}~\bibnamefont {Taniguchi}}, \bibinfo {author}
  {\bibfnamefont {B.}~\bibnamefont {Yan}}, \bibinfo {author} {\bibfnamefont
  {A.}~\bibnamefont {Stern}}, \bibinfo {author} {\bibfnamefont
  {E.}~\bibnamefont {Berg}}, \bibinfo {author} {\bibfnamefont {D.~K.}\
  \bibnamefont {Efetov}},\ and\ \bibinfo {author} {\bibfnamefont
  {E.}~\bibnamefont {Zeldov}},\ }\href
  {https://doi.org/10.1038/s41567-022-01635-7} {\bibfield  {journal} {\bibinfo
  {journal} {Nat. Phys.}\ }\textbf {\bibinfo {volume} {18}},\ \bibinfo {pages}
  {885} (\bibinfo {year} {2022})}\BibitemShut {NoStop}%
\bibitem [{\citenamefont {Zhou}\ \emph {et~al.}(2023)\citenamefont {Zhou},
  \citenamefont {Liu}, \citenamefont {Hao}, \citenamefont {Yan}, \citenamefont
  {Zheng}, \citenamefont {Ren}, \citenamefont {Zhao}, \citenamefont {Watanabe},
  \citenamefont {Taniguchi},\ and\ \citenamefont {He}}]{ZHO23a}%
  \BibitemOpen
  \bibfield  {author} {\bibinfo {author} {\bibfnamefont {X.-F.}\ \bibnamefont
  {Zhou}}, \bibinfo {author} {\bibfnamefont {Y.-W.}\ \bibnamefont {Liu}},
  \bibinfo {author} {\bibfnamefont {C.-Y.}\ \bibnamefont {Hao}}, \bibinfo
  {author} {\bibfnamefont {C.}~\bibnamefont {Yan}}, \bibinfo {author}
  {\bibfnamefont {Q.}~\bibnamefont {Zheng}}, \bibinfo {author} {\bibfnamefont
  {Y.-N.}\ \bibnamefont {Ren}}, \bibinfo {author} {\bibfnamefont {Y.-X.}\
  \bibnamefont {Zhao}}, \bibinfo {author} {\bibfnamefont {K.}~\bibnamefont
  {Watanabe}}, \bibinfo {author} {\bibfnamefont {T.}~\bibnamefont
  {Taniguchi}},\ and\ \bibinfo {author} {\bibfnamefont {L.}~\bibnamefont
  {He}},\ }\href {https://doi.org/10.1103/PhysRevB.107.125410} {\bibfield
  {journal} {\bibinfo  {journal} {Phys. Rev. B}\ }\textbf {\bibinfo {volume}
  {107}},\ \bibinfo {pages} {125410} (\bibinfo {year} {2023})}\BibitemShut
  {NoStop}%
\bibitem [{\citenamefont {Chen}\ \emph
  {et~al.}(2019{\natexlab{a}})\citenamefont {Chen}, \citenamefont {Jiang},
  \citenamefont {Wu}, \citenamefont {Lyu}, \citenamefont {Li}, \citenamefont
  {Chittari}, \citenamefont {Watanabe}, \citenamefont {Taniguchi},
  \citenamefont {Shi}, \citenamefont {Jung}, \citenamefont {Zhang},\ and\
  \citenamefont {Wang}}]{CHE19}%
  \BibitemOpen
  \bibfield  {author} {\bibinfo {author} {\bibfnamefont {G.}~\bibnamefont
  {Chen}}, \bibinfo {author} {\bibfnamefont {L.}~\bibnamefont {Jiang}},
  \bibinfo {author} {\bibfnamefont {S.}~\bibnamefont {Wu}}, \bibinfo {author}
  {\bibfnamefont {B.}~\bibnamefont {Lyu}}, \bibinfo {author} {\bibfnamefont
  {H.}~\bibnamefont {Li}}, \bibinfo {author} {\bibfnamefont {B.~L.}\
  \bibnamefont {Chittari}}, \bibinfo {author} {\bibfnamefont {K.}~\bibnamefont
  {Watanabe}}, \bibinfo {author} {\bibfnamefont {T.}~\bibnamefont {Taniguchi}},
  \bibinfo {author} {\bibfnamefont {Z.}~\bibnamefont {Shi}}, \bibinfo {author}
  {\bibfnamefont {J.}~\bibnamefont {Jung}}, \bibinfo {author} {\bibfnamefont
  {Y.}~\bibnamefont {Zhang}},\ and\ \bibinfo {author} {\bibfnamefont
  {F.}~\bibnamefont {Wang}},\ }\href
  {https://doi.org/10.1038/s41567-018-0387-2} {\bibfield  {journal} {\bibinfo
  {journal} {Nat. Phys.}\ }\textbf {\bibinfo {volume} {15}},\ \bibinfo {pages}
  {237} (\bibinfo {year} {2019}{\natexlab{a}})}\BibitemShut {NoStop}%
\bibitem [{\citenamefont {Chen}\ \emph
  {et~al.}(2019{\natexlab{b}})\citenamefont {Chen}, \citenamefont {Sharpe},
  \citenamefont {Gallagher}, \citenamefont {Rosen}, \citenamefont {Fox},
  \citenamefont {Jiang}, \citenamefont {Lyu}, \citenamefont {Li}, \citenamefont
  {Watanabe}, \citenamefont {Taniguchi}, \citenamefont {Jung}, \citenamefont
  {Shi}, \citenamefont {{Goldhaber-Gordon}}, \citenamefont {Zhang},\ and\
  \citenamefont {Wang}}]{CHE19a}%
  \BibitemOpen
  \bibfield  {author} {\bibinfo {author} {\bibfnamefont {G.}~\bibnamefont
  {Chen}}, \bibinfo {author} {\bibfnamefont {A.~L.}\ \bibnamefont {Sharpe}},
  \bibinfo {author} {\bibfnamefont {P.}~\bibnamefont {Gallagher}}, \bibinfo
  {author} {\bibfnamefont {I.~T.}\ \bibnamefont {Rosen}}, \bibinfo {author}
  {\bibfnamefont {E.~J.}\ \bibnamefont {Fox}}, \bibinfo {author} {\bibfnamefont
  {L.}~\bibnamefont {Jiang}}, \bibinfo {author} {\bibfnamefont
  {B.}~\bibnamefont {Lyu}}, \bibinfo {author} {\bibfnamefont {H.}~\bibnamefont
  {Li}}, \bibinfo {author} {\bibfnamefont {K.}~\bibnamefont {Watanabe}},
  \bibinfo {author} {\bibfnamefont {T.}~\bibnamefont {Taniguchi}}, \bibinfo
  {author} {\bibfnamefont {J.}~\bibnamefont {Jung}}, \bibinfo {author}
  {\bibfnamefont {Z.}~\bibnamefont {Shi}}, \bibinfo {author} {\bibfnamefont
  {D.}~\bibnamefont {{Goldhaber-Gordon}}}, \bibinfo {author} {\bibfnamefont
  {Y.}~\bibnamefont {Zhang}},\ and\ \bibinfo {author} {\bibfnamefont
  {F.}~\bibnamefont {Wang}},\ }\href
  {https://doi.org/10.1038/s41586-019-1393-y} {\bibfield  {journal} {\bibinfo
  {journal} {Nature}\ }\textbf {\bibinfo {volume} {572}},\ \bibinfo {pages}
  {215} (\bibinfo {year} {2019}{\natexlab{b}})}\BibitemShut {NoStop}%
\bibitem [{\citenamefont {Park}\ \emph
  {et~al.}(2021{\natexlab{b}})\citenamefont {Park}, \citenamefont {Cao},
  \citenamefont {Watanabe}, \citenamefont {Taniguchi},\ and\ \citenamefont
  {{Jarillo-Herrero}}}]{PAR21}%
  \BibitemOpen
  \bibfield  {author} {\bibinfo {author} {\bibfnamefont {J.~M.}\ \bibnamefont
  {Park}}, \bibinfo {author} {\bibfnamefont {Y.}~\bibnamefont {Cao}}, \bibinfo
  {author} {\bibfnamefont {K.}~\bibnamefont {Watanabe}}, \bibinfo {author}
  {\bibfnamefont {T.}~\bibnamefont {Taniguchi}},\ and\ \bibinfo {author}
  {\bibfnamefont {P.}~\bibnamefont {{Jarillo-Herrero}}},\ }\href
  {https://doi.org/10.1038/s41586-021-03192-0} {\bibfield  {journal} {\bibinfo
  {journal} {Nature}\ }\textbf {\bibinfo {volume} {590}},\ \bibinfo {pages}
  {249} (\bibinfo {year} {2021}{\natexlab{b}})}\BibitemShut {NoStop}%
\bibitem [{\citenamefont {Hao}\ \emph {et~al.}(2021)\citenamefont {Hao},
  \citenamefont {Zimmerman}, \citenamefont {Ledwith}, \citenamefont {Khalaf},
  \citenamefont {Najafabadi}, \citenamefont {Watanabe}, \citenamefont
  {Taniguchi}, \citenamefont {Vishwanath},\ and\ \citenamefont {Kim}}]{HAO21}%
  \BibitemOpen
  \bibfield  {author} {\bibinfo {author} {\bibfnamefont {Z.}~\bibnamefont
  {Hao}}, \bibinfo {author} {\bibfnamefont {A.~M.}\ \bibnamefont {Zimmerman}},
  \bibinfo {author} {\bibfnamefont {P.}~\bibnamefont {Ledwith}}, \bibinfo
  {author} {\bibfnamefont {E.}~\bibnamefont {Khalaf}}, \bibinfo {author}
  {\bibfnamefont {D.~H.}\ \bibnamefont {Najafabadi}}, \bibinfo {author}
  {\bibfnamefont {K.}~\bibnamefont {Watanabe}}, \bibinfo {author}
  {\bibfnamefont {T.}~\bibnamefont {Taniguchi}}, \bibinfo {author}
  {\bibfnamefont {A.}~\bibnamefont {Vishwanath}},\ and\ \bibinfo {author}
  {\bibfnamefont {P.}~\bibnamefont {Kim}},\ }\href
  {https://doi.org/10.1126/science.abg0399} {\bibfield  {journal} {\bibinfo
  {journal} {Science}\ }\textbf {\bibinfo {volume} {371}},\ \bibinfo {pages}
  {1133} (\bibinfo {year} {2021})}\BibitemShut {NoStop}%
\bibitem [{\citenamefont {Cao}\ \emph {et~al.}(2021{\natexlab{b}})\citenamefont
  {Cao}, \citenamefont {Park}, \citenamefont {Watanabe}, \citenamefont
  {Taniguchi},\ and\ \citenamefont {{Jarillo-Herrero}}}]{CAO21b}%
  \BibitemOpen
  \bibfield  {author} {\bibinfo {author} {\bibfnamefont {Y.}~\bibnamefont
  {Cao}}, \bibinfo {author} {\bibfnamefont {J.~M.}\ \bibnamefont {Park}},
  \bibinfo {author} {\bibfnamefont {K.}~\bibnamefont {Watanabe}}, \bibinfo
  {author} {\bibfnamefont {T.}~\bibnamefont {Taniguchi}},\ and\ \bibinfo
  {author} {\bibfnamefont {P.}~\bibnamefont {{Jarillo-Herrero}}},\ }\href
  {https://doi.org/10.1038/s41586-021-03685-y} {\bibfield  {journal} {\bibinfo
  {journal} {Nature}\ }\textbf {\bibinfo {volume} {595}},\ \bibinfo {pages}
  {526} (\bibinfo {year} {2021}{\natexlab{b}})}\BibitemShut {NoStop}%
\bibitem [{\citenamefont {Li}\ \emph {et~al.}(2022{\natexlab{a}})\citenamefont
  {Li}, \citenamefont {Zhang}, \citenamefont {Chen}, \citenamefont {Wei},
  \citenamefont {Zhang}, \citenamefont {Xiao}, \citenamefont {Gao},
  \citenamefont {Chen}, \citenamefont {Liang}, \citenamefont {Pei},
  \citenamefont {Xu}, \citenamefont {Watanabe}, \citenamefont {Taniguchi},
  \citenamefont {Yang}, \citenamefont {Miao}, \citenamefont {Liu},
  \citenamefont {Cheng}, \citenamefont {Wang}, \citenamefont {Chen},\ and\
  \citenamefont {Liu}}]{LI22d}%
  \BibitemOpen
  \bibfield  {author} {\bibinfo {author} {\bibfnamefont {Y.}~\bibnamefont
  {Li}}, \bibinfo {author} {\bibfnamefont {S.}~\bibnamefont {Zhang}}, \bibinfo
  {author} {\bibfnamefont {F.}~\bibnamefont {Chen}}, \bibinfo {author}
  {\bibfnamefont {L.}~\bibnamefont {Wei}}, \bibinfo {author} {\bibfnamefont
  {Z.}~\bibnamefont {Zhang}}, \bibinfo {author} {\bibfnamefont
  {H.}~\bibnamefont {Xiao}}, \bibinfo {author} {\bibfnamefont {H.}~\bibnamefont
  {Gao}}, \bibinfo {author} {\bibfnamefont {M.}~\bibnamefont {Chen}}, \bibinfo
  {author} {\bibfnamefont {S.}~\bibnamefont {Liang}}, \bibinfo {author}
  {\bibfnamefont {D.}~\bibnamefont {Pei}}, \bibinfo {author} {\bibfnamefont
  {L.}~\bibnamefont {Xu}}, \bibinfo {author} {\bibfnamefont {K.}~\bibnamefont
  {Watanabe}}, \bibinfo {author} {\bibfnamefont {T.}~\bibnamefont {Taniguchi}},
  \bibinfo {author} {\bibfnamefont {L.}~\bibnamefont {Yang}}, \bibinfo {author}
  {\bibfnamefont {F.}~\bibnamefont {Miao}}, \bibinfo {author} {\bibfnamefont
  {J.}~\bibnamefont {Liu}}, \bibinfo {author} {\bibfnamefont {B.}~\bibnamefont
  {Cheng}}, \bibinfo {author} {\bibfnamefont {M.}~\bibnamefont {Wang}},
  \bibinfo {author} {\bibfnamefont {Y.}~\bibnamefont {Chen}},\ and\ \bibinfo
  {author} {\bibfnamefont {Z.}~\bibnamefont {Liu}},\ }\href
  {https://doi.org/10.1002/adma.202205996} {\bibfield  {journal} {\bibinfo
  {journal} {Adv. Mater.}\ }\textbf {\bibinfo {volume} {34}},\ \bibinfo {pages}
  {2205996} (\bibinfo {year} {2022}{\natexlab{a}})}\BibitemShut {NoStop}%
\bibitem [{\citenamefont {Turkel}\ \emph {et~al.}(2022)\citenamefont {Turkel},
  \citenamefont {Swann}, \citenamefont {Zhu}, \citenamefont {Christos},
  \citenamefont {Watanabe}, \citenamefont {Taniguchi}, \citenamefont {Sachdev},
  \citenamefont {Scheurer}, \citenamefont {Kaxiras}, \citenamefont {Dean},\
  and\ \citenamefont {Pasupathy}}]{TUR22}%
  \BibitemOpen
  \bibfield  {author} {\bibinfo {author} {\bibfnamefont {S.}~\bibnamefont
  {Turkel}}, \bibinfo {author} {\bibfnamefont {J.}~\bibnamefont {Swann}},
  \bibinfo {author} {\bibfnamefont {Z.}~\bibnamefont {Zhu}}, \bibinfo {author}
  {\bibfnamefont {M.}~\bibnamefont {Christos}}, \bibinfo {author}
  {\bibfnamefont {K.}~\bibnamefont {Watanabe}}, \bibinfo {author}
  {\bibfnamefont {T.}~\bibnamefont {Taniguchi}}, \bibinfo {author}
  {\bibfnamefont {S.}~\bibnamefont {Sachdev}}, \bibinfo {author} {\bibfnamefont
  {M.~S.}\ \bibnamefont {Scheurer}}, \bibinfo {author} {\bibfnamefont
  {E.}~\bibnamefont {Kaxiras}}, \bibinfo {author} {\bibfnamefont {C.~R.}\
  \bibnamefont {Dean}},\ and\ \bibinfo {author} {\bibfnamefont {A.~N.}\
  \bibnamefont {Pasupathy}},\ }\href {https://doi.org/10.1126/science.abk1895}
  {\bibfield  {journal} {\bibinfo  {journal} {Science}\ }\textbf {\bibinfo
  {volume} {376}},\ \bibinfo {pages} {193} (\bibinfo {year}
  {2022})}\BibitemShut {NoStop}%
\bibitem [{\citenamefont {Liu}\ \emph {et~al.}(2022)\citenamefont {Liu},
  \citenamefont {Zhang}, \citenamefont {Watanabe}, \citenamefont {Taniguchi},\
  and\ \citenamefont {Li}}]{LIU22b}%
  \BibitemOpen
  \bibfield  {author} {\bibinfo {author} {\bibfnamefont {X.}~\bibnamefont
  {Liu}}, \bibinfo {author} {\bibfnamefont {N.~J.}\ \bibnamefont {Zhang}},
  \bibinfo {author} {\bibfnamefont {K.}~\bibnamefont {Watanabe}}, \bibinfo
  {author} {\bibfnamefont {T.}~\bibnamefont {Taniguchi}},\ and\ \bibinfo
  {author} {\bibfnamefont {J.~I.~A.}\ \bibnamefont {Li}},\ }\href
  {https://doi.org/10.1038/s41567-022-01515-0} {\bibfield  {journal} {\bibinfo
  {journal} {Nat. Phys.}\ }\textbf {\bibinfo {volume} {18}},\ \bibinfo {pages}
  {522} (\bibinfo {year} {2022})}\BibitemShut {NoStop}%
\bibitem [{\citenamefont {Kim}\ \emph {et~al.}(2022)\citenamefont {Kim},
  \citenamefont {Choi}, \citenamefont {Lewandowski}, \citenamefont {Thomson},
  \citenamefont {Zhang}, \citenamefont {Polski}, \citenamefont {Watanabe},
  \citenamefont {Taniguchi}, \citenamefont {Alicea},\ and\ \citenamefont
  {{Nadj-Perge}}}]{KIM22}%
  \BibitemOpen
  \bibfield  {author} {\bibinfo {author} {\bibfnamefont {H.}~\bibnamefont
  {Kim}}, \bibinfo {author} {\bibfnamefont {Y.}~\bibnamefont {Choi}}, \bibinfo
  {author} {\bibfnamefont {C.}~\bibnamefont {Lewandowski}}, \bibinfo {author}
  {\bibfnamefont {A.}~\bibnamefont {Thomson}}, \bibinfo {author} {\bibfnamefont
  {Y.}~\bibnamefont {Zhang}}, \bibinfo {author} {\bibfnamefont
  {R.}~\bibnamefont {Polski}}, \bibinfo {author} {\bibfnamefont
  {K.}~\bibnamefont {Watanabe}}, \bibinfo {author} {\bibfnamefont
  {T.}~\bibnamefont {Taniguchi}}, \bibinfo {author} {\bibfnamefont
  {J.}~\bibnamefont {Alicea}},\ and\ \bibinfo {author} {\bibfnamefont
  {S.}~\bibnamefont {{Nadj-Perge}}},\ }\href
  {https://doi.org/10.1038/s41586-022-04715-z} {\bibfield  {journal} {\bibinfo
  {journal} {Nature}\ }\textbf {\bibinfo {volume} {606}},\ \bibinfo {pages}
  {494} (\bibinfo {year} {2022})}\BibitemShut {NoStop}%
\bibitem [{\citenamefont {Zhang}\ \emph
  {et~al.}(2022{\natexlab{a}})\citenamefont {Zhang}, \citenamefont {Polski},
  \citenamefont {Lewandowski}, \citenamefont {Thomson}, \citenamefont {Peng},
  \citenamefont {Choi}, \citenamefont {Kim}, \citenamefont {Watanabe},
  \citenamefont {Taniguchi}, \citenamefont {Alicea}, \citenamefont {{von
  Oppen}}, \citenamefont {Refael},\ and\ \citenamefont
  {{Nadj-Perge}}}]{ZHA22d}%
  \BibitemOpen
  \bibfield  {author} {\bibinfo {author} {\bibfnamefont {Y.}~\bibnamefont
  {Zhang}}, \bibinfo {author} {\bibfnamefont {R.}~\bibnamefont {Polski}},
  \bibinfo {author} {\bibfnamefont {C.}~\bibnamefont {Lewandowski}}, \bibinfo
  {author} {\bibfnamefont {A.}~\bibnamefont {Thomson}}, \bibinfo {author}
  {\bibfnamefont {Y.}~\bibnamefont {Peng}}, \bibinfo {author} {\bibfnamefont
  {Y.}~\bibnamefont {Choi}}, \bibinfo {author} {\bibfnamefont {H.}~\bibnamefont
  {Kim}}, \bibinfo {author} {\bibfnamefont {K.}~\bibnamefont {Watanabe}},
  \bibinfo {author} {\bibfnamefont {T.}~\bibnamefont {Taniguchi}}, \bibinfo
  {author} {\bibfnamefont {J.}~\bibnamefont {Alicea}}, \bibinfo {author}
  {\bibfnamefont {F.}~\bibnamefont {{von Oppen}}}, \bibinfo {author}
  {\bibfnamefont {G.}~\bibnamefont {Refael}},\ and\ \bibinfo {author}
  {\bibfnamefont {S.}~\bibnamefont {{Nadj-Perge}}},\ }\href
  {https://doi.org/10.1126/science.abn8585} {\bibfield  {journal} {\bibinfo
  {journal} {Science}\ }\textbf {\bibinfo {volume} {377}},\ \bibinfo {pages}
  {1538} (\bibinfo {year} {2022}{\natexlab{a}})}\BibitemShut {NoStop}%
\bibitem [{\citenamefont {Zhang}\ \emph
  {et~al.}(2022{\natexlab{b}})\citenamefont {Zhang}, \citenamefont {Wang},
  \citenamefont {Watanabe}, \citenamefont {Taniguchi}, \citenamefont {Vafek},\
  and\ \citenamefont {Li}}]{ZHA22c}%
  \BibitemOpen
  \bibfield  {author} {\bibinfo {author} {\bibfnamefont {N.~J.}\ \bibnamefont
  {Zhang}}, \bibinfo {author} {\bibfnamefont {Y.}~\bibnamefont {Wang}},
  \bibinfo {author} {\bibfnamefont {K.}~\bibnamefont {Watanabe}}, \bibinfo
  {author} {\bibfnamefont {T.}~\bibnamefont {Taniguchi}}, \bibinfo {author}
  {\bibfnamefont {O.}~\bibnamefont {Vafek}},\ and\ \bibinfo {author}
  {\bibfnamefont {J.~I.~A.}\ \bibnamefont {Li}},\ }\href
  {https://doi.org/10.48550/arXiv.2211.01352} {\bibinfo {title} {Electronic
  anisotropy in magic-angle twisted trilayer graphene}} (\bibinfo {year}
  {2022}{\natexlab{b}}),\ \Eprint {https://arxiv.org/abs/2211.01352}
  {arxiv:2211.01352 [cond-mat]} \BibitemShut {NoStop}%
\bibitem [{\citenamefont {Shen}\ \emph {et~al.}(2023)\citenamefont {Shen},
  \citenamefont {Ledwith}, \citenamefont {Watanabe}, \citenamefont {Taniguchi},
  \citenamefont {Khalaf}, \citenamefont {Vishwanath},\ and\ \citenamefont
  {Efetov}}]{SHE23}%
  \BibitemOpen
  \bibfield  {author} {\bibinfo {author} {\bibfnamefont {C.}~\bibnamefont
  {Shen}}, \bibinfo {author} {\bibfnamefont {P.~J.}\ \bibnamefont {Ledwith}},
  \bibinfo {author} {\bibfnamefont {K.}~\bibnamefont {Watanabe}}, \bibinfo
  {author} {\bibfnamefont {T.}~\bibnamefont {Taniguchi}}, \bibinfo {author}
  {\bibfnamefont {E.}~\bibnamefont {Khalaf}}, \bibinfo {author} {\bibfnamefont
  {A.}~\bibnamefont {Vishwanath}},\ and\ \bibinfo {author} {\bibfnamefont
  {D.~K.}\ \bibnamefont {Efetov}},\ }\href
  {https://doi.org/10.1038/s41563-022-01428-6} {\bibfield  {journal} {\bibinfo
  {journal} {Nat. Mater.}\ }\textbf {\bibinfo {volume} {22}},\ \bibinfo {pages}
  {316} (\bibinfo {year} {2023})}\BibitemShut {NoStop}%
\bibitem [{\citenamefont {Kim}\ \emph {et~al.}(2023)\citenamefont {Kim},
  \citenamefont {Choi}, \citenamefont {{Lantagne-Hurtubise}}, \citenamefont
  {Lewandowski}, \citenamefont {Thomson}, \citenamefont {Kong}, \citenamefont
  {Zhou}, \citenamefont {Baum}, \citenamefont {Zhang}, \citenamefont {Holleis},
  \citenamefont {Watanabe}, \citenamefont {Taniguchi}, \citenamefont {Young},
  \citenamefont {Alicea},\ and\ \citenamefont {{Nadj-Perge}}}]{KIM23}%
  \BibitemOpen
  \bibfield  {author} {\bibinfo {author} {\bibfnamefont {H.}~\bibnamefont
  {Kim}}, \bibinfo {author} {\bibfnamefont {Y.}~\bibnamefont {Choi}}, \bibinfo
  {author} {\bibfnamefont {{\'E}.}~\bibnamefont {{Lantagne-Hurtubise}}},
  \bibinfo {author} {\bibfnamefont {C.}~\bibnamefont {Lewandowski}}, \bibinfo
  {author} {\bibfnamefont {A.}~\bibnamefont {Thomson}}, \bibinfo {author}
  {\bibfnamefont {L.}~\bibnamefont {Kong}}, \bibinfo {author} {\bibfnamefont
  {H.}~\bibnamefont {Zhou}}, \bibinfo {author} {\bibfnamefont {E.}~\bibnamefont
  {Baum}}, \bibinfo {author} {\bibfnamefont {Y.}~\bibnamefont {Zhang}},
  \bibinfo {author} {\bibfnamefont {L.}~\bibnamefont {Holleis}}, \bibinfo
  {author} {\bibfnamefont {K.}~\bibnamefont {Watanabe}}, \bibinfo {author}
  {\bibfnamefont {T.}~\bibnamefont {Taniguchi}}, \bibinfo {author}
  {\bibfnamefont {A.~F.}\ \bibnamefont {Young}}, \bibinfo {author}
  {\bibfnamefont {J.}~\bibnamefont {Alicea}},\ and\ \bibinfo {author}
  {\bibfnamefont {S.}~\bibnamefont {{Nadj-Perge}}},\ }\href
  {https://doi.org/10.1038/s41586-023-06663-8} {\bibfield  {journal} {\bibinfo
  {journal} {Nature}\ }\textbf {\bibinfo {volume} {623}},\ \bibinfo {pages}
  {942} (\bibinfo {year} {2023})}\BibitemShut {NoStop}%
\bibitem [{\citenamefont {Uchida}\ \emph {et~al.}(2014)\citenamefont {Uchida},
  \citenamefont {Furuya}, \citenamefont {Iwata},\ and\ \citenamefont
  {Oshiyama}}]{UCH14}%
  \BibitemOpen
  \bibfield  {author} {\bibinfo {author} {\bibfnamefont {K.}~\bibnamefont
  {Uchida}}, \bibinfo {author} {\bibfnamefont {S.}~\bibnamefont {Furuya}},
  \bibinfo {author} {\bibfnamefont {J.-I.}\ \bibnamefont {Iwata}},\ and\
  \bibinfo {author} {\bibfnamefont {A.}~\bibnamefont {Oshiyama}},\ }\href
  {https://doi.org/10.1103/PhysRevB.90.155451} {\bibfield  {journal} {\bibinfo
  {journal} {Phys. Rev. B}\ }\textbf {\bibinfo {volume} {90}},\ \bibinfo
  {pages} {155451} (\bibinfo {year} {2014})}\BibitemShut {NoStop}%
\bibitem [{\citenamefont {van Wijk}\ \emph {et~al.}(2015)\citenamefont {van
  Wijk}, \citenamefont {Schuring}, \citenamefont {Katsnelson},\ and\
  \citenamefont {Fasolino}}]{WIJ15}%
  \BibitemOpen
  \bibfield  {author} {\bibinfo {author} {\bibfnamefont {M.~M.}\ \bibnamefont
  {van Wijk}}, \bibinfo {author} {\bibfnamefont {A.}~\bibnamefont {Schuring}},
  \bibinfo {author} {\bibfnamefont {M.~I.}\ \bibnamefont {Katsnelson}},\ and\
  \bibinfo {author} {\bibfnamefont {A.}~\bibnamefont {Fasolino}},\ }\href
  {https://doi.org/10.1088/2053-1583/2/3/034010} {\bibfield  {journal}
  {\bibinfo  {journal} {2D Mater.}\ }\textbf {\bibinfo {volume} {2}},\ \bibinfo
  {pages} {034010} (\bibinfo {year} {2015})}\BibitemShut {NoStop}%
\bibitem [{\citenamefont {Dai}\ \emph {et~al.}(2016)\citenamefont {Dai},
  \citenamefont {Xiang},\ and\ \citenamefont {Srolovitz}}]{DAI16}%
  \BibitemOpen
  \bibfield  {author} {\bibinfo {author} {\bibfnamefont {S.}~\bibnamefont
  {Dai}}, \bibinfo {author} {\bibfnamefont {Y.}~\bibnamefont {Xiang}},\ and\
  \bibinfo {author} {\bibfnamefont {D.~J.}\ \bibnamefont {Srolovitz}},\ }\href
  {https://doi.org/10.1021/acs.nanolett.6b02870} {\bibfield  {journal}
  {\bibinfo  {journal} {Nano Lett.}\ }\textbf {\bibinfo {volume} {16}},\
  \bibinfo {pages} {5923} (\bibinfo {year} {2016})}\BibitemShut {NoStop}%
\bibitem [{\citenamefont {Jain}\ \emph {et~al.}(2016)\citenamefont {Jain},
  \citenamefont {Juri{\v c}i{\'c}},\ and\ \citenamefont {Barkema}}]{JAI16}%
  \BibitemOpen
  \bibfield  {author} {\bibinfo {author} {\bibfnamefont {S.~K.}\ \bibnamefont
  {Jain}}, \bibinfo {author} {\bibfnamefont {V.}~\bibnamefont {Juri{\v
  c}i{\'c}}},\ and\ \bibinfo {author} {\bibfnamefont {G.~T.}\ \bibnamefont
  {Barkema}},\ }\href {https://doi.org/10.1088/2053-1583/4/1/015018} {\bibfield
   {journal} {\bibinfo  {journal} {2D Mater.}\ }\textbf {\bibinfo {volume}
  {4}},\ \bibinfo {pages} {015018} (\bibinfo {year} {2016})}\BibitemShut
  {NoStop}%
\bibitem [{\citenamefont {Nam}\ and\ \citenamefont {Koshino}(2017)}]{NAM17}%
  \BibitemOpen
  \bibfield  {author} {\bibinfo {author} {\bibfnamefont {N.~N.~T.}\
  \bibnamefont {Nam}}\ and\ \bibinfo {author} {\bibfnamefont {M.}~\bibnamefont
  {Koshino}},\ }\href {https://doi.org/10.1103/PhysRevB.96.075311} {\bibfield
  {journal} {\bibinfo  {journal} {Phys. Rev. B}\ }\textbf {\bibinfo {volume}
  {96}},\ \bibinfo {pages} {075311} (\bibinfo {year} {2017})}\BibitemShut
  {NoStop}%
\bibitem [{\citenamefont {Efimkin}\ and\ \citenamefont
  {MacDonald}(2018)}]{EFI18}%
  \BibitemOpen
  \bibfield  {author} {\bibinfo {author} {\bibfnamefont {D.~K.}\ \bibnamefont
  {Efimkin}}\ and\ \bibinfo {author} {\bibfnamefont {A.~H.}\ \bibnamefont
  {MacDonald}},\ }\href {https://doi.org/10.1103/PhysRevB.98.035404} {\bibfield
   {journal} {\bibinfo  {journal} {Phys. Rev. B}\ }\textbf {\bibinfo {volume}
  {98}},\ \bibinfo {pages} {035404} (\bibinfo {year} {2018})}\BibitemShut
  {NoStop}%
\bibitem [{\citenamefont {Kang}\ and\ \citenamefont {Vafek}(2018)}]{KAN18}%
  \BibitemOpen
  \bibfield  {author} {\bibinfo {author} {\bibfnamefont {J.}~\bibnamefont
  {Kang}}\ and\ \bibinfo {author} {\bibfnamefont {O.}~\bibnamefont {Vafek}},\
  }\href {https://doi.org/10.1103/PhysRevX.8.031088} {\bibfield  {journal}
  {\bibinfo  {journal} {Phys. Rev. X}\ }\textbf {\bibinfo {volume} {8}},\
  \bibinfo {pages} {031088} (\bibinfo {year} {2018})}\BibitemShut {NoStop}%
\bibitem [{\citenamefont {Zou}\ \emph {et~al.}(2018)\citenamefont {Zou},
  \citenamefont {Po}, \citenamefont {Vishwanath},\ and\ \citenamefont
  {Senthil}}]{ZOU18}%
  \BibitemOpen
  \bibfield  {author} {\bibinfo {author} {\bibfnamefont {L.}~\bibnamefont
  {Zou}}, \bibinfo {author} {\bibfnamefont {H.~C.}\ \bibnamefont {Po}},
  \bibinfo {author} {\bibfnamefont {A.}~\bibnamefont {Vishwanath}},\ and\
  \bibinfo {author} {\bibfnamefont {T.}~\bibnamefont {Senthil}},\ }\href
  {https://doi.org/10.1103/PhysRevB.98.085435} {\bibfield  {journal} {\bibinfo
  {journal} {Phys. Rev. B}\ }\textbf {\bibinfo {volume} {98}},\ \bibinfo
  {pages} {085435} (\bibinfo {year} {2018})}\BibitemShut {NoStop}%
\bibitem [{\citenamefont {Po}\ \emph {et~al.}(2019)\citenamefont {Po},
  \citenamefont {Zou}, \citenamefont {Senthil},\ and\ \citenamefont
  {Vishwanath}}]{PO19}%
  \BibitemOpen
  \bibfield  {author} {\bibinfo {author} {\bibfnamefont {H.~C.}\ \bibnamefont
  {Po}}, \bibinfo {author} {\bibfnamefont {L.}~\bibnamefont {Zou}}, \bibinfo
  {author} {\bibfnamefont {T.}~\bibnamefont {Senthil}},\ and\ \bibinfo {author}
  {\bibfnamefont {A.}~\bibnamefont {Vishwanath}},\ }\href
  {https://doi.org/10.1103/PhysRevB.99.195455} {\bibfield  {journal} {\bibinfo
  {journal} {Phys. Rev. B}\ }\textbf {\bibinfo {volume} {99}},\ \bibinfo
  {pages} {195455} (\bibinfo {year} {2019})}\BibitemShut {NoStop}%
\bibitem [{\citenamefont {Liu}\ \emph {et~al.}(2019{\natexlab{a}})\citenamefont
  {Liu}, \citenamefont {Ma}, \citenamefont {Gao},\ and\ \citenamefont
  {Dai}}]{LIU19a}%
  \BibitemOpen
  \bibfield  {author} {\bibinfo {author} {\bibfnamefont {J.}~\bibnamefont
  {Liu}}, \bibinfo {author} {\bibfnamefont {Z.}~\bibnamefont {Ma}}, \bibinfo
  {author} {\bibfnamefont {J.}~\bibnamefont {Gao}},\ and\ \bibinfo {author}
  {\bibfnamefont {X.}~\bibnamefont {Dai}},\ }\href
  {https://doi.org/10.1103/PhysRevX.9.031021} {\bibfield  {journal} {\bibinfo
  {journal} {Phys. Rev. X}\ }\textbf {\bibinfo {volume} {9}},\ \bibinfo {pages}
  {031021} (\bibinfo {year} {2019}{\natexlab{a}})}\BibitemShut {NoStop}%
\bibitem [{\citenamefont {Tarnopolsky}\ \emph {et~al.}(2019)\citenamefont
  {Tarnopolsky}, \citenamefont {Kruchkov},\ and\ \citenamefont
  {Vishwanath}}]{TAR19}%
  \BibitemOpen
  \bibfield  {author} {\bibinfo {author} {\bibfnamefont {G.}~\bibnamefont
  {Tarnopolsky}}, \bibinfo {author} {\bibfnamefont {A.~J.}\ \bibnamefont
  {Kruchkov}},\ and\ \bibinfo {author} {\bibfnamefont {A.}~\bibnamefont
  {Vishwanath}},\ }\href {https://doi.org/10.1103/PhysRevLett.122.106405}
  {\bibfield  {journal} {\bibinfo  {journal} {Phys. Rev. Lett.}\ }\textbf
  {\bibinfo {volume} {122}},\ \bibinfo {pages} {106405} (\bibinfo {year}
  {2019})}\BibitemShut {NoStop}%
\bibitem [{\citenamefont {Mora}\ \emph {et~al.}(2019)\citenamefont {Mora},
  \citenamefont {Regnault},\ and\ \citenamefont {Bernevig}}]{MOR19}%
  \BibitemOpen
  \bibfield  {author} {\bibinfo {author} {\bibfnamefont {C.}~\bibnamefont
  {Mora}}, \bibinfo {author} {\bibfnamefont {N.}~\bibnamefont {Regnault}},\
  and\ \bibinfo {author} {\bibfnamefont {B.~A.}\ \bibnamefont {Bernevig}},\
  }\href {https://doi.org/10.1103/PhysRevLett.123.026402} {\bibfield  {journal}
  {\bibinfo  {journal} {Phys. Rev. Lett.}\ }\textbf {\bibinfo {volume} {123}},\
  \bibinfo {pages} {026402} (\bibinfo {year} {2019})}\BibitemShut {NoStop}%
\bibitem [{\citenamefont {Li}\ \emph {et~al.}(2019)\citenamefont {Li},
  \citenamefont {Wu},\ and\ \citenamefont {MacDonald}}]{LI19}%
  \BibitemOpen
  \bibfield  {author} {\bibinfo {author} {\bibfnamefont {X.}~\bibnamefont
  {Li}}, \bibinfo {author} {\bibfnamefont {F.}~\bibnamefont {Wu}},\ and\
  \bibinfo {author} {\bibfnamefont {A.~H.}\ \bibnamefont {MacDonald}},\
  }\href@noop {} {\bibfield  {journal} {\bibinfo  {journal} {arXiv:1907.12338
  [cond-mat]}\ } (\bibinfo {year} {2019})},\ \Eprint
  {https://arxiv.org/abs/1907.12338} {arxiv:1907.12338 [cond-mat]} \BibitemShut
  {NoStop}%
\bibitem [{\citenamefont {Fang}\ \emph {et~al.}(2019)\citenamefont {Fang},
  \citenamefont {Carr}, \citenamefont {Zhu}, \citenamefont {Massatt},\ and\
  \citenamefont {Kaxiras}}]{FAN19}%
  \BibitemOpen
  \bibfield  {author} {\bibinfo {author} {\bibfnamefont {S.}~\bibnamefont
  {Fang}}, \bibinfo {author} {\bibfnamefont {S.}~\bibnamefont {Carr}}, \bibinfo
  {author} {\bibfnamefont {Z.}~\bibnamefont {Zhu}}, \bibinfo {author}
  {\bibfnamefont {D.}~\bibnamefont {Massatt}},\ and\ \bibinfo {author}
  {\bibfnamefont {E.}~\bibnamefont {Kaxiras}},\ }\href
  {https://doi.org/10.48550/arXiv.1908.00058} {\bibinfo {title}
  {Angle-{{Dependent}} {{{\emph{Ab}}}}{\emph{ initio}} {{Low-Energy
  Hamiltonians}} for a {{Relaxed Twisted Bilayer Graphene Heterostructure}}}}
  (\bibinfo {year} {2019}),\ \Eprint {https://arxiv.org/abs/1908.00058}
  {arxiv:1908.00058 [cond-mat]} \BibitemShut {NoStop}%
\bibitem [{\citenamefont {Khalaf}\ \emph {et~al.}(2019)\citenamefont {Khalaf},
  \citenamefont {Kruchkov}, \citenamefont {Tarnopolsky},\ and\ \citenamefont
  {Vishwanath}}]{KHA19}%
  \BibitemOpen
  \bibfield  {author} {\bibinfo {author} {\bibfnamefont {E.}~\bibnamefont
  {Khalaf}}, \bibinfo {author} {\bibfnamefont {A.~J.}\ \bibnamefont
  {Kruchkov}}, \bibinfo {author} {\bibfnamefont {G.}~\bibnamefont
  {Tarnopolsky}},\ and\ \bibinfo {author} {\bibfnamefont {A.}~\bibnamefont
  {Vishwanath}},\ }\href {https://doi.org/10.1103/PhysRevB.100.085109}
  {\bibfield  {journal} {\bibinfo  {journal} {Phys. Rev. B}\ }\textbf {\bibinfo
  {volume} {100}},\ \bibinfo {pages} {085109} (\bibinfo {year}
  {2019})}\BibitemShut {NoStop}%
\bibitem [{\citenamefont {Carr}\ \emph
  {et~al.}(2019{\natexlab{a}})\citenamefont {Carr}, \citenamefont {Fang},
  \citenamefont {Zhu},\ and\ \citenamefont {Kaxiras}}]{CAR19a}%
  \BibitemOpen
  \bibfield  {author} {\bibinfo {author} {\bibfnamefont {S.}~\bibnamefont
  {Carr}}, \bibinfo {author} {\bibfnamefont {S.}~\bibnamefont {Fang}}, \bibinfo
  {author} {\bibfnamefont {Z.}~\bibnamefont {Zhu}},\ and\ \bibinfo {author}
  {\bibfnamefont {E.}~\bibnamefont {Kaxiras}},\ }\href
  {https://doi.org/10.1103/PhysRevResearch.1.013001} {\bibfield  {journal}
  {\bibinfo  {journal} {Phys. Rev. Res.}\ }\textbf {\bibinfo {volume} {1}},\
  \bibinfo {pages} {013001} (\bibinfo {year} {2019}{\natexlab{a}})}\BibitemShut
  {NoStop}%
\bibitem [{\citenamefont {Carr}\ \emph
  {et~al.}(2019{\natexlab{b}})\citenamefont {Carr}, \citenamefont {Fang},
  \citenamefont {Po}, \citenamefont {Vishwanath},\ and\ \citenamefont
  {Kaxiras}}]{CAR19}%
  \BibitemOpen
  \bibfield  {author} {\bibinfo {author} {\bibfnamefont {S.}~\bibnamefont
  {Carr}}, \bibinfo {author} {\bibfnamefont {S.}~\bibnamefont {Fang}}, \bibinfo
  {author} {\bibfnamefont {H.~C.}\ \bibnamefont {Po}}, \bibinfo {author}
  {\bibfnamefont {A.}~\bibnamefont {Vishwanath}},\ and\ \bibinfo {author}
  {\bibfnamefont {E.}~\bibnamefont {Kaxiras}},\ }\href
  {https://doi.org/10.1103/PhysRevResearch.1.033072} {\bibfield  {journal}
  {\bibinfo  {journal} {Phys. Rev. Res.}\ }\textbf {\bibinfo {volume} {1}},\
  \bibinfo {pages} {033072} (\bibinfo {year} {2019}{\natexlab{b}})}\BibitemShut
  {NoStop}%
\bibitem [{\citenamefont {Rademaker}\ \emph {et~al.}(2019)\citenamefont
  {Rademaker}, \citenamefont {Abanin},\ and\ \citenamefont {Mellado}}]{RAD19}%
  \BibitemOpen
  \bibfield  {author} {\bibinfo {author} {\bibfnamefont {L.}~\bibnamefont
  {Rademaker}}, \bibinfo {author} {\bibfnamefont {D.~A.}\ \bibnamefont
  {Abanin}},\ and\ \bibinfo {author} {\bibfnamefont {P.}~\bibnamefont
  {Mellado}},\ }\href {https://doi.org/10.1103/PhysRevB.100.205114} {\bibfield
  {journal} {\bibinfo  {journal} {Phys. Rev. B}\ }\textbf {\bibinfo {volume}
  {100}},\ \bibinfo {pages} {205114} (\bibinfo {year} {2019})}\BibitemShut
  {NoStop}%
\bibitem [{\citenamefont {Kwan}\ \emph {et~al.}(2020)\citenamefont {Kwan},
  \citenamefont {Parameswaran},\ and\ \citenamefont {Sondhi}}]{KWA20}%
  \BibitemOpen
  \bibfield  {author} {\bibinfo {author} {\bibfnamefont {Y.~H.}\ \bibnamefont
  {Kwan}}, \bibinfo {author} {\bibfnamefont {S.~A.}\ \bibnamefont
  {Parameswaran}},\ and\ \bibinfo {author} {\bibfnamefont {S.~L.}\ \bibnamefont
  {Sondhi}},\ }\href {https://doi.org/10.1103/PhysRevB.101.205116} {\bibfield
  {journal} {\bibinfo  {journal} {Phys. Rev. B}\ }\textbf {\bibinfo {volume}
  {101}},\ \bibinfo {pages} {205116} (\bibinfo {year} {2020})}\BibitemShut
  {NoStop}%
\bibitem [{\citenamefont {Carr}\ \emph
  {et~al.}(2020{\natexlab{a}})\citenamefont {Carr}, \citenamefont {Li},
  \citenamefont {Zhu}, \citenamefont {Kaxiras}, \citenamefont {Sachdev},\ and\
  \citenamefont {Kruchkov}}]{CAR20}%
  \BibitemOpen
  \bibfield  {author} {\bibinfo {author} {\bibfnamefont {S.}~\bibnamefont
  {Carr}}, \bibinfo {author} {\bibfnamefont {C.}~\bibnamefont {Li}}, \bibinfo
  {author} {\bibfnamefont {Z.}~\bibnamefont {Zhu}}, \bibinfo {author}
  {\bibfnamefont {E.}~\bibnamefont {Kaxiras}}, \bibinfo {author} {\bibfnamefont
  {S.}~\bibnamefont {Sachdev}},\ and\ \bibinfo {author} {\bibfnamefont
  {A.}~\bibnamefont {Kruchkov}},\ }\href
  {https://doi.org/10.1021/acs.nanolett.9b04979} {\bibfield  {journal}
  {\bibinfo  {journal} {Nano Lett.}\ }\textbf {\bibinfo {volume} {20}},\
  \bibinfo {pages} {3030} (\bibinfo {year} {2020}{\natexlab{a}})}\BibitemShut
  {NoStop}%
\bibitem [{\citenamefont {Tritsaris}\ \emph {et~al.}(2020)\citenamefont
  {Tritsaris}, \citenamefont {Carr}, \citenamefont {Zhu}, \citenamefont {Xie},
  \citenamefont {Torrisi}, \citenamefont {Tang}, \citenamefont {Mattheakis},
  \citenamefont {Larson},\ and\ \citenamefont {Kaxiras}}]{TRI20}%
  \BibitemOpen
  \bibfield  {author} {\bibinfo {author} {\bibfnamefont {G.~A.}\ \bibnamefont
  {Tritsaris}}, \bibinfo {author} {\bibfnamefont {S.}~\bibnamefont {Carr}},
  \bibinfo {author} {\bibfnamefont {Z.}~\bibnamefont {Zhu}}, \bibinfo {author}
  {\bibfnamefont {Y.}~\bibnamefont {Xie}}, \bibinfo {author} {\bibfnamefont
  {S.~B.}\ \bibnamefont {Torrisi}}, \bibinfo {author} {\bibfnamefont
  {J.}~\bibnamefont {Tang}}, \bibinfo {author} {\bibfnamefont {M.}~\bibnamefont
  {Mattheakis}}, \bibinfo {author} {\bibfnamefont {D.~T.}\ \bibnamefont
  {Larson}},\ and\ \bibinfo {author} {\bibfnamefont {E.}~\bibnamefont
  {Kaxiras}},\ }\href {https://doi.org/10.1088/2053-1583/ab8f62} {\bibfield
  {journal} {\bibinfo  {journal} {2D Mater.}\ }\textbf {\bibinfo {volume}
  {7}},\ \bibinfo {pages} {035028} (\bibinfo {year} {2020})}\BibitemShut
  {NoStop}%
\bibitem [{\citenamefont {Wilson}\ \emph {et~al.}(2020)\citenamefont {Wilson},
  \citenamefont {Fu}, \citenamefont {Das~Sarma},\ and\ \citenamefont
  {Pixley}}]{WIL20}%
  \BibitemOpen
  \bibfield  {author} {\bibinfo {author} {\bibfnamefont {J.~H.}\ \bibnamefont
  {Wilson}}, \bibinfo {author} {\bibfnamefont {Y.}~\bibnamefont {Fu}}, \bibinfo
  {author} {\bibfnamefont {S.}~\bibnamefont {Das~Sarma}},\ and\ \bibinfo
  {author} {\bibfnamefont {J.~H.}\ \bibnamefont {Pixley}},\ }\href
  {https://doi.org/10.1103/PhysRevResearch.2.023325} {\bibfield  {journal}
  {\bibinfo  {journal} {Phys. Rev. Research}\ }\textbf {\bibinfo {volume}
  {2}},\ \bibinfo {pages} {023325} (\bibinfo {year} {2020})}\BibitemShut
  {NoStop}%
\bibitem [{\citenamefont {Park}\ \emph {et~al.}(2020)\citenamefont {Park},
  \citenamefont {Chittari},\ and\ \citenamefont {Jung}}]{PAR20}%
  \BibitemOpen
  \bibfield  {author} {\bibinfo {author} {\bibfnamefont {Y.}~\bibnamefont
  {Park}}, \bibinfo {author} {\bibfnamefont {B.~L.}\ \bibnamefont {Chittari}},\
  and\ \bibinfo {author} {\bibfnamefont {J.}~\bibnamefont {Jung}},\ }\href
  {https://doi.org/10.1103/PhysRevB.102.035411} {\bibfield  {journal} {\bibinfo
   {journal} {Phys. Rev. B}\ }\textbf {\bibinfo {volume} {102}},\ \bibinfo
  {pages} {035411} (\bibinfo {year} {2020})}\BibitemShut {NoStop}%
\bibitem [{\citenamefont {Carr}\ \emph
  {et~al.}(2020{\natexlab{b}})\citenamefont {Carr}, \citenamefont {Fang},\ and\
  \citenamefont {Kaxiras}}]{CAR20a}%
  \BibitemOpen
  \bibfield  {author} {\bibinfo {author} {\bibfnamefont {S.}~\bibnamefont
  {Carr}}, \bibinfo {author} {\bibfnamefont {S.}~\bibnamefont {Fang}},\ and\
  \bibinfo {author} {\bibfnamefont {E.}~\bibnamefont {Kaxiras}},\ }\href
  {https://doi.org/10.1038/s41578-020-0214-0} {\bibfield  {journal} {\bibinfo
  {journal} {Nat. Rev. Mater.}\ }\textbf {\bibinfo {volume} {5}},\ \bibinfo
  {pages} {748} (\bibinfo {year} {2020}{\natexlab{b}})}\BibitemShut {NoStop}%
\bibitem [{\citenamefont {Fu}\ \emph {et~al.}(2020)\citenamefont {Fu},
  \citenamefont {K{\"o}nig}, \citenamefont {Wilson}, \citenamefont {Chou},\
  and\ \citenamefont {Pixley}}]{FU20}%
  \BibitemOpen
  \bibfield  {author} {\bibinfo {author} {\bibfnamefont {Y.}~\bibnamefont
  {Fu}}, \bibinfo {author} {\bibfnamefont {E.~J.}\ \bibnamefont {K{\"o}nig}},
  \bibinfo {author} {\bibfnamefont {J.~H.}\ \bibnamefont {Wilson}}, \bibinfo
  {author} {\bibfnamefont {Y.-Z.}\ \bibnamefont {Chou}},\ and\ \bibinfo
  {author} {\bibfnamefont {J.~H.}\ \bibnamefont {Pixley}},\ }\href
  {https://doi.org/10.1038/s41535-020-00271-9} {\bibfield  {journal} {\bibinfo
  {journal} {npj Quantum Mater.}\ }\textbf {\bibinfo {volume} {5}},\ \bibinfo
  {pages} {1} (\bibinfo {year} {2020})}\BibitemShut {NoStop}%
\bibitem [{\citenamefont {Huang}\ \emph {et~al.}(2020)\citenamefont {Huang},
  \citenamefont {Hosur},\ and\ \citenamefont {Pal}}]{HUA20a}%
  \BibitemOpen
  \bibfield  {author} {\bibinfo {author} {\bibfnamefont {Y.}~\bibnamefont
  {Huang}}, \bibinfo {author} {\bibfnamefont {P.}~\bibnamefont {Hosur}},\ and\
  \bibinfo {author} {\bibfnamefont {H.~K.}\ \bibnamefont {Pal}},\ }\href
  {https://doi.org/10.1103/PhysRevB.102.155429} {\bibfield  {journal} {\bibinfo
   {journal} {Phys. Rev. B}\ }\textbf {\bibinfo {volume} {102}},\ \bibinfo
  {pages} {155429} (\bibinfo {year} {2020})}\BibitemShut {NoStop}%
\bibitem [{\citenamefont {Calder{\'o}n}\ and\ \citenamefont
  {Bascones}(2020)}]{CAL20}%
  \BibitemOpen
  \bibfield  {author} {\bibinfo {author} {\bibfnamefont {M.~J.}\ \bibnamefont
  {Calder{\'o}n}}\ and\ \bibinfo {author} {\bibfnamefont {E.}~\bibnamefont
  {Bascones}},\ }\href {https://doi.org/10.1103/PhysRevB.102.155149} {\bibfield
   {journal} {\bibinfo  {journal} {Phys. Rev. B}\ }\textbf {\bibinfo {volume}
  {102}},\ \bibinfo {pages} {155149} (\bibinfo {year} {2020})}\BibitemShut
  {NoStop}%
\bibitem [{\citenamefont {Wu}\ \emph {et~al.}(2021{\natexlab{b}})\citenamefont
  {Wu}, \citenamefont {Zhan},\ and\ \citenamefont {Yuan}}]{WU21d}%
  \BibitemOpen
  \bibfield  {author} {\bibinfo {author} {\bibfnamefont {Z.}~\bibnamefont
  {Wu}}, \bibinfo {author} {\bibfnamefont {Z.}~\bibnamefont {Zhan}},\ and\
  \bibinfo {author} {\bibfnamefont {S.}~\bibnamefont {Yuan}},\ }\href
  {https://doi.org/10.1007/s11433-020-1690-4} {\bibfield  {journal} {\bibinfo
  {journal} {Sci. China Phys. Mech. Astron.}\ }\textbf {\bibinfo {volume}
  {64}},\ \bibinfo {pages} {267811} (\bibinfo {year}
  {2021}{\natexlab{b}})}\BibitemShut {NoStop}%
\bibitem [{\citenamefont {Ren}\ \emph {et~al.}(2021)\citenamefont {Ren},
  \citenamefont {Gao}, \citenamefont {MacDonald},\ and\ \citenamefont
  {Niu}}]{REN21}%
  \BibitemOpen
  \bibfield  {author} {\bibinfo {author} {\bibfnamefont {Y.}~\bibnamefont
  {Ren}}, \bibinfo {author} {\bibfnamefont {Q.}~\bibnamefont {Gao}}, \bibinfo
  {author} {\bibfnamefont {A.~H.}\ \bibnamefont {MacDonald}},\ and\ \bibinfo
  {author} {\bibfnamefont {Q.}~\bibnamefont {Niu}},\ }\href
  {https://doi.org/10.1103/PhysRevLett.126.016404} {\bibfield  {journal}
  {\bibinfo  {journal} {Phys. Rev. Lett.}\ }\textbf {\bibinfo {volume} {126}},\
  \bibinfo {pages} {016404} (\bibinfo {year} {2021})}\BibitemShut {NoStop}%
\bibitem [{\citenamefont {Hejazi}\ \emph {et~al.}(2021)\citenamefont {Hejazi},
  \citenamefont {Chen},\ and\ \citenamefont {Balents}}]{HEJ21}%
  \BibitemOpen
  \bibfield  {author} {\bibinfo {author} {\bibfnamefont {K.}~\bibnamefont
  {Hejazi}}, \bibinfo {author} {\bibfnamefont {X.}~\bibnamefont {Chen}},\ and\
  \bibinfo {author} {\bibfnamefont {L.}~\bibnamefont {Balents}},\ }\href
  {https://doi.org/10.1103/PhysRevResearch.3.013242} {\bibfield  {journal}
  {\bibinfo  {journal} {Phys. Rev. Research}\ }\textbf {\bibinfo {volume}
  {3}},\ \bibinfo {pages} {013242} (\bibinfo {year} {2021})}\BibitemShut
  {NoStop}%
\bibitem [{\citenamefont {C{\u a}lug{\u a}ru}\ \emph
  {et~al.}(2021)\citenamefont {C{\u a}lug{\u a}ru}, \citenamefont {Xie},
  \citenamefont {Song}, \citenamefont {Lian}, \citenamefont {Regnault},\ and\
  \citenamefont {Bernevig}}]{CAL21}%
  \BibitemOpen
  \bibfield  {author} {\bibinfo {author} {\bibfnamefont {D.}~\bibnamefont {C{\u
  a}lug{\u a}ru}}, \bibinfo {author} {\bibfnamefont {F.}~\bibnamefont {Xie}},
  \bibinfo {author} {\bibfnamefont {Z.-D.}\ \bibnamefont {Song}}, \bibinfo
  {author} {\bibfnamefont {B.}~\bibnamefont {Lian}}, \bibinfo {author}
  {\bibfnamefont {N.}~\bibnamefont {Regnault}},\ and\ \bibinfo {author}
  {\bibfnamefont {B.~A.}\ \bibnamefont {Bernevig}},\ }\href
  {https://doi.org/10.1103/PhysRevB.103.195411} {\bibfield  {journal} {\bibinfo
   {journal} {Phys. Rev. B}\ }\textbf {\bibinfo {volume} {103}},\ \bibinfo
  {pages} {195411} (\bibinfo {year} {2021})}\BibitemShut {NoStop}%
\bibitem [{\citenamefont {Bernevig}\ \emph
  {et~al.}(2021{\natexlab{a}})\citenamefont {Bernevig}, \citenamefont {Song},
  \citenamefont {Regnault},\ and\ \citenamefont {Lian}}]{BER21}%
  \BibitemOpen
  \bibfield  {author} {\bibinfo {author} {\bibfnamefont {B.~A.}\ \bibnamefont
  {Bernevig}}, \bibinfo {author} {\bibfnamefont {Z.-D.}\ \bibnamefont {Song}},
  \bibinfo {author} {\bibfnamefont {N.}~\bibnamefont {Regnault}},\ and\
  \bibinfo {author} {\bibfnamefont {B.}~\bibnamefont {Lian}},\ }\href
  {https://doi.org/10.1103/PhysRevB.103.205411} {\bibfield  {journal} {\bibinfo
   {journal} {Phys. Rev. B}\ }\textbf {\bibinfo {volume} {103}},\ \bibinfo
  {pages} {205411} (\bibinfo {year} {2021}{\natexlab{a}})}\BibitemShut
  {NoStop}%
\bibitem [{\citenamefont {Bernevig}\ \emph
  {et~al.}(2021{\natexlab{b}})\citenamefont {Bernevig}, \citenamefont {Song},
  \citenamefont {Regnault},\ and\ \citenamefont {Lian}}]{BER21a}%
  \BibitemOpen
  \bibfield  {author} {\bibinfo {author} {\bibfnamefont {B.~A.}\ \bibnamefont
  {Bernevig}}, \bibinfo {author} {\bibfnamefont {Z.-D.}\ \bibnamefont {Song}},
  \bibinfo {author} {\bibfnamefont {N.}~\bibnamefont {Regnault}},\ and\
  \bibinfo {author} {\bibfnamefont {B.}~\bibnamefont {Lian}},\ }\href
  {https://doi.org/10.1103/PhysRevB.103.205413} {\bibfield  {journal} {\bibinfo
   {journal} {Phys. Rev. B}\ }\textbf {\bibinfo {volume} {103}},\ \bibinfo
  {pages} {205413} (\bibinfo {year} {2021}{\natexlab{b}})}\BibitemShut
  {NoStop}%
\bibitem [{\citenamefont {Wang}\ \emph
  {et~al.}(2021{\natexlab{a}})\citenamefont {Wang}, \citenamefont {Zheng},
  \citenamefont {Millis},\ and\ \citenamefont {Cano}}]{WAN21a}%
  \BibitemOpen
  \bibfield  {author} {\bibinfo {author} {\bibfnamefont {J.}~\bibnamefont
  {Wang}}, \bibinfo {author} {\bibfnamefont {Y.}~\bibnamefont {Zheng}},
  \bibinfo {author} {\bibfnamefont {A.~J.}\ \bibnamefont {Millis}},\ and\
  \bibinfo {author} {\bibfnamefont {J.}~\bibnamefont {Cano}},\ }\href
  {https://doi.org/10.1103/PhysRevResearch.3.023155} {\bibfield  {journal}
  {\bibinfo  {journal} {Phys. Rev. Research}\ }\textbf {\bibinfo {volume}
  {3}},\ \bibinfo {pages} {023155} (\bibinfo {year}
  {2021}{\natexlab{a}})}\BibitemShut {NoStop}%
\bibitem [{\citenamefont {Ramires}\ and\ \citenamefont {Lado}(2021)}]{RAM21}%
  \BibitemOpen
  \bibfield  {author} {\bibinfo {author} {\bibfnamefont {A.}~\bibnamefont
  {Ramires}}\ and\ \bibinfo {author} {\bibfnamefont {J.~L.}\ \bibnamefont
  {Lado}},\ }\href {https://doi.org/10.1103/PhysRevLett.127.026401} {\bibfield
  {journal} {\bibinfo  {journal} {Phys. Rev. Lett.}\ }\textbf {\bibinfo
  {volume} {127}},\ \bibinfo {pages} {026401} (\bibinfo {year}
  {2021})}\BibitemShut {NoStop}%
\bibitem [{\citenamefont {Shin}\ \emph {et~al.}(2021)\citenamefont {Shin},
  \citenamefont {Chittari},\ and\ \citenamefont {Jung}}]{SHI21}%
  \BibitemOpen
  \bibfield  {author} {\bibinfo {author} {\bibfnamefont {J.}~\bibnamefont
  {Shin}}, \bibinfo {author} {\bibfnamefont {B.~L.}\ \bibnamefont {Chittari}},\
  and\ \bibinfo {author} {\bibfnamefont {J.}~\bibnamefont {Jung}},\ }\href
  {https://doi.org/10.1103/PhysRevB.104.045413} {\bibfield  {journal} {\bibinfo
   {journal} {Phys. Rev. B}\ }\textbf {\bibinfo {volume} {104}},\ \bibinfo
  {pages} {045413} (\bibinfo {year} {2021})}\BibitemShut {NoStop}%
\bibitem [{\citenamefont {Lei}\ \emph {et~al.}(2021)\citenamefont {Lei},
  \citenamefont {Linhart}, \citenamefont {Qin}, \citenamefont {Libisch},\ and\
  \citenamefont {MacDonald}}]{LEI21}%
  \BibitemOpen
  \bibfield  {author} {\bibinfo {author} {\bibfnamefont {C.}~\bibnamefont
  {Lei}}, \bibinfo {author} {\bibfnamefont {L.}~\bibnamefont {Linhart}},
  \bibinfo {author} {\bibfnamefont {W.}~\bibnamefont {Qin}}, \bibinfo {author}
  {\bibfnamefont {F.}~\bibnamefont {Libisch}},\ and\ \bibinfo {author}
  {\bibfnamefont {A.~H.}\ \bibnamefont {MacDonald}},\ }\href
  {https://doi.org/10.1103/PhysRevB.104.035139} {\bibfield  {journal} {\bibinfo
   {journal} {Phys. Rev. B}\ }\textbf {\bibinfo {volume} {104}},\ \bibinfo
  {pages} {035139} (\bibinfo {year} {2021})}\BibitemShut {NoStop}%
\bibitem [{\citenamefont {Cao}\ \emph {et~al.}(2021{\natexlab{c}})\citenamefont
  {Cao}, \citenamefont {Wang}, \citenamefont {Qian}, \citenamefont {Liu},\ and\
  \citenamefont {Yao}}]{CAO21a}%
  \BibitemOpen
  \bibfield  {author} {\bibinfo {author} {\bibfnamefont {J.}~\bibnamefont
  {Cao}}, \bibinfo {author} {\bibfnamefont {M.}~\bibnamefont {Wang}}, \bibinfo
  {author} {\bibfnamefont {S.-F.}\ \bibnamefont {Qian}}, \bibinfo {author}
  {\bibfnamefont {C.-C.}\ \bibnamefont {Liu}},\ and\ \bibinfo {author}
  {\bibfnamefont {Y.}~\bibnamefont {Yao}},\ }\href
  {https://doi.org/10.1103/PhysRevB.104.L081403} {\bibfield  {journal}
  {\bibinfo  {journal} {Phys. Rev. B}\ }\textbf {\bibinfo {volume} {104}},\
  \bibinfo {pages} {L081403} (\bibinfo {year}
  {2021}{\natexlab{c}})}\BibitemShut {NoStop}%
\bibitem [{\citenamefont {Sheffer}\ and\ \citenamefont {Stern}(2021)}]{SHE21}%
  \BibitemOpen
  \bibfield  {author} {\bibinfo {author} {\bibfnamefont {Y.}~\bibnamefont
  {Sheffer}}\ and\ \bibinfo {author} {\bibfnamefont {A.}~\bibnamefont
  {Stern}},\ }\href {https://doi.org/10.1103/PhysRevB.104.L121405} {\bibfield
  {journal} {\bibinfo  {journal} {Phys. Rev. B}\ }\textbf {\bibinfo {volume}
  {104}},\ \bibinfo {pages} {L121405} (\bibinfo {year} {2021})}\BibitemShut
  {NoStop}%
\bibitem [{\citenamefont {Wu}\ \emph {et~al.}(2021{\natexlab{c}})\citenamefont
  {Wu}, \citenamefont {Kuang}, \citenamefont {Zhan},\ and\ \citenamefont
  {Yuan}}]{WU21c}%
  \BibitemOpen
  \bibfield  {author} {\bibinfo {author} {\bibfnamefont {Z.}~\bibnamefont
  {Wu}}, \bibinfo {author} {\bibfnamefont {X.}~\bibnamefont {Kuang}}, \bibinfo
  {author} {\bibfnamefont {Z.}~\bibnamefont {Zhan}},\ and\ \bibinfo {author}
  {\bibfnamefont {S.}~\bibnamefont {Yuan}},\ }\href
  {https://doi.org/10.1103/PhysRevB.104.205104} {\bibfield  {journal} {\bibinfo
   {journal} {Phys. Rev. B}\ }\textbf {\bibinfo {volume} {104}},\ \bibinfo
  {pages} {205104} (\bibinfo {year} {2021}{\natexlab{c}})}\BibitemShut
  {NoStop}%
\bibitem [{\citenamefont {Koshino}\ \emph {et~al.}(2018)\citenamefont
  {Koshino}, \citenamefont {Yuan}, \citenamefont {Koretsune}, \citenamefont
  {Ochi}, \citenamefont {Kuroki},\ and\ \citenamefont {Fu}}]{KOS18}%
  \BibitemOpen
  \bibfield  {author} {\bibinfo {author} {\bibfnamefont {M.}~\bibnamefont
  {Koshino}}, \bibinfo {author} {\bibfnamefont {N.~F.~Q.}\ \bibnamefont
  {Yuan}}, \bibinfo {author} {\bibfnamefont {T.}~\bibnamefont {Koretsune}},
  \bibinfo {author} {\bibfnamefont {M.}~\bibnamefont {Ochi}}, \bibinfo {author}
  {\bibfnamefont {K.}~\bibnamefont {Kuroki}},\ and\ \bibinfo {author}
  {\bibfnamefont {L.}~\bibnamefont {Fu}},\ }\href
  {https://doi.org/10.1103/PhysRevX.8.031087} {\bibfield  {journal} {\bibinfo
  {journal} {Phys. Rev. X}\ }\textbf {\bibinfo {volume} {8}},\ \bibinfo {pages}
  {031087} (\bibinfo {year} {2018})}\BibitemShut {NoStop}%
\bibitem [{\citenamefont {Ledwith}\ \emph {et~al.}(2021)\citenamefont
  {Ledwith}, \citenamefont {Khalaf}, \citenamefont {Zhu}, \citenamefont {Carr},
  \citenamefont {Kaxiras},\ and\ \citenamefont {Vishwanath}}]{LED21}%
  \BibitemOpen
  \bibfield  {author} {\bibinfo {author} {\bibfnamefont {P.~J.}\ \bibnamefont
  {Ledwith}}, \bibinfo {author} {\bibfnamefont {E.}~\bibnamefont {Khalaf}},
  \bibinfo {author} {\bibfnamefont {Z.}~\bibnamefont {Zhu}}, \bibinfo {author}
  {\bibfnamefont {S.}~\bibnamefont {Carr}}, \bibinfo {author} {\bibfnamefont
  {E.}~\bibnamefont {Kaxiras}},\ and\ \bibinfo {author} {\bibfnamefont
  {A.}~\bibnamefont {Vishwanath}},\ }\href
  {https://doi.org/10.48550/arXiv.2111.11060} {\bibinfo {title} {{{TB}} or not
  {{TB}}? {{Contrasting}} properties of twisted bilayer graphene and the
  alternating twist $n$-layer structures ($n=3, 4, 5, \dots$)}} (\bibinfo
  {year} {2021}),\ \Eprint {https://arxiv.org/abs/2111.11060} {arxiv:2111.11060
  [cond-mat]} \BibitemShut {NoStop}%
\bibitem [{\citenamefont {Guerci}\ \emph {et~al.}(2022)\citenamefont {Guerci},
  \citenamefont {Simon},\ and\ \citenamefont {Mora}}]{GUE22}%
  \BibitemOpen
  \bibfield  {author} {\bibinfo {author} {\bibfnamefont {D.}~\bibnamefont
  {Guerci}}, \bibinfo {author} {\bibfnamefont {P.}~\bibnamefont {Simon}},\ and\
  \bibinfo {author} {\bibfnamefont {C.}~\bibnamefont {Mora}},\ }\href
  {https://doi.org/10.1103/PhysRevResearch.4.L012013} {\bibfield  {journal}
  {\bibinfo  {journal} {Phys. Rev. Res.}\ }\textbf {\bibinfo {volume} {4}},\
  \bibinfo {pages} {L012013} (\bibinfo {year} {2022})}\BibitemShut {NoStop}%
\bibitem [{\citenamefont {Davydov}\ \emph {et~al.}(2022)\citenamefont
  {Davydov}, \citenamefont {Choo}, \citenamefont {Fischer},\ and\ \citenamefont
  {Neupert}}]{DAV22}%
  \BibitemOpen
  \bibfield  {author} {\bibinfo {author} {\bibfnamefont {A.}~\bibnamefont
  {Davydov}}, \bibinfo {author} {\bibfnamefont {K.}~\bibnamefont {Choo}},
  \bibinfo {author} {\bibfnamefont {M.~H.}\ \bibnamefont {Fischer}},\ and\
  \bibinfo {author} {\bibfnamefont {T.}~\bibnamefont {Neupert}},\ }\href
  {https://doi.org/10.1103/PhysRevB.105.165153} {\bibfield  {journal} {\bibinfo
   {journal} {Phys. Rev. B}\ }\textbf {\bibinfo {volume} {105}},\ \bibinfo
  {pages} {165153} (\bibinfo {year} {2022})}\BibitemShut {NoStop}%
\bibitem [{\citenamefont {Classen}\ \emph {et~al.}(2022)\citenamefont
  {Classen}, \citenamefont {Pixley},\ and\ \citenamefont {K{\"o}nig}}]{CLA22}%
  \BibitemOpen
  \bibfield  {author} {\bibinfo {author} {\bibfnamefont {L.}~\bibnamefont
  {Classen}}, \bibinfo {author} {\bibfnamefont {J.~H.}\ \bibnamefont
  {Pixley}},\ and\ \bibinfo {author} {\bibfnamefont {E.~J.}\ \bibnamefont
  {K{\"o}nig}},\ }\href {https://doi.org/10.1088/2053-1583/ac6e71} {\bibfield
  {journal} {\bibinfo  {journal} {2D Mater.}\ }\textbf {\bibinfo {volume}
  {9}},\ \bibinfo {pages} {031001} (\bibinfo {year} {2022})}\BibitemShut
  {NoStop}%
\bibitem [{\citenamefont {Lin}\ \emph {et~al.}(2022)\citenamefont {Lin},
  \citenamefont {Li}, \citenamefont {Su},\ and\ \citenamefont {Ni}}]{LIN22}%
  \BibitemOpen
  \bibfield  {author} {\bibinfo {author} {\bibfnamefont {X.}~\bibnamefont
  {Lin}}, \bibinfo {author} {\bibfnamefont {C.}~\bibnamefont {Li}}, \bibinfo
  {author} {\bibfnamefont {K.}~\bibnamefont {Su}},\ and\ \bibinfo {author}
  {\bibfnamefont {J.}~\bibnamefont {Ni}},\ }\href
  {https://doi.org/10.1103/PhysRevB.106.075423} {\bibfield  {journal} {\bibinfo
   {journal} {Phys. Rev. B}\ }\textbf {\bibinfo {volume} {106}},\ \bibinfo
  {pages} {075423} (\bibinfo {year} {2022})}\BibitemShut {NoStop}%
\bibitem [{\citenamefont {Samajdar}\ \emph {et~al.}(2022)\citenamefont
  {Samajdar}, \citenamefont {Teng},\ and\ \citenamefont {Scheurer}}]{SAM22}%
  \BibitemOpen
  \bibfield  {author} {\bibinfo {author} {\bibfnamefont {R.}~\bibnamefont
  {Samajdar}}, \bibinfo {author} {\bibfnamefont {Y.}~\bibnamefont {Teng}},\
  and\ \bibinfo {author} {\bibfnamefont {M.~S.}\ \bibnamefont {Scheurer}},\
  }\href {https://doi.org/10.1103/PhysRevB.106.L201403} {\bibfield  {journal}
  {\bibinfo  {journal} {Phys. Rev. B}\ }\textbf {\bibinfo {volume} {106}},\
  \bibinfo {pages} {L201403} (\bibinfo {year} {2022})}\BibitemShut {NoStop}%
\bibitem [{\citenamefont {Kang}\ and\ \citenamefont {Vafek}(2023)}]{KAN23b}%
  \BibitemOpen
  \bibfield  {author} {\bibinfo {author} {\bibfnamefont {J.}~\bibnamefont
  {Kang}}\ and\ \bibinfo {author} {\bibfnamefont {O.}~\bibnamefont {Vafek}},\
  }\href {https://doi.org/10.1103/PhysRevB.107.075408} {\bibfield  {journal}
  {\bibinfo  {journal} {Phys. Rev. B}\ }\textbf {\bibinfo {volume} {107}},\
  \bibinfo {pages} {075408} (\bibinfo {year} {2023})}\BibitemShut {NoStop}%
\bibitem [{\citenamefont {Vafek}\ and\ \citenamefont {Kang}(2023)}]{VAF23}%
  \BibitemOpen
  \bibfield  {author} {\bibinfo {author} {\bibfnamefont {O.}~\bibnamefont
  {Vafek}}\ and\ \bibinfo {author} {\bibfnamefont {J.}~\bibnamefont {Kang}},\
  }\href {https://doi.org/10.1103/PhysRevB.107.075123} {\bibfield  {journal}
  {\bibinfo  {journal} {Phys. Rev. B}\ }\textbf {\bibinfo {volume} {107}},\
  \bibinfo {pages} {075123} (\bibinfo {year} {2023})}\BibitemShut {NoStop}%
\bibitem [{\citenamefont {Shin}\ \emph {et~al.}(2023)\citenamefont {Shin},
  \citenamefont {Jang}, \citenamefont {Shin}, \citenamefont {Jung},\ and\
  \citenamefont {Min}}]{SHI23}%
  \BibitemOpen
  \bibfield  {author} {\bibinfo {author} {\bibfnamefont {K.}~\bibnamefont
  {Shin}}, \bibinfo {author} {\bibfnamefont {Y.}~\bibnamefont {Jang}}, \bibinfo
  {author} {\bibfnamefont {J.}~\bibnamefont {Shin}}, \bibinfo {author}
  {\bibfnamefont {J.}~\bibnamefont {Jung}},\ and\ \bibinfo {author}
  {\bibfnamefont {H.}~\bibnamefont {Min}},\ }\href
  {https://doi.org/10.1103/PhysRevB.107.245139} {\bibfield  {journal} {\bibinfo
   {journal} {Phys. Rev. B}\ }\textbf {\bibinfo {volume} {107}},\ \bibinfo
  {pages} {245139} (\bibinfo {year} {2023})}\BibitemShut {NoStop}%
\bibitem [{\citenamefont {Hejazi}\ \emph
  {et~al.}(2019{\natexlab{a}})\citenamefont {Hejazi}, \citenamefont {Liu},
  \citenamefont {Shapourian}, \citenamefont {Chen},\ and\ \citenamefont
  {Balents}}]{HEJ19}%
  \BibitemOpen
  \bibfield  {author} {\bibinfo {author} {\bibfnamefont {K.}~\bibnamefont
  {Hejazi}}, \bibinfo {author} {\bibfnamefont {C.}~\bibnamefont {Liu}},
  \bibinfo {author} {\bibfnamefont {H.}~\bibnamefont {Shapourian}}, \bibinfo
  {author} {\bibfnamefont {X.}~\bibnamefont {Chen}},\ and\ \bibinfo {author}
  {\bibfnamefont {L.}~\bibnamefont {Balents}},\ }\href
  {https://doi.org/10.1103/PhysRevB.99.035111} {\bibfield  {journal} {\bibinfo
  {journal} {Phys. Rev. B}\ }\textbf {\bibinfo {volume} {99}},\ \bibinfo
  {pages} {035111} (\bibinfo {year} {2019}{\natexlab{a}})}\BibitemShut
  {NoStop}%
\bibitem [{\citenamefont {Ahn}\ \emph {et~al.}(2019)\citenamefont {Ahn},
  \citenamefont {Park},\ and\ \citenamefont {Yang}}]{AHN19}%
  \BibitemOpen
  \bibfield  {author} {\bibinfo {author} {\bibfnamefont {J.}~\bibnamefont
  {Ahn}}, \bibinfo {author} {\bibfnamefont {S.}~\bibnamefont {Park}},\ and\
  \bibinfo {author} {\bibfnamefont {B.-J.}\ \bibnamefont {Yang}},\ }\href
  {https://doi.org/10.1103/PhysRevX.9.021013} {\bibfield  {journal} {\bibinfo
  {journal} {Phys. Rev. X}\ }\textbf {\bibinfo {volume} {9}},\ \bibinfo {pages}
  {021013} (\bibinfo {year} {2019})}\BibitemShut {NoStop}%
\bibitem [{\citenamefont {Song}\ \emph {et~al.}(2019)\citenamefont {Song},
  \citenamefont {Wang}, \citenamefont {Shi}, \citenamefont {Li}, \citenamefont
  {Fang},\ and\ \citenamefont {Bernevig}}]{SON19}%
  \BibitemOpen
  \bibfield  {author} {\bibinfo {author} {\bibfnamefont {Z.}~\bibnamefont
  {Song}}, \bibinfo {author} {\bibfnamefont {Z.}~\bibnamefont {Wang}}, \bibinfo
  {author} {\bibfnamefont {W.}~\bibnamefont {Shi}}, \bibinfo {author}
  {\bibfnamefont {G.}~\bibnamefont {Li}}, \bibinfo {author} {\bibfnamefont
  {C.}~\bibnamefont {Fang}},\ and\ \bibinfo {author} {\bibfnamefont {B.~A.}\
  \bibnamefont {Bernevig}},\ }\href
  {https://doi.org/10.1103/PhysRevLett.123.036401} {\bibfield  {journal}
  {\bibinfo  {journal} {Phys. Rev. Lett.}\ }\textbf {\bibinfo {volume} {123}},\
  \bibinfo {pages} {036401} (\bibinfo {year} {2019})}\BibitemShut {NoStop}%
\bibitem [{\citenamefont {Hejazi}\ \emph
  {et~al.}(2019{\natexlab{b}})\citenamefont {Hejazi}, \citenamefont {Liu},\
  and\ \citenamefont {Balents}}]{HEJ19a}%
  \BibitemOpen
  \bibfield  {author} {\bibinfo {author} {\bibfnamefont {K.}~\bibnamefont
  {Hejazi}}, \bibinfo {author} {\bibfnamefont {C.}~\bibnamefont {Liu}},\ and\
  \bibinfo {author} {\bibfnamefont {L.}~\bibnamefont {Balents}},\ }\href
  {https://doi.org/10.1103/PhysRevB.100.035115} {\bibfield  {journal} {\bibinfo
   {journal} {Phys. Rev. B}\ }\textbf {\bibinfo {volume} {100}},\ \bibinfo
  {pages} {035115} (\bibinfo {year} {2019}{\natexlab{b}})}\BibitemShut
  {NoStop}%
\bibitem [{\citenamefont {Xie}\ \emph {et~al.}(2020)\citenamefont {Xie},
  \citenamefont {Song}, \citenamefont {Lian},\ and\ \citenamefont
  {Bernevig}}]{XIE20}%
  \BibitemOpen
  \bibfield  {author} {\bibinfo {author} {\bibfnamefont {F.}~\bibnamefont
  {Xie}}, \bibinfo {author} {\bibfnamefont {Z.}~\bibnamefont {Song}}, \bibinfo
  {author} {\bibfnamefont {B.}~\bibnamefont {Lian}},\ and\ \bibinfo {author}
  {\bibfnamefont {B.~A.}\ \bibnamefont {Bernevig}},\ }\href
  {https://doi.org/10.1103/PhysRevLett.124.167002} {\bibfield  {journal}
  {\bibinfo  {journal} {Phys. Rev. Lett.}\ }\textbf {\bibinfo {volume} {124}},\
  \bibinfo {pages} {167002} (\bibinfo {year} {2020})}\BibitemShut {NoStop}%
\bibitem [{\citenamefont {Lian}\ \emph {et~al.}(2020)\citenamefont {Lian},
  \citenamefont {Xie},\ and\ \citenamefont {Bernevig}}]{LIA20}%
  \BibitemOpen
  \bibfield  {author} {\bibinfo {author} {\bibfnamefont {B.}~\bibnamefont
  {Lian}}, \bibinfo {author} {\bibfnamefont {F.}~\bibnamefont {Xie}},\ and\
  \bibinfo {author} {\bibfnamefont {B.~A.}\ \bibnamefont {Bernevig}},\ }\href
  {https://doi.org/10.1103/PhysRevB.102.041402} {\bibfield  {journal} {\bibinfo
   {journal} {Phys. Rev. B}\ }\textbf {\bibinfo {volume} {102}},\ \bibinfo
  {pages} {041402} (\bibinfo {year} {2020})}\BibitemShut {NoStop}%
\bibitem [{\citenamefont {Song}\ \emph {et~al.}(2021)\citenamefont {Song},
  \citenamefont {Lian}, \citenamefont {Regnault},\ and\ \citenamefont
  {Bernevig}}]{SON21}%
  \BibitemOpen
  \bibfield  {author} {\bibinfo {author} {\bibfnamefont {Z.-D.}\ \bibnamefont
  {Song}}, \bibinfo {author} {\bibfnamefont {B.}~\bibnamefont {Lian}}, \bibinfo
  {author} {\bibfnamefont {N.}~\bibnamefont {Regnault}},\ and\ \bibinfo
  {author} {\bibfnamefont {B.~A.}\ \bibnamefont {Bernevig}},\ }\href
  {https://doi.org/10.1103/PhysRevB.103.205412} {\bibfield  {journal} {\bibinfo
   {journal} {Phys. Rev. B}\ }\textbf {\bibinfo {volume} {103}},\ \bibinfo
  {pages} {205412} (\bibinfo {year} {2021})}\BibitemShut {NoStop}%
\bibitem [{\citenamefont {Ochi}\ \emph {et~al.}(2018)\citenamefont {Ochi},
  \citenamefont {Koshino},\ and\ \citenamefont {Kuroki}}]{OCH18}%
  \BibitemOpen
  \bibfield  {author} {\bibinfo {author} {\bibfnamefont {M.}~\bibnamefont
  {Ochi}}, \bibinfo {author} {\bibfnamefont {M.}~\bibnamefont {Koshino}},\ and\
  \bibinfo {author} {\bibfnamefont {K.}~\bibnamefont {Kuroki}},\ }\href
  {https://doi.org/10.1103/PhysRevB.98.081102} {\bibfield  {journal} {\bibinfo
  {journal} {Phys. Rev. B}\ }\textbf {\bibinfo {volume} {98}},\ \bibinfo
  {pages} {081102} (\bibinfo {year} {2018})}\BibitemShut {NoStop}%
\bibitem [{\citenamefont {Thomson}\ \emph {et~al.}(2018)\citenamefont
  {Thomson}, \citenamefont {Chatterjee}, \citenamefont {Sachdev},\ and\
  \citenamefont {Scheurer}}]{THO18}%
  \BibitemOpen
  \bibfield  {author} {\bibinfo {author} {\bibfnamefont {A.}~\bibnamefont
  {Thomson}}, \bibinfo {author} {\bibfnamefont {S.}~\bibnamefont {Chatterjee}},
  \bibinfo {author} {\bibfnamefont {S.}~\bibnamefont {Sachdev}},\ and\ \bibinfo
  {author} {\bibfnamefont {M.~S.}\ \bibnamefont {Scheurer}},\ }\href
  {https://doi.org/10.1103/PhysRevB.98.075109} {\bibfield  {journal} {\bibinfo
  {journal} {Phys. Rev. B}\ }\textbf {\bibinfo {volume} {98}},\ \bibinfo
  {pages} {075109} (\bibinfo {year} {2018})}\BibitemShut {NoStop}%
\bibitem [{\citenamefont {Xu}\ \emph {et~al.}(2018)\citenamefont {Xu},
  \citenamefont {Law},\ and\ \citenamefont {Lee}}]{XU18b}%
  \BibitemOpen
  \bibfield  {author} {\bibinfo {author} {\bibfnamefont {X.~Y.}\ \bibnamefont
  {Xu}}, \bibinfo {author} {\bibfnamefont {K.~T.}\ \bibnamefont {Law}},\ and\
  \bibinfo {author} {\bibfnamefont {P.~A.}\ \bibnamefont {Lee}},\ }\href
  {https://doi.org/10.1103/PhysRevB.98.121406} {\bibfield  {journal} {\bibinfo
  {journal} {Phys. Rev. B}\ }\textbf {\bibinfo {volume} {98}},\ \bibinfo
  {pages} {121406} (\bibinfo {year} {2018})}\BibitemShut {NoStop}%
\bibitem [{\citenamefont {Po}\ \emph {et~al.}(2018{\natexlab{a}})\citenamefont
  {Po}, \citenamefont {Zou}, \citenamefont {Vishwanath},\ and\ \citenamefont
  {Senthil}}]{PO18a}%
  \BibitemOpen
  \bibfield  {author} {\bibinfo {author} {\bibfnamefont {H.~C.}\ \bibnamefont
  {Po}}, \bibinfo {author} {\bibfnamefont {L.}~\bibnamefont {Zou}}, \bibinfo
  {author} {\bibfnamefont {A.}~\bibnamefont {Vishwanath}},\ and\ \bibinfo
  {author} {\bibfnamefont {T.}~\bibnamefont {Senthil}},\ }\href
  {https://doi.org/10.1103/PhysRevX.8.031089} {\bibfield  {journal} {\bibinfo
  {journal} {Phys. Rev. X}\ }\textbf {\bibinfo {volume} {8}},\ \bibinfo {pages}
  {031089} (\bibinfo {year} {2018}{\natexlab{a}})}\BibitemShut {NoStop}%
\bibitem [{\citenamefont {Venderbos}\ and\ \citenamefont
  {Fernandes}(2018)}]{VEN18}%
  \BibitemOpen
  \bibfield  {author} {\bibinfo {author} {\bibfnamefont {J.~W.~F.}\
  \bibnamefont {Venderbos}}\ and\ \bibinfo {author} {\bibfnamefont {R.~M.}\
  \bibnamefont {Fernandes}},\ }\href
  {https://doi.org/10.1103/PhysRevB.98.245103} {\bibfield  {journal} {\bibinfo
  {journal} {Phys. Rev. B}\ }\textbf {\bibinfo {volume} {98}},\ \bibinfo
  {pages} {245103} (\bibinfo {year} {2018})}\BibitemShut {NoStop}%
\bibitem [{\citenamefont {Yuan}\ and\ \citenamefont {Fu}(2018)}]{YUA18}%
  \BibitemOpen
  \bibfield  {author} {\bibinfo {author} {\bibfnamefont {N.~F.~Q.}\
  \bibnamefont {Yuan}}\ and\ \bibinfo {author} {\bibfnamefont {L.}~\bibnamefont
  {Fu}},\ }\href {https://doi.org/10.1103/PhysRevB.98.045103} {\bibfield
  {journal} {\bibinfo  {journal} {Phys. Rev. B}\ }\textbf {\bibinfo {volume}
  {98}},\ \bibinfo {pages} {045103} (\bibinfo {year} {2018})}\BibitemShut
  {NoStop}%
\bibitem [{\citenamefont {Dodaro}\ \emph {et~al.}(2018)\citenamefont {Dodaro},
  \citenamefont {Kivelson}, \citenamefont {Schattner}, \citenamefont {Sun},\
  and\ \citenamefont {Wang}}]{DOD18}%
  \BibitemOpen
  \bibfield  {author} {\bibinfo {author} {\bibfnamefont {J.~F.}\ \bibnamefont
  {Dodaro}}, \bibinfo {author} {\bibfnamefont {S.~A.}\ \bibnamefont
  {Kivelson}}, \bibinfo {author} {\bibfnamefont {Y.}~\bibnamefont {Schattner}},
  \bibinfo {author} {\bibfnamefont {X.~Q.}\ \bibnamefont {Sun}},\ and\ \bibinfo
  {author} {\bibfnamefont {C.}~\bibnamefont {Wang}},\ }\href
  {https://doi.org/10.1103/PhysRevB.98.075154} {\bibfield  {journal} {\bibinfo
  {journal} {Phys. Rev. B}\ }\textbf {\bibinfo {volume} {98}},\ \bibinfo
  {pages} {075154} (\bibinfo {year} {2018})}\BibitemShut {NoStop}%
\bibitem [{\citenamefont {Padhi}\ \emph {et~al.}(2018)\citenamefont {Padhi},
  \citenamefont {Setty},\ and\ \citenamefont {Phillips}}]{PAD18}%
  \BibitemOpen
  \bibfield  {author} {\bibinfo {author} {\bibfnamefont {B.}~\bibnamefont
  {Padhi}}, \bibinfo {author} {\bibfnamefont {C.}~\bibnamefont {Setty}},\ and\
  \bibinfo {author} {\bibfnamefont {P.~W.}\ \bibnamefont {Phillips}},\ }\href
  {https://doi.org/10.1021/acs.nanolett.8b02033} {\bibfield  {journal}
  {\bibinfo  {journal} {Nano Lett.}\ }\textbf {\bibinfo {volume} {18}},\
  \bibinfo {pages} {6175} (\bibinfo {year} {2018})}\BibitemShut {NoStop}%
\bibitem [{\citenamefont {Kennes}\ \emph {et~al.}(2018)\citenamefont {Kennes},
  \citenamefont {Lischner},\ and\ \citenamefont {Karrasch}}]{KEN18}%
  \BibitemOpen
  \bibfield  {author} {\bibinfo {author} {\bibfnamefont {D.~M.}\ \bibnamefont
  {Kennes}}, \bibinfo {author} {\bibfnamefont {J.}~\bibnamefont {Lischner}},\
  and\ \bibinfo {author} {\bibfnamefont {C.}~\bibnamefont {Karrasch}},\ }\href
  {https://doi.org/10.1103/PhysRevB.98.241407} {\bibfield  {journal} {\bibinfo
  {journal} {Phys. Rev. B}\ }\textbf {\bibinfo {volume} {98}},\ \bibinfo
  {pages} {241407} (\bibinfo {year} {2018})}\BibitemShut {NoStop}%
\bibitem [{\citenamefont {Rademaker}\ and\ \citenamefont
  {Mellado}(2018)}]{RAD18}%
  \BibitemOpen
  \bibfield  {author} {\bibinfo {author} {\bibfnamefont {L.}~\bibnamefont
  {Rademaker}}\ and\ \bibinfo {author} {\bibfnamefont {P.}~\bibnamefont
  {Mellado}},\ }\href {https://doi.org/10.1103/PhysRevB.98.235158} {\bibfield
  {journal} {\bibinfo  {journal} {Phys. Rev. B}\ }\textbf {\bibinfo {volume}
  {98}},\ \bibinfo {pages} {235158} (\bibinfo {year} {2018})}\BibitemShut
  {NoStop}%
\bibitem [{\citenamefont {Liu}\ \emph {et~al.}(2019{\natexlab{b}})\citenamefont
  {Liu}, \citenamefont {Liu},\ and\ \citenamefont {Dai}}]{LIU19}%
  \BibitemOpen
  \bibfield  {author} {\bibinfo {author} {\bibfnamefont {J.}~\bibnamefont
  {Liu}}, \bibinfo {author} {\bibfnamefont {J.}~\bibnamefont {Liu}},\ and\
  \bibinfo {author} {\bibfnamefont {X.}~\bibnamefont {Dai}},\ }\href
  {https://doi.org/10.1103/PhysRevB.99.155415} {\bibfield  {journal} {\bibinfo
  {journal} {Phys. Rev. B}\ }\textbf {\bibinfo {volume} {99}},\ \bibinfo
  {pages} {155415} (\bibinfo {year} {2019}{\natexlab{b}})}\BibitemShut
  {NoStop}%
\bibitem [{\citenamefont {Huang}\ \emph {et~al.}(2019)\citenamefont {Huang},
  \citenamefont {Zhang},\ and\ \citenamefont {Ma}}]{HUA19}%
  \BibitemOpen
  \bibfield  {author} {\bibinfo {author} {\bibfnamefont {T.}~\bibnamefont
  {Huang}}, \bibinfo {author} {\bibfnamefont {L.}~\bibnamefont {Zhang}},\ and\
  \bibinfo {author} {\bibfnamefont {T.}~\bibnamefont {Ma}},\ }\href
  {https://doi.org/10.1016/j.scib.2019.01.026} {\bibfield  {journal} {\bibinfo
  {journal} {Sci. Bull.}\ }\textbf {\bibinfo {volume} {64}},\ \bibinfo {pages}
  {310} (\bibinfo {year} {2019})}\BibitemShut {NoStop}%
\bibitem [{\citenamefont {Wu}\ \emph {et~al.}(2019{\natexlab{a}})\citenamefont
  {Wu}, \citenamefont {Jian},\ and\ \citenamefont {Xu}}]{WU19}%
  \BibitemOpen
  \bibfield  {author} {\bibinfo {author} {\bibfnamefont {X.-C.}\ \bibnamefont
  {Wu}}, \bibinfo {author} {\bibfnamefont {C.-M.}\ \bibnamefont {Jian}},\ and\
  \bibinfo {author} {\bibfnamefont {C.}~\bibnamefont {Xu}},\ }\href
  {https://doi.org/10.1103/PhysRevB.99.161405} {\bibfield  {journal} {\bibinfo
  {journal} {Phys. Rev. B}\ }\textbf {\bibinfo {volume} {99}},\ \bibinfo
  {pages} {161405} (\bibinfo {year} {2019}{\natexlab{a}})}\BibitemShut
  {NoStop}%
\bibitem [{\citenamefont {Classen}\ \emph {et~al.}(2019)\citenamefont
  {Classen}, \citenamefont {Honerkamp},\ and\ \citenamefont {Scherer}}]{CLA19}%
  \BibitemOpen
  \bibfield  {author} {\bibinfo {author} {\bibfnamefont {L.}~\bibnamefont
  {Classen}}, \bibinfo {author} {\bibfnamefont {C.}~\bibnamefont {Honerkamp}},\
  and\ \bibinfo {author} {\bibfnamefont {M.~M.}\ \bibnamefont {Scherer}},\
  }\href {https://doi.org/10.1103/PhysRevB.99.195120} {\bibfield  {journal}
  {\bibinfo  {journal} {Phys. Rev. B}\ }\textbf {\bibinfo {volume} {99}},\
  \bibinfo {pages} {195120} (\bibinfo {year} {2019})}\BibitemShut {NoStop}%
\bibitem [{\citenamefont {Kang}\ and\ \citenamefont {Vafek}(2019)}]{KAN19}%
  \BibitemOpen
  \bibfield  {author} {\bibinfo {author} {\bibfnamefont {J.}~\bibnamefont
  {Kang}}\ and\ \bibinfo {author} {\bibfnamefont {O.}~\bibnamefont {Vafek}},\
  }\href {https://doi.org/10.1103/PhysRevLett.122.246401} {\bibfield  {journal}
  {\bibinfo  {journal} {Phys. Rev. Lett.}\ }\textbf {\bibinfo {volume} {122}},\
  \bibinfo {pages} {246401} (\bibinfo {year} {2019})}\BibitemShut {NoStop}%
\bibitem [{\citenamefont {Seo}\ \emph {et~al.}(2019)\citenamefont {Seo},
  \citenamefont {Kotov},\ and\ \citenamefont {Uchoa}}]{SEO19}%
  \BibitemOpen
  \bibfield  {author} {\bibinfo {author} {\bibfnamefont {K.}~\bibnamefont
  {Seo}}, \bibinfo {author} {\bibfnamefont {V.~N.}\ \bibnamefont {Kotov}},\
  and\ \bibinfo {author} {\bibfnamefont {B.}~\bibnamefont {Uchoa}},\ }\href
  {https://doi.org/10.1103/PhysRevLett.122.246402} {\bibfield  {journal}
  {\bibinfo  {journal} {Phys. Rev. Lett.}\ }\textbf {\bibinfo {volume} {122}},\
  \bibinfo {pages} {246402} (\bibinfo {year} {2019})}\BibitemShut {NoStop}%
\bibitem [{\citenamefont {Da~Liao}\ \emph {et~al.}(2019)\citenamefont
  {Da~Liao}, \citenamefont {Meng},\ and\ \citenamefont {Xu}}]{DA19}%
  \BibitemOpen
  \bibfield  {author} {\bibinfo {author} {\bibfnamefont {Y.}~\bibnamefont
  {Da~Liao}}, \bibinfo {author} {\bibfnamefont {Z.~Y.}\ \bibnamefont {Meng}},\
  and\ \bibinfo {author} {\bibfnamefont {X.~Y.}\ \bibnamefont {Xu}},\ }\href
  {https://doi.org/10.1103/PhysRevLett.123.157601} {\bibfield  {journal}
  {\bibinfo  {journal} {Phys. Rev. Lett.}\ }\textbf {\bibinfo {volume} {123}},\
  \bibinfo {pages} {157601} (\bibinfo {year} {2019})}\BibitemShut {NoStop}%
\bibitem [{\citenamefont {Angeli}\ \emph {et~al.}(2019)\citenamefont {Angeli},
  \citenamefont {Tosatti},\ and\ \citenamefont {Fabrizio}}]{ANG19}%
  \BibitemOpen
  \bibfield  {author} {\bibinfo {author} {\bibfnamefont {M.}~\bibnamefont
  {Angeli}}, \bibinfo {author} {\bibfnamefont {E.}~\bibnamefont {Tosatti}},\
  and\ \bibinfo {author} {\bibfnamefont {M.}~\bibnamefont {Fabrizio}},\ }\href
  {https://doi.org/10.1103/PhysRevX.9.041010} {\bibfield  {journal} {\bibinfo
  {journal} {Phys. Rev. X}\ }\textbf {\bibinfo {volume} {9}},\ \bibinfo {pages}
  {041010} (\bibinfo {year} {2019})}\BibitemShut {NoStop}%
\bibitem [{\citenamefont {Xie}\ and\ \citenamefont {MacDonald}(2020)}]{XIE20b}%
  \BibitemOpen
  \bibfield  {author} {\bibinfo {author} {\bibfnamefont {M.}~\bibnamefont
  {Xie}}\ and\ \bibinfo {author} {\bibfnamefont {A.~H.}\ \bibnamefont
  {MacDonald}},\ }\href {https://doi.org/10.1103/PhysRevLett.124.097601}
  {\bibfield  {journal} {\bibinfo  {journal} {Phys. Rev. Lett.}\ }\textbf
  {\bibinfo {volume} {124}},\ \bibinfo {pages} {097601} (\bibinfo {year}
  {2020})}\BibitemShut {NoStop}%
\bibitem [{\citenamefont {Bultinck}\ \emph
  {et~al.}(2020{\natexlab{a}})\citenamefont {Bultinck}, \citenamefont
  {Chatterjee},\ and\ \citenamefont {Zaletel}}]{BUL20b}%
  \BibitemOpen
  \bibfield  {author} {\bibinfo {author} {\bibfnamefont {N.}~\bibnamefont
  {Bultinck}}, \bibinfo {author} {\bibfnamefont {S.}~\bibnamefont
  {Chatterjee}},\ and\ \bibinfo {author} {\bibfnamefont {M.~P.}\ \bibnamefont
  {Zaletel}},\ }\href {https://doi.org/10.1103/PhysRevLett.124.166601}
  {\bibfield  {journal} {\bibinfo  {journal} {Phys. Rev. Lett.}\ }\textbf
  {\bibinfo {volume} {124}},\ \bibinfo {pages} {166601} (\bibinfo {year}
  {2020}{\natexlab{a}})}\BibitemShut {NoStop}%
\bibitem [{\citenamefont {Chatterjee}\ \emph {et~al.}(2020)\citenamefont
  {Chatterjee}, \citenamefont {Bultinck},\ and\ \citenamefont
  {Zaletel}}]{CHA20}%
  \BibitemOpen
  \bibfield  {author} {\bibinfo {author} {\bibfnamefont {S.}~\bibnamefont
  {Chatterjee}}, \bibinfo {author} {\bibfnamefont {N.}~\bibnamefont
  {Bultinck}},\ and\ \bibinfo {author} {\bibfnamefont {M.~P.}\ \bibnamefont
  {Zaletel}},\ }\href {https://doi.org/10.1103/PhysRevB.101.165141} {\bibfield
  {journal} {\bibinfo  {journal} {Phys. Rev. B}\ }\textbf {\bibinfo {volume}
  {101}},\ \bibinfo {pages} {165141} (\bibinfo {year} {2020})}\BibitemShut
  {NoStop}%
\bibitem [{\citenamefont {Repellin}\ \emph {et~al.}(2020)\citenamefont
  {Repellin}, \citenamefont {Dong}, \citenamefont {Zhang},\ and\ \citenamefont
  {Senthil}}]{REP20}%
  \BibitemOpen
  \bibfield  {author} {\bibinfo {author} {\bibfnamefont {C.}~\bibnamefont
  {Repellin}}, \bibinfo {author} {\bibfnamefont {Z.}~\bibnamefont {Dong}},
  \bibinfo {author} {\bibfnamefont {Y.-H.}\ \bibnamefont {Zhang}},\ and\
  \bibinfo {author} {\bibfnamefont {T.}~\bibnamefont {Senthil}},\ }\href
  {https://doi.org/10.1103/PhysRevLett.124.187601} {\bibfield  {journal}
  {\bibinfo  {journal} {Phys. Rev. Lett.}\ }\textbf {\bibinfo {volume} {124}},\
  \bibinfo {pages} {187601} (\bibinfo {year} {2020})}\BibitemShut {NoStop}%
\bibitem [{\citenamefont {Cea}\ and\ \citenamefont {Guinea}(2020)}]{CEA20}%
  \BibitemOpen
  \bibfield  {author} {\bibinfo {author} {\bibfnamefont {T.}~\bibnamefont
  {Cea}}\ and\ \bibinfo {author} {\bibfnamefont {F.}~\bibnamefont {Guinea}},\
  }\href {https://doi.org/10.1103/PhysRevB.102.045107} {\bibfield  {journal}
  {\bibinfo  {journal} {Phys. Rev. B}\ }\textbf {\bibinfo {volume} {102}},\
  \bibinfo {pages} {045107} (\bibinfo {year} {2020})}\BibitemShut {NoStop}%
\bibitem [{\citenamefont {Zhang}\ \emph {et~al.}(2020)\citenamefont {Zhang},
  \citenamefont {Jiang}, \citenamefont {Wang},\ and\ \citenamefont
  {Zhang}}]{ZHA20}%
  \BibitemOpen
  \bibfield  {author} {\bibinfo {author} {\bibfnamefont {Y.}~\bibnamefont
  {Zhang}}, \bibinfo {author} {\bibfnamefont {K.}~\bibnamefont {Jiang}},
  \bibinfo {author} {\bibfnamefont {Z.}~\bibnamefont {Wang}},\ and\ \bibinfo
  {author} {\bibfnamefont {F.}~\bibnamefont {Zhang}},\ }\href
  {https://doi.org/10.1103/PhysRevB.102.035136} {\bibfield  {journal} {\bibinfo
   {journal} {Phys. Rev. B}\ }\textbf {\bibinfo {volume} {102}},\ \bibinfo
  {pages} {035136} (\bibinfo {year} {2020})}\BibitemShut {NoStop}%
\bibitem [{\citenamefont {Kang}\ and\ \citenamefont {Vafek}(2020)}]{KAN20a}%
  \BibitemOpen
  \bibfield  {author} {\bibinfo {author} {\bibfnamefont {J.}~\bibnamefont
  {Kang}}\ and\ \bibinfo {author} {\bibfnamefont {O.}~\bibnamefont {Vafek}},\
  }\href {https://doi.org/10.1103/PhysRevB.102.035161} {\bibfield  {journal}
  {\bibinfo  {journal} {Phys. Rev. B}\ }\textbf {\bibinfo {volume} {102}},\
  \bibinfo {pages} {035161} (\bibinfo {year} {2020})}\BibitemShut {NoStop}%
\bibitem [{\citenamefont {Bultinck}\ \emph
  {et~al.}(2020{\natexlab{b}})\citenamefont {Bultinck}, \citenamefont {Khalaf},
  \citenamefont {Liu}, \citenamefont {Chatterjee}, \citenamefont {Vishwanath},\
  and\ \citenamefont {Zaletel}}]{BUL20a}%
  \BibitemOpen
  \bibfield  {author} {\bibinfo {author} {\bibfnamefont {N.}~\bibnamefont
  {Bultinck}}, \bibinfo {author} {\bibfnamefont {E.}~\bibnamefont {Khalaf}},
  \bibinfo {author} {\bibfnamefont {S.}~\bibnamefont {Liu}}, \bibinfo {author}
  {\bibfnamefont {S.}~\bibnamefont {Chatterjee}}, \bibinfo {author}
  {\bibfnamefont {A.}~\bibnamefont {Vishwanath}},\ and\ \bibinfo {author}
  {\bibfnamefont {M.~P.}\ \bibnamefont {Zaletel}},\ }\href
  {https://doi.org/10.1103/PhysRevX.10.031034} {\bibfield  {journal} {\bibinfo
  {journal} {Phys. Rev. X}\ }\textbf {\bibinfo {volume} {10}},\ \bibinfo
  {pages} {031034} (\bibinfo {year} {2020}{\natexlab{b}})}\BibitemShut
  {NoStop}%
\bibitem [{\citenamefont {Chichinadze}\ \emph
  {et~al.}(2020{\natexlab{a}})\citenamefont {Chichinadze}, \citenamefont
  {Classen},\ and\ \citenamefont {Chubukov}}]{CHI20b}%
  \BibitemOpen
  \bibfield  {author} {\bibinfo {author} {\bibfnamefont {D.~V.}\ \bibnamefont
  {Chichinadze}}, \bibinfo {author} {\bibfnamefont {L.}~\bibnamefont
  {Classen}},\ and\ \bibinfo {author} {\bibfnamefont {A.~V.}\ \bibnamefont
  {Chubukov}},\ }\href {https://doi.org/10.1103/PhysRevB.102.125120} {\bibfield
   {journal} {\bibinfo  {journal} {Phys. Rev. B}\ }\textbf {\bibinfo {volume}
  {102}},\ \bibinfo {pages} {125120} (\bibinfo {year}
  {2020}{\natexlab{a}})}\BibitemShut {NoStop}%
\bibitem [{\citenamefont {Soejima}\ \emph {et~al.}(2020)\citenamefont
  {Soejima}, \citenamefont {Parker}, \citenamefont {Bultinck}, \citenamefont
  {Hauschild},\ and\ \citenamefont {Zaletel}}]{SOE20}%
  \BibitemOpen
  \bibfield  {author} {\bibinfo {author} {\bibfnamefont {T.}~\bibnamefont
  {Soejima}}, \bibinfo {author} {\bibfnamefont {D.~E.}\ \bibnamefont {Parker}},
  \bibinfo {author} {\bibfnamefont {N.}~\bibnamefont {Bultinck}}, \bibinfo
  {author} {\bibfnamefont {J.}~\bibnamefont {Hauschild}},\ and\ \bibinfo
  {author} {\bibfnamefont {M.~P.}\ \bibnamefont {Zaletel}},\ }\href
  {https://doi.org/10.1103/PhysRevB.102.205111} {\bibfield  {journal} {\bibinfo
   {journal} {Phys. Rev. B}\ }\textbf {\bibinfo {volume} {102}},\ \bibinfo
  {pages} {205111} (\bibinfo {year} {2020})}\BibitemShut {NoStop}%
\bibitem [{\citenamefont {Christos}\ \emph {et~al.}(2020)\citenamefont
  {Christos}, \citenamefont {Sachdev},\ and\ \citenamefont {Scheurer}}]{CHR20}%
  \BibitemOpen
  \bibfield  {author} {\bibinfo {author} {\bibfnamefont {M.}~\bibnamefont
  {Christos}}, \bibinfo {author} {\bibfnamefont {S.}~\bibnamefont {Sachdev}},\
  and\ \bibinfo {author} {\bibfnamefont {M.~S.}\ \bibnamefont {Scheurer}},\
  }\href {https://doi.org/10.1073/pnas.2014691117} {\bibfield  {journal}
  {\bibinfo  {journal} {PNAS}\ }\textbf {\bibinfo {volume} {117}},\ \bibinfo
  {pages} {29543} (\bibinfo {year} {2020})}\BibitemShut {NoStop}%
\bibitem [{\citenamefont {Eugenio}\ and\ \citenamefont {Dag}(2020)}]{EUG20}%
  \BibitemOpen
  \bibfield  {author} {\bibinfo {author} {\bibfnamefont {P.}~\bibnamefont
  {Eugenio}}\ and\ \bibinfo {author} {\bibfnamefont {C.}~\bibnamefont {Dag}},\
  }\href {https://doi.org/10.21468/SciPostPhysCore.3.2.015} {\bibfield
  {journal} {\bibinfo  {journal} {SciPost Physics Core}\ }\textbf {\bibinfo
  {volume} {3}},\ \bibinfo {pages} {015} (\bibinfo {year} {2020})}\BibitemShut
  {NoStop}%
\bibitem [{\citenamefont {Wu}\ and\ \citenamefont {Das~Sarma}(2020)}]{WU20}%
  \BibitemOpen
  \bibfield  {author} {\bibinfo {author} {\bibfnamefont {F.}~\bibnamefont
  {Wu}}\ and\ \bibinfo {author} {\bibfnamefont {S.}~\bibnamefont {Das~Sarma}},\
  }\href {https://doi.org/10.1103/PhysRevLett.124.046403} {\bibfield  {journal}
  {\bibinfo  {journal} {Phys. Rev. Lett.}\ }\textbf {\bibinfo {volume} {124}},\
  \bibinfo {pages} {046403} (\bibinfo {year} {2020})}\BibitemShut {NoStop}%
\bibitem [{\citenamefont {Vafek}\ and\ \citenamefont {Kang}(2020)}]{VAF20}%
  \BibitemOpen
  \bibfield  {author} {\bibinfo {author} {\bibfnamefont {O.}~\bibnamefont
  {Vafek}}\ and\ \bibinfo {author} {\bibfnamefont {J.}~\bibnamefont {Kang}},\
  }\href {https://doi.org/10.1103/PhysRevLett.125.257602} {\bibfield  {journal}
  {\bibinfo  {journal} {Phys. Rev. Lett.}\ }\textbf {\bibinfo {volume} {125}},\
  \bibinfo {pages} {257602} (\bibinfo {year} {2020})}\BibitemShut {NoStop}%
\bibitem [{\citenamefont {Xie}\ \emph {et~al.}(2021{\natexlab{b}})\citenamefont
  {Xie}, \citenamefont {Cowsik}, \citenamefont {Song}, \citenamefont {Lian},
  \citenamefont {Bernevig},\ and\ \citenamefont {Regnault}}]{XIE21}%
  \BibitemOpen
  \bibfield  {author} {\bibinfo {author} {\bibfnamefont {F.}~\bibnamefont
  {Xie}}, \bibinfo {author} {\bibfnamefont {A.}~\bibnamefont {Cowsik}},
  \bibinfo {author} {\bibfnamefont {Z.-D.}\ \bibnamefont {Song}}, \bibinfo
  {author} {\bibfnamefont {B.}~\bibnamefont {Lian}}, \bibinfo {author}
  {\bibfnamefont {B.~A.}\ \bibnamefont {Bernevig}},\ and\ \bibinfo {author}
  {\bibfnamefont {N.}~\bibnamefont {Regnault}},\ }\href
  {https://doi.org/10.1103/PhysRevB.103.205416} {\bibfield  {journal} {\bibinfo
   {journal} {Phys. Rev. B}\ }\textbf {\bibinfo {volume} {103}},\ \bibinfo
  {pages} {205416} (\bibinfo {year} {2021}{\natexlab{b}})}\BibitemShut
  {NoStop}%
\bibitem [{\citenamefont {Kang}\ \emph {et~al.}(2021)\citenamefont {Kang},
  \citenamefont {Bernevig},\ and\ \citenamefont {Vafek}}]{KAN21}%
  \BibitemOpen
  \bibfield  {author} {\bibinfo {author} {\bibfnamefont {J.}~\bibnamefont
  {Kang}}, \bibinfo {author} {\bibfnamefont {B.~A.}\ \bibnamefont {Bernevig}},\
  and\ \bibinfo {author} {\bibfnamefont {O.}~\bibnamefont {Vafek}},\ }\href
  {https://doi.org/10.1103/PhysRevLett.127.266402} {\bibfield  {journal}
  {\bibinfo  {journal} {Phys. Rev. Lett.}\ }\textbf {\bibinfo {volume} {127}},\
  \bibinfo {pages} {266402} (\bibinfo {year} {2021})}\BibitemShut {NoStop}%
\bibitem [{\citenamefont {Liu}\ \emph {et~al.}(2021{\natexlab{b}})\citenamefont
  {Liu}, \citenamefont {Khalaf}, \citenamefont {Lee},\ and\ \citenamefont
  {Vishwanath}}]{LIU21}%
  \BibitemOpen
  \bibfield  {author} {\bibinfo {author} {\bibfnamefont {S.}~\bibnamefont
  {Liu}}, \bibinfo {author} {\bibfnamefont {E.}~\bibnamefont {Khalaf}},
  \bibinfo {author} {\bibfnamefont {J.~Y.}\ \bibnamefont {Lee}},\ and\ \bibinfo
  {author} {\bibfnamefont {A.}~\bibnamefont {Vishwanath}},\ }\href
  {https://doi.org/10.1103/PhysRevResearch.3.013033} {\bibfield  {journal}
  {\bibinfo  {journal} {Phys. Rev. Research}\ }\textbf {\bibinfo {volume}
  {3}},\ \bibinfo {pages} {013033} (\bibinfo {year}
  {2021}{\natexlab{b}})}\BibitemShut {NoStop}%
\bibitem [{\citenamefont {Da~Liao}\ \emph {et~al.}(2021)\citenamefont
  {Da~Liao}, \citenamefont {Kang}, \citenamefont {Brei{\o}}, \citenamefont
  {Xu}, \citenamefont {Wu}, \citenamefont {Andersen}, \citenamefont
  {Fernandes},\ and\ \citenamefont {Meng}}]{DA21}%
  \BibitemOpen
  \bibfield  {author} {\bibinfo {author} {\bibfnamefont {Y.}~\bibnamefont
  {Da~Liao}}, \bibinfo {author} {\bibfnamefont {J.}~\bibnamefont {Kang}},
  \bibinfo {author} {\bibfnamefont {C.~N.}\ \bibnamefont {Brei{\o}}}, \bibinfo
  {author} {\bibfnamefont {X.~Y.}\ \bibnamefont {Xu}}, \bibinfo {author}
  {\bibfnamefont {H.-Q.}\ \bibnamefont {Wu}}, \bibinfo {author} {\bibfnamefont
  {B.~M.}\ \bibnamefont {Andersen}}, \bibinfo {author} {\bibfnamefont {R.~M.}\
  \bibnamefont {Fernandes}},\ and\ \bibinfo {author} {\bibfnamefont {Z.~Y.}\
  \bibnamefont {Meng}},\ }\href {https://doi.org/10.1103/PhysRevX.11.011014}
  {\bibfield  {journal} {\bibinfo  {journal} {Phys. Rev. X}\ }\textbf {\bibinfo
  {volume} {11}},\ \bibinfo {pages} {011014} (\bibinfo {year}
  {2021})}\BibitemShut {NoStop}%
\bibitem [{\citenamefont {Liu}\ and\ \citenamefont {Dai}(2021)}]{LIU21a}%
  \BibitemOpen
  \bibfield  {author} {\bibinfo {author} {\bibfnamefont {J.}~\bibnamefont
  {Liu}}\ and\ \bibinfo {author} {\bibfnamefont {X.}~\bibnamefont {Dai}},\
  }\href {https://doi.org/10.1103/PhysRevB.103.035427} {\bibfield  {journal}
  {\bibinfo  {journal} {Phys. Rev. B}\ }\textbf {\bibinfo {volume} {103}},\
  \bibinfo {pages} {035427} (\bibinfo {year} {2021})}\BibitemShut {NoStop}%
\bibitem [{\citenamefont {Thomson}\ and\ \citenamefont {Alicea}(2021)}]{THO21}%
  \BibitemOpen
  \bibfield  {author} {\bibinfo {author} {\bibfnamefont {A.}~\bibnamefont
  {Thomson}}\ and\ \bibinfo {author} {\bibfnamefont {J.}~\bibnamefont
  {Alicea}},\ }\href {https://doi.org/10.1103/PhysRevB.103.125138} {\bibfield
  {journal} {\bibinfo  {journal} {Phys. Rev. B}\ }\textbf {\bibinfo {volume}
  {103}},\ \bibinfo {pages} {125138} (\bibinfo {year} {2021})}\BibitemShut
  {NoStop}%
\bibitem [{\citenamefont {Kwan}\ \emph
  {et~al.}(2021{\natexlab{a}})\citenamefont {Kwan}, \citenamefont {Hu},
  \citenamefont {Simon},\ and\ \citenamefont {Parameswaran}}]{KWA21a}%
  \BibitemOpen
  \bibfield  {author} {\bibinfo {author} {\bibfnamefont {Y.~H.}\ \bibnamefont
  {Kwan}}, \bibinfo {author} {\bibfnamefont {Y.}~\bibnamefont {Hu}}, \bibinfo
  {author} {\bibfnamefont {S.~H.}\ \bibnamefont {Simon}},\ and\ \bibinfo
  {author} {\bibfnamefont {S.~A.}\ \bibnamefont {Parameswaran}},\ }\href
  {https://doi.org/10.1103/PhysRevLett.126.137601} {\bibfield  {journal}
  {\bibinfo  {journal} {Phys. Rev. Lett.}\ }\textbf {\bibinfo {volume} {126}},\
  \bibinfo {pages} {137601} (\bibinfo {year} {2021}{\natexlab{a}})}\BibitemShut
  {NoStop}%
\bibitem [{\citenamefont {Lian}\ \emph {et~al.}(2021)\citenamefont {Lian},
  \citenamefont {Song}, \citenamefont {Regnault}, \citenamefont {Efetov},
  \citenamefont {Yazdani},\ and\ \citenamefont {Bernevig}}]{LIA21}%
  \BibitemOpen
  \bibfield  {author} {\bibinfo {author} {\bibfnamefont {B.}~\bibnamefont
  {Lian}}, \bibinfo {author} {\bibfnamefont {Z.-D.}\ \bibnamefont {Song}},
  \bibinfo {author} {\bibfnamefont {N.}~\bibnamefont {Regnault}}, \bibinfo
  {author} {\bibfnamefont {D.~K.}\ \bibnamefont {Efetov}}, \bibinfo {author}
  {\bibfnamefont {A.}~\bibnamefont {Yazdani}},\ and\ \bibinfo {author}
  {\bibfnamefont {B.~A.}\ \bibnamefont {Bernevig}},\ }\href
  {https://doi.org/10.1103/PhysRevB.103.205414} {\bibfield  {journal} {\bibinfo
   {journal} {Phys. Rev. B}\ }\textbf {\bibinfo {volume} {103}},\ \bibinfo
  {pages} {205414} (\bibinfo {year} {2021})}\BibitemShut {NoStop}%
\bibitem [{\citenamefont {Zhang}\ \emph {et~al.}(2021)\citenamefont {Zhang},
  \citenamefont {Pan}, \citenamefont {Zhang}, \citenamefont {Kang},\ and\
  \citenamefont {Meng}}]{ZHA21}%
  \BibitemOpen
  \bibfield  {author} {\bibinfo {author} {\bibfnamefont {X.}~\bibnamefont
  {Zhang}}, \bibinfo {author} {\bibfnamefont {G.}~\bibnamefont {Pan}}, \bibinfo
  {author} {\bibfnamefont {Y.}~\bibnamefont {Zhang}}, \bibinfo {author}
  {\bibfnamefont {J.}~\bibnamefont {Kang}},\ and\ \bibinfo {author}
  {\bibfnamefont {Z.~Y.}\ \bibnamefont {Meng}},\ }\href
  {https://doi.org/10.1088/0256-307X/38/7/077305} {\bibfield  {journal}
  {\bibinfo  {journal} {Chinese Phys. Lett.}\ }\textbf {\bibinfo {volume}
  {38}},\ \bibinfo {pages} {077305} (\bibinfo {year} {2021})}\BibitemShut
  {NoStop}%
\bibitem [{\citenamefont {Parker}\ \emph {et~al.}(2021)\citenamefont {Parker},
  \citenamefont {Soejima}, \citenamefont {Hauschild}, \citenamefont {Zaletel},\
  and\ \citenamefont {Bultinck}}]{PAR21a}%
  \BibitemOpen
  \bibfield  {author} {\bibinfo {author} {\bibfnamefont {D.~E.}\ \bibnamefont
  {Parker}}, \bibinfo {author} {\bibfnamefont {T.}~\bibnamefont {Soejima}},
  \bibinfo {author} {\bibfnamefont {J.}~\bibnamefont {Hauschild}}, \bibinfo
  {author} {\bibfnamefont {M.~P.}\ \bibnamefont {Zaletel}},\ and\ \bibinfo
  {author} {\bibfnamefont {N.}~\bibnamefont {Bultinck}},\ }\href
  {https://doi.org/10.1103/PhysRevLett.127.027601} {\bibfield  {journal}
  {\bibinfo  {journal} {Phys. Rev. Lett.}\ }\textbf {\bibinfo {volume} {127}},\
  \bibinfo {pages} {027601} (\bibinfo {year} {2021})}\BibitemShut {NoStop}%
\bibitem [{\citenamefont {Vafek}\ and\ \citenamefont {Kang}(2021)}]{VAF21}%
  \BibitemOpen
  \bibfield  {author} {\bibinfo {author} {\bibfnamefont {O.}~\bibnamefont
  {Vafek}}\ and\ \bibinfo {author} {\bibfnamefont {J.}~\bibnamefont {Kang}},\
  }\href {https://doi.org/10.1103/PhysRevB.104.075143} {\bibfield  {journal}
  {\bibinfo  {journal} {Phys. Rev. B}\ }\textbf {\bibinfo {volume} {104}},\
  \bibinfo {pages} {075143} (\bibinfo {year} {2021})}\BibitemShut {NoStop}%
\bibitem [{\citenamefont {Kwan}\ \emph
  {et~al.}(2021{\natexlab{b}})\citenamefont {Kwan}, \citenamefont {Wagner},
  \citenamefont {Chakraborty}, \citenamefont {Simon},\ and\ \citenamefont
  {Parameswaran}}]{KWA21b}%
  \BibitemOpen
  \bibfield  {author} {\bibinfo {author} {\bibfnamefont {Y.~H.}\ \bibnamefont
  {Kwan}}, \bibinfo {author} {\bibfnamefont {G.}~\bibnamefont {Wagner}},
  \bibinfo {author} {\bibfnamefont {N.}~\bibnamefont {Chakraborty}}, \bibinfo
  {author} {\bibfnamefont {S.~H.}\ \bibnamefont {Simon}},\ and\ \bibinfo
  {author} {\bibfnamefont {S.~A.}\ \bibnamefont {Parameswaran}},\ }\href
  {https://doi.org/10.1103/PhysRevB.104.115404} {\bibfield  {journal} {\bibinfo
   {journal} {Phys. Rev. B}\ }\textbf {\bibinfo {volume} {104}},\ \bibinfo
  {pages} {115404} (\bibinfo {year} {2021}{\natexlab{b}})}\BibitemShut
  {NoStop}%
\bibitem [{\citenamefont {Chen}\ \emph {et~al.}(2021)\citenamefont {Chen},
  \citenamefont {Liao}, \citenamefont {Chen}, \citenamefont {Vafek},
  \citenamefont {Kang}, \citenamefont {Li},\ and\ \citenamefont
  {Meng}}]{CHE21}%
  \BibitemOpen
  \bibfield  {author} {\bibinfo {author} {\bibfnamefont {B.-B.}\ \bibnamefont
  {Chen}}, \bibinfo {author} {\bibfnamefont {Y.~D.}\ \bibnamefont {Liao}},
  \bibinfo {author} {\bibfnamefont {Z.}~\bibnamefont {Chen}}, \bibinfo {author}
  {\bibfnamefont {O.}~\bibnamefont {Vafek}}, \bibinfo {author} {\bibfnamefont
  {J.}~\bibnamefont {Kang}}, \bibinfo {author} {\bibfnamefont {W.}~\bibnamefont
  {Li}},\ and\ \bibinfo {author} {\bibfnamefont {Z.~Y.}\ \bibnamefont {Meng}},\
  }\href {https://doi.org/10.1038/s41467-021-25438-1} {\bibfield  {journal}
  {\bibinfo  {journal} {Nat. Commun.}\ }\textbf {\bibinfo {volume} {12}},\
  \bibinfo {pages} {5480} (\bibinfo {year} {2021})}\BibitemShut {NoStop}%
\bibitem [{\citenamefont {Potasz}\ \emph {et~al.}(2021)\citenamefont {Potasz},
  \citenamefont {Xie},\ and\ \citenamefont {MacDonald}}]{POT21}%
  \BibitemOpen
  \bibfield  {author} {\bibinfo {author} {\bibfnamefont {P.}~\bibnamefont
  {Potasz}}, \bibinfo {author} {\bibfnamefont {M.}~\bibnamefont {Xie}},\ and\
  \bibinfo {author} {\bibfnamefont {A.~H.}\ \bibnamefont {MacDonald}},\ }\href
  {https://doi.org/10.1103/PhysRevLett.127.147203} {\bibfield  {journal}
  {\bibinfo  {journal} {Phys. Rev. Lett.}\ }\textbf {\bibinfo {volume} {127}},\
  \bibinfo {pages} {147203} (\bibinfo {year} {2021})}\BibitemShut {NoStop}%
\bibitem [{\citenamefont {Xie}\ \emph {et~al.}(2021{\natexlab{c}})\citenamefont
  {Xie}, \citenamefont {Regnault}, \citenamefont {C{\u a}lug{\u a}ru},
  \citenamefont {Bernevig},\ and\ \citenamefont {Lian}}]{XIE21b}%
  \BibitemOpen
  \bibfield  {author} {\bibinfo {author} {\bibfnamefont {F.}~\bibnamefont
  {Xie}}, \bibinfo {author} {\bibfnamefont {N.}~\bibnamefont {Regnault}},
  \bibinfo {author} {\bibfnamefont {D.}~\bibnamefont {C{\u a}lug{\u a}ru}},
  \bibinfo {author} {\bibfnamefont {B.~A.}\ \bibnamefont {Bernevig}},\ and\
  \bibinfo {author} {\bibfnamefont {B.}~\bibnamefont {Lian}},\ }\href
  {https://doi.org/10.1103/PhysRevB.104.115167} {\bibfield  {journal} {\bibinfo
   {journal} {Phys. Rev. B}\ }\textbf {\bibinfo {volume} {104}},\ \bibinfo
  {pages} {115167} (\bibinfo {year} {2021}{\natexlab{c}})}\BibitemShut
  {NoStop}%
\bibitem [{\citenamefont {Xie}\ and\ \citenamefont {MacDonald}(2021)}]{XIE21a}%
  \BibitemOpen
  \bibfield  {author} {\bibinfo {author} {\bibfnamefont {M.}~\bibnamefont
  {Xie}}\ and\ \bibinfo {author} {\bibfnamefont {A.~H.}\ \bibnamefont
  {MacDonald}},\ }\href {https://doi.org/10.1103/PhysRevLett.127.196401}
  {\bibfield  {journal} {\bibinfo  {journal} {Phys. Rev. Lett.}\ }\textbf
  {\bibinfo {volume} {127}},\ \bibinfo {pages} {196401} (\bibinfo {year}
  {2021})}\BibitemShut {NoStop}%
\bibitem [{\citenamefont {Cha}\ \emph {et~al.}(2021)\citenamefont {Cha},
  \citenamefont {Patel},\ and\ \citenamefont {Kim}}]{CHA21}%
  \BibitemOpen
  \bibfield  {author} {\bibinfo {author} {\bibfnamefont {P.}~\bibnamefont
  {Cha}}, \bibinfo {author} {\bibfnamefont {A.~A.}\ \bibnamefont {Patel}},\
  and\ \bibinfo {author} {\bibfnamefont {E.-A.}\ \bibnamefont {Kim}},\ }\href
  {https://doi.org/10.1103/PhysRevLett.127.266601} {\bibfield  {journal}
  {\bibinfo  {journal} {Phys. Rev. Lett.}\ }\textbf {\bibinfo {volume} {127}},\
  \bibinfo {pages} {266601} (\bibinfo {year} {2021})}\BibitemShut {NoStop}%
\bibitem [{\citenamefont {Kwan}\ \emph
  {et~al.}(2021{\natexlab{c}})\citenamefont {Kwan}, \citenamefont {Wagner},
  \citenamefont {Soejima}, \citenamefont {Zaletel}, \citenamefont {Simon},
  \citenamefont {Parameswaran},\ and\ \citenamefont {Bultinck}}]{KWA21}%
  \BibitemOpen
  \bibfield  {author} {\bibinfo {author} {\bibfnamefont {Y.~H.}\ \bibnamefont
  {Kwan}}, \bibinfo {author} {\bibfnamefont {G.}~\bibnamefont {Wagner}},
  \bibinfo {author} {\bibfnamefont {T.}~\bibnamefont {Soejima}}, \bibinfo
  {author} {\bibfnamefont {M.~P.}\ \bibnamefont {Zaletel}}, \bibinfo {author}
  {\bibfnamefont {S.~H.}\ \bibnamefont {Simon}}, \bibinfo {author}
  {\bibfnamefont {S.~A.}\ \bibnamefont {Parameswaran}},\ and\ \bibinfo {author}
  {\bibfnamefont {N.}~\bibnamefont {Bultinck}},\ }\href
  {https://doi.org/10.1103/PhysRevX.11.041063} {\bibfield  {journal} {\bibinfo
  {journal} {Phys. Rev. X}\ }\textbf {\bibinfo {volume} {11}},\ \bibinfo
  {pages} {041063} (\bibinfo {year} {2021}{\natexlab{c}})}\BibitemShut
  {NoStop}%
\bibitem [{\citenamefont {Hofmann}\ \emph {et~al.}(2022)\citenamefont
  {Hofmann}, \citenamefont {Khalaf}, \citenamefont {Vishwanath}, \citenamefont
  {Berg},\ and\ \citenamefont {Lee}}]{HOF22}%
  \BibitemOpen
  \bibfield  {author} {\bibinfo {author} {\bibfnamefont {J.~S.}\ \bibnamefont
  {Hofmann}}, \bibinfo {author} {\bibfnamefont {E.}~\bibnamefont {Khalaf}},
  \bibinfo {author} {\bibfnamefont {A.}~\bibnamefont {Vishwanath}}, \bibinfo
  {author} {\bibfnamefont {E.}~\bibnamefont {Berg}},\ and\ \bibinfo {author}
  {\bibfnamefont {J.~Y.}\ \bibnamefont {Lee}},\ }\href
  {https://doi.org/10.1103/PhysRevX.12.011061} {\bibfield  {journal} {\bibinfo
  {journal} {Phys. Rev. X}\ }\textbf {\bibinfo {volume} {12}},\ \bibinfo
  {pages} {011061} (\bibinfo {year} {2022})}\BibitemShut {NoStop}%
\bibitem [{\citenamefont {Wagner}\ \emph {et~al.}(2022)\citenamefont {Wagner},
  \citenamefont {Kwan}, \citenamefont {Bultinck}, \citenamefont {Simon},\ and\
  \citenamefont {Parameswaran}}]{WAG22}%
  \BibitemOpen
  \bibfield  {author} {\bibinfo {author} {\bibfnamefont {G.}~\bibnamefont
  {Wagner}}, \bibinfo {author} {\bibfnamefont {Y.~H.}\ \bibnamefont {Kwan}},
  \bibinfo {author} {\bibfnamefont {N.}~\bibnamefont {Bultinck}}, \bibinfo
  {author} {\bibfnamefont {S.~H.}\ \bibnamefont {Simon}},\ and\ \bibinfo
  {author} {\bibfnamefont {S.~A.}\ \bibnamefont {Parameswaran}},\ }\href
  {https://doi.org/10.1103/PhysRevLett.128.156401} {\bibfield  {journal}
  {\bibinfo  {journal} {Phys. Rev. Lett.}\ }\textbf {\bibinfo {volume} {128}},\
  \bibinfo {pages} {156401} (\bibinfo {year} {2022})}\BibitemShut {NoStop}%
\bibitem [{\citenamefont {Christos}\ \emph {et~al.}(2022)\citenamefont
  {Christos}, \citenamefont {Sachdev},\ and\ \citenamefont {Scheurer}}]{CHR22}%
  \BibitemOpen
  \bibfield  {author} {\bibinfo {author} {\bibfnamefont {M.}~\bibnamefont
  {Christos}}, \bibinfo {author} {\bibfnamefont {S.}~\bibnamefont {Sachdev}},\
  and\ \bibinfo {author} {\bibfnamefont {M.~S.}\ \bibnamefont {Scheurer}},\
  }\href {https://doi.org/10.1103/PhysRevX.12.021018} {\bibfield  {journal}
  {\bibinfo  {journal} {Phys. Rev. X}\ }\textbf {\bibinfo {volume} {12}},\
  \bibinfo {pages} {021018} (\bibinfo {year} {2022})}\BibitemShut {NoStop}%
\bibitem [{\citenamefont {Song}\ and\ \citenamefont {Bernevig}(2022)}]{SON22}%
  \BibitemOpen
  \bibfield  {author} {\bibinfo {author} {\bibfnamefont {Z.-D.}\ \bibnamefont
  {Song}}\ and\ \bibinfo {author} {\bibfnamefont {B.~A.}\ \bibnamefont
  {Bernevig}},\ }\href {https://doi.org/10.1103/PhysRevLett.129.047601}
  {\bibfield  {journal} {\bibinfo  {journal} {Phys. Rev. Lett.}\ }\textbf
  {\bibinfo {volume} {129}},\ \bibinfo {pages} {047601} (\bibinfo {year}
  {2022})}\BibitemShut {NoStop}%
\bibitem [{\citenamefont {Brillaux}\ \emph {et~al.}(2022)\citenamefont
  {Brillaux}, \citenamefont {Carpentier}, \citenamefont {Fedorenko},\ and\
  \citenamefont {Savary}}]{BRI22}%
  \BibitemOpen
  \bibfield  {author} {\bibinfo {author} {\bibfnamefont {E.}~\bibnamefont
  {Brillaux}}, \bibinfo {author} {\bibfnamefont {D.}~\bibnamefont
  {Carpentier}}, \bibinfo {author} {\bibfnamefont {A.~A.}\ \bibnamefont
  {Fedorenko}},\ and\ \bibinfo {author} {\bibfnamefont {L.}~\bibnamefont
  {Savary}},\ }\href {https://doi.org/10.1103/PhysRevResearch.4.033168}
  {\bibfield  {journal} {\bibinfo  {journal} {Phys. Rev. Res.}\ }\textbf
  {\bibinfo {volume} {4}},\ \bibinfo {pages} {033168} (\bibinfo {year}
  {2022})}\BibitemShut {NoStop}%
\bibitem [{\citenamefont {Hong}\ \emph {et~al.}(2022)\citenamefont {Hong},
  \citenamefont {Soejima},\ and\ \citenamefont {Zaletel}}]{HON22}%
  \BibitemOpen
  \bibfield  {author} {\bibinfo {author} {\bibfnamefont {J.~P.}\ \bibnamefont
  {Hong}}, \bibinfo {author} {\bibfnamefont {T.}~\bibnamefont {Soejima}},\ and\
  \bibinfo {author} {\bibfnamefont {M.~P.}\ \bibnamefont {Zaletel}},\ }\href
  {https://doi.org/10.1103/PhysRevLett.129.147001} {\bibfield  {journal}
  {\bibinfo  {journal} {Phys. Rev. Lett.}\ }\textbf {\bibinfo {volume} {129}},\
  \bibinfo {pages} {147001} (\bibinfo {year} {2022})}\BibitemShut {NoStop}%
\bibitem [{\citenamefont {Zhang}\ \emph {et~al.}(2023)\citenamefont {Zhang},
  \citenamefont {Pan}, \citenamefont {Chen}, \citenamefont {Li}, \citenamefont
  {Sun},\ and\ \citenamefont {Meng}}]{ZHA23a}%
  \BibitemOpen
  \bibfield  {author} {\bibinfo {author} {\bibfnamefont {X.}~\bibnamefont
  {Zhang}}, \bibinfo {author} {\bibfnamefont {G.}~\bibnamefont {Pan}}, \bibinfo
  {author} {\bibfnamefont {B.-B.}\ \bibnamefont {Chen}}, \bibinfo {author}
  {\bibfnamefont {H.}~\bibnamefont {Li}}, \bibinfo {author} {\bibfnamefont
  {K.}~\bibnamefont {Sun}},\ and\ \bibinfo {author} {\bibfnamefont {Z.~Y.}\
  \bibnamefont {Meng}},\ }\href {https://doi.org/10.1103/PhysRevB.107.L241105}
  {\bibfield  {journal} {\bibinfo  {journal} {Phys. Rev. B}\ }\textbf {\bibinfo
  {volume} {107}},\ \bibinfo {pages} {L241105} (\bibinfo {year}
  {2023})}\BibitemShut {NoStop}%
\bibitem [{\citenamefont {Blason}\ and\ \citenamefont
  {Fabrizio}(2022)}]{BLA22}%
  \BibitemOpen
  \bibfield  {author} {\bibinfo {author} {\bibfnamefont {A.}~\bibnamefont
  {Blason}}\ and\ \bibinfo {author} {\bibfnamefont {M.}~\bibnamefont
  {Fabrizio}},\ }\href {https://doi.org/10.1103/PhysRevB.106.235112} {\bibfield
   {journal} {\bibinfo  {journal} {Phys. Rev. B}\ }\textbf {\bibinfo {volume}
  {106}},\ \bibinfo {pages} {235112} (\bibinfo {year} {2022})}\BibitemShut
  {NoStop}%
\bibitem [{\citenamefont {Xie}\ \emph {et~al.}(2023)\citenamefont {Xie},
  \citenamefont {Kang}, \citenamefont {Bernevig}, \citenamefont {Vafek},\ and\
  \citenamefont {Regnault}}]{XIE23a}%
  \BibitemOpen
  \bibfield  {author} {\bibinfo {author} {\bibfnamefont {F.}~\bibnamefont
  {Xie}}, \bibinfo {author} {\bibfnamefont {J.}~\bibnamefont {Kang}}, \bibinfo
  {author} {\bibfnamefont {B.~A.}\ \bibnamefont {Bernevig}}, \bibinfo {author}
  {\bibfnamefont {O.}~\bibnamefont {Vafek}},\ and\ \bibinfo {author}
  {\bibfnamefont {N.}~\bibnamefont {Regnault}},\ }\href
  {https://doi.org/10.1103/PhysRevB.107.075156} {\bibfield  {journal} {\bibinfo
   {journal} {Phys. Rev. B}\ }\textbf {\bibinfo {volume} {107}},\ \bibinfo
  {pages} {075156} (\bibinfo {year} {2023})}\BibitemShut {NoStop}%
\bibitem [{\citenamefont {Kwan}\ \emph {et~al.}(2023)\citenamefont {Kwan},
  \citenamefont {Wagner}, \citenamefont {Bultinck}, \citenamefont {Simon},
  \citenamefont {Berg},\ and\ \citenamefont {Parameswaran}}]{KWA23}%
  \BibitemOpen
  \bibfield  {author} {\bibinfo {author} {\bibfnamefont {Y.~H.}\ \bibnamefont
  {Kwan}}, \bibinfo {author} {\bibfnamefont {G.}~\bibnamefont {Wagner}},
  \bibinfo {author} {\bibfnamefont {N.}~\bibnamefont {Bultinck}}, \bibinfo
  {author} {\bibfnamefont {S.~H.}\ \bibnamefont {Simon}}, \bibinfo {author}
  {\bibfnamefont {E.}~\bibnamefont {Berg}},\ and\ \bibinfo {author}
  {\bibfnamefont {S.~A.}\ \bibnamefont {Parameswaran}},\ }\href
  {https://doi.org/10.48550/arXiv.2303.13602} {\bibinfo {title}
  {Electron-phonon coupling and competing {{Kekul{\'e}}} orders in twisted
  bilayer graphene}} (\bibinfo {year} {2023}),\ \Eprint
  {https://arxiv.org/abs/2303.13602} {arxiv:2303.13602 [cond-mat]} \BibitemShut
  {NoStop}%
\bibitem [{\citenamefont {Yu}\ \emph {et~al.}(2023{\natexlab{b}})\citenamefont
  {Yu}, \citenamefont {Xie}, \citenamefont {Bernevig},\ and\ \citenamefont
  {Das~Sarma}}]{YU23a}%
  \BibitemOpen
  \bibfield  {author} {\bibinfo {author} {\bibfnamefont {J.}~\bibnamefont
  {Yu}}, \bibinfo {author} {\bibfnamefont {M.}~\bibnamefont {Xie}}, \bibinfo
  {author} {\bibfnamefont {B.~A.}\ \bibnamefont {Bernevig}},\ and\ \bibinfo
  {author} {\bibfnamefont {S.}~\bibnamefont {Das~Sarma}},\ }\href
  {https://doi.org/10.1103/PhysRevB.108.035129} {\bibfield  {journal} {\bibinfo
   {journal} {Phys. Rev. B}\ }\textbf {\bibinfo {volume} {108}},\ \bibinfo
  {pages} {035129} (\bibinfo {year} {2023}{\natexlab{b}})}\BibitemShut
  {NoStop}%
\bibitem [{\citenamefont {Wang}\ \emph
  {et~al.}(2023{\natexlab{a}})\citenamefont {Wang}, \citenamefont {Kwan},
  \citenamefont {Wagner}, \citenamefont {Bultinck}, \citenamefont {Simon},\
  and\ \citenamefont {Parameswaran}}]{WAN23c}%
  \BibitemOpen
  \bibfield  {author} {\bibinfo {author} {\bibfnamefont {Z.}~\bibnamefont
  {Wang}}, \bibinfo {author} {\bibfnamefont {Y.~H.}\ \bibnamefont {Kwan}},
  \bibinfo {author} {\bibfnamefont {G.}~\bibnamefont {Wagner}}, \bibinfo
  {author} {\bibfnamefont {N.}~\bibnamefont {Bultinck}}, \bibinfo {author}
  {\bibfnamefont {S.~H.}\ \bibnamefont {Simon}},\ and\ \bibinfo {author}
  {\bibfnamefont {S.~A.}\ \bibnamefont {Parameswaran}},\ }\href
  {https://doi.org/10.48550/arXiv.2310.16094} {\bibinfo {title} {Kekul{\'e}
  spirals and charge transfer cascades in twisted symmetric trilayer graphene}}
  (\bibinfo {year} {2023}{\natexlab{a}}),\ \Eprint
  {https://arxiv.org/abs/2310.16094} {arxiv:2310.16094 [cond-mat]} \BibitemShut
  {NoStop}%
\bibitem [{\citenamefont {Fernandes}\ and\ \citenamefont
  {Venderbos}(2020)}]{FER20}%
  \BibitemOpen
  \bibfield  {author} {\bibinfo {author} {\bibfnamefont {R.~M.}\ \bibnamefont
  {Fernandes}}\ and\ \bibinfo {author} {\bibfnamefont {J.~W.~F.}\ \bibnamefont
  {Venderbos}},\ }\href {https://doi.org/10.1126/sciadv.aba8834} {\bibfield
  {journal} {\bibinfo  {journal} {Sci. Adv.}\ }\textbf {\bibinfo {volume}
  {6}},\ \bibinfo {pages} {eaba8834} (\bibinfo {year} {2020})}\BibitemShut
  {NoStop}%
\bibitem [{\citenamefont {Datta}\ \emph {et~al.}(2023)\citenamefont {Datta},
  \citenamefont {Calder{\'o}n}, \citenamefont {Camjayi},\ and\ \citenamefont
  {Bascones}}]{DAT23}%
  \BibitemOpen
  \bibfield  {author} {\bibinfo {author} {\bibfnamefont {A.}~\bibnamefont
  {Datta}}, \bibinfo {author} {\bibfnamefont {M.~J.}\ \bibnamefont
  {Calder{\'o}n}}, \bibinfo {author} {\bibfnamefont {A.}~\bibnamefont
  {Camjayi}},\ and\ \bibinfo {author} {\bibfnamefont {E.}~\bibnamefont
  {Bascones}},\ }\href {https://doi.org/10.1038/s41467-023-40754-4} {\bibfield
  {journal} {\bibinfo  {journal} {Nat. Commun.}\ }\textbf {\bibinfo {volume}
  {14}},\ \bibinfo {pages} {5036} (\bibinfo {year} {2023})}\BibitemShut
  {NoStop}%
\bibitem [{\citenamefont {Rai}\ \emph {et~al.}(2023)\citenamefont {Rai},
  \citenamefont {Crippa}, \citenamefont {C{\u a}lug{\u a}ru}, \citenamefont
  {Hu}, \citenamefont {de'{\aftergroup\ignorespaces} Medici}, \citenamefont
  {Georges}, \citenamefont {Bernevig}, \citenamefont {Valent{\'i}},
  \citenamefont {Sangiovanni},\ and\ \citenamefont {Wehling}}]{RAI23a}%
  \BibitemOpen
  \bibfield  {author} {\bibinfo {author} {\bibfnamefont {G.}~\bibnamefont
  {Rai}}, \bibinfo {author} {\bibfnamefont {L.}~\bibnamefont {Crippa}},
  \bibinfo {author} {\bibfnamefont {D.}~\bibnamefont {C{\u a}lug{\u a}ru}},
  \bibinfo {author} {\bibfnamefont {H.}~\bibnamefont {Hu}}, \bibinfo {author}
  {\bibfnamefont {L.}~\bibnamefont {de'{\aftergroup\ignorespaces} Medici}},
  \bibinfo {author} {\bibfnamefont {A.}~\bibnamefont {Georges}}, \bibinfo
  {author} {\bibfnamefont {B.~A.}\ \bibnamefont {Bernevig}}, \bibinfo {author}
  {\bibfnamefont {R.}~\bibnamefont {Valent{\'i}}}, \bibinfo {author}
  {\bibfnamefont {G.}~\bibnamefont {Sangiovanni}},\ and\ \bibinfo {author}
  {\bibfnamefont {T.}~\bibnamefont {Wehling}},\ }\href
  {https://doi.org/10.48550/arXiv.2309.08529} {\bibinfo {title} {Dynamical
  correlations and order in magic-angle twisted bilayer graphene}} (\bibinfo
  {year} {2023}),\ \Eprint {https://arxiv.org/abs/2309.08529} {arxiv:2309.08529
  [cond-mat, physics:quant-ph]} \BibitemShut {NoStop}%
\bibitem [{\citenamefont {Abouelkomsan}\ \emph {et~al.}(2020)\citenamefont
  {Abouelkomsan}, \citenamefont {Liu},\ and\ \citenamefont
  {Bergholtz}}]{ABO20}%
  \BibitemOpen
  \bibfield  {author} {\bibinfo {author} {\bibfnamefont {A.}~\bibnamefont
  {Abouelkomsan}}, \bibinfo {author} {\bibfnamefont {Z.}~\bibnamefont {Liu}},\
  and\ \bibinfo {author} {\bibfnamefont {E.~J.}\ \bibnamefont {Bergholtz}},\
  }\href {https://doi.org/10.1103/PhysRevLett.124.106803} {\bibfield  {journal}
  {\bibinfo  {journal} {Phys. Rev. Lett.}\ }\textbf {\bibinfo {volume} {124}},\
  \bibinfo {pages} {106803} (\bibinfo {year} {2020})}\BibitemShut {NoStop}%
\bibitem [{\citenamefont {Ledwith}\ \emph {et~al.}(2020)\citenamefont
  {Ledwith}, \citenamefont {Tarnopolsky}, \citenamefont {Khalaf},\ and\
  \citenamefont {Vishwanath}}]{LED20}%
  \BibitemOpen
  \bibfield  {author} {\bibinfo {author} {\bibfnamefont {P.~J.}\ \bibnamefont
  {Ledwith}}, \bibinfo {author} {\bibfnamefont {G.}~\bibnamefont
  {Tarnopolsky}}, \bibinfo {author} {\bibfnamefont {E.}~\bibnamefont
  {Khalaf}},\ and\ \bibinfo {author} {\bibfnamefont {A.}~\bibnamefont
  {Vishwanath}},\ }\href {https://doi.org/10.1103/PhysRevResearch.2.023237}
  {\bibfield  {journal} {\bibinfo  {journal} {Phys. Rev. Research}\ }\textbf
  {\bibinfo {volume} {2}},\ \bibinfo {pages} {023237} (\bibinfo {year}
  {2020})}\BibitemShut {NoStop}%
\bibitem [{\citenamefont {Repellin}\ and\ \citenamefont
  {Senthil}(2020)}]{REP20a}%
  \BibitemOpen
  \bibfield  {author} {\bibinfo {author} {\bibfnamefont {C.}~\bibnamefont
  {Repellin}}\ and\ \bibinfo {author} {\bibfnamefont {T.}~\bibnamefont
  {Senthil}},\ }\href {https://doi.org/10.1103/PhysRevResearch.2.023238}
  {\bibfield  {journal} {\bibinfo  {journal} {Phys. Rev. Research}\ }\textbf
  {\bibinfo {volume} {2}},\ \bibinfo {pages} {023238} (\bibinfo {year}
  {2020})}\BibitemShut {NoStop}%
\bibitem [{\citenamefont {Khalaf}\ \emph {et~al.}(2020)\citenamefont {Khalaf},
  \citenamefont {Bultinck}, \citenamefont {Vishwanath},\ and\ \citenamefont
  {Zaletel}}]{KHA20}%
  \BibitemOpen
  \bibfield  {author} {\bibinfo {author} {\bibfnamefont {E.}~\bibnamefont
  {Khalaf}}, \bibinfo {author} {\bibfnamefont {N.}~\bibnamefont {Bultinck}},
  \bibinfo {author} {\bibfnamefont {A.}~\bibnamefont {Vishwanath}},\ and\
  \bibinfo {author} {\bibfnamefont {M.~P.}\ \bibnamefont {Zaletel}},\ }\href
  {https://doi.org/10.48550/arXiv.2009.14827} {\bibinfo {title} {Soft modes in
  magic angle twisted bilayer graphene}} (\bibinfo {year} {2020}),\ \Eprint
  {https://arxiv.org/abs/2009.14827} {arxiv:2009.14827 [cond-mat]} \BibitemShut
  {NoStop}%
\bibitem [{\citenamefont {Bernevig}\ \emph
  {et~al.}(2021{\natexlab{c}})\citenamefont {Bernevig}, \citenamefont {Lian},
  \citenamefont {Cowsik}, \citenamefont {Xie}, \citenamefont {Regnault},\ and\
  \citenamefont {Song}}]{BER21b}%
  \BibitemOpen
  \bibfield  {author} {\bibinfo {author} {\bibfnamefont {B.~A.}\ \bibnamefont
  {Bernevig}}, \bibinfo {author} {\bibfnamefont {B.}~\bibnamefont {Lian}},
  \bibinfo {author} {\bibfnamefont {A.}~\bibnamefont {Cowsik}}, \bibinfo
  {author} {\bibfnamefont {F.}~\bibnamefont {Xie}}, \bibinfo {author}
  {\bibfnamefont {N.}~\bibnamefont {Regnault}},\ and\ \bibinfo {author}
  {\bibfnamefont {Z.-D.}\ \bibnamefont {Song}},\ }\href
  {https://doi.org/10.1103/PhysRevB.103.205415} {\bibfield  {journal} {\bibinfo
   {journal} {Phys. Rev. B}\ }\textbf {\bibinfo {volume} {103}},\ \bibinfo
  {pages} {205415} (\bibinfo {year} {2021}{\natexlab{c}})}\BibitemShut
  {NoStop}%
\bibitem [{\citenamefont {Kumar}\ \emph {et~al.}(2021)\citenamefont {Kumar},
  \citenamefont {Xie},\ and\ \citenamefont {MacDonald}}]{KUM21}%
  \BibitemOpen
  \bibfield  {author} {\bibinfo {author} {\bibfnamefont {A.}~\bibnamefont
  {Kumar}}, \bibinfo {author} {\bibfnamefont {M.}~\bibnamefont {Xie}},\ and\
  \bibinfo {author} {\bibfnamefont {A.~H.}\ \bibnamefont {MacDonald}},\ }\href
  {https://doi.org/10.1103/PhysRevB.104.035119} {\bibfield  {journal} {\bibinfo
   {journal} {Phys. Rev. B}\ }\textbf {\bibinfo {volume} {104}},\ \bibinfo
  {pages} {035119} (\bibinfo {year} {2021})}\BibitemShut {NoStop}%
\bibitem [{\citenamefont {Kwan}\ \emph {et~al.}(2022)\citenamefont {Kwan},
  \citenamefont {Wagner}, \citenamefont {Bultinck}, \citenamefont {Simon},\
  and\ \citenamefont {Parameswaran}}]{KWA22}%
  \BibitemOpen
  \bibfield  {author} {\bibinfo {author} {\bibfnamefont {Y.~H.}\ \bibnamefont
  {Kwan}}, \bibinfo {author} {\bibfnamefont {G.}~\bibnamefont {Wagner}},
  \bibinfo {author} {\bibfnamefont {N.}~\bibnamefont {Bultinck}}, \bibinfo
  {author} {\bibfnamefont {S.~H.}\ \bibnamefont {Simon}},\ and\ \bibinfo
  {author} {\bibfnamefont {S.~A.}\ \bibnamefont {Parameswaran}},\ }\href
  {https://doi.org/10.1103/PhysRevX.12.031020} {\bibfield  {journal} {\bibinfo
  {journal} {Phys. Rev. X}\ }\textbf {\bibinfo {volume} {12}},\ \bibinfo
  {pages} {031020} (\bibinfo {year} {2022})}\BibitemShut {NoStop}%
\bibitem [{\citenamefont {Guo}\ \emph {et~al.}(2018)\citenamefont {Guo},
  \citenamefont {Zhu}, \citenamefont {Feng},\ and\ \citenamefont
  {Scalettar}}]{GUO18}%
  \BibitemOpen
  \bibfield  {author} {\bibinfo {author} {\bibfnamefont {H.}~\bibnamefont
  {Guo}}, \bibinfo {author} {\bibfnamefont {X.}~\bibnamefont {Zhu}}, \bibinfo
  {author} {\bibfnamefont {S.}~\bibnamefont {Feng}},\ and\ \bibinfo {author}
  {\bibfnamefont {R.~T.}\ \bibnamefont {Scalettar}},\ }\href
  {https://doi.org/10.1103/PhysRevB.97.235453} {\bibfield  {journal} {\bibinfo
  {journal} {Phys. Rev. B}\ }\textbf {\bibinfo {volume} {97}},\ \bibinfo
  {pages} {235453} (\bibinfo {year} {2018})}\BibitemShut {NoStop}%
\bibitem [{\citenamefont {Xu}\ and\ \citenamefont {Balents}(2018)}]{XU18}%
  \BibitemOpen
  \bibfield  {author} {\bibinfo {author} {\bibfnamefont {C.}~\bibnamefont
  {Xu}}\ and\ \bibinfo {author} {\bibfnamefont {L.}~\bibnamefont {Balents}},\
  }\href {https://doi.org/10.1103/PhysRevLett.121.087001} {\bibfield  {journal}
  {\bibinfo  {journal} {Phys. Rev. Lett.}\ }\textbf {\bibinfo {volume} {121}},\
  \bibinfo {pages} {087001} (\bibinfo {year} {2018})}\BibitemShut {NoStop}%
\bibitem [{\citenamefont {Liu}\ \emph {et~al.}(2018)\citenamefont {Liu},
  \citenamefont {Zhang}, \citenamefont {Chen},\ and\ \citenamefont
  {Yang}}]{LIU18a}%
  \BibitemOpen
  \bibfield  {author} {\bibinfo {author} {\bibfnamefont {C.-C.}\ \bibnamefont
  {Liu}}, \bibinfo {author} {\bibfnamefont {L.-D.}\ \bibnamefont {Zhang}},
  \bibinfo {author} {\bibfnamefont {W.-Q.}\ \bibnamefont {Chen}},\ and\
  \bibinfo {author} {\bibfnamefont {F.}~\bibnamefont {Yang}},\ }\href
  {https://doi.org/10.1103/PhysRevLett.121.217001} {\bibfield  {journal}
  {\bibinfo  {journal} {Phys. Rev. Lett.}\ }\textbf {\bibinfo {volume} {121}},\
  \bibinfo {pages} {217001} (\bibinfo {year} {2018})}\BibitemShut {NoStop}%
\bibitem [{\citenamefont {Isobe}\ \emph {et~al.}(2018)\citenamefont {Isobe},
  \citenamefont {Yuan},\ and\ \citenamefont {Fu}}]{ISO18}%
  \BibitemOpen
  \bibfield  {author} {\bibinfo {author} {\bibfnamefont {H.}~\bibnamefont
  {Isobe}}, \bibinfo {author} {\bibfnamefont {N.~F.~Q.}\ \bibnamefont {Yuan}},\
  and\ \bibinfo {author} {\bibfnamefont {L.}~\bibnamefont {Fu}},\ }\href
  {https://doi.org/10.1103/PhysRevX.8.041041} {\bibfield  {journal} {\bibinfo
  {journal} {Phys. Rev. X}\ }\textbf {\bibinfo {volume} {8}},\ \bibinfo {pages}
  {041041} (\bibinfo {year} {2018})}\BibitemShut {NoStop}%
\bibitem [{\citenamefont {Peltonen}\ \emph {et~al.}(2018)\citenamefont
  {Peltonen}, \citenamefont {Ojaj{\"a}rvi},\ and\ \citenamefont
  {Heikkil{\"a}}}]{PEL18}%
  \BibitemOpen
  \bibfield  {author} {\bibinfo {author} {\bibfnamefont {T.~J.}\ \bibnamefont
  {Peltonen}}, \bibinfo {author} {\bibfnamefont {R.}~\bibnamefont
  {Ojaj{\"a}rvi}},\ and\ \bibinfo {author} {\bibfnamefont {T.~T.}\ \bibnamefont
  {Heikkil{\"a}}},\ }\href {https://doi.org/10.1103/PhysRevB.98.220504}
  {\bibfield  {journal} {\bibinfo  {journal} {Phys. Rev. B}\ }\textbf {\bibinfo
  {volume} {98}},\ \bibinfo {pages} {220504} (\bibinfo {year}
  {2018})}\BibitemShut {NoStop}%
\bibitem [{\citenamefont {Wu}\ \emph {et~al.}(2018)\citenamefont {Wu},
  \citenamefont {MacDonald},\ and\ \citenamefont {Martin}}]{WU18}%
  \BibitemOpen
  \bibfield  {author} {\bibinfo {author} {\bibfnamefont {F.}~\bibnamefont
  {Wu}}, \bibinfo {author} {\bibfnamefont {A.~H.}\ \bibnamefont {MacDonald}},\
  and\ \bibinfo {author} {\bibfnamefont {I.}~\bibnamefont {Martin}},\ }\href
  {https://doi.org/10.1103/PhysRevLett.121.257001} {\bibfield  {journal}
  {\bibinfo  {journal} {Phys. Rev. Lett.}\ }\textbf {\bibinfo {volume} {121}},\
  \bibinfo {pages} {257001} (\bibinfo {year} {2018})}\BibitemShut {NoStop}%
\bibitem [{\citenamefont {Guinea}\ and\ \citenamefont {Walet}(2018)}]{GUI18}%
  \BibitemOpen
  \bibfield  {author} {\bibinfo {author} {\bibfnamefont {F.}~\bibnamefont
  {Guinea}}\ and\ \bibinfo {author} {\bibfnamefont {N.~R.}\ \bibnamefont
  {Walet}},\ }\href {https://doi.org/10.1073/pnas.1810947115} {\bibfield
  {journal} {\bibinfo  {journal} {PNAS}\ }\textbf {\bibinfo {volume} {115}},\
  \bibinfo {pages} {13174} (\bibinfo {year} {2018})}\BibitemShut {NoStop}%
\bibitem [{\citenamefont {Gonz{\'a}lez}\ and\ \citenamefont
  {Stauber}(2019)}]{GON19}%
  \BibitemOpen
  \bibfield  {author} {\bibinfo {author} {\bibfnamefont {J.}~\bibnamefont
  {Gonz{\'a}lez}}\ and\ \bibinfo {author} {\bibfnamefont {T.}~\bibnamefont
  {Stauber}},\ }\href {https://doi.org/10.1103/PhysRevLett.122.026801}
  {\bibfield  {journal} {\bibinfo  {journal} {Phys. Rev. Lett.}\ }\textbf
  {\bibinfo {volume} {122}},\ \bibinfo {pages} {026801} (\bibinfo {year}
  {2019})}\BibitemShut {NoStop}%
\bibitem [{\citenamefont {Roy}\ and\ \citenamefont {Juri{\v
  c}i{\'c}}(2019)}]{ROY19}%
  \BibitemOpen
  \bibfield  {author} {\bibinfo {author} {\bibfnamefont {B.}~\bibnamefont
  {Roy}}\ and\ \bibinfo {author} {\bibfnamefont {V.}~\bibnamefont {Juri{\v
  c}i{\'c}}},\ }\href {https://doi.org/10.1103/PhysRevB.99.121407} {\bibfield
  {journal} {\bibinfo  {journal} {Phys. Rev. B}\ }\textbf {\bibinfo {volume}
  {99}},\ \bibinfo {pages} {121407} (\bibinfo {year} {2019})}\BibitemShut
  {NoStop}%
\bibitem [{\citenamefont {Wu}\ \emph {et~al.}(2019{\natexlab{b}})\citenamefont
  {Wu}, \citenamefont {Hwang},\ and\ \citenamefont {Das~Sarma}}]{WU19a}%
  \BibitemOpen
  \bibfield  {author} {\bibinfo {author} {\bibfnamefont {F.}~\bibnamefont
  {Wu}}, \bibinfo {author} {\bibfnamefont {E.}~\bibnamefont {Hwang}},\ and\
  \bibinfo {author} {\bibfnamefont {S.}~\bibnamefont {Das~Sarma}},\ }\href
  {https://doi.org/10.1103/PhysRevB.99.165112} {\bibfield  {journal} {\bibinfo
  {journal} {Phys. Rev. B}\ }\textbf {\bibinfo {volume} {99}},\ \bibinfo
  {pages} {165112} (\bibinfo {year} {2019}{\natexlab{b}})}\BibitemShut
  {NoStop}%
\bibitem [{\citenamefont {You}\ and\ \citenamefont {Vishwanath}(2019)}]{YOU19}%
  \BibitemOpen
  \bibfield  {author} {\bibinfo {author} {\bibfnamefont {Y.-Z.}\ \bibnamefont
  {You}}\ and\ \bibinfo {author} {\bibfnamefont {A.}~\bibnamefont
  {Vishwanath}},\ }\href {https://doi.org/10.1038/s41535-019-0153-4} {\bibfield
   {journal} {\bibinfo  {journal} {npj Quantum Mater.}\ }\textbf {\bibinfo
  {volume} {4}},\ \bibinfo {pages} {1} (\bibinfo {year} {2019})}\BibitemShut
  {NoStop}%
\bibitem [{\citenamefont {Lian}\ \emph {et~al.}(2019)\citenamefont {Lian},
  \citenamefont {Wang},\ and\ \citenamefont {Bernevig}}]{LIA19}%
  \BibitemOpen
  \bibfield  {author} {\bibinfo {author} {\bibfnamefont {B.}~\bibnamefont
  {Lian}}, \bibinfo {author} {\bibfnamefont {Z.}~\bibnamefont {Wang}},\ and\
  \bibinfo {author} {\bibfnamefont {B.~A.}\ \bibnamefont {Bernevig}},\ }\href
  {https://doi.org/10.1103/PhysRevLett.122.257002} {\bibfield  {journal}
  {\bibinfo  {journal} {Phys. Rev. Lett.}\ }\textbf {\bibinfo {volume} {122}},\
  \bibinfo {pages} {257002} (\bibinfo {year} {2019})}\BibitemShut {NoStop}%
\bibitem [{\citenamefont {Hu}\ \emph {et~al.}(2019)\citenamefont {Hu},
  \citenamefont {Hyart}, \citenamefont {Pikulin},\ and\ \citenamefont
  {Rossi}}]{HU19a}%
  \BibitemOpen
  \bibfield  {author} {\bibinfo {author} {\bibfnamefont {X.}~\bibnamefont
  {Hu}}, \bibinfo {author} {\bibfnamefont {T.}~\bibnamefont {Hyart}}, \bibinfo
  {author} {\bibfnamefont {D.~I.}\ \bibnamefont {Pikulin}},\ and\ \bibinfo
  {author} {\bibfnamefont {E.}~\bibnamefont {Rossi}},\ }\href
  {https://doi.org/10.1103/PhysRevLett.123.237002} {\bibfield  {journal}
  {\bibinfo  {journal} {Phys. Rev. Lett.}\ }\textbf {\bibinfo {volume} {123}},\
  \bibinfo {pages} {237002} (\bibinfo {year} {2019})}\BibitemShut {NoStop}%
\bibitem [{\citenamefont {Julku}\ \emph {et~al.}(2020)\citenamefont {Julku},
  \citenamefont {Peltonen}, \citenamefont {Liang}, \citenamefont
  {Heikkil{\"a}},\ and\ \citenamefont {T{\"o}rm{\"a}}}]{JUL20}%
  \BibitemOpen
  \bibfield  {author} {\bibinfo {author} {\bibfnamefont {A.}~\bibnamefont
  {Julku}}, \bibinfo {author} {\bibfnamefont {T.~J.}\ \bibnamefont {Peltonen}},
  \bibinfo {author} {\bibfnamefont {L.}~\bibnamefont {Liang}}, \bibinfo
  {author} {\bibfnamefont {T.~T.}\ \bibnamefont {Heikkil{\"a}}},\ and\ \bibinfo
  {author} {\bibfnamefont {P.}~\bibnamefont {T{\"o}rm{\"a}}},\ }\href
  {https://doi.org/10.1103/PhysRevB.101.060505} {\bibfield  {journal} {\bibinfo
   {journal} {Phys. Rev. B}\ }\textbf {\bibinfo {volume} {101}},\ \bibinfo
  {pages} {060505} (\bibinfo {year} {2020})}\BibitemShut {NoStop}%
\bibitem [{\citenamefont {Chichinadze}\ \emph
  {et~al.}(2020{\natexlab{b}})\citenamefont {Chichinadze}, \citenamefont
  {Classen},\ and\ \citenamefont {Chubukov}}]{CHI20a}%
  \BibitemOpen
  \bibfield  {author} {\bibinfo {author} {\bibfnamefont {D.~V.}\ \bibnamefont
  {Chichinadze}}, \bibinfo {author} {\bibfnamefont {L.}~\bibnamefont
  {Classen}},\ and\ \bibinfo {author} {\bibfnamefont {A.~V.}\ \bibnamefont
  {Chubukov}},\ }\href {https://doi.org/10.1103/PhysRevB.101.224513} {\bibfield
   {journal} {\bibinfo  {journal} {Phys. Rev. B}\ }\textbf {\bibinfo {volume}
  {101}},\ \bibinfo {pages} {224513} (\bibinfo {year}
  {2020}{\natexlab{b}})}\BibitemShut {NoStop}%
\bibitem [{\citenamefont {{Lopez-Bezanilla}}\ and\ \citenamefont
  {Lado}(2020)}]{LOP20}%
  \BibitemOpen
  \bibfield  {author} {\bibinfo {author} {\bibfnamefont {A.}~\bibnamefont
  {{Lopez-Bezanilla}}}\ and\ \bibinfo {author} {\bibfnamefont {J.~L.}\
  \bibnamefont {Lado}},\ }\href
  {https://doi.org/10.1103/PhysRevResearch.2.033357} {\bibfield  {journal}
  {\bibinfo  {journal} {Phys. Rev. Research}\ }\textbf {\bibinfo {volume}
  {2}},\ \bibinfo {pages} {033357} (\bibinfo {year} {2020})}\BibitemShut
  {NoStop}%
\bibitem [{\citenamefont {K{\"o}nig}\ \emph {et~al.}(2020)\citenamefont
  {K{\"o}nig}, \citenamefont {Coleman},\ and\ \citenamefont {Tsvelik}}]{KON20}%
  \BibitemOpen
  \bibfield  {author} {\bibinfo {author} {\bibfnamefont {E.~J.}\ \bibnamefont
  {K{\"o}nig}}, \bibinfo {author} {\bibfnamefont {P.}~\bibnamefont {Coleman}},\
  and\ \bibinfo {author} {\bibfnamefont {A.~M.}\ \bibnamefont {Tsvelik}},\
  }\href {https://doi.org/10.1103/PhysRevB.102.104514} {\bibfield  {journal}
  {\bibinfo  {journal} {Phys. Rev. B}\ }\textbf {\bibinfo {volume} {102}},\
  \bibinfo {pages} {104514} (\bibinfo {year} {2020})}\BibitemShut {NoStop}%
\bibitem [{\citenamefont {Wang}\ \emph
  {et~al.}(2021{\natexlab{b}})\citenamefont {Wang}, \citenamefont {Kang},\ and\
  \citenamefont {Fernandes}}]{WAN21}%
  \BibitemOpen
  \bibfield  {author} {\bibinfo {author} {\bibfnamefont {Y.}~\bibnamefont
  {Wang}}, \bibinfo {author} {\bibfnamefont {J.}~\bibnamefont {Kang}},\ and\
  \bibinfo {author} {\bibfnamefont {R.~M.}\ \bibnamefont {Fernandes}},\ }\href
  {https://doi.org/10.1103/PhysRevB.103.024506} {\bibfield  {journal} {\bibinfo
   {journal} {Phys. Rev. B}\ }\textbf {\bibinfo {volume} {103}},\ \bibinfo
  {pages} {024506} (\bibinfo {year} {2021}{\natexlab{b}})}\BibitemShut
  {NoStop}%
\bibitem [{\citenamefont {Khalaf}\ \emph {et~al.}(2021)\citenamefont {Khalaf},
  \citenamefont {Chatterjee}, \citenamefont {Bultinck}, \citenamefont
  {Zaletel},\ and\ \citenamefont {Vishwanath}}]{KHA21}%
  \BibitemOpen
  \bibfield  {author} {\bibinfo {author} {\bibfnamefont {E.}~\bibnamefont
  {Khalaf}}, \bibinfo {author} {\bibfnamefont {S.}~\bibnamefont {Chatterjee}},
  \bibinfo {author} {\bibfnamefont {N.}~\bibnamefont {Bultinck}}, \bibinfo
  {author} {\bibfnamefont {M.~P.}\ \bibnamefont {Zaletel}},\ and\ \bibinfo
  {author} {\bibfnamefont {A.}~\bibnamefont {Vishwanath}},\ }\href
  {https://doi.org/10.1126/sciadv.abf5299} {\bibfield  {journal} {\bibinfo
  {journal} {Sci. Adv.}\ }\textbf {\bibinfo {volume} {7}},\ \bibinfo {pages}
  {eabf5299} (\bibinfo {year} {2021})}\BibitemShut {NoStop}%
\bibitem [{\citenamefont {Lewandowski}\ \emph {et~al.}(2021)\citenamefont
  {Lewandowski}, \citenamefont {Chowdhury},\ and\ \citenamefont
  {Ruhman}}]{LEW21}%
  \BibitemOpen
  \bibfield  {author} {\bibinfo {author} {\bibfnamefont {C.}~\bibnamefont
  {Lewandowski}}, \bibinfo {author} {\bibfnamefont {D.}~\bibnamefont
  {Chowdhury}},\ and\ \bibinfo {author} {\bibfnamefont {J.}~\bibnamefont
  {Ruhman}},\ }\href {https://doi.org/10.1103/PhysRevB.103.235401} {\bibfield
  {journal} {\bibinfo  {journal} {Phys. Rev. B}\ }\textbf {\bibinfo {volume}
  {103}},\ \bibinfo {pages} {235401} (\bibinfo {year} {2021})}\BibitemShut
  {NoStop}%
\bibitem [{\citenamefont {Fernandes}\ and\ \citenamefont {Fu}(2021)}]{FER21}%
  \BibitemOpen
  \bibfield  {author} {\bibinfo {author} {\bibfnamefont {R.~M.}\ \bibnamefont
  {Fernandes}}\ and\ \bibinfo {author} {\bibfnamefont {L.}~\bibnamefont {Fu}},\
  }\href {https://doi.org/10.1103/PhysRevLett.127.047001} {\bibfield  {journal}
  {\bibinfo  {journal} {Phys. Rev. Lett.}\ }\textbf {\bibinfo {volume} {127}},\
  \bibinfo {pages} {047001} (\bibinfo {year} {2021})}\BibitemShut {NoStop}%
\bibitem [{\citenamefont {Qin}\ and\ \citenamefont {MacDonald}(2021)}]{QIN21}%
  \BibitemOpen
  \bibfield  {author} {\bibinfo {author} {\bibfnamefont {W.}~\bibnamefont
  {Qin}}\ and\ \bibinfo {author} {\bibfnamefont {A.~H.}\ \bibnamefont
  {MacDonald}},\ }\href {https://doi.org/10.1103/PhysRevLett.127.097001}
  {\bibfield  {journal} {\bibinfo  {journal} {Phys. Rev. Lett.}\ }\textbf
  {\bibinfo {volume} {127}},\ \bibinfo {pages} {097001} (\bibinfo {year}
  {2021})}\BibitemShut {NoStop}%
\bibitem [{\citenamefont {Phong}\ \emph {et~al.}(2021)\citenamefont {Phong},
  \citenamefont {Pantale{\'o}n}, \citenamefont {Cea},\ and\ \citenamefont
  {Guinea}}]{PHO21}%
  \BibitemOpen
  \bibfield  {author} {\bibinfo {author} {\bibfnamefont {V.~T.}\ \bibnamefont
  {Phong}}, \bibinfo {author} {\bibfnamefont {P.~A.}\ \bibnamefont
  {Pantale{\'o}n}}, \bibinfo {author} {\bibfnamefont {T.}~\bibnamefont {Cea}},\
  and\ \bibinfo {author} {\bibfnamefont {F.}~\bibnamefont {Guinea}},\ }\href
  {https://doi.org/10.1103/PhysRevB.104.L121116} {\bibfield  {journal}
  {\bibinfo  {journal} {Phys. Rev. B}\ }\textbf {\bibinfo {volume} {104}},\
  \bibinfo {pages} {L121116} (\bibinfo {year} {2021})}\BibitemShut {NoStop}%
\bibitem [{\citenamefont {Choi}\ and\ \citenamefont {Choi}(2021)}]{CHO21d}%
  \BibitemOpen
  \bibfield  {author} {\bibinfo {author} {\bibfnamefont {Y.~W.}\ \bibnamefont
  {Choi}}\ and\ \bibinfo {author} {\bibfnamefont {H.~J.}\ \bibnamefont
  {Choi}},\ }\href {https://doi.org/10.1103/PhysRevLett.127.167001} {\bibfield
  {journal} {\bibinfo  {journal} {Phys. Rev. Lett.}\ }\textbf {\bibinfo
  {volume} {127}},\ \bibinfo {pages} {167001} (\bibinfo {year}
  {2021})}\BibitemShut {NoStop}%
\bibitem [{\citenamefont {Lake}\ and\ \citenamefont {Senthil}(2021)}]{LAK21}%
  \BibitemOpen
  \bibfield  {author} {\bibinfo {author} {\bibfnamefont {E.}~\bibnamefont
  {Lake}}\ and\ \bibinfo {author} {\bibfnamefont {T.}~\bibnamefont {Senthil}},\
  }\href {https://doi.org/10.1103/PhysRevB.104.174505} {\bibfield  {journal}
  {\bibinfo  {journal} {Phys. Rev. B}\ }\textbf {\bibinfo {volume} {104}},\
  \bibinfo {pages} {174505} (\bibinfo {year} {2021})}\BibitemShut {NoStop}%
\bibitem [{\citenamefont {Chou}\ \emph {et~al.}(2021)\citenamefont {Chou},
  \citenamefont {Wu}, \citenamefont {Sau},\ and\ \citenamefont
  {Das~Sarma}}]{CHO21c}%
  \BibitemOpen
  \bibfield  {author} {\bibinfo {author} {\bibfnamefont {Y.-Z.}\ \bibnamefont
  {Chou}}, \bibinfo {author} {\bibfnamefont {F.}~\bibnamefont {Wu}}, \bibinfo
  {author} {\bibfnamefont {J.~D.}\ \bibnamefont {Sau}},\ and\ \bibinfo {author}
  {\bibfnamefont {S.}~\bibnamefont {Das~Sarma}},\ }\href
  {https://doi.org/10.1103/PhysRevLett.127.217001} {\bibfield  {journal}
  {\bibinfo  {journal} {Phys. Rev. Lett.}\ }\textbf {\bibinfo {volume} {127}},\
  \bibinfo {pages} {217001} (\bibinfo {year} {2021})}\BibitemShut {NoStop}%
\bibitem [{\citenamefont {Li}\ \emph {et~al.}(2022{\natexlab{b}})\citenamefont
  {Li}, \citenamefont {Zheng},\ and\ \citenamefont {Huang}}]{LI22c}%
  \BibitemOpen
  \bibfield  {author} {\bibinfo {author} {\bibfnamefont {S.}~\bibnamefont
  {Li}}, \bibinfo {author} {\bibfnamefont {G.}~\bibnamefont {Zheng}},\ and\
  \bibinfo {author} {\bibfnamefont {J.}~\bibnamefont {Huang}},\ }\href
  {https://doi.org/10.48550/arXiv.2111.04451} {\bibinfo {title} {Induced
  superconductivity in magic-angle twisted trilayer graphene through
  graphene-metal contacts}} (\bibinfo {year} {2022}{\natexlab{b}}),\ \Eprint
  {https://arxiv.org/abs/2111.04451} {arxiv:2111.04451 [cond-mat]} \BibitemShut
  {NoStop}%
\bibitem [{\citenamefont {Fischer}\ \emph {et~al.}(2022)\citenamefont
  {Fischer}, \citenamefont {Goodwin}, \citenamefont {Mostofi}, \citenamefont
  {Lischner}, \citenamefont {Kennes},\ and\ \citenamefont {Klebl}}]{FIS22}%
  \BibitemOpen
  \bibfield  {author} {\bibinfo {author} {\bibfnamefont {A.}~\bibnamefont
  {Fischer}}, \bibinfo {author} {\bibfnamefont {Z.~A.~H.}\ \bibnamefont
  {Goodwin}}, \bibinfo {author} {\bibfnamefont {A.~A.}\ \bibnamefont
  {Mostofi}}, \bibinfo {author} {\bibfnamefont {J.}~\bibnamefont {Lischner}},
  \bibinfo {author} {\bibfnamefont {D.~M.}\ \bibnamefont {Kennes}},\ and\
  \bibinfo {author} {\bibfnamefont {L.}~\bibnamefont {Klebl}},\ }\href
  {https://doi.org/10.1038/s41535-021-00410-w} {\bibfield  {journal} {\bibinfo
  {journal} {npj Quantum Mater.}\ }\textbf {\bibinfo {volume} {7}},\ \bibinfo
  {pages} {1} (\bibinfo {year} {2022})}\BibitemShut {NoStop}%
\bibitem [{\citenamefont {Yu}\ \emph {et~al.}(2022)\citenamefont {Yu},
  \citenamefont {Chen},\ and\ \citenamefont {Das~Sarma}}]{YU22}%
  \BibitemOpen
  \bibfield  {author} {\bibinfo {author} {\bibfnamefont {J.}~\bibnamefont
  {Yu}}, \bibinfo {author} {\bibfnamefont {Y.-A.}\ \bibnamefont {Chen}},\ and\
  \bibinfo {author} {\bibfnamefont {S.}~\bibnamefont {Das~Sarma}},\ }\href
  {https://doi.org/10.1103/PhysRevB.105.104515} {\bibfield  {journal} {\bibinfo
   {journal} {Phys. Rev. B}\ }\textbf {\bibinfo {volume} {105}},\ \bibinfo
  {pages} {104515} (\bibinfo {year} {2022})}\BibitemShut {NoStop}%
\bibitem [{\citenamefont {Scammell}\ \emph {et~al.}(2022)\citenamefont
  {Scammell}, \citenamefont {Li},\ and\ \citenamefont {Scheurer}}]{SCA22}%
  \BibitemOpen
  \bibfield  {author} {\bibinfo {author} {\bibfnamefont {H.~D.}\ \bibnamefont
  {Scammell}}, \bibinfo {author} {\bibfnamefont {J.~I.~A.}\ \bibnamefont
  {Li}},\ and\ \bibinfo {author} {\bibfnamefont {M.~S.}\ \bibnamefont
  {Scheurer}},\ }\href {https://doi.org/10.1088/2053-1583/ac5b16} {\bibfield
  {journal} {\bibinfo  {journal} {2D Mater.}\ }\textbf {\bibinfo {volume}
  {9}},\ \bibinfo {pages} {025027} (\bibinfo {year} {2022})}\BibitemShut
  {NoStop}%
\bibitem [{\citenamefont {Chatterjee}\ \emph {et~al.}(2022)\citenamefont
  {Chatterjee}, \citenamefont {Ippoliti},\ and\ \citenamefont
  {Zaletel}}]{CHA22}%
  \BibitemOpen
  \bibfield  {author} {\bibinfo {author} {\bibfnamefont {S.}~\bibnamefont
  {Chatterjee}}, \bibinfo {author} {\bibfnamefont {M.}~\bibnamefont
  {Ippoliti}},\ and\ \bibinfo {author} {\bibfnamefont {M.~P.}\ \bibnamefont
  {Zaletel}},\ }\href {https://doi.org/10.1103/PhysRevB.106.035421} {\bibfield
  {journal} {\bibinfo  {journal} {Phys. Rev. B}\ }\textbf {\bibinfo {volume}
  {106}},\ \bibinfo {pages} {035421} (\bibinfo {year} {2022})}\BibitemShut
  {NoStop}%
\bibitem [{\citenamefont {Wagner}\ \emph
  {et~al.}(2023{\natexlab{a}})\citenamefont {Wagner}, \citenamefont {Kwan},
  \citenamefont {Bultinck}, \citenamefont {Simon},\ and\ \citenamefont
  {Parameswaran}}]{WAG23a}%
  \BibitemOpen
  \bibfield  {author} {\bibinfo {author} {\bibfnamefont {G.}~\bibnamefont
  {Wagner}}, \bibinfo {author} {\bibfnamefont {Y.~H.}\ \bibnamefont {Kwan}},
  \bibinfo {author} {\bibfnamefont {N.}~\bibnamefont {Bultinck}}, \bibinfo
  {author} {\bibfnamefont {S.~H.}\ \bibnamefont {Simon}},\ and\ \bibinfo
  {author} {\bibfnamefont {S.~A.}\ \bibnamefont {Parameswaran}},\ }\href
  {https://doi.org/10.48550/arXiv.2302.00682} {\bibinfo {title}
  {Superconductivity from repulsive interactions in {{Bernal-stacked}} bilayer
  graphene}} (\bibinfo {year} {2023}{\natexlab{a}}),\ \Eprint
  {https://arxiv.org/abs/2302.00682} {arxiv:2302.00682 [cond-mat]} \BibitemShut
  {NoStop}%
\bibitem [{\citenamefont {Gonz{\'a}lez}\ and\ \citenamefont
  {Stauber}(2023)}]{GON23}%
  \BibitemOpen
  \bibfield  {author} {\bibinfo {author} {\bibfnamefont {J.}~\bibnamefont
  {Gonz{\'a}lez}}\ and\ \bibinfo {author} {\bibfnamefont {T.}~\bibnamefont
  {Stauber}},\ }\href {https://doi.org/10.1038/s41467-023-38250-w} {\bibfield
  {journal} {\bibinfo  {journal} {Nat. Commun.}\ }\textbf {\bibinfo {volume}
  {14}},\ \bibinfo {pages} {2746} (\bibinfo {year} {2023})}\BibitemShut
  {NoStop}%
\bibitem [{\citenamefont {Wagner}\ \emph
  {et~al.}(2023{\natexlab{b}})\citenamefont {Wagner}, \citenamefont {Kwan},
  \citenamefont {Bultinck}, \citenamefont {Simon},\ and\ \citenamefont
  {Parameswaran}}]{WAG23}%
  \BibitemOpen
  \bibfield  {author} {\bibinfo {author} {\bibfnamefont {G.}~\bibnamefont
  {Wagner}}, \bibinfo {author} {\bibfnamefont {Y.~H.}\ \bibnamefont {Kwan}},
  \bibinfo {author} {\bibfnamefont {N.}~\bibnamefont {Bultinck}}, \bibinfo
  {author} {\bibfnamefont {S.~H.}\ \bibnamefont {Simon}},\ and\ \bibinfo
  {author} {\bibfnamefont {S.~A.}\ \bibnamefont {Parameswaran}},\ }\href
  {https://doi.org/10.48550/arXiv.2308.10938} {\bibinfo {title} {Coulomb-driven
  band unflattening suppresses $k$-phonon pairing in moir{\'e} graphene}}
  (\bibinfo {year} {2023}{\natexlab{b}}),\ \Eprint
  {https://arxiv.org/abs/2308.10938} {arxiv:2308.10938 [cond-mat]} \BibitemShut
  {NoStop}%
\bibitem [{\citenamefont {Wang}\ \emph {et~al.}(2024)\citenamefont {Wang},
  \citenamefont {Zhou}, \citenamefont {Peng}, \citenamefont {Lian},\ and\
  \citenamefont {Song}}]{WAN24}%
  \BibitemOpen
  \bibfield  {author} {\bibinfo {author} {\bibfnamefont {Y.-J.}\ \bibnamefont
  {Wang}}, \bibinfo {author} {\bibfnamefont {G.-D.}\ \bibnamefont {Zhou}},
  \bibinfo {author} {\bibfnamefont {S.-Y.}\ \bibnamefont {Peng}}, \bibinfo
  {author} {\bibfnamefont {B.}~\bibnamefont {Lian}},\ and\ \bibinfo {author}
  {\bibfnamefont {Z.-D.}\ \bibnamefont {Song}},\ }\href
  {https://doi.org/10.48550/arXiv.2402.00869} {\bibinfo {title} {Molecular
  {{Pairing}} in {{Twisted Bilayer Graphene Superconductivity}}}} (\bibinfo
  {year} {2024}),\ \Eprint {https://arxiv.org/abs/2402.00869} {arxiv:2402.00869
  [cond-mat]} \BibitemShut {NoStop}%
\bibitem [{\citenamefont {Moon}\ and\ \citenamefont {Koshino}(2013)}]{MOO13}%
  \BibitemOpen
  \bibfield  {author} {\bibinfo {author} {\bibfnamefont {P.}~\bibnamefont
  {Moon}}\ and\ \bibinfo {author} {\bibfnamefont {M.}~\bibnamefont {Koshino}},\
  }\href {https://doi.org/10.1103/PhysRevB.87.205404} {\bibfield  {journal}
  {\bibinfo  {journal} {Phys. Rev. B}\ }\textbf {\bibinfo {volume} {87}},\
  \bibinfo {pages} {205404} (\bibinfo {year} {2013})}\BibitemShut {NoStop}%
\bibitem [{\citenamefont {Liu}\ and\ \citenamefont {Dai}(2020)}]{LIU20b}%
  \BibitemOpen
  \bibfield  {author} {\bibinfo {author} {\bibfnamefont {J.}~\bibnamefont
  {Liu}}\ and\ \bibinfo {author} {\bibfnamefont {X.}~\bibnamefont {Dai}},\
  }\href {https://doi.org/10.1038/s41524-020-0299-4} {\bibfield  {journal}
  {\bibinfo  {journal} {npj Comput. Mater.}\ }\textbf {\bibinfo {volume} {6}},\
  \bibinfo {pages} {1} (\bibinfo {year} {2020})}\BibitemShut {NoStop}%
\bibitem [{\citenamefont {Padhi}\ \emph {et~al.}(2020)\citenamefont {Padhi},
  \citenamefont {Tiwari}, \citenamefont {Neupert},\ and\ \citenamefont
  {Ryu}}]{PAD20}%
  \BibitemOpen
  \bibfield  {author} {\bibinfo {author} {\bibfnamefont {B.}~\bibnamefont
  {Padhi}}, \bibinfo {author} {\bibfnamefont {A.}~\bibnamefont {Tiwari}},
  \bibinfo {author} {\bibfnamefont {T.}~\bibnamefont {Neupert}},\ and\ \bibinfo
  {author} {\bibfnamefont {S.}~\bibnamefont {Ryu}},\ }\href
  {https://doi.org/10.1103/PhysRevResearch.2.033458} {\bibfield  {journal}
  {\bibinfo  {journal} {Phys. Rev. Research}\ }\textbf {\bibinfo {volume}
  {2}},\ \bibinfo {pages} {033458} (\bibinfo {year} {2020})}\BibitemShut
  {NoStop}%
\bibitem [{\citenamefont {{Garc{\'i}a-Ruiz}}\ \emph {et~al.}(2020)\citenamefont
  {{Garc{\'i}a-Ruiz}}, \citenamefont {Thompson}, \citenamefont
  {{Mucha-Kruczy{\'n}ski}},\ and\ \citenamefont {Fal'ko}}]{GAR20}%
  \BibitemOpen
  \bibfield  {author} {\bibinfo {author} {\bibfnamefont {A.}~\bibnamefont
  {{Garc{\'i}a-Ruiz}}}, \bibinfo {author} {\bibfnamefont {J.~J.~P.}\
  \bibnamefont {Thompson}}, \bibinfo {author} {\bibfnamefont {M.}~\bibnamefont
  {{Mucha-Kruczy{\'n}ski}}},\ and\ \bibinfo {author} {\bibfnamefont {V.~I.}\
  \bibnamefont {Fal'ko}},\ }\href
  {https://doi.org/10.1103/PhysRevLett.125.197401} {\bibfield  {journal}
  {\bibinfo  {journal} {Phys. Rev. Lett.}\ }\textbf {\bibinfo {volume} {125}},\
  \bibinfo {pages} {197401} (\bibinfo {year} {2020})}\BibitemShut {NoStop}%
\bibitem [{\citenamefont {Kruchkov}(2023)}]{KRU23}%
  \BibitemOpen
  \bibfield  {author} {\bibinfo {author} {\bibfnamefont {A.}~\bibnamefont
  {Kruchkov}},\ }\href {https://doi.org/10.1103/PhysRevB.107.L241102}
  {\bibfield  {journal} {\bibinfo  {journal} {Phys. Rev. B}\ }\textbf {\bibinfo
  {volume} {107}},\ \bibinfo {pages} {L241102} (\bibinfo {year}
  {2023})}\BibitemShut {NoStop}%
\bibitem [{\citenamefont {Wang}\ \emph
  {et~al.}(2023{\natexlab{b}})\citenamefont {Wang}, \citenamefont {Finney},
  \citenamefont {Sharpe}, \citenamefont {Rodenbach}, \citenamefont {Hsueh},
  \citenamefont {Watanabe}, \citenamefont {Taniguchi}, \citenamefont {Kastner},
  \citenamefont {Vafek},\ and\ \citenamefont {{Goldhaber-Gordon}}}]{WAN23b}%
  \BibitemOpen
  \bibfield  {author} {\bibinfo {author} {\bibfnamefont {X.}~\bibnamefont
  {Wang}}, \bibinfo {author} {\bibfnamefont {J.}~\bibnamefont {Finney}},
  \bibinfo {author} {\bibfnamefont {A.~L.}\ \bibnamefont {Sharpe}}, \bibinfo
  {author} {\bibfnamefont {L.~K.}\ \bibnamefont {Rodenbach}}, \bibinfo {author}
  {\bibfnamefont {C.~L.}\ \bibnamefont {Hsueh}}, \bibinfo {author}
  {\bibfnamefont {K.}~\bibnamefont {Watanabe}}, \bibinfo {author}
  {\bibfnamefont {T.}~\bibnamefont {Taniguchi}}, \bibinfo {author}
  {\bibfnamefont {M.~A.}\ \bibnamefont {Kastner}}, \bibinfo {author}
  {\bibfnamefont {O.}~\bibnamefont {Vafek}},\ and\ \bibinfo {author}
  {\bibfnamefont {D.}~\bibnamefont {{Goldhaber-Gordon}}},\ }\href
  {https://doi.org/10.1073/pnas.2307151120} {\bibfield  {journal} {\bibinfo
  {journal} {PNAS}\ }\textbf {\bibinfo {volume} {120}},\ \bibinfo {pages}
  {e2307151120} (\bibinfo {year} {2023}{\natexlab{b}})}\BibitemShut {NoStop}%
\bibitem [{\citenamefont {Liu}\ \emph {et~al.}(2020)\citenamefont {Liu},
  \citenamefont {Hao}, \citenamefont {Khalaf}, \citenamefont {Lee},
  \citenamefont {Ronen}, \citenamefont {Yoo}, \citenamefont {Haei~Najafabadi},
  \citenamefont {Watanabe}, \citenamefont {Taniguchi}, \citenamefont
  {Vishwanath},\ and\ \citenamefont {Kim}}]{LIU20a}%
  \BibitemOpen
  \bibfield  {author} {\bibinfo {author} {\bibfnamefont {X.}~\bibnamefont
  {Liu}}, \bibinfo {author} {\bibfnamefont {Z.}~\bibnamefont {Hao}}, \bibinfo
  {author} {\bibfnamefont {E.}~\bibnamefont {Khalaf}}, \bibinfo {author}
  {\bibfnamefont {J.~Y.}\ \bibnamefont {Lee}}, \bibinfo {author} {\bibfnamefont
  {Y.}~\bibnamefont {Ronen}}, \bibinfo {author} {\bibfnamefont
  {H.}~\bibnamefont {Yoo}}, \bibinfo {author} {\bibfnamefont {D.}~\bibnamefont
  {Haei~Najafabadi}}, \bibinfo {author} {\bibfnamefont {K.}~\bibnamefont
  {Watanabe}}, \bibinfo {author} {\bibfnamefont {T.}~\bibnamefont {Taniguchi}},
  \bibinfo {author} {\bibfnamefont {A.}~\bibnamefont {Vishwanath}},\ and\
  \bibinfo {author} {\bibfnamefont {P.}~\bibnamefont {Kim}},\ }\href
  {https://doi.org/10.1038/s41586-020-2458-7} {\bibfield  {journal} {\bibinfo
  {journal} {Nature}\ }\textbf {\bibinfo {volume} {583}},\ \bibinfo {pages}
  {221} (\bibinfo {year} {2020})}\BibitemShut {NoStop}%
\bibitem [{\citenamefont {Shen}\ \emph {et~al.}(2020)\citenamefont {Shen},
  \citenamefont {Chu}, \citenamefont {Wu}, \citenamefont {Li}, \citenamefont
  {Wang}, \citenamefont {Zhao}, \citenamefont {Tang}, \citenamefont {Liu},
  \citenamefont {Tian}, \citenamefont {Watanabe}, \citenamefont {Taniguchi},
  \citenamefont {Yang}, \citenamefont {Meng}, \citenamefont {Shi},
  \citenamefont {Yazyev},\ and\ \citenamefont {Zhang}}]{SHE20}%
  \BibitemOpen
  \bibfield  {author} {\bibinfo {author} {\bibfnamefont {C.}~\bibnamefont
  {Shen}}, \bibinfo {author} {\bibfnamefont {Y.}~\bibnamefont {Chu}}, \bibinfo
  {author} {\bibfnamefont {Q.}~\bibnamefont {Wu}}, \bibinfo {author}
  {\bibfnamefont {N.}~\bibnamefont {Li}}, \bibinfo {author} {\bibfnamefont
  {S.}~\bibnamefont {Wang}}, \bibinfo {author} {\bibfnamefont {Y.}~\bibnamefont
  {Zhao}}, \bibinfo {author} {\bibfnamefont {J.}~\bibnamefont {Tang}}, \bibinfo
  {author} {\bibfnamefont {J.}~\bibnamefont {Liu}}, \bibinfo {author}
  {\bibfnamefont {J.}~\bibnamefont {Tian}}, \bibinfo {author} {\bibfnamefont
  {K.}~\bibnamefont {Watanabe}}, \bibinfo {author} {\bibfnamefont
  {T.}~\bibnamefont {Taniguchi}}, \bibinfo {author} {\bibfnamefont
  {R.}~\bibnamefont {Yang}}, \bibinfo {author} {\bibfnamefont {Z.~Y.}\
  \bibnamefont {Meng}}, \bibinfo {author} {\bibfnamefont {D.}~\bibnamefont
  {Shi}}, \bibinfo {author} {\bibfnamefont {O.~V.}\ \bibnamefont {Yazyev}},\
  and\ \bibinfo {author} {\bibfnamefont {G.}~\bibnamefont {Zhang}},\ }\href
  {https://doi.org/10.1038/s41567-020-0825-9} {\bibfield  {journal} {\bibinfo
  {journal} {Nat. Phys.}\ }\textbf {\bibinfo {volume} {16}},\ \bibinfo {pages}
  {520} (\bibinfo {year} {2020})}\BibitemShut {NoStop}%
\bibitem [{\citenamefont {Hu}\ \emph {et~al.}(2023{\natexlab{a}})\citenamefont
  {Hu}, \citenamefont {Rai}, \citenamefont {Crippa}, \citenamefont
  {{Herzog-Arbeitman}}, \citenamefont {C{\u a}lug{\u a}ru}, \citenamefont
  {Wehling}, \citenamefont {Sangiovanni}, \citenamefont {Valent{\'i}},
  \citenamefont {Tsvelik},\ and\ \citenamefont {Bernevig}}]{HU23}%
  \BibitemOpen
  \bibfield  {author} {\bibinfo {author} {\bibfnamefont {H.}~\bibnamefont
  {Hu}}, \bibinfo {author} {\bibfnamefont {G.}~\bibnamefont {Rai}}, \bibinfo
  {author} {\bibfnamefont {L.}~\bibnamefont {Crippa}}, \bibinfo {author}
  {\bibfnamefont {J.}~\bibnamefont {{Herzog-Arbeitman}}}, \bibinfo {author}
  {\bibfnamefont {D.}~\bibnamefont {C{\u a}lug{\u a}ru}}, \bibinfo {author}
  {\bibfnamefont {T.}~\bibnamefont {Wehling}}, \bibinfo {author} {\bibfnamefont
  {G.}~\bibnamefont {Sangiovanni}}, \bibinfo {author} {\bibfnamefont
  {R.}~\bibnamefont {Valent{\'i}}}, \bibinfo {author} {\bibfnamefont {A.~M.}\
  \bibnamefont {Tsvelik}},\ and\ \bibinfo {author} {\bibfnamefont {B.~A.}\
  \bibnamefont {Bernevig}},\ }\href
  {https://doi.org/10.1103/PhysRevLett.131.166501} {\bibfield  {journal}
  {\bibinfo  {journal} {Phys. Rev. Lett.}\ }\textbf {\bibinfo {volume} {131}},\
  \bibinfo {pages} {166501} (\bibinfo {year} {2023}{\natexlab{a}})}\BibitemShut
  {NoStop}%
\bibitem [{\citenamefont {Zhou}\ \emph {et~al.}(2024)\citenamefont {Zhou},
  \citenamefont {Wang}, \citenamefont {Tong},\ and\ \citenamefont
  {Song}}]{ZHO24}%
  \BibitemOpen
  \bibfield  {author} {\bibinfo {author} {\bibfnamefont {G.-D.}\ \bibnamefont
  {Zhou}}, \bibinfo {author} {\bibfnamefont {Y.-J.}\ \bibnamefont {Wang}},
  \bibinfo {author} {\bibfnamefont {N.}~\bibnamefont {Tong}},\ and\ \bibinfo
  {author} {\bibfnamefont {Z.-D.}\ \bibnamefont {Song}},\ }\href
  {https://doi.org/10.1103/PhysRevB.109.045419} {\bibfield  {journal} {\bibinfo
   {journal} {Phys. Rev. B}\ }\textbf {\bibinfo {volume} {109}},\ \bibinfo
  {pages} {045419} (\bibinfo {year} {2024})}\BibitemShut {NoStop}%
\bibitem [{\citenamefont {Lau}\ and\ \citenamefont {Coleman}(2023)}]{LAU23}%
  \BibitemOpen
  \bibfield  {author} {\bibinfo {author} {\bibfnamefont {L.~L.~H.}\
  \bibnamefont {Lau}}\ and\ \bibinfo {author} {\bibfnamefont {P.}~\bibnamefont
  {Coleman}},\ }\href {https://doi.org/10.48550/arXiv.2303.02670} {\bibinfo
  {title} {Topological {{Mixed Valence Model}} for {{Twisted Bilayer
  Graphene}}}} (\bibinfo {year} {2023}),\ \Eprint
  {https://arxiv.org/abs/2303.02670} {arxiv:2303.02670 [cond-mat]} \BibitemShut
  {NoStop}%
\bibitem [{\citenamefont {C{\u a}lug{\u a}ru}\ \emph
  {et~al.}(2023)\citenamefont {C{\u a}lug{\u a}ru}, \citenamefont {Borovkov},
  \citenamefont {Lau}, \citenamefont {Coleman}, \citenamefont {Song},\ and\
  \citenamefont {Bernevig}}]{CAL23}%
  \BibitemOpen
  \bibfield  {author} {\bibinfo {author} {\bibfnamefont {D.}~\bibnamefont {C{\u
  a}lug{\u a}ru}}, \bibinfo {author} {\bibfnamefont {M.}~\bibnamefont
  {Borovkov}}, \bibinfo {author} {\bibfnamefont {L.~L.~H.}\ \bibnamefont
  {Lau}}, \bibinfo {author} {\bibfnamefont {P.}~\bibnamefont {Coleman}},
  \bibinfo {author} {\bibfnamefont {Z.-D.}\ \bibnamefont {Song}},\ and\
  \bibinfo {author} {\bibfnamefont {B.~A.}\ \bibnamefont {Bernevig}},\ }\href
  {https://doi.org/10.1063/10.0019421} {\bibfield  {journal} {\bibinfo
  {journal} {Low Temp. Phys.}\ }\textbf {\bibinfo {volume} {49}},\ \bibinfo
  {pages} {640} (\bibinfo {year} {2023})}\BibitemShut {NoStop}%
\bibitem [{\citenamefont {Chou}\ and\ \citenamefont {Das~Sarma}(2023)}]{CHO23}%
  \BibitemOpen
  \bibfield  {author} {\bibinfo {author} {\bibfnamefont {Y.-Z.}\ \bibnamefont
  {Chou}}\ and\ \bibinfo {author} {\bibfnamefont {S.}~\bibnamefont
  {Das~Sarma}},\ }\href {https://doi.org/10.1103/PhysRevLett.131.026501}
  {\bibfield  {journal} {\bibinfo  {journal} {Phys. Rev. Lett.}\ }\textbf
  {\bibinfo {volume} {131}},\ \bibinfo {pages} {026501} (\bibinfo {year}
  {2023})}\BibitemShut {NoStop}%
\bibitem [{\citenamefont {Hu}\ \emph {et~al.}(2023{\natexlab{b}})\citenamefont
  {Hu}, \citenamefont {Bernevig},\ and\ \citenamefont {Tsvelik}}]{HU23i}%
  \BibitemOpen
  \bibfield  {author} {\bibinfo {author} {\bibfnamefont {H.}~\bibnamefont
  {Hu}}, \bibinfo {author} {\bibfnamefont {B.~A.}\ \bibnamefont {Bernevig}},\
  and\ \bibinfo {author} {\bibfnamefont {A.~M.}\ \bibnamefont {Tsvelik}},\
  }\href {https://doi.org/10.1103/PhysRevLett.131.026502} {\bibfield  {journal}
  {\bibinfo  {journal} {Phys. Rev. Lett.}\ }\textbf {\bibinfo {volume} {131}},\
  \bibinfo {pages} {026502} (\bibinfo {year} {2023}{\natexlab{b}})}\BibitemShut
  {NoStop}%
\bibitem [{\citenamefont {Li}\ \emph {et~al.}(2023)\citenamefont {Li},
  \citenamefont {Fregoso},\ and\ \citenamefont {Dzero}}]{LI23a}%
  \BibitemOpen
  \bibfield  {author} {\bibinfo {author} {\bibfnamefont {Y.}~\bibnamefont
  {Li}}, \bibinfo {author} {\bibfnamefont {B.~M.}\ \bibnamefont {Fregoso}},\
  and\ \bibinfo {author} {\bibfnamefont {M.}~\bibnamefont {Dzero}},\ }\href
  {https://doi.org/10.48550/arXiv.2309.03416} {\bibinfo {title} {Topological
  {{Mixed Valence Model}} in {{Magic-Angle Twisted Bilayer Graphene}}}}
  (\bibinfo {year} {2023}),\ \Eprint {https://arxiv.org/abs/2309.03416}
  {arxiv:2309.03416 [cond-mat]} \BibitemShut {NoStop}%
\bibitem [{\citenamefont {Merino}\ \emph {et~al.}(2024)\citenamefont {Merino},
  \citenamefont {Calugaru}, \citenamefont {Hu}, \citenamefont {{Diez-Merida}},
  \citenamefont {{Diez-Carlon}}, \citenamefont {Taniguchi}, \citenamefont
  {Watanabe}, \citenamefont {Seifert}, \citenamefont {Bernevig},\ and\
  \citenamefont {Efetov}}]{MER24}%
  \BibitemOpen
  \bibfield  {author} {\bibinfo {author} {\bibfnamefont {R.~L.}\ \bibnamefont
  {Merino}}, \bibinfo {author} {\bibfnamefont {D.}~\bibnamefont {Calugaru}},
  \bibinfo {author} {\bibfnamefont {H.}~\bibnamefont {Hu}}, \bibinfo {author}
  {\bibfnamefont {J.}~\bibnamefont {{Diez-Merida}}}, \bibinfo {author}
  {\bibfnamefont {A.}~\bibnamefont {{Diez-Carlon}}}, \bibinfo {author}
  {\bibfnamefont {T.}~\bibnamefont {Taniguchi}}, \bibinfo {author}
  {\bibfnamefont {K.}~\bibnamefont {Watanabe}}, \bibinfo {author}
  {\bibfnamefont {P.}~\bibnamefont {Seifert}}, \bibinfo {author} {\bibfnamefont
  {B.~A.}\ \bibnamefont {Bernevig}},\ and\ \bibinfo {author} {\bibfnamefont
  {D.~K.}\ \bibnamefont {Efetov}},\ }\href
  {https://doi.org/10.48550/arXiv.2402.11749} {\bibinfo {title} {Evidence of
  heavy fermion physics in the thermoelectric transport of magic angle twisted
  bilayer graphene}} (\bibinfo {year} {2024}),\ \Eprint
  {https://arxiv.org/abs/2402.11749} {arxiv:2402.11749 [cond-mat]} \BibitemShut
  {NoStop}%
\bibitem [{\citenamefont {{Batlle-Porro}}\ \emph {et~al.}(2024)\citenamefont
  {{Batlle-Porro}}, \citenamefont {Calugaru}, \citenamefont {Hu}, \citenamefont
  {Kumar}, \citenamefont {Hesp}, \citenamefont {Watanabe}, \citenamefont
  {Taniguchi}, \citenamefont {Bernevig}, \citenamefont {Stepanov},\ and\
  \citenamefont {Koppens}}]{BAT24}%
  \BibitemOpen
  \bibfield  {author} {\bibinfo {author} {\bibfnamefont {S.}~\bibnamefont
  {{Batlle-Porro}}}, \bibinfo {author} {\bibfnamefont {D.}~\bibnamefont
  {Calugaru}}, \bibinfo {author} {\bibfnamefont {H.}~\bibnamefont {Hu}},
  \bibinfo {author} {\bibfnamefont {R.~K.}\ \bibnamefont {Kumar}}, \bibinfo
  {author} {\bibfnamefont {N.~C.~H.}\ \bibnamefont {Hesp}}, \bibinfo {author}
  {\bibfnamefont {K.}~\bibnamefont {Watanabe}}, \bibinfo {author}
  {\bibfnamefont {T.}~\bibnamefont {Taniguchi}}, \bibinfo {author}
  {\bibfnamefont {B.~A.}\ \bibnamefont {Bernevig}}, \bibinfo {author}
  {\bibfnamefont {P.}~\bibnamefont {Stepanov}},\ and\ \bibinfo {author}
  {\bibfnamefont {F.~H.~L.}\ \bibnamefont {Koppens}},\ }\href
  {https://doi.org/10.48550/arXiv.2402.12296} {\bibinfo {title}
  {Cryo-{{Near-Field Photovoltage Microscopy}} of {{Heavy-Fermion Twisted
  Symmetric Trilayer Graphene}}}} (\bibinfo {year} {2024}),\ \Eprint
  {https://arxiv.org/abs/2402.12296} {arxiv:2402.12296 [cond-mat]} \BibitemShut
  {NoStop}%
\bibitem [{\citenamefont {C{\u a}lug{\u a}ru}\ and\ \citenamefont
  {{others}}(2024)}]{CAL23b}%
  \BibitemOpen
  \bibfield  {author} {\bibinfo {author} {\bibfnamefont {D.}~\bibnamefont {C{\u
  a}lug{\u a}ru}}\ and\ \bibinfo {author} {\bibnamefont {{others}}},\
  }\href@noop {} {\bibfield  {journal} {\bibinfo  {journal} {To be published}\
  } (\bibinfo {year} {2024})}\BibitemShut {NoStop}%
\bibitem [{\citenamefont {Mahan}(2000)}]{MAH00}%
  \BibitemOpen
  \bibfield  {author} {\bibinfo {author} {\bibfnamefont {G.~D.}\ \bibnamefont
  {Mahan}},\ }\href@noop {} {\emph {\bibinfo {title} {Many Particle
  Physics}}},\ \bibinfo {edition} {3rd}\ ed.,\ Physics of Solids and Liquids\
  (\bibinfo  {publisher} {{Springer Science + Business Media, LLC}},\ \bibinfo
  {address} {{New York}},\ \bibinfo {year} {2000})\BibitemShut {NoStop}%
\bibitem [{\citenamefont {Peskin}\ and\ \citenamefont
  {Schroeder}(1995)}]{PES95}%
  \BibitemOpen
  \bibfield  {author} {\bibinfo {author} {\bibfnamefont {M.~E.}\ \bibnamefont
  {Peskin}}\ and\ \bibinfo {author} {\bibfnamefont {D.~V.}\ \bibnamefont
  {Schroeder}},\ }\href@noop {} {\emph {\bibinfo {title} {An Introduction to
  Quantum Field Theory}}}\ (\bibinfo  {publisher} {{Addison-Wesley Pub. Co}},\
  \bibinfo {address} {{Reading, Mass}},\ \bibinfo {year} {1995})\BibitemShut
  {NoStop}%
\bibitem [{\citenamefont {Moreno}\ and\ \citenamefont {Coleman}(1996)}]{MOR96}%
  \BibitemOpen
  \bibfield  {author} {\bibinfo {author} {\bibfnamefont {J.}~\bibnamefont
  {Moreno}}\ and\ \bibinfo {author} {\bibfnamefont {P.}~\bibnamefont
  {Coleman}},\ }\href {https://doi.org/10.48550/arXiv.cond-mat/9603079}
  {\bibinfo {title} {Thermal currents in highly correlated systems}} (\bibinfo
  {year} {1996}),\ \Eprint {https://arxiv.org/abs/cond-mat/9603079}
  {arxiv:cond-mat/9603079} \BibitemShut {NoStop}%
\bibitem [{\citenamefont {Paul}\ and\ \citenamefont {Kotliar}(2003)}]{PAU03}%
  \BibitemOpen
  \bibfield  {author} {\bibinfo {author} {\bibfnamefont {I.}~\bibnamefont
  {Paul}}\ and\ \bibinfo {author} {\bibfnamefont {G.}~\bibnamefont {Kotliar}},\
  }\href {https://doi.org/10.1103/PhysRevB.67.115131} {\bibfield  {journal}
  {\bibinfo  {journal} {Phys. Rev. B}\ }\textbf {\bibinfo {volume} {67}},\
  \bibinfo {pages} {115131} (\bibinfo {year} {2003})}\BibitemShut {NoStop}%
\bibitem [{\citenamefont {Schweitzer}\ and\ \citenamefont
  {Czycholl}(1990)}]{SCH90}%
  \BibitemOpen
  \bibfield  {author} {\bibinfo {author} {\bibfnamefont {H.}~\bibnamefont
  {Schweitzer}}\ and\ \bibinfo {author} {\bibfnamefont {G.}~\bibnamefont
  {Czycholl}},\ }\href {https://doi.org/10.1007/BF01437647} {\bibfield
  {journal} {\bibinfo  {journal} {Z. Physik B - Condensed Matter}\ }\textbf
  {\bibinfo {volume} {79}},\ \bibinfo {pages} {377} (\bibinfo {year}
  {1990})}\BibitemShut {NoStop}%
\bibitem [{\citenamefont {Georges}\ \emph {et~al.}(1996)\citenamefont
  {Georges}, \citenamefont {Kotliar}, \citenamefont {Krauth},\ and\
  \citenamefont {Rozenberg}}]{GEO96}%
  \BibitemOpen
  \bibfield  {author} {\bibinfo {author} {\bibfnamefont {A.}~\bibnamefont
  {Georges}}, \bibinfo {author} {\bibfnamefont {G.}~\bibnamefont {Kotliar}},
  \bibinfo {author} {\bibfnamefont {W.}~\bibnamefont {Krauth}},\ and\ \bibinfo
  {author} {\bibfnamefont {M.~J.}\ \bibnamefont {Rozenberg}},\ }\href
  {https://doi.org/10.1103/RevModPhys.68.13} {\bibfield  {journal} {\bibinfo
  {journal} {Rev. Mod. Phys.}\ }\textbf {\bibinfo {volume} {68}},\ \bibinfo
  {pages} {13} (\bibinfo {year} {1996})}\BibitemShut {NoStop}%
\bibitem [{\citenamefont {Luttinger}(1964)}]{LUT64}%
  \BibitemOpen
  \bibfield  {author} {\bibinfo {author} {\bibfnamefont {J.~M.}\ \bibnamefont
  {Luttinger}},\ }\href {https://doi.org/10.1103/PhysRev.135.A1505} {\bibfield
  {journal} {\bibinfo  {journal} {Phys. Rev.}\ }\textbf {\bibinfo {volume}
  {135}},\ \bibinfo {pages} {A1505} (\bibinfo {year} {1964})}\BibitemShut
  {NoStop}%
\bibitem [{\citenamefont {{Herzog-Arbeitman}}\ \emph
  {et~al.}(2024)\citenamefont {{Herzog-Arbeitman}}, \citenamefont {Yu},
  \citenamefont {C{\u a}lug{\u a}ru}, \citenamefont {Hu}, \citenamefont
  {Vafek}, \citenamefont {Regnault}, \citenamefont {Kang},\ and\ \citenamefont
  {Bernevig}}]{HER24a}%
  \BibitemOpen
  \bibfield  {author} {\bibinfo {author} {\bibfnamefont {J.}~\bibnamefont
  {{Herzog-Arbeitman}}}, \bibinfo {author} {\bibfnamefont {J.}~\bibnamefont
  {Yu}}, \bibinfo {author} {\bibfnamefont {D.}~\bibnamefont {C{\u a}lug{\u
  a}ru}}, \bibinfo {author} {\bibfnamefont {H.}~\bibnamefont {Hu}}, \bibinfo
  {author} {\bibfnamefont {O.}~\bibnamefont {Vafek}}, \bibinfo {author}
  {\bibfnamefont {N.}~\bibnamefont {Regnault}}, \bibinfo {author}
  {\bibfnamefont {J.}~\bibnamefont {Kang}},\ and\ \bibinfo {author}
  {\bibfnamefont {B.~A.}\ \bibnamefont {Bernevig}},\ }\href@noop {} {\bibfield
  {journal} {\bibinfo  {journal} {To be published}\ } (\bibinfo {year}
  {2024})}\BibitemShut {NoStop}%
\bibitem [{\citenamefont {Po}\ \emph {et~al.}(2018{\natexlab{b}})\citenamefont
  {Po}, \citenamefont {Watanabe},\ and\ \citenamefont {Vishwanath}}]{PO18c}%
  \BibitemOpen
  \bibfield  {author} {\bibinfo {author} {\bibfnamefont {H.~C.}\ \bibnamefont
  {Po}}, \bibinfo {author} {\bibfnamefont {H.}~\bibnamefont {Watanabe}},\ and\
  \bibinfo {author} {\bibfnamefont {A.}~\bibnamefont {Vishwanath}},\ }\href
  {https://doi.org/10.1103/PhysRevLett.121.126402} {\bibfield  {journal}
  {\bibinfo  {journal} {Phys. Rev. Lett.}\ }\textbf {\bibinfo {volume} {121}},\
  \bibinfo {pages} {126402} (\bibinfo {year} {2018}{\natexlab{b}})}\BibitemShut
  {NoStop}%
\bibitem [{\citenamefont {Czycholl}(2008)}]{CZY08}%
  \BibitemOpen
  \bibfield  {author} {\bibinfo {author} {\bibfnamefont {G.}~\bibnamefont
  {Czycholl}},\ }\href {https://doi.org/10.1007/978-3-540-74790-1} {\emph
  {\bibinfo {title} {{Theoretische Festk{\"o}rperphysik}}}},\
  {Springer-Lehrbuch}\ (\bibinfo  {publisher} {{Springer}},\ \bibinfo {address}
  {{Berlin, Heidelberg}},\ \bibinfo {year} {2008})\BibitemShut {NoStop}%
\bibitem [{\citenamefont {{Wikipedia contributors}}(2023)}]{WIK23}%
  \BibitemOpen
  \bibfield  {author} {\bibinfo {author} {\bibnamefont {{Wikipedia
  contributors}}},\ }\href@noop {} {\emph {\bibinfo {title} {Matsubara
  Frequency}}}\ (\bibinfo {year} {2023})\BibitemShut {NoStop}%
\bibitem [{\citenamefont {Sommerfeld}(1928)}]{SOM28}%
  \BibitemOpen
  \bibfield  {author} {\bibinfo {author} {\bibfnamefont {A.}~\bibnamefont
  {Sommerfeld}},\ }\href {https://doi.org/10.1007/BF01391052} {\bibfield
  {journal} {\bibinfo  {journal} {Z. Physik}\ }\textbf {\bibinfo {volume}
  {47}},\ \bibinfo {pages} {1} (\bibinfo {year} {1928})}\BibitemShut {NoStop}%
\bibitem [{\citenamefont {{Chien-Lih}}(2005)}]{CHI05}%
  \BibitemOpen
  \bibfield  {author} {\bibinfo {author} {\bibfnamefont {H.}~\bibnamefont
  {{Chien-Lih}}},\ }\href@noop {} {\bibfield  {journal} {\bibinfo  {journal}
  {The Mathematical Gazette}\ }\textbf {\bibinfo {volume} {89}},\ \bibinfo
  {pages} {469} (\bibinfo {year} {2005})},\ \Eprint
  {https://arxiv.org/abs/3621947} {3621947} \BibitemShut {NoStop}%
\bibitem [{\citenamefont {Jonson}\ and\ \citenamefont {Mahan}(1980)}]{JON80}%
  \BibitemOpen
  \bibfield  {author} {\bibinfo {author} {\bibfnamefont {M.}~\bibnamefont
  {Jonson}}\ and\ \bibinfo {author} {\bibfnamefont {G.~D.}\ \bibnamefont
  {Mahan}},\ }\href {https://doi.org/10.1103/PhysRevB.21.4223} {\bibfield
  {journal} {\bibinfo  {journal} {Phys. Rev. B}\ }\textbf {\bibinfo {volume}
  {21}},\ \bibinfo {pages} {4223} (\bibinfo {year} {1980})}\BibitemShut
  {NoStop}%
\bibitem [{\citenamefont {Abramowitz}\ and\ \citenamefont
  {Stegun}(1965)}]{ABR65}%
  \BibitemOpen
  \bibfield  {author} {\bibinfo {author} {\bibfnamefont {M.}~\bibnamefont
  {Abramowitz}}\ and\ \bibinfo {author} {\bibfnamefont {I.~A.}\ \bibnamefont
  {Stegun}},\ }\href@noop {} {\emph {\bibinfo {title} {Handbook of
  {{Mathematical Functions}}: {{With Formulas}}, {{Graphs}}, and {{Mathematical
  Tables}}}}}\ (\bibinfo  {publisher} {{Courier Corporation}},\ \bibinfo {year}
  {1965})\BibitemShut {NoStop}%
\bibitem [{\citenamefont {Mehew}\ \emph {et~al.}(2024)\citenamefont {Mehew},
  \citenamefont {Merino}, \citenamefont {Ishizuka}, \citenamefont {Block},
  \citenamefont {M{\'e}rida}, \citenamefont {Carl{\'o}n}, \citenamefont
  {Watanabe}, \citenamefont {Taniguchi}, \citenamefont {Levitov}, \citenamefont
  {Efetov},\ and\ \citenamefont {Tielrooij}}]{MEH24}%
  \BibitemOpen
  \bibfield  {author} {\bibinfo {author} {\bibfnamefont {J.~D.}\ \bibnamefont
  {Mehew}}, \bibinfo {author} {\bibfnamefont {R.~L.}\ \bibnamefont {Merino}},
  \bibinfo {author} {\bibfnamefont {H.}~\bibnamefont {Ishizuka}}, \bibinfo
  {author} {\bibfnamefont {A.}~\bibnamefont {Block}}, \bibinfo {author}
  {\bibfnamefont {J.~D.}\ \bibnamefont {M{\'e}rida}}, \bibinfo {author}
  {\bibfnamefont {A.~D.}\ \bibnamefont {Carl{\'o}n}}, \bibinfo {author}
  {\bibfnamefont {K.}~\bibnamefont {Watanabe}}, \bibinfo {author}
  {\bibfnamefont {T.}~\bibnamefont {Taniguchi}}, \bibinfo {author}
  {\bibfnamefont {L.~S.}\ \bibnamefont {Levitov}}, \bibinfo {author}
  {\bibfnamefont {D.~K.}\ \bibnamefont {Efetov}},\ and\ \bibinfo {author}
  {\bibfnamefont {K.-J.}\ \bibnamefont {Tielrooij}},\ }\href
  {https://doi.org/10.1126/sciadv.adj1361} {\bibfield  {journal} {\bibinfo
  {journal} {Sci. Adv.}\ }\textbf {\bibinfo {volume} {10}},\ \bibinfo {pages}
  {eadj1361} (\bibinfo {year} {2024})}\BibitemShut {NoStop}%
\end{thebibliography}
\end{document}